\documentclass[11pt,titlepage,twoside,a4paper]{report}
\usepackage{a4}                        
\usepackage{graphics}                  
\usepackage{epsfig}
\usepackage{cite}                      
\usepackage[nottoc]{tocbibind}         
\usepackage[usenames,dvipsnames]
                           {color}     
 \definecolor{darkred}{rgb}{0.5,0,0}
 \definecolor{darkgreen}{rgb}{0,0.5,0}
 \definecolor{darkblue}{rgb}{0,0,0.5}
 \definecolor{w}{rgb}{1,1,1}
 \definecolor{b}{cmyk}{0,0,0,1}
 \definecolor{r}{rgb}{1,0,0}
 \definecolor{g}{rgb}{0,1,0}
 \definecolor{bl}{rgb}{0,0,1}
 \definecolor{c}{rgb}{0,1.,1.}
 \definecolor{m}{cmyk}{0.0,1,0,0}
 \definecolor{v}{rgb}{0.3,0,1}
 \definecolor{gr}{gray}{0.85}
 \definecolor{gr2}{gray}{0.5}
 \definecolor{skyblue}{rgb}{0.196078,0.6, 0.8}
 \definecolor{midnightblue}{rgb}{0.184314,0.184314,0.309804}
 \definecolor{orange}{rgb}{1.0, 0.5, 0.0}
  \ifx\pdfoutput\undefined   
   \usepackage[pdftitle={Dissertation},
               pdfauthor={Lars Goerlich},
               pdfsubject={N=1 and non-supersymmetric open string theories
                  in six and four space-time dimensions},
               pdfkeywords={string theory, orientifolds, asymmetric orbifolds,
               open strings, D-branes, non-supersymmetric string theories, 
               supersymmetric models, string phenomenology}, 
               hypertexnames=false, 
               hyperindex=true,
               citebordercolor= 0 0 0,
               filebordercolor= 0 0 0,
               linkbordercolor= 0 0 0,
               menubordercolor= 0 0 0,
               pagebordercolor= 0 0 0,
               urlbordercolor= 0 0 0,
               colorlinks=b,
               anchorcolor=b,
               linkcolor=b,
               filecolor=b,
               urlcolor=b,
               citecolor=b,
               menucolor=b,
               pagecolor=b,
               breaklinks=true,
               linktocpage=true,
               ]{hyperref}
  \else
    \usepackage[pdftex=true,               
                pdftitle={Dissertation},
                pdfauthor={Lars Görlich},
                pdfsubject={N=1 and non-supersymmetric open string theories
                   in six and four space-time dimensions},
                pdfkeywords={String theory, Orientifolds, Open strings, 
                D-branes, Supersymmetric models, Asymmetric Orbifolds, 
                open strings, D-branes,
                Non-supersymmetric string theories, String phenomenology}, 
                plainpages=false,
                hypertexnames=false, 
                pdfpagelabels=true,
                pdfborder= 0 0 0,  
                pdfpagemode=UseOutlines,
                hyperindex=true,
                citebordercolor= 0 0 0,
                filebordercolor= 0 0 0,
                linkbordercolor= 0 0 0,
                menubordercolor= 0 0 0,
                pagebordercolor= 0 0 0,
                urlbordercolor= 0 0 0,
                colorlinks=true,
                anchorcolor=blue,
                linkcolor=darkblue,
                filecolor=darkgreen,
                urlcolor=darkred,
                citecolor=darkblue,
                menucolor=blue,
                pagecolor=blue,
                breaklinks=true,
               linktocpage=true,
                ]{hyperref}
   \fi

\usepackage{amsmath}      
\usepackage{amssymb}      
\usepackage{dsfont}       

\newcommand{\vt}{\vartheta}
\newcommand{\oh}{\frac{1}{2}}
\newcommand{\beqn}{\begin{eqnarray}} 
\newcommand{\eeqn}{\end{eqnarray}}
\newcommand{\be}{\begin{equation}}
\newcommand{\ee}{\end{equation}}
\newcommand{\non}{\nonumber}
\newcommand{\al}{{\alpha^{\scriptscriptstyle \prime}}}
\newcommand{\bra}[1]{\left<\left. #1\right.\right|}
\newcommand{\ket}[1]{\left|\left. #1\right.\right>}
\newcommand{\braket}[3]{\big< #1\big| #2 \big| #3 \big>}
\newcommand{\vev}[1]{\big< #1  \big>}
\newcommand{\vevs}[1]{\langle #1  \rangle}
\newcommand{\thef}[2]{\vartheta\left[{#1\atop #2}\right]}
\newlength{\neglength}
\newcommand{\negphantom}[1]{\text{\settowidth{\neglength}{$#1$}$
                             \hspace{-\neglength}$}}

\newtheorem{defin}{Definition}
\newtheorem{theorem}{Theorem}

\DeclareMathOperator*{\image}{im}
\DeclareMathOperator*{\tr}{tr}
\DeclareMathOperator*{\Tr}{Tr}
\DeclareMathOperator{\rk}{rank}
\DeclareMathOperator{\pf}{Pf}
\DeclareMathOperator{\id}{Id}
\DeclareMathOperator{\diag}{diag}
\DeclareMathOperator{\ch}{ch}

 \newcommand{\dckeyena}{string theory}
 \newcommand{\dckeyenb}{orientifolds}
 \newcommand{\dckeyenc}{asymmetric orbifolds}
 \newcommand{\dckeyend}{open strings}
 \newcommand{\dckeyene}{D-branes}
 \newcommand{\dckeyenf}{background fluxes}
 \newcommand{\dckeyeng}{non-supersymmetric string theories}
 \newcommand{\dckeyenh}{supersymmetric models}
 \newcommand{\dckeyeni}{string phenomenology}
 \newcommand{\dckeywordsen}{\vspace*{2cm} \\{\bf{Keywords:}}\\ \dckeyena, 
   \dckeyenb, \dckeyenc, \dckeyend, \dckeyene, \dckeyenf, \dckeyeng,
            \dckeyenh, \dckeyeni \\}

\begingroup\makeatletter\ifx\SetFigFont\undefined%
 \gdef\SetFigFont#1#2#3#4#5{%
   \reset@font\fontsize{#1}{#2pt}%
   \fontfamily{#3}\fontseries{#4}\fontshape{#5}%
   \selectfont}%
 \fi\endgroup%

\begin{titlepage}
       \title{
        \begin{flushright} 
          \vspace{-2cm} 
          \small  hep-th/0401040                   \\
          \small  HUB-EP-04/02                     \\
          \small  TIFR/TH/04-01 
        \end{flushright}
         \vspace{3cm}
           $\textstyle{\cal N}=1$
          and non-supersymmetric open string theories
                in six and four space-time dimensions\footnote{PhD thesis
           written  under the supervision of Prof.~D.~L \"ust,  
           handed in at ``Humboldt Univerisit\"at zu Berlin'' in August 2003}}
\date{}
\author{\rule{0cm}{2cm}\Large Lars G\"orlich\footnote{Email: \sf goerlich@tifr.res.in} \\ \sl
        \rule{0cm}{1cm} Tata Institute of Fundamental Research, \\
      \sl       Homi Bhabha Rd, Mumbai 400005, India}
\end{titlepage}

\begin{document}
 
\pagenumbering{roman}
 \maketitle
 \thispagestyle{empty}
 \clearpage
\pdfbookmark[1]{Abstract}{b}
\abstract
This thesis contains an introductory chapter on orbifolds.
Besides rudimentary basics we discuss more advanced topics like {\it discrete 
torsion} and {\it asymmetric} orbifold groups. As examples we investigate 
torus compactifications and an asymmetric 
$T^4/\mathbb{Z}_3^\text{L}\times\mathbb{Z}_3^\text{R}$  orbifold.

The following chapter explains  the foundations of orientifolds, including
open strings with  {\it Chan-Paton} degrees of freedom.

Chapters \ref{strbg}-\ref{z4} present own research.

In chapter \ref{strbg} we quantize open strings with linear boundary
conditions, as they show up in electro-magnetic fields. We quantize 
the zero- and momentum-modes for toroidal compactifications, too. 
As an application we calculate the commutator of the coordinate fields in the
case of general constant Neveu-Schwarz $U(1)$-field strengths. Thereby  
we confirm previous results on {\it non-commutativity} of open string 
theories in Neveu-Schwarz backgrounds.

Chapter \ref{ncg} reviews the results of a former publication 
\cite{Blumenhagen:2000fp} on asymmetric orientifolds, supplemented by 
some recent insights in connection with the preceeding chapter.

Chapter \ref{magbf} is a summary of \cite{Blumenhagen:2000wh}. In this
publication we investigated to what extend  one can build phenomenologically
interesting models from toroidal orientifolds. By turning on magnetic fluxes
on D$9$-branes we induce chiral fermions. Most calculations are performed 
in an (equivalent) T-dual picture. Here the number of chiral fermions is 
given by the topological intersection number of $D$-branes.

In orientifolds of toroidal compactifications one obtains either non-chiral
or non-supersymmetric orientifold solutions. However both properties can
be reconciled in orientifolds that are obtained from specific supersymmetric
orbifold compactifications. In chapter \ref{z4} we present
 the $\Bar{\sigma}\Omega$-Orientifold  on a  $T^6/\mathbb{Z}_4$ orbifold.
As a very attractive example we investigate a supersymmetric
$U(4)\times U(2)^3_\text{L}\times U(2)^3_\text{R}$ model that is broken
to an {\it MSSM}\;\footnote{ ``MSSM''$=$ minimal supersymmetric Standard Model}
-like model by switching on suitable background fields in the {\sl low energy
effective action}.
This chapter is based on our publication \cite{Blumenhagen:2002gw}.

The thesis is supplemented by an appendix with formul\ae\; applied in the
text, as well as proofs to two theorems  that were used as well.
\\
\dckeywordsen

\setcounter{page}{2} 
\clearpage
\pdfbookmark[1]{Contents}{d}
\tableofcontents
\vfill
\pagebreak
\listoffigures
\vfill
\pagebreak
\listoftables
\vfill
\pagebreak

\pagestyle{myheadings}
\pagenumbering{arabic}

\chapter{Introduction}
In this chapter we will give a short motivation for 
string theory and its supersymmetric extension.
By doing so we will expose some of the basic ideas underlying this theory.
As the main part of the thesis and the whole part of our own research 
presented here deals with open strings, we devote a section to this topic as well.
In chapter \ref{magbf} and \ref{z4} we will investigate the chance to find
realistic models in a special class of unoriented open-string
theories. Therefore we will make some comments on these orientifolds as
well. Since we have included two more extended chapters on orbifolds and
orientifolds in this thesis, this introductory chapter is rather 
condensed, only concentrating on the rudiments without going into details.
General texts on string theory are the classical work of Green, Schwarz and
Witten \cite{Green:1987sp,Green:1987mn}, the book of L\"ust and  
Theisen \cite{Lust:1989tj} and the two volumes by Polchinski 
\cite{Polchinski:1998v1,Polchinski:1998v2}.
The latter reference is especially interesting as it 
includes a chapter on D-branes and gives a D-brane interpretation of 
orientifolds. The book of Bailin and Love provides a 
good introduction to both supersymmetric field theory
and superstring theory (cf.\ \cite{Bailin:1994qt}).

\section{String theory}
String theory is a quantum theory of string-like (i.e. one-dimensional)
objects. Even though it has many similarities to quantum mechanics of
point-like particles and known quantum field-theories, there are many
 striking differences as well. We will recall some principles used to quantize
 classical systems and try to apply those to the string as well.
It turns out that string theory is in some respect much more restrictive, but
offers a lot of promising features at the same time.
 The most interesting one is surely 
that string theory is automatically a (presumably) consistent  theory
of quantum gravity, already at the perturbative level. Other interesting
features are gauge-symmetries in the low-energy effective action 
of a huge class of string theories.
A third feature is that chiral fermions appear in many string theories,
thereby making string theory a good candidate for a unified theory of
nature.

\section{Quantization of classical strings}
In many approaches to quantum theory one starts with a classical system in
canonical formalism. A main ingredient in this formulation is the symplectic
phase space of the system, which is the cotangent bundle
$T^\ast({\cal M})$ of a manifold ${\cal M}$. For example in classical
mechanics 
 ${\cal M}$ is the $3n$-dimensional manifold describing the positions of $n$
point-like particles. In a system with infinitely many degrees of freedom
 (dofs)
one often does not bother about the precise structure of ${\cal M}$, which is
in this case infinite, too. On the (finite-dimensional) 
manifold $T^\ast({\cal M})$ an algebra of  $C^\infty$-functions exists, which
we denote by ${\cal T}\big(T^\ast({\cal M})\big)$.
What is now  important, is that the symplectic structure of the phase-space
induces a
bilinear map from the space of $C^\infty$-function to itself. This map is
commonly known as the {\it Poisson-bracket}:
\begin{equation}
 \begin{aligned}
   {\cal T}\big(T^\ast({\cal M})\big)
    \times  {\cal T}\big(T^\ast({\cal M}) \big)
       &\to {\cal T}\big(T^\ast({\cal M})\big)            \\
  (f,g) & \mapsto \{f,g\}_\text{PB} 
 \end{aligned}
\end{equation}
By quantizing the system the classical algebra of observables
${\cal T}\big(T^\ast({\cal M})\big)$ gets exchanged by some
operator algebra.\footnote{These statements should be taken with a grain of
  salt. We are not very precise about the operator algebra involved, and
  especially not about the map form ${\cal T}\big(T^\ast({\cal M})\big)$ to 
  this algebra. Furthermore there might appear additional subtleties.}
Since both the Poisson-bracket and the {\it commutator} of an operator algebra 
share three important properties (bilinearity, anti-symmetry and Jacobian 
identity) it is natural to map the Poisson-bracket of two
 functions $f,g$  to the commutator of
the corresponding operators $\hat{f},\hat{g}$ in the operator
algebra.\footnote{Usually one maps the Poisson bracket to $\hbar/i$
 times the  commutator, but the normalization is somehow redundant.}  

In contrast to point-particles, strings are one-dimensional objects, but they
admit a classical description in terms of a {\it Lagrangian} (density),
a derived symplectic form and a Hamiltonian as well.
Functions on its  phase space, especially the coordinate functions
 of the string and  the canonical momentum might be substituted by operators
 as well, thereby preparing the grounds for a quantum theory of strings.
Quantum mechanics has still a richer structure.  The
probability interpretation of quantum mechanics requires that the operator
algebra has to act on some vector space which admits a  {\sl hermitian}
 scalar product. In the best case the vector space is closed (w.r.t.\
 the scalar product) i.e.\ a {\sl Hilbert space}.
Observable quantities are the spectra (Eigenvalues)
of operators corresponding to classical quantities, and as these are real, one
requires these operators to be self-adjoint. 

While in the simplest case of point particles such a Hilbert space and
operator algebra are relatively easy to find,\footnote{Even though one
  encounters already ambiguities in the map from
  the function- to the operator-algebra (e.g.\ ordering of operators).}
it turns out that more complicated systems (i.e. infinitely many degrees of
freedom, like quantum field-theories) a direct map from the classical- to the
quantum-system is often problematic. This may have many reasons and up to now
there exists no prescription, how to quantize an arbitrary classical system.
For example in classical field theory there is no obstruction in multiplying
two fields $\phi(x)$, $\psi(y)$, even if both fields have the same argument
$x=y$.
 In quantum field theory the product of the corresponding  fields 
(which are in a naive approach  operator valued functions) 
with coinciding  arguments $x=y$  
 will in general be singular, i.e.\ not properly defined.
In many quantum field theories, these infinities can be ``regularized''
and a program called renormalization expresses the parameters that are
introduced in the regularization procedure by physical, i.e.\ measurable
quantities.

While the problem discussed above usually becomes important if one considers
some kind of interaction, it might already be a challenge to find the correct
Hilbert space even if one neglects interactions. This is the case for the
electromagnetic Maxwell field,
 but also for the string.
It is well known that a naive quantization of the electromagnetic field $A^\mu$
induces states of negative norm due to the minkowskian scalar product in 
space-time. Some classical equalities can lead to contradictions if directly
translated into operator equations.
This problem can be solved by the so called {\sl Gupta-Bleuler}
quantization. The idea is to split the state-space into a physical one, and
a redundant space. The physical Hilbert-space is obtained by requiring that 
the classical conditions are fulfilled by the positive frequency (or
annihilation-) part of the corresponding quantum fields:\footnote{As the
creation-part $F^{(-)}$ is the hermitian conjugate of  $F^{(+)}$  matrix
elements between physical states involving {\sl normal-ordered} combinations
of $F^{(\pm)}$ will vanish.}
\begin{equation}
 F_\text{class}=0 \quad\Rightarrow F^{(+)}_\text{qm} |\psi,\text{phys}\rangle=0
\end{equation}
For Gupta-Bleuler quantization of the electromagnetic field,
 $F$ equals the (four-dimensional)
divergence of the vector potential: $F=\partial_\mu A^\mu=0$.
These conditions are
linear, therefore the physical space is a linear subspace of the vector space from
which one starts.
This physical subspace has a positive {\sl semi}-definite norm. In
  electrodynamics there is still a redundancy in this subspace. 
  Physical states belong to equivalence classes of the subspace and 
  measurable quantities are not affected by the representative chosen.
  The redundancy corresponds to the (unphysical) longitudinal and time-like 
  polarization part whose {\it non-zero} excitations are of
  zero-norm. Requiring that the longitudinal part admits a zero-excitation
  contribution with non-vanishing norm
  ensures that the state is normalizable, thereby turning the 
  space of equivalence classes into a (pre-) Hilbert space, i.e.\ a vector
  space of positive definite scalar product. 
  It is very assuring, that the (non-physical) longitudinal excitations
  decouple from  the $S$-matrix.  
The  gauge invariance of  electromagnetism might be regarded as the 
origin of the split.
 Analogous features are encountered if one quantizes more general quantum field
 theories (QFTs)
like non-abelian gauge theories and even more interesting for us:
 Something similar occurs for the string as well.

The classical action for the string is proportional to the area of the
string {\it world sheet}. The string world sheet is the two-dimensional analog
of the world line for point-particles. 
The name  {\it Nambu-Goto action} is devoted to its inventors:
\begin{equation} \label{nambu}
 S_{\text{NG}}=-\frac{1}{2\pi\al}\int d^2\sigma \sqrt{
    -\det_{\alpha\beta} \Big(  \big(\partial_{\sigma^\alpha} X^\mu\big)
              \big(\partial_{\sigma^\beta} X_\mu\big) \Big)}
\end{equation}
$\al$ is the so called {\it Regge-slope}.\footnote{In general the Regge-slope
 is defined as the maximal angular momentum per energy\raisebox{1ex}{\tiny 2}.}
The solutions of the equations of motion (eoms) for the embedding fields 
$X^\mu$  justify this identification. 
The eoms of the Nambu-Goto action are highly nonlinear, even if the
background on which the string propagates is flat and consequently
difficult to solve. Therefore the following more tractable 
action ({\it Polyakov-action}) was proposed:\footnote{This action was found by
  Brink, Di Vecchia, Howe, Deser and Zumino. Polyakov used it to
  perform  path-integral quantization.}
\begin{equation} \label{polyakov}
 S_{\text{P}}=-\frac{1}{4\pi\al}\int  d^2\sigma  \sqrt{-\det h}\, 
             h^{\alpha\beta}\,   
      \big(\partial_{\sigma^\alpha} X^\mu\big) 
      \big(\partial_{\sigma^\beta} X_\mu\big)
\end{equation}
$h^{\alpha\beta}$ is a metric defined on the world sheet. Solving the eoms
for $h^{\alpha\beta}$ and reinserting the (formal) solution into the Polyakov
action \eqref{polyakov} results in the  Nambu-Goto action \eqref{nambu}.
At this classical level the two actions are therefore equivalent. It is however
an open issue to show that they coincide if one takes the
quantum fluctuations in $h^{\alpha\beta}$  into account as well.
Taking the Polyakov-action as our starting point, there are several
possibilities to quantize the theory, all (of the explicitly known) leading
to the same result. We restrict to the case of flat space-time metric, 
which implies 
the maximal  (i.e. $D$-dimensional) Poincar{\'e}-invariance of the string 
Lagrangian-density and action. Like the Nambu-Goto action, $S_{\text{P}}$ is
invariant under diffeomorphisms of the two-dimensional world-sheet.
In addition, the Polyakov-action is invariant under a Weyl rescaling of the 
world-sheet metric:\footnote{The Weyl-invariance would not be present for
  higher or lower dimensional objects like membranes or point-particles.}
\begin{equation} \label{weyltransform}
 \begin{gathered}
  X'(\tau,\sigma)=X(\tau,\sigma) \\
  h_{\alpha\beta}= e^{2\omega(\tau,\sigma)} h_{\alpha\beta},
                 \quad\text{with }\omega(\tau,\sigma)\text{ arbitrary}
 \end{gathered}
\end{equation}
Reparameterization invariance
is sufficient to transform the metric $h$ (at least locally) to diagonal
form proportional to $\diag(-1,1)$.
Using in addition Weyl invariance allows one to obtain the following
gauge:
\begin{equation} \label{confgauge}
 h_{\alpha\beta}=\eta_{\alpha\beta}\qquad \eta_{\alpha\beta}\equiv \diag(-1,1) 
\end{equation}
The gauge \eqref{confgauge} is called {\it conformal gauge}
because it is preserved by a combination of general conformal transformations
(which leave $ h_{\alpha\beta}$ invariant up to a scale factor) and a
subsequent Weyl
transformation, that rescales the metric to its original form
\eqref{confgauge}. 
The quantization procedure might be performed as follows:
First one solves the eoms for the $X^\mu$ coordinate fields which become
wave equations for flat space-time metric:
\begin{equation} 
  \partial^2_{\sigma^0} X^\mu= \partial^2_{\sigma^1} X^\mu\quad \mu=0\ldots D
\end{equation}
Furthermore the $X$-fields are subjected to boundary conditions. 
The most common  boundary conditions are periodic ones in the world sheet
coordinate $\sigma\equiv\sigma^1$ ($\tau\equiv\sigma^0$):
\begin{equation} 
 X^\mu(\tau,\sigma+2\pi)= X^\mu(\tau,\sigma)
\end{equation}
and open string boundary conditions which are of Neumann-type:
\begin{equation}
  \partial_{\sigma}X^\mu(\tau,\sigma)\big|_{\sigma=0,\pi}=0
\end{equation}
Classical closed and open strings fulfilling these eoms and 
boundary conditions are depicted in figure \ref{closedstrpic} and  
\ref{opstringpic}.\footnote{However the depicted world-sheet does
           not fulfill the classical constraint equations. These would imply
           that  no oscillator modes are excited for a light-like
           center of mass
           momentum $p$. The normal-ordering of the quantum theory
           enforces however that oscillators are excited for $p^2=0$.}
 \begin{figure} 
  \setlength{\unitlength}{0.1in}
  \addtocounter{footnote}{-2}
  \begin{minipage}[t]{\textwidth}
   \begin{minipage}[t]{6cm}
   \begin{center}
   \begin{picture}(25,29)
   \SetFigFont{14}{20.4}{\rmdefault}{\mddefault}{\updefault}
   \put(0,0.2){\scalebox{0.22}{\includegraphics{picint1.EPS2}}}
    \put(5,27){$X_0$} 
   \end {picture}
   \caption[Closed-string evolving in time]
           {\label{closedstrpic}Closed-string (\textcolor{bl}{blue}) evolving in 
            time. The world-sheet, which is a classical solution is indicated 
            in transparent orange.\footnotemark}
  \end{center}
  \end{minipage} 
   \hfill
   \addtocounter{footnote}{-2}
   \begin{minipage}[t]{6cm}
   \begin{center}
   \begin{picture}(25,29)
   \SetFigFont{14}{20.4}{\rmdefault}{\mddefault}{\updefault}
   \put(0,0){\scalebox{0.22}{\includegraphics{picint2.EPS2}}}
    \put(3.85,27){$X_0$}
   \end{picture}
    \caption[Open-string evolving in time]
             {\label{opstringpic}Open-string (\textcolor{bl}{blue}) evolving in time.
              Both world-sheet boundaries (\textcolor{g}{green}) belong to the 
              same stack of D-branes.
              This classical solution can be associated with a gauge-boson of
              the quantized theory.\footnotemark}
   \end{center}
   \end{minipage}
  \end{minipage}
 \end{figure}
Then one quantizes the classical degrees of freedom. The world-sheet
Hamiltonian suggests a splitting into creation and annihilation operators.
Like in the case of electrodynamics one encounters however necessarily states
of zero-, and even worse: negative-norm. Much alike in the case of
electrodynamics one now tries to impose further classical constraints. 
The additional classical conditions stem from the eoms of the world-sheet
metric $h_{\alpha\beta}$ i.e.\ the vanishing of the world sheet energy-momentum
tensor $T$. Its trace vanishes already by the Weyl-invariance of the
classical action.\footnote{However the Weyl-symmetry might get spoiled by
  quantum effects. Actually in order that Weyl-anomalies are absent in the
  path integral formalism, one is
  restricted  to $D=26$ space-time dimensions  
  under certain assumptions on the background. This is especially true for 
 constant background fields. 
 }
$T$ can be expressed in terms of $X$ and consequently in terms
of operators. The Fourier components are the famous Virasoro generators $L_n$.
In calculating the Poisson brackets of the classical  Virasoro
generators\footnote{The Virasoro generators together with its commutators are
  called the (central extension of the) {\it Virasoro algebra}.} and 
comparing it with  the quantum mechanical commutator obtained by 
expressing the $L_n$ in terms of operators encountered in quantizing $X$, one
discovers a $c$-number anomaly, the so called  {\it Virasoro anomaly}.
Unlike to electrodynamics, the condition that the positive frequency part 
of the energy-momentum tensor $T^{(+)}$ 
has to vanish on the physical Hilbert-space ${\cal H}_\text{phys}$, 
does not remove negative-norm states from ${\cal H}_\text{phys}$. It does so
if a previously obtained normal ordering constant $a$ equals one
 and if the space-time
dimensions  $D$ equals $26$. Higher space-time dimensions are not possible,
while there are examples for   $a<1$ and $D<26$ that emerge 
from projecting  the $26$-dimensional theory to lower dimensions.
Looking at the massless spectrum, one can however not single out the
case $D$ = $26$ and $a=1$. This might be done by looking at vertex operators,
or probably more conveniently: to choose the method of {\it light-cone 
quantization}.

In light-cone quantization one transforms the time-coordinate $X_0$ and one
 arbitrary space-coordinate, say $X_1$ to new coordinates 
$X_\pm=1/\sqrt{2}(X_0\pm X_1)$.
The constraints $T=0$ take the following simple form in light-cone coordinates:
\begin{equation}
 \begin{aligned}  
  \partial_\tau X_+  \partial_\tau X_- +  
   (\tau\leftrightarrow \sigma)
 &=\frac{1}{2} \sum_{i=2}^D (\partial_\tau X_i)^2 + \tau\leftrightarrow \sigma
  \\
   \partial_\tau X_+ \partial_\sigma X_- + (\tau\leftrightarrow \sigma) 
 &=
   \frac{1}{2} \sum_{i=2}^D (\partial_\tau X_i)(\partial_\sigma X_i)
  \end{aligned} 
\end{equation}
The interesting observation is that $X_-$ appears only linear in the above 
constraint equations. If we would be able to bring $X_+$ to a particular simple
form, i.e.\ one which is linear in $\tau$ (or $\sigma$),  we could solve these
equations directly. Due to the formerly mentioned residual conformal symmetry
(which leaves the form of the gauge fixed action and metric $h$ invariant)
 and due to the
fact that the $X$-fields have the same periodicity
as the conformal transformations, which are harmonic functions 
on the world-sheet as well,
this is indeed possible. The resulting spectrum can be shown to be ghost-free.
However Lorentz-symmetry is no longer manifest. It turns out that in general
the  Lorentz-symmetry is plagued with anomalies, except for the case of 
space-time dimension $D=26$.

Another method to quantize strings is the {\sl path-integral\,}
approach. It is relatively complicated, although leading 
to most insights in mathematical respects.
In path-integral formalism  the absence of  quantum anomaly
in Weyl-transformations \eqref{weyltransform} restricts the space time
dimension to  $D=26$ which also removes a possible 
BRST-anomaly.\footnote{Anomalies in symmetries that are used to split
   the Hilbert space into a physical and an unphysical part (and this is
   exactly what the BRST-symmetry is used for) would indicate  
  that this split is ruined by quantum corrections.}

\section{String theory as a theory of quantum gravity}
Up to now we explained how string theory is quantized in principle, and how the
corresponding Hilbert space can  be obtained. We saw that 
this  Hilbert space only exists for bosonic strings moving in  $D=26$
space-time dimensions, which already puts surprisingly many constraints on the
geometry. (For supersymmetric strings the number of flat dimensions turns out
to be $10$.)  Up to now we restricted to a flat target space. However the form
of the string action suggests some generalization. 
If one computes the spectrum, one sees that it is quantized due to the
constraint on the energy-momentum tensor, or its Fourier-components, the
$L_n$'s. The linear mode $p_\text{com}$ of the fields 
\begin{equation} 
 X(\tau,\sigma)
  \sim x_\text{com}+\tau\cdot p_\text{com}+ \text{oscillator modes}   
\end{equation}
is interpreted as the center of mass (com.) momentum. Its (minkowskian) 
square determines the mass of the state. It turns out that for the bosonic
string on flat space-time there exist tachyons for both open and closed
strings. The mass\raisebox{1ex}{\tiny 2}$=0$ level consist of an excitation
that has exactly the degrees of freedom of a $U(1)$-gauge field for the open
string. The closed string  mass\raisebox{1ex}{\tiny 2}$=0$ can be identified
with a scalar (the dilaton), an antisymmetric tensor, and a traceless
symmetric tensor, the latter interpreted as the graviton. 
This makes string theory particularly interesting. String theory gives further
evidence that this identification is justified. According to the massless
 particle content, it is suggestive to include further terms in the Polyakov
 action, which are compatible with two-dimensional diffeomorphism and 
Weyl invariance at the classical level:\footnote{We neglect for the moment a
  possible boundary action that would include a vector-potential $A^\mu$
  corresponding to the open string massless mode.}
\begin{equation}  \label{nonlinsigmamod}
S_\sigma = -\frac{1}{4\pi\al}\int  d^2\sigma  \sqrt{-|h|} 
  \Big(\big(h^{\alpha\beta}G(X)+\epsilon^{\alpha\beta}B(X)\big)_{\mu\nu}\,
       \frac{X^\mu}{ \partial {\sigma^\alpha} }
       \frac{\partial X^\nu}{\partial {\sigma^\beta}} 
          +\al R(h) \Phi(X)
  \Big)
\end{equation}
$G$ is the space-time dependent $D$-dimensional metric, $B$ the antisymmetric
tensor, $\Phi$ the dilaton field, while $R$ is the two dimensional
Ricci-scalar.
The background fields in the above action might be interpreted as coherent
states of strings, which might be represented by insertions of vertex operators
into the path-integral.\footnote{States can be created by so called
 {\it vertex  operators}. This is similar to the case of QFT, where {\it in}
 and {\it out} states are created by corresponding fields. Vertex operators
 play an essential role in calculating string interactions.} 
The action \eqref{nonlinsigmamod} describes a coupled two-dimensional field
theory with the couplings $G$, $B$ and $\Phi$ depending on the fields $X^\mu$
in a possibly non-linear way. (Such an action is therefore called a non-linear
$\sigma$-model.) These coupling functionals will admit $\beta$ functions like
any coupling in a QFT. Weyl invariance at the quantum level requires, 
that these
$\beta$-functions vanish. It is possible to obtain the $\beta$ functions (of
the two dimensional world-sheet theory)
corresponding to the three fields  $G$, $B$ and $\Phi$ as eoms
of the following $D$-dimensional action:\footnote{The $\beta$-function leading
  to this action were obtained by expanding the background field up to first
  order in coordinate fields $X$. Higher order corrections are included in
  $O(\al)$.}
\begin{equation} \label{stringfraction}
 S_\text{S}=\frac{1}{2\kappa^2_0}\int
 d^D x\sqrt{-G}e^{-2\Phi}\Big(-\tfrac{2(D-26)}{3\al}+R(G)-\frac{1}{12}
 H\wedge\ast H+ 4d\Phi\wedge\ast d\Phi +O(\al) \Big) 
\end{equation} 
$H$ is the field strength of the antisymmetric tensor: $H=dB$.
Upon a Weyl rescaling of the metric $\widetilde{G}(x)=\exp(2\omega(x))G(x)$,
$\omega(x)=2(\Phi_0-\Phi(x))/(D-2)$ 
together with the induced transformation of the Ricci scalar $R(G)$ and
a further field redefinition of the dilaton
$\widetilde{\Phi}=\Phi(x)-\Phi_0$ the
action \eqref{stringfraction} becomes ($\kappa=\kappa_0\exp(\Phi_0)$):
\begin{multline} \label{einsteinfr}
 S_\text{E}=\frac{1}{2\kappa^2}\int
 d^D x\sqrt{-\widetilde{G}}\Big(-\tfrac{2(D-26)}{3\al}
  e^{\tfrac{4\widetilde{\Phi}}{D-2}}+R(\widetilde{G}) 
  \\-
  \tfrac{1}{12}
 e^{-\tfrac{8\widetilde{\Phi}}{D-2}}H\wedge\tilde{\ast} H
  -\tfrac{4}{D-2}d\Phi\wedge\tilde{\ast} d\Phi +O(\al) \Big) 
\end{multline} 
Because the action \eqref{einsteinfr} is the Einstein-Hilbert action of
gravity supplemented with some additional fields, the metric 
$\widetilde{G}$ is denoted as
the {\it Einstein metric}, while $G$ is called {\it string metric}.
The action  \eqref{einsteinfr} governing the background fields is the most
impressive justification for identifying the symmetric traceless mode of
the perturbative closed string with the quantum excitation of the graviton
field.

\begin{figure}
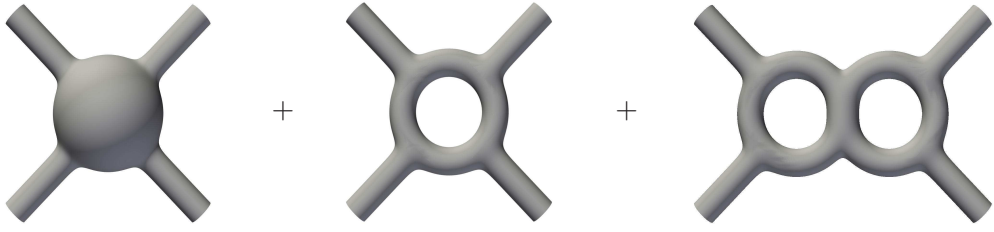

  \setlength{\unitlength}{0.1in}
  \begin{picture}(60,12)
   \put(0,0){\scalebox{0.09}{\includegraphics{picint31.EPS2}}}
   \put(14,5.5){$+$}
   \put(16,0){\scalebox{0.082}{\includegraphics{picint32.EPS2}}}
   \put(32.,5.5){$+$}
   \put(36,0){\scalebox{0.09}{\includegraphics{picint33.EPS2}}}
  \end{picture}
 \caption[String perturbation series]
              {\label{stringpertseries}First three terms of the string
            perturbation series with four external closed string states 
            involved}
\end{figure}

\begin{figure}[h]
\begin{center}
\includegraphics{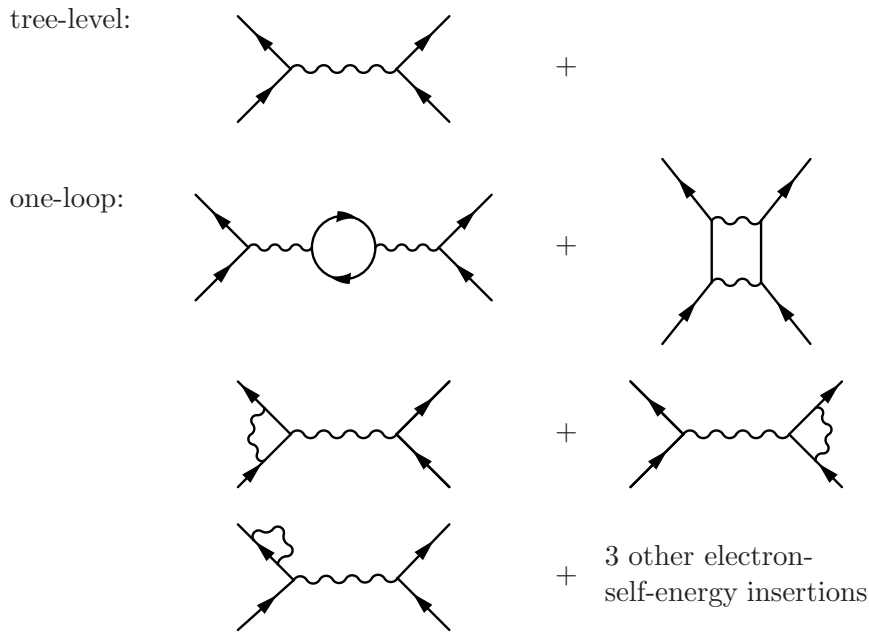}
\caption[QED perturbation series]{\label{qftoneloop}Perturbative expansion of
                                  electron-electron scattering in QED with one
                                  fermion generation.}
\end{center}
\end{figure}

String theory perturbation series are  defined as integrals over the moduli
space of Riemann surfaces with insertions of vertex operators (whose positions
are also moduli).\footnote{To be more precise, one
  only integrates over a region in moduli space, which is not connected to
  another one by an holomorphic transormation.}
The vertex operators correspond to
external (i.e.\ incoming and outcoming) particles. For closed strings there
exists only one diagram  at a given genus. It includes implicitly all possible 
string excitations in the internal part of the diagram. A four-particle 
closed string scattering process is depicted up to third order 
in figure \ref{stringpertseries}. Besides the sphere it includes a torus with
one and another torus with two handles.
From the aspect of simplicity (i.e.\ one diagram at each level of
closed-string perturbation series, and still comparatively few, if one
includes open strings and unoriented diagrams) string theory is very economic.
If one considers for example
 all diagrams contributing to the one-loop level of
electron-electron scattering ($e^-e^-\to e^-e^-$) one gets a variety of
diagrams which are shown in figure \ref{qftoneloop}.
We even suppressed the different combinations of legs, which would lead to a
multiple of the depicted diagrams. Similar combinatorics occur however in 
string theory as well, if several open strings participate as external states.
We also have to admit that the actual calculation of the few string-diagrams 
is a highly non-trivial task, at least at higher loop orders, or for many 
external strings participating. 

As string theory  has the graviton in its spectrum, we can (at least formally) calculate
scattering amplitudes that include this excitation as an external state.
In loop diagrams the graviton is implicitly included as an internal state as 
well. If one tries to include the graviton in conventional QFT, one is lead to
serious problems in performing the
 perturbative expansion (while little is known
about a non-perturbative treatment of a QFT of gravitation). This is due to
the fact that the usual renormalization program, that allows to absorb 
all divergencies in a {\sl finite} number of (measurable) constants, fails.
This kind of non-renormalizibility can be traced back
to the fact that the gravitational constant has negative mass dimension 
(in units where $\hbar =1$).

In string theory the problems with (UV-) divergencies are circumvented in a 
very elegant way: In field theory the dangerous divergencies are
UV-divergencies, i.e.\ divergencies that appear for high momenta. In
coordinate space an one-loop UV divergence would correspond to the limit,
where the loop size shrinks to zero. In principle this divergence is also seen
in string theory, if one considers the limit, when the modular parameter
(or complex structure) that describes the shape of the torus, approaches zero.
However the symmetries of string theory, in this case: modular invariance,
require that one only integrates the modular parameter over a region that
describes inequivalent tori. A convenient integration region for the torus
modulus is given by the shaded region 
in figure \ref{fundregion1} on page \pageref{fundregion1}. The regions 
including possible singularities are explicitly excluded by this choice.
Therefore one-loop torus amplitudes are UV-finite 
(Shapiro \cite{Shapiro:1972ph}). 
Modular invariance extends to higher loop-levels as well. 
Even though not strictly proven yet, it is believed that the
finiteness of string-scattering amplitudes extends to all orders of string
perturbation theory.
It would imply that string theory describes perturbative quantum
gravitation.
This is one (maybe even the strongest) motivation to consider string theory
as a unifying theory.

Up to now we concentrated on the massless modes of string theory. 
There is still an
infinite tower of massive states. For flat backgrounds the different
mass-levels are equally spaced.
Bosonic string theory contains a tachyon, which seems to indicate an 
instability.
Some researchers undertake however considerably effort in order to stabilize
the theory via some kind of tachyon condensation. Another way out of this
problem is to look for a string theory where tachyons are  manifestly absent.
This is the case for:

\section{\label{susystrings}Supersymmetric string theories}
An obvious shortcoming of the bosonic string is the absence of 
space-time fermi\-ons in its spectrum.\footnote{It has however been suggested 
that the bosonic string includes even the supersymmetric string-theories in a
rather subtle manner (cf.\
 \cite{Freund:1985nd,Casher:1985ra,Englert:1986na}).}
 Furthermore it is desirable to have a supersymmetric
theory in space time, at least to a good approximation. In any
 phenomenologically relevant theory this supersymmetry has to be broken at some
 scale, and if string-theory serves as a unifying theory, this breaking must
 be compatible with the underlying principles of string theory. In this
 thesis, we will not concentrate on the supersymmetry breaking mechanism.

To include space-time fermions, the bosonic string action
\ref{polyakov} is extended by terms that involve fermionic dofs. 
There are two common ways to achieve this. One is the {\sl Green-Schwarz}
  (GS) superstring(-formalism) \cite{Green:1981zg,Green:1982yb}.
 Instead of the purely bosonic action one considers ($\al$ set to $1/2$):
\begin{equation} 
 S_1=-\frac{1}{2\pi}\int  d^2\sigma    \sqrt{-\det h}\, 
             h^{\alpha\beta}\,  \Pi^\mu_\alpha (\Pi_\beta)_\mu,\qquad
 \Pi^\mu_\alpha=\partial_\alpha X^\mu-
                   i\bar{\theta}^A\Gamma^\mu\partial_\alpha \theta^A
\end{equation}
In this approach $\theta^A, A=1\ldots N$ are $N$ space-time spinors. 
Each of the spinor components is a world sheet
scalar. This action  is reparameterization invariant. Requiring a so called
$\kappa$-symmetry in order 
to reduce the number of fermionic dofs
 lets one introduce an additional action piece
$S_2$. $\kappa$-symmetry restricts then the maximal number of spacetime 
spinors to $N=2$. Requiring $S_2$ to be supersymmetric reduces the possible 
space-time dimensions considerably. The quantized version singles out 
$D=10$ in which case already supersymmetry of the action $S_2$ requires that 
both spinors $\theta^1$ and $\theta^2$ are of {\it Majorana-Weyl} type.
The Green-Schwarz formalism has the advantage to be manifestly
supersymmetric in space-time.
 However the resulting eoms are extremely complicated, as they
are non-linear. They can be drastically simplified by choosing light-cone
gauge, and simplifying the $X_+$-coordinate as in the bosonic case. 
As the Lorentz-algebra can only be realized in $D=10$ space-time dimensions
this case is considered as the only consistent. 
Depending on the relative handedness of $\theta^1$ and $\theta^2$ one obtains
either the Type IIA ($\theta^1$ and $\theta^2$ have opposite chirality) or 
Type IIB ($\theta^1$ and $\theta^2$ have equal chirality).
In the case of Type I both spinors are identified (by moding out the world
sheet-parity $\Omega$). We come back to this case in section 
\ref{orientiintro}.
There is still a third kind of ten-dimensional superstring known, the 
{\it heterotic string}. As its name suggests, the construction of the
heterotic string is composed from several pieces. The heterotic string takes
advantage from the fact that the closed string-states 
can be decomposed into left and right moving parts. 
The same is true for the fields. Roughly speaking, the
theories considered so far are constructed from a tensor product of left- and
right moving degrees of freedom. This does not mean that the resulting
theories are tensor products as well, since in general some additional
conditions have to be imposed. 
The probably most famous construction of the
heterotic string starts from  ten-dimensional superstring of one (say the
right-moving) sector and the 26-dimensional bosonic string in the other (here:
left-moving) sector. In order to have a sensible space-time interpretation,
one compactifies the sixteen surplus bosonic dimensions. 
Especially one considers flat toroidal compactifications that are obtained 
by identifying points $x\sim x+2\pi\gamma$ with $\gamma$ a vector of a
 sixteen dimensional lattice $\Lambda^{16}$. Associated with 
the torus lattice $\Lambda$ is an even self dual lattice, the so  called 
{\it Narain-lattice} $\Gamma^{16}$, which is in general not unique
 (cf.\ chapter  \ref{orbifolds}). However there are only two
 16-dimensional Narain-lattices:  
\begin{enumerate}
 \item $\Gamma^{16}$ is the weight-lattice of the 
                        $Spin(32)/\mathbb{Z}_2$\,\footnote{By this we mean the
                          sub-lattice of the weight-lattice of $Spin(32)$ which
                          is generated by  the weights of {\sl one}
                          spinor representation together with the
                          the roots of $SO(32)$. (Remember that  the weight
                          lattice of $Spin(32)$ consist of four conjugacy
                          classes: adjoint (i.e.\ roots), vector, spinor,
                          spinor'.)}\\
                        $\big(\Gamma_\text{root}(SO(32))
                  \subset \Gamma_\text{weight}(Spin(32)/\mathbb{Z}_2)\big)$
 \item $\Gamma^{16}=\Gamma^{8}\times\Gamma^{8}$ with $\Gamma^{8}$ 
                    the root-lattice of the $E_8$ Lie-algebra 
\end{enumerate} 
What makes heterotic string-theories so extremely interesting is the fact 
that they admit non-abelian Lie-algebras as gauge-symmetries of their 
low-energy-effective field theory. 
In the first case  this symmetry is $SO(32)$ 
 while in the second it is $E_8\times E_8$.
  These symmetries are also manifest in the operator product
expansion (OPE) of the formerly mentioned vertex operators.\footnote{In 
 the heterotic string there exist currents on the world-sheet
 (associated with
 charges) that build up the corresponding  {\it Kac-Moody algebras}.\,A 
  Kac-Moody algebra  is an infinite
  dimensional extension of a Lie algebra.} 
 Compactifications of the heterotic string to four space-time dimensions 
 have led to many interesting  models, especially such that come pretty close
  the Standard Model (SM) of electro-weak and strong interactions. 
 
 In parallel to the Green-Schwarz superstring there exists the 
 so called {\it Neveu-Schwarz-Ramond} (NSR) superstring.\footnote{The NSR
 formalism  was developed before the GS-formalism} 
 It turns out that the GS- and the NSR-superstring describe the same physics,
 though they use other formalisms.
 While the GS-superstring exhibits manifest space-time supersymmetry, its
 covariant quantization is not at all obvious.
 The NSR superstring however can be quantized by path-integral formalism in
 parallel with the bosonic string, up to some generalizations. It becomes
 space-time supersymmetric, if one imposes the so called {\sl
  Gliozzi-Scherk-Olive} (GSO) projection (which is absent in GS-formalism)
  \cite{Gliozzi:1976jf,Gliozzi:1977qd}.
 From the world-sheet point of view the bosonic string action \ref{polyakov} 
 might be considered as a two dimensional gravity theory (after inclusion 
 of the $R(h)$-term like in \eqref{nonlinsigmamod}) coupled to $D$  world-sheet
 scalars $X^\mu$. It is now quite natural (and actually 
 necessary in order to use
 path-integral formalism in a subsequent analysis) to extend this theory to
 $N=1$ local supersymmetry (or: $N=1$ supergravity) on the world sheet.
 Supersymmetrizing the scalar part
of the bosonic action  is achieved by adding the following term to the 
 Polyakov action \eqref{polyakov}:
\begin{equation}\label{fermionicext}
 S_F= -\frac{1}{2\pi\al}\int  d^2\sigma  \sqrt{-\det h}\,
   \big(i\bar{\psi}^\mu\rho^\alpha\partial_{\sigma^\alpha} \psi_\mu 
        +F^\mu F_\mu
   \big)
\end{equation}
Some comments are in order: Each $\psi_\mu$, $\mu=0,\ldots, D-1$ is a two
 dimensional (world-sheet) Majorana-spinor. The $D$  world-sheet spinors 
$\psi_\mu$ make up a space-time Lorentz-vector. 
(This is in contrast to the GS-formalism, where the $\theta^A$ are space-time
 spinors as well as world-sheet scalars.) The $F^\mu$ are auxiliary fields
 which are needed to realize the off-shell supersymmetry-algebra. Their eoms 
however require them to vanish on-shell. Each of the $F^\mu$ is a 
world-sheet scalar while in total they  make up a $D$ dimensional 
space-time Lorentz-vector. The metric $h$ can be expressed by world-sheet 
Vielbeins $e^\alpha_{\phantom{\alpha}a}$ 
(or more precisely as they live in {\sl two} dimensions: by {\sl Zwei}beins.):
\begin{equation} 
  e^\alpha_{\phantom{\alpha}a}  e^\beta_{\phantom{\beta}b}
      h_{\alpha\beta}=\eta_{ab} \qquad a,b,\alpha,\beta\in\{0,1\},\qquad
     \eta=\diag(-1,1)
\end{equation}
As $GL(d,\mathbb{R})$ does not admit finite-dimensional
 spinor-representations,
 Vielbeins are a way to define spinors on curved space-time,
 in our case: on a curved  world-sheet. The two dimensional matrices 
$\rho^\alpha$ are obtained from the  two dimensional Dirac matrices (cf.\ eq.\
\eqref{2ddiracmat},\eqref{2ddiracalg}, p.\ \pageref{2ddiracmat}) by:
\begin{equation}
  \rho^\alpha\equiv  e^\alpha_{\phantom{\alpha}a}\rho^a
\end{equation}
The sum of the bosonic action \eqref{polyakov} and the fermionic action
 \eqref{fermionicext} does not yet admit {\it local} supersymmetry. This goal
 is achieved by adding a third piece to the action:
\begin{equation} 
 S_3= \frac{i}{4\pi\al}\int  d^2\sigma  \sqrt{-\det h}\, 
    \bar{\chi}_\alpha \rho^\beta \rho^\alpha \psi^\mu 
 \big(\partial_{\sigma^\beta} X_\mu-\tfrac{i}{4}\bar{\chi}_{\beta}\psi_\mu\big)
\end{equation}
$\chi_\alpha$ is the superpartner of the world-sheet metric
 $h_{\alpha\beta}$ (or of the Zweibein $e^\alpha_{\phantom{\alpha}a}$).
It has a world-sheet vector- and a world-sheet spinor-index.  
The resulting action has a variety of symmetries:
\begin{itemize}
 \item Local world-sheet supersymmetry
 \item Local Weyl-invariance (The Weyl transformation rescales also the
   Ma\-jo\-ra\-na-fermions $\psi^\mu$ and the gravitino $\chi_\alpha$ besides 
the Zweibein $e^\alpha_{\phantom{\alpha}a}$)
 \item Local super-Weyl-invariance ($\lambda(\tau,\sigma)$ 
    a Majorana spinor parameter): 
   \begin{equation}
     \delta_\lambda \chi_\alpha=\rho_\alpha\lambda\qquad 
     \delta_\lambda (\text{others}) =0
   \end{equation}
 \item World-sheet (or: two-dimensional) Lorentz-invariance
 \item World-sheet reparameterization- (or: diffeomorphism-) invariance
\end{itemize}
 Very similar to the purely bosonic case, one could use some of the symmetries 
 to eliminate some degrees of freedom. Using local supersymmetry,
 reparameterization and Lorentz-invariance, one can reduce the two-dimensional
 supergravity action to a much simpler action (cf.\ eq.\
 \eqref{opensuperstaction}, \eqref{susywsactionconf}, 
 p.\ \pageref{opensuperstaction}). The corresponding gauge where the 
 gravitino is efficiently eliminated while $h$ is brought to the standard 
 minkowskian form is called {\it superconformal-gauge}.
 Besides the conformal symmetry encountered in the bosonic conformal gauge,
 this action admits a further symmetry, generated by the fermionic current $T_F$.
 $T_F$ is determined  by varying the  (non-gauge fixed)
   action with respect to the gravitino $\bar{\chi}^\alpha$:
\begin{equation}
 T_F= \frac{2\pi}{i\det e} \cdot\frac{\delta S}{\delta\bar{\chi}}
\end{equation}
Along the lines of light-cone gauge in the bosonic case, one can eliminate in
 addition the $\psi^+$ component from the world-sheet Majorana
 spinor.\footnote{$\psi^+\propto \psi^0+\psi^1$ should not be confused with
 the spinor component $\psi^\mu_+$ which will be introduced below.}
In contrast to the bosonic case the critical space-time dimension $D$ turns
 out to be {\it ten}, rather than $26$. (The formerly mentioned normal
 ordering constant $a$ equals now {\it one half}\, in the bosonic 
(Neveu-Schwarz) sector
instead of one, while it is {\it zero} in the fermionic (Ramond) sector. These
sectors will be explained below.)
Equivalent results can also be derived via path-integral quantization.

There is still a peculiar feature in the NSR superstring which we now want to
address: So far we have not specified the boundary conditions of the Majorana
spinor $\psi$.  For several reasons, and the most striking one is 
{\sl modular invariance} (to be explained in chapter \ref{orbifolds}), one
is forced to allow $\psi$ to be both periodic and antiperiodic for closed
strings:\footnote{In calculating partition functions, a fermion-field is
  anti-periodic in time (if no further trace-insertion acts on this field in
  operator formalism).
  The modular group (which is a
  important symmetry in string theory) maps sectors in the partition 
  that correspond to certain 
  periodicities to other sectors. Thereby a spinor $\psi$ that is periodic in
  $\sigma$ and anti-periodic in time will get mapped to a different
  sector. This explains the presence of different boundary (or periodicity)
  conditions as well as the presence of the GSO-projection.}  
\begin{equation} \label{GSOprojclos}
  \psi^\mu_\pm(\tau,\sigma)=\kappa_\pm  \psi^\mu_\pm(\tau,\sigma+2\pi)
  \qquad \kappa_\pm\in\{ -1,+1\}
\end{equation}
Here we have denoted the two components of the Majorana spinor 
$\psi^\mu$ by $\psi^\mu_+$ and $\psi^\mu_-$ which is suggested by
 the fact that after
solving the eoms the first component only depends on $\tau+\sigma$ while the 
second only depends on $\tau-\sigma$.
A similar freedom like in \eqref{GSOprojclos} exists also 
for open strings, where the supersymmetric partner
 of the bosonic boundary condition
(eg.\ Neumann-type: $\partial_\sigma X^\mu=0\Rightarrow\partial_+ X^\mu=\partial_-X^\mu$, 
  $\partial_\pm \equiv 1/2(\partial{\tau}\pm\partial_\sigma)$)
  becomes:\footnote{There is still a redundancy in the following equations:
By a field redefinition one can set $\kappa(0)=+1$.}
\begin{equation} 
    \psi_+ (\tau,\sigma)=\kappa(\sigma)\psi_- (\tau,\sigma)
    \qquad\text{for } \sigma\in\{0,\pi\}, \;\kappa(\sigma)\in\{\pm 1\}
\end{equation}
Depending on the sign $\kappa$ one distinguishes between Ramond (R) ($\kappa=1$) and 
Neveu-Schwarz (NS) ($\kappa=-1$) fermions. It turns out that unless one
    imposes a projection, the GSO-projection, the NS-sector contains a
    tachyon. The GSO-projection is also needed for modular invariance.
Even though defined on the whole string-spectrum, the action of the
    GSO-projection on the R-ground-states is particularly interesting. 
Solving the equations of motion subjected to the boundary conditions, one
discovers a zero-mode in each $\psi^\mu$-coordinate. In light-cone-gauge one
has therefore $8$ zero-modes $b^i$ which are anti-commuting and fulfill a
Clifford-algebra:
\begin{equation}
  \{b^i,b^j\}=\eta^{ij}\qquad i,j\in\{2\ldots D-1\}
\end{equation}
Thus one can represent the above algebra on a vector space 
with the following basis: $|s_1,s_2,s_3,s_4\rangle$ with $s_i=\pm\oh$.
The resulting vector space can then be described as the sum of a vector 
space  of positive chirality and another one of negative chirality. 
Performing the GSO-projection   eliminates
one chirality from the massless ground-state.
In the closed string sector there are two sectors containing fermions
  depending on the combination of left- and right-moving sectors:
These are the NSR and the RNS sectors, while the NSNS- and the RR-sector 
make up  space-time bosons. In the open string there are just two sectors,
 the NS- and the R-sector, the latter containing the space-time fermions.
As the GSO projection picks up one chirality, there is still the freedom to
  choose equal or opposite chiralities on left- and right-movers. Equal 
chiralities lead to the Type IIB superstring, while opposite chiralities 
yield Type IIA. Upon compactification on a circle, this does not make big
  difference, since both theories are then related by a perturbative duality,
  the so called {\sl T-duality}. The massless spectrum of Type IIA theory 
  can be found
  in table \ref{clspectypeIIA}, its Type IIB pendant 
  is given in table \ref{clspecIIaI}, page \pageref{clspecIIaI}.
   While the resulting spectrum is
  supersymmetric, it is much harder to show that the interacting theory is
  supersymmetric as well. We will not investigate this topic.
\begin{table}
 \begin{center}
 \begin{tabular}{|c|c|}
  \hline
  \multicolumn{2}{|c|}{bosons} \\
  \hline 
   NS-NS& R-R \\
   metric $g_{ij}$, 2-form $B_{ij}$ &  \\
  dilaton $\phi$ & \raisebox{1.5ex}[-1.5ex]{vector $A_i$, 3-form $C_{ijk}$}
 \\
 \hline
 \hline
  \multicolumn{2}{|c|}{fermions}   \\
  \hline 
   NSR &RNS   \\
   gravitino $\psi_{i\dot{a}}$ & gravitino $\tilde{\psi}_{jb}$ \\
 \hline 
 \end{tabular}
 \caption{\label{clspectypeIIA}Massless closed-string  spectra of Type IIA theory}
 \end{center}
\end{table}

 It is possible to build modular-invariant partition functions that
  consist only
  of RR and NSNS sectors. The resulting theories are called Type 0A and Type
  0B. They do not contain any fermions in the closed string sectors and are
  plagued with tachyons. However there exist interesting generalizations of
  these Type 0A/B by performing an {\it orientifold}\,\footnote{We will introduce
   orientifolds in section \ref{orientiintro}. In addition we devoted a 
  whole chapter to these constructions (cf.\ chap.\ \ref{orientifolds}).} projection of these theories.
This removes the closed-string tachyon and introduces fermions via a necessary open string
  sector (cf.\ \cite{Angelantonj:2002ct} and references therein). 
  There exist non-supersymmetric orientifolds of Type 0B that are completely
  tachyon free \cite{Sagnotti:1995ga,Sagnotti:1997qj,Angelantonj:2002ct}.
  Something similar is known for the heterotic string
  as well: If one constructs the heterotic string in the NSR
  formalism one discovers that by changing the GSO-projection one can obtain 
  a tachyon-free  non-supersymmetric $O(16)\times O(16)$ string theory in 
  ten space-time dimensions
  (cf.\
  \cite{Alvarez-Gaume:1986jb,Seiberg:1986by}).
  Several other non-supersymmetric modular-invariant variants of the heterotic
  string (which contain however tachyons) are known.

 It is a natural task to consider $N>1$ world-sheet supergravities. 
 However it turns out that for $N=2$ the critical space-time dimension would
 be 4 with a $(2,2)$ space-time signature
 (which is phenomenologically uninteresting),
  while for $N=4$ the dimension is even negative, and thus unacceptable for a
 reasonable space-time interpretation. 

\subsection*{Space-time supersymmetry}
Space-time supersymmetry is a desirable feature for physical theories.
This has several reasons. The probably most important one  
is the {\it hierarchy problem}:
In electroweak-theory the big difference between electroweak-scale (which is
about $246$ GeV, the {\it vacuum expectation value} (VEV) of the Standard
Model Higgs field) and Planck-scale ($1.22\cdot10^{19}$ GeV) is believed to be
very unnatural. Furthermore the parameters describing the 
Higgs-boson (which is the only scalar particle
of the Standard Model) receive enormous contributions from radiative 
corrections up to the Planck scale. In order that these parameters take
exactly those values required by measurements at typical ``high-energy''
 experiments, the values have to be met within enormous precision (something
 like one part in $10^{30}$) at the Planck scale. Furthermore this fine-tuning has to be repeated 
 at each order of perturbation theory. In parallel the higher order
 corrections 
exceed in general the lower order approximations.     
\subsubsection*{Grand unified theories}
In (most) {\sl grand unified theories}  in general a 
second hierarchy problem comes along which is due to an additional Higgs 
particle. The underlying idea of grand unified theories is the following:
Each lepton generation comes up with a
quark-generation (or flavor) which however sits in a separate representation.
One could now try to unify leptons with quarks in  multiplets of the gauge
group. This is  achieved for example in the {\it Pati-Salam} (PS) $SU(4)\times
SU(2)_\text{R}\times SU(2)_{L}$-model where the leptons correspond to a fourth
color (cf.\ \cite{Pati:1973uk}). Each generation of  matter 
transforms in a $(4,2,0)$ and  $(4,0,2)$ representation of the gauge group. 
This Pati-Salam model  has two interesting features, that are common to most
other GUTs as well: 
\begin{itemize}
\item Additional matter that is absent in the (``minimal'')
  Standard Model  (In Pati-Salam $SU(4)\times
SU(2)_\text{R}\times SU(2)_{L}$: right-handed neutrinos)
\item The electric-charge is quantized    
\end{itemize} 
Quantization of electric-charge is in general true for models with simple
gauge-group but also for this semi-simple example. In unifications with
simple gauge-group the SM gauge-group is embedded into a larger, simple
Lie-group $G$:
\begin{equation}
  SU(3)\times SU(2)\times U(1)_Y\hookrightarrow G
\end{equation}
Thus not only  leptons and quarks become unified, but gauge-bosons of different
gauge-groups as well.
Well known  examples for GUTs with simple gauge group $G$ 
are $SU(5)$, $SO(32)$ and even $E(6)$ GUTs, the
latter based on the exceptional group $E(6)$.\footnote{The $SU(5)$ model was
proposed by Georgi and Glashow \cite{Georgi:1974sy}, the SO$(32)$ theory by
Georgi \cite{Carlson:1975gu}
in parallel to Fritzsch and Minkowski \cite{Fritzsch:1975nn}. The $E(6)$ model was found by
Gursey, Ramond and Sikivie \cite{Gursey:1976ki}.} 
Among several interesting and attractive features of GUTs we want to mention
the probably best known: GUTs in general predict {\it proton decay}.
Proton decay, if present, can be measured  (up to a certain bound) by
experiments. Several GUTs have already been ruled out by experimental data.
Supersymmetry suppresses the decay rate considerably. For example the
non-supersymmetric $SU(5)$
GUT is forbidden, while its supersymmetric extension is still in accord with
the bound given by current
proton decay experiments. Analogous statements can be made for $SO(10)$.

Now we address a second hierarchy problem that comes along with most GUTs.
 What is important in  GUTs, is that the unifying gauge-symmetry has
to be broken at some scale, which is of course above the electro-weak scale.
This will be done in general by some Higgs mechanism with the corresponding
Higgs-field  acquiring a VEV 
$\langle 0|\Phi|0\rangle=w$  which is of the
order of the unification scale. We assume that the unification scale  {\it a
 priori} does not 
 coincide with the Planck scale.
 The running of the couplings strongly suggests that it is of the order of
 $10^{15}$ to $10^{16}$ GeV.\footnote{The first value is
 already excluded by experiment, and assuming solely the SM particle content
 will not lead to gauge coupling unification.}
 The second gauge-breaking is the usual
electro-weak symmetry breaking which occurs at a VEV of
$\langle 0|\phi_\text{e.w.}|0\rangle=v\approx 246$ GeV.
A generic Higgs potential looks like:\footnote{We have suppressed 
 group indices which are present since the Higgs fields transform 
under the gauge group.}
\begin{equation} 
V = -\frac{A}{2}\Phi^2+\frac{B}{4}\Phi^4
 -\frac{a}{2}\phi^2+\frac{b}{4}\phi^4+ \frac{\lambda}{2}\Phi^2\phi^2
\end{equation}
The term proportional to $\lambda$ is generic and thus has to be included.
The GUT scale value is obtained, if we tune $A$ and $B$ such that: $w^2=A/B$.
The problem occurs for the VEV of the second (i.e.\ the electroweak) Higgs:
Since $v^2=(a-\lambda w^2)/b$ has to be obeyed this requires a fine tuning of
 $a$ to one part in $10^{26}$. Radiative corrections will require this fine
 tuning at each order in perturbation theory. If present, supersymmetry
 ensures however
 that radiative corrections do not destroy the hierarchy and parameters do not
 have to be retuned. On the other hand supersymmetry has to be
 broken. Requiring the
 hierarchy to be preserved by this breaking leads to the prediction, that
 supersymmetric partners of the known particles should show up at $1$ TeV.

Other mechanism like 
composite Higgs-particles have been proposed to circumvent the hierarchy
problem without the use of supersymmetry. However these approaches are plagued
with other difficulties.

Inspired from string-theory, it has been suggested that {\it extra large
  dimensions} 
could solve the hierarchy problem as well. In these scenarios the known gauge
interactions are restricted to a lower dimensional subspace (a {\it brane})
while gravity propagates in the entire space (often denoted by ``bulk''),
  which in most models has relatively large,\footnote{By ``large'' we mean much
  bigger than the Planck length, and in order to solve the hierarchy problem:
 in the region up to a few TeV.} but compact directions.
Future experiments can put severe constraints on the size of possible extra 
large dimensions, which might sustain or rule out these proposals.

As a third argument for supersymmetry, we mention the 
unification of the Standard Model couplings at a scale of
$10^{16}$ GeV if one assumes the 
supersym\-me\-try-breaking scale at about one TeV.

\section{Compactifications}
It goes back to the early twenties of the 20\raisebox{1ex}{\tiny th}
century that {\it Kaluza} suggested a theory with an
additional small dimension. Even though this dimension might not be discovered
directly due to its smallness, it influences the four dimensional physics
indirectly. As string theory on flat backgrounds has too many dimensions of
unrestricted size, one has to figure out some explanation, why only four
space-time dimensions are seen. A very fruitful idea is to compactify string
theory on some tiny space $ {\cal X}_{d}$:
\begin{equation}
  {\cal M} = \mathbb{R}^{(1,D-d-1)}\times {\cal X}_{d}
\end{equation}
By this we obtain effectively a theory with one time and $D-d-1$ space
  dimensions.
The exact form of the space ${\cal X}_{d}$ has big influence on the theory
  seen in uncompactified space. If we compactify a 10-dimensional ${\cal N}=1$
  superstring theory on a {\sl Calabi-Yau} (CY) space, ${\cal N}=1$ will be
  present in $10-d$ dimensional space time.\footnote{To be precise, one has to
  deform the CY space to take the $\al$-string corrections to the supersymmetry algebra
 into account.} 
Furthermore the chiral massless spectrum is determined by topological data of 
 ${\cal X}_{d}$. The Calabi-Yau space admits in general additional
  structures like gauge-bundles. Physical requirements like
  anomaly-cancellation put further constraints on the geometry. 
  The topic is too extended in order to enter into details. For some geometric
  aspects of compactifications we refer the reader to the book of GSW \cite{Green:1987mn}. 
  
 A special case of compactification spaces are
 {\sl orbifolds} to which we have devoted the next chapter. Roughly speaking,
 an orbifold is
 the orbit-space of some discrete group $G$ that acts on a manifold ${\cal B}$:
\begin{equation}
 {\cal X}_{d} = {\cal B}/G
\end{equation}
The action of $G$ may admit fixed-points, which usually result in
singularities on ${\cal X}_d$. If the string theory on ${\cal B}$ is known, it
is comparatively easy to construct the orbifold by $G$.
Even though ${\cal X}_d$ might be singular in some points, string propagation 
turns out to be regular (in most cases). In all of our thesis we encounter either
tori (that can also be interpreted as fixed-point free orbifolds) or toroidal 
orbifolds of $\mathbb{Z}_N$-groups or products thereof.  

\section{\label{orientiintro}Open strings and unoriented string theories}
We have already seen that the perturbative spectrum of the heterotic string 
leads to a non-abelian gauge-symmetry in the low-energy effective action.
However both Type II theories do not show this gauge-symmetry. 
If one does not insist on 10-dimensional Lorentz-invariance, one can 
include gauge-symmetries in Type II theories as well.
One way to achieve this is to include open strings, 
and in general: world-sheets with
boundaries. One can assign charges to the end-points of open-strings in the
way proposed by Chan and Paton (cf.\ \cite{Paton:1969je}). 
\begin{figure}[h]
   \begin{minipage}[t]{\textwidth}
    \begin{minipage}[t]{5cm}
     \setlength{\unitlength}{1mm}
      \begin{center}
     \begin{picture}(45,20)
      \begin{picture}(0,0)%
       \includegraphics{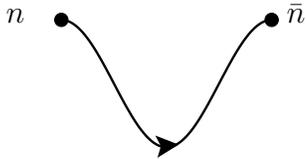}%
      \end{picture}%
      \setlength{\unitlength}{987sp}%
      \begingroup\makeatletter\ifx\SetFigFont\undefined%
      \gdef\SetFigFont#1#2#3#4#5{%
      \reset@font\fontsize{#1}{#2pt}%
      \fontfamily{#3}\fontseries{#4}\fontshape{#5}%
      \selectfont}%
      \fi\endgroup%
     \begin{picture}(7050,3627)(1151,-5904)
      \put(1151,-2472){\makebox(0,0)[lb]{\smash{\SetFigFont{12}{14.4}{\rmdefault}{\mddefault}{\updefault}{\color[rgb]{0,0,0}$n$}%
}}}
      \put(8201,-2472){\makebox(0,0)[lb]{\smash{\SetFigFont{12}{14.4}{\rmdefault}{\mddefault}{\updefault}{\color[rgb]{0,0,0}$\bar{n}$}%
}}}
     \end{picture}
     \end{picture}
     \end{center}
    \end{minipage}
     \hfill
     \raisebox{2.4cm}{\begin{minipage}[t]{8cm}
      \begin{center}
      \caption[Open string with Chan-Paton charges]
              {\label{opscpcharge}Open-string
                with Chan-Paton charges $n$ and $\bar{n}$}
      \end{center}
     \end{minipage}}
   \end{minipage}
\end{figure}
In figure
\ref{opscpcharge} we have depicted an open string with two charges $n$ and
$\bar{n}$ at its endpoints. It can be shown that one can define consistently
string perturbation theory if one assumes that both endpoints are represented 
by $n$-dimensional vectors in $\mathbb{C}^n$. The resulting string theory
admits a global $U(n)$ symmetry, which is promoted to a local (i.e.\ gauge-)
symmetry 
of the low-energy effective action. The $\bar{n}$ signals that the
right-endpoint transforms in the complex conjugate representation with respect
to the left one. 
\subsection*{D-branes}
The loci of open string endpoints can be associated to so called {\it
  D$p$-branes}, where $p+1$ is the space-time dimension of these loci. 
 In supersymmetric string-theories there exists a connection between
 D$p$-branes  and the supercharges which are preserved by these 
 objects.\footnote{Not every possible locus (more precisely: submanifold)
 the D$p$-brane wraps  
  can be associated with supercharges. The submanifold has to fulfill
  additional condition, eg.\ the sLag condition.} If several D-branes are
  present they may or may not preserve some or several supercharges.
  The number of  supercharges preserved by a D-brane configuration
  determines the amount of supersymmetry of the particular string model. 
  Supersymmetric D$p$-branes are usually {\it Bogomolnyi-Prasad-Sommerfield
  (BPS)
  states}, i.e.\ states with a reduced
  amount of supersymmetry that saturate the BPS-bound. BPS-states carry a 
  {\it central charge}
  $Z$ of the super-symmetry algebra which is a conserved charge. 
  This fact made D-branes so important in the 
  second ``string revolution''.  String-theory was defined so far by a
  perturbative expansion, very similar to the way in which Feynman rules may
  be introduced by hand in electrodynamics. 
  Whatever the correct non-perturbative definition of string-theory would be,
  it is extremely likely that it preserves the BPS-property, especially in the
  process of renormalization. 
  On the other hand BPS-solutions were known
  to appear in the form of p-dimensional soliton-like solutions in the
  low-energy effective actions of string theories. Besides D-branes there 
  exist other BPS-states in string theory as well.
  By identifying BPS states
  in perturbatively inequivalent theories the notion of an {\it M-theory} was
  born. M-theory is considered to be the unifying theory which includes all
  superstring theories as special limits of the M-theory
  moduli-space.\footnote{11-dimensional supergravity is another limit in the 
   M-theory moduli-space. }
  D-branes have also proven extremely useful in explaining  
  Bekenstein-Hawking entropy at a microscopic level.\footnote{At least for
  some supersymmetric black hole configurations.} 
  
\subsection*{Type I and orientifolds}
  We have claimed so far that Type II theories, if they contain open-strings,
 will break 10-dimensional Lorentz-symmetry. This is not disastrous, since for
  phenomenological reasons we will break this symmetry anyway at some point. 
  So far we did not explain why we are not allowed to introduce freely some
  D9-branes in Type II, thereby maintaining Lorentz-invariance. 
  It will become soon clear, that D-branes carry a special type of charge,
  a so called {\it Ramond-Ramond (RR) charge},
  which is of topological type, and
  that this charge has to cancel in total. The RR-charge of a D-brane
 constitutes its central charge.
  
  The low energy limit of 
  Type I theory is a 10-dimensional ${\cal N}=1$ supergravity coupled to 
  10-dimensional ${\cal N}=1$ supersymmetric Yang-Mills with gauge group
 $SO(32)$. Both open and closed strings admit a further symmetry, which is
  world-sheet parity. 
  World-sheet parity reverses the orientation of the world
  sheet, while leaving the action invariant. This has two effects: 
  \begin{itemize}
  \item Only 
  closed string-states that are invariant under the world-sheet parity
  $\Omega$ are kept in the spectrum. 
  \item Due to the formula $\chi=2-2h-b-c$ for the Euler-character $\chi$ 
   we need to include 
   the Klein-bottle ($h=0$ handles, $b=0$ boundaries, $c=2$ cross-caps) as the
   second closed-string one-loop vacuum amplitude.
   \end{itemize}
It turns out that the Klein-bottle amplitude has severe divergences. They are
interpreted as uncanceled RR-charges under which the so called {\sl
  orientifold plane} (O-plane) is charged. (The O-plane corresponds to the
cross-caps in the Klein-bottle). In analogy to field theory these
divergences are called RR-{\sl tadpoles}. As D-branes carry RR-charges as well they may
serve as a neutralizer of the O-plane charge, provided that their charge has
the right sign and value. This is indeed the case. In Type I the RR-charge is
exactly canceled by 32 D$9$-branes. In computing the open string partition
function we have to include the parity projection $\Omega$ as well. This
implies that we have to introduce the {\it M\"obius-strip} ($b=1$, $c=1$)
besides the cylinder ($b=2$, $c=0$). 
The projection together with the RR-tadpole
cancellation conditions implies that the $U(32)$-symmetry 
gets broken to $SO(32)$.
The only gauge groups which can be obtained in the perturbative spectrum of
Type I and compactifications thereof 
are orthogonal, symplectic and (under certain circumstances)
unitary groups.    

The Type I construction can be generalized. On one hand Type I can be
compactified on some space ${\cal X}_d$. As before ${\cal X}_d$ might be an
orbifold. We can also gauge a combination  $s\Omega$, where $s$ acts on
space-time such that $s\Omega$ is a symmetry of the string theory under
consideration. We can even try to include several such elements. However we
will show in section \ref{orientifolds} that this does not lead
to new consistent models in most cases. Given a projection via $s\Omega$ and a
compactification space ${\cal X}_d$ there might be several inequivalent ways
to cancel the RR-tadpole of the O-plane(s).\footnote{In more complicated 
spaces ${\cal X}_d$ the O-planes consist of several parts. Therefore, we refer
to ``several O-planes''.} All these generalizations which include the world
sheet-parity in some way are summarized by the term: ``orientifold''.

\section{Chiral fermions in open string theories}
As half of the thesis deals with chiral fermions from the open string sector
in one way or the other, we want to make some comments here.
Chiral-fermions are an essential feature of the SM. In string theory they can
be obtained in many ways (cf.\ the introduction to chap.\ \ref{magbf}, p.\
\pageref{magbf}). For open string theories three mechanism are very prominent:
  \begin{figure}[t] 
  \begin{minipage}[t]{\textwidth}
   \setlength{\unitlength}{0.1in}
   \begin{minipage}[t]{6cm}
  \begin{center}
  \begin{picture}(25,27.0)
  \SetFigFont{14}{20.4}{\rmdefault}{\mddefault}{\updefault}
  \put(0,0.0){\scalebox{0.20}{\includegraphics{picint6.EPS2}}}
   \put(25,22){$X_0$} 
  \end{picture}
  \caption[Open-string with endpoints located on intersecting D-branes]
          {\label{angledbranesfig}Time evolution of an open-string 
           with endpoints located on D-branes 
           intersecting at an angle. The classical string oscillates around the
           intersection point. Upon toroidal 
         compactification on a $T^2$  angled D-branes are T-dual to 
         magnetic backgrounds (right figure).}
  \end{center}
  \end{minipage} 
 \hfill
    \begin{minipage}[t]{6cm}
    \begin{center}
    \begin{picture}(25,27.0)
    \SetFigFont{14}{20.4}{\rmdefault}{\mddefault}{\updefault}
    \put(0,2){\scalebox{0.20}{\includegraphics{picint7.EPS2}}}
     \put(20.7,22){$X_0$}
    \end{picture}
    \caption[Open-string in magnetic background fields]
          {\label{magstringpic}
           Time evolution of a bosonic open-string in constant 
            magnetic background
             fields. The classical string rotates around a point, whose 
             position is however {\sl not} determined by the NS-fields on its
             boundaries.
            }
    \end{center}
    \end{minipage}
 \\  The \rule{0ex}{3.2ex} string is a \textcolor{bl}{blue} line, 
  while the world-sheet
  boundaries are in \textcolor{r}{red} and ``\textcolor{skyblue}{skyblue}''.
  The string depicted obeys the classical eoms. Its
  lowest (non-zero) mode is excited. In fig.\ \ref{angledbranesfig} 
  the D-branes are drawn in transparent colors, while 
  in fig.\ \ref{magstringpic} the branes are two-dimensional.
  The world sheet is in transp.\ \textcolor{orange}{orange}.
\end{minipage}
\end{figure}
\begin{enumerate}
 \item Open strings with endpoints on D-branes with non-trivial topological
   intersection number
 \item Open strings with endpoints on D-branes which carry different magnetic
   background fields
 \item Open strings stuck to a singularity
\end{enumerate}
For flat space-time,
the first method was discovered in \cite{Berkooz:1996km}. The second method
was (to our knowledge) first applied to model building in
\cite{Bachas:1995ik}.
Both methods are related by T-duality, if the branes intersect as lines when
restricted to a $T^2$. T-duality acts on each $T^2$ in {\sl one} coordinate 
by $R/\sqrt{\al}\to\sqrt{\al}/R$.
 The classical solutions for both scenarios are depicted
in figure \ref{angledbranesfig} and \ref{magstringpic}. (These figures show the
string, its boundaries and the world sheet, as well as the D-branes for the
intersecting scenario.)  The quantized version 
has some features in common with the classical solution. In the case of
intersecting D-branes (fig.\ \ref{angledbranesfig}), one sees that the string
oscillates around the intersection point. The string which is coupled to the
magnetized D-branes circulates around some point as well. This point is
classically not restricted. In the quantized version it corresponds to a Landau
level. The infinite Landau degeneracy gets finite after compactification,
e.g.\ compactification on a torus. The easiest way to see the appearance
of chiral fermions is first to note that by the altered boundary conditions
the number of Ramond-zero-modes $b^i$ (cf.\ section \ref{susystrings}) is
reduced, such that (for suitable boundary condition) only one Ramond-state
survives:
\begin{equation}
 \begin{array} {ccc} 
 \text{homogenous} &  &\text{inhomogenous}
 \\  
 \text{ boundary conditions} &  &\text{boundary conditions}
 \\ \rule{0ex}{3ex} |s_1,s_2,s_3,s_4\rangle\big|_{\text{GSO-proj.}} 
  &\longrightarrow & \big|+\tfrac{1}{2}\big>
 \end{array}
\end{equation}
By ``homogenous'' we mean that there are identical boundary conditions on the
 left- and right-endpoint of the string (at least concerning the derivatives).
Each chiral fermion obtained this way appears with a multiplicity that is
determined by the bosonic zero-modes, where the sign has to be properly taken
into account. This multiplicity is the topological intersection number or 
the Landau degeneracy which might be calculate via the {\sl Atiyah-Singer
  index theorem} for twisted spin-complexes.
Lower dimensional D-branes which have exactly half the dimension of the
embedding space like in figure \ref{angledbranesfig} are encountered in 
$\bar{\sigma}\Omega$-orientifolds. In these orientifold $\bar{\sigma}$ acts as
complex conjugation on each $T^2$. (The compactification space is a product of
$T^2$`s or an orbifold thereof.) We use $\bar{\sigma}\Omega$-orientifold
constructions in chapter \ref{magbf} and \ref{z4} and obtain interesting
 chiral spectra. In chapter \ref{magbf} we
alternatively consider the T-dual magnetized situation as well.
 \begin{figure} 
 \begin{minipage}[t]{\textwidth}
  \begin{minipage}[t]{5cm}
  \begin{center}
  \setlength{\unitlength}{0.1in}
  \begin{picture}(27,36.4)
  \put(0,0){\scalebox{0.22}{\includegraphics{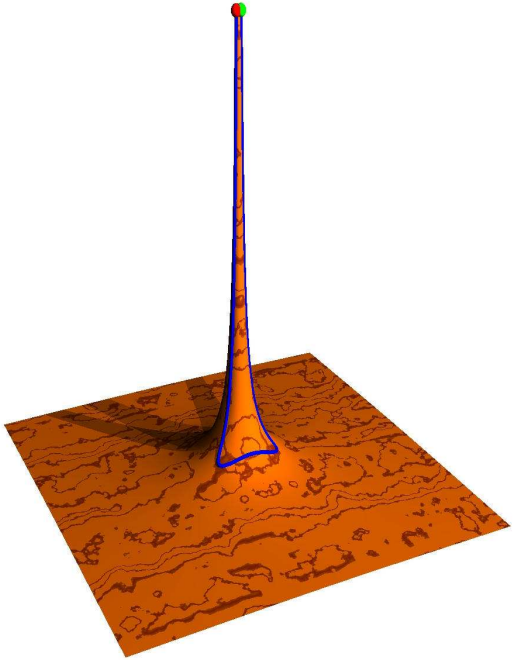}}}
  \end{picture}
  \end{center}
 \end{minipage} 
  \hfill 
  \begin{minipage}[t]{5.1cm}
    \caption[Open-string located at a singularity]
      {\label{openstringsingpic}
       Open-string located at a singularity (schematic):
       A D-brane which is bound to a
       singularity in compactification space is the locus of 
       open-string end-points. (The string is painted in  
       \textcolor{bl}{blue}, its 
        endpoints are  \textcolor{r}{red} and \textcolor{g}{green}). 
       Open superstrings in this sector 
         might admit chiral fermions. If the string end-points  belong to
        different stacks  
       of D-branes (denoted by $a$ and $b$), this can lead to
       chiral fermions in bifundamental representation $(n_a,\bar{n}_b)$ of the
       associated gauge group $U(n_a)\times U(n_b)$.}
    \end{minipage}
 \end{minipage}
 \end{figure}
Even though we do not apply it in this thesis, we want to mention that chiral 
fermions can be obtained from D-branes which are located at singularities
(cf.\ \cite{Douglas:1996sw,Klebanov:1998hh}). The orbifold case, especially the
$T^2/\mathbb{Z}_3$ case has been exhaustively explored (cf.\
\cite{Douglas:1996sw,Douglas2001:tr} and references therein). 
A D-brane that is stuck to an (orbifold) singularity is called a {\it fractional brane}.
This is due to the fact that it carries only a fractional amount of
untwisted RR-charge in comparison to an ordinary D-brane. Open string states 
in the Ramond-sector are described by:\footnote{For untwisted
  boundary conditions.}
\begin{equation} 
 |s_1,s_2,s_3,s_4;\Lambda_{ij}\rangle\big|_{\text{GSO-proj.}}
\end{equation}
$\Lambda_{ij}$ is the Chan-Paton matrix that encodes the CP-dofs.
In symmetrizing the state with respect to an orbifold group $G$ one encounters 
conditions like:
\begin{equation}
 |g(s_1,s_2,s_3,s_4);(g^{-1}\Lambda g)_{ij}\rangle\big|_{\text{GSO-proj.}}=
 |s_1,s_2,s_3,s_4;\Lambda_{ij}\rangle\big|_{\text{GSO-proj.}}\qquad \forall
 g\in G
\end{equation}
This will in general reduce the number of CP-dofs such that the gauge group
gets broken. In addition many zero-modes $|s_1,s_2,s_3,s_4\rangle$ will be
projected out. Depending on the orbifold group $G$ this can result in a single
chiral fermion. We have illustrated the situation where an open string is stuck
to a singularity in figure \ref{openstringsingpic}.\footnote{Of course such a
  singularity can not exist in one dimension. In order to get chiral
  fermions in four space-time dimensions, the singularity has to be complex
  three-dimensional.}  
 
\chapter[Orbifolds]{\centerline{\label{orbifolds} Orbifolds}}
In this chapter we will introduce the notion of a  (string-theoretic)
orbifold. While we give several refenrences during this chapter,
the fundamental publications for string-theoretic orbifolds
are the two papers by Dixon, Harvey, Vafa and Witten
\cite{Dixon:1985jw,Dixon:1986jc}.
 
We will present  {\sl orientifolds} in the next chapter.
Orientifolds are  string theories with an
 orbifold group containing elements which interchange left- and right-moving
sectors of the theory. 

It is assumed that the reader is familiar with the basic concepts
 of string theory.
After giving the ideas of orbifold constructions we will present
the $d$ dimensional torus $T^d$ as a concrete example. The formally introduced
orbifold torsion can  be identified with the exponential
of closed NSNS two-form fluxes $B_{\mu\nu}$ in this case. 
We will explain the T-duality group $SO(2,2,\mathbb{Z})$ for the
two-torus, because we will use these results in later chapters.
A good review for T-duality is the report by 
Giveon, Porrati and Rabinovici \cite{Giveon:1994fu}.
Finally, we will present the asymmetric 
$T^4/\mathbb{Z}^L_3\times\mathbb{Z}^R_3$ orbifold in some detail.  
This introductory chapter on orbifolds is far from being
exhaustive. Even though the notion of an orbifold is introduced,
 it is  impossible to enter into the details. This chapter  
is meant as a tool  to  understand 
 the concrete models which are presented in the following chapters,
especially chapter \ref{ncg} and \ref{z4}.  

\section{General construction of orbifolds\label{orbifoldsintro}}
Compactifications in superstring theory are usually of the form:
\begin{equation}\label{orbi1} 
 {\cal M} = \mathbb{R}^{(1,9-d)}\times {\cal X}_{d} 
\end{equation}
Whereas  $\mathbb{R}^{(1,9-d)}$ is the flat Minkowski space, ${\cal X}_{d}$
is a small $d$ dimensional, compact space. 
Even though an orbifold can be of any dimension we 
will concentrate on the dimension $d\leq 6$ case since it 
seems to be the most relevant
one for superstring compactifications to $10-d$ dimensional space-time.
In common orbifold constructions ${\cal X}_{d}$ is obtained
as a quotient of a manifold ${\cal B}$ by a group G acting in a
discrete way on ${\cal B}$:
\begin{equation}\label{orbidef1}
 {\cal X}_{d} = {\cal B}/G
\end{equation}
String theory is usually defined on spaces admitting a metric. Especially this
is the case for $\cal B$. As the metric appears in several quantities like
the Hamiltonian it is an essential structure of the theory. We 
require it to be invariant under the action of $G$.
In order for $G$ to be a symmetry of the theory on $\cal B$ we require that
all physical quantities like transition amplitudes and especially the 
Hamiltonian stay invariant under $G$.

As G can admit fixed-points in ${\cal B}$ (more generally:
fixed-sets) the orbifold might get singular
at these points. In going from the geometrical space to string theory one
is especially interested in the Hilbert space of the string theory living
on ${\cal X}_{d}$, or more precisely, in the Hilbert space associated with
${\cal M}$  in \eqref{orbi1} (We call this Hilbert space 
${\cal H}_{\cal X}$). The states in ${\cal H}_{\cal X}$ can partially be 
obtained by projecting on the $G$-invariant subspace of ${\cal H}_{\cal B}$ 
(${\cal H}_{\cal B}$ the Hilbert space of the string-theory on $\cal B$). 
 It turns out that there are additional states in the Hilbert space coming
from so called {\it twisted sectors} ${\cal H}_g$, $g\in G$ which form
 subspaces of ${\cal H}_{\cal X}$. These states stem from closed-strings
which are closed on $\cal X$ but on $\cal B$ only by an element $g$ of $G$:
\begin{equation} \label{twisted1}
X(\tau,\sigma+1) = g X(\tau,\sigma),\quad g\in G
\end{equation}

If the general solution of the equations of motion (eoms)
 for  $X$ on the space $\cal B$ is
known, it is often quite easy to implement the modified (i.e.\ twisted) 
boundary (or: periodicity) condition \eqref{twisted1}. After quantizing the
fields in this new sector, one can construct the resulting Hilbert space 
by known methods. Especially, one has to ensure that the states in the
${\cal H}_g$ sectors are invariant under all $h\in G$. A more detailed 
investigation reveals that states in ${\cal H}_g$ have to be invariant only 
 under the centralizer $C$ of $g$ ($h\in C \Leftrightarrow hg = gh$). 
As the information on the particle spectrum is encoded in the partition 
function, this quantity is extremely important. The perturbative 
spectrum
is encoded in the one-loop 
partition function.
The latter enters the one-loop vacuum amplitude as an integrand. 
For closed oriented strings the one-loop amplitude is the {\it torus amplitude}.

\begin{figure}
   \setlength{\unitlength}{0.1in}
   \begin{picture}(50,30)
   \put(10,0){
    \begin{picture}(50,30)
     \SetFigFont{14}{20.4}{\rmdefault}{\mddefault}{\updefault}
      \put(0,0){\scalebox{0.5}{\includegraphics{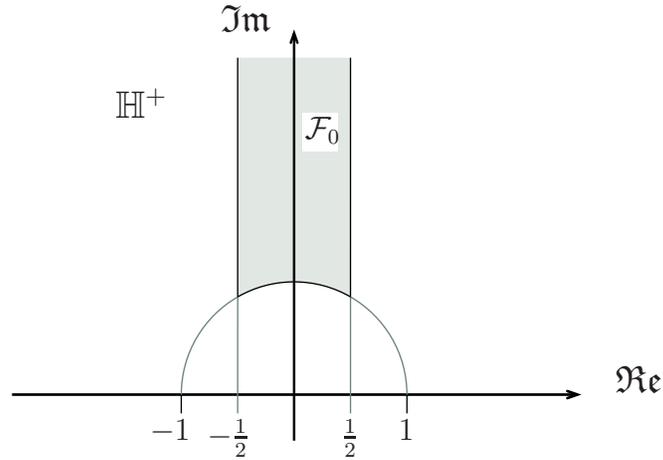}}}
     \put(11,21.5){$\mathfrak{Im}$}
     \put(31,2.6){  $\mathfrak{Re}$}
      \put(5.5,17){$\mathbb{H}^+$}
     \SetFigFont{12}{20.4}{\rmdefault}{\mddefault}{\updefault}
     \put(7.3,2.2){\makebox(1,-1)[tl]{$-1$}}
     \put(10.3,2.2){\makebox(1,-1)[tl]{$-\tfrac{1}{2}$}}
     \put(20.3,2.2){\makebox(1,-1)[tl]{$1$}}
     \put(17.3,2.2){\makebox(1,-1)[tl]{$\tfrac{1}{2}$}}
     \put(15.4,15.9){${\cal F}_0$}
    \end{picture}
   }
   \end{picture}
 \caption[Fundamental region ${\cal F}_0$ of the complex structure]
         {\label{fundregion1} The upper half plane $\mathbb{H}^+$ and the
          fundamental region ${\cal F}_0$ of the complex structure.
         }
\end{figure}

The torus amplitude can be  written as a path-integral with integration
over  fields of definite periodicity. Equivalently it can 
be calculated in the operator formalism 
as a trace over states corresponding to these periodicities in the $\sigma$ 
direction and trace insertions corresponding to the $\tau$ (world sheet time)
direction.\footnote{No trace insertion corresponds to periodicity of bosonic
  fields in the time (or: $\tau$) direction and {\sl anti}-periodicity
 of fermionic fields in this direction.}
 The torus amplitude (including the world sheet fermions) takes
then the form:
\beqn\label{torusamp}
T=V_{10}\int_{\cal F} \frac{d^2\tau}{4\mathfrak{Im}\tau}\int\frac{d^{10}p}{(2\pi)^{10}}
\Tr  {\bf P}_{\text{GSO}} (-1)^F q^H\bar{q}^{\widetilde{H}}
\eeqn 
$\tau$ is the modular parameter of the torus, $q=\exp(i2\pi\tau)$ and $\cal F$ 
is one fundamental region of the torus. $V_{10}$ denotes the ten dimensional 
regulated space time volume and 
${\bf P}_{\text{GSO}}=\tfrac{1}{2}\big(1+(-1)^f\big)$ the GSO 
projection ($f$ the world sheet fermion number). $F$ is the space time Fermion
number ($(-1)^F = -1$ in the RNS and NSR sector, otherwise $=1$). The trace in
\eqref{torusamp} is over the world sheet bosonic and over the fermionic
sector. The fermionic sector divides into a Neveu-Schwarz (NS) 
sector (corresponding to world
sheet fermions anti-periodic in $\sigma$) and a Ramond (R)
 sector (corresponding to periodic fermions).  
\begin{figure}
 \setlength{\unitlength}{0.1in}
  \begin{picture}(50,30)
   \SetFigFont{14}{20.4}{\rmdefault}{\mddefault}{\updefault}
   \put(0,0){\scalebox{0.5}{\includegraphics{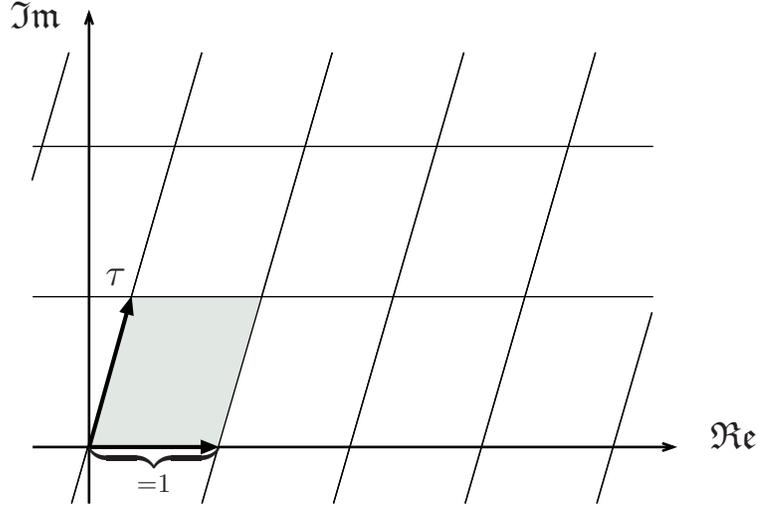}}}
   \put(9,25){$\mathfrak{Im}$}
   \put(45,3){  $\mathfrak{Re}$}
   \put(13.15,3.){\makebox(0,0)[tl]{$\underbrace{\rule{17.1mm}{0mm}}_{=1}$}}
   \put(14,11.5){$\tau$}
  \end{picture}
\caption[Torus lattice $\Gamma$ with complex structure $\tau$]
        {\label{twotorus} The Torus lattice $\Gamma$ with complex 
          structure $\tau$.
          One fundamental region of the two-torus is shaded.
        }
\end{figure}
Integrating in \eqref{torusamp} over $p^0$ and one component of
 momentum $p^i$ (which are part of $H+\widetilde{H}$)
 leads to an additional factor 
$(\al\mathfrak{Im}\tau)^{-1}$.\footnote{To regularize the momentum integral one 
has to perform a Wick rotation.}
We notice that the resulting measure 
$\big(4\al(\mathfrak{Im}\tau)^2\big)^{-1}d^2\tau$ 
is invariant  under modular
transformations $\tau\rightarrow\tfrac{a\tau+b}{c\tau+d}$ with $\bigl(
\begin{smallmatrix} a& b \\ c & d\end{smallmatrix}\bigr)\in SL(2,\mathbb{Z})$.
The torus is defined as the orbit space of a two dimensional lattice
acting additively on $\mathbb{C}\simeq\mathbb{R}^2$ (cf.\ figure
 \ref{twotorus}): 
\begin{align}\label{twotorusformula}
T^2_\tau &\equiv\mathbb{C}/\Gamma & \Gamma_\tau&= \left\{
             m+n\cdot\tau \,|\, m,n\in \mathbb{Z} \right\}
\end{align}
$g\in SL(2,\mathbb{Z})$ acts on $\tau$ as described above, or equivalently
on a vector $\vec{v}=(m,n)^T\in\Gamma$ as matrix multiplication from the
left by a matrix described above. It is therefore obvious that $\Gamma$
and as a consequence $T^2_\tau$ is invariant under $SL(2,\mathbb{Z})$.
This modular invariance should also be reflected in the torus partition 
function. The torus amplitude is modular invariant if the integrand of 
the $d^2\tau$-integration (the trace including the remaining momentum integral)
is modular invariant. Since the integrand is 
essentially the partition function, its
modular invariance is commonly referred to  as 
{\it modular invariance} of the partition function. With the explicit 
modular invariance of the integrand one is free to choose a fixed 
fundamental
region $\cal F$ which under the action of $G$ is mapped to the complete
upper half plane $\mathbb{H}^+$. The choice ${\cal F}_0=\{|\tau|>1,\,
|\mathfrak{Re}\tau| < 1/2,\,\mathfrak{Im}\tau>0 \}$ (cf.\ 
fig.\ \ref{fundregion1}) eliminates
explicitly potential divergencies in the region $\tau\rightarrow 0$.
This is in contrast to field theory where this limit corresponds to
UV-divergencies. Therefore modular invariance is essential for the finiteness
of string theory. Anomalies in field theory have several interpretations.
They signal a breakdown of classical symmetries at the quantum level. 
Gauge symmetries in field theory play an important role in decoupling 
unphysical states in physical quantities like transition amplitudes.
It can be shown that in order to decouple unphysical states (i.e.\ unphysical
vertex operators) in string theory, modular invariance is needed. 
It ensures also 
the absence  of anomalies in the low energy effective field theory limit of
 the corresponding string theory. In field theory anomalies in gauge symmetries
ruin the renormalizibility of a theory. Therefore modular invariance in
string
theory is intimately connected to the finiteness of the theory, the absence
of anomalies and the decoupling of unphysical states.

 In orbifolds the torus vacuum amplitude \eqref{torusamp} gets modified. 
One has to sum up the traces over sectors ${\cal H}_g$ representing
states with $g$-twisted boundary conditions \eqref{twisted1} in 
the $\sigma$-direction. One will also have to insert projectors
of the form
\begin{equation}
\frac{1}{|G|}\sum_{h\in G} h
\end{equation}
in the trace over states in ${\cal H}_g$, thereby projecting onto $G$-invariant
states. 
An insertion of $h$ into the trace corresponds to integrating over 
fields with periodicity 
\begin{equation} \label{twisted2}
X(\tau+1,\sigma) = h X(\tau,\sigma)
\end{equation} 
in the path-integral formalism.
This only makes sense for $h$ and $g$ commuting (i.e.\
$hg=gh$). 
Commonly a sector of the partition function that corresponds to fields with
periodicity \eqref{twisted1} and \eqref{twisted2} is represented by:
\begin{equation} \label{twisted3}
{ h}\underset{\mbox{$ g$}}{\square}
\end{equation}
A modular transformation of the parameter 
$\tau\rightarrow\tfrac{a\tau+b}{c\tau+d}$ has the same effect
as transforming
\begin{equation} \label{twisted4}
 { h}\underset{\mbox{$ g$}}{\fbox{\phantom{\rule{1.8ex}{1.8ex}}}}\rightarrow
 {\textstyle h^a g^b}\underset{\mbox{$\textstyle h^c
    g^d$}}{\fbox{\phantom{\rule{1.8ex}{1.8ex}}}},\qquad \text{for }hg=gh
\end{equation}
with $\tau$ unchanged. If one knows all trace insertions in the
$\sigma$ untwisted sectors, i.e.\ all contributions 
${ h}\underset{\mbox{$ 1$}}{\square}$ one can construct a big part of the
  twisted 
sectors ${ h^n}\underset{\mbox{$h^m$}}{\square}$ by applying eq.\ 
\eqref{twisted4}. This proves extremely useful in the construction
of so called  {\it left-right asymmetric orbifolds}, where solving of the
boundary conditions \eqref{twisted1} is not problematic.
This is due to the fact that the asymmetric action on the center of mass
(com.) coordinate
of the string is not well defined.  
World sheet fermions which correspond to anti-commuting
fields come in four different types for each $h$, $g\in G$:
The NS fermions have an additional twist of $-1$ in the $\sigma$ direction and
the $(-1)^f$ trace insertion of the GSO projection corresponds to an 
additional twist of $-1$ in the world sheet time direction that is also 
denoted by $\tau$.   
The whole one-loop partition function of the closed-string sector 
is then given by:
\begin{equation}\label{partittionf1}
Z_{\cal X}(q,\bar{q}) = \frac{1}{|G|}\sum_{\substack{h,g\in G \\ hg =gh}}
       { h}\underset{\mbox{$ g$}}{\square}
\end{equation}
Although this expression is formally modular invariant there are some 
subtleties. They appear especially in so called asymmetric orbifolds.
Eq.\ \eqref{orbidef1} defines an orbifold as a geometric space. 
 Typically a string theory admits more symmetries than the 
background on which the string propagates. As the string splits into left-
and right-moving parts, a symmetry can interchange these parts (like in 
orientifolds) or act differently on left- and right-movers. A 
generalized orbifold group might also contain elements which do both.
It has been observed that in asymmetric orbifolds the naive partition
function \eqref{partittionf1} might be ill defined. If certain conditions
are not fulfilled\footnote{These obstructions are described for instance in
\cite{Vafa:1986wx}.} a sector which should be transformed
 to an equivalent sector by some element of the modular group, might however
gain  a non-trivial phase. (The modular group is represented only projectively
on the partition function.)
One remark is in order: If one wants to extract the spectrum from 
$Z(q,\bar{q})$ one still has to impose the condition that the number of
left-moving excitations equals the number of right-moving excitations
($N=\widetilde{N}$). If $G$ is a product of groups, i.e. $G=G_1\times G_2$,  
$Z_{\cal X}$ in eq.\ \eqref{partittionf1} is not the only modular
invariant partition function.
In general $Z_{\cal X}$ can split into  sectors which are not related 
by modular transformations. These sectors are allowed to aquire $U(1)$ phases
$\epsilon(g,h)$ where $g$ is the twist in the $\tau$-, 
$h$ the twist in the $\sigma$-direction. However higher loop consistency of
string theory puts further constraints (cf.\ \cite{Vafa:1986wx}) (in form
of co-cycle conditions) on $\epsilon(c,d)$:
\begin{align} \label{cocycle1}
 \epsilon(g_1g_2,h) &= \epsilon(g_1,h)\epsilon(g_2,h) \\\label{cocycle2}
 \epsilon(g,h)      &= \epsilon(h,g)^{-1} \\\label{cocycle3}
 \epsilon(g,g)      &= 1 
\end{align}
Imposing $\epsilon(h,g)$ to be invariant under the modular transformation
 $\epsilon(h,g)\rightarrow\epsilon(h^ag^b,h^cg^d)$ the sectors 
${\scriptstyle h}\underset{{ g}}{\square}$ in the partition function 
\eqref{partittionf1} get multiplied by $\epsilon(h,g)$.
The phase $\epsilon$ is commonly called {\it discrete torsion}.

 We will now consider toroidal
compactifications which are the starting manifold
 $\cal B$  (eq.\ \eqref{orbidef1}) for so called toroidal orbifolds. Toroidal
orbifolds and compactifications 
play an essential role in chapters \ref{ncg}-\ref{z4}.

\section{\label{toruscomp}Torus compactification as an orbifold}
If one wants to build orbifolds which descent form string theory on 
flat ten dimensional Minkowski space (with constant NSNS
two-from potential $B$),
 where the string can be
explicitly quantized, $G$ is allowed to be a subgroup of the euclidian
group acting on the base space ${\cal B}=\mathbb{R}^{d}$ 
(cf.\ eq.\ \eqref{orbi1}). This 
group leaves the Hamiltonian of the string theory on $\cal B$ invariant.
As the Hamiltonian on $\cal B$ splits into independent left- and right-moving
parts (in light cone gauge: 
$H_{\text{bos.\ , L/R}}= \left\|\partial_\pm X\right\|_{\text{transv.}
}^2+ a_{\text{L/R}}$) one can mod out
independent subgroups of $\text{Euc}(\mathbb{R}^{d})$ in the left- and
 right-moving
sector of the theory. However in constructing the twisted sector of an
asymmetric orbifold one  faces the question what the fixed points 
are. We will discuss this topic in section \ref{toroidorbs}.
As toroidal orbifolds (i.e.\ ${\cal X}_{d} = T^{d}/G$ with $T^{d}$ a 
flat $d$ dimensional torus) are of special interest, we will first consider
strings living on 
\begin{equation}
 {\cal M} = \mathbb{R}^{(1,9-d)}\times T^{d}
\end{equation}
The bosonic string action on this background has the form:
\begin{equation}
S_{\text{bos}}=-\frac{1}{4\pi \alpha^\prime} 
                         \int_{\cal{M}} d^2\sigma \,
                            \big( \partial_\alpha X^\mu \partial^\alpha X_\mu
                           -  B_{\mu\nu} \epsilon^{\alpha\beta}
                                   \partial_\alpha X^\mu
                                         \partial_\beta X^\nu\big)
\end{equation} 
We note that the equations of motion
of a closed-string are not affected by  a constant $B$-field 
as the variation of the action w.r.t.\ $X^\mu$ only contributes a boundary
piece which is assumed to vanish. (The $\tau\rightarrow\pm\infty$ pieces
of the boundary for an infinite cylinder are assumed to give no contribution. 
Quantizing the closed-string on a torus circumvents this problem because
the torus has no boundary.) However the canonical momentum gets modified
by the constant $B$-field:
\begin{equation}\label{canmomcl}
 P^\mu (\tau,\sigma) =\frac{\partial}{\partial \Dot{X}_\mu} L(X,\partial X)
                =
 \frac{1}{2\pi\alpha^\prime}\bigl( 
                          \Dot{X}^\mu  + B^\mu_{\phantom{\mu}\nu}
                                                  X^{\prime\,\nu} 
                           \bigr) 
\end{equation}
The integrated momentum $\int d\sigma P^\mu$ is $\tau$-independent
which follows from the equation of motion and the fact that 
$\partial{\cal M}=\emptyset$.
Since
\begin{align} \label{dtorusdef}
 T^{d} &= \mathbb{R}^{d}/2\pi\Gamma \\ \nonumber
 \Gamma &=\big\{ n^i e_i| \; n^i \in\mathbb{Z},
 \; e_i,\;i=1\ldots d\,\text{ a fixed basis of }\mathbb{R}^{d} \big\}
\end{align}
the torus is an orbifold itself, with a discrete, but fix-point free acting
group $G=\Gamma\simeq\mathbb{Z}^n$. As $\Gamma$ acts additively the $g=n^i$
twisted sector is characterized as follows:\footnote{$X$, $e_i$, 
$P$ and $\Pi$ should be understood as vectors, not just as numbers.}
\raisebox{0.8ex}{\tiny\negphantom{\;\;}, }\footnote{Here we set the closed-string length to
$2\pi$ in world sheet coordinates, because this choice seems to be in common
 use for concrete mode expansions. The periodicities in formul\ae\, like
\eqref{twisted1}, \eqref{twisted2} then changes from $1$ to $2\pi$.}
\begin{equation} \label{torusbdy1}
 X(\tau,\sigma+2\pi)= 2\pi n^i e_i + X(\tau,\sigma)
\end{equation}
This sector is usually called the {\it winding sector} since a string in
a twisted sector with $g=n^i$ corresponds to a string winding $n^i$ times 
around the $i$\raisebox{1ex}{\tiny th} one-cycle.
A solution in flat Euclidian space for the $d$ $X^j(\tau,\sigma)$ 
coordinates 
involved in \eqref{torusbdy1} has an oscillator part which is 
unchanged compared to the $2\pi$ periodic sector. In addition a piece linear 
in $\sigma$ is needed to accomplish equation \eqref{torusbdy1}. Altogether the
bosonic string coordinate in the $n^i$-twisted sector takes the following
form:
\begin{equation}\label{closstring1}
 X(\tau,\sigma) = x +\sqrt{2\al}
           p\cdot\tau+n^ie_i\cdot\sigma
           + i\sqrt{\frac{\al}{2}}
           \sum_{k,\in \mathbb{Z}^\ast} 
           \Big(\frac{\alpha_k}{k}e^{ik(\sigma-\tau)} 
           +\frac{\Tilde{\alpha}_k}{k}e^{-ik(\sigma+\tau)} \Big) 
\end{equation}
Whereas the separation into left- and right-movers for the oscillators
in the above expression is completely obvious, we will have a closer look
at the parts linear in $\sigma$ and $\tau$. The condition that momentum 
states are
invariant under $s^i$, i.e.\ \footnote{By acting with $s^j$ we mean a
  translation by $2\pi s^je_j$. This condition puts no constraints on
the oscillator part of the state (both the bosonic and fermionic) since
$n^i$ acts trivially on the oscillators.}
\begin{equation}
 s^i\ket{p} = \ket{p}
\end{equation}
puts constraints on the allowed spectrum for $p$.
With the momentum \eqref{canmomcl}
one finds the following commutators for the modes in 
\eqref{closstring1}:\footnote{$G^{ij }$ is the metric of the torus dual
to $T^d$.}
\begin{align}\label{commutrel1}
 [x^i,p^j] &= i\sqrt{\alpha^\prime/2} G^{ij}& 
 [\alpha^i_l,\alpha^j_m] &=
    [\tilde{\alpha}^i_l,\tilde{\alpha}^j_m]=l\cdot\delta_{l+m,0}G^{ij}
\end{align}
Since\footnote{$B_{ij}\equiv e_i^\mu B_{\mu\nu}e_j^\nu$} 
$i\Pi_j\equiv\int d\sigma P_j=i\bigl(\sqrt{2/\alpha^\prime}G_{jk}p^k
+(1/\al)B_{jk} n^k
\bigr)$
is the generator of
translations in the $e_j$ direction, $s^j$ acts
as:\footnote{The metric $G_{ij}$ of $T^d$ which is dual to $G^{ij}$ satisfies: 
$G^{ij}G_{jk}= \delta^i_k$.}
\begin{equation}
 s^j\ket{p} = \exp\bigl(i2\pi s^j\Pi_j\bigr)\ket{p}
\end{equation} 
Therefore $s^j$-invariant states have to fulfill:\footnote{The
 following  product $\vec{s}\cdot\Pi$ is defined by
 $(\vec{s}\cdot\Pi)^\mu\equiv\sum_j s^j \Pi^j e_j^\mu$
  and $\Pi_j=G_{jk}\Pi^k$.}
\begin{equation}
 \vec{s}\cdot\Pi\in \Gamma^\ast, \quad 
 \Gamma^\ast=\big\{m_i e^i | \; m_i \in\mathbb{Z},
 \; e^i e_j=\delta^i_{\phantom{i}j},
 \;i,j=1\ldots d \big\}
\end{equation}
$\Gamma^\ast$ is the lattice dual to $\Gamma$ in the sense that its elements
$\vec{v}^\ast$ have integer scalar products with vectors $\vec{w}\in\Gamma$ and
that the fundamental cells have inverse volumes. Invariance of the states
under the full translation group $\Gamma$ requires of course that 
$\Pi\in \Gamma^\ast$. 
With this information the $p$'s in the $n^i$-twisted and $s^j$-invariant 
sector of
eq.\ \eqref{closstring1} can be expressed as:
\begin{equation}\label{latticemom1} 
 p(\vec{m})_{\vec{s},\vec{n}}=e_kp^k =  e_k \sqrt{\frac{\al}{2}}
                 \Bigl( G^{kj} \frac{m_j}{ s^j}
                       - \frac{B^k_{\phantom{k}i}}{\al}n^i\Bigr),
                   \qquad \vec{m}\in\mathbb{Z}^d
\end{equation}
The Hamiltonian of the theory is not explicitly $B$ dependent when expressed
in terms of $\dot{X}$ and $X^\prime$, or equivalently in terms of 
$\partial_\pm X$:
\begin{equation}\label{clhamilton1}
H_{\text{bos}}= \frac{1}{4\pi\al} \int d\sigma \,
   \big(\dot{X}^2 + X^{\prime\, 2}\big)
=\frac{1}{2\pi\al}\int d\sigma  \,
          \Big(\bigl(\partial_+ X\bigr)^2+\bigl(\partial_- X\bigr)^2\Big)
\end{equation}
where we have defined:
\begin{equation}\label{partialdef}
\partial_\pm \equiv \tfrac{1}{2}\big(\partial_{\tau}\pm\partial_\sigma\big)
 \end{equation}
The bosonic oscillator part of $H$ is:\footnote{The dot product
is meant to be the product w.r.t.\ the (dual) metric  $G_{ij}$}
\begin{equation}\label{clhamiltonosc2}
 \begin{aligned}H_{\text{bos, osc}}= H+\widetilde{H}&=
  \frac{1}{2}\sum_{k\in \mathbb{Z}^\ast} (\alpha_{-k}\cdot\alpha_k+
  \tilde{\alpha}_{-k}\cdot\tilde{\alpha}_k )\\ 
  &=\sum_{k\in \mathbb{N}^\ast} (\alpha_{-k}\cdot\alpha_k- a_\text{bos}
    + \tilde{\alpha}_{-k}\cdot\tilde{\alpha}_k 
         -\tilde{a}_\text{bos})
 \end{aligned}
\end{equation}
with $a$ the normal ordering constant which is $1/24$ for each transverse
bosonic coordinate in light cone gauge. The oscillator part of the Hamiltonian
is not affected by a constant $B$. It is the same as in the non-compactified
theory.
 Requiring that the states $\ket{p}$ to be 
invariant under the whole lattice group $\Gamma$ restricts the $s^j$
in eq.\ \eqref{latticemom1}:
 $s^j=1\,\forall j$. 
We call the $\vec{m}$ excitations the
{\it Kaluza Klein} (KK) modes because they correspond to
the quantized momentum modes of a point particle compactified on
the torus $T^d$. In contrast to the oscillator part, the part linear in
$\tau$ and $\sigma$ gets affected by the torus compactification. We will
call it the {\it lattice part} since it depends on $\Gamma$. 
With the definition
\begin{equation}
 \begin{aligned}\label{momentapm}
  p(\vec{m},\vec{n})_\pm 
  &\equiv\frac{1}{\sqrt{2}}p(\vec{m})_{s^j=1,\,\vec{n}}
    \pm \frac{n^ie_i}{2\sqrt{\al}}\\
  &=\frac{e^k}{2}\biggl(\sqrt{\al} m_k
    + \frac{1}{\sqrt{\al}}\bigl(\pm G-B\bigr)_{kj}
    n^j\biggr)
 \end{aligned} 
\end{equation} 
the lattice part of the bosonic field can be rewritten:
\begin{equation}
  X_{\Gamma}(\tau,\sigma) = x +\sqrt{\al}
           \bigl(p_+\cdot(\tau+\sigma) 
               + p_-(\tau-\sigma)
           \bigr)
\end{equation}
The lattice Hamiltonian $H_\Gamma$ takes the form:
\begin{equation}
 \begin{aligned}\label{latticeham1}
 H_{\Gamma}  
 &= p(\vec{m},\vec{n})_+^{\phantom{+}2}
  + p(\vec{m},\vec{n})_-^{\phantom{-}2} \\
  &=\frac{1}{2}
   \begin{pmatrix}
    m_i , n^j
   \end{pmatrix}
 \begin{pmatrix}
 \al G^{ik}  & 
    -B^i_{\phantom{i}l} \\
     B_j^{\phantom{j}k}
   & \frac{1}{\al}\bigl(G - B^2 \bigr)_{jl}
 \end{pmatrix}
   \begin{pmatrix}
    m_k \\ n^l
    \end{pmatrix}
 \end{aligned}
\end{equation} 
and
\begin{equation} \label{lrmomenta}
p(\vec{m},\vec{n})_\pm^{\phantom{+}2}
  =\tfrac{1}{4}
   \begin{pmatrix}
    m_i,n^j
    \end{pmatrix}
 \begin{pmatrix}
    \al G^{ik}  & \bigl(  \pm G- B\bigr)^i_{\phantom{i}l} \\
    \bigl(\pm G + B\bigr)_j^{\phantom{j}k}  &
                  \frac{1}{\al}\bigl(G - B^2 \bigr)_{jl}
 \end{pmatrix}
   \begin{pmatrix}
    m_k \\ n^l
    \end{pmatrix}
\end{equation}
As the splitting of the linear part of $P(\tau,\sigma)$ into
 $p(\vec{m},\vec{n})_\pm$ is unambiguous we can  embed naturally
  $p(\vec{m},\vec{n})$ into a $2d$ dimensional lattice
 $\Gamma^{(d,d)}$ by the map:\footnote{The product structure
$\Gamma^d\times\Gamma^{d\,\ast}$ should be understood set theoretically.
It is not of physical relevance because the Hamiltonian \eqref{latticeham1}
couples vectors in both lattices.}
\begin{equation}
 \begin{array}{ccc}
  \Gamma^d\times\Gamma^{d\,\ast} &\stackrel{\Upsilon}{\longrightarrow} 
        &\Gamma^{(d,d)} \\
  (n^ie_i,m_j e^j)            &\longmapsto &
     \bigl(p(\vec{m},\vec{n})_+,p(\vec{m},\vec{n})_-\bigr)  
 \end{array}
\end{equation}
$\Upsilon$ has the following matrix representation:
\begin{equation}\label{ups}
 \Upsilon
  =\frac{1}{2}
  \begin{pmatrix}
     \sqrt{\al}G_k^{\phantom{k}i}
     & \frac{1}{\sqrt{\al}}\bigl( G- B\bigr)_{kj} \\
     \sqrt{\al}G_l^{\phantom{k}i}  
     & -\frac{1}{\sqrt{\al}}\bigl(G +  B\bigr)_{kj}
  \end{pmatrix}
 \end{equation}
For later use in asymmetric orbifolds we calculate the inverse of $\Upsilon$:
\begin{equation} \label{upsinv}
  \begin{pmatrix}
    m_{i} \\ n^j
  \end{pmatrix}
  =
   \Upsilon^{-1}
   \begin{pmatrix}
    ( p_+) \\ (p_-)
   \end{pmatrix}
  =
   \begin{pmatrix}
    \frac{1}{\sqrt{\al}}\bigl(G +  B\bigr)_i^{\phantom{i}k}  
     & \frac{1}{\sqrt{\al}}\bigl( G- B\bigr)_i^{\phantom{i}l} \\
      \sqrt{\al}G^{jk}  & - \sqrt{\al} G^{jl}
  \end{pmatrix}
  \begin{pmatrix}
   ( p_+)_{k} \\ (p_-)_{l}
  \end{pmatrix}
\end{equation}
Besides a positive definite and non degenerate quadratic form given
by the Hamiltonian \eqref{latticeham1} we can define another 
non degenerate but non definite quadratic form of signature $(d,d)$
 which is of physical importance:
\begin{equation}\label{xidef1}
\frac{1}{2}\Xi\bigl(p(\vec{n},\vec{m})\bigr)
 \equiv H_{\Gamma\,+}-H_{\Gamma\,-} 
 =\begin{pmatrix}
    m_i ,\, n^j
   \end{pmatrix}
 \begin{pmatrix}
   0 & 
     \tfrac{1}{2}\delta^i_{\phantom{i}l} \\
     \tfrac{1}{2}\delta_j^{\phantom{j}k}  & 0
 \end{pmatrix}
 \begin{pmatrix}
  m_k \\ n^l
 \end{pmatrix} = m_k n^k
\end{equation}
As $\Xi$ naturally induces a nondegenerate metric (also denoted by
$\Xi$) of signature $(d,d)$ 
the $2d$ dimensional lattice is denoted by $\Gamma^{(d,d)}$:
\begin{equation}
\Xi\bigl(p(\vec{k},\vec{l}),\,p(\vec{n},\vec{m})\bigr)
 =\begin{pmatrix}
    k_i ,\, l^j
   \end{pmatrix}
 \begin{pmatrix}
   0 & 
     \delta^i_{\phantom{i}l} \\
     \delta_j^{\phantom{j}k}  & 0
 \end{pmatrix}
 \begin{pmatrix}
  m_k \\ n^l
 \end{pmatrix} 
\end{equation}
We note that lattice $\Gamma^{(d,d)}$ is self dual w.r.t.\ $\Xi$.
The scalar product $\Xi$ is clearly $\mathbb{Z}$-valued, and furthermore 
the norm $\Xi$ of a vector is an even number.  A lattice with this 
property is called 
called {\it even}. A part of the physical relevance of $\Xi$ is clear by
the definition \eqref{xidef1}: In  order to fulfill the physical 
state condition $H-\widetilde{H}\ket{\text{phys}}=0$  we require:
\begin{equation}
 H_\text{bos, osc}-\widetilde{H}_\text{bos, osc} 
  = -\frac{\Xi\bigl(p(\vec{n},\vec{m})\bigr)}{2}
   \qquad\text{on physical states}
\end{equation}
\subsection[Moduli-space of toroidal compactifications,
            T-duality group and symmetries] 
           {\centerline{
            Moduli-space of toroidal compactifications,} 
             \centerline{T-duality group and symmetries}
            }
For the modular invariance of the partition function $Z_{T^d}(q,\bar{q})$ the
above equality needs to hold only $\mod\mathbb{Z}$.
 Therefore the fact that
$\Gamma^{(d,d)}$ is even ensures the modular invariance of 
the partition function.  Since any other even self-dual lattice of signature
$(d,d)$ can be reached by performing an $O(d,d,\mathbb{R})$ rotation
on a given  lattice $\Gamma^{(d,d)}$, the moduli space should be
 locally isomorphic to this Lorentz group. However separate 
$O(d,\mathbb{R})$ rotations 
(which implicitly transform both the $d$-dimensional lattice $\Gamma^d$, i.e.\
the metric $G_{ij}$, as
well as the $B$-field) on the left- and the right-movers do not change
the spectrum (cf.\ eq.\ \eqref{clhamiltonosc2} and \eqref{latticeham1})
and are therefore (at this level, i.e.\
one-loop vacuum) physically irrelevant. These $O(d,\mathbb{R})$ rotations 
leave not only the spectrum, but also the mass of an individual state
$\ket{m,n}$ invariant. Furthermore there are rotations, which leave the
whole spectrum, but not necessarily the mass of the individual 
states $\ket{m,n}$ 
invariant, thereby leading to an equivalent theory (at this level again), 
too.  These are exactly the elements of $O(d,d,\mathbb{Z})$, the so called
{\it T-duality group} (or Target space duality group) of the $d$-torus.
 $O(d,d,\mathbb{Z})$ transformations
  only permute the basis vectors of $\Gamma^{d,d}$ (possibly changing the 
orientation of a given, ordered basis). Therefore the moduli space 
of toroidal compactifications with constant $G$ and $B$ takes 
the form:
\begin{equation}
{\cal M}_{T^d} \simeq\frac{O(d,d,\mathbb{R})}{O(d,\mathbb{R})
            \times O(d,\mathbb{R})\times O(d,d,\mathbb{Z})}
\end{equation}
It has been shown that the T-duality group $O(d,d,\mathbb{Z})$ has 
a well defined action on the oscillators $\alpha_k$, $\tilde{\alpha}_l$, too
(which respects the commutation relations \eqref{commutrel1}).
We shall mention that under world sheet parity $\Omega:\sigma\mapsto -\sigma$
which has the effect:
\begin{align}
\alpha_n &\stackrel{\Omega}{\longleftrightarrow} \tilde{\alpha}_n &
 \begin{pmatrix}
  m_k \\ n^l  
 \end{pmatrix}\stackrel{\Omega}{\longrightarrow}
 \begin{pmatrix}
   \delta_k^{\phantom{k}i} &  -2 B_{kj}/\al \\
     0 & -\delta^l_{\phantom{l}j}
 \end{pmatrix}
\begin{pmatrix}
   m_i \\ n^j  
 \end{pmatrix}
\end{align}
the scalar product $\Xi$ changes its sign. Even though the mass formula
\eqref{latticeham1} is invariant under the above transformation,
for $\Omega$ to be a symmetry (and not just a duality), $B$ is quantized 
(cf.\ \cite{Bianchi:1992eu},\cite{Kakushadze:1998bw}): 
\begin{equation}\label{quantizedB}
 B_{ki}/\al \in \frac{1}{2}\cdot\mathbb{Z}
\end{equation}  
such that the lattice $\Gamma^{(d,d)}$ is mapped to itself. The world sheet
parity $\Omega$ should not be confused with the following kind of 
$SO(d,d,\mathbb{Z})$ transformation:
\begin{equation}
\Theta:\,
\begin{pmatrix}
  m_k \\ n^l  
 \end{pmatrix}\longrightarrow
 \begin{pmatrix}
   \delta_k^{\phantom{k}i} &  \theta_{kj}\\
     0 & \delta^l_{\phantom{l}j}
 \end{pmatrix}
\begin{pmatrix}
   m_i \\ n^j  
 \end{pmatrix} 
 \, ,\quad \theta_{kj}\in \mathbb{Z}, \quad \theta=-\theta^T
\end{equation}
which is equivalent to shifting $B_{ij}/\al\rightarrow B_{ij}/\al-\theta_{ij}$.
Even though the spectrum is unchanged under $\Theta$, it is in general
not a symmetry of the theory since states $\ket{p}$ are  mapped
 to states of different masses. If
we want to mod out by the world sheet parity $\Omega$ by using this 
duality, we only need to distinguish the cases:
\begin{equation}
 B_{kl} \in \bigl\{0,\,\tfrac{1}{2}\bigr\},\qquad l,i=1\ldots d
\end{equation}
We will consider these constructions later (cf.\ section \ref{wsparityt2} and
 chap.\ \ref{ncg}-\ref{z4}).

The partition function is constructed according to eq.\ \eqref{partittionf1}.
The translation group $G\simeq\Gamma^d\simeq\mathbb{Z}^d$ 
is abelian but infinite. We have to be careful with the
regularization of the projector $P_{\Gamma^d}$. $V_d$ is the regularized $d$-dimensional
volume (cf.\ \eqref{torusamp}) 
while $N$ is the order of $\Gamma$, which is infinite as well, 
but the ratio is just the volume $V_{\Gamma}$ of the elementary $d$-cycle:
\begin{align} \label{torusproj1}
 P_{\Gamma^d}&=\frac{V_d}{N}
    \sum_{2\pi s^j\in\Gamma^d} \exp\bigl(i2\pi s^j\Pi_j\bigr)\\
 \label{torusproj2}&=V_{\Gamma}
               \sum_{2\pi s^j\in\Gamma^d}
           \exp\bigl(i2\pi s^j\bigl(\sqrt{2/\al}G_{jk}p^k
            +(1/\al) B_{jk}n^k\bigr)\bigr)\\
 \label{torusproj3}&= V_{\Gamma}
               \sum_{2\pi s^j\in\Gamma^d}
  \epsilon(s^j,n^k)\cdot\exp\bigl(i2\pi s^j\sqrt{2/\al}G_{jk}p^k\bigr) \\
  \label{torustorsion}
  \text{with }\epsilon(s^j,n^k)&\equiv\exp\bigl(i(2\pi/\al) s^j B_{jk}n^k\bigr)
\end{align}
\eqref{torusproj3} is just a rewriting of \eqref{torusproj2}. However
$\epsilon(s^j,n^k)$ is recovered as the discrete torsion introduced at the end
of section \ref{orbifolds}. 
Consistency condition \eqref{cocycle1} is fulfilled 
due to 
the defining properties of $\exp$. Taking into account that $B_{ij}$ 
is antisymmetric,  \eqref{cocycle2} and \eqref{cocycle3} are obeyed.
The integral $\int d^dp$ over the projector \eqref{torusproj1} 
restricts the {\sl canonical} momenta
 to lie on the dual lattice $\Gamma^{d\,\ast}$
whereas in the torsion form \eqref{torusproj3} the {\sl kinematical} momentum
$p$ is restricted to the dual torus lattice.
After we perform the $p$ integration,
the partition function for  the $d$ real bosons takes the form 
($\tau=\tau_1+i\tau_2$):
\begin{align}
 Z_{T^d}(q,\bar{q})
 &=\tr\bigl( q^{H_{\text{osc}}}+ \bar{q}^{\widetilde{H}_{\text{osc}}}\bigr)
 \cdot\sum_{\vec{n}\in\mathbb{Z}^d}\sum_{\vec{m}\in\mathbb{Z}^d} 
 q^{H_{\Gamma}}+\bar{q}^{\widetilde{H}_{\Gamma}} \\ \label{toruspartf1}
  &= |\eta(q)|^{-2d} \cdot \sum_{\vec{n},\,\vec{m}\in\mathbb{Z}^d} 
    \exp\bigl(-2\pi\tau_2(H_{\Gamma}+\widetilde{H}_{\Gamma})\bigr) 
    \exp\bigl(2i\pi\tau_1(H_{\Gamma}-\widetilde{H}_{\Gamma})\bigr)   
\end{align}
We will prove its modular invariance. We will first consider invariance
under $T:\,\tau\mapsto \tau+1$. From \eqref{Ttransf} (p.~\pageref{Ttransf})
we see that the oscillator part transforms trivially (because of $|\eta |^2$ 
which eliminates the twelfth root of unity). The shift of $\tau_1$
in the lattice part introduces a phase, that is however trivial as
 the lattice is even. 
To investigate  the transformation of $Z_{T^d}(q,\bar{q})$ under
 $S:\,\tau\rightarrow-1/\tau$ 
($\Rightarrow \tau_1\rightarrow -\tau_1/|\tau|^2$ and 
$\tau_2\rightarrow \tau_2/|\tau|^2$) we note 
that we can split the exponential of the lattice part as follows:
 \begin{align}
  \frac{\tau_2}{|\tau |^2} 2(H_{\Gamma}+\widetilde{H}_{\Gamma})&= 
   \frac{\tau_2}{|\tau |^2}\bigl((\vec{m}-B\vec{n})^T \al 
   G^\ast(\vec{m}-B\vec{n})
                      +\vec{n}\frac{G}{\al}\vec{n} \bigr)  \\  \label{trasc1}
 -i\frac{\tau_1}{|\tau |^2} (H_{\Gamma}-\widetilde{H}_{\Gamma})&=
 -i\frac{\tau_1}{|\tau |^2}\vec{n}\cdot(\vec{m} -B\vec{n})
\end{align}
In \eqref{trasc1} we have added a term which vanishes because of 
 $B$ being anti-symmet\-ric. Therefore we can apply the Poisson-resummation
 formula \eqref{poissonf1} (p.~\pageref{poissonf1}) for the $\vec{m}$ 
resummation. For the lattice part we obtain after this resummation:
\begin{align}
  \biggl(\frac{|\tau|^2}{\tau_2}\biggr)^{\tfrac{d}{2}}\sqrt{\frac{|G|}{\al}}
 \sum_{\vec{n}\in\mathbb{Z}^d}\sum_{\vec{w}\in\mathbb{Z}^d}
 e^{-i2\pi \vec{w}B\vec{n}}
 e^{\frac{\pi}{\tau_2}(\vec{n}+\tau_1\vec{w})^T \frac{G}{\al} 
  (\vec{n}+\tau_1\vec{w})
   +\tau_2 \vec{w} \frac{G}{\al}\vec{w} }
\end{align}
A second Poisson resummation (now in the opposite direction) with respect
to the $\vec{n}$ sum turns the above expression into:
\begin{align}
 {|\tau|^d}
 \sum_{\vec{v}\in\mathbb{Z}^d}\sum_{\vec{w}\in\mathbb{Z}^d}
 e^{i2\pi \tau_1\vec{v}\vec{w}}
 e^{-\pi\tau_2(\vec{v}-B\vec{w})^T \al G^\ast (\vec{v}-B\vec{w})
   +\tau_2 \vec{w} \frac{G}{\al}\vec{w} }
\end{align}
where we surpressed a vanishing expression $\propto i\pi\tau_1\vec{w}B\vec{w}$.
Taking into account that $|\eta (q)|^{2d}$ with $q=\exp(i2\pi (-1/\tau))$
equals $|\tau|^{-d}|\eta (q)|^{2d}$ with $q=\exp(i2\pi \tau)$ (cf.\
\eqref{Stransf}, p.~\pageref{Stransf}), we have
proven the modular invariance of a bosonic string compactified on a torus 
that is described by constant background fields $G$ and $B$.
 The fermionic
part is untouched by the torus compactification. This is due to the fact
that the world sheet fermions (in the RNS-formalism) are insensitive
to space-time translations. The complete partition function of the super string
is merely a product of the toroidal bosonic partition function times the
unchanged fermionic partition function.

 Before we present as a concrete example
the two-torus $T^2$ we want to mention that it can be
shown that the duality group $O(d,d,\mathbb{Z})$ of 
string theory compactified on $T^d$ is also preserved by 
string interactions and
at higher loop orders. By considering higher loop vacuum amplitudes
 it turns out that also the dilaton VEV is transformed under the duality
 group.
First work on the lattice $\Gamma^{(d,d)}$ has been done by Narain, Sarmadi
and Witten \cite{Narain:1986jj, Narain:1987am}. 
 Therefore $\Gamma^{(d,d)}$  is commonly 
called a {\it Narain lattice}. The construction  generalizes naturally
to heterotic compactifications. There the  Narain lattice 
$\Gamma^{(d+16,d)}$ is $2d+16$ dimensional and has the indicated signature.
\subsection{Compactification on \texorpdfstring{$T^2$}{T\texttwosuperior},
            T-duality and symmetries}
The main part of the work in this thesis consists of orientifolds
which are derived  either from:
\begin{equation}
\text{Type II on } T^2\times T^2 (\times T^2)
\end{equation} 
or from orbifolds of this special torus. Thus we will have a closer look on
string theory on $T^2$, especially the lattice part. 
The Narain lattice  $\Gamma^{(2,2)}$ of $T^2$ is four dimensional. 
Its Lorentz group $SO(2,2,\mathbb{R})\simeq SL(2,\mathbb{R})\times 
SL(2,\mathbb{R})$ has dimension $3+3=6$. Excluding world sheet parity for the
moment, which changes the sign of $\Xi$, we note that the T-duality group
is a semi-direct product of the normal subgroup 
$N=SL(2,\mathbb{Z})_1\times SL(2,\mathbb{Z})_2$ and an $H=\mathbb{Z}_2\times
\mathbb{Z}_2^\prime$ subgroup sharing only the identity with $N$:
\begin{equation}
 SO(2,2,\mathbb{Z})=\bigl(SL(2,\mathbb{Z})_1\times SL(2,\mathbb{Z})_2\bigr)
  \rtimes\bigl( \mathbb{Z}_2\times \mathbb{Z}_2^\prime\bigr)
\end{equation}
The (real two-dimensional)
 fundamental representation of $SL(2,\mathbb{Z})$ is equivalent to its
complex one-dimensional representation. We embed the two
$SL(2,\mathbb{Z})$ in the following way: $SL(2,\mathbb{Z})_1$ acts on 
$\tau\hookrightarrow SO(2,2,\mathbb{R})$ by:\footnote{$\tau$ parametrizes
$SL(2,\mathbb{R})/SO(2,\mathbb{R})$ and can therefore
be embedded into $SO(2,2,\mathbb{R})$.}
\begin{equation}\label{tautransf}
 \begin{array}{ccc}
  SL(2,\mathbb{Z})_1 \times SO(2,2,\mathbb{R}) 
  & \longrightarrow & SO(2,2,\mathbb{R}) \\
 \bigl(\begin{smallmatrix}
  a & b \\ c & d
 \end{smallmatrix}\bigr) (\tau) &\longmapsto & \frac{a\tau+b}{c\tau+d}
 \end{array}  \\
\end{equation}
with
\be \label{taudef}
\tau\equiv \tau_1+i\tau_2 \equiv \frac{G_{12}}{G_{11}}
                  + i  \frac{\sqrt{\det G}}{G_{11}}
\ee
The action of the second $SL(2,\mathbb{Z})$ is defined analogously
by the following embedding of the modular parameter 
$\rho\hookrightarrow SO(2,2,\mathbb{R})$:
\be  \label{rhodef}
\rho\equiv \rho_1+i\rho_2 \equiv B_{12}/\al + i\sqrt{\det (G/\al)}
\ee
The two $\mathbb{Z}_2$ subgroups are:
\begin{align} \label{Dduality}
 \mathbb{Z}_2 &=\big\{\id, D| 
                 \tau\stackrel{D}{\longleftrightarrow}\rho\big\}\\
 \mathbb{Z}^\prime_2 &=\big\{\id, R| 
                 (\tau,\rho)\stackrel{R}{\longleftrightarrow}
                 -(\bar{\tau},\bar{\rho}) \big\}
\end{align}
In order to discover the role of the subgroups in terms of their action
on the background fields we will rewrite the transformations.
The $SL(2,\mathbb{Z})_1$ is purely geometric. It just describes 
the change to a new basis of the torus lattice $\Gamma^2$. 
We define the  mass matrix $M^2$  by 
\begin{equation}\label{msquaremat}
 \frac{M^2}{2} \equiv \frac{1}{2}
 \biggl(
  \begin{smallmatrix}
  \al G^{ik}  & 
    - B^i_{\phantom{i}l} \\
      B_j^{\phantom{j}k}
   &\frac{1}{\al} \bigl(G -B^2 \bigr)_{jl}
  \end{smallmatrix}
 \biggr)
\end{equation}
Then $SL(2,\mathbb{Z})_1$ transforms $M^2$ under \eqref{tautransf}
in the following way (not
distinguishing lower and upper indices in $M^2$ this time):
\begin{equation}
M^2\mapsto S^T M^2S,\qquad 
 S=\biggl(\begin{array} {c|c}
     \begin{smallmatrix} 
      d & -c \\ -b & a
     \end{smallmatrix}  & 0 
    \\ \hline
     0 & 
     \begin{smallmatrix} \rule[0ex]{0.ex}{1.4ex} 
      a & b\\ c & d
     \end{smallmatrix}
   \end{array}
   \biggr)  ,\qquad 
     \bigl(\begin{smallmatrix} 
            a & b\\ c & d
           \end{smallmatrix}
     \bigr) \in SL(2,\mathbb{Z})
\end{equation}
where we have taken into account that a linear map transforms the 
dual basis with the transposed inverse.
An alternative expression of the mass\raisebox{1ex}{\tiny 2}
matrix \eqref{msquaremat} in terms of the parameters $\tau$
(called {\it complex structure}\,\footnote{The flat torus is a {\sl
    Calabi-Yau} space. 
     $\tau$ is called the complex structure as it parametrizes 
    the different complex structures of the 
     two-torus (cf.~fig.~\ref{twotorus}). 
    The origin of the name {\sl K\"ahler
    structure} comes from the observation  that $\rho$ can be interpreted as
     a complex $(1,1)$-form.  In CY-spaces $(1,1)$ forms correspond 
    to deformations of the
     {\sl K\"ahler structure} (that preserve the CY-property). This point
    is a bit subtle as the $(1,1)$ forms that describe the  CY-preserving 
    deformations of the metric are real. However they can be complexified
    by combining them with the  NSNS $B$-field (cf.\ \cite{Candelas1990}).})
    and $\rho$ (commonly denoted as
{\it K\"ahler structure}\,\addtocounter{footnote}{-1}\footnotemark) is (c.f.\
eq.\ \eqref{lrmomenta}):
\begin{gather}
 \frac{M^2}{2} =\bigl(p_{+}^2+p_{-}^2\bigr) \\
\begin{aligned}
 p_{+}^2 &= \frac{1}{4}\frac{1}{\tau_2\rho_2} 
            | (\tau m_1 -m_2) - \Bar{\rho}(n_1+\tau n_2) |^2 \\
 p_{-}^2 &= \frac{1}{4}\frac{1}{\tau_2\rho_2} 
            | (\tau m_1 -m_2) - \rho(n_1+\tau n_2) |^2 
\end{aligned}\qquad n_i,m_i \in \mathbb{Z}
\end{gather}
 The $SL(2,\mathbb{Z})_2$ is stringy.
 The generators have the following correspondences:
\begin{align}
  \rho&\rightarrow\rho +1 &\rho&\rightarrow -\frac{1}{\rho} \\
  B_{12}/\al&\rightarrow B_{12}/\al+1  
    &   B_{12}/\al &\rightarrow -\frac{\al B_{12}}{B_{12}^2+\det G}
   \\\nonumber
  & &\sqrt{\det (G/\al)}&\rightarrow \frac{\al\sqrt{\det G}}{B_{12}^2+\det G} 
   \\
  S&= \biggl( 
     \begin{array}{c|c}
      \mathds{1}_2
      &  
     \begin{smallmatrix} 
      0 & 1\\  
      \rule[-.45ex]{0.ex}{0.1ex} -1& 0
     \end{smallmatrix}
    \\ \hline
     0 &  \mathds{1}_2
   \end{array} \biggr)
  & S&= \biggl( 
     \begin{array}{c|c}
      0 &  \mathds{1}_2  
      \\ \hline   \mathds{1}_2 & 0     
   \end{array} \biggr)
\end{align}  
However, the $\det g=1$-condition ($g\in SL(\mathbb{Z})$) is not so easily 
imposed, so that we have presented only the action of the generators 
of the latter
$SL(2,\mathbb{Z})_2$ transformation on the background fields and not the 
action of a general element of $SL(2,\mathbb{Z})_2$.
$\rho\rightarrow\rho +1$ represents an integer shift on $B$.
$\rho\rightarrow -1/\rho$ corresponds to T-duality along both
directions of the torus. The two nontrivial elements $D$ and $R$ 
of the $\mathbb{Z}_2\times\mathbb{Z}^\prime_2$ subgroup are
 a stringy T-duality   respectively a reflection
in either the $e_1$ or $e_2$ direction:\footnote{$S_D$ represents usual
  T-duality along $e_1$ times a reflection in the same direction, while $S_R$ 
  is a reflection along $e_1$}
\begin{equation}
 \label{z2duality}
 \begin{aligned}
 &D & &R\\
   \frac{G_{12}}{G_{11}}&\leftrightarrow B_{12}/\al
   & G_{12}&\leftrightarrow -G_{12}  \\
  \frac{\sqrt{\det G}}{G_{11}}&\leftrightarrow\sqrt{(\det G)/\al} 
   & \qquad B_{12}&\leftrightarrow -B_{12} \\ 
  S_D&= \biggl( 
     \begin{array}{c|c}
      \begin{smallmatrix} 
      0 & 0\\  
      \rule[-.45ex]{0.ex}{0.1ex} 0& 1
     \end{smallmatrix} 
      &       
     \begin{smallmatrix} 
                                  -1 & 0\\  
      \rule[-.45ex]{0.ex}{0.1ex}   0 & 0
     \end{smallmatrix} 
    \\ \hline
       \begin{smallmatrix}  
      \rule[0ex]{0.ex}{1.4ex}      -1 & 0\\  
       \rule[-.45ex]{0.ex}{0.1ex}   0 & 0
      \end{smallmatrix}
      &
     \begin{smallmatrix} 
      \rule[0ex]{0.ex}{1.4ex}    0 & 0\\  
      \rule[-.45ex]{0.ex}{0.1ex} 0 & 1
     \end{smallmatrix}     
   \end{array} \biggr)
  & S_R&= \biggl( 
     \begin{array}{c|c}
      \begin{smallmatrix} 
      -1 & 0\\  
      \rule[-.45ex]{0.ex}{0.1ex} 0& 1
     \end{smallmatrix} 
      & 
      0     
      \\ \hline
       0 &
       \begin{smallmatrix}
         \rule[0ex]{0.ex}{1.4ex} -1 & 0\\  
       \rule[-.45ex]{0.ex}{0.1ex} 0& 1
      \end{smallmatrix}
      \end{array}
     \biggr)
 \end{aligned}
\end{equation}

\subsubsection{\label{enhsym} Points of enhanced symmetry in the moduli space 
$SO(2,2,\mathbb{R})$}
As toroidal orbifolds are obtained by modding out symmetries of the torus,
we are especially interested in points of the moduli space which are fixed
by certain elements of the T-duality group. In these cases the dualities
are enhanced to symmetries of the string compactification, so that they can
be modded out. In chapter \ref{z4}  we consider a $Z_4$ orbifold
which is obtained from a special $T^6$:
\be
\frac{T^2\times T^2 \times T^2}{\mathbb{Z}_4} 
\ee
The $\mathbb{Z}_4$  acts on each of the first two $T^2$s as
$\exp(i2\pi/4)$ and as $\exp(-i2\pi/2)$ on the last torus (written in terms
of complexified coordinates). 
The $\exp(i2\pi/4)$ rotation restricts the moduli in the following way:
\begin{align}\label{z4metric}
 \mathbb{Z}_4:\qquad
 \tau &=i &\Longrightarrow\;  G_{11}&=G_{22},& G_{12}&=0 
\end{align}
The corresponding matrix acting on $M^2$ (leaving it invariant) and 
consequently on the vectors of $\Gamma^{(d,d)}_{\mathbb{Z}_4}$ is:
\be
 S_{\mathbb{Z}_4} = \biggl( 
     \begin{array}{c|c}
      \begin{smallmatrix} 
      0 & -1\\  
      \rule[-.45ex]{0.ex}{0.1ex} 1& 0
     \end{smallmatrix} 
      & 
      0     
    \\ \hline
     0 &
     \begin{smallmatrix} 
      \rule[0ex]{0.ex}{1.4ex}0 & -1\\  
      \rule[-.45ex]{0.ex}{0.1ex} 1 & 0
     \end{smallmatrix}     
   \end{array} \biggr)
\ee
This means that the torus lattice has a basis consisting of two vectors of
equal but unrestricted length, making an angle of $\pi/2$.
The $\mathbb{Z}_2$ symmetry $\exp(-i2\pi/2)$ does not restrict the values
of $G$ and $B$. It multiplies all vectors in the Narain lattice by $-1$:
 $S_{\mathbb{Z}_2}=- \mathds{1}_4$.
In section \ref{asz3orbi} we will consider an orbifold where we divide out a
special four torus $T^2\times T^2$ by the product 
$\mathbb{Z}^L_3\times\mathbb{Z}^R_3$ with one $\mathbb{Z}_3$ only acting on
the left-moving degrees of freedom and the other only on the right-moving
part of the string. Since this direct product contains the symmetric 
$\mathbb{Z}_3$ as a subgroup with the generator acting as $\exp(i2\pi/3)$,
we will first look at the lattice having this symmetry:
\begin{align}\label{z3metric}
 \mathbb{Z}_3:\qquad\tau &=\frac{1}{2}+i\frac{\sqrt{3}}{2} 
    &\Longrightarrow\; G_{11}&=G_{22}, & G_{12}&=\frac{1}{2}G_{11}
\end{align}
This describes a lattice admitting a basis with vectors of equal length
and a mutual angle of $2\pi/3$. 
It is up to a scale factor  the root lattice
of the $SU(3)$ Lie algebra. The action of the symmetric $\mathbb{Z}_3$
is given by (in terms of the basis which was described above):
\begin{equation}\label{Z3action}
 S_{\theta} =S_{\mathbb{Z}_3} = \left( 
     \begin{array}{c|c}
      \begin{smallmatrix} 
      0& -1 \\
      1& -1
     \end{smallmatrix} 
      & 
      0     
    \\ \hline
     0 &
     \begin{smallmatrix} 
       \rule[0ex]{0.ex}{1.4ex} -1 & -1 \\
      1 & 0
     \end{smallmatrix}     
   \end{array} \right)
\end{equation}
We note that the two-torus described by \eqref{Z3action} 
admits in addition a $\mathbb{Z}_6$ symmetry, namely the geometric rotation
by $\pi/3$.
Since the asymmetric $\mathbb{Z}^L_3\times\mathbb{Z}^R_3$ can be
generated by the symmetric $\exp(i2\pi/3)$ and an element 
$\hat{\theta}$ rotating the left-movers by $\exp(i2\pi/3)$  and
the right-movers by the reversed angle, we will search for Narain
lattices $\Gamma^{(d,d)}$ admitting this latter symmetry.
Especially the associated matrix $S_{\hat{\theta}}$ acting on the $(m_i,n^j)$
has to have integer entries. In principle we could determine the
 form of $S_{\hat{\theta}}$
by mapping $(m_i,n^j)$ to $(p_+,p_-)$ (where the form of the asymmetric
$\hat{\theta}$ is explicitly known), performing the rotation $\hat{\theta}$ 
and mapping back to the $(m_i,n^j)$ basis. The map between the two basis 
is described by $\Upsilon$ and $\Upsilon^{-1}$ (eq.\ \eqref{ups} and 
\eqref{upsinv}):
\begin{equation}
 S_{\theta}(p_+,p_-) = \Upsilon^{-1}
     S_{\theta}(m,n)
      \Upsilon
\end{equation}
However, we will proceed differently. We know 
(from \eqref{momentapm}, \eqref{z2duality})
that at the self dual radius with vanishing $B$-field, the T-Duality $D$
from \eqref{z2duality} reflects the $X_2$ coordinate on the right-movers,
leaving the rest unchanged. By acting with $D^{-1}\theta D$ we achieve 
(on the $T^2$ at the self dual radius, $B=0$) that the right-movers get 
rotated in the inverse direction w.r.t.\ the left-movers (if we choose
the metric of the dual $T^2$ to admit the symmetric $\mathbb{Z}_3$ action).
Via the $D$-duality \eqref{Dduality}, \eqref{z2duality} the metric 
\eqref{z3metric} of the symmetric $\mathbb{Z}_3$ maps to the background
fields of the asymmetric $\widehat{\mathbb{Z}}_3$ (in terms of the old background 
fields):
\begin{align} \label{z3hatmetric}
 \widehat{\mathbb{Z}}_3:\qquad
 \frac{G_{ij}^{\widehat{\mathbb{Z}}_3}}{\al}
 &= 
 \begin{pmatrix}
   \frac{\al}{G_{11}}
            &  \frac{B_{12}}{G_{11}} \\
   \frac{B_{12}}{G_{11}} &\al\Bigl(\frac{3}{4}G_{11}
                                   +\frac{(B_{12})^2}{G_{11}}
                             \Bigr)  
  \end{pmatrix} 
  &  \frac{B_{12}^{\widehat{\mathbb{Z}}_3}}{\al}&= 1/2  \\
  \rho&=\frac{1}{2}+i\frac{\sqrt{3}}{2}
 \end{align}
The $\widehat{\mathbb{Z}}_3$ action  takes
the following form in the momentum and winding number basis:
\be  
  \label{Z3asymaction}
   S_{\hat{\theta}}=S_{D}^{-1}S_\theta S_D
 =  \left( 
     \begin{array}{c|c}
      -\mathds{1}
      & 
      \begin{smallmatrix} 
       0 &  1 \\
      -1 &  0
     \end{smallmatrix}     
    \\ \hline
     \begin{smallmatrix} 
       0 &  1 \rule[0ex]{0.ex}{1.4ex}\\
      -1 &  0
     \end{smallmatrix}
       &
      \rule[0ex]{0.ex}{1.4ex} 0
   \end{array} \right)
\ee
The $\widehat{\mathbb{Z}}_3$ action on the left- and right-moving momenta 
 $(e^k (p_{+})_k,(e^l p_{-})_l$  is:
\begin{equation}
  S_{\hat{\theta}}(p_+,p_-) = \Upsilon_{\widehat{\mathbb{Z}}_3} 
                 S_{\theta}\,\Upsilon^{-1}_{\widehat{\mathbb{Z}}_3}
   =\left(\begin{array}{c|c}
     \theta & 0  \\ \hline
    0 & \theta^{-1}\rule[0ex]{0.ex}{2.2ex}
    \end{array}\right),\quad
 \theta_i^{\phantom{i}j} 
  =\left(
   \begin{smallmatrix}
     -\bigl(\frac{B_{12}}{G_{11}}+\oh\bigr)             & \frac{\al}{G_{11}}  \\
     \frac {4 B_{12}^{2}+ 3 G_{11}^2}{4\al G_{11}} & \frac{B_{12}}{G_{11}}-\oh
    \end{smallmatrix}
   \right) 
\end{equation}
Therefore the action $S_{\hat{\theta}}$ represents for all allowed backgrounds
(even those with $B\neq 0$ in the $D$ dual geometry) an asymmetric rotation
of the form $(\theta_L,\theta_R)=(\theta,\theta^{-1})$.
If we want to have the full 
$\widehat{\mathbb{Z}}_3^L\times\widehat{\mathbb{Z}}_3^R$ symmetry, we are restricted
to backgrounds of the form:
\begin{align}\label{z3z3bg1}
\mathbb{Z}_3^L\times\mathbb{Z}_3^R:\qquad
\tau=\rho=\frac{1}{2}+i\frac{\sqrt{3}}{2}
\end{align}

In the same way  we get for the asymmetric $\widehat{\mathbb{Z}}_4$:
\begin{align}
\widehat{\mathbb{Z}}_4:\qquad
\rho&=i
\end{align}
If we want to maintain the asymmetric $\widehat{\mathbb{Z}}_4$ as well as
symmetric $\mathbb{Z}_4$ action we need:
\begin{align}
 \mathbb{Z}_4\times\widehat{\mathbb{Z}}_4:\qquad
 \tau&=\rho=i
\end{align}
However $ \mathbb{Z}_4\times\widehat{\mathbb{Z}}_4 $ does not generate
the full $\mathbb{Z}_4^L\times\mathbb{Z}_4^R$ since it does not
contain elements like $(\theta_L,\theta_R)=(\theta,\id)$.
In chapter \ref{ncg} we will investigate the
 orientifold of the asymmetric $\mathbb{Z}_3^L\times\mathbb{Z}_3^R$ 
orbifold.\footnote{The {\sl four} 
dimensional
$T^6/(\mathbb{Z}_4\times\widehat{\mathbb{Z}}_4)$  orientifold is  
presumably fraught with the same problems as the  four dimensional
$T^6/{\mathbb{Z}}_4 $ $\Omega$-orientifold of \cite{Aldazabal:1998mr}.
We will make some comments about this in the following chapter.}

\subsubsection{\label{wsparityt2}The world-sheet-parity on
               \texorpdfstring{$T^2$}{T\texttwosuperior}}
 We mentioned already that the world-sheet parity $\Omega: \sigma\to -\sigma$
 is a symmetry
 of the compactified theory, iff the $B$-field obeys condition 
\eqref{quantizedB}. This means especially that $\Omega$ is a symmetry
of the  $\mathbb{Z}^L_3\times\mathbb{Z}^R_3$-symmetric background 
\eqref{z3z3bg1}. In the next chapter we will see how symmetries involving
$\Omega$ are gauged, leading to so called {\it orientifolds}. One can 
also combine $\Omega$  with an element $s$ which acts on the space(-time) 
and more generally, on the Narain lattice, s.th.~the resulting  $s\Omega$
is still a symmetry of the theory.
In chapter \ref{magbf} and \ref{z4} 
we will gauge by $\Bar{\sigma}\Omega$ with $\Bar{\sigma}$
acting as:
\begin{equation} \label{sigmabaraction}
  \begin{pmatrix}  X_1 \\ X_2
  \end{pmatrix}
  \xrightarrow{\Bar{\sigma}}
  \begin{pmatrix}  X_1 \\ -X_2
  \end{pmatrix}
  \qquad\text{or in complex coords.:}\quad Z \xrightarrow{\Bar{\sigma}}\Bar{Z}
\end{equation}
This action is a  symmetry of bosonic string theory on $T^2$ iff the complex
structure $\tau$ fulfills either
\begin{equation}
  \tau_1 = 0 \quad\text{or } \quad\tau_1 = 1/2
\end{equation}
For  $\Bar{\sigma}\Omega$ to be a symmetry of the superstring, it has
to be compatible with the GSO projection: $\Bar{\sigma}$ acts on a Ramond zero
mode $\ket{s}$ by $s=\pm1/2\xrightarrow{\Bar{\sigma}}s=\mp1/2$. If the 
GSO-projection takes the form:\footnote{This is true for sectors
which are untwisted or twisted by a left-right symmetric twist.
$a$ is an integer which can be chosen to be one.}
\be 
 \ket{s_0\ldots s_n}_\text{L}:
  \qquad\sum_{i=0}^n \varepsilon_i s^\text{L}_i = a\mod 2,
   \qquad \varepsilon_i\in \{-1,1\}
\ee 
on left-movers, it is for right-movers:
\be 
  \ket{s_0\ldots s_n}_\text{R}:
  \qquad\sum_{i=0}^n \varepsilon_i \Bar{\sigma}\big(s^\text{R}_i\big) 
     =a\mod 2,\qquad \varepsilon_i\in \{-1,1\}
\ee 
 $\Bar{\sigma}$  multiplies the $s_i$ with $-1$ on the complex 
planes on which it acts by \eqref{sigmabaraction}. We can absorb this
action in a redefinition of the $\varepsilon_i$. For $\Bar{\sigma}\Omega$ 
orientifolds with even number of planes with  $\Bar{\sigma}$ action
 \eqref{sigmabaraction}, left- and right-movers have the same GSO-projection.
Therefore in this case, $\Bar{\sigma}\Omega$ is a symmetry of  Type IIB 
theory. The converse is true for an odd number of complex planes
on which $\Bar{\sigma}$ acts by complex conjugation. In the latter case
$\Bar{\sigma}\Omega$ is a symmetry of Type IIA theory.
We will make one comment on the $\mathbb{Z}_3$-symmetric
torus \eqref{z3metric}. There is an (equivalent) torus obtained from
 \eqref{z3metric} by transforming the background by the element
\be\label{tsttransf}
 TST:\,\tau\to\tfrac{\tau}{\tau +1}
\ee
 of the T-duality group
 $SL(2,\mathbb{Z})_1$.
It leads to:
\begin{align}
 \tau_B &= \frac{1}{2}+i\frac{1}{2\sqrt{3}}
  &\Longrightarrow\; G_{11}&=3 G_{22}, & G_{12}&=\frac{1}{2}G_{11}
\end{align}
 while leaving $\rho$ unchanged. 
Even though the orbifold theory is completely
equivalent, the gauging of $\Bar{\sigma}\Omega$ leads to inequivalent models. 
We call the torus obtained from \eqref{tsttransf} the $\bf B$ torus and
the  ``usual''  $\mathbb{Z}_3$-torus \eqref{z3metric} the $\bf A$ torus.
\begin{figure}
\begin{picture}(0,0)%
\includegraphics{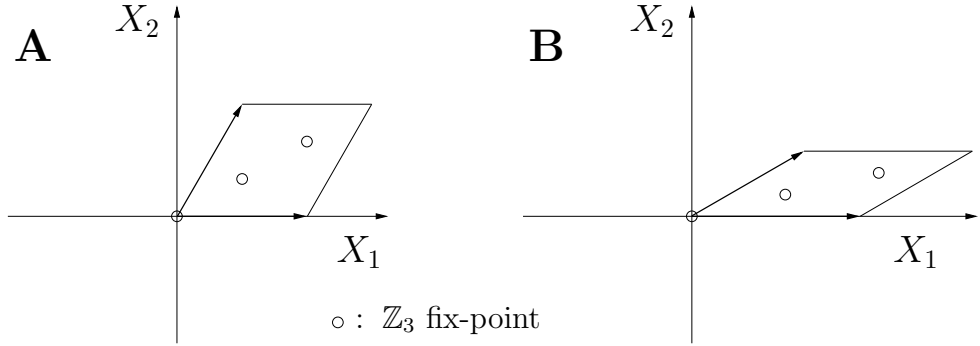}%
\end{picture}%
\setlength{\unitlength}{2072sp}%
\begingroup\makeatletter\ifx\SetFigFont\undefined%
\gdef\SetFigFont#1#2#3#4#5{%
  \reset@font\fontsize{#1}{#2pt}%
  \fontfamily{#3}\fontseries{#4}\fontshape{#5}%
  \selectfont}%
\fi\endgroup%
\begin{picture}(11622,4064)(-9,-3258)
\put(4186,-3031){\makebox(0,0)[lb]{\smash{\SetFigFont{12}{14.4}{\rmdefault}{\mddefault}{\updefault}{\color[rgb]{0,0,0}$:\;\mathbb{Z}_3$ fix-point}%
}}}
\put(10621,-2241){\makebox(0,0)[lb]{\smash{\SetFigFont{14}{16.8}{\rmdefault}{\mddefault}{\updefault}{\color[rgb]{0,0,0}$X_1$}%
}}}
\put(6256,164){\makebox(0,0)[lb]{\smash{\SetFigFont{17}{20.4}{\rmdefault}{\mddefault}{\updefault}{\color[rgb]{0,0,0}{\bf B}}%
}}}
\put(1306,524){\makebox(0,0)[lb]{\smash{\SetFigFont{14}{16.8}{\rmdefault}{\mddefault}{\updefault}{\color[rgb]{0,0,0}$X_2$}%
}}}
\put(3961,-2241){\makebox(0,0)[lb]{\smash{\SetFigFont{14}{16.8}{\rmdefault}{\mddefault}{\updefault}{\color[rgb]{0,0,0}$X_1$}%
}}}
\put(7466,524){\makebox(0,0)[lb]{\smash{\SetFigFont{14}{16.8}{\rmdefault}{\mddefault}{\updefault}{\color[rgb]{0,0,0}$X_2$}%
}}}
\put( 96,164){\makebox(0,0)[lb]{\smash{\SetFigFont{17}{20.4}{\rmdefault}{\mddefault}{\updefault}{\color[rgb]{0,0,0}{\bf A}}%
}}}
\end{picture}
\caption[$\bf A$ torus and $\bf B$ torus]
    {\label{AsndBz3torus} Left: $\bf A$ torus with complex structure
        $\tau=\frac{1}{2}+i\frac{\sqrt{3}}{2}$. Right:
                             $\bf B$ torus with 
                      $\tau=\frac{1}{2}+i\frac{1}{2\sqrt{3}}$. 
       The $\mathbb{Z}_3$ fix points are depicted as well.}
\end{figure}
Both $\mathbb{Z}_3$ symmetric tori are depicted in figure \ref{AsndBz3torus}.
The asymmetric  $\mathbb{Z}_3$ rotation $\hat{\theta}$ is now again obtained
by $D$-duality \eqref{Dduality}. It leads to the background:
\begin{equation}
 \begin{aligned}\label{z3hatmetricB}
\widehat{\mathbb{Z}}_3:\qquad
 \frac{G_{ij}^{\widehat{\mathbb{Z}}_3}}{\al}
 &= 
 \begin{pmatrix}
   \frac{\al}{G_{11}}
            &  \frac{B_{12}}{G_{11}} \\
   \frac{B_{12}}{G_{11}} &\al\Bigl(\frac{1}{12}G_{11}
                                   +\frac{(B_{12})^2}{G_{11}}
                             \Bigr) \rule[-2.5ex]{0.0ex}{2ex} 
  \end{pmatrix} 
  &  \frac{B_{12}^{\widehat{\mathbb{Z}}_3}}{\al}&= 1/2  \\      
     \tau&= \frac{B_{12}}{G_{11}}  +i\frac{G_{11}}{2\sqrt{3}}   
      & \rho&=\frac{1}{2}+i\frac{1}{2\sqrt{3}}
 \end{aligned}
\end{equation}
Symmetry under the asymmetric $\mathbb{Z}^L_3\times\mathbb{Z}^R_3$ rotation
group then requires the background:
\begin{align}\label{z3z3bg2}
\mathbb{Z}_3^L\times\mathbb{Z}_3^R:\qquad
\tau_B=\rho_B=\frac{1}{2}+i\frac{1}{2\sqrt{3}}
\end{align}
Instead of acting with $\Bar{\sigma}\Omega$ (or $\Omega$)
 on the $\bf B$ torus, we can alternatively implement  the action of 
\eqref{tsttransf}   directly into $\Bar{\sigma}\Omega$ (or $\Omega$):
\be
 \Bar{\sigma}\Omega_B \equiv  (TST)^{-1}\, \Bar{\sigma}\Omega \,(TST)
\ee
The action of $\Bar{\sigma}\Omega$ takes the following form on the
$(m,n)$-basis
of  the $\mathbb{Z}_3$ background \eqref{z3metric}: 
\begin{equation}
 \label{Omegasigmactionz3}
 \mathbb{Z}_3:\qquad\begin{array}{ccc}
  \Bar{\sigma}\Omega_{\bf A}= \biggl( 
     \begin{array}{c|c}
      \begin{smallmatrix} 
      -1 & 0\\  
      \rule[-.45ex]{0.ex}{0.1ex} 
      -1& 1
     \end{smallmatrix} 
      &       
      0
    \\ \hline
      0
      &
     \begin{smallmatrix} 
      \rule[0ex]{0.ex}{1.4ex}    1 &  1\\  
      \rule[-.45ex]{0.ex}{0.1ex} 0 & -1
     \end{smallmatrix}     
   \end{array} \biggr)
  & \qquad &
   \Bar{\sigma}\Omega_{\bf B}= \biggl( 
     \begin{array}{c|c}
      \begin{smallmatrix} 
      0 & 1\\  
      \rule[-.45ex]{0.ex}{0.1ex} 
      1 & 0
     \end{smallmatrix} 
      & 
      0     
      \\ \hline
       0 &
       \begin{smallmatrix}
         \rule[0ex]{0.ex}{1.4ex}  0 & -1 \\  
       \rule[-.45ex]{0.ex}{0.1ex}-1 &  0
      \end{smallmatrix}
      \end{array}
     \biggr)
 \end{array}
\end{equation} 
Under the T-duality $D$ \eqref{Dduality} these symmetries become symmetries 
of the $\widehat{\mathbb{Z}}_3$ background \eqref{z3hatmetric}:\footnote{
$\Omega_{\bf A}$ is indeed a symmetry of the $\mathbb{Z}_3$ 
background \eqref{z3metric} if we impose in addition: $B_{12}/\al=1/2$ 
(cf.~\eqref{quantizedB}).
$\Omega_{\bf B}$ requires however the full
$\mathbb{Z}_3^L\times\mathbb{Z}_3^R$ symmetric point \eqref{z3z3bg1} in order
to be a symmetry.}
\begin{equation}
 \label{Omegasigmactionz3B}
 \widehat{\mathbb{Z}}_3:\qquad\begin{array}{ccc}
\phantom{\Bar{\sigma}}\Omega_{\bf A}= \biggl( 
     \begin{array}{c|c}
      \mathds{1}
      &       
      \begin{smallmatrix} 
      \rule[0ex]{0.ex}{1.4ex}    0 & -1\\  
      \rule[-.45ex]{0.ex}{0.1ex} 1 &  0
     \end{smallmatrix}
    \\ \hline
      0
      &
      -\mathds{1}    
   \end{array} \biggr)
  & \qquad &
  \phantom{\Bar{\sigma}}\Omega_{\bf B}
  = \biggl( 
     \begin{array}{c|c}
       0
       &
       \begin{smallmatrix} 
        0 & -1\\  
        \rule[-.45ex]{0.ex}{0.1ex} 
        1 & 0
       \end{smallmatrix} 
       \\ \hline
        \begin{smallmatrix}
         \rule[0ex]{0.ex}{1.4ex}    0 &  1 \\  
         \rule[-.45ex]{0.ex}{0.1ex}-1 &  0
      \end{smallmatrix}
      &
      0
      \end{array}
     \biggr)
 \end{array}
\end{equation} 
At the  $\mathbb{Z}_3^L\times\mathbb{Z}_3^R$ symmetric point
\eqref{z3z3bg1} all four actions of $(\Bar{\sigma})\Omega$ 
\eqref{Omegasigmactionz3},\eqref{Omegasigmactionz3B}
 are symmetries of the theory and can be gauged.
These symmetries extend to the sector of the zero- and oscillator-modes as 
well. While the bosonic zero-modes (i.e. the center of mass coordinates) are
sensitive to the distinction between $\bf A$ and $\bf B$ lattices 
(equivalently:  sensitive to $..\Omega_{\bf A}$ and $..\Omega_{\bf B}$)
as well, fermionic and oscillator parts are unaffected in the end.\footnote{Of 
course the coordinate representation is different even for the oscillator
modes. However in physical quantities like partition functions the 
oscillator and w.s.-fermionic parts 
are unaffected by the distinction between $\bf A$ and $\bf B$ lattices (or
$..\Omega$-action).}
 
The $\Bar{\sigma}\Omega$ action has the interesting property that it maps
a $g$-twisted sector onto itself ($g$ a geometric rotation in the complex
  plane, on that
 $\Bar{\sigma}$ acts by complex conjugation). This is in contrast to the
 pure world sheet parity $\Omega$ that maps a $g$-twisted sector to the
 $g^{-1}$ twisted sector (which is different if $g\notin \mathbb{Z}_2$).
 Another interesting feature of the $\Bar{\sigma}\Omega$ action is, that
 its invariant closed-string oscillator excitations on each $T^2$ are of the 
form:\footnote{We present only bosonic degrees of freedom
 schematically, but the
 world sheet fermions are treated analogously, implementing the fact that
 fermionic occupation numbers for individual modes are either one or zero.}
\be  \label{sigmaomegainv}
  \ket{\alpha^{k_1}_{i_1}\alpha^{k_2}_{i_2}\ldots\alpha^{k_n}_{i_n},
       \Bar{\alpha}^{l_1}_{j_1}\Bar{\alpha}^{l_2}_{j_2}\ldots
        \Bar{\alpha}^{l_n}_{j_m}}
  \otimes
  \ket{\Bar{\Tilde{\alpha}}^{k_1}_{i_1}\Bar{\Tilde{\alpha}}^{k_2}_{i_2}
   \ldots\Bar{\Tilde{\alpha}}^{k_n}_{i_n},
   \Tilde{ \alpha}^{l_1}_{j_1}\Tilde{\alpha}^{l_2}_{j_2}
        \ldots\Tilde{\alpha}^{l_n}_{j_m}}
\ee
We used here a complexified basis for the oscillators.
In this basis a (symmetric) rotation $g\in U(1)\simeq SO(2)$ acts as 
\be \label{complexrotation}
 g:\,  Z \mapsto e^{i2\pi \phi_g}Z\qquad  
  \Bar{Z}\mapsto e^{-i2\pi \phi_g} \Bar{Z}
\ee
on the complex coordinates \eqref{sigmabaraction}. This implies that
 $\Bar{\sigma}\Omega$-invariant oscillator states \eqref{sigmaomegainv}
 and its fermionic counterparts
 are automatically invariant under geometric $U(1)$ rotations $g$.
 Remark: this is in general not 
 true for linear excitations (i.e.\ excitations in the Narain lattice). These
 properties extend naturally to direct products of two-tori.
 
\section{\label{toroidorbs}Toroidal orbifolds}
In this section we will consider orbifolds of the type:
\begin{equation}
 {\cal X}_d= T^d/G
\end{equation}
with $G$ a symmetry of the torus $T^d$. For the sake of simplicity, we will
restrict to the case of {\it abelian} groups $G$.
However, we will also consider the case where $G$ acts differently on left- and
right-movers, such that the action of $G$ is well defined on the Narain lattice
$\Gamma^{(d,d)}$ and on the left- and right-moving parts.
 The group $G$ can consist of rotations 
$\in O(d,d,\mathbb{Z})$ and translations.
We will restrict in this work to the case where G is purely rotational, even 
though translations (or shifts) can give rise to interesting effects. 
The untwisted sector of the orbifold is obtained by projection on  
$G$-invariant states. The untwisted partition function consists of 
all trace insertions of elements $g\in G$.  
We will have a closer look at 
the sector twisted by $gs$, ($g\in G$, $s\in \Gamma$) 
in the $\sigma$ direction (c.f.\ eq. 
\eqref{twisted1}).  These boundary conditions are solved by the
following mode expansion (lattice momenta and center of mass
 coordinate are left out):
\begin{equation}\label{closstrtwist1}
 \begin{aligned}
 X^{\text{osc}}_+(\tau+\sigma)^\mu&= 
            i\sqrt{\frac{\al}{2}}
           \sum_j\, 
           \sideset{}{'}
            \sum_{n_j\in \mathbb{Z}+\beta_j}
               C^\mu_j\frac{\tilde{\alpha}^j_{{-n}_j}}{n_j}
            e^{in_j(\tau+\sigma)}  \\
 X^{\text{osc}}_-(\tau-\sigma)^\mu&= 
            i\sqrt{\frac{\al}{2}}
           \sum_k\, 
           \sideset{}{'}
            \sum_{n_k\in \mathbb{Z}+\gamma_k}
               D^\mu_k\frac{\alpha^k_{{-n}_k}}{n_k}
            e^{-in_k(\tau-\sigma)}
 \end{aligned}
\end{equation}
We split $g$ into $(g_+,g_-)$, a part acting on $X_+$ which is $g_+$
and $g_-$ acting on $X_-$. Then the $C^{\mu}_j$ are defined as
Eigenvectors of $g_+$ with Eigenvalue $\kappa_j=\exp i2\pi\beta_j$. 
(Analogously, the $D^{\mu}_j$ are Eigenvectors of $g_-$ with Eigenvalue
$\lambda_j=\exp i2\pi\gamma_j$).
Since $g_+$ is a rotation, the $C_j$ form an
orthogonal system (which can be normalized) w.r.t.\ the hermitian form that is 
induced by the euclidian scalar product on 
$\mathbb{R}^d$. They can be interpreted as Vielbeins. This is analogous
to the discussion of open strings in constant background fields (cf.\ 
chapter \ref{strbg}). For each complex $C_j$ there exists a complex conjugate 
$C_{-j}$ with c.c.\ Eigenvalue $\lambda_{-j}=\bar{\lambda}_j$.
The $D_k$ fulfill analogous properties w.r.t.\ the rotation $g_-$.
The oscillators obey the following commutation relations:
\begin{align}
 \big[\alpha^i_{l_i},\alpha^k_{m_k}\big] 
   &=l_i\cdot\delta_{(l_i+m_k),0}\bigl<C^{i}, C^{k}\bigr>, &
    \big[\tilde{\alpha}^i_{l_i},\tilde{\alpha}^k_{m_k}\big]
     =l_i\cdot\delta_{(l_i+m_k),0}
     \bigl<D^{i}, D^{k}\bigr>
\end{align}
In this case $\bigl<C^{i}, C^{k}\bigr>$ is the inverse of 
$\bigl<C_{i}, C_{k}\bigr>$. 
The lattice part obeys:
\begin{equation}\label{closstrtwist2}
 \begin{aligned}
 X^{\Gamma}_+(\tau,\sigma)&=
           \sqrt{\al}\bigl(p_+(\tau+\sigma)+ p_-(\tau-\sigma)\bigr) \\
\text{with }\quad p_+&=g_+p_+\quad \text{and }\quad 
             p_-=g_-p_-,\quad \bigl(p_+,p_-\bigr) \in \Gamma^{(d,d)}
 \end{aligned}
\end{equation}
The center of mass $x_{\text{com}}$ has {\it a priori} no well defined
splitting into left- and right-movers. Therefore $g$ has a clear 
interpretation in terms of a geometric action only in the
symmetric case: $(g_L,g_R)=(g,g)$. Then the boundary condition reads:
\begin{equation}\label{closstrtwist3}
 (\id-g) x_{\text{com}} =  - 2\pi\sqrt{\al}(p_+-p_-) + s,\quad s\in\Gamma^d
\end{equation} 
If $I_g\subset\Gamma^d$ is the lattice invariant under $g$ and
$N_g=\{u\in \Gamma | u\cdot v=0 \,\forall v\in I_g\}$ the lattice
perpendicular to $I_g$, the right hand side of \eqref{closstrtwist3}
is also contained in $N_g$. (Proof: multiply both sides by $v\in I_g$. The left
side vanishes since $v\cdot (\id-g)x=(\id-g^{-1})v\cdot x$.) For
$g=\id$ we get exactly the result of the toroidal compactification. 
For $g\neq\id$ the  com.\ coordinate is restricted to be fixed up to
addition of a lattice vector.   
$x_{\text{com}}+w$, with $w$ an arbitrary lattice vector, fulfills
eq.\ \eqref{closstrtwist3} if $s$ is shifted by $(1-g)w$ which is a vector 
lying in the sub-lattice $N_g$. Therefore such a shifted com.\ coordinate
is equivalent to the non shifted one. In \cite{Narain:1987qm} the number
of inequivalent fixed points (more general: fixed planes)
 was thereby determined to be:\footnote{By $(1-g)\Gamma$ we mean the lattice
 $\{(\vec{r}-g\vec{r})\,|\,\vec{r}\in\Gamma  \}$.}
\begin{equation}
n_{\text{fix},g}= \left|\frac{N}{(1-g)\Gamma^d}\right|
\end{equation}
This formula  has a simple generalization for left-right asymmetric
twists $g=(g_+,g_-)$: In \cite{Narain:1987qm} the authors gave a 
definition that only refers to the Narain lattice:\footnote{In fact they 
consider heterotic compactifications where the Narain lattice is of the more 
general form $\Gamma^{d,p}$.}
\begin{equation}\label{fixform1}
n_{\text{fix},g}= \sqrt{ \left|\frac{N}{(1-g)\Gamma^{d,d}}\right|}
\end{equation}
$N$ from our former definition has to be replaced by the lattice which
is orthogonal to the invariant sub-lattice $I$ of the Narain lattice 
$\Gamma^{d,d}$. In the same publication the authors proved
that \eqref{fixform1} is always an integer. The proof is rather lengthy.
It involves the embedding of the Narain lattice $\Gamma^{d,d}$  into a
lattice of doubled dimension. Especially in the path-integral formalism
the splitting into holomorphic and anti-holomorphic fields is {\it a priori}
not possible. The number of $h$-invariant
fixed-points in the $g$-twisted sector is of big importance, too. It appears
as an overall constant in the sector  ${ h}\underset{\mbox{$ g$}}{\square}$.
In a subsequent publication
 (cf.\ \cite{Narain:1991mw}) the same authors determined this number to be 
 the {\sl square root} of:
\begin{equation}\label{fixform2}
C_{h,g}=  \left|\frac{N}{(1-g)N^\ast\cup (1-h)N}\right|
\end{equation}
$N^\ast$ is the lattice dual to $N$. In the example which we present
in the following,  most of the numbers $C_{h,g}$
are determined by  modular transformations of untwisted parts in
 the partition function.
(By applying \eqref{fixform2} in combination with modular transformations
we will  relate in fact all these numbers to partition functions in 
the  ($\sigma$-) untwisted sector of the 
$T^4\big/\big(\mathbb{Z}_3^{\text{L}}\times\mathbb{Z}_3^{\text{R}}\big)$
 orbifold.)

For the superstring there exists an analog of the mode
expansion \eqref{closstrtwist2} for the twisted world-sheet fermions
 (in the
NSR formalism). World sheet fermions do not have extra com. degrees of
freedom. As the Hilbert space is a tensor product of the fermionic
and the bosonic sector, the bosonic zero-modes (i.e.\ the com.\ coordinates)
determine the multiplicity of fermionic states as well. If the space-time
fermionic sector (e.g.\ the R-sector of the heterotic string or the
$\Omega$-symmetrized NSR-sector of an orientifold) has only
one massless excitation, the number of fixed points (=inequivalent 
 com.\ coordinates) in this sector determines the number of chiral fermions.
This is similar to the case of intersecting $D$-branes, where the number of
chiral fermions in the respective open string sector is determined by the
intersection number of the two $D$-branes to which the string is attached
 (c.f.\ chapter \ref{magbf}, \ref{z4}, and \cite{Blumenhagen:2002wn}).
There  are additional conditions for a group $G$ to be a symmetry of the
superstring: $G$ must be a symmetry not only of the bosonic and
fermionic Hamiltonian (or: world-sheet energy-momentum tensor)
(i.e. $T_B$ ), but also of the world sheet supercurrent ($T_F$). Of course
$G$ must preserve interactions, especially the $OPE$, as well.  
In addition, the  partition function has to be modular invariant.

Before we will turn to an (asymmetric) example, we will summarize some well
known facts about space-time supersymmetric, geometric toroidal orbifolds.

\subsection{\label{susyorbs}Space-time supersymmetric (toroidal) orbifolds}
In \cite{Dixon:1985jw,Dixon:1986jc} orbifolds were introduced 
in the context of superstring theory. However this was not the first time
orbifolds appeared in physics (cf.\ references in \cite{Dixon:1985jw}).
In mathematics they go back to Satake \cite{Satake:1956}. Our intention 
is to restate (sufficient) conditions for the 
orbifold to be supersymmetric.
Even though an orbifold is not a manifold, certain orbifolds 
can be deformed into a manifold. An interesting class of obifolds
are those which can be deformed into a Calabi-Yau manifold, as superstrings
compactified on such an orbifold yield space-time supersymmetry. 
A complex $n$-dimensional compact  manifold $\cal M$ that is Calabi-Yau 
 (i.e. K\"ahler and first Chern class
$c_1=0$) admits a unique Ricci-flat metric for a given K\"ahler class and
complex structure.\footnote{Sometimes CY manifolds are defined as the triple 
$({\cal M},J,g)$ with ${\cal M}$ a complex 
$n$-dimensional K\"ahler manifold with $c_1=0$,
 $J$ its complex  structure, and $g$ the  K\"ahler metric, which in 
addition should be Ricci-flat. In section \ref{slagsec} we will give a third 
definition making explicit use of a holomorphic $(n,0)$-form $\Omega$
which always exist on a CY$n$-fold.}
 This does not mean that every metric
on  $\cal M$ is Ricci flat, but it means that such a metric  exists.
The property of a complex $n$-dimensional manifold to be k\"ahler 
restricts its holonomy to be at most $U(n)$. 
Ricci flatness implies that the holonomy group with respect to the
Ricci-flat metric  is contained even in 
$SU(n)$, iff $c_1({\cal M})=0$.\footnote{In K\"ahler manifolds, 
$SU(n)$ holonomy implies
also Ricci-flatness. Conversely Ricci-flat K\"ahler manifolds
admit $SU(n)$ holonomy, if they are simply-connected.}
As a consequence the manifold as a spin-manifold admits one 
Killing-spinor (=covariantly constant spinor) of each chirality.
 Unbroken supersymmetry requires the supersymmetry variation of
 the gravitino to vanish. In the absence of an NSNS 3-form field strength $H$ 
 this is equivalent to the statement
 that the covariant derivative of the supersymmetry
 parameter $\eta$ vanishes: $D\eta =0$ (i.e.\ $\eta$ is a Killing spinor).
 If one covariant constant spinor exists on the compactification space
  $\cal M$, it can serve as the supersymmetry parameter $\eta$. 
 Supersymmetry requires in addition that  the variation of the 
 gluino vanishes for each gauge group. We will not pursue this second
 question. We note however that in the absence of non-trivial field strength
 $H$, the low energy supersymmetry-conditions 
 require the compactification manifold $\cal M$  to be of CY-type to first
 order in $\al$. ($\al$-corrections from string theory deform the CY-condition
only continiously, s.th. it is justified to neglect them in a 
first approximation.) 

 Like a smooth manifold, an orbifold
 admits a holonomy group. The holonomy group  of an orbifold is
closely related to the orbifold group. The holonomy group
of a general real four dimensional manifold is 
$SO(4)\simeq SU(2)\times SU(2)$. In general, such a manifold will 
admit no global Killing spinors.
 If however the holonomy is 
 contained in  an $SU(2)$ subgroup, a single Killing spinor exists 
(for each chirality) and one supersymmetry will survive.
Orbifold groups that are discrete subgroups of $SU(2)$, and 
 which admit  geometric action  on four-dimensional tori,
have been listed in \cite{Dixon:1985jw,Dixon:1986jc}. We list them in
table \ref{sixdorbif}. If a ten dimensional string-theory with 
${\cal N}=1$ supersymmetry in ten space dimensions is compactified on
such  real four dimensional orbifold space, it will lead to ${\cal N}=1$
in six dimensions. This is the case for heterotic string and for Type I.  
\begin{table}
\begin{center}
\begin{tabular}{|c|c|}
 \hline
 \rule{0cm}{2.4ex}
 $\mathbb{Z}_2  : \vec{v}=(1,-1)/2 $ &
 $\mathbb{Z}_4 : \vec{v}=(1,-1)/4$ \\
   $\mathbb{Z}_3 : \vec{v}=(1,-1)/3$ & 
  $\mathbb{Z}_6 : \vec{v}=(1,-1)/6$  \\
 \hline
\end{tabular}
 \caption{\label{sixdorbif}$\mathbb{Z}_N$ groups preserving
                   ${\cal N}=1$ supersymmetry in $D=6$.}
 \end{center}
 \end{table}
Table \ref{sixdorbif} should be understood as follows: The Eigenvalues 
of the rotations $\theta\simeq g\in G$ are $\exp(\pm2i\pi v_1)$ and
 $\exp(\pm2i\pi v_2)$. A general orbifold-rotation $\theta\in G$ can 
then be described in a suitable (complexified) basis by:
\begin{align}
   \theta : Z_i  \mapsto \exp \left( 2\pi i v_i \right)       Z_i & &
       \bar{Z}_i \mapsto \exp \left( -2\pi i v_i \right) \bar{Z}_i 
\end{align}
Four dimensional theories with ${\cal N}=1$ in four dimensions are obtained
by compactifying a ten-dimensional  ${\cal N}=1$ supersymmetric theory
on a complex three dimensional manifold with $SU(3)$ holonomy. 
(The general $SO(6)\simeq SU(4)$
  holonomy of a real six dimensional manifold 
  is reduced to $SU(3)$, leaving one Killing spinor.) Possible orbifold
actions, that lead to four dimensional  ${\cal N}=1$ supersymmetry  (and
that in addition are geometric symmetries of some 
six-tori\footnote{The torus is not completely determined by the symmetry.
In general several tori exist for a $\mathbb{Z}_N$ group, that lead
to different spectra of the orbifolded theory (cf.\ 
\cite{Erler:1993ki}).}) are listed
in table \ref{tablezn}.
\begin{table}
\begin{center} 
\begin{tabular}{|l|l|l|}
\hline \rule{0cm}{2.4ex}
 $\mathbb{Z}_3 : \vec{v}= (1,1,-2)/3$ & $\mathbb{Z}^\prime_6 : 
 \vec{v}= (1,2,-3)/6$ &
 $\mathbb{Z}_8^\prime : \vec{v}= (1,2,-3)/8$ \\
 $\mathbb{Z}_4 : \vec{v}= (1,1,-2)/4$ & $\mathbb{Z}_7 :
  \vec{v}= (1,2,-3)/7$ &
 $\mathbb{Z}_{12} : \vec{v}= (1,4,-5)/12$ \\
 $\mathbb{Z}_6 : \vec{v}= (1,1,-2)/6$ & $\mathbb{Z}_8 : \vec{v}= (1,3,-4)/8$ &
 $\mathbb{Z}_{12}^\prime : \vec{v}= (1,5,-6)/12$ \\
\hline
\end{tabular}
\caption{\label{tablezn} $\mathbb{Z}_N$ groups preserving
                   ${\cal N}=1$ supersymmetry in $D=4$.}
\end{center}
\end{table}
We will note however, that this classification of supersymmetric orbifold
actions is far from being exhaustive. For example one
can also build products of the above groups. Another possibility
are non-abelian orbifolds.  In addition to geometric orbifolds, string-theory
offers
the chance to build asymmetric  orbifolds, many of them supersymmetric as 
well. In these cases, the supersymmetry is recovered in the spectrum.
There are also combinations of translations and rotations possible.
The Scherk-Schwarz mechanism \cite{Scherk:1979ta} is an example of 
such an orbifold. Scherk-Schwarz orbifolds, that generically break 
supersymmetry admit nevertheless points in  parameter space (corresponding
to decompactification) where 
supersymmetry is restored. It is difficult to give 
a general simple rule which states
if supersymmetry exists for an orbifold 
or not. We also neglected
conditions for preserving supersymmetry in the gauge-sector.
The restriction to vanishing NSNS field strength $H$ can be weakened. Many
recent and some older work elaborated the obstructions in the more 
general case 
(cf.\ \cite{Candelas:1985en,Strominger:1986uh,Polchinski:1996sm,Gukov:1999gr, 
Taylor:1999ii,Gurrieri:2002wz,Kachru:2002sk,Becker:2003yv,Cardoso:2003af}). 
For our purpose the material presented here is sufficient
and we will turn to a non-trivial example, featuring asymmetry and the
freedom of a $\mathbb{Z}_3$-valued torsion.

\section{\label{asz3orbi}
         The asymmetric 
         \texorpdfstring{$(T^2\times T^2)/(\mathbb{Z}^{\text{L}}_3
         \times\mathbb{Z}^{\text{R}}_3)$}
         {(T\texttwosuperior T\textmultiply\texttwosuperior)/
           (Z(3)L\textmultiply Z(3)R}) 
         orbifold}
We explored in section \ref{enhsym} that a two-torus $T^2$ exists where
a $G= \mathbb{Z}^{\text{L}}_3\times\mathbb{Z}^{\text{R}}_3$ subgroup
of the $SO(2,2,\mathbb{Z})$ duality group is enhanced
to a symmetry such that it can be gauged
 (i.e.~modded out as an orbifold group). This point in moduli space
is given by \eqref{z3z3bg1}. It can be rewritten in terms of the 
mass\raisebox{1ex}{\tiny 2} matrix \eqref{msquaremat}:

\begin{align}\label{msqasz3}
\frac{M^2}{2} \equiv \frac{1}{2}
 \biggl(
  \begin{array}{cc}
  \al G^{ik}  & 
    - B^i_{\phantom{i}l} \\
      B_j^{\phantom{j}k}
   &\frac{1}{\al} \bigl(G -B^2 \bigr)_{jl}
  \end{array}
 \biggr)
 =\frac{1}{3}
 \left(
  \begin{array}{c|c}
   \begin{smallmatrix}
    \rule[0ex]{0.ex}{1.4ex}  2 & -1 \\
      -1 & 2\rule[-.55ex]{0.ex}{0.1ex}
    \end{smallmatrix} &
    \begin{smallmatrix}
      -1/2 &-1 \\
      1 & 1/2
    \end{smallmatrix}    
  \\
   \hline
    \begin{smallmatrix}
      \rule[0ex]{0.ex}{1.4ex}-1/2 & 1 \\
      -1   & 1/2
    \end{smallmatrix} &
    \begin{smallmatrix}
      2 & \phantom{+}1 \\
      1 & \phantom{+}2
    \end{smallmatrix} 
  \end{array}
 \right)
\end{align} 
From this equation we easily see that the compactification scale is of the
order of the string scale. We denote the lattice part of the partition
function without trace insertions by $\Lambda_{SU(3)^2}$. However this part
of the partition function does not factorize into a purely left- and a purely 
right-moving part. The action on the KK and winding modes $\vec{v}=
(m_1,m_2,n_1,n_2)$ is given by \eqref{Z3action} and 
\eqref{Z3asymaction}
Besides the $\mathds{1}$-trace insertion 
only  two other classes of rotations, namely
$(\theta(\hat{\theta}^2)$,$\theta^2\hat{\theta})$  and
$(\theta\hat{\theta}$, 
$\theta^2\hat{\theta}^2)$, 
 have Eigenvectors on the Narain lattice with
Eigenvalue one and can therefore contribute to the lattice-trace.
They span the following invariant lattices (the $\vec{r}_i$ are given
in the same basis as $M^2/2$ in \eqref{msqasz3}):
\begin{align}
 I_{\theta\hat{\theta}}=I_{\theta^2\hat{\theta}^2} 
 &=\{n\vec{v}_1+m\vec{v}_2\,|\, n,m \in \mathbb{Z};\,
    \vec{v}_1=(1, 1, 0, 1),\, \vec{v}_2=(1, 0, 1, 0) \} \\
 I_{\theta(\hat{\theta}^2)}=I_{\theta^2\hat{\theta}}
 &=\{n\vec{v}_1+m\vec{v}_2\,|\, n,m \in \mathbb{Z};\,
    \vec{v}_1=(0, -1, 0, 1),\, \vec{v}_2=(1, 1,-1, 0) \}
\end{align} 
Notifying that the respective normal lattices fulfill the relations:
\begin{align} 
 N_{\theta\hat{\theta}}=N_{\theta^2\hat{\theta}^2}
 &=I_{\theta(\hat{\theta}^2)}=I_{\theta^2\hat{\theta}}
 &
 N_{\theta(\hat{\theta}^2)}=N_{\theta^2\hat{\theta}}
 &=I_{\theta\hat{\theta}}=I_{\theta^2\hat{\theta}^2}  
\end{align}
 and using that $I_{\theta(\hat{\theta}^2)}\perp I_{\theta\hat{\theta}}$ 
(w.r.t.\ the Lorentzian scalar product $\Xi$)
we derive the useful identities for the (squared) multiplicities
$C_{h,g}$ (c.f.~\eqref{fixform2}):
\begin{equation} \label{invasymfpts}
  \begin{aligned} 
  C_{h=(\theta^\prime,\id),g=(\id,\theta)} 
  &=C_{h=(\id,\id),g=(\id,\theta)}
  \\
  C_{h=(\id,\theta^\prime),g=(\theta,id)} 
  &=C_{h=(\id,\id),g=(\theta,\id)}
 \end{aligned}
\end{equation}
This means that the fix-point (precisely: fix-plane) multiplicity 
is unaffected by inserting a purely left-moving twist into the trace
of  a purely  right-moving twisted sector. As the fix-point degeneracy
 in the $g$-twisted sector is obtained by modular $S$-transformation
of the untwisted sector with $g$-insertion, we fix again  many 
(in fact: all) prefactors by requiring modular invariance (using
\eqref{fixform2} only indirectly). We would not be able to determine
the numbers $C_{h=(\theta^\prime,\id),g=(\id,\theta)}$ by modular
transformation of trace inserted untwisted sectors because 
$k\underset{\mbox{$ 1$}}{\square}$ and $h\underset{\mbox{$ g$}}{\square}$
lie in different modular orbits for all $k\in G$.  

We get the following bosonic {\sl lattice} partition functions for each
$T^2$:\footnote{The $\vt$-functions are given in appendix 
\ref{thetafunctions} (p.~\pageref{thetafunctions}). Its modular
properties and  some identities can be found there as well.}
\begin{align}\label{trins1}
 {\theta\hat{\theta} }\underset{\mbox{$ 1$}}{\square}&=
  \biggl(\frac{1}{\eta^2(q)}\sum_{\vec{v}\in\mathbb{Z}^2} q^{\vec{v}^T
  S\vec{v}} \biggr)
  \cdot \frac{2\sin(2\pi/3)\eta(\bar{q})}{\thef{1/2}{1/2+2/3}(\bar{q})}
  \\
 {\theta^2\hat{\theta}^2 }\underset{\mbox{$ 1$}}{\square}
 &=\biggl(\frac{1}{\eta^2(q)}
    \sum_{\vec{v}\in\mathbb{Z}^2} q^{\vec{v}^T S\vec{v}}\biggr)
   \cdot \frac{2\sin(\pi/3)\eta(\bar{q})}{\thef{1/2}{1/2+1/3}(\bar{q})}
  \\ 
 {\theta\hat{\theta} }\underset{\mbox{$ 1$}}{\square}
  =\overline{{\theta(\hat{\theta}^2) }\underset{\mbox{$ 1$}}{\square}} 
  &\qquad\qquad
   \theta^2\hat{\theta}^2  \underset{\mbox{$ 1$}}{\square}=
  \overline{\theta^2\hat{\theta}  \underset{\mbox{$ 1$}}{\square}} 
\end{align}
where $ S= \Bigl(\begin{smallmatrix} 
       \rule[0ex]{0.ex}{1.4ex} 1& 1/2 \\
       1/2& 1\rule[-.45ex]{0.ex}{0.1ex}
      \end{smallmatrix} 
      \Bigr)$.
The $q$-dependent part in \eqref{trins1} multiplies just by a phase $i^{1/6}$
 under a modular $T$-transformation
 ($\tau\to\tau+1$) but the $S$-transformed left-moving part is slightly
more complicated:
\begin{align}\label{latsumas1}
 1 \underset{\mbox{$\theta\hat{\theta} $}}
 {\fbox{\phantom{\rule{1.8ex}{1.8ex}}}} 
   &=\frac{1}{\sqrt{3}\eta^2}\sum_{\vec{v}\in\mathbb{Z}^2} 
           q^{\vec{v}^T S^{-1}\vec{v}}
   & 
   S^{-1} &= \frac{1}{3}
      \begin{pmatrix} 
       \rule[0ex]{0.ex}{1.4ex} 1& 1/2 \\
       1/2& 1\rule[-.45ex]{0.ex}{0.1ex}
      \end{pmatrix} 
\end{align}
To calculate further  $T$- and $S$-transformed partition functions with
the help of the Poisson resummation formula \eqref{poissonf1}
we have to rewrite  the $q$-dependent part of the above function 
as a sum over {\it shifted} 
lattices:\footnote{The
coordinates of the $\mathbb{Z}_3$-fix points are written in the basis
 which defines the metric $S$ of the lattice $\Gamma_S$.}
\begin{gather}\label{latsumas2}
 1 \underset{\mbox{$\theta^2\hat{\theta} $}}
 {\fbox{\phantom{\rule{1.8ex}{1.8ex}}}} 
 = \frac{1}{\sqrt{3}}\sum_{i=0}^2 \chi_i
 \\ \label{su3char}
 \begin{aligned}
  \chi_i &=\frac{1}{\eta^2}
          \Biggl(  \sum_{\vec{v}\in\mathbb{Z}^2} 
           q^{(\vec{v}+\vec{r}_i)^T S(\vec{v}+\vec{r}_i)}
          \Biggr)\\
    \{r_i| i=0, 1, 2\}
    &=\bigl\{ (0,0),\bigl(\tfrac{1}{3},\tfrac{1}{3}\bigr),
              \bigl(\tfrac{2}{3},\tfrac{2}{3}\bigr)\bigr\}
      \quad(=\mathbb{Z}_3\text{-fix points})
 \end{aligned}
\end{gather}
The sums \eqref{latsumas1} and \eqref{latsumas2} are easily seen to be equal:
The KK and winding lattice in  \eqref{latsumas1} has one third of the volume 
 of the original lattice $\Gamma_S$ which is defined by the metric $S$. 
This lattice also
admits  $\mathbb{Z}_3$ symmetry. 
In addition the  fixed-points  span a 
fundamental cell of a $\mathbb{Z}_3$ symmetric lattice 
{\sl modulo} a lattice vector of $\Gamma_S$. This fundamental cell has exactly
one third of the volume of the $S$ lattice. Therefore the lattice associated 
with $S^{-1}$
equals the direct sum:
\begin{equation}\label{latticedecomp1}
 \Gamma_{S^{-1}} 
  = \bigoplus_{r_i\in\mathbb{Z}_3\text{-fix-pts}\negphantom{fix}}
  \bigl(\Gamma_{S} +\vec{r}_i\bigr)
\end{equation} 
Decompositions of the above type  appear in compactifications on 
lattices which are associated with Lie algebras. In our case it is the
lattice of the Lie algebra $A_2$ (or equivalently: $SU(3)$). 
$\chi_0$ 
equals the left partition function in \eqref{trins1} and multiplies
with a phase under a  modular $T$-transformation. In total we can describe
the mapping of the  so called {\it characters} $\chi_i$ under $T$ and $S$
by matrices\footnote{We will not explain the relation between general 
characters and partition functions. A short introduction can be found in
\cite{Lust:1989tj}, chap.~11.}:
\begin{align}\label{modtransfTz3}
       \begin{pmatrix} 
       \chi_0(q) \\ \chi_1(q) \\ \chi_2(q)
      \end{pmatrix} \xrightarrow{T}
       e^{-i\pi/6}
       \begin{pmatrix} 
       \rule[0ex]{0.ex}{1.4ex} 1& 0 & 0  \\
       0& e^{i2\pi/3}\rule[-.45ex]{0.ex}{0.1ex} &0\\
       0 &0& e^{-i2\pi/3} 
      \end{pmatrix} 
      \begin{pmatrix} 
       \chi_0(q) \\ \chi_1(q) \\ \chi_2(q)
      \end{pmatrix} 
\end{align}
The $S$-transformation is a bit more involved. Application of the 
Poisson resummation formula \eqref{poissonf1} leads to:
\begin{equation}\label{latsumas3}
  \frac{1}{\eta^2}\sum_{\vec{v}\in\mathbb{Z}^2} 
           q^{(\vec{v}+\vec{r}_j)^T S(\vec{v}+\vec{r}_j)}
          \xrightarrow{S}
   \frac{1}{\eta^2}\frac{1}{\sqrt{\det S}} \sum_{\vec{w}\in\mathbb{Z}^2} 
        e^{i2\pi\vec{r}_j\cdot\vec{w}}
           q^{(\vec{w})^T S^{-1}\vec{w}}
\end{equation}
The r.h.s.\ can be split again as a sum according to the decomposition
\eqref{latticedecomp1} of $\Gamma_{S^{-1}}$. The phase in the
summation sector \eqref{latsumas3} is then constant in each shifted lattice
$\Gamma_S+\vec{r}_k$. It equals: $\exp(i2\pi\vec{r}_{k}\cdot\vec{r}_{j})/3)
=\exp(i2\pi(k+j)/3)$. With this information 
  we express the action of $S$ on the characters $\chi_j$ by:
\begin{align}\label{modtransfSz3}
      \begin{pmatrix} 
       \chi_0(q) \\ \chi_1(q) \\ \chi_2(q)
      \end{pmatrix} \xrightarrow{S}
       \frac{1}{\sqrt{3}}
       \begin{pmatrix} 
       \rule[0ex]{0.ex}{1.4ex} 
       1 &           1                           & 1  \\
       1 & e^{i2\pi/3}\rule[-.45ex]{0.ex}{0.1ex} & e^{-i2\pi /3}\\
       1 &        e^{-i2\pi /3}                  & e^{i2\pi /3} 
      \end{pmatrix} 
      \begin{pmatrix} 
       \chi_0(q) \\ \chi_1(q) \\ \chi_2(q)
      \end{pmatrix} 
\end{align}
The above characters $\chi_i$ describe a {\sl free} boson compactified on a
 torus given by the root lattice of $SU(3)$ (or $A_2$). Because the boson is
free, the characters are called {\sl level one}. (In total:
 ``$SU(3)$ characters at level one".) There are further connections to 
Kac-Moody algebras which are however not the aim of this thesis.
  An introduction is given in \cite{Lust:1989tj} together with
 further references.
\begin{table}
\begin{center}
\includegraphics{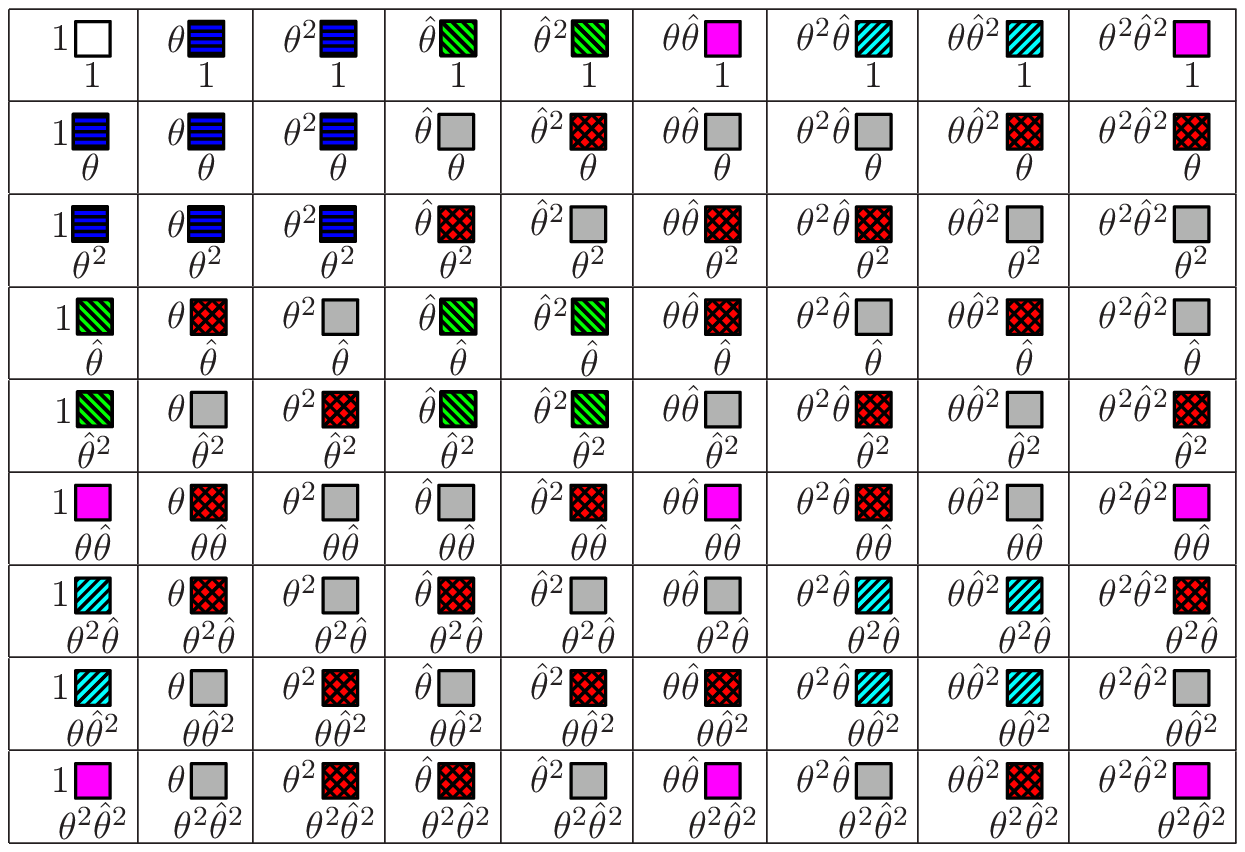}
\caption[Different traces in the partition function]
          {\label{partfz3ass}Different traces in the partition function. 
  Sectors that belong to the same modular orbit are painted in the same 
  color and style. }
  \end{center}
 \end{table}
Now we are able to derive the orbits under 
the modular group. The different orbits under the modular group
are marked by boxes with different colours in table \ref{partfz3ass}.
In total we have {\sl seven} different orbits which are not connected
by modular transformations. In principle we could add phases
to  different orbits. These phases are the discrete
torsion introduced at the end of section \ref{orbifoldsintro}.
However  higher loop consistency imposes further conditions
onto the phases which we have summarized in \eqref{cocycle1} to \eqref{cocycle3}
(p.\ \pageref{cocycle3}). Condition  \eqref{cocycle3} forbids nontrivial
phases for all modular orbits except the one marked by \textcolor{r}{red} 
(and doubly striped) boxes and 
the one marked by \textcolor{gr2}{gray} boxes. These two are exactly the
orbits that contain elements of the type discussed in \eqref{invasymfpts} 
(i.e.\ partition functions that are  not derived from 
any $\sigma$-untwisted partition function by a modular transformation).
We further read off from table \ref{partfz3ass} and the second co-cycle
condition \eqref{cocycle2} that these two orbits (each of them containing
24 traces) have complex conjugate torsions $\epsilon$ w.r.t.\ each other.
In addition, \eqref{cocycle1} and table \ref{partfz3ass}
 tells us that $\epsilon^{3}=1$. This leaves two inequivalent 
choices:\footnote{The choice $\epsilon =e^{-i2\pi/3}$ turns out to be 
equivalent to $\epsilon =e^{+i2\pi/3}$.}
\be
 \epsilon =1\qquad\text{and}\qquad\epsilon =e^{i2\pi/3}
\ee
We will schematically present one orbit explicitly. To be economical with
space we introduce the notation of \cite{Bianchi:1999uq}. It is given in
appendix \ref{confblocks}.
The same notation is used in the discussion of the orientifold
of this orbifold in chapter \ref{ncg}, section \ref{z3z3asymorienti}, too.
The orbit we choose is the one containing\, 
\raisebox{-2.4ex}
  {\includegraphics{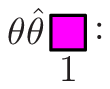}}
\begin{equation}
  \begin{aligned}
 \includegraphics{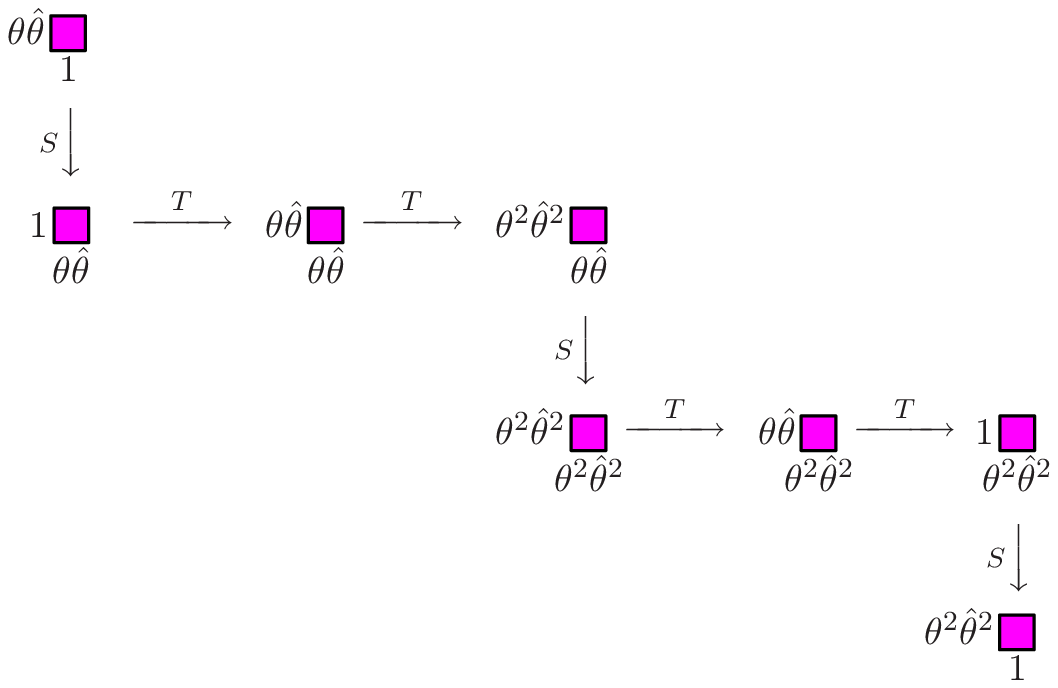}
  \end{aligned}
\end{equation}
Explicitly:
\be 
\begin{array}{rlrlrlrlrl}
 \renewcommand{\arraycolsep}{-2ex}
  {\cal T_{(\theta\hat{\theta})}}
=\phantom{+}\rho_{00}\Lambda_R&\hspace{-2.ex}\Bar{\rho}_{01} & &   && &&&&\\
   \hspace{-2.ex}\,+\,\rho_{00}\Lambda_W&\hspace{-2.ex}\Bar{\rho}_{10} 
 &\hspace{-2.ex} \,+\, \rho_{00}\Lambda^{(1)}_W&\hspace{-2.ex}\Bar{\rho}_{11}
 &\hspace{-2.ex}
   \,+\, \rho_{00}\Lambda^{(-1)}_W&\hspace{-2.ex}\Bar{\rho}_{12} && & &\\
  && & & \hspace{-2.ex}  \,+\,\rho_{00}\Lambda^{(1)}_W
  &\hspace{-2.ex}\Bar{\rho}_{22} &
          \hspace{-2.ex}  \,+\,\rho_{00}\Lambda^{(-{1})}_W
  &\hspace{-2.ex}\Bar{\rho}_{21} &
           \hspace{-2.ex} \,+\,\rho_{00}\Lambda_W
     &\hspace{-2.ex}\Bar{\rho}_{20} \\
  && && && && \hspace{-2.ex}\,+\,\rho_{00}\Lambda_R
            &\hspace{-2.ex}\Bar{\rho}_{02}
\end{array}
\ee
To obtain the contribution to the torus amplitude we have to integrate the
above expression (cf.\ eq.\ \eqref{torusamp}):
\begin{align}
 T_{(\theta\hat{\theta})}&=V_{6}\int_{\cal F}
 \frac{d^2\tau}{4\tau_2}
  \overbrace{\int\frac{d^{6}p}{(2\pi)^{6}} e^{-\pi\al\tau_2 \|\vec{p}\|^2}}^
            {(\ast)}
  \bigg(\frac{1}{\eta\Bar{\eta}}\bigg)^4{\cal T_{(\theta\hat{\theta})}}
  ,\quad (\ast)= \bigg(\frac{1}{\al \tau_2}\bigg)^{6/2} \\
  \label{torusintz3z3orb}
 &=v_d\int_{\cal F}
 \frac{d^2\tau}{4(\tau_2)^2} 
    \bigg(\frac{1}{\sqrt{\tau_2}\eta\Bar{\eta}}\bigg)^4
{\cal T_{(\theta\hat{\theta})}}
\end{align}
with $v_d\equiv\tfrac{V_d}{(4\pi^2\alpha^\prime)^{d/2}}$ being the regularized
$d$-dimensional space-time volume.  The advantage in the definition
 of the functions $\rho_{gh}$ (cf.\ \ref{rhos}, appendix \ref{confblocks})
 is their simple modular transformation behaviour 
(eq.\ \eqref{rhosS} and \eqref{rhosT}). Both the measure 
$\big(4(\tau_2)^2)^{-1}$
and the space-time contribution 
$\big(\sqrt{\tau_2}\eta\Bar{\eta}\big)^{-4}$ of the world-sheet bosons
are modular invariant. 
For completeness we present the complete torus partition function in terms
of quantities ${\cal T}_{gh}$ that have to be inserted in place of 
${\cal T}_{(\theta\Hat{\theta})}$ into the integral 
\eqref{torusintz3z3orb}.\footnote{This torus partition function
 was presented in \cite{Bianchi:1999uq}. We put boxes of the colors
 used in table \ref{partfz3ass} around the individual contributions,
 to indicate the  modular orbit they belong to.}
\be
 \begin{aligned}
  \mbox{\raisebox{2.21ex}{${\cal T}_{00}=$}} &
    \begin{split}\frac{1}{9}
      \Big[ \rho_{00}\Lambda_{SU(3)^2}\bar{\rho}_{00} 
    +&\, \setlength{\fboxrule}{0.8pt}
     \fcolorbox{bl}{w}{$\rho_{01}\bar{\rho}_{01}+\rho_{02}\bar{\rho}_{02}$}
    +\fcolorbox{g}{w}{$\rho_{01}\bar{\rho}_{02}+\rho_{02}\bar{\rho}_{01}$} \\
    +&\,\setlength{\fboxrule}{0.8pt}
    \fcolorbox{c}{w}{$(\rho_{01}+\rho_{02})\bar{\rho}_{00}\bar{\Lambda}_{R}$}
     +\fcolorbox{m}{w}{$\rho_{00}\Lambda_{R}(\bar{\rho}_{01}+\bar{\rho}_{02})$}
     \,\Big]
     \end{split} \\
  \mbox{\raisebox{2.21ex}{${\cal T}_{01}=$}} &\,    
    \begin{split}\frac{1}{9}\Big[\,
      \setlength{\fboxsep}{1pt}\setlength{\fboxrule}{0.8pt}
      \raisebox{-1.5pt}{\fcolorbox{m}{w}{$\rule{3.pt}{0pt}\rho_{00}
       (\Lambda_{W}\bar{\rho}_{10}+\Lambda^{(1)}_{W}\bar{\rho}_{11}
        +\Lambda^{{(-1)}}_{W}\bar{\rho}_{12})\rule{3.pt}{0pt}$}}
       & \,\raisebox{-1.5pt}{$+$}\raisebox{-1.5pt}{\setlength{\fboxsep}{3pt} 
                         \setlength{\fboxrule}{0.8pt}
          \hspace{-2pt}\fcolorbox{r}{w}{$\epsilon\rho_{01}(\bar{\rho}_{10}+
        \bar{\rho}_{11}+\bar{\rho}_{12})$}}  \\
       & +\,\fcolorbox{gr}{w}{$\bar{\epsilon}\rho_{02}(\bar{\rho}_{10}+
         \bar{\rho}_{11}+\bar{\rho}_{12})$}\, \Big]
     \end{split} \\
  \mbox{\raisebox{2.21ex}{${\cal T}_{02}=$}}&\setlength{\fboxrule}{0.8pt}
     \begin{split}\frac{1}{9}\Big[ \,
       \setlength{\fboxsep}{1pt}\setlength{\fboxrule}{0.8pt}
      \raisebox{-1.5pt}{\fcolorbox{m}{w}{$\rule{3.pt}{0pt}\rho_{00}
       (\Lambda_{W}\bar{\rho}_{20}+\Lambda^{(1)}_{W}\bar{\rho}_{22}
       +\Lambda^{{(-1)}}_{W}\bar{\rho}_{21})\rule{3.pt}{0pt}$}}
      & \,\raisebox{-1.5pt}{$+$}\raisebox{-1.5pt}{\setlength{\fboxsep}{3pt}
         \setlength{\fboxrule}{0.8pt}
       \hspace{-2pt}\fcolorbox{r}{w}{$\epsilon\rho_{02}(\bar{\rho}_{20}+
      \bar{\rho}_{22}+\bar{\rho}_{21})$}} \\ 
       & +\,\setlength{\fboxrule}{0.8pt}
        \fcolorbox{gr}{w}{$\bar{\epsilon}\rho_{01}(\bar{\rho}_{20}+
        \bar{\rho}_{22}+\bar{\rho}_{21})$} \, \Big]
     \end{split} \\
 \mbox{\raisebox{2.21ex}{${\cal T}_{11}=$}}&
     \begin{split}\setlength{\fboxrule}{0.8pt}
     \frac{1}{9}\Big[ \,
       \setlength{\fboxrule}{0.8pt}
       \fcolorbox{bl}{w}{$9(\rho_{10}\bar{\rho}_{10}
      + \rho_{11}\bar{\rho}_{11}+\rho_{12}\bar{\rho}_{12})$} 
      &\;\setlength{\fboxrule}{0.8pt}\fcolorbox{r}{w}{$ -3\epsilon
      (\rho_{10}\bar{\rho}_{12}+\rho_{11}\bar{\rho}_{10}+
      \rho_{12}\bar{\rho}_{11})$}\\
      &\; \setlength{\fboxrule}{0.8pt}\fcolorbox{gr}{w}{$-3\bar{\epsilon}
      (\rho_{11}\bar{\rho}_{12}+\rho_{12}\bar{\rho}_{10}+
      \rho_{10}\bar{\rho}_{11})$}\, \Big]
     \end{split} \\
 \mbox{\raisebox{2.21ex}{${\cal T}_{22}=$}}&
     \begin{split}\frac{1}{9}\Big[ \,
     \setlength{\fboxrule}{0.8pt}
     \fcolorbox{bl}{w}{$9(\rho_{20}\bar{\rho}_{20}
     +\rho_{22}\bar{\rho}_{22}+\rho_{21}\bar{\rho}_{21})$} 
      &\; \setlength{\fboxrule}{0.8pt}\fcolorbox{r}{w}{$-3{\epsilon}
     (\rho_{20}\bar{\rho}_{21}+\rho_{22}\bar{\rho}_{20}+
     \rho_{21}\bar{\rho}_{22})$}\\
       &\;\setlength{\fboxrule}{0.8pt}\fcolorbox{gr}{w}{$-3\bar{\epsilon}
     (\rho_{22}\bar{\rho}_{21}+\rho_{21}\bar{\rho}_{20}+
     \rho_{20}\bar{\rho}_{22})$}\, \Big]
     \end{split} \\
 \mbox{\raisebox{2.21ex}{${\cal T}_{12}=$}}&
     \begin{split}\frac{1}{9}\Big[ \,\setlength{\fboxrule}{0.8pt}
       \fcolorbox{g}{w}{$9(\rho_{10}\bar{\rho}_{20}
       +\rho_{11}\bar{\rho}_{22}+\rho_{12}\bar{\rho}_{21})$}
       &\;\setlength{\fboxrule}{0.8pt}\fcolorbox{r}{w}{$ -3\epsilon
       (\rho_{10}\bar{\rho}_{22}+\rho_{11}\bar{\rho}_{21}+
       \rho_{12}\bar{\rho}_{20})$}\\
       &\;\setlength{\fboxrule}{0.8pt}\fcolorbox{gr}{w}{$ -3\bar{\epsilon}
       (\rho_{11}\bar{\rho}_{20}+\rho_{12}\bar{\rho}_{22}+
       \rho_{10}\bar{\rho}_{21})$}\, \Big]
     \end{split} 
\label{torus3}
 \end{aligned}
\ee
The remaining integrands for the torus vacuum amplitude 
(cf.\ eq.\ \eqref{torusintz3z3orb}) are obtained by complex conjugation:
\begin{align}
 {\cal T}_{10}&=\overline{{\cal T}}_{01} 
 & {\cal T}_{20}&=\overline{{\cal T}}_{02} &
 {\cal T}_{21}&=\overline{{\cal T}}_{12}
\end{align}
The spectrum can be read off from the partition function, if one imposes in
addition the condition $H=\widetilde{H}$. One also has to distinguish if
individual
excitations belong to the compact or the space-time part. The spectrum
depends on the torsion $\epsilon$. It is listed in table \ref{z3z3clspec}. 
\begin{table}
 \renewcommand{\arraystretch}{1.2}
  \begin{center}
   \begin{tabular}{|c|c|}
  \hline
  $\epsilon$          & Spectrum       \\
  \hline\hline
  $1$                 & ${\cal N}=(2,2)$: Supergravity multiplet \\
 \hline
  $ e^{\pm i2\pi /3} $ & $ {\cal N}=(2,0) $:  Supergravity 
                       + $ 21\times $ Tensor multiplet \\
 \hline
   \end{tabular}
 \caption[Closed-string spectra of the $\big(T^2\times T^2\big)\big/
            \big(\mathbb{Z}^{\text{L}}_3
         \times\mathbb{Z}^{\text{R}}_3\big)$ orbifold]
       {\label{z3z3clspec}Closed-string spectra of the asymmetric
            $(T^2\times T^2)/(\mathbb{Z}^{\text{L}}_3
         \times\mathbb{Z}^{\text{R}}_3)$ orbifold in dependence of the torsion
         $\epsilon$.}
  \end{center}
\end{table}
We will come back to the  $(T^2\times T^2)/(\mathbb{Z}^{\text{L}}_3
         \times\mathbb{Z}^{\text{R}}_3)$ orbifold in chapter \ref{ncg} 
 where we  build an orientifold of the model. Orientifolds 
will be introduced in the next chapter. In addition
to gauging  world sheet parity $\Omega$ (or a combination
 $s\Omega$ with a space group element $s$) they will introduce 
open strings. We will find out that the asymmetric action 
$\mathbb{Z}^{\text{R, L}}_N$ contains $D$-branes with magnetic fluxes
in its orbit.  $D$-branes with electro-magnetic fluxes are the topic
of chapter \ref{strbg}.

 \chapter[Orientifolds]{\centerline{\label{orientifolds}Orientifolds}}
String theories which contain
non-orientable  world-sheets  are called orientifolds.
  Closed-string theories containing  non-orientable  
diagrams (like the Klein bottle at Euler-characteristic $\chi=0$)
admit tadpoles which lead to inconsistencies.
These inconsistencies can often be cured by adding 
world-sheets with holes and thereby open-strings. At the $\chi=0$-level
these are the M\"obius-strip and the Cylinder. By introducing open-strings
via D-branes of appropriate number and type, all tadpoles can be eliminated
in many cases.\footnote{In order to cancel both the NSNS and the RR tadpoles
  the parent closed-string theory should be supersymmetric. Even though this
  is not strictly proven, this  is likely to be true. There exist also cases
 (e.g.\ the $\Omega$ orientifold of Type IIB on $T^6/\mathbb{Z}_N$, 
$N\in {4,8,12}$ c.f.\ \cite{Aldazabal:1998mr}) where the Klein bottle is 
supersymmetric but so far no attempt to cancel its tadpoles by D-branes
in a supersymmetric manner has been successful and there are hints that
this might be impossible.} Often there exist different
D-bane spectra which lead to tadpole cancellation. By this many (attractive)
spectra can arise in orientifold models as we will see in the following 
chapters.
After presenting a heuristic definition of a  general orbifold,
we will enter into the details and consequences like tadpoles, open-strings
and associated particle spectra. References and overview articles can
be found in the ``concluding remarks'' on page \pageref{conclremori}.

\section{Basic concepts} 
All of our orientifold models are based on orbifold constructions, 
where an element $s\Omega$, with $\Omega$ the world sheet parity and $s$
an element acting on space time but {\sl not} on the world-sheet,
is added to the orbifold group. More generally we define:\footnote{The idea
that an  orientifold is an orbifold of the two-dimensional world-sheet
theory goes back to Sagnotti \cite{Sagnotti:1987tw}.
An orientifold in which the world-sheet parity-reversal $\Omega$ only
appears in combination $s\Omega$
 with some space-time action $s$ is described for the first time 
in \cite{Horava:1989vt}.}
\begin{defin}\label{Orientidef}  
An {\bf orientifold} is a string theory, which is obtained by
modding out a symmetry group $O$ of the original theory:
\begin{equation}
O=\overline{G\cup S\Omega }
\end{equation}
$G$ is a group which does not mix left- and right-movers. $\Omega$ is the 
world sheet parity (which exchanges left- and right-movers).
 $S$ is  a set, which defines a symmetry 
of the underlying CFT if it gets multiplied by $\Omega$. The elements
of $S$ should not mix the left- and right-moving Hilbert spaces.
\end{defin} 
The line over the union indicates algebraic closure. That means that
 $O$ is a group (and not a half group). As $s\Omega s^\prime\Omega\in O$
we can choose the $s$-dependent decomposition ($s\in S$):
\begin{equation}\label{odecomp}
O=G\times \{\id,s\Omega \}
\end{equation}
where we have redefined G to contain all elements  of $O$ which
do not mix left- and right-movers. 
In this thesis we  will restrict $G$ to be a toroidal orbifold group.
That means that $G=\Gamma\rtimes E$, $E\subset SO(\mathbb{Z})$  
(chapter \ref{magbf} and \ref{z4}) or 
$E\subset \Gamma\rtimes\big(SO(\mathbb{Z})_L\times SO(\mathbb{Z})_R\big)$ 
for the models of chapter \ref{ncg}. 
An orientifold is in some respect quite similar to an orbifold, but there
are also striking differences. In this chapter we will consider several
aspects, namely:
\begin{itemize}
 \item non-oriented spectra (i.e.\ $S\Omega$-invariant spectra) 
 \item non-orientable world-sheets
 \item closed-string tadpoles
 \item D-branes, open-strings and  Chan Paton factors
\end{itemize} 
 All these aspects are related to each other.
 We consider the first item which is treated very  similar as in pure 
  orbifolds.
\section{\texorpdfstring{$S\Omega$}{S Omega}-invariant closed-string spectra and Klein bottle 
             amplitude}
\begin{figure}
  \setlength{\unitlength}{0.1in}
  \begin{picture}(60,50)
    \SetFigFont{14}{20.4}{\rmdefault}{\mddefault}{\updefault}
   \put(5,3){\scalebox{0.5}{\includegraphics{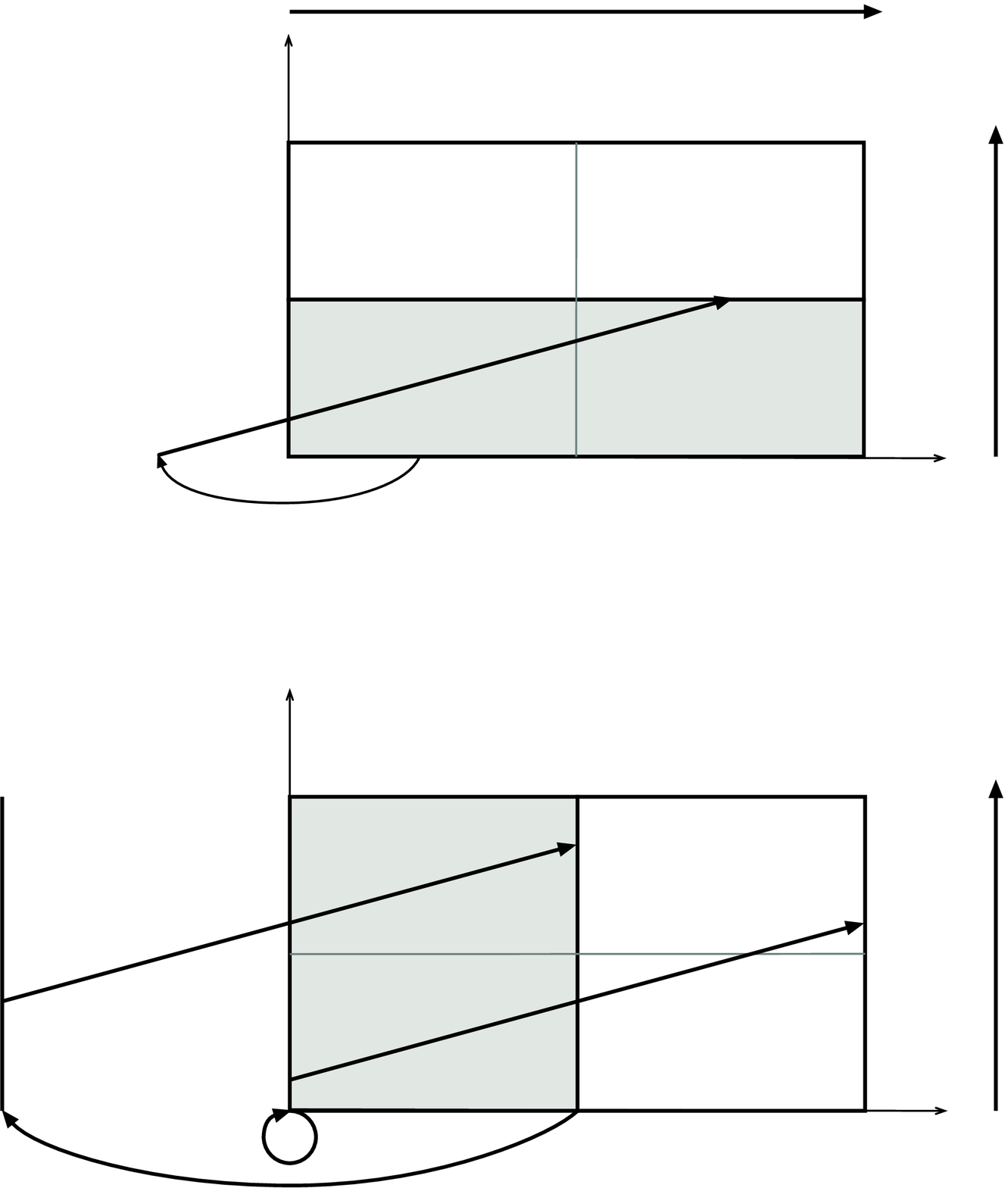}}}
   \put(0,48){loop-channel:}
    \put(26,48){$g$}
    \put(12.7,45.){$\mathfrak{Im}$}
    \put(13,42){$i\tau_2$}
    \put(14.,35.9){$it$}
    \put(44,35.5){$h$} 
    \put(16,36.3){\SetFigFont{12}{20.4}{\rmdefault}{\mddefault}{\updefault}
              \makebox(1,-1)[tl]{$ 1+i\tau_2/2$}}
     \put(12.5,29.3){\SetFigFont{12}{20.4}{\rmdefault}{\mddefault}{\updefault}
                \makebox(1,-1)[tl]{$ z\rightarrow -\bar{z}$}}
     \put(38,28.5){$\mathfrak{Re}$}
    \put(0,23.5){tree-channel:}
    \put(12.7,20.5){$\mathfrak{Im}$}
    \put(13,17.5){$i\tau_2$} 
    \put(44,11.0){$h$}
    \put(16,11.6){\SetFigFont{12}{20.4}{\rmdefault}{\mddefault}{\updefault}
               \makebox(1,-1)[tl]{$ 1+i\tau_2/2$}}
    \put(26.2,4.){$l$}
    \put(12.5,3.7){\SetFigFont{12}{20.4}{\rmdefault}{\mddefault}{\updefault}
               \makebox(1,-1)[tl]{$ z\rightarrow -\bar{z}$}}
     \put(38,4){$\mathfrak{Re}$}
  \end{picture}
 \caption[Periodicities of the Klein bottle]{\label{kleinbotfund} 
                             The periodicities of the Klein bottle embedded
                                in the underlying torus (-cell) and the 
                                two fundamental regions of the Klein bottle
                                (shaded areas).}
\end{figure}

 As in an orbifold, we require the states of the orientifold Hilbert-space
 $\ket{\psi}\in{\cal H}_{\cal O}$ to be invariant under the whole group
 $O$:
 \begin{equation}
   o\ket{\psi}=\ket{\psi} \quad\forall\, o\in O
 \end{equation}
The closed-string partition function therefore has to include
the projection $1/|O|\sum_{o \in O} o$. As we will discuss, this
partition function does in general not need to be modular invariant,
since the trace taken in the operator formalism (or equivalently the
path-integral) does not only correspond to the torus but to   
the sum of a torus partition function and a so called {\it Klein bottle}
 partition function. While the two dimensional
world sheet torus can be obtained by dividing the complex plane
by a lattice group $\Gamma$ (cf.\ eq.\ \eqref{twotorusformula}) the Klein bottle
is obtained by dividing the two-torus by a $\mathbb{Z}_2$-symmetry:
\begin{equation}\label{omegaaction1}
   \Omega:\; z\mapsto -\bar{z}+(1+i\tau_2/2)
 \end{equation} 
$\tau_2$ is again the imaginary part of the complex structure,
which is purely imaginary, because we require $\Omega$ to leave the
world sheet metric $h_{\alpha\beta}$ invariant. Note that $\tau$ with
real parts (i.e. $\tau_1\neq 0$) would define a $\mathbb{Z}_2$ action
leaving the lattice $\Gamma^2$ invariant (as a set)
if we replace the complex conjugation in \eqref{omegaaction1} by a reflection
in the $\tau$-direction. However this $\mathbb{Z}_2$ is not a symmetry of
the string theory as it changes the world-sheet metric. 
The Klein bottle with two 
choices of its fundamental region is depicted in figure \ref{kleinbotfund}.
We have separated the action of $\Omega$ \eqref{omegaaction1} into
the reflection along the $\mathfrak{Re}$-direction and the translation
by $1+i\tau/2$. The first figure corresponds to the loop-channel for the
following reason: The string path-integral with string fields integrated over
the shaded area (in the upper picture of fig.\ \ref{kleinbotfund}) and having
 the depicted
periodicities ($\Omega$ from \eqref{omegaaction1})\footnote{The world
sheet time $\tau$ is assumed to be along the $\mathfrak{Im}$ direction
while the world sheet coordinate $\sigma$ is along the $\mathfrak{Re}$-axis.}
\begin{equation}\label{omegaperiod}
  X(\tau,\sigma) = \Omega X(\tau,\sigma) =  X(\tau+t,-\sigma+1)
\end{equation}
corresponds in operator formalism to the trace-insertion 
($q=\exp(-2\pi t)$):
 \begin{align}\label{kbampl} 
  \text{KB}&= \frac{V_{10}}{(2\pi)^{10}}
   \,\lim\limits_{\epsilon\to 0}
    \int_\epsilon^\infty\frac{dt}{t} Z_{\text{KB}}(q)
       \\ \label{omegatr}
  Z_{\text{KB}}(q)&={ \Omega^\prime}\underset{\mbox{$1$}}{\square} 
   = \frac{1}{|O|}\tr \Omega^\prime e^{-2\pi t(H+\widetilde{H})}
\end{align}
This follows from the common argument that trace-insertions correspond
to periodicity conditions along the time direction in the path-integral
picture. $\Omega^\prime$ means that 
in the operator formalism left- and right-movers are exchanged 
($\Omega^\prime: \sigma\leftrightarrow 2\pi-\sigma$). The shift in the
time coordinate $\tau\rightarrow\tau+t$ is implicitly included in the trace:
 \eqref{omegatr} corresponds to fields in the path-integral which
admit the periodicity $X(\sigma,\tau)=\Omega X(\sigma,\tau)$.
  As the partition function is associated with loop diagrams, we
 call the associated fundamental region the {\it loop-channel}.
The name {\it direct channel} is often used, too.

\begin{figure}
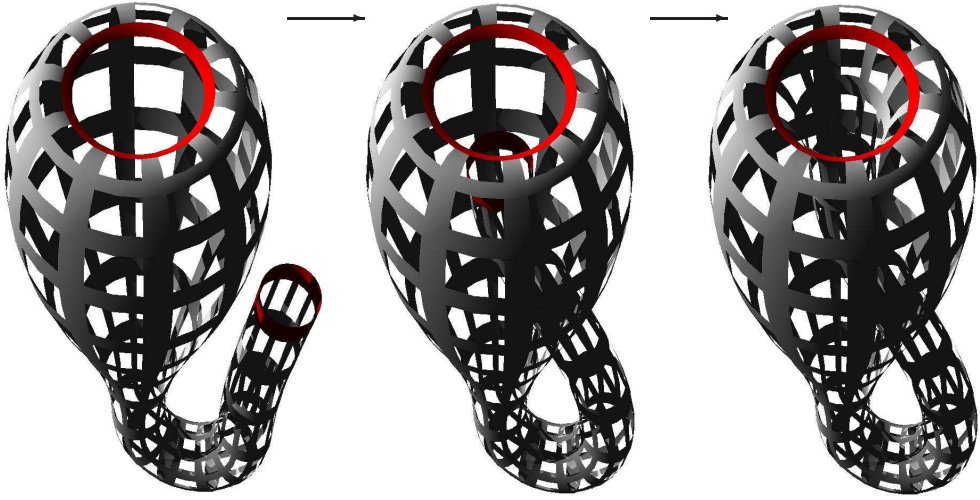

    \setlength{\unitlength}{0.1in}
   \begin{picture}(60,27)
    \put(0,0){\scalebox{0.2}{\includegraphics{picori21.EPS2}}}
    \put(19,0){\scalebox{0.2}{\includegraphics{picori22.EPS2}}}
    \put(37,0){\scalebox{0.2}{\includegraphics{picori23.EPS2}}}
    \put(15,25){\vector(1, 0){4}} \put(34,25){\vector(1, 0){4}}
   \end{picture}
 \caption[Construction of the Klein bottle]
              {\label{kleinbotfig} Construction of the Klein bottle by 
               joining two ends of a cylinder in the way depicted.}
\end{figure}

\begin{figure}
\begin{center}
\begin{picture}(0,0)%
\put(22,10){\epsfig{file=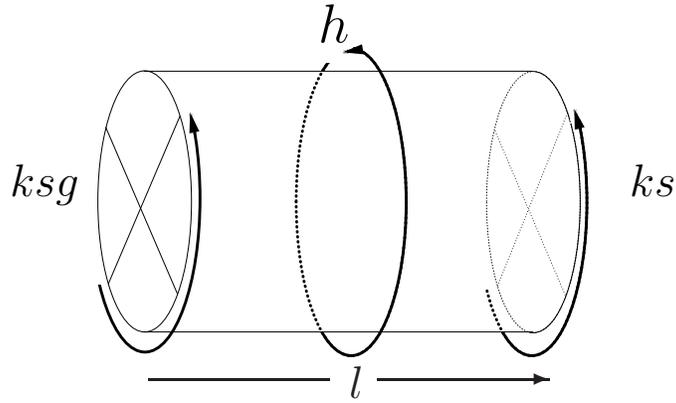}}%
\end{picture}
\setlength{\unitlength}{1184sp}%
\begingroup\makeatletter\ifx\SetFigFont\undefined%
\gdef\SetFigFont#1#2#3#4#5{%
  \reset@font\fontsize{#1}{#2pt}%
  \fontfamily{#3}\fontseries{#4}\fontshape{#5}%
  \selectfont}%
\fi\endgroup%
\begin{picture}(13626,7529)(1001,-6989)
\put(14476,-3061){\makebox(0,0)[lb]{\smash{\SetFigFont{17}{20.4}{\rmdefault}{\mddefault}{\updefault}
$ks$
}}}
\put(7951,164){\makebox(0,0)[lb]{\smash{\SetFigFont{20}{24.0}{\rmdefault}{\mddefault}{\updefault}
$h$
}}}
\put(1501,-3061){\makebox(0,0)[lb]{\smash{\SetFigFont{17}{20.4}{\rmdefault}{\mddefault}{\updefault}
$k s g$
}}}
\thicklines\put(4400,-6900){\line(1,0){3800}}
\put(9200,-6900){\vector(1,0){3600}}
\put(8600,-7300){\SetFigFont{17}{20.4}{\rmdefault}{\mddefault}{\updefault}$l$}
\end{picture}
\end{center}
 \caption[Klein bottle in tree-channel]{\label{Kleinbottletreech} 
    The Klein bottle in tree-channel. The twist $h$ of the closed-string
    is restricted to: $h= (ks\Omega)^2=(ks\Omega g)^2$.}
\end{figure}

The closed-string
which propagates in the loop sweeps out the surface of a  Klein bottle.
A Klein bottle can be obtained (topologically) by identifying the sides
of the shaded rectangle as depicted in figure \ref{kleinbotfund}. 
This is topological the same as joining the two ends of a cylinder
as described in the sequence of figure \ref{kleinbotfig}. We have painted 
a grid (even though the Klein bottle admits of course no holes) to 
make the situation more transparent. The Klein bottle has inevitably 
self-intersections when embedded into three dimensions. 
Equation \eqref{omegaperiod} can easily be generalized to cases where
$\Omega$ is combined with some element $g$ of $G$ and $s\in S$ from 
definition
\ref{Orientidef}. The closed-string can also have boundary conditions
twisted by $h$ and $g$ in the $\tau$- resp.\ $\sigma$-direction 
on the underlying  torus as depicted in figure \ref{kleinbotfund}.
However consistency of the $\mathbb{Z}_2$ involution $\Omega$ 
\eqref{omegaaction1} puts some constraints on the combination of
$h,g,k\in G$ and $s\in S$ if we consider the trace-insertion $ks\Omega$:
\footnote{$\Omega$ can be omitted
if $g,k$ and $s$ act left-right symmetric.}
\begin{align}\label{kbrel1}
  (ks\Omega)^2 = h \;&,\; (ks\Omega g)^2 = h \\
   \label{kbrel2}
   \Longrightarrow h=(ks\Omega)^2 \;&,\; g\,ks\Omega\,g = ks\Omega   
\end{align}
The last equivalence in \eqref{kbrel2} defines exactly the fundamental
group $\pi_1$ of the Klein bottle if we take $g$ and $ks\Omega$ as 
its generators.
In addition to these relations (which are realized on the fields)
the derivatives $\partial_{\pm}$ (cf.\ \eqref{partialdef}) get
exchanged:
\begin{align}
 \Omega\partial_\pm&=\partial_\mp\Omega \\
 \Rightarrow ks\Omega \partial_\pm X(\tau,\sigma) 
  &=  k s\partial_\mp\Omega X(\tau,\sigma)
   =  k s\partial_\mp X(\tau+t,-\sigma+1)
\end{align}
One has to distinguish between world sheet time $\tau$ and the modular
parameter $i t=i\tau_2/2$ of the underlying Torus $T^2$. 
In the {\it tree-channel} (sometimes called: {\it transverse channel})
\begin{align}\label{kbbdycond1}
  X(\tau,\sigma)&= k s\Omega X(\tau,\sigma)  &
  \Rightarrow 
   \left\{
    \begin{array}{cc} X(\tau,\sigma)\big|_{\sigma=\oh} 
        & = ks X(\tau+t,\sigma)\big|_{\sigma=\oh} \\
        \partial_\pm X(\tau,\sigma)\big|_{\sigma=\oh} 
        & = ks \partial_\mp X(\tau+t,\sigma)\big|_{\sigma=\oh} 
    \end{array}
   \right.
\end{align} 
The line at $\sigma=0$ is mapped to the line at $\sigma=2l=1$
in a similar way. If we take the twist $X(\tau,\sigma+1)=g X(\tau,\sigma)$
on the underlying torus into account, we get analogously:
\begin{equation}
 \begin{aligned}\label{kbbdycond2}
    X(\tau,\sigma)\big|_{\sigma=0} 
        & = ks X(\tau+t,\sigma=1)= ksg X(\tau+t,\sigma)|_{\sigma=0} \\
   \partial_\pm X(\tau,\sigma)\big|_{\sigma=0} 
        & = ks \partial_\mp X(\tau+t,\sigma=1)
   = ksg\partial_\mp  X(\tau+t,\sigma)\big|_{\sigma=0}
 \end{aligned}
\end{equation}
We notice that in the case of $k=h=g=s=\id$ the boundary conditions
\eqref{kbbdycond1} and \eqref{kbbdycond2} define topologically what is called
a {\it cross-cap} as they identify opposite points on the ends of a cylinder.
The formul\ae\, \eqref{kbbdycond1} and \eqref{kbbdycond2} give a hint to
an alternative calculation of the Klein bottle amplitude. We could
as well calculate the path-integral with the fields argument lying
in the fundamental region which we called the tree-channel in
figure \ref{kleinbotfund}. This channel is depicted in figure 
\ref{Kleinbottletreech}
as a cylinder with two cross-caps (twisted by $ks$ and $ksg$).
 These fields have to fulfill the
periodicity (or boundary) conditions described by \eqref{kbbdycond1} and
 \eqref{kbbdycond2}. The tree-channel corresponds to $h$-twisted 
closed-strings which travel from the $\tau=0$ line to $\tau=l$ if we 
Weyl-rescale the world sheet by $\lambda=1/(2t)$, such that 
$\tau_2=2t\mapsto 1$ and
the propagation length $l=1/2$ rescales accordingly to $l=1/(4t)$.
 The rescaling
prescription is summarized in table \ref{transftreeloop}
 (p.~\pageref{transftreeloop}) together with the  
corresponding values for the annulus and M{\"o}bius strip which we will
introduce in section \ref{openstringssec}. In the operator formalism
the tree-channel amplitude
 is described as a transition amplitude between two {\it cross-cap states}
$\ket{{\cal C}(ks\Omega)}$ and  $\ket{{\cal C}(ks\Omega g)}$ 
which obey the  boundary condition
 \eqref{kbbdycond1} resp.\ \eqref{kbbdycond2}:\footnote{A ``cross-cap state''
   is a special kind of boundary state. ``Cross-cap states'' only appear
   in orientifolds, whereas general boundary states are present in any
   theory with open-strings.}
\begin{gather} \label{BoundstateKB}
\widetilde{\text{KB}}_{(ks, ks g)}=
      \int_{\mathbb{R}^+} dl\,\braket{{\cal C}(ks)}
            {e^{-2\pi l (H+\widetilde{H})_{h\text{-twisted}}}}{{\cal C}(ksg)}
             \text{ with }   h=(ks\Omega)^2 \\
 \begin{aligned}\label{BoundstateKB2}
   \bigl(X(\tau=0,\sigma) - a X(\tau=0,\sigma+1/2)\bigr)_{h\text{-twisted}}
        \ket{{\cal C}(a)}&=0 
  \\ \bigl(\partial_\pm X(\tau=0,\sigma) + a \partial_\mp
              X(\tau=0,\sigma+1/2)\bigr)_{h\text{-twisted}}
        \ket{{\cal C}(a)}&=0   
 \end{aligned} 
\end{gather}

The boundary conditions in \eqref{BoundstateKB2} are defined at $\tau=0$.
However the time evolution operator $\exp\bigl(-2\pi l (H+\widetilde{H})\bigr)$
appearing in the integrand of \eqref{BoundstateKB} maps one of them to
 $\tau=l$.
The boundary conditions do not determine the normalization of the
cross-cap states. They do not fix the phase, too. 
The normalization is fixed if we compute the amplitude in  loop-channel,
which has to lead to an identical result, since we are choosing a different
but equivalent fundamental region of the Klein bottle:
\begin{align}\label{kbampl2} 
  \text{KB}_{(ks, ksg)}&= \frac{V_{10}}{(2\pi)^{10}}
   \,\lim\limits_{\epsilon\to 0}\int_\epsilon^\infty\frac{dt}{t} 
                Z_{\text{KB}_{(ks, ksg)}}(q)
       \\ \label{omegatr2}
  Z_{\text{KB}_{(ks, ks g)}}(q)
   &= \frac{1}{|O|}\tr ks \Omega^\prime  
               e^{-2\pi t(H+\widetilde{H})_{g\text{-twisted}}}
\end{align}
Summing over all allowed twists $h$ and $g$ and inserting all allowed 
trace-inser\-tions containing $k\Omega^\prime s$ gives the complete 
non-orientable closed-string contribution to the one-loop 
amplitude:\footnote{The tilde over $Z$ indicates that its modular parameter is
          expressed as $q=\exp(-2\pi l)$. In the following we will skip
          the $Z$ if it is clear from the context that we consider only the 
          integrand (which is the partition function) 
          but not the whole expression (which is the amplitude).} 
\begin{align} \label{kbcomplete1}
\widetilde{Z}_{\text{KB}}&= \sum_{\begin{smallmatrix}
                         g,k\in G   \\
                         g\,ks\Omega\,g =ks\Omega 
                       \end{smallmatrix}}
                        \widetilde{Z}_{\text{KB}_{(ks, ksg)}} \\ 
                       &= \sum_{\begin{smallmatrix}
                         g,k\in G \\
                          g\,ks\,g =ks 
                       \end{smallmatrix}}
                           \braket{{\cal C}(ks)}
         {e^{-2\pi l (H+\widetilde{H})_{h^2\text{-twisted}}}}{{\cal C}(ksg)}\\
       \label{completesq} 
       &= \sum_{h\in G} 
          \braket{{\cal C}|s;h}
         {e^{-2\pi l (H+\widetilde{H})_{h^2\text{-twisted}}}}{{\cal C}|s;h}
\end{align}
where we have defined the complete cross-cap state in the $h$-twisted tree
channel (with orientifold projection $s\Omega$, $s\in S$) by:        
\begin{equation}
    \ket{{\cal C}|s;h}\equiv \sum_{a\in G\atop (as)^2=h} \ket{{\cal C}(as)} 
\end{equation}
In \eqref{completesq} the Klein bottle tree channel integrand 
$\widetilde{Z}_{\text{KB}}$ is written as a sum of
{\it complete squares} with a product defined by 
$(a,b)=\braket{a}{\exp -2\pi l (H+\widetilde{H})}{b}$. (Taking into account that
a) the closed-string propagator does not mix differently twisted sectors and
that b) the scalar product of states in differently twisted sectors vanishes,
we can write \eqref{completesq} as a single complete square.) 
The situation is very similar to the case of orbifolds: In orbifolds the
knowledge of traces in the untwisted sectors is sufficient to determine
many  twisted sectors by modular transformations (cf.\ eq.\ \eqref{twisted4}).
In the orientifold we can determine cross-cap state normalizations
 from loop amplitudes
in the untwisted sector of the loop-channel by considering amplitudes
of the form:
\begin{equation}
      \widetilde{\text{KB}}_{(ks, ks)}=  \text{KB}_{(ks, ks)}  
\end{equation}
for all $k\in G,\;s\in S$. 
 This leaves still a phase for the
cross-cap state which we can not determine in this way.
 In principle
we have shown, how one can calculate the one-loop vacuum amplitudes for 
the closed-string sector of an orientifold. The two diagrams are the
torus from the orbifold theory and the Klein bottle amplitude which is
unique to orientifolds. We should be aware that the Klein bottle amplitude
describes a geometrically closed surface only on the orbifold space 
and in general
not in the ambient space. In addition for the Klein bottle amplitude to
describe a closed world-sheet in space time, one has to require $S=\{\id\}$.
The Euler character for general surfaces with $h$ handles, $b$ boundaries 
and $c$ cross-caps
 is given by: 
\begin{equation} \label{eulerch}
\chi=2-2h-b-c,
\end{equation}
In this sense both the torus and the Klein bottle are $\chi=0$ amplitudes,
i.e.\ first order in perturbation theory. Even though we can formally calculate
the amplitude we have so far
neglected one important problem which we will treat in the 
following section.
\subsection{Closed-string tadpoles} 
We recall that by choosing an appropriate fundamental region ${\cal F}_0$
 for the integration region of $\tau$ (cf.\ fig.\ \ref{fundregion1}) we 
can avoid singularities in the integrand of the torus amplitude
\eqref{torusamp}. However the modular group of the Klein bottle is trivial
and we will in general encounter a singularity (or multiple ones) 
in the integrand of
  \eqref{kbampl2} for $t\rightarrow 0$.\footnote{However 
there exist examples with finite KB amplitude.} This or these 
divergences are mapped
to a divergence (or multiple divergences)
 $\propto dl\,q^0\times(\text{volume factors})$
 for $l\rightarrow\infty$ in tree-channel.
There could be multiple divergencies with different volume dependencies 
(e.g.\ in orientifolds of (toroidal) orbifolds). However, we will say
``the'' divergence in the following. We can  give a physical interpretation of
the divergence if we observe that the $l\to\infty$ , $q^0$ part  of the 
tree-channel corresponds to 
massless closed-string states traveling an infinite distance in space 
(cf.\ fig. \ref{Kleinbottletreech}). In field theory there is an analogous
phenomenon called  {\it tadpole}.  We take as a simple  example the following  
action for a real scalar field $\phi$:
\begin{equation}
\int d^dx \bigr(\tfrac{1}{2}\partial_\mu\phi\partial^\mu\phi+ Q\phi\bigl)
\end{equation}
The equation of motion is: 
\begin{equation}\label{vaccum1}
 \partial_\mu\partial^\mu\phi=Q
\end{equation}
If we expand around $Q=0$ we  will encounter Feynman diagrams like 
\begin{equation}\label{tadpole1}
 \begin{aligned}
  \rule[4ex]{0ex}{5.1ex}
  \includegraphics{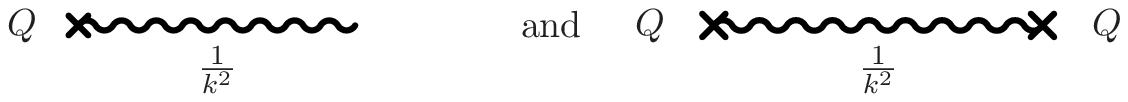}
 \end{aligned}
\end{equation}
Both diagrams have divergencies at vanishing momentum $k^2\to 0$.
If we rewrite the propagator $1/k^2$ as
\begin{equation}
 \frac{1}{k^2}=\int_0^\infty dl\,e^{-k^2l}
\end{equation}
we inspect that  this
divergence originates from huge times $l$ in the time evolution 
operator $\exp(-k^2 l)$.
We could have avoided the divergence if we would have expanded
around the true vacuum \eqref{vaccum1}. 
The divergence in the Klein bottle amplitude can also be traced back to
an ill defined vacuum. There are several plausible solutions to the problem:
one is to continously deform the values of background fields 
(in a space time dependent manner) such that the resulting theory is 
tadpole-free and consistent. 
This would be the so called {\it Fischler-Susskind mechanism}
\cite{Fischler:1986ci, Fischler:1986tb} or a generalization thereof.
However there exist situations where this continuous deformation does not
remove the tadpole. The Feynman graphs in \eqref{tadpole1} would be the
same if we  have introduced (uncanceled) background charges $Q$ to which
a gauge field couples (represented by the wiggled line in the diagram).
 These charges can be of topological nature and
they can not be canceled by deforming the background continously.  This
is usually the case in orientifolds.
As an example we take a closer look at the 10-dimensional Type I
string which can be derived from  Type IIB  superstring by 
constructing the orientifold with $O=\{1,\Omega\}$.
The spectra of both  Type IIB and Type I (closed-strings only) superstring
are listed in table \ref{clspecIIaI}.
\begin{table}
 \begin{center}
 \begin{tabular}{|c|c|c|}
  \hline
  Type IIB &\multicolumn{2}{c|}{bosons} \\
  \hline 
  & NS-NS& R-R \\
  & metric $g_{ij}$, 2-form $B_{ij}$,  & scalar $\chi$,
  2-form $B^\prime_{ij}$, \\ 
  \rule[-1.4ex]{0.0ex}{1ex}
  & dilaton $\phi$ & self-dual 4-form $D_{ijkl}$ \\
 \hline
 \hline
 & \multicolumn{2}{c|}{fermions}   \\
  \hline 
  & NSR &RNS   \\
 & gravitino $\psi_{i\dot{a}}$ & gravitino $\tilde{\psi}_{j\dot{b}}$ \\
 \hline \hline
  Type I &\multicolumn{2}{c|}{bosons} \\
  \hline 
  &NS-NS& R-R \\
  \rule[-1.4ex]{0.0ex}{1ex}
  &Metric $g_{ij}$, dilaton $\phi$ & 
  2-form $B^\prime_{ij}$,  \\
 \hline
 \hline
  &\multicolumn{2}{c|}{fermions}   \\
  \hline 
 &  \multicolumn{2}{c|}{1 gravitino \rule[-1.4ex]{0.0ex}{3.8ex} 
    $\psi_{j\dot{b}}$} \\
 \hline 
 \end{tabular}
 \caption{\label{clspecIIaI}Massless closed-string  spectra of Type IIB and
 Type I}
 \end{center}
\end{table} 
The spectrum of the (un-oriented) Type I theory is obtained by projecting
onto $\Omega$-invariant states. When one  computes  the
tree-channel KB amplitude a tadpole stemming  from an NSNS state and 
another tadpole from the  RR sector shows up.
 Both tadpoles have the same magnitude
 but opposite sign. The sum of both vanishes of course, since the tree-channel
amplitude can be rewritten as an integral over the 
partition function which has to vanish due to
supersymmetry. However both tadpoles are associated with physically 
distinguishable fields, which lead to equations of motion for both
 fields. 
The tadpole means that both field equations are not fulfilled in our
background.
The 10-dimensional superstring has a peculiar problem in the RR sector.
 The tadpole is Lorentz invariant. It should be transmitted by a scalar field
(or its Poincar{\'e} dual: in 10 dim.\ this is an 8-form).
 However no scalars exist in the (perturbative) RR sector of Type I
 (table \ref{clspecIIaI}).  But a 10-form potential $C_{10}$ would give
raise to a Lorentz invariant tadpole, if it appears via  
\begin{equation}\label{CStype1}
S_\text{C.S.}=\mu_{10}
\int d^{10}x\,C_{10}
\end{equation}
 in the action. The action would be invariant under
the gauge transformation $C_{10}\mapsto C_{10}+ d\Lambda_9$. So $C_{10}$
is in principle charged under this $U(1)$ symmetry. Dimensional reasons
forbid a kinetic term $F_{11}=dC_{10}$. 
So the eom takes the form $\mu_{10}=0$. This can not be fulfilled because
$\mu_{10}$ is a constant.  However the action \eqref{CStype1} should
be interpreted as the coupling of the form $C_{10}$ to a 10-dimensional 
object. The action \eqref{CStype1} is of topological type. It is a so
called {\it Chern Simons action}. We could now add other objects 
charged under $C_{10}$ s.th.\ the total action for $C_{10}$  vanishes.
These objects are {\cal D9-branes} in the case of the Type I string.
Thereby we automatically introduce open-strings and open-string diagrams.
Before we will have a closer look at the open-string sector of orientifolds,
we will make a comment about other possible tadpoles.\\
The eoms of a $p+1$ form field strength $H_{p+1}$ can 
be written as a generalization
of the Maxwell equations in 10 space-time dimensions:
\begin{align}
 d H_{p+1}&=\ast J_{8-p}& d\ast H_{p+1}&= \ast J_{p}
\end{align}
$J_{8-p}$ is the magnetic source and $J_{p}$ the electric one.
(The $J$'s depend in general on other fields of the theory as well.)
For $p=10$ like in Type I theory no field strength exists. As a consequence the
associated total charge $J_{10}$ has to vanish pointwise.
For $p<10$  and a spacetime without boundaries 
the eoms imply that the following integral vanishes by Stoke's theorem:
\begin{equation}
\int\limits_{{\cal M},\,\partial{\cal M }=\emptyset}\ast J_{k}=0
\end{equation}
In the low energy field theory limit the orientifold projection in the
closed-string sectors is described by adding expressions of the type (c.f.\
\cite{Morales:1998ux,Stefanski:1998yx}):
\begin{equation}\label{csoplane}
S_{\text{C.S.}} = Q_q\int_{O_q}\sqrt{
   \frac{\hat{{\cal L}}({\cal R}_T/4)}{\hat{{\cal L}}({\cal R}_N/4)}}\wedge
   \sum_{p\in \text{RR}} C_p
\end{equation}
These actions are of Chern Simons type and therfore topological. 
In \eqref{csoplane} we have introduced the Hirzebruch polynomial:
\begin{equation}
  \hat{{\cal L}}({\cal R}) =1+ \frac{p_1({\cal R})}{3}+
                   \frac{p_2({\cal R})-\bigl(p_1({\cal R})\bigr)^2}{45}+\ldots
\end{equation}
with $p_i$ the $i$\raisebox{1ex}{\tiny th} Pontrjagin class. ${\cal R}_T$
and ${\cal R}_N$ are the pull-backs of the curvature two-form to the tangent
and normal bundle of the subspace $O_q$. $O_q$ is called the
 {\it orientifold }$q$ 
{\it plane} (short: O$q$ plane). The O$q$ planes are in direct correspondence
to cross-cap states $\ket{\cal C}$: In geometric orientifolds, the $q$-planes
are $q$ dimensional subspaces which are left   point-wisely invariant
by the element
$gs$ ($g\in G$, $s\in S$) appearing in the definition \eqref{BoundstateKB2}
 of the cross-cap state $\ket{{\cal C}(gs)}$ ($O_q(gs)=\{x 
 \;|\;x\in {\cal X},\; x = gs\,x\}$). If we are able to
normalize the action \eqref{csoplane} correctly, we can in principle determine
the O-plane charge without doing the explicit CFT calculation. This
program has been carried out in \cite{Blumenhagen:2002wn} for a class
of orientifolds descending from supersymmetric Type IIA closed-string theories
  (compactified on Calabi-Yau spaces). In these models the set $S$ consists
 of a $\mathbb{Z}_2$-involution $\bar{\sigma}$ leaving invariant 
a $d/2$-dimensional subspace of the CY$d$ space. 
In addition the $\bar{\sigma}$ considered in this
publication exchanges  forms with complex  conjugate forms of the
 CY space (i.e. it maps a $(p,q)$-form to a $(q,p)$-form). The resulting
 O-(hyper)planes are {\it special Lagrangian submanifolds} (short:
 sLags).\footnote{A definition of a sLag is given in section \ref{slagsec}.}
 In the CFT description this leads in general to a supersymmetric
 closed-string sector (if one starts with a supersymmetric string theory).
Orientifolds of this kind have been also considered in 
\cite{Blumenhagen:1999md,Blumenhagen:1999ev,Blumenhagen:1999db,Blumenhagen:2001te}.
In \cite{Blumenhagen:2002gw} we investigated a $\mathbb{Z}_4$ orientifold
of this kind in four space-time dimensions which is supersymmetric in both 
closed- and open-string sectors and which is in addition chiral
in the open-string sector. Furthermore, it has other 
phenomenologically appealing features.
 We will return to this orientifold in chapter \ref{z4}.
The NSNS tadpole can be derived 
from a {\it Dirac Born Infeld}-type action which
for the O-plane is proportional to the volume of the hyperplane:
\begin{equation}\label{dbioplane}
 S_{\text{DBI}} = T_q\int_{O_q}e^{-\Phi} \sqrt{-\det G}
\end{equation}
$G$ is the pullback of the space-time metric to the O$q$-plane,  
$\Phi$ the Dilaton and the constant $T_q$ is the so called {\it tension} 
of the 
O$q$-plane. 
The action \ref{dbioplane} is {\sl not} topological in contrast to
 the Chern Simons action \eqref{csoplane}. 
In principle the NSNS tadpole can be removed by a continuous
deformation of the background. However this deformation can lead to a
degenerate space (e.g.\ zero or infinite volume in compactifications).  

\begin{table}
 \begin{center}
 \begin{tabular}{c|c|c|c}
   \multicolumn{1}{c|}{} &
  Klein bottle & Annulus & M\"obius strip \\ \hline
  loop-channel (direct channel) &$t$&$t$&$t$\\
  tree-channel (transverse channel)  \rule[-1.4ex]{0.0ex}{3.8ex}&
        $\tfrac{1}{4l}$&$\tfrac{1}{2l}$&$\tfrac{1}{8l}$\\
 \end{tabular}
 \caption{\label{transftreeloop} Relation between the parameters $t$ 
  and $l$ in the loop- and tree-channel.}
 \end{center}
\end{table}
\section{\label{openstringssec}Open-strings}
 Open-strings are very similar to closed-strings. However
open-string diagrams have boundaries. Their Lagrangian can be written 
as a sum of a bulk and of a boundary Lagrangian.\footnote{We consider
for simplicity only string theories which can be described by Lagrangians.}
 The bulk Lagrangian is the same for closed- and open-strings.
Though it is  only integrated from $\sigma=0$ to $\sigma=\pi$. In other
words: The open-string has half the length of a closed-string. This 
normalization is important for the comparison of different closed- and
open-string amplitudes and has direct impact on the tadpole cancellation
conditions. 
We will have a closer look at the details of the boundary terms
in the next chapter. 
There also the open-string Lagrangian can be found.
Here we only summarize some results.   The boundary conditions for 
the open-string take the form:
\begin{equation}\label{bdyc1}
 \partial_+ X(\tau,\sigma)= V_{(i)}\partial_- X(\tau,\sigma)\, ,
          \, \sigma\in\partial{\cal M}_i
\end{equation}
${\cal M}_i$ is the $i$\raisebox{1ex}{\tiny th} connected part of the
 world sheet boundary. We can associate it with an object 
usually called a {\it D-brane}. Sometimes the word D-brane implies
more: e.g. a special amount of space time supersymmetry
which is preserved by this kind of boundary condition. In order
to consider space-time supersymmetry one has to generalize these boundary
conditions to the fermionic sector of the world-sheet fields (if one 
applies  RNS formalism, cf.\ section \ref{susyori}).
 We will consider Chan-Paton degrees of freedom (dofs)
 later
which lead to a slightly different characterization of a D-brane. 
The $V_{(i)}$ are matrices that specify the boundary conditions.
In all of our concrete examples they are orthogonal matrices.
As will be shown in chapter \ref{strbg}, 
electric fields coupling to the boundary will invoke
$V_{(i)}\in SO(1,d-1)$. There the concrete form of the $V_{(i)}$ can be
found (cf.\ eq.\ \eqref{defV}, p.\ \pageref{defV}). 
 
Equation \eqref{bdyc1} does not include zero-modes. Zero-modes
are nevertheless important as they govern multiplicities in the spectrum.
 We will not enter in the details here. In the examples presented
in this work the zero-mode contribution can be derived consistently. 
 We are especially interested in
$\chi=0$ diagrams. According to eq.\ \eqref{eulerch} the remaining diagrams 
are:
\begin{figure}
  \setlength{\unitlength}{0.1in}
  \begin{picture}(60,21)
    \SetFigFont{14}{20.4}{\rmdefault}{\mddefault}{\updefault}
   \put(15.5,6){\scalebox{0.5}{\includegraphics{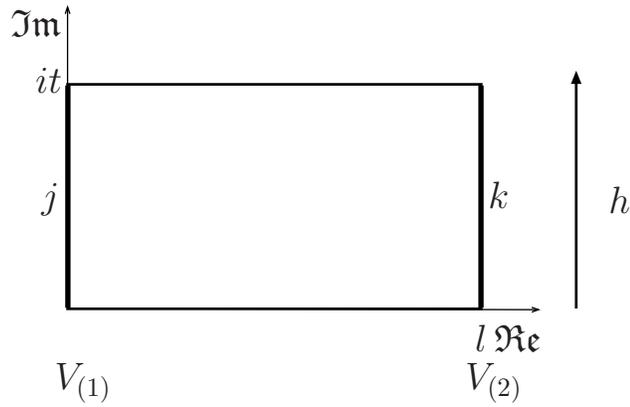}}}
    \put(12.7,20.5){$\mathfrak{Im}$}
    \put(14,17.5){$it$}
     \put(14.3,11.5){$j$}
     \put(37.7,11.5){$k$}
    \put(44,11.2){$h$}
    \put(37,4.){$l$}
     \put(38,4){$\mathfrak{Re}$}
    \put(15.,2.){$V_{(1)}$}
    \put(36.5,2.){$V_{(2)}$}
  \end{picture}
 \caption[Periodicities of the cylinder with
                           boundary conditions \ldots]
             {\label{cylfund} The periodicities of the cylinder with
                           boundary conditions $V_{(1)}$ and  $V_{(2)}$
                           as well as Chan Paton labels $j$, $k$.}
\end{figure}
\begin{itemize}
 \item Cylinder (or annulus) ($b=2,\, h=c=0$)
 \item M\"obius strip ($b=c=1$,\, $h=0$)
\end{itemize}
\begin{figure}
\vspace{0.5cm}
\hspace{1.65cm}
\begin{picture}(0,0)%
\put(45,10){\epsfig{file=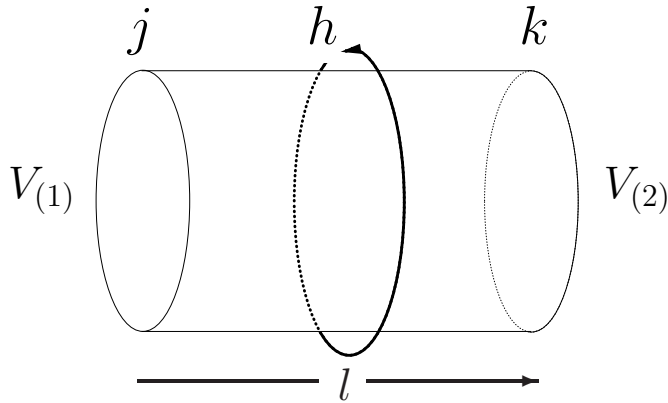}}%
\end{picture}%
\setlength{\unitlength}{1184sp}%
\begingroup\makeatletter\ifx\SetFigFont\undefined%
\gdef\SetFigFont#1#2#3#4#5{%
  \reset@font\fontsize{#1}{#2pt}%
  \fontfamily{#3}\fontseries{#4}\fontshape{#5}%
  \selectfont}%
\fi\endgroup%
\begin{picture}(13626,7529)(1001,-6989)
\put(14176,-3061){\makebox(0,0)[lb]{\smash{\SetFigFont{17}{20.4}{\rmdefault}{\mddefault}{\updefault}
$V_{(2)}$
}}}
\put(7951,164){\makebox(0,0)[lb]{\smash{\SetFigFont{20}{24.0}{\rmdefault}{\mddefault}{\updefault}
$h$
}}}
\put(4200,164){\makebox(0,0)[lb]{\smash{\SetFigFont{20}{24.0}
        {\rmdefault}{\mddefault}{\updefault}
$j$
}}}
\put(12400,164){\makebox(0,0)[lb]{\smash{\SetFigFont{20}{24.0}
        {\rmdefault}{\mddefault}{\updefault}
$k$
}}}
\put(1701,-3061){\makebox(0,0)[lb]{\smash{\SetFigFont{17}{20.4}{\rmdefault}{\mddefault}{\updefault}
$V_{(1)}$
}}}
\thicklines\put(4400,-6900){\line(1,0){3800}}
\put(9200,-6900){\vector(1,0){3600}}
\put(8600,-7300){\SetFigFont{17}{20.4}{\rmdefault}{\mddefault}{\updefault}$l$}
\end{picture}
\caption[Cylinder diagram]{\label{cylfig} 
The cylinder diagram. The boundary conditions are represented by
$V_{(1)}$ and $V_{(2)}$. The twist in the closed-string channel is denoted by 
$h$. As it can be also viewed as a loop diagram, we have included the  Chan
 Paton indices $j$ and $k$. These have to fulfill
the condition $h(j)=j$ and $h(k) = k$. $h$ acts differently 
 on the left and right CP index (cf.\ text). }
\end{figure}
The annulus is depicted in figure \ref{cylfig}. 
We have already included possible
twists in the boundary conditions which appear if we consider the CFT on 
this cylinder. The cylinder is not obtained by modding out a torus CFT
by a finite group. The world sheet of a torus is simply a square with
two opposite sides identified in an  orientation preserving way. 
(In figure \ref{cylfund} the horizontal lines are identified.)
 Therefore the torus has only one fundamental region. The complex structure
is however fixed to be purely imaginary due to the boundary conditions 
\eqref{bdyc1}.
Nevertheless, 
there exists a notion of a loop- and a tree-channel as well. The loop-channel
corresponds to open-strings of length $\pi$ with end-points $k$ and $j$
(cf.\ fig.\ \ref{cylfig}) propagating in a loop. In the path-integral
the corresponding fields are periodic in the world-sheet $\tau$-direction
up to a twist $h\in G$. This implies that  the  boundary conditions are 
invariant as well as the CP-labels (in figure  \ref{cylfig}: $j$ and $k$),
 which we consider in section \ref{chanp}.
The tree-channel is just another parameterization of the one cylinder (fig.\ 
\ref{cylfund}). It is interpreted as closed-strings (of length $2\pi$) 
propagating from (closed-string) world sheet time $\sigma=0$ to $\sigma=l$. 
   The correspondence 
between tree- and loop-channel parameters is computed in the same way as in
the Klein bottle by Weyl rescaling with $\lambda= 1/t$ such that 
$l=1/2{\mapsto}l=1/(2t)$ (taking into account the length
of the open-string). This result is listed in table \ref{transftreeloop}.
In analogy to cross-cap states the so called {\it boundary states}
are defined by (up to normalization and phase):
\begin{align}
  \label{bdystatedef1}
 \bigl(\partial_+ X(\tau=0,\sigma) + V \partial_-
              X(\tau=0,\sigma)\bigr)_{h\text{-twisted}}
        \ket{{\cal D} \,(V);h}=0  
\end{align}
We also have to impose analogous conditions on the zero-modes.
In addition the boundary condition specified by $V$ (c.f.\ \eqref{bdyc1})
has to be invariant under $h$ (c.f.\ section \ref{osonorb},
 eq.\ \eqref{bdycondg}).
The world-sheet coordinates in \eqref{bdystatedef1} correspond
to the tree-channel parameterization.
\subsection{\label{chanp}Chan Paton factors and gauge symmetries}
In fig.\ \ref{cylfig} we have already included so called {\it Chan Paton}
indices (short: CP-indices). Their name is devoted to its inventors 
\cite{Paton:1969je}.
 It has been observed that additional dofs can be added to the endpoints
of open-strings. In general a string vacuum (on which the perturbative
Fock-space is built by creation operators) looks like:
\begin{equation}\label{cpwvfc}
  \ket{\theta(i,j);a;b}
\end{equation}
$a$ and $b$ are the CP-indices associated with the $\sigma=0$ and
$\sigma=\pi$ ends of the open-string, $\theta(i,j)$  labels sectors 
which are given by boundary
conditions including \eqref{bdyc1} and conditions for the zero modes as well.
The first entry in $\theta$ refers to the   $\sigma=0$ point (which we
call the left end-point), the
other to the $\sigma=\pi$ boundary (the right end-point)
of the open-string, in analogy with the CP indices.
We have assumed here that the $V_{(i)}$ of \eqref{bdyc1} do not act on the
CP degrees of freedom. However we do not want the states \eqref{cpwvfc} to
  factorize. This means that the CP-Hilbert spaces are in general 
 distinct in different sectors $\theta(i,j)$ of the ambient
 theory:\footnote{By ambient we mean that we have not performed any orbifold
   projection so far. As we will see in section \ref{osonorb},
 the structure of ${\cal H}$ might change on orbifolds.}
 \begin{equation}\label{hilbspacebd}
 {\cal H} = \bigoplus_{i\in \text{U}}\bigoplus_{j\in \text{U}}
                          \bigl({\cal H}^{\text{L}}_{(i)}
                              \otimes {\cal H}^{\text{R}}_{(j)}\bigr)
                          \otimes {\cal H}_{\theta(i,j)}
 \end{equation}
 ${\cal H}^{\text{L}}_{(i)} $ and ${\cal H}^{\text{R}}_{(j)}$
 correspond to the CP-Hilbert
space associated with the left and with the right string end-point. 
On the other hand the string end-points  are 
associated with (a set of) D-branes obeying the boundary  condition 
\eqref{bdyc1}. The left and right Hilbert spaces are not independent. In order
to define a consistent perturbation theory, left and right spaces have
to be identified.  
 For string theory with CP dofs to
 be consistent, the corresponding vertex operators of  the states
 \eqref{cpwvfc}
have to exist. The interactions described with the help of these 
operators have also to be consistent. This is possible with the
Hilbert space structure given by
\eqref{hilbspacebd} (though we give no rigorous proof). We assume
that ${\cal H}^{\text{L/R}}_{(i)}$ is a
 $n(i)$  dimensional vector spaces over $\mathbb{C}$ 
($n<\infty$).
The Chan Paton states  can be represented by $n\times
 m$ matrices $\Lambda^{(i,j)}_{ab}\in\mathbb{C}^{n(i)}
  \otimes\mathbb{C}^{m(j)}$.
 The inner product  induced by the hermitian product on $\mathbb{C}^{n(i)}$
and $\mathbb{C}^{m(j)}$ takes the form:
\begin{equation}\label{innerprodcp}
 \big<\theta(i,j);\Lambda_1^{(i,j)}\big| \theta(k,l);\Lambda_2^{(k,l)}\big>
 =\delta^{ik}\delta^{jl}\, \tr\bigl({\Lambda_1^{(i,j)}}^\dagger 
      \,\Lambda_2^{(k,l)}\bigr)
\end{equation}
 Different states of a quantum theory should be normalized. Furthermore,
 the (overall) phase of the (complete) state is not important for the
 amplitudes that we measure. For identical left and right
 Hilbert spaces (which we require  in the case of identical
 boundary conditions $(i,j)=(i,i)$)
 the set of hermitian matrices with the following normalization:
\begin{equation}
 \tr\Big(\Lambda^{(i,i)}_{(a)}\Lambda^{(i,i)}_{(b)}\Big) =\delta_{ab}
 \qquad a,b=1\ldots \bigl(n^{(i)}\bigr)^2
\end{equation} 
forms a complete set of states in the $n^{(i)}\times n^{(i)}$-dimensional 
Hilbert space 
${\cal H}^{\text{L}}_{(i)}\otimes {\cal H}^{\text{R}}_{(i)}$.
The $\mathbb{C}$-vector space of these matrices forms a representation space
for the irreducible adjoint representation of $U(n)$. The 
adjoint representation of $SU(n)$ acts however only reducibly on this space. 

In scattering amplitudes two such CP matrices $\Lambda$
get multiplied if they belong to adjacent string endpoints
 of open-string vertex operators (no sum over
upper indices which specify the boundary conditions):
\begin{equation}\label{cpampl1}
  {\cal A}_{1,2,\ldots n}\propto\tr
  \bigl( \Lambda^{(i,j)}_1\Lambda^{(j,k)}_2 \ldots \Lambda^{(l,i)}_n\bigr)
\end{equation}
By this form we have implicitly identified left with right CP-Hilbert spaces:
 ${\cal H}^{\text{L}}_{(i)}\sim{\cal H}^{\text{R}}_{(i)}$.
Of course one will also encounter sums over expressions of the type 
\eqref{cpampl1} if one considers permutations of vertex operators. 
Inner boundaries (i.e.\ without external string states attached) get 
multiplied just by:
\begin{equation}\label{innersum}
\sum_i \sum_a \tr \Lambda^{(i,i)}_{(a)} \qquad i\text{: inner boundary} 
\end{equation}
The expressions \eqref{cpampl1} and \eqref{innersum} 
 are clearly invariant if we perform the following combined transformations:
\begin{equation}
\begin{aligned}
  \Lambda^{(i,j)}&\mapsto  \Lambda^{(i,j)}U^{(j)}\\
  \Lambda^{(j,k)}&\mapsto  \bigl(U^{(j)}\bigr)^{-1}\Lambda^{(j,k)}
\end{aligned}  
\end{equation}
$ U^{(j)}$ is associated with the $j$\raisebox{1ex}{\tiny th} stack of
D-branes. To be a symmetry of the full quantum theory, $U^{(j)}$ should
preserve the inner product \eqref{innerprodcp} (or at least its modulus)
 so that it is restricted
to be a unitary (or anti-unitary)   
map of dimension $n(j)$. 
(We restrict ourself to the unitary case for simplicity in which case 
$U^{(j)}$ is described by a unitary matrix.)
Note that in this sense the left Hilbert space ${\cal H}^{\text{L}}_{(i)}$ 
transforms in the {\sl complex conjugate} representation in 
comparison to the right Hilbert space ${\cal H}^{\text{R}}_{(i)}$ :
\begin{equation}
 \begin{array}{rccc}
 {\cal H}^{\text{L}}_{(i)}\otimes{\cal H}^{\text{R}}_{(j)} & 
 \ni v^{\text{L}}\otimes w^{\text{R}}
 &\xrightarrow{U^{(i)} \times U^{(j)}}
  &\bigl(U^{-1}_{(i)}v\bigr)\otimes\bigl(U^{T}_{(j)}w\bigr) \\
 &\,\Lambda^{(i,j)}& \xrightarrow{\phantom{U^{(i)}\times U^{(j)}}}
              &U^{-1}_{(i)} \Lambda^{(i,j)}U_{(j)} 
 \end{array}
\end{equation}
 It turns out that strings
in the sector $\theta(j,j)$ give rise to vectors\footnote{In most
compactifications
this sector contributes scalars transforming in the adjoint
 representation as well.}. The global $U(n(j))$ symmetry
is then lifted to a local one, i.e.\ a {\it gauge symmetry}.
Strings in sectors $\theta(i,j)$ with $i\neq j$ usually contain  massless
matter as moding and vacuum energy are lifted (cf.\ chapter \ref{strbg}). In
addition Lorentz symmetry gets broken by this kind of boundary conditions.
The open-string massless matter states then transform in the bifundamental 
$n(i)\times \Bar{n}(j)$ of $U(n(i))\times U(n(j))$.
It is natural to ask whether the Chan-Paton Hilbert-space 
${\cal H}^{\text{L}}\otimes {\cal H}^{\text{R}}$
which we assumed to be a tensor product $\mathbb{C}^{n}
  \otimes\mathbb{C}^{m}$ might be projected to a smaller space, thereby
  reducing the $U(n)\times U(m)$ to some smaller symmetry group. It turns
  out that this is indeed possible. However not all gauge groups are possible
  and there are constraints. Related with these issues is the question if
  unitarity conditions like {\it factorization of tree level amplitudes} are
  respected. For unitary gauge groups these conditions are obeyed. 
  Unitary gauge groups or products thereof are the only perturbative 
  gauge-symmetries of oriented closed string Type II spectra.

We will now consider the case where additional symmetries are modded out, such
that the Hilbert space $\cal H$, eq.\ \eqref{hilbspacebd} gets modified.

\subsubsection{\label{osonorb}Open-strings on orbifolds}
In building the orbifold we have first to symmetrize our set of boundary
conditions represented by  the $V_{(i)}$ i.e.\ we consider all boundary 
conditions of the type $gV_{(i)}$. To be more specific: we write $g$ 
as $(g_+,g_-)$, thereby keeping  the possibility of left-right
 asymmetric twists. 
(The symmetric case is represented by: $g_+=g_-$.) We deduce with the help
of \eqref{bdyc1} that under $\partial_+X\mapsto g_+\partial_+X$, 
$\partial_-X\mapsto g_-\partial_-X$, the $V_{(i)}$ map as:
\begin{equation} \label{bdycondg}
 (g_+,g_-):\, V_{(i)}\mapsto V_{g(i)}\equiv g_+^{-1}V_{(i)}g_-
\end{equation}  
The zero modes map as well. However there is no {\it a priori} well defined
action on the zero modes for asymmetric $g$. In addition, $g$ as a symmetry
admits a unitary (possibly projective) representation  on
the CP factors: 
$\Lambda^{(i,j)}\in{\cal H}^{\text{L}}_{(i)}\otimes {\cal H}^{\text{R}}_{(j)}$ 
with  
\begin{equation}\label{gorbaction}
 \Lambda^{(i,j)}\mapsto 
  \bigl(\gamma_{g}^{(i)}\bigr)^{\dagger}\Lambda^{\left(g(i),g(j)\right)}
   \gamma^{(j)}_g
\end{equation}
$\gamma$ is the matrix representation of g.
The $G$-invariant open-string states have to obey:
\begin{equation}\label{ginvopst}
g \ket{\psi}=\ket{\psi} \qquad \forall g \in G
\end{equation}
The states in \eqref{ginvopst} are in 
general linear combinations of states in the ambient space. However
$g$ does not mix or exchange left and right end-points. As a result each state 
will still be
built on a CP-Hilbert space of the form ${\cal H}^{\text{L}}_{(i,n)} \otimes
{\cal H}^{\text{R}}_{(j,n)}$. Now the indices  
$i$ and $j$ need no longer refer to a single pair of boundary conditions
$\theta(i,j)$ but they can belong to g-invariant linear combinations.
 The index
$n$  indicates that in general the CP-space depends also on the oscillator 
excitations. We will not step into a classification of all possible 
resulting Hilbert-spaces. We only state that the  the gauge
group of the orbifolded theory is a product
\begin{equation}
 \prod_{i=1}^{n} U(n_i)
\end{equation}
 of unitary groups $U(n_i)$. The breaking by $G$ 
 occurs in general when the boundary conditions $V_{(j)}$
   are $g$-invariant, and the $U(n_j)$ group is broken 
  down to a product of groups $\prod_{i} U(n_i)$ by the action of $g$ .

 Our analysis was restricted to 
 tree-level. Up to this point it is fully consistent to mod out by twists
 which act
 only on the CP space while being trivial in space-time (i.e.\ not only
 trivial on the bdy.-condition but trivial on $\ket{\psi_{\text{osc}}}$,
  too). However such kind of gauge symmetry breaking 
 would be  inconsistent at one-loop order of string
 perturbation theory. 
 
 To see this,
 we have a look at the non-planar one-loop diagram with four external states 
in figure \ref{cpcons}. We assume
 that the orbifold group $G$ splits (or more appropriately: projects) the
  CP-Hilbert space such that 
 \begin{equation}
  U(n_1+n_2)\xrightarrow{{\cal P}_g} U(n_1)\times U(n_2)
 \end{equation}
 Without the projection ${\cal P}_g$ there would be (massless) 
 states with CP indices
 in the off-diagonal blocks, while after the projection, they have vanished
 (seemingly):
 \begin{equation}
  \Lambda^{(1\cup 2,1\cup 2)}
   =\biggl(\begin{array} {c|c}
     \Lambda^{(1,1)} & \Lambda^{(1,2)}
    \\ \hline
     \rule[0.ex]{0.ex}{2.55ex} \Lambda^{(2,1)} & \Lambda^{(2,2)}     
   \end{array}
   \biggr) 
   \xrightarrow{{\cal P}_g}\Lambda^{\prime\,(1\cup 2,1\cup 2)}
    =\biggl(\begin{array} {c|c}
     \Lambda^{(1,1)} & 0
    \\ \hline
     \rule[0.ex]{0.ex}{2.55ex} 0  & \Lambda^{(2,2)}     
   \end{array}
   \biggr) 
 \end{equation}
 \begin{figure}
   \setlength{\unitlength}{0.1in}
   \begin{minipage}[t]{\textwidth}
   \begin{minipage}[t]{6cm}
    \begin{center}
    \begin{picture}(25,30.)
    \SetFigFont{14}{20.4}{\rmdefault}{\mddefault}{\updefault}
    \put(1,0.2){\scalebox{0.22}{\includegraphics{picori7.EPS2}}}
     \put(3.3,27.7){${\cal V}_a$}  \put(0.5,1){${\cal V}_b$}
      \put(26.,26.5){${\cal V}_c$}  \put(28,.0){${\cal V}_d$} 
    \end{picture}
   \end{center}
   \end{minipage}
   \hfill\vspace{1cm}
    \begin{minipage}[t]{5cm}
    \begin{center}\caption[Non-planar open-string one-loop diagram with four
            external states]
            {\label{cpcons}Non-planar open-string one-loop diagram with four
            external states. We assume the states 
              ${\cal V}_a$ and ${\cal V}_b$ to transform in the $U(n_1)$
            while the states ${\cal V}_c$ and ${\cal V}_d$ transform in the
            second gauge group, the $U(n_2)$.}
    \end{center}
    \end{minipage}
  \end{minipage}
   \rule{0cm}{1cm}
   \setlength{\unitlength}{0.1in}
   \begin{minipage}[t]{\textwidth}
   \begin{minipage}[t]{6cm}
   \begin{center}
   \begin{picture}(25,29.4)
   \SetFigFont{14}{20.4}{\rmdefault}{\mddefault}{\updefault}
   \put(1.5,0.4){\scalebox{0.20}{\includegraphics{picori8.EPS2}}}
      \put(8.,26.5){${\cal V}_a$}  \put(0.,15.){${\cal V}_b$}
      \put(29.,22.5){${\cal V}_c$}  \put(33.3,9.){${\cal V}_d$}
       \put(6.5,2){${\cal V}_i$} \put(23.4,-1.5){${\cal V}_i$}
   \end{picture}
    \end{center}
    \end{minipage}
   \hfill
     \begin{minipage}[t]{3.6cm}
     \begin{center}
     \caption[Cut of the non-planar open-string one-loop diagram with
            external states]
             {\label{cpconscut}After cutting the above non-planar diagram between
              the external states ${\cal V}_b$ and ${\cal V}_d$ we obtain the
             diagram on the left. It has the topology of a disk-diagram. \\ 
               ${\cal V}_i$ corresponds to internal states. ${\cal
             V}_i$ takes Chan-Paton values in the bifundamental of $U(n_1)\times
             U(n_2)$.}
    \end{center}
    \end{minipage}
  \end{minipage}
  \end{figure}
 In non-planar loop diagrams the states with values in the off-diagonal
 components $\Lambda^{(1\cup 2,1\cup 2)}$ re-enter:
 In figure \ref{cpcons} The states ${\cal V}_a$ and ${\cal V}_b$ 
 transform in the $U(n_1)$ while the two other states 
  ${\cal V}_c$ and ${\cal V}_d$ 
 transform in the second factor, i.e.\ the $U(n_2)$. We make this transparent 
 by putting different colors on the two boundaries in the figure.
 According to the rules, this amplitude is therefore proportional to
 \be \label{planaramp}
   {\cal A}_{a,b,c,d}\propto
  \tr\bigl( \Lambda^{(1,1)}_a\Lambda^{(1,1)}_b\bigr)
  \cdot\tr\bigl( \Lambda^{(2,2)}_c\Lambda^{(2,2)}_d\bigr)
 \ee  
 Unitarity of string theory is manifest in cutting rules. 
 In figure \ref{cpconscut} we cut the non-planar diagram between the 
 external states (or: strips) ${\cal V}_b$ and ${\cal V}_d$. The resulting
 diagram represents a disk-amplitude with six vertex-operator insertions, 
two of the stemming from the cut, which we denote by ${\cal V}_i$.
 According to our rules, this  amplitude is proportional to 
 \be 
   {\cal A}_{a,b,i,c,d,i}\propto
  \tr\bigl( \Lambda^{(1,1)}_a\Lambda^{(1,1)}_b\Lambda^{(1,2)}_i
   \Lambda^{(2,2)}_c\Lambda^{(2,2)}_d\Lambda^{(2,1)}_i\bigr)
 \ee 
 For
 unitarity reasons, by summing over internal states
 ${\cal V}_i$ we should obtain the planar amplitude
 \eqref{planaramp}. This is only possible, if the matrices 
 $\Lambda^{(1,2)}_i$ are non-vanishing for the internal states.
 As we require unitarity these internal states have to show up in the
 spectrum.
 If the space times twist in the orbifold group $G$ are trivial, we get 
 massless vectors in the bifundamental representation of 
$U(n_1)\times U(n_2)$.  
 These massless vectors can be combined with the massless 
 vectors  that transform in the adjoint representation of
  $U(n_1)$ and $U(n_2)$  to 
  fill the representation adjoint representation of $U(n_1+n_2)$. This
  indicates that the full $U(n_1+n_2)$ gauge symmetry
 is still present.\footnote{I want to thank Stephan Stieberger for clarifying
 the above discussion.}
 Thus gauge symmetry breaking by pure gauge twist (i.e.\ a group $G$ 
 that solely acts on the CP Hilbert space) is forbidden by unitarity.  
We require a consistent representation of the twists $g$ 
 on both the Chan-Paton wave functions and on the oscillator part
 $\ket{\psi_{\text{osc}}}$.
 Other restrictions on the representation stem from tadpole 
 cancellation conditions.

 In a similar spirit, we demand the representation of 
 $g\in G$ on the CP-Hilbert space to depend only on the boundary 
 condition $V_{(i)}$ of the corresponding left or right string-endpoint.
 In other words: it is not possible to split {\it a priory}  the set of
 D-branes with identical boundary conditions into several sets 
  on which $g$ then acts differently: $N_i$ branes with boundary condition
 $V_{(i)}$ give always rise to a $U(n_i)$ gauge group which might 
 get broken by a consistently acting orbifold group. However one
 can modify the boundary conditions by introducing different
 {\it Wilson lines} on the respective set of branes, thereby evading
 the restriction (cf.\ chapter \ref{z4}).

 There is another subtlety 
 concerning the zero modes appearing in the context of so called 
 {\it fractional branes} (also: {\cal twisted branes}). These are branes
 coming from sectors with $V_{(i)}=V_{g(i)}$: The boundary condition associated
 with the {\sl single} brane is  already symmetric 
 under the group $G$. Therefore one does not need to introduce its
 $G$-pictures. As a consequence they are ``smaller'' (in the ambient space of 
 the orbifold) than the D-branes which originate from non $G$-invariant 
 boundary conditions. As branes carry charges, the fractional branes will carry
 a smaller amount of charge than the objects which are obtained by explicitly
 symmetrizing over $g\in G$. 
 
 So far we have not established a link
 between the CP-degrees of freedom and the boundary-states. We know however
 that  loop- and tree-channel are only different parameterizations of the
 same world sheet, and therefore the amplitudes are identical. However closed
 strings do not carry CP-degrees of freedom. They also do not appear
in the   definition of the $V_{(i)}$. 
Keeping in mind the tensor product structure of the CP Hilbert space of the
 non-orbifolded theory (eq.\ \eqref{hilbspacebd}) we notice that the 
loop-channel trace (in the operator formalism) splits into a product of
 traces over the CP-Hilbert space
times a trace over the boundary conditions. Suppressing the trace over 
the CP-Hilbert space we get for each CP-sector $\Lambda^{(i,j)}$:\footnote{As 
we have assumed that the $V_{(i)}$  act trivially on the CP-Hilbert space of
the ambient space, we assume that this is the case for the orbifold, too. 
 (I.e.\ the $V_{g(i)}$, $g\in G$ do not carry CP-indices.)}
 \begin{align}
  \text{A}^{ij}_{ab}(h) 
  &=  \frac{V_{10}}{(2\pi)^d}
   \,\lim\limits_{\epsilon\to 0}
   \int_\epsilon^\infty\frac{dt}{t} Z_{\text{A}^{ij}_{ab}(h)}\\
  &=\int dl\,
   \braket{{\cal D}(V_{(i)});h;a}
   {e^{-2\pi l (H+\widetilde{H})_{h\text{-twisted}}}}{{\cal D}(V_{(j)});h;b}\\
   Z_{\text{A}^{ij}(h)}&=Z_{\text{A}^{ij}_{ab}(h)}\equiv\frac{1}{|O|}
   \tr h\, e^{-2\pi t(H_{\theta(j,l)})} 
 \end{align}
 Tracing both expressions over the CP-Hilbert space 
 ${\cal H}^{\text{L}}_{(i)}\otimes {\cal H}^{\text{R}}_{(j)}$ and taking
 into account that $h$   acts on it by 
 $\gamma_{h}^{(i)}\otimes\bar{\gamma}_{h}^{(j)}$
 we can rewrite the complete annulus amplitude in the sector given 
 by $\theta(i,j)$ as:
 \begin{align}\label{annf1}
 \text{A}^{ij}(h)
 &\equiv  
  \Bigl(\tr\gamma^{(i)}_h\Bigr)
  \Bigl(\tr\bar{\gamma}_{h}^{(j)}\Bigr)
  \frac{V_{10}}{(2\pi)^d}
  \,\lim\limits_{\epsilon\to 0}
  \int_\epsilon^\infty\frac{dt}{t} Z_{\text{A}^{ij}(h)}\\ 
 &= \Bigl(\tr\gamma_{h}^{(i)}\Bigr)
  \Bigl(\tr\bar{\gamma}^{(j)}_h\Bigr) \int dl\,
  \braket{{\cal D}(V_{(i)});h;b}
  {e^{-2\pi l (H+\widetilde{H})_{h\text{-twisted}}}}{{\cal D}(V_{(j)});h;b}\\
 &\ket{{{\cal D}(V_{(j)});h}}\equiv 
  \tr\Bar{\gamma}^{(j)}_h\cdot\ket{{\cal D}(V_{(j)});h;b} 
 \end{align} 
 Therefore the trace-insertion in  loop-channel is mapped to the
 normalization and to the closed-string twist $h$
 of the boundary state. (Note that $\ket{{\cal D}(V_{(j)});h;b}$ does not 
 depend  on the CP-index $b$ due to  the assumption 
  that $h$ preserves the structure 
 ${\cal H}^{\text{CP}}\otimes {\cal H}_{\theta(ij)}$.)
 The whole annulus contribution can be written  as a sum over
 perfect squares:
 \begin{align} \label{anntrace}
  \text{A}
  &\equiv\sum_{h\in G}\sum_i\sum_j\text{A}^{ij}(h)\\
  &=\sum_{h\in G}
   \int dl\,
   \bra{{\cal D};h}
   e^{-2\pi l (H+\widetilde{H})_{h\text{-twisted}}}\ket{{\cal D};h}
  \\
  \ket{{\cal D};h}&\equiv\sum_j\ket{{\cal D}(V_{(j)});h}
 \end{align} 
 $j$ runs over a set of $G$-symmetrized boundary conditions.
 Taking into account that vacua of different $h$-twists have 
 vanishing scalar products and observing that the closed-string propagator
 ${\cal P}_{\text{cl}}=\sum_h\int dl\,
 \exp\bigr({-2\pi l (H+\widetilde{H})_{h\text{-twisted}}}\bigr)$
 does not mix the twisted sectors, we can rewrite:
 \begin{align}
 \text{A}&= \int dl\,
  \sum_{i\in G\text{-inv.}\rule{3.5ex}{0.ex}\atop\text{set}}
  \negphantom{\rule{2.5ex}{0.ex}}
  \sum_{j\in G\text{-inv.}\rule{3.5ex}{0.ex}\atop\text{set}}
   \bra{{\cal D}(V_{(i)})}
  e^{-2\pi l (H+\widetilde{H})}\ket{{\cal D}(V_{(j)})}
 \\
  \ket{{\cal D}(V_{(j)})}&\equiv{\cal S}_j\sum_{h\in G}
   \ket{{\cal D}(V_{(j)});h}
   \qquad\text{no sum over } j 
\end{align}    
  ${\cal S}_j$ defines a symmetrization of the boundary conditions
 in the sector described by $V_{(j)}$. The choice of ${\cal S}_j$ leaves
 some freedom: if a brane is symmetric under $h\in G$ its image may 
 (which would
 be just a doubling of dofs) or may not be included.
 Invariance
 under $h$ includes also the invariance of the zero modes.
 By the doubling, a non-trivial action $\gamma_h$  on the CP dofs can be
 chosen (cf.\ eq.\ \eqref{gorbaction}). As the amplitude is proportional
 to the product of the traces of both $\gamma_h$-matrices 
 (cf.\ eq.\ \eqref{annf1}), the $h$-twisted part of the corresponding
 boundary state  vanishes, if the corresponding trace $\tr\gamma_h$ equals
 zero. In geometric orbifolds the massless closed-string states of the
 twisted sectors can be associated with blowing up modes of the singularity.
 (For  Calabi-Yau orbifolds, these  fields are contained in 
 $H^{(1,1)}({\cal M})$.) $\tr\gamma_h=0$ would mean that the D$p$-brane does
 not wrap an (exceptional) cycle of the blow-up whereas for $\tr\gamma_h\neq 0$
 the contrary is true.  As we mentioned, without the doubling and
 $\gamma_h=\id$ the $h$-invariant branes (or: boundary states) 
  are called fractional branes.   
 For the D$6$-branes discussed in chapter \ref{z4} this means that fractional
 branes have to intersect $\mathbb{Z}_2$-fix-points. As they are stuck to 
 fix-points which in turn are associated with certain twisted closed-string 
 sectors, we also refer to these branes as {\it twisted} branes. 
Like the cross-cap 
states, also the boundary states couple to RR and NSNS fields. The coupling
to the untwisted (i.e.\ $h=0$) closed-string fields for a fractional D-brane
is a fraction  of what it would be for a brane which was originally
 not $G$-invariant in ambient space. 

As we have noticed that D-branes have similar couplings to closed-string
fields as O-planes, we could check if the addition of D-branes could
cancel the tadpole of the Klein bottle. We are especially interested
in the cancellation  of RR tadpoles as they can not be cured by a Fischler
Susskind mechanism. We would also be glad to cancel the NSNS tadpoles as
well, because the NSNS tadpoles can also lead to a deformation of the theory
to a singular limit. If the model is supersymmetric in both the open- and
the closed-string sector, we also have other phenomenologically appealing
features like possible solution to the hierarchy problem etc. We will
have a closer look at a $\mathbb{Z}_4$-orientifold in chapter \ref{z4} which
admits supersymmetric solutions.\footnote{Condiditions for a D-brane to be
supersymmetric are given in section \ref{susycondop} and in section 
\ref{slagsec}.}
Similar to the O$p$-planes, the D$p$-branes have a low energy effective 
action, too. 
The Chern Simons action looks like (cf.\ \cite{Douglas:1995bn,Green:1997dd,Morales:1998ux,Stefanski:1998yx}):
\begin{equation}\label{csdbrane}
S^{(\text{D}_p)}_{\text{C.S.}} = \mu_p\int_{\text{D}_p}
  \ch({\cal F})\sqrt{
   \frac{\hat{{\cal A}}({\cal R}_T)}{\hat{{\cal A}}({\cal R}_N)}}\wedge
   \sum_{q\in \text{RR}} C_q
\end{equation}
$\mu_p$ is the D$p$-brane charge.
$\ch$ is the Chern character, ${\cal F}=F+B$ the sum of the electro-magnetic
$U(1)$ NS-gauge field $F$ and $B$ is the NSNS two-form. 
$\hat{{\cal A}}$ denotes the $A$-roof (or Dirac) genus:
\begin{equation}
 \hat{{\cal A}}({\cal R}) = 1 -\frac{p_1({\cal R})}{24}
    + \frac{7\bigl(p_1({\cal R})\bigr)^2 -4p_2({\cal R})}{5760}+\ldots
\end{equation}
Like for the O$p$-plane, ${\cal R}_T$ (and  ${\cal R}_N$) are the pull-backs 
of the curvature two-form to the tangent- (and normal-) bundle of the 
D$p$-brane (and $p_i$ are Pontrjagin classes).
 The Dirac-Born-Infeld action also contains a term that
 couples to the combination of NS and NSNS fields ${\cal F}$:
\begin{equation}\label{dbdbrane}
 S^{(\text{D}p)}_{\text{DBI}} = T_p\int_{{\text D}q}
  e^{-\Phi} \sqrt{-\det (G+\cal F)}
\end{equation}
$T_p$ is the D-brane tension.
However it is still a field of research how the non-abelian 
gauge degrees of freedom are
correctly incorporated. One method, which is sufficient for all of our tadpole
considerations, is simply to trace over the gauge degrees of freedom. Several
non-abelian extensions of the above actions have been suggested, motivated 
by different approaches (\cite{Tseytlin:1997cs,Tseytlin:1999dj,
Myers:1999ps,Koerber:2002zb,Stieberger:2002fh,Stieberger:2002wk}).  

For consistency we also have to project on $S\Omega$-invariant states
in the open-string sector. This leads to non-orientable diagrams with
boundaries in the
perturbation series. At one-loop level this is the M\"obius strip.

\subsection{\texorpdfstring{$s\Omega$}{s Omega}-invariant open-string sector}  
The $s\Omega$-projection must be also imposed in the open-string sector.
We assume that we have already created the orbifold including oriented
open-strings as discussed in the last section. We do not expect that the
 orbifolded theory is fully consistent at this stage. It will in general still
suffer from tadpoles. Consistency of the string perturbation expansion 
will force us to include non-orientable diagrams with boundaries. 
We will first have a closer look at the spectrum. 
For $s\Omega$ to be a symmetry, we have to include all $s\Omega$ images of
the brane. In the case without $U(1)$-valued electro-magnetic NSNS fields
$F^{(i)}$,
we only have to make sure that the configuration (i.e. the
Hilbert space) is $s$-invariant.
With  non-trivial  $U(1)$-fields we note that $\Omega:\, 
(F_{(i)},F_{(j)})\mapsto -(F_{(j)},F_{(i)})$ (cf.\ chapter \ref{strbg}).\\
Since $\Omega$ acts as
an orientation reversal on $X(\tau,\sigma)$ there are two possibilities:
\begin{enumerate}
 \item $s\Omega$ interchanges the sector $(i)$ with a different sector   
       $s\Omega (i)$ ($(i)$ describes  a
       $G$-invariant combination of boundary conditions). 
       In this case the $s\Omega$ projection breaks the gauge group:\\
 \begin{minipage}[t]{12cm} 
 \begin{equation}
  U(n_i)\times U(n_{s\Omega(i)})\xrightarrow{{\cal P}_{s\Omega}}  U(n_i)
 \end{equation}
 \end{minipage}
 \item The sector $(i)$ is mapped to itself by $s\Omega$. We will consider
       this case in more detail:
 \end{enumerate} 
As $\Omega$ acts as orientation-reversal, left
 and right CP-degrees also have to get exchanged in the sector $(ii)$ under
the $s\Omega$ action.  $\Omega$ includes a transposition of $\Lambda$:
\begin{equation}  
 \ket{\Lambda^{(i,i)}}\xrightarrow{\Omega}\ket{\bigl(\Lambda^{(i,i)}\bigr)^T}
\end{equation}
(A hermitian conjugation would leave the hermitian matrix $\Lambda^{(i,i)}$
invariant. Hermitian or anti-hermitian matrices do however not form a
$\mathbb{C}$-vector space.)
$s\Omega$ could also 
 contain an additional $U(n)$-rotation $U_{s\Omega}$. 
In total we have for $s\Omega$ in this case:
\begin{equation}
 s\Omega:\, \ket{\psi (i,i),\text{osc}}\otimes\ket{\Lambda^{(i,i)}}
 \mapsto \ket{s\Omega(\psi) (i,i),\text{osc}}\otimes
 \ket{\bigl(U^{(i)}_{s\Omega}\bigr)^{-1}\bigl(\Lambda^{(i,i)}\bigr)^T 
   U^{(i)}_{s\Omega}}
\end{equation}
We note that $\big(U^{(i)}_{s\Omega}\big)^{-1}
=\big(U^{(i)}_{s\Omega}\big)^{\dagger}
 =\big(\overline{U}^{(i)}_{s\Omega}\big)^{T}$.
Under a unitary basis-change $V^{(i)}$ in the left or right CP-Hilbert space
${\cal H}_{\text{CP}}^{(i)}$
(belonging to the boundary condition $(i)$), $U^{(i)}_{s\Omega}$ will 
{\sl not} transform  by conjugation:
\begin{align}\label{gammaomtr}
 \Lambda^{(i,j)}&\mapsto \Lambda^{\prime\,(i,j)}
 =\bigl(V^{(i)}\bigr)^{-1}\Lambda^{(i,j)}V^{(j)} &\Rightarrow
 U^{(i)}_{s\Omega}&\mapsto U^{\prime\,(i)}_{s\Omega} 
 = \bigl(V^{(i)}\bigr)^{T}U^{(i)}_{s\Omega} V^{(i)}
\end{align}
Hence a basis  of 
${\cal H}_{\text{CP}}^{(i)}$  in which $U^{(i)}_{s\Omega}$ is
diagonal, does not exist in generic cases.
 However it turns out that $U^{(i)}_{s\Omega}$ is either 
symmetric (or anti-symmetric) in which case simple representations exist:\\ 
The relation
$(s\Omega)^2=h$, $h\in G$ has
to be obeyed at least up to a phase in both the oscillator and the CP-part
of the state (and in total without a phase).
 Because we have already performed the $G$-projection ({\sl and assuming that
 the resulting state can be written as a direct product} of a CP-part and 
 a part $|\psi (i,j), \text{osc}\rangle$)  this reduces
 to:\footnote{By this we assume that $|\psi\rangle$ is $G$-invariant. The
   split of a general $G$-invariant state into a product of a CP and an
   oscillator part is in general not possible. However there should exist a
   basis of $G$-invariant states that admits this product structure. 
   Relation \eqref{phirel} holds for each of these basis vectors.}
\begin{align}
 \mbox{$(s\Omega)^2\ket{\psi (i,j), \text{osc}}=$}
 \begin{split}
   &\exp(i\phi^{(i,j)}_{\text{osc}})\ket{\psi(i,j), \text{osc}}
   \end{split} \\
   \raisebox{2.5ex}{\mbox{$(s\Omega)^2\big|\Lambda^{(i,j)}\big>=$}}
   \begin{split}
        &\exp\big(i(\phi^{(i)}_{\text{CP}}-\phi^{(j)}_{\text{CP}})\big) \\  
   &\,\cdot\Bigl|
 \Bigl(\bigl( (U^{(i)}_{s\Omega})^{-1} \bigr)^T U^{(i)}_{s\Omega}\Bigl)^{-1}
 \Lambda^{(i,j)}
 \Bigl(\bigl( (U^{(i)}_{s\Omega})^{-1} \bigr)^T U^{(i)}_{s\Omega}\Bigl)
 \Bigr> 
 \end{split}
\end{align}
with the phase depending on the  super-selection sector $(i,j)$. 
We have used that the right CP space transforms in the complex conjugate
representation with respect to the left one. 
We also deduce that for identical boundary conditions on left- and 
right-movers (including the effect of the GSO-projection):
\begin{equation}\label{phirel}
 \phi^{(i,i)}_{\text{osc}}=0 \mod 2\pi
\end{equation}
If $(s\Omega)^2$ equals the identity,\footnote{In the bosonic string, and
on GSO-projected states of the superstring.}
 $U^{(i)}_{s\Omega}$ is
either symmetric or anti-symmetric in sectors in which 
$s\Omega$ leaves the boundary condition invariant:
\begin{align}
 U^{(i)}_{s\Omega}&=(U^{(i)}_{s\Omega})^{T}\qquad\text{or} &
 U^{(i)}_{s\Omega}&=-(U^{(i)}_{s\Omega})^{T} 
 \qquad\text{for } \quad s\Omega(i)=i
\end{align}
By a unitary base change of the form $\eqref{gammaomtr}$ we can achieve
$U^{(i)}_{s\Omega}$ to be:
\begin{align} \label{symform}
 U^{(i)}_{s\Omega}\qquad&  \text{\phantom{ anti-}symmetric: }
 \qquad U^{(i)}_{s\Omega} =\mathds{1}_{n_i}  \\ \label{asymform}
 U^{(i)}_{s\Omega}\qquad& \text{ anti-symmetric: }
 \qquad U^{(i)}_{s\Omega} = 
  \biggl(\begin{array} {c|c}
     0 & i\mathds{1}_{n_i/2}\rule[-1.05ex]{0.0ex}{0ex}
    \\ \hline
     -i\mathds{1}_{n_i/2} & 0
   \end{array}
   \biggr) 
\end{align}
This fact is proven in appendix \ref{omegstform}. The situation
is more complicated if $(s\Omega)^2=h\neq \id$. To the knowledge
of the author, this case is not  classified in  physics
literature. If an element $(s\Omega)^2=\id$ is contained
in the orientifold-group $O$, we take this element as the representative in
eq.\ \eqref{odecomp} (p.\ \pageref{odecomp}). This element does not need to
be unique, of course. If we know the form of $U^{(i)}_{s\Omega}$ for this
element, we may determine the form of the remaining $U^{(i)}_{s\Omega g}$,
$g\in G$ by requiring that the $U$ matrices (often called $\gamma$-matrices
 as well, which should not be confused with the generators of the
 Dirac-algebra) form a representation of the orientifold group $O$. 

Even though all models considered in this thesis belong to the class
$(s\Omega)^2=\id$, we explain a problem in the case that such an element
is not included in $O$. This case implies
that the Klein bottle only leads to twisted-sector tadpoles. The twist
corresponds to $(ks\Omega)^2=h$, with $k,h\in G$. It is not clear, if 
such kind of twisted tadpoles can be canceled. However it is obvious that
the cancellation of purely twisted tadpoles by adding D-branes and 
leaving the background otherwise unmodified, is impossible. D-branes always
couple to untwisted closed-string fields as the partition function can be 
written in terms of traces (cf.\ eq.\ \eqref{anntrace}). The 
$\id$-trace-insertion (which is not allowed to vanish)
corresponds to untwisted closed-string exchange in the 
tree-channel. Therefore each individual annulus amplitude has non-vanishing
untwisted closed-string contribution. As the Klein bottle does not contribute
to this untwisted closed-string exchange (in the case at hand), the total 
untwisted annulus tadpole has to vanish by itself. For the RR-tadpole
this would imply that both branes and anti-branes (which have the opposite
coupling to the RR-fields) are present. As anti-branes have identical
boundary conditions as branes, except for the GSO-projection, which is
reversed, supersymmetry gets broken. The annulus NSNS tadpole however can 
not be eliminated since it has the same sign for branes and anti-branes.
As a consequence we do not expect that supersymmetric orientifolds
exist, with $(hs\Omega)^2\neq\id\;\forall h\in G$.   

Assuming from {\sl now  onwards}, that an {\sl order two element} $s\Omega$ is
 contained in the orientifold group $O$,
 we still have not derived if $s\Omega$ is represented on the
CP dofs by a symmetric \eqref{symform} or anti-symmetric matrix 
\eqref{asymform}. This is in general not easy to decide. It may get derived
from the tadpole-cancellation conditions. However it is often possible
to derive relations between different  $U^{(i)}_{s\Omega}$ acting
on different boundary conditions by the use of the vertex operator algebra.
We will sketch one method. We assume that we have a (GSO-invariant)
vertex operator $V^{(i,i)}$  that corresponds to the boundary 
 condition $(i,i)$ on which we assume $s\Omega$ to act
 trivially: $s\Omega(i)=i$. The same we assume for a second boundary
 condition: $s\Omega(j)=j$.
 The CP factors are not yet
included in the vertex operators. We further assume that we know the
explicit form of the vertex operators in the
 following OPE:
\footnote{For illustrative reasons, we assume to have operators with this
simple OPE. Of course, an asymptotic expansion of the OPE involves in general
a sum over vertex operators on the right hand side which are multiplied by 
different (not necessarily constant) coefficients.}
\begin{equation}\label{ope1}
  {\cal V}^{(i,j)}{\cal V}^{(j,i)}\sim {\cal V}^{(i,i)}
\end{equation}
In addition we require ${\cal V}^{(i,i)}$  and ${\cal V}^{(j,j)}$ to be  
$s\Omega$-Eigenstates with known Eigenvalue $\lambda_i$ resp.\ $\lambda_j$. 
 We are  then able to determine the relative sign in the 
$(s\Omega)^2$ projection on the CP-dofs:
Since $s\Omega$ interchanges the left boundary condition $i$ 
with the right boundary condition $j$
without changing the level (mass) of the vertex operator, we deduce:
\begin{equation} \label{relv1}
  {\cal V}^{(i,j)}= \xi s\Omega \bigl({\cal V}^{(j,i)}\bigr),
  \qquad\xi\in\mathbb{C}
\end{equation}
(The proof relies very much on this fact, i.e.\ on $s\Omega$ invariant
boundary conditions $i$ {\sl and} $j$.)
Now we use that:\footnote{Here we made the assumption that $s\Omega$
exchanges two vertex operators. In principle it could also 
exchange the vertex operators and multiply the resulting product by $-1$.
This second possibility would reverse the conclusions in such a way
that both matrices $ U^{(i)}_{s\Omega}$ and 
 $U^{(j)}_{s\Omega}$ would have the  same symmetry properties in 
 \eqref{somegasymmetry}.}
\begin{equation}
  s\Omega \bigl({\cal V}^{(i,j)}{\cal V}^{(j,i)}\bigr)
  =s\Omega \bigl({\cal V}^{(j,i)}\bigr)s\Omega \bigl({\cal V}^{(i,j)}\bigr)
   \sim s\Omega\bigl({\cal V}^{(i,i)}\bigr)= \lambda_i {\cal V}^{(i,i)}
\end{equation}
Inserting  relation \eqref{relv1} and denoting the $(s\Omega)^2$ Eigenvalue
of ${\cal V}^{(j,i)}$ by $\epsilon$ we get:
\begin{equation}
 \begin{aligned}
\epsilon\xi s\Omega  \bigl({\cal V}^{(j,i)}\bigr) {\cal V}^{(j,i)}=
  \epsilon {\cal V}^{(i,j)}{\cal V}^{(j,i)}
 \end{aligned}
   \sim s\Omega\bigl({\cal V}^{(i,i)}\bigr)= \lambda_i {\cal V}^{(i,i)}
\end{equation}
If \eqref{ope1} and $s\Omega(i)=i$ holds, we  directly deduce: 
\begin{equation}
 \begin{aligned}
   (s\Omega)^2 \bigl({\cal V}^{(i,j)}\bigr) 
     = \epsilon \bigl({\cal V}^{(i,j)}\bigr) \\
   s\Omega  \bigl({\cal V}^{(i,i)}\bigr) 
    = \lambda_i {\cal V}^{(i,i)}
  \end{aligned}
 \Longrightarrow  \epsilon= \lambda_i
\end{equation}
Given an $s\Omega$ invariant sector $i$ with 
 $s\Omega$ Eigenvalue $\lambda_i=-1$
  for a specific boundary vertex operator 
${\cal V}^{(i,i)}$ 
 we require that $(s\Omega)^2$
acts as the identity in the $(i,j)$ sector (i.e.\ on the whole state including
the CP dofs). This
 imposes opposite $(s\Omega)^2$ projections on the CP Hilbert space
in the $i$\raisebox{1ex}{\tiny th} and $j$\raisebox{1ex}{\tiny th} sector:
\begin{align}\label{somegasymmetry}
 U^{(i)}_{s\Omega}&=\pm\bigl(U^{(i)}_{s\Omega}\bigr) &
 U^{(j)}_{s\Omega}&=\mp\bigl(U^{(j)}_{s\Omega}\bigr) 
\end{align}
(In other words: The action on the CP dofs has to compensate the
phase $-1$ of $(s\Omega)^2$ acting on the oscillators.) 
This method was used in \cite{Gimon:1996rq} to derive opposite 
$\Omega$-projections on D$9$- and D$5$-branes. 
Even though we might have reduced the choices of $U^{(i)}_{s\Omega}$ in
this way, we cannot deduce the spectrum directly. First we have
to determine the tadpole cancellation conditions, which in addition
to the algebraic  restrictions
further constrain the form of the  $U^{(i)}_{s\Omega}$
and $U^{(i)}_{g}$. We will however state the result, that the
only gauge groups that can be obtained in the perturbative
spectrum of orientifold theories are the $SO(n)$, $USp(n)$ and
$U(n)$ groups, as well as direct products of these groups.
These restrictions arise if one imposes factorization of 
open-string amplitudes (which is needed in order 
that the theory is consistent with unitarity) 
\cite{Marcus:1982fr,Marcus:1987cm}.
No simple rule is known to deduce the spectrum of
generic orientifolds  directly 
(except for some classes of orientifolds like the models considered in
\cite{Blumenhagen:2002wn}). For the models presented in this thesis,
 we will 
find consistent actions of the orientifold group $O$ on the
CP Hilbert space, which allow a projection of the open-string spectrum
onto an $O$-invariant subspace. 

\subsection{M\"obius amplitude}
\begin{figure}
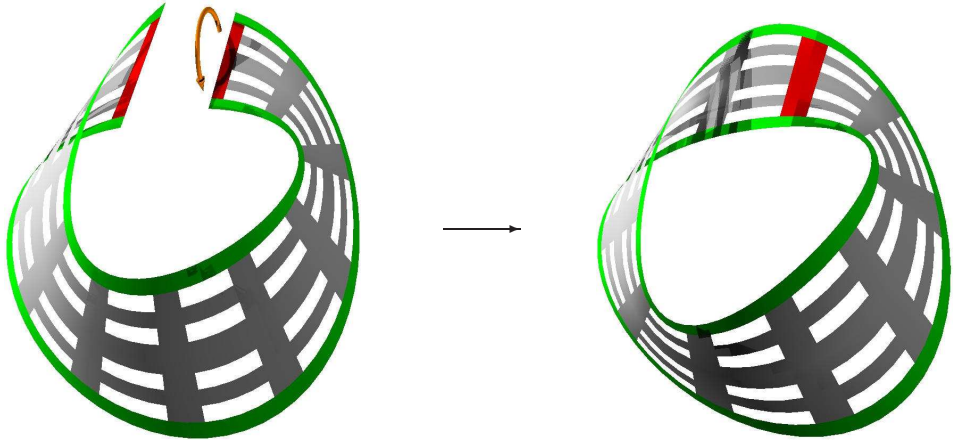

\vspace{1cm}
  \setlength{\unitlength}{0.1in}
  \begin{picture}(60,18)
   \put(2,0){\scalebox{0.2}{\includegraphics{picori91.EPS2}}}
   \put(33,0){\scalebox{0.2}{\includegraphics{picori92.EPS2}}}
 \put(25,11){\vector(1, 0){4}}
  \end{picture}
\caption[Construction of the M{\"o}bius strip]
  {\label{moebfig} Construction of the M{\"o}bius strip  by gluing
  two ends of a twisted strip  in the way depicted.}
\end{figure}

\begin{figure}
  \setlength{\unitlength}{0.1in}
  \begin{picture}(60,50)
    \SetFigFont{14}{20.4}{\rmdefault}{\mddefault}{\updefault}
   \put(5,3){\scalebox{0.5}{\includegraphics{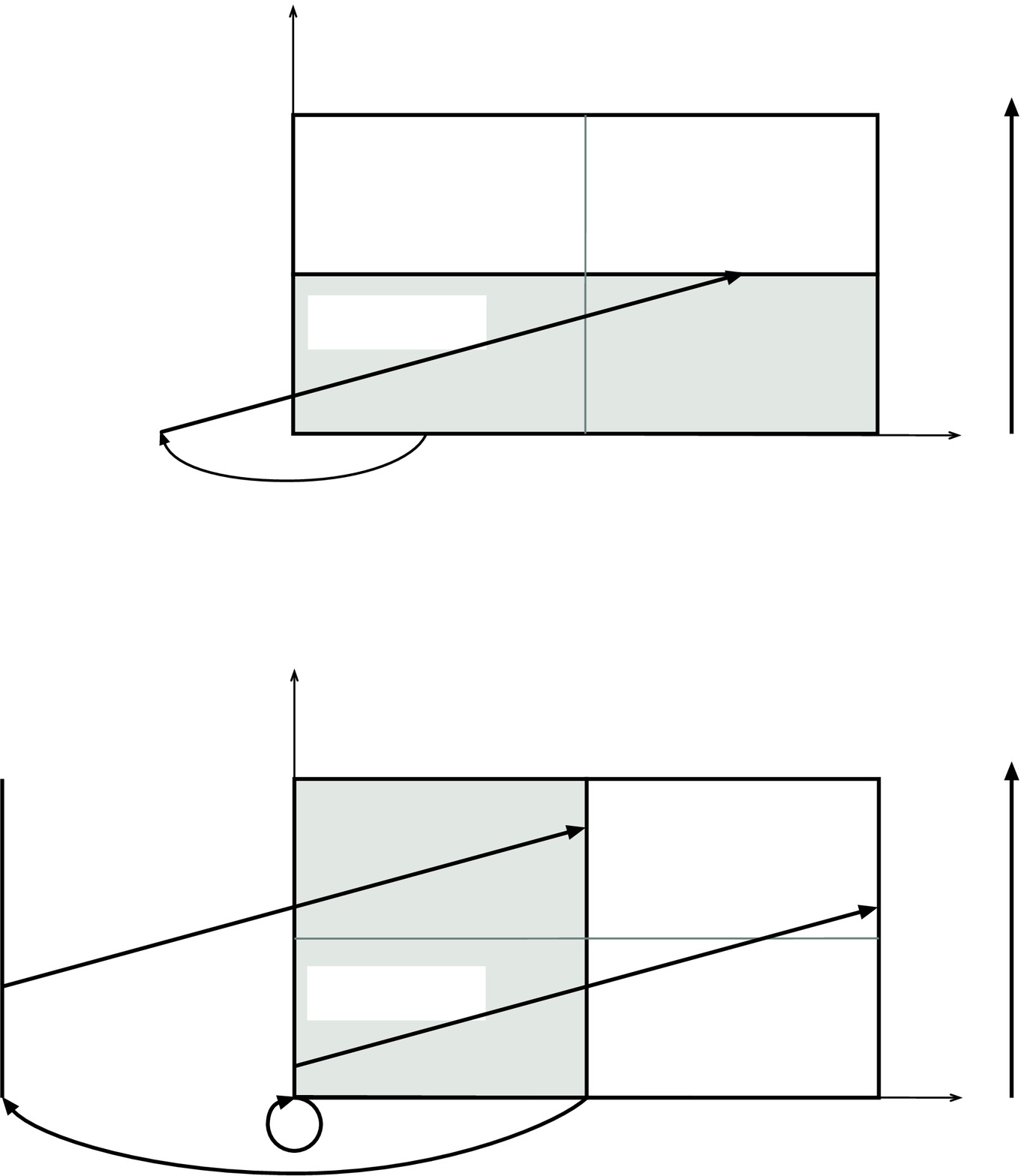}}}
   \put(0,48){loop-channel:}
    \put(12.7,45.){$\mathfrak{Im}$}
    \put(13,42){$i\tau_2$}
    \put(14.,35.9){$it$}
    \put(44,35.5){$h$} 
    \put(16,36.3){\SetFigFont{12}{20.4}{\rmdefault}{\mddefault}{\updefault}
              \makebox(1,-1)[tl]{$ 1+i\tau_2/2$}}
     \put(12.5,29.3){\SetFigFont{12}{20.4}{\rmdefault}{\mddefault}{\updefault}
                \makebox(1,-1)[tl]{$ z\rightarrow -\bar{z}$}}
     \put(38,28.5){$\mathfrak{Re}$}
   
    \put(0,23.5){tree-channel:}
    \put(12.7,20.5){$\mathfrak{Im}$}
    \put(13,17.5){$i\tau_2$}
    \put(44,11.0){$h$}
    \put(16,11.6){\SetFigFont{12}{20.4}{\rmdefault}{\mddefault}{\updefault}
               \makebox(1,-1)[tl]{$ 1+i\tau_2/2$}}
    \put(26.2,4.){$l$}
    \put(12.5,3.7){\SetFigFont{12}{20.4}{\rmdefault}{\mddefault}{\updefault}
               \makebox(1,-1)[tl]{$ z\rightarrow -\bar{z}$}}
     \put(38,4){$\mathfrak{Re}$}
    \put(14.5,39.2){$j$}
    \put(37.8,39.2){$k$}
   \put(15.,25.6){$V_{(1)}$} \put(36.5,25.6){$V_{(2)}$}
     \put(14.5,14.7){$j$}
     \put(37.8,14.7){$k$}
    \put(15.,0){$V_{(1)}$} \put(36.5,0){$V_{(2)}$}
  \end{picture}
     \caption[Periodicities of the M\"obius strip embedded
                                 in the underlying annulus]
         {\label{moebiusfund} Periodicities of the M\"obius strip embedded
                                 in the underlying annulus and the 
                                 two fundamental regions 
                                 (shaded areas). The boundaries correspond 
                                 to  boundary conditions $V_{(1)}$ and
                                 $V_{(2)}$.
                                   Chan Paton
                                 labels $j$ and $k$ are also included in the
                                 diagrams.}
\end{figure}

The remaining $\chi=0$ diagram is the M\"obius strip. Topologically, it
is obtained form a strip, with the two ends twisted and then glued together,
such that the resulting object is non-orientable (cf.\ figure \ref{moebfig}).
This picture corresponds to the loop-channel in which an open-string 
circulates in a loop. Like the Klein bottle, the M\"obius strip is obtained
by moding out another world sheet by a $\mathbb{Z}_2$-involution. For the 
M\"obius strip this ambient world sheet is the annulus. 
Like for the Klein bottle we paint a diagram, from which we will read off the
periodicities (figure \eqref{moebiusfund}). 

\begin{figure}
\hspace{2.5cm}
\begin{center}
\begin{picture}(0,0)%
\put(45,10){\epsfig{file=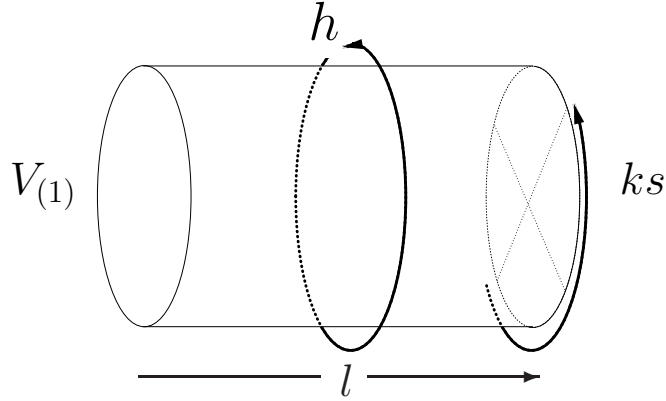}}%
\end{picture}%
\setlength{\unitlength}{1184sp}%
\begingroup\makeatletter\ifx\SetFigFont\undefined%
\gdef\SetFigFont#1#2#3#4#5{%
  \reset@font\fontsize{#1}{#2pt}%
  \fontfamily{#3}\fontseries{#4}\fontshape{#5}%
  \selectfont}%
\fi\endgroup%
\begin{picture}(13626,7529)(1001,-6989)
\put(14476,-3061){\makebox(0,0)[lb]{\smash{\SetFigFont{17}{20.4}{\rmdefault}{\mddefault}{\updefault}
$ks$
}}}
\put(7951,164){\makebox(0,0)[lb]{\smash{\SetFigFont{20}{24.0}{\rmdefault}{\mddefault}{\updefault}
$h$
}}}
\put(1701,-3061){\makebox(0,0)[lb]{\smash{\SetFigFont{17}{20.4}{\rmdefault}{\mddefault}{\updefault}
$V_{(1)}$
}}}
\thicklines\put(4400,-6900){\line(1,0){3800}}
\put(9200,-6900){\vector(1,0){3600}}
\put(8600,-7300){\SetFigFont{17}{20.4}{\rmdefault}{\mddefault}{\updefault}$l$}
\end{picture}
 \caption[M\"obius strip in tree-channel]{\label{moebfigtree} 
  The M\"obius strip in the tree-channel. It describes a closed-string that
  is emitted from a D-brane described by boundary condition $V_{(1)}$ and
  absorbed by the cross-cap state (or O-plane) $\ket{{\cal C}(ks)}$ after
  world sheet time $l$.}
 \end{center}
\end{figure}

The involution for the M\"obius strip is the same as for the Klein bottle
(cf.\ \eqref{omegaaction1}, p.\ \pageref{omegaaction1}). 
Similarly the relations
\eqref{kbrel1} and \eqref{kbrel2} (p.\ \pageref{kbrel1}) between
the trace-insertion and the twist in the tree-channel are valid for the
M\"obius strip as well. In tree-channel only the cross-cap condition 
\eqref{kbbdycond1} (p.~\pageref{kbbdycond1}) is valid, if we take $\sigma$
to be half the length of the open and {\sl not} of the closed-string.
The periodicity
w.r.t.\ the $\mathbb{Z}_2$-involution relates both boundary  conditions
$V_{(1)}$ and $V_{(2)}$, if we have a trace-insertion $ks\Omega$
in the corresponding partition function:
\begin{align}
 \partial_\mp X(\tau,\sigma=0) &= ks_\pm\partial_\pm X(\tau+t,\sigma=1) \\
              &= ks_\pm V^{\pm1}_{(2)}\partial_\mp X(\tau+t,\sigma=1) \\
              &= ks_\pm V^{\pm1}_{(2)}ks_\pm\partial_\pm X(\tau+2t,\sigma=0) \\
 ks_\pm ks_\mp\partial_\mp X(\tau+2t,\sigma=0) 
  &= ks_\pm V^{\pm1}_{(2)}ks_\pm V^{\mp1}_{(1)}\partial_\mp X(\tau+2t,\sigma=0)
\end{align}
Thus we found the necessary condition for the boundary conditions of the
M\"obius strip:\footnote{The index $\pm$ on $ks$ takes into account that $ks$
might act differently on left- and right-movers in the case of asymmetric
orientifolds.}
\begin{equation}\label{moebcond2}
  ks_- V_{(1)}=V_{(2)}ks_+
\end{equation}
Similar conditions hold for the zero modes as well.
In addition the CP states $\Lambda^{(i,j)}_{a;b}$ have to obey:
\begin{equation}
   \Lambda^{(i,j)}_{ab} = 
        \biggl(\Bigl(U^{(j)}_{s\Omega}U^{(j)}_{k}\Bigr)^{-1}
        \bigl(\Lambda^{(i,j)}\bigr)^T 
         U^{(i)}_{s\Omega}U^{(i)}_{k}\biggr)_{ab}
\end{equation}
The tree-channel of the M\"obius-amplitude is depicted in figure
\ref{moebfigtree}. The relation between loop- and tree-channel parameter 
$t$ and $l$ is listed in table \ref{transftreeloop} 
(p.~\pageref{transftreeloop}). It is obtained by the same reasoning as
for the Klein bottle, except that the open-string length is half of the 
closed-string length.
Like the other diagrams, we write the M\"obius amplitude in loop-channel
 and in tree-channel, where it corresponds to a
closed-string exchange between a boundary and a cross-cap state.
\begin{align}
  \text{M}^{i}(ks)   
  &=  \tr\Bigl(\bar{\gamma}^{(i)}_{ks\Omega}
  \bigl(\gamma^{(i)}_{ks}\bigr)^{-1}
  \gamma^{(i)}_{ks\Omega}\bigl(\gamma^{(i)}_{ks}\bigr)\Bigr)
   \cdot   \frac{V_{10}}{(2\pi)^d}
   \,\lim\limits_{\epsilon\to 0}
   \int_\epsilon^\infty\frac{dt}{t} Z_{\text{M}^{i}}\\
  \label{treembtransf}
  &= \int dl\,
   \braket{{\cal C}(ks)}
   {e^{-2\pi l (H+\widetilde{H})_{h=(k\Omega s)^2\text{-twisted}}}}
   {{\cal D}(V_{(i)});h}+  \text{c.c.}\\
   Z_{\text{M}^{i}(h)}&=Z_{\text{M}^{i}_{ab}(ks\Omega)}
   \equiv\frac{1}{|O|}
   \tr_{(V_{i},(ks)^{-1}V_{i}(ks)) \atop \setminus{\cal H}_\text{CP}} 
   ks\Omega\, e^{-2\pi t(H_{\theta(jl)})} 
 \end{align}
The subscript under the $\tr$ indicates that the CP-trace is excluded in this
expression in accordance with eq.\ \eqref{annf1}. Condition \eqref{moebcond2} 
 is already imposed. We have used the common notation 
$U^{(i)}_g=\gamma^{(i)}_{g}$ for the representation of $O$ on the CP Hilbert
space. Summing $\text{M}^{i}(ks)$ over all $i\in\{\text{bdy-conditions}\}$ as
well as over all trace-insertions $ks\Omega$ compatible with 
\eqref{moebcond2}, we obtain the whole M\"obius amplitude. Written in 
tree-channel it takes the form:
\begin{equation} \label{treembtransf2}
  \text{M}=\int dl\,\sum_{h\in G} 
          \braket{{\cal C}|s;h}
        {e^{-2\pi l (H+\widetilde{H})_{h^2\text{-twisted}}}}{{\cal D};h} 
      + \text{c.c.}
\end{equation}
The fact that eq.\ \eqref{treembtransf} and  \eqref{treembtransf2} take this
 special form is highly non-trivial (and we do not prove it). It
follows from the normalization of the cross-cap and boundary states,
which is determined by rewriting the Klein bottle and annulus loop-amplitudes
 as tree-channel amplitudes. This form (especially the prefactor)
 is deeply linked to the dimension of the target space.
For general CFTs it can be different.
 However in all models presented in this paper, this 
factor is present. It is important for rewriting the whole amplitude 
as a perfect square:
\begin{multline}\label{tadptot}
\text{KB}+ \text{M}+\text{A}  \\
 = \int dl\,\sum_{h\in G} 
   \bigl( \bra{{\cal C}|s;h} + \bra{{\cal D};h} 
   \bigr)
    e^{-2\pi l (H+\widetilde{H})_{h^2\text{-twisted}}}
   \bigl( \ket{{\cal C}|s;h} + \ket{{\cal D};h} 
   \bigr)
\end{multline}
In addition, this factorization imposes conditions on the 
representation matrices $\gamma^{(i)}_{\ldots}$. 
The ability to write this amplitude as a perfect square leads to
its factorization in the limit of huge world-sheet-times $l$.
The cancellation of the overall tadpole requires the $l\to\infty$ limit
of the  integrand in \eqref{tadptot} to vanish separately for each physically 
distinguishable closed-string excitation. 
In this limit only
the closed-string $q^0$-term, i.e.\ massless modes can contribute. 
Each independent tadpole  must be canceled separately.
In the
$\mathbb{Z}_4$-orientifold of chapter \ref{z4} this means that branes wrapped
around blown up fix-points have to cancel their twisted sector charges on 
each fix-point individually.
 Furthermore, physically different tadpoles might be 
distinguished by their dependence on geometrical data like the volume 
(complex structure, etc.~). The tadpole cancellation conditions impose
constraints both on the allowed boundary conditions $V_{(i)}$ and
on the representation matrices $\gamma$ of the orientifold group
$O$. These constraints still leave some freedom in many cases. 
Requiring supersymmetry can further reduce this freedom and leads in
some cases to a unique solution. 
To be more specific: The $\id$ trace-insertion determines the dimension(s)
of the CP-Hilbert space(s). This is usually identified with the number of
D-branes (though different authors sometimes include or exclude
 the $s\Omega$ and/or $G$ images of the branes in their counting).
The $s\Omega$-insertion (i.e.~the M\"obius amplitude without further
insertions) also fixes the form of the $U_{s\Omega}^{(i)}$
 matrices. Usually  the $\id$ sector leads   to a binomial
 formula of the type $(N-C)^2\stackrel{!}{=}0$. $C$ is associated with the
  coupling of the cross-cap to the field that generates the tadpole. In the
simplest case $N$ counts the number of branes. More generally it 
includes topological data for the RR sector and data which depends
on differential (i.e.\ not purely topological) properties of the brane for 
NSNS sector tadpoles.
 If different untwisted sectors are present one gets several of
these binomial formul\ae. Non trivial trace-insertions lead generically 
to twisted sector tadpoles. They constrain the form of the $U^{(i)}_{g}$ 
(or $\gamma^{(i)}_{g}$)
matrices. Often a geometrical interpretation of some invariants like
$\tr U^{(i)}_{g}$ is accessible. In the $\mathbb{Z}_4$ models of chapter
\ref{z4} this trace is interpreted as the wrapping number of the brane $(i)$
around (blown-up) $g\in\mathbb{Z}_2$-twisted cycles.
As boundary states have often an 
interpretation in terms of geometrical D-branes, which are (connected)
 sub-manifolds, one could also determine
the tadpole conditions via the low-energy effective actions  like 
\eqref{csoplane} (p.~\pageref{csoplane})  and \eqref{csdbrane} 
(p.~\pageref{csdbrane}).
 However one should have in
mind, that an interpretation in terms of partition functions always has to 
exist in order to give a sensible string interpretation. 

Infinities caused by tadpoles can be seen in other than one-loop vacuum
amplitudes, too.
Green and Schwarz proved UV-finiteness of one-loop scattering amplitudes 
with four external massless open-string states in $D=10$ type I 
superstring with gauge-group $SO(32)$.
They also guessed that closed string tree-amplitudes become
finite if {\it disk} and {\it projective plane}-diagrams are combined (c.f.\
\cite{Green:1985ed}). The 
absence of the tree-level dilaton-tadpole in $SO(32)$ type I superstring has 
been proven 
in \cite{Itoyama:1987qz} by Itoyama and Moxhay.  

Instead of modifying the open-string background, one could also try to modify
 the closed string background, thereby solving the eoms \cite{Uranga:2002vk}. 
 For example the NSNS 3-form field strength couples naturally to the 
 RR 4-form potential as well as to the RR 3-form field strength in the
 CS-action of Type IIB theory.\footnote{This fact was used in 
 \cite{Blumenhagen:2003vr} to construct orientifolds with background fluxes}
 One could also combine both possibilities. However the 
stringy description (in form of a CFT solution) of a non-trivially modified
closed-string background is often not known. This is also true for most
non-linear, i.e. $X$-dependent boundary conditions $V_{(i)}$, that
relate the left- and right-moving parts of the open-string in addition to
specifying its zero-modes. However linear boundary conditions are solved 
more easily. By this one obtains usually a great variety of different, 
and often phenomenological appealing solutions including chiral fermions
transforming in interesting gauge-groups.\footnote{However exceptional groups
are not contained in perturbative orientifold spectra.}

\subsection{\label{susyori}Orientifolds of supersymmetric strings} 
 In this section we will make some comments about orientifolds of 
supersymmetric string theories. In addition to the bosonic string sector,
world-sheet fermions appear. 
The fermionic term of the gauge fixed fermionic action looks like:
\begin{equation}
S_{\text{ferm}} =  -\frac{1}{4\pi \alpha^\prime} \int_{\cal{M}} d^2\sigma
    \, 2 i\, \Bar{\psi}^\mu\rho^\alpha\partial_\alpha\psi_\mu 
\end{equation}
The $\psi$´s are world-sheet majorana spinors. 
The two dimensional Dirac matrices $\rho^\alpha$ are in this gauge 
($h= \Bigl(\begin{smallmatrix} 
            h_{\tau\tau} & h_{\tau\sigma}  \\
            h_{\sigma\tau} & h_{\sigma\sigma}
          \end{smallmatrix}  \Bigr)
  = \diag (-1,1)$) :
\begin{equation}\label{diracmata}
   \rho^0= \begin{pmatrix}
            0 & 1  \\
           -1 & 0
           \end{pmatrix} ,\qquad
   \rho^1= \begin{pmatrix}
            0 & 1  \\
            1 & 0
           \end{pmatrix}
\end{equation}

For open-strings it is well known, that one 
has to include altered boundary conditions (cf.\ \eqref{bdyc1}):
\begin{equation}\label{bdycferm}
 \psi_+(\tau,\sigma)= \kappa_i V_{(i)} \psi_-(\tau,\sigma)\, ,
          \, \sigma\in\partial{\cal M}_i
\end{equation}
The relative sign $\kappa_i\kappa_j=\pm 1$ 
of a string in the $(i,j)$ sector 
determines whether the open-string belongs to the Ramond  $(+)$
or Neveu Schwarz sector $(-)$. As we have to fix the $\kappa_i$, this
 choice is obviously asymmetric in the Neveu Schwarz sector.
In the following we write a spinor as: 
\begin{equation}
 \psi(\tau,\sigma)=
  \begin{pmatrix}
    \psi_+ \\ \psi_-
  \end{pmatrix} (\tau,\sigma)
\end{equation}
 We have
 not yet determined how
the $\mathbb{Z}_2$-involution $s\Omega$ (cf.\ eq.\ \eqref{omegaaction1} 
, p.\ \pageref{omegaaction1}) acts on the world sheet fermions.
The trace includes a GSO projection as well. For open-strings:
\begin{equation}\label{gsoop}
 {\cal P}_{\text{GSO}} = \frac{1+(-1)^{f}}{2}
\end{equation}
with $f$ the world sheet fermion number and for closed-strings:
\begin{equation}\label{gsocl}
 {\cal P}_{\text{GSO}} = \frac{1+(-1)^{f_{\text{L}}}}{2}
                         \cdot\frac{1+(-1)^{f_{\text{R}}}}{2}
\end{equation}
In order to have a sensible mapping between path-integral and operator
 formalism we have to be able to describe the action of $s\Omega$ on
the fermionic fields in such a way, that they are compatible with allowed
boundary conditions in the path-integral. 
We will consider the three amplitudes seperately:
\subsubsection{Fermionic sector of the Klein bottle}
In the ferionic sector, additional signs can be inserted in the
boundary conditions compared to the bosonic sector. For example a fermion
is usually anti-periodic in $\tau$-direction. The GSO projection 
adds however the time-periodic boundary condition as well. In operator
formalism, this corresponds to a $(-1)^f$ insertion, $f$ being the world sheet
fermion number (possibly restricted to the left- or right-moving sector).
A Ramond fermion is periodic in $\sigma$, while a Neveu-Schwarz fermion is
antiperiodic in this direction. This describes the situation for the torus
amplitude. For the Klein bottle and the two remaining $\chi=0$ amplitudes,
we will proceed in the spirit of  \cite{Polchinski:1988tu}: We will first
add  all  possible signs in to the bosonic boundary conditions and determine
further restrictions on these signs afterwards.  For the following discussion
we refer again to figure \ref{kleinbotfund} (p.\ \pageref{kleinbotfund}).
There are two possible signs in the $\tau$-direction:
\begin{align}\label{tausign}
   h\psi (\tau,\sigma) = \diag (\epsilon_1,\epsilon_2)\,\psi(\tau+2t,\sigma)
\end{align}
Furthermore, we know that $ks\Omega$ exchanges left- and right-movers.
In the fermionic sector there are possibly additional signs. We will write the
fermionic analog of  $ks\Omega$ as
\begin{align}\label{ksomegferm}
  ks\Omega=
 ks\begin{pmatrix}
     0      & \chi_1 \\
     \chi_2 &  0
   \end{pmatrix}
\end{align}
Furthermore there is a twist that determines if the string is 
in the NS or R (precisely NSNS, RR, NSR or RNS) sector of the loop-channel:
\begin{align}
   g\psi (\tau,\sigma) = \diag (\kappa_1,\kappa_2)\,\psi (\tau,\sigma+1)
\end{align}
From  condition \eqref{kbrel1} we derive:
\begin{align} \label{kbrel1ferm}
  (ks\Omega)^2&=h  &\Longrightarrow \chi_1\chi_2&=\epsilon_1=\epsilon_2 \\
  \label{kbrel2ferm}
  (gks\Omega)^2&=h &\Longrightarrow \chi_1\chi_2\kappa_1\kappa_2&=\epsilon_1
\end{align}
With \eqref{kbrel1ferm} we get from \eqref{kbrel2ferm}:
\begin{equation}
 \kappa_1=\kappa_2
\end{equation}
The above formula  states that in the Klein bottle fields have to be 
 either of RR- or NSNS-type in the loop-channel. 
The same is also true  for the tree-channel
due to \eqref{kbrel1ferm}. One sign in \eqref{ksomegferm},
 e.g.~$\chi_2$ can be eliminated by
a field redefinition.  
The fermionic part of the two crosscaps is now determined by the conditions:
\begin{gather} \label{BoundstateKBferm}
\widetilde{\text{KB}}_{(ks, ks g)}=
      \int_{\mathbb{R}^+} dl\,\braket{{\cal C}(ks)}
            {e^{-2\pi l (H+\widetilde{H})_{h\text{-twisted}}}}{{\cal C}(ksg)}
             \text{ with }   h=(ks\Omega)^2 \\
 \begin{aligned}\label{BoundstateKB2ferm}
   \bigl(\psi_- (\tau=0,\sigma)- ks\psi_+(\tau=0,\sigma+1/2)\bigr)
     _{h\epsilon\text{-twisted}}
        \ket{{\cal C}(ks)}&=0 
 \\ \bigl(\psi_+ (\tau=0,\sigma)-\chi ks\psi_-(\tau=0,\sigma+1/2)\bigr)
     _{h\epsilon\text{-twisted}}
        \ket{{\cal C}(ks)}&=0   
 \end{aligned} 
\end{gather}
Similarly we get at the other end: 
\begin{gather} 
\begin{aligned}\label{BoundstateKB2lferm}
   \bigl(\psi_- (\tau=l,\sigma)- \kappa g ks\psi_+(\tau=l,\sigma+1/2)\bigr)
     _{h\epsilon\text{-twisted}}
        \ket{{\cal C}(ks)}&=0 
 \\\bigl(\psi_+ (\tau=l,\sigma)-(\chi\kappa) g ks\psi_-(\tau=l,\sigma+1/2)\bigr)
     _{h\epsilon\text{-twisted}}
        \ket{{\cal C}(gks)}&=0   
 \end{aligned} 
\end{gather}
The interpretation is as follows: $\chi$ determines the GSO projection
in the loop-channel. As $\chi=\epsilon$ its sign determines whether
a state belongs to the RR or NSNS sector in tree-channel. 
The sign of $\kappa$ determines whether a state belongs to the NSNS or RR
sector
in loop-channel. In tree-channel it shows up in an additional sign
in the $l$-direction. This sign is the analog of a trace-insertion in
the torus amplitude. We will therefore distinguish the tree-channel sector
by a sign as well. In table \ref{sectormappingkb}, where all tree-loop channel 
relations are listed, $\kappa$ is denoted by $+1$ or $-1$. A general crosscap
state is now denoted as follows:
\begin{gather} 
\begin{aligned}\label{BoundstateKBgenferm}
   \bigl(\psi_- (\tau=0,\sigma)- \kappa a\psi_+(\tau=0,\sigma+1/2)\bigr)
     _{a^2\epsilon\text{-twisted}}
        \ket{{\cal C}(a)|\chi,\kappa}&=0 
 \\\bigl(\psi_+ (\tau=0,\sigma)-(\chi\kappa) a\psi_-(\tau=0,\sigma+1/2)\bigr)
     _{a^2\epsilon\text{-twisted}}
        \ket{{\cal C}(a)|\chi,\kappa}&=0   
 \end{aligned} 
\end{gather}
$\chi=+1$ is the NSNS,   $\chi=-1$ the RR sector. The bosonic part fulfills 
the same conditions as in \eqref{BoundstateKB2} (p.\
\pageref{BoundstateKB2}). 
The complete boundary state is a sum over all four possible choices   
for $\chi,\kappa$. 
\begin{table}
\renewcommand{\tabcolsep}{3pt}
\begin{minipage}[t]{6.5cm}
\begin{center}
\begin{tabular}{rr|c|c}
 $\chi$    &  $\kappa$ &     loop-channel &  tree-channel \\
  \hline
 $-1$  & $-1$   &        (NSNS,$1$)             & (NSNS,$+$) \\
 $ 1$  & $-1$   &        (NSNS,$(-1)^f$)        & (RR,$+$)    \\
 $-1$  & $ 1$   &        (RR,  $1$)             & (NSNS,$-$) \\
 $ 1$  & $ 1$   &        (RR,  $(-1)^f$ )       & (RR,$-$ )
\end{tabular}
 \caption[Klein bottle, Cylinder: 
          Relation between  fermionic sectors in tree- and loop-channel]
         {\label{sectormappingkb} Klein bottle, Cylinder: 
          Relation between the fermionic sectors in tree- and loop-channel.
          For the cylinders´ loop-channel
          NSNS and RR mean the NS- and R-sector respectively.
         } 
 \end{center}
 \end{minipage}\hfill
 \begin{minipage}[t]{6.0cm}
\begin{center}
\begin{tabular}{rr|c|c}
 $\chi$    &  $\kappa$ &     loop-channel &  tree-channel \\
  \hline
 $+1$ & $-1$   &        (NS,$1$)             & (NS,$-$) \\
 $-1$ & $-1$   &        (NS,$(-1)^f$)        & (NS,$+$)    \\
 $ 1$ & $ 1$   &        (R,  $1$)             & (R,$-$) \\
 $-1$ & $ 1$   &        (R,  $(-1)^f$ )       & (R,$-$ )
 \end{tabular}
  \caption{\label{sectormapms} M\"obius strip:
 Relation between fermionic sectors in tree- and loop-channel.}
 \end{center} 
 \end{minipage}
\end{table}

\subsubsection{Fermionic sector of the Cylinder}
The same program can be applied to the cylinder as well.  
Equation \eqref{tausign} is valid (with the signs free), but there is no
condition from $ks\Omega$. However we get two conditions from 
\eqref{bdycferm}. From \eqref{tausign} and \eqref{bdycferm} we get
like for the Klein bottle:
\begin{equation}
 \epsilon\equiv\epsilon_1=\epsilon_2
\end{equation}
However both signs $\kappa_1$ and $\kappa_2$ in \eqref{bdycferm} from both
boundaries are free. One sign, e.g. $\kappa_2$ can be fixed to one by a
field redefinition s.th.\ $\kappa\equiv\kappa_1$ is the second free parameter.
The interpretation of these signs is the same as for the Klein bottle.
However in loop-channel
 the NSNS sector is the NS sector for the cylinder. The same is valid
for the RR sector. Therefore the tree-loop channel relations can be read off 
from table  \ref{sectormappingkb} as well.
The fermionic boundary condition in one fermionic sector specified by
$\epsilon$ and $\kappa$ reads:
\begin{equation}
 \bigl(\psi_+(\tau=0,\sigma)  -  \kappa V \psi_-(\tau=0,\sigma)\bigr)
         _{\epsilon h\text{-twisted}}
        \ket{{\cal D} \,(V);h|\epsilon,\kappa}=0
\end{equation}
For the bosons the corresponding boundary condition is given by
\eqref{bdystatedef1} (p.\ \pageref{bdystatedef1}).

\subsubsection{Fermionic sector of the M\"obius strip}
For the M\"obius strip the cylinder relations  are inherited. 
However the $ks\Omega$ action imposes additional conditions.
The fermionic action of $ks\Omega$ is again given by \eqref{ksomegferm}.
We can apply \eqref{kbrel1ferm} for the M\"obius strip, too. 
We are lead to the same result: $\chi_1\chi_2=\epsilon$. 
The M\"obius strip condition \eqref{moebcond2} (p.\ \pageref{moebcond2}) that
relates the boundary conditions $V_1$ and $V_2$, leads in the fermionic sector
to:
\begin{equation}\label{moebfermrel1}
 \chi_1\chi_2=\kappa_1\kappa_2 \quad(=\epsilon)
\end{equation}
Field redefinition can fix one sign of  $\{\chi_1,\chi_2,\kappa_1,\kappa_2\}$.
We fix $\kappa_2=1$ leaving $\kappa\equiv\kappa_1$ as a free parameter.
$\kappa$ determines whether the fermion belongs to NS- ($\kappa=1$)
or to R-sector ($\kappa=-1$) in loop-channel. In contrast
 to Klein bottle and cylinder, the NS-sector in loop-channel corresponds
 to the NSNS sector in tree-channel (c.f.\ \eqref{moebfermrel1}).
There is still the freedom to choose an overall sign in $ks\Omega$. 
 We define $\chi\equiv\chi_1$ (with  $\chi_2= \kappa/\chi$ imposed by 
\eqref{moebfermrel1}).
The relations between different sectors in tree- and loop-channel are given 
in table \ref{sectormapms}.

\addcontentsline{toc}{section}{Concluding remarks}
\section*{\label{conclremori}Concluding remarks}
We presented basic notions of orientifolds, including subjects like tadpoles,
cross-cap- and more generally: boundary-states,  Chan Paton factors and
non-orientable world-sheets. Special emphasis was put on the open-closed
string correspondence of the one-loop vacuum amplitudes.
Of course, we could not cover all subjects. Boundary states for examples can
be defined via more general symmetries than the ones derived from the chiral
fields $\partial_\pm X^\mu$ (cf.\ \cite{Ishibashi:1989kg,Onogi:1989qk}).
We also excluded the ghost sector from our discussion, as we work in  
light-cone quantization in most of the following chapters.
We have not been very precise in specifying the boundary states, i.e.\
solving the  boundary state conditions. This will however be done in
some examples in the following chapters. 

We also restricted to the case of orientifolds that stem from orbifolds of
smooth manifolds. We want to mention that other interesting constructions
exist including for example orientifolds on WZNW-models (describing smooth
manifolds as well) and Coset-spaces 
(cf.\ \cite{Pradisi:1995qy,Pradisi:1995pp,Hikida:2002ws,Couchoud:2002eg}). The spectrum of strings and D-branes can be
investigated by advanced means of the corresponding CFT. Also more geometrical
approaches to D-branes and open strings  have been undertaken  
(cf.\ \cite{Douglas2001:tr} and references given there).
 The relatively simple  orbifolds (though interesting as they usually contain
 singularities) serve as  valuable examples and give
 hints to  more general (mathematical) descriptions of D-branes.

There exists a huge amount of literature on orientifolds and it is impossible
to cite all of the publications.
 Some articles with major contributions to the field
have been cited in the text, many others not. In the remaining chapters  we
will mention additional publications, many 
 dealing directly with orientifolds.

As stated at the beginning of this chapter, the idea to define an 
orientifold as an orbifold of the two-dimensional
world-sheet theory was first described in an article by A.~Sagnotti's from 1987
 \cite{Sagnotti:1987tw}. Sagnotti and many of his collaborators have 
contributed a substantial part of research to the field of unoriented string
 theories from that time onwards.

 Not many survey articles have been written on 
orientifolds however, so it is not too hard to mention some. Polchinski,
who did a lot of important work, introduces orientifolds in his 
two books on string theory \cite{Polchinski:1998v1,Polchinski:1998v2}. 
Even though the name ``orientifold'' was not in use around that time, some
basics about open (oriented and non-oriented) superstring theories can
be found in the two volumes written by Green, Schwarz and Witten 
\cite{Green:1987sp,Green:1987mn}. In 1997, Atish Dabholkar gave a
 lecture on orientifolds in Trieste. The notes can be found  in the internet 
 \cite{Dabholkar:1997zd}. More recently Angelantonj and Sagnotti 
published an overview article \cite{Angelantonj:2002ct} that is devoted
to the CFT-oriented approach to orientifolds, developed by Sagnotti and
his collaborators. 

Even though the
material is not complete, we hope that it gives the reader necessary tools
to follow the sucsessive text.

\chapter[Open Strings in  Electro-Magnetic Background-Fields]
   {\label{strbg}\centerline{Open Strings in Electro-}
                 \centerline{Magnetic Background-Fields}
   }
In this chapter we will quantize the bosonic open string with
linear boundary conditions in flat space-time. These
are given both by the fact that D-branes are lower dimensional hypersurfaces
the string end-points are confined to and by the constant NSNS two-form $B$ 
in combination with the NS $U(1)$-field strength $F$.
The generalization to 
superstrings is straightforward and simpler than the bosonic
case.\footnote{There is a subtlety in deriving the boundary conditions
for the world sheet fermions from the action. Instead of coupling the
fermions to the boundary-$U(1)$-field strength $F$ via a boundary term
in the action, which might over-constrain the problem, we impose 
fermionic boundary conditions {\sl by hand}. We demand these fermionic 
boundary conditions to be compatible with super-symmetry transformations which
 also have to be reduced (to one half of the bulk super-symmetry)
  at the boundary (i.e. super-symmetry invariance under
the full Majorana spinor $\epsilon$ (cf.~eq.~\eqref{susytransf}) 
 over-constrains the
problem, too). See also \cite{Albertsson:2001dv,Albertsson:2002qc}.} 
We will generalize the result of \cite{Seiberg:1999vs,Chu:1999ta}
 on the non-commutativity
of string fields $X^\mu(\tau,\sigma)$ located at the boundary to the one
loop case (in comparison with \cite{Seiberg:1999vs})
and to the case where the boundary conditions on both boundaries are given
by  NS field-strengths $F_1$ and $F_2$ that are constant,
but completely independent form each other 
(in comparison with \cite{Chu:1999ta}). There are many other approaches 
to derive the commutator as well. A prominent one is by deformation
quantization \cite{Schomerus:1999ug}, others are guided by
constrained (or Dirac) quantization  \cite{Chu:1999gi,Ardalan:1999av}.
Laidlaw calculated the propagator for the cylinder with independent, constant
$U(1)$ $F$-fields at the boundaries and reproduced the result for the 
commutator as well  \cite{Laidlaw:2000kb}.
 The list is surely not complete.
We want to mention that the approach to solve the string boundary conditions,
which is actually a variant of a doubling trick, was motivated by 
\cite{Roy:1986hq,Roy:1987nv,Roy:1988xn} where the boundary condition
problem for open strings
on intersecting $D$-branes of arbitrary dimension {\sl without} NS
and NSNS background fields was solved.\footnote{Of course the term 
{\it D-brane} was not used around that time. The authors
also gave no space-time interpretation of the boundary conditions in the
spirit of \cite{Polchinski:1995mt}.}
 However we adopted another quantization
method here: We first calculate the canonical two-form in terms of the
individual modes. Then we restrict to its
 invertible part. The inverse is (up to a factor) the Poisson-bracket. 

We present a new and (to our knowledge) first direct derivation
of the open string mass formula for toroidally compactified D-branes
with magnetic fields (section \ref{toroidaldbranes}). 

\section[Action and boundary conditions of the open string]
        {\centerline{Action and boundary conditions}
         \centerline{of the open string}
        }
We consider the following superconformal-gauge action (space-time metric 
$G_{\mu\nu}$ is of signature $(1,d-1)$ ):
\begin{equation} \label{opensuperstaction}
  S = S_{\text{bos}} + S_{\text{ferm}}
\end{equation}
with the convention $\epsilon^{\tau \sigma}=1 $:

\begin{align}
  S_{\text{bos}} &= S_{\text{bos}}^{\text{bulk}}+S_{\text{bos}}^{\text{bdy}} 
                    \nonumber \\
  \label{bosonicaction} 
                 &=   -\frac{1}{4\pi \alpha^\prime} 
                         \left(\int_{\cal{M}} d^2\sigma \,
                             \partial_\alpha X^\mu \partial^\alpha X_\mu
                           -  B_{\mu\nu} \epsilon^{\alpha\beta}
                                   \partial_\alpha X^\mu
                                         \partial_\beta X^\nu
                           +   2\int_{\partial\cal{M}} d\tau \,   
                                     \Dot{X}^\mu A_{\mu}
                         \right)
  \\  \label{susywsactionconf} 
S_{\text{ferm}} &=  -\frac{1}{4\pi \alpha^\prime} \int_{\cal{M}} d^2\sigma
    \, 2 i\, \Bar{\psi}^\mu\rho^\alpha\partial_\alpha\psi_\mu 
\end{align}
In case of a constant $U(1)$-field strength with the gauge 
$A_\nu = \tfrac{X^\mu}{2}F_{\mu\nu}$,
($ F_{\mu\nu}= \partial_\mu A_\nu-\partial_\nu A_\mu $) and
constant $B_{\mu\nu}$ equation \eqref{bosonicaction}
reduces to:\footnote{Then $B_{\mu\nu} 
                  \epsilon^{\alpha\beta}\partial_\alpha X^\mu
                  \partial_\beta X^\nu = B_{\mu\nu} d(X^\mu\cdot dX^\nu)=
                  d( B_{\mu\nu} (X^\mu\cdot dX^\nu))$ with $d$ the 
                  exterior derivative on the world sheet.}
\begin{equation} \label{actionb2}
 S_{\text{bos}} = -\frac{1}{4\pi \alpha^\prime}
                   \left( \int_{\cal{M}} d^2\sigma \,
                               \partial_\alpha X^\mu \partial^\alpha X_\mu
                   - 2\Dot{X}^\mu (B)_{\mu\nu}X^{\prime\,\nu}
                   -      \int_{\partial\cal{M}}d\tau \,
                               \Dot{X}^\mu (F)_{\mu\nu}X^\nu 
                   \right)
\end{equation}
The $\psi$`s are world-sheet Majorana spinors. 
The two dimensional Dirac matrices $\rho^\alpha$ are in this gauge 
($h= \bigl(\begin{smallmatrix} 
            h_{\tau\tau} & h_{\tau\sigma}  \\
            h_{\sigma\tau} & h_{\sigma\sigma}
          \end{smallmatrix}  \bigr)
  = \diag (-1,1)$) :
\begin{equation}\label{2ddiracmat}
   \rho^0= \begin{pmatrix}
            0 & 1  \\
           -1 & 0
           \end{pmatrix} , \,
   \rho^1= \begin{pmatrix}
            0 & 1  \\
            1 & 0
           \end{pmatrix}
\end{equation}
They satisfy the algebra:
\begin{equation} \label{2ddiracalg}
   \bigl\{ \rho^\alpha ,\rho^\beta \bigr\} = 2h^{\alpha\,\beta}
\end{equation}
The spinor conjugate to $\lambda := \bigl(\begin{smallmatrix} 
            \lambda_+ \\
            \lambda_-
          \end{smallmatrix}  \bigr)$ is $\bar{\lambda} := \lambda^\dagger\rho^0
          = (
              -\lambda_- ,
              \lambda_+
            ) $.
The charge conjugation matrix $C$ is defined as $\rho^0$. Then a Majorana
spinor is real. The action \ref{opensuperstaction} is invariant under 
the following bulk super-symmetry transformation:
\begin{align}\label{susytransf}
  \delta_\epsilon X^\mu &= i\Bar{\epsilon}\psi^\mu
   & \delta_\epsilon\psi^\mu 
       &= \frac{1}{2} \rho^\alpha\big(\partial_\alpha X^\mu\big)\epsilon
\end{align}
In the case of constant $G_{\mu\nu}$ the variation of the bosonic action
 with respect to $X^\mu$ gives the bulk equation of motion 
\begin{eqnarray}\label{boseom}
\left(\partial_{\tau}^2-\partial_{\sigma}^2 \right) X^\mu =
4\partial_+\partial_- X^\mu = 0
\end{eqnarray} 
with $ \partial_\pm \equiv
      \tfrac{1}{2} ( \partial_\tau \pm \partial_\sigma ) $
plus a boundary condition.
The boundary contribution to $\delta S_{\text{bos}}$ is:
\begin{align}
 \delta S_{\text{bos, bdy.}} =& \nonumber \\ \nonumber
                     -\frac{1}{2\pi \alpha^\prime}&
                    \int_{\cal{\partial M}} d\tau \, \delta X^\mu \cdot
                    \big( \partial_\sigma X_\mu 
                    + B_{\mu\nu} \partial_\tau X^\nu +
                    \partial_\mu A_{\nu}\partial_\tau X^\nu - 
                    \partial_\nu A_{\mu}\partial_\tau X^\mu
                   \big) \\
                            = -\frac{1}{2\pi \alpha^\prime} &
                    \int_{\cal{\partial M}} d\tau \, \delta X^\mu \cdot
                    \big( \partial_\sigma X_\mu 
                           + {\cal F}_{\mu\nu}\partial_\tau X^\nu 
                    \big)    \text{ with }\;{\cal F}_{\mu\nu}\equiv 
                   (B+F)_{\mu\nu}
\end{align}

\begin{figure}
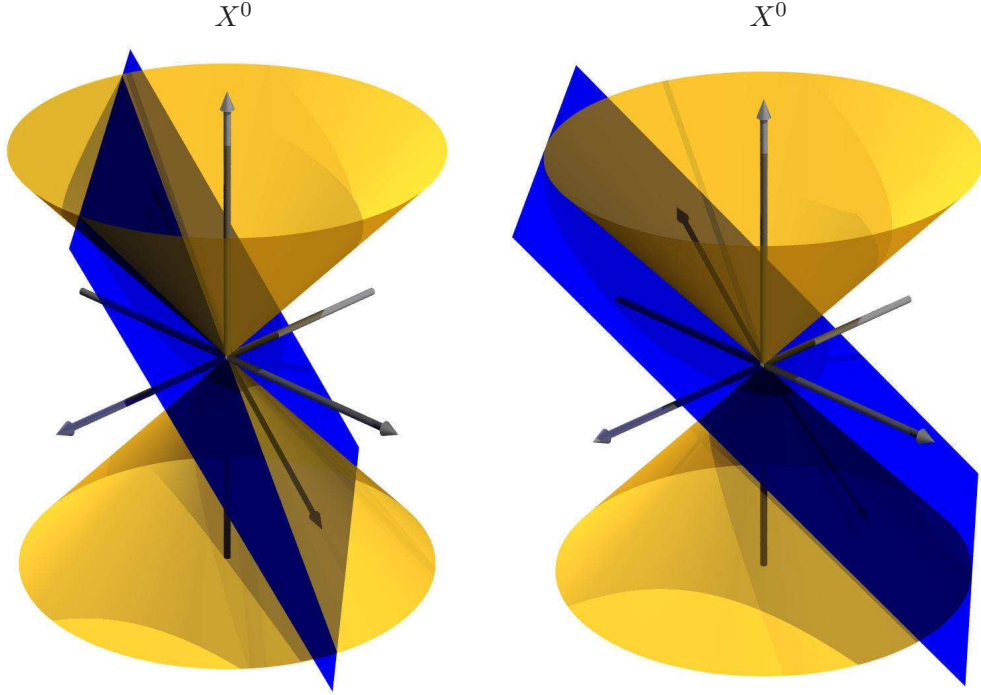

   \setlength{\unitlength}{0.1in}
   \begin{picture}(60,36)
   \put(0,0){\scalebox{0.25}{\includegraphics{picstrbg1.EPS2}}}
   \put(26.5,0.3){\scalebox{0.25}{\includegraphics{picstrbg2.EPS2}}}
  \put(11,35){$X^0$}\put(39,35){$X^0$}
  \end{picture}
\caption[Time-like and light-like branes]{
   \label{lightconepic}Time-like and light-like branes:
   D-branes can intersect the light-cone 
  (left) or only touch it (right) . 
  We will not consider the case of space-like
  branes. }
\end{figure}

We will consider {\sl flat} D-branes of arbitrary dimension with 
{\sl constant} but otherwise completely general
$B$ and $U(1)$ background flux $F$. Then these D-branes are hyperplanes. 
The $\mathbb{R}^d$ can be decomposed as $D_p\oplus V_{d-(p+1)}$ with 
$V_{d-(p+1)}$ the orthogonal compliment of the $Dp$-brane $D_p$. 
${\cal P}_\parallel$ and ${\cal P}_\perp$ denote the  parallel
resp.\ tranverse projections with respect to the brane. They can be defined as
follows:
Let the D-brane
be spanned by a set of vectors $d^\mu_i$, $i=0\ldots p$ and V by $c^\mu_j$, 
$j=p+1\ldots d-1$.
It turns out useful to distinguish light-like and non-light-like branes (c.f.\
figure \ref{lightconepic})
\begin{enumerate}
\item If the brane is not tangential to the light-cone,
choose the $c_i$ and $d_j$ s.th.~:
 \begin{equation}
  \begin{aligned}
    d^\mu_iG_{\mu\nu}d^\nu_j &= \eta^{\parallel}_{ij} &
    c^\mu_iG_{\mu\nu}c^\nu_j &= \eta^{\perp}_{ij} 
  \end{aligned}
 \end{equation} 
 By definition, $d^\mu_iG_{\mu\nu}c^\nu_j = 0$. $\eta^{(\cdot)}
 =\diag(-1,1\ldots1)$ or $= \id$, if the subspace contains a space-like
 direction or not.  
 \begin{equation}
  \begin{aligned}
   \big({\cal P}_\parallel\big)^\mu_{\phantom{\mu}\nu} &\equiv  
     \sum_{i\in D_p} \big(\eta^{\parallel}\big)^{ii} 
      d^\mu_id^\lambda_i G_{\lambda\nu} &
     \big({\cal P}_\perp\big)^\mu_{\phantom{\mu}\nu} &\equiv  
     \sum_{i\in V_{d-{p+1}}}^d \big(\eta^{\perp}\big)^{ii}
            c^\mu_ic^\lambda_i G_{\lambda\nu}
  \end{aligned}
 \end{equation}
\item 
  If the brane is tangential to the light-cone let 
  $d^\mu_0$ $\in D_p$ be light-like. 
  Then we choose $d^\nu_0$ and $c^\nu_{p+1}$ s.th.\ 
  $d^\mu_{0} G_{\mu\nu}c^\nu_{p+1}=1$ which is always possible. All other
  inner products involving one of the light-like $d^\nu_0$ and $c^\nu_{p+1}$
  with the other basis-vectors should vanish. This means that
  the other vectors lie in a subspace that is perpendicular to the one spanned
  by $d_0$ and $c_{p+1}$ 
  For the
  remaining vectors we can achieve:
  \begin{equation}
  \begin{aligned}
    d^\mu_iG_{\mu\nu}d^\nu_j &= \delta_{ij}\quad i,j={\bf 1}\ldots p; &
    c^\mu_iG_{\mu\nu}c^\nu_j &= \delta_{ij}\quad i,j={\bf p+2}\ldots d 
  \end{aligned}
 \end{equation} 
  In this case:
  \begin{equation} \label{lightlikeproj}
   \begin{aligned}
    \big({\cal P}_\parallel\big)^\mu_{\phantom{\mu}\nu} &\equiv  
     d^\mu_0 c^\lambda_{p+1} G_{\lambda\nu} +
       \sum_{i=\bf 1}^p d^\mu_id^\lambda_i G_{\lambda\nu} \\
       \big({\cal P}_\perp\big)^\mu_{\phantom{\mu}\nu} &\equiv  
     c^\mu_{p+1}d^\lambda_0  G_{\lambda\nu} +
       \sum_{i=\bf p+2}^d c^\mu_i c^\lambda_i G_{\lambda\nu} 
  \end{aligned}
 \end{equation}
 We further choose $c_{p+1}$ to be light-like as well (which is achieved
 by adding $\lambda d_0$ with appropriate $\lambda$). This will ensure 
 the relation  $R=R^{-1}$ in \eqref{Rmatrix}.
\end{enumerate}
The two cases are not continuously connected.
{\sl In some of the following formul\ae \;we will omit the indices.} 
They can be added
by taking into account that in matrix-multiplication an upper index is
contracted with a lower one. In addition, matrix inversion changes an upper
to a lower index and vice versa.   Indices are raised and lowered by 
 $G^{\mu\nu}$ and $G_{\mu\nu}$. Then 
\begin{align}
  &{\cal P_\parallel}+{\cal P_\perp} = G \\ 
  \label{Rmatrix}
  &{\cal P_\parallel}-{\cal P_\perp} \equiv R,\quad R
    \text{ invertible, } R=R^{-1}
\end{align}
The resulting bdy.-conditions are valid
even for non-constant $G$ and $\cal F$ (${\cal F}_\parallel\equiv 
   {\cal P}^T_\parallel {\cal F}{\cal P}_\parallel$):
\begin{align} \label{neumann}
          \begin{split}  \big({\cal P}^T_\parallel G\,\partial_\sigma   + 
                         {\cal F}_\parallel \,\partial_\tau \big) 
                          X(\tau,\sigma)
                         \big|_{\sigma\in\partial{\cal M}} =&  \\
                          \big({\cal P}^T_\parallel G\,
                          (\partial_+ -\partial_-)   + 
                         {\cal F}_\parallel \,
                            (\partial_+ + \partial_-)\big) 
                          X(\tau,\sigma)
                         \big|_{\sigma\in\partial{\cal M}} =& 0 
             \end{split} \\
\label{dirichlet} 
{\cal P_\perp} X(\tau,\sigma)\big|_{\sigma\in\partial{\cal M}} 
    = U^i c_i  \quad i=p+1\ldots d \quad U^i\in\mathbb{R} \text{, const.} 
\end{align}
${\cal P_\perp}$ is constant for a hyperplane. We can differentiate 
\eqref{dirichlet} w.r.t.\ $\tau$. 
The bdy.-condition can be reformulated as a relation between the left- and
right-moving part of the open string. One has to distinguish light-like and
other branes. In the non-light-like case ${\cal P}^T G = G{\cal P}$. 
Therefore:
\begin{equation}\label{bdy0}
\begin{split}
\partial_- X(\tau,\sigma)&= \big(GR-{\cal F}_\parallel\big)^{-1} 
                 \big(G+ {\cal F}_\parallel\big)
                                   \partial_+ X (\tau,\sigma)\\
                         &= \big(G-{\cal F}_\parallel\big)^{-1} 
                 \big(GR+{\cal F}_\parallel\big)
                                   \partial_+ X (\tau,\sigma)
\end{split}
                \quad\text{ for } \sigma\in{\partial\cal M}
\end{equation}
If $(G\mp{\cal F}_\parallel)$ is not invertible one has a critical
 case corresponding
to  vanishing DBI action.\footnote{For non-light-like branes 
$\det(G\mp{\cal F}_\parallel)\neq0$ is equivalent to 
$\det(GR\mp{\cal F}_\parallel)\neq0$.}
This means that $(G\pm{\cal F}_\parallel)$ has
 a non-trivial kernel. With 
$v^\mu {\cal F_{\mu\nu}} v^\nu=0$ it is trivial that two such coordinates
$v,w\neq 0$  with $(G+{\cal F}_\parallel)v=0$ and $(G-{\cal F_\parallel})w=0$ 
are light-like and 
distinct.  One can modify the condition \eqref{bdy0} by inserting suitable 
projectors, projecting out the light-like $v$ resp.\ $w$ direction  s.th.\ 
the matrices $(G\pm{\cal F})$ get invertible on the remaining subspaces. In 
this case the remaining boundary conditions put no obstructions on the
left-moving part of the $v$- and the right-moving part of the $w$-direction.
So the corresponding states will have  continuous spectra. 
This could lead to problems with 
positivity and unitarity in the quantum theory.
 The critical case will not be pursued further 
in this investigation (even though interesting in its own right). 
\\
Defining 
\begin{equation}\label{defV}
 V\equiv
     \big(G +{\cal F}_\parallel\big)^{-1}\big(GR - {\cal F}_\parallel\big) = 
 G^{-1} \big(GR - {\cal F}_\parallel\big)\big(G +{\cal F}_\parallel\big)^{-1}G
\end{equation} 
we note that V is in $O(1,d-1)$ (i.e.\ $V^TGV=G$) and \eqref{bdy0} reduces to:
\begin{equation}
\partial_- X(\tau,\sigma)= V^{-1}\partial_+ X(\tau,\sigma)\, ,
          \, \sigma\in\partial{\cal M}
\end{equation}

\subsection{Open strings with two boundaries}
In this subsection we will consider the open string with two boundaries 
(at $\sigma=0$ and $\sigma=\pi$), i.e.\ with two constant but otherwise 
completely independent U(1) gauge fields $F_1$ (at $\sigma=0$) and 
$F_2$ (at $\sigma=\pi$). Using \eqref{actionb2} we will absorb the constant 
$B$-field into  boundary terms.
Defining\footnote{The minus sign in front of $B$ takes into account that the
direction of the $\tau$ derivative at the $\sigma=0$ end-point
is reverse to the derivative at $\sigma=\pi$.}
\begin{align}
{\cal F}_1 &\equiv -B+ F_1 & {\cal F}_2 &\equiv B+ F_2
\end{align}
$V_1$ and $V_2$ are given by \eqref{defV} with ${\cal F}$ substituted
by ${\cal F}_1$ res.\ ${\cal F}_2$ and $R$ by $R_1$, $R_2$.
Taking into account the opposite
orientation of the $\sigma =0$ and the $\sigma =\pi$ boundary the two
relations between the right and the left moving part of the string are 
valid: 
\begin{align}  
\label{bdy1}
 \partial_- X^\nu(\tau,0) &=V_{\bf 1} \partial_+ X^\nu (\tau,0) \\
 \label{bdy2}
 \partial_- X^\mu(\tau,\pi) &= V_{\bf 2}^{-1}\partial_+ X^\nu (\tau,\pi).
\end{align}  
Because of the eom \eqref{boseom} $X^\mu$ can be expanded
in the following way:
\begin{equation}
\label{decomp}
X^\mu = H^\mu + X_L^\mu(\tau+\sigma) + X_R^\mu(\tau-\sigma) 
\end{equation}
with $H^\mu$ a constant vector.
$X_L^\mu$ depends only on $\tau+\sigma$ and $\widetilde{X}_R$ only on 
$\tau-\sigma$.
Let us consider the periodicity properties of say $\partial_+ X^\nu$. 
Note that $V$ is in $O(1,n-1)$ that means it is a  
Lorentz-transformation. $O(1,n-1)$ consists of four disconnected pieces.
 For the non-light-like branes one can use
$GR-{\cal F}_\parallel= (G-{\cal F}_\parallel)R$ to see that branes
with an even number of transverse dimensions correspond to the $SO(1,n-1)$
 subgroup.
\subsection{Solution to linear boundary conditions for the cylinder}
Defining another w.s.-time $\tilde{\tau}=\tau+ \pi$ ($\pi$ the open string
length) the first bdy.-condition \eqref{bdy1} reads:
\begin{equation}
\underbrace{\partial_-X^\mu(\tilde{\tau}-\pi,0)}
          _{=\partial_-X^\mu(\tilde{\tau},\pi) } 
          = \bigl(V_{\bf{1}}\bigr)^{\mu}_{\;\,\nu} \partial_+ X^\nu(\tau,0)
\end{equation}
Plugging the left hand side into the second condition \eqref{bdy2} one gets:
\begin{equation}
     \partial_-X^\mu(\tilde{\tau},\pi) = 
           \bigl(V^{-1}_{\bf{2}}\bigr)^{\mu}_{\;\,\nu} 
           \underbrace{ \partial_+ X^\nu(\tilde{\tau},\pi)}
           _{=\partial_+X^\mu(\tau,2\pi) }   
\end{equation}
Combining the last two equations we see that the following quasi-periodicity
holds for $\partial_+X^\mu$:
\begin{equation}\label{bdy3}
\partial_+X^\mu(\tau+\sigma+2\pi) = 
  \bigl(V_{\bf{2}}V_{\bf{1}}\bigr)^{\mu}_{\;\,\nu} 
                \partial_+ X^\nu(\tau+\sigma)
\end{equation}
As $V_{\bf{1,2}}$ (and therefore their product) are in $O(1,n-1)\subset 
U(1,n-1)$, 
$V_{\bf{2}}V_{\bf{1}}$  admits in general $n-1$
complex Eigenvectors $C^\mu_{\lambda_i}$ with Eigenvalues 
$\lambda$ ($|\lambda |=1$) and two real Eigenvalues $\xi$, $\xi^{-1}$ which 
belong to two real Eigenvectors $C^\mu_{\xi}$ and $C^\mu_{\xi^{-1}}$.\footnote{This is derived in appendix \ref{loerentzform},
 where also the other stated facts on
  the Eigenvectors are proven. We exclude a degenerate case (in which our
statement about the Eigenvectors would be wrong and) that might show
up for some special light-like Eigenvectors with Eigenvalue $\lambda=\pm 1$,
from our further analysis. We shed some light on this case in appendix
 \ref{loerentzform} as well. For a  purely {\sl magnetic} $F_{\bf{i}}$-field
  the transformation $V_{\bf{i}}$ is actually a rotation (i.e.\ $\in O(n)$).}
 We assume the $C^\mu_{\lambda_i}$ to be "ortho-normalized" 
(always possible in the case
at hand) with respect to the hermitian scalar product  
$\left<C_1,C_2\right>\equiv\sum_\mu {C^\mu_1}^\ast G_{\mu\nu}C^\nu_2$. As
$V_{\bf{2}}V_{\bf{1}}$ is real, for every $\lambda_i$,  $\lambda_i^\ast$ is
 Eigenvalue (with Eigenvector ${C^\mu_{\lambda^\ast}=C^\mu_\lambda}^\ast$),
 too. For $\xi\neq \pm 1$, $C^\mu_{\xi}$ and $C^\mu_{\xi^{-1}}$ are
 light-like.
 As their scalar product is non-vanishing, we will normalize $C^\mu_{\xi}$ and 
 $C^\mu_{\xi^{-1}}$ 
 such that $\left<C_{\xi^{-1}},C_\xi\right>=1$. $C_{\xi^{-1}},\, 
C_\xi$ are perpendicular to the $C_{\lambda_i}$. Let us represent 
$\lambda_i$ as
\begin{eqnarray}
 \lambda_1 =& \exp(i2\pi(\tfrac{-i}{2\pi}\ln\xi)    
           =  \exp(i2\pi\theta_1)               \\
 \lambda_2 =& \exp(i2\pi(\tfrac{i}{2\pi}\ln\xi)
           =  \exp(i2\pi\theta_2)               \\
 \lambda_i =& \exp(i2\pi\theta_i)      
\end{eqnarray}
$-1/2<\theta_i\le1/2,\, i=3\ldots n$ and $\mathfrak{Im}\ni\theta_{1,2}= 
\pm\tfrac{i}{2\pi}\ln\xi$.
Denote $\lambda_i^\ast$ by $\lambda_{-i}$ for $i=3\ldots n$ and 
$\lambda_{-1}= \lambda_2$. Similarly $\theta_{-i}= -\theta_{i}$ 
represents $\lambda_{-i}$. Note however that there are cases without
light-like Eigenvectors. For example a pure space rotation could
lead (in odd space dimensions) to a time-like Eigenvector with Eigenvalue
$\lambda=1$.
The following useful identity holds:
\begin{equation}
\sum_i C_{i}^\mu C_{-i}^\nu = G^{\mu\nu}
\end{equation} 
We will abbreviate the $\lambda_\ell=1$ Eigenvectors as $C_\ell$,
 $C_\jmath$,
 etc.
With $\partial_\pm X^\mu = \partial_\pm X^\mu_{L/R} $
we see that $X_{L/R}$ is of the form:
\begin{equation}
X_L^\mu = \text{const} + \bigg(\sum_{\ell : \lambda_\ell = 1} 
                        C^\mu_\ell p^\ell (\tau+\sigma)
                  \bigg) + 
           \frac{\sqrt{\alpha^\prime}}{i\sqrt{2}}\sum_j\, 
           C^\mu_j\sideset{}{'}
            \sum_{n_j\in \mathbb{Z}+\theta_j}
            \frac{a^j_{{-n}_j}}{n_j}e^{in_j(\tau+\sigma)}
\end{equation}
In the following we will absorb the constant part of the $X_L$ and $X_R$-field
into the common constant $H^\mu$ (eq.\ \eqref{decomp}).
Boundary condition \eqref{bdy1}  prescribes what $X_R$ has to be:
\begin{eqnarray}
 \partial_- X^\mu_R(\tau-\sigma,0) 
   =&  \bigl(V_{\bf{1}}\bigr)^{\mu}_{\;\,\nu}\partial_+
                           X^\nu_L(\tau-\sigma,0) \\
 \partial_- X^\mu_R(\tau-\sigma) 
   =&\bigl(V_{\bf{1}}\bigr)^{\mu}_{\;\,\nu} \partial_+ X^\nu_L(\tau+(-\sigma)) 
\end{eqnarray}
In the last equation the differentials act on the functions argument.
This then leads to:
\begin{equation}
X_R^\mu = 
         \bigg(\sum_{\ell : \lambda_\ell = 1} 
               \bigl(V_{\bf{1}}\bigr)^{\mu}_{\;\,\nu}
               C^\nu_\ell p^\ell (\tau-\sigma)
         \bigg)
          +
         \frac{\sqrt{\alpha^\prime}}{i\sqrt{2}}\sum_j\, 
           \bigl(V_{\bf{1}}\bigr)^{\mu}_{\;\,\nu}
            C^\nu_j\sideset{}{'}\sum_{n_j\in \mathbb{Z}+\theta_j}
            \frac{a^j_{{-n}_j}}{n_j}e^{in_j(\tau-\sigma)} 
\end{equation} 
Thus we have the following mode expansion of $X^\mu(\tau,\sigma)$:
\begin{multline}\label{oscillatorexpansionX}
X^\mu =
       \overbrace{ H^\mu}^{0\text{-modes}} + 
       \overbrace{\sum_{\ell : \lambda_\ell = 1}
                   \left( (\tau +\sigma)\id + 
                  (\tau -\sigma)V_{\bf{1}}\right)^{\mu}_{\;\,\nu}
                  C^\nu_\ell  p^\ell
                 }^{\text{linear modes}}  \\
         + \underbrace{\frac{\sqrt{\alpha^\prime}}{i\sqrt{2}}
         \sum_j\, \sideset{}{'} \sum_{n_j\in \mathbb{Z}+\theta_j}
            \frac{a^j_{{-n}_j}}{n_j}
             \left( e^{in_j(\tau+\sigma)} C^\mu_j +
                   \bigl(V_{\bf{1}}\bigr)^{\mu}_{\;\,\nu}C^\nu_j
                   e^{in_j(\tau-\sigma)} 
             \right)}_{\text{oscillator-modes}}
\end{multline} 
The Dirichlet condition \eqref{dirichlet} imposes further restriction on the
zero-mode part of the string. At $\sigma = 0$ the brane
(hyperplane)  is located at ${\cal P}_{\perp,\mathbf 1}U=U_{\mathbf 1}$. Then
${\cal P}_{\perp,\mathbf 2}U_{\mathbf 2}=U_{\mathbf 2}$ specifies 
the position of the $\sigma=\pi$ brane.
We become aware that
\begin{align}
{\cal P}_{\perp,\mathbf 1} V^{\pm 1}_1 &= -{\cal P}_{\perp,\mathbf 1} &
{\cal P}_{\perp,\mathbf 2} V^{\pm 1}_2 &= -{\cal P}_{\perp,\mathbf 2} 
\end{align}
as well as 
\begin{equation}
 V_1C_{\lambda_\ell}=V^{-1}_2C_{\lambda_\ell}\quad\text{ for } 
 \lambda_\ell=1
\end{equation}
with $\lambda_\ell$ the Eigenvalue of $V_2V_1$.
Let us rewrite the zero- and linear-modes of $X$:
\begin{equation}
X^\mu_{0} = H^\mu + 
           \bigg(\sum_{\ell : \lambda_\ell = 1}
                 \left(  \tau\left( \id + V_{\bf{1}}\right)^{\mu}_{\;\,\nu}
                  +\sigma\left(\id - V_{\bf{1}}\right)^{\mu}_{\;\,\nu}
                 \right)
                   C^\nu_\ell  p^\ell
           \bigg)    
\end{equation}
\subsubsection{World sheet momentum and Hamiltonian}
In order to quantize the string one has to know the (gauge-dependent)
canonical momentum:
 \begin{align} \nonumber
 P^\mu (\sigma) &=\frac{\partial}{\partial \Dot{X}_\mu} L(X,\partial X) \\
    \label{canmomg}
               &=
   \frac{1}{2\pi\alpha^\prime}\left( 
               \Dot{X}^\mu  + B^\mu_{\phantom{\mu}\nu}X^{\prime\,\nu}  -
                   \big( 
                       \delta({\sigma}) A_{\bf{1}}^\mu
                     + \delta({\sigma}-\pi) A_{\bf{2}}^\mu
                   \big)
                  \right) 
\\\label{canmom}
 &=\frac{1}{2\pi\alpha^\prime}\Big( 
                    \Dot{X}^\mu + B^\mu_{\phantom{\mu}\nu}X^{\prime\,\nu}
  +\frac{1}{2}\big( 
    \delta({\sigma}) F^{\phantom{\bf{\;}}\mu}_{\bf{1}\phantom{\bf{\;}}\nu}X^\nu
  +\delta({\sigma}-\pi)
                    F^{\phantom{\bf{\;}}\mu}_{\bf{2}\phantom{\bf{\;}}\nu}X^\nu
                   \big)
                                   \Big)                   
\end{align}
Equation \eqref{canmomg} is valid for all $U(1)$-field strengths,  
\eqref{canmom} only for constant field strengths in our particular gauge.
The momentum is not conserved since the Lagrangian varies under pure
translations.\footnote{By momentum we mean also the integral 
$\int d\sigma P$.} 

Another important quantity is the  world sheet Hamiltonian:
\begin{multline}
{ H}= \int_0^\pi d\sigma\, P^\mu(\sigma)\Dot{X}_\mu - L(\sigma) = 
\frac{1}{4\pi\alpha^\prime}\int_0^\pi d\sigma\, \Dot{X}^\mu\Dot{X}_\mu
+ X^{\prime\,\mu}X^\prime_\mu   \\
= \frac{1}{2\pi\alpha^\prime}\int_0^\pi d\sigma\, 
  \partial_+ X^\mu\partial_+ X_\mu + \partial_- X^\mu\partial_- X_\mu =
\frac{1}{\pi\alpha^\prime}\int_0^\pi d\sigma\, 
\partial_+ X^\mu\partial_+ X_\mu
\end{multline}
$ H$ is gauge invariant in terms of the $X^\mu$'s and its derivatives.
With 
\begin{equation}
  \partial_+ X^\mu_0 =   C^\mu_\ell  p^\ell
\end{equation}
the momentum-mode part of the Hamiltonian is:
\begin{equation}\label{Hamiltonianmom}
{ H}_{\text{lin}} = \frac{1}{\alpha^\prime} 
              \underbrace{C_\jmath^\mu G_{\mu\nu} C_{\ell}^\nu}
               _{\equiv G_{\jmath\ell}}\, p^\jmath p^\ell 
\end{equation}
Similarly one obtains the oscillator-part 
(which still has to be normal ordered):
\begin{equation}\label{Hamiltonianosc}
{ H}_\text{osc.} = \frac{1}{2}
    \sum_k\sideset{}{'}\sum_{n_k\in \mathbb{Z}+\theta_k}
     \underbrace{C_{-k}^\mu G_{\mu\nu}C_{k}^\nu}
      _{\equiv G_{-k\,k}}\, a^{-k}_{n_k} a^k_{{-n}_k}    
\end{equation}
\section{Quantization of open strings with linear boundary conditions}
In the following we will quantize the classical solution. To do this, we 
first calculate the {\it canonical two-form} $\Omega(P,X)$ 
in terms of the classical solution which we restrict in a second
step to a subspace on which  $\Omega(P,X)$ is nondegenerate. 
The inverse of $\Omega(P,X)$ (on this subspace) defines the 
{\it Poisson-bracket}. By substitution of the Poisson bracket by $-i$ times
the commutator we perform the transition to the quantized string. 
Alternatively one could have tried to implement the boundary conditions 
via a Dirac-bracket. However we think that the method applied here is more
direct and less complicated.

\subsection{Canonical two-form and canonical quantization}
In order to quantize the system lets look at the canonical two-form\footnote{
$D\,P$ and $D\,X$ means the derivative with respect to the target space. 
Therefore $\Omega$ is manifestly $U(1)$-gauge invariant since $P\rightarrow
P+D\, \xi$ and $X$ is invariant.}
\begin{align}
\Omega(P,X) &= \int_0^\pi d\sigma\ DP^\mu\wedge DX_\mu \\ \nonumber
            &=\frac{1}{2\pi\alpha^\prime}\int_0^\pi d\sigma\ 
              \Big(D\Dot{X}^\mu \wedge\, DX_\mu  \\ \nonumber
           &+\tfrac{1}{2}(\delta (\sigma)F_1 + 
                         \delta (\sigma-\pi)F_2)^{\mu}_{\;\,\nu} DX^\nu
              \wedge DX_\mu 
                    +\tfrac{1}{2} B^{\mu}_{\;\,\nu} 
                      \frac{\partial}{\partial\sigma}
                        \big(D X^\nu\wedge DX_\mu\big)\Big) \\
          &=\frac{1}{2\pi\alpha^\prime}\int_0^\pi d\sigma\ \nonumber
              \Big(D\Dot{X}^\mu \wedge\, DX_\mu  \\
           &\phantom{\frac{1}{2\pi\alpha^\prime}\int_0^\pi d\sigma\ 
              \Big(D\,} 
             +\tfrac{1}{2}\left(\delta(\sigma){\cal F}_{\mathbf{1}}
                 +\delta (\sigma-\pi)
          {\cal F}_{\mathbf{2}}\right)^{\mu}_{\;\,\nu} 
               DX^\nu \wedge DX_\mu \Big)
\end{align}
which is  time independent  in case of  constant field strengths,
two-form potential $B$ and on-shell ${X^\mu}$:\footnote{In the 
last line only parallel components of the ${\cal F}$ fields contribute,
as the perpendicular components of $\Dot{X}^{\mu}$ vanish at boundary.}
\begin{multline}
\frac{d}{d\tau}\Omega(P,X) = \\
  \nonumber
   \frac{1}{2\pi\alpha^\prime}
    \int_0^\pi d\sigma\big(\overbrace{D\Ddot{X}^\mu}^{=DX^{\prime\prime\,\mu}}
              \wedge\, DX_\mu + (\delta (\sigma){\cal F}_1 + 
                         \delta (\sigma-\pi){\cal F}_2)^{\mu}_{\;\,\nu} 
                    D\Dot{X}^\nu \wedge DX_\mu \big)\\
 \nonumber
   =  \frac{1}{2\pi\alpha^\prime}     \int_0^\pi d\sigma \,
          d\left(DX^{\prime\,\mu}\wedge\, DX_\mu\right) +
          \underbrace{DX^{\prime\,\mu}\wedge\, DX^\prime_\mu}_{=0}   \\ 
  \nonumber
       +\big(
         \delta (\sigma){\cal F}_{\parallel\,\mathbf{1}} +\delta (\sigma-\pi)
          {\cal F}_{\parallel\,\mathbf{2}}
         \big)^{\mu}_{\;\,\nu} 
               D\Dot{X}^\nu \wedge DX_\mu = 0
\end{multline}
Therefore one can neglect the $\tau$-dependent parts of $\Omega$.
\subsubsection{Quantization of zero- and linear-modes}
The following two expressions are useful to follow the discussion:
\begin{align}\label{zeromodeexpansion}
       \Dot{X}^\mu_{0}    &= (G+V_1)^{\mu}_{\;\,\nu}C^\nu_\ell p^\ell &
  X^\mu_{\tau\text{-indep.}} &= H^\mu + \sigma(G-V_1)^{\mu}_{\;\,\nu}
               C^\nu_\ell p^\ell 
\end{align}
The Poisson bracket is the inverse of the restriction $\Omega|_U$
of the
 canonical two-form $\Omega$ to a subspace $U$ s.th.\ $\Omega|_U$ 
is invertible.
To shorten the notation, we will not write down the derivative 
$D$ explicitly. We will
now quantize the individual (zero) modes:\footnote{The perpendicular
components of $B_{\mu\nu}$ in the term $\propto \big({\cal F}_{\mathbf{1}} +
       {\cal F}_{\mathbf{2}}\big)_{\mu\nu} H^\nu \wedge H^\mu$ cancel.}
\begin{multline}
\nonumber
\Omega_0(P,X)  \\=\int_0^\pi d\sigma\, 
         \Biggl(\overbrace{\frac{1}{2\pi\alpha^\prime}\left(
         (G+V_1)_{\mu\nu}C^\nu_\ell p^\ell\wedge H^\mu 
        + \sigma C^\mu_\ell(V^T_1-V_1)_{\mu\nu}
               C^\nu_\jmath p^\ell\wedge   p^\jmath  \right)}
                 ^{\text{from}\, \Dot{X}^\mu\wedge X_\mu}\Biggr)
\\
+ \frac{1}{4\pi\alpha^\prime}\big({\cal F}_{\parallel\,\mathbf{1}} +
       {\cal F}_{\parallel\,\mathbf{2}}\big)_{\mu\nu} H^\nu \wedge H^\mu
\\
    +\frac{1}{2\al}\big({\cal F}_{\mathbf{2}}(G-V_1)\big)
                  _{\mu\nu}C^\nu_\ell p^\ell\wedge H^\mu
    +\frac{\pi}{4\al}
     C^{\mu}_{\jmath}\big((G-V^T_1){\cal F}_{\mathbf{2}}(G-V_1)
                     \big)
           _{\mu\nu} C^{\nu}_\ell  \, p^\ell\wedge p^\jmath
\\
\end{multline}
Summarizing,
\begin{multline}
 \label{omegazerowedge}
 \Omega_0(P,X)  =\frac{1}{4\pi\al} \big({\cal F}_{\parallel\,\mathbf{1}} +
       {\cal F}_{\parallel\,\mathbf{2}}\big)_{\mu\nu} H^\nu \wedge H^\mu
   \hfill
\\
  +\frac{1}{2\al}\Big( (G+V_1)
                        +\big({\cal F}_{\mathbf{2}}(G-V_1)\big)
                 \Big)_{\mu\nu}C^\nu_\ell p^\ell\wedge H^\mu
\\
  +\frac{\pi}{4\alpha^\prime}
    C^{\mu}_{\jmath}\big((G-V^T_1){\cal F}_{\mathbf{2}}(G-V_1)
                              + (V_1-V_1^T)
                    \big)
           _{\mu\nu} C^{\nu}_\ell  \, p^\ell\wedge p^\jmath
\end{multline}
Now we will restrict to an invertible subspace.
We introduce a system of vectors $e^\mu_i$ 
such that
\begin{equation}  \label{aonedef}
  (A_1)_{ij}    \equiv  
   \frac{1}{4\pi} e^\mu_i\big({\cal F}_{\parallel\,\mathbf{1}} +
       {\cal F}_{\parallel\,\mathbf{2}}\big)_{\mu\nu}e^\nu_j   \qquad
   i,\,j = 1\ldots p 
\end{equation}
is invertible (such a system, (including possibly the empty set) always 
exists as ${\cal F}_{\parallel\,\mathbf{i}}$ is anti-symmetric).
Then we define $H^i$ and  $H^a$ by:
\begin{equation}
  H^\mu = e^\mu_i H^i +  d^\mu_a H^a
\end{equation}
The  $d^\mu_a$ are made orthogonal to the $e^\mu_i$. $H^\mu$ is real as
the string coordinate $X^\mu$ is real. This leads to some restrictions
on the phase of the $H^i$ and   $H^a$. In the quantized version of the string
$X^\mu$ becomes a hermitian operator. The classical restrictions on the
$H^i$ and   $H^a$ lead to some restrictions on their properties as operators.

In some special cases (i.e.~when Dirichlet conditions are absent), the
 $e^\mu_i$ are chosen to be $(G+{\cal F}_1)C_i$ with $\lambda_i\neq 1$.
This leads to the simple relation $H^{-i}=H^{i\,\dagger}$. The analogous
definition $d_a\equiv(G+{\cal F}_1)C_a$ with  $\lambda_a=1$ forces the
corresponding $H^a$ to be hermitian in this case.

With this choice and taking into account 
\eqref{omegazerowedge}
the general form of $\Omega_0|_u$ is (in matrix form):
\begin{equation}
\Omega_0|_U = \frac{1}{\alpha^\prime}
\begin{pmatrix} \label{omegamatrix}
   A_1        &        0          &     K          \\
    0         &        0          &     N     \\
   -K^T       &       -N^T        &     A_2         \\
\end{pmatrix}
\end{equation}
By comparison with \eqref{omegazerowedge} we identify in addition to
\eqref{aonedef}:
\begin{gather} \label{KNA2def}
 \begin{aligned} 
   (K)_{i\ell}   &\equiv\frac{1}{4}
     e^\mu_i \Big( (G+V_1)+\big({\cal F}_{\mathbf{2}}(G-V_1)\big)
             \Big)_{\mu\nu}C^\nu_\ell     \\
    (N)_{a\ell}       &\equiv\frac{1}{4}
     d^\mu_a \Big( (G+V_1)+\big({\cal F}_{\mathbf{2}}(G-V_1)\big)
             \Big)_{\mu\nu}C^\nu_\ell     \\
    (A_2)_{\ell\jmath} 
                 &\equiv  \frac{\pi}{4}
    C^{\mu}_{\jmath}\big((G-V^T_1){\cal F}_{\mathbf{2}}(G-V_1)
                         + (V_1-V_1^T)\big)
           _{\mu\nu} C^{\nu}_\ell  
 \end{aligned}                   \\
 \label{valueofN}
 \text{with } i= 1\ldots p;\qquad a,\,\ell,\, \jmath  = p+1\ldots r
\end{gather}
While $p$ is the dimension of $\image \big({\cal F}_{\parallel\,\mathbf{1}} +
       {\cal F}_{\parallel\,\mathbf{2}}\big)$ (or
equivalently: the dimension of $A_1$), $(r-p)$ is
the dimension of its kernel.
In \eqref{valueofN} we already used that the matrix $N$ has to be a
square matrix. This is true for the following reasons:
\begin{enumerate}
\item In order for $\Omega_0|_U$ to be invertible,  $N^T$ has to
   be of maximal rank (i.e. $\dim A_2$).
   Therefore the column number of $N$ equals the number
   of $p^\ell$ dofs (or equivalently: the number of Eigenvalue $\lambda=1$
   Eigenvectors of $V_2V_1$).
\item In order for \eqref{omegamatrix} to be invertible, its determinant has
 to be $\neq 0$. If the number of rows in $N$ exceeds the dimension
   of the square matrix $A_2$, at least two column vectors of
    $\Omega_0|_U$  would be linear dependent. As a consequence the
    determinant would vanish which contradicts the
    invertibility of $\Omega_0|_U$. As $\Omega_0|_U$ is invertible by
   definition,  $N$ must be a    square matrix. 
\item  As $N$ has maximal rank, it is invertible.
\end{enumerate}
In other words: the space spanned by the $d^\mu_a$ has to be reduced
such that $N$ gets invertible which is needed to ensure invertibility
of $\Omega_0|_U$.
While $A_{1}$  and $A_2$ are antisymmetric, $K$ is in general not
 a square matrix.\footnote{We found that the dimensions of the
spaces $\langle d_a\rangle$ and $\langle C^\jmath;\lambda_\jmath=1\rangle$ 
is equal. 
For the specific situation where the Dirichlet directions are the same
for both branes we can even identify: $d^\jmath=C^\jmath$.}
In this notation 
\begin{equation}
  \Omega_0(P,X)|_U = x_i (\Omega_0)_{ij} \wedge  x_j , \quad 
  \vec{x} = (H^i, H^a , p^\jmath )
\end{equation}
To get the Poisson bracket of the bosonic zero-modes one has to invert 
$\Omega_0$:
\begin{equation}\label{pbrack1}
  \left\{x^i,x^j\right\}_\text{P.B.}= \frac{1}{2}(\Omega_0^{-1})^{ij}
\end{equation}
The solution is
\begin{equation}
 \Omega_0^{-1} = \alpha^\prime
\begin{pmatrix} \label{pbrack2}
   A_1^{-1 }          &       -A_1^{-1 }KN^{-1}              &      0    \\
 -\big(N^T\big)^{-1} K^TA_1^{-1 } 
                      &  \big(N^T\big)^{-1}( A_2-K^TA_1^{-1 }K )N^{-1} 
                                                     & -\big(N^T\big)^{-1}\\
   0                  &           N^{-1}                    &      0    \\
\end{pmatrix} 
\end{equation}
As $\Omega_0$ is a symplectic form it can be transformed into a more 
convenient form by a general linear transformation of the zero modes
$H^i$, $H^a$, $p^\jmath$.
\begin{equation}
\Tilde{\Omega}_0 = S^T\Omega_0 S = \frac{1}{\alpha^\prime}
\begin{pmatrix} 
    A_1            &        0        &     0           \\
    0              &        0        &     N          \\
    0              &      -N^T       &     0           \\
\end{pmatrix} 
\end{equation}
with
\begin{equation}
 S = 
\begin{pmatrix} 
    1                       &        0          &       0                  \\
  -\big(N^T\big)^{-1} K^T   &        1          
                                                &\big(N^T\big)^{-1} A_2/2 \\
    0                       &        0          &       1                  \\
\end{pmatrix} 
\end{equation}
The transformed zero modes
$\tilde{H}^i$, $\tilde{H}^\ell$, $\tilde{p}^\jmath$ are defined by:
\begin{equation}
 \begin{pmatrix} 
\tilde{H}^i  \\
\tilde{H}^a  \\
\tilde{p}^\jmath  \\
\end{pmatrix}
 = S^{-1} 
\begin{pmatrix} 
H^i  \\
H^a  \\
p^\jmath  \\
\end{pmatrix}
=
\begin{pmatrix} 
    1                    &  0          &     0                           \\
  \big(N^T\big)^{-1} K^T &        1          &  -\big(N^T\big)^{-1} A_2/2 \\
    0                    &  0          &     1                           \\
\end{pmatrix} 
\begin{pmatrix} 
H^i  \\
H^a  \\
p^\jmath  \\
\end{pmatrix}
\end{equation}
The $\tilde{H}^i$, $\tilde{H}^a$, $\tilde{p}^\jmath$ have rather simple
Poisson-brackets and commutators (all others vanishing):\footnote{$N^{a\jmath}$ is the inverse
 transposed of $N_{\imath b}$ }
\begin{equation} \label{poissonbckt2}
 \begin{aligned}
    \big\{\tilde{H}^i,\tilde{H}^k\big\}  =&  
             \frac{\al}{2} \big(A_1^{-1}\big)^{ik}   \\
               \big\{\tilde{H}^a,\tilde{p}^\jmath \big\}  =& 
                      -\frac{\al}{2}    N^{a\jmath}
  \end{aligned}
  \quad\xrightarrow{\{\,.\, ,\,.\, \}\to -i[\,.\, ,\,.\, ]}\quad
  \begin{aligned}
    \big[\tilde{H}^i,\tilde{H}^k\big]  =&  
             \frac{\al}{2i} \big(A_1^{-1}\big)^{ik}   \\
               \big[\tilde{H}^a,\tilde{p}^\jmath \big]  =& 
                      i\frac{\al}{2}    \big(N^{-1}\big)^{a\jmath}
  \end{aligned}
\end{equation}
\subsubsection{Quantization of oscillator modes}
Now we look at the oscillator part of $\Omega$, which is also time
independent.
For simplicity we split $\Omega_{\text{osc.}}$ into $\Omega_1+\Omega_2$ 
and $\Omega_3$
where $\Omega_1+\Omega_2\propto\Dot{X}\wedge X$ is from the bulk integral and 
$\Omega_3$ is the boundary term. 
\begin{multline}
\frac{1}{2\pi\alpha^\prime} \Dot{X}_{\text{osc.}}^\mu(\tau,\sigma) \wedge  
X_{\text{osc.}\, _\mu}(\tau,\sigma) \\=    
\frac{1}{2\pi\alpha^\prime} \frac{-i \alpha^\prime}{2}  
\cdot
\sum_{j,\, l} \,  \sideset{}{'}\sum_{n_j\in \mathbb{Z}+\theta_j \atop
                     m_l\in \mathbb{Z}+\theta_l    } 
 e^{i\tau (n_j+m_l)} 
   \big(G e^{in_j\sigma} + V_1 e^{-i n_j\sigma}\big)
             ^{\mu}_{\phantom{\mu}\epsilon}
       C_j^\epsilon
    \\
   G_{\mu\nu}\big(G e^{i m_l\sigma} + 
        V_1 e^{-i m_l\sigma}\big)^{\nu}_{\phantom{\nu}\kappa} 
    C_l^\kappa
  \frac{a^j_{-n_j}\wedge a^l_{-m_l}}{m_l}
\end{multline}
As we consider only the $\tau$-independent terms, we impose $n_j=-m_l$, $j=-l$.
\begin{multline}
\frac{1}{2\pi\alpha^\prime} \Dot{X}_{\text{osc.}}^\mu(\tau,\sigma) \wedge  
X_{\text{osc.}\, _\mu}(\tau,\sigma) =  \frac{-i}{4\pi} 
\sum_{l} \,  \sideset{}{'}\sum_{m_l\in \mathbb{Z}+\theta_l}        \\
\Big( C_{-l}^\epsilon 
\underbrace{
\big(G e^{-i m_l\sigma} + V_1^T e^{i m_l\sigma}\big)
                                            _{\epsilon}^{\phantom{\mu}\mu}  
\big(G e^{i m_l\sigma} + V_1 e^{-i m_l\sigma}\big)_{\mu\kappa}
    }_{= (2G +  V_1^{\,-1}\cdot\exp(i2m_l\sigma)+
          V_1\cdot\exp(-i2m_l\sigma))_{\epsilon\kappa}}
C_l^\kappa\Big) 
 \frac{a^{-l}_{m_l}\wedge a^l_{-m_l}}{m_l}
\end{multline} 
From this we get $\Omega_1+\Omega_2 $:
\begin{multline}
 \Omega_1+\Omega_2 = \frac{1}{2\pi\alpha^\prime}\int_0^\pi  d\sigma
 \Dot{X}_{\text{osc.}}^\mu \wedge  X_{\text{osc.}\, \mu} \\
 =
   \overbrace{\frac{-i}{2} \sum_{l} G_{-l,l} 
   \sideset{}{'}\sum_{m_l\in \mathbb{Z}+\theta_l}
   \frac{a^{-l}_{m_l}\wedge a^l_{-m_l}}{m_l}}^{\equiv\Omega_1} \\ 
  \underbrace{- \sum_{l} 
   \sideset{}{'}\sum_{m_l\in \mathbb{Z}+\theta_l}
   \frac{1}{8\pi}  C_{-l}^\mu        
     \big(\overbrace{V_1^{\,-1} (\lambda_l -G)}^
         {\negphantom{\rule{6ex}{0ex}}
             \lambda_l C^T_{-l}= ((V_2V_1)^{-1}C_{-l})^T
          \negphantom{\rule{2ex}{0ex}} }
    - \overbrace{ V_1 (\lambda_{-l} -G)}^
         {V_1\lambda_{-l} C_l = V_2^{\,-1} C_l\negphantom{\rule{2ex}{0ex}}}
     \big)_{\mu\nu}C_{l}^\nu
          \frac{a^{-l}_{m_l}\wedge a^l_{-m_l}}{m_l^2}}_{\equiv\Omega_2}
\end{multline}
\begin{equation}
\Omega_2= -\frac{1}{8\pi}
  \sideset{}{'}\sum_{l \atop m_l\in \mathbb{Z}+\theta_l}
   C_{-l}^\mu
   \Big( 
    \big(V_2 - V_2^{\,-1}
        \big)
   +\big(V_1 - V_1^{\,-1}
        \big)
   \Big)_{\mu\nu}   
   C_{l}^\nu 
   \frac{a^{-l}_{m_l}\wedge a^l_{-m_l}}{m_l^2}
\end{equation}
Next we will show that $\Omega_2 + \Omega_3$ vanishes. If we assume that this is
true we immediately obtain the Poisson-bracket  for the modes (all others
vanishing):
\begin{equation}\label{poissinbrosz}
\big\{a^l_{-m_l} , a^{-l}_{m_l}\big\} = i m_l G^{-l\,l}
\end{equation}
$\Omega_3$ is the $d\sigma$-integral over:
\begin{multline}
 \frac{1}{4\pi\alpha^\prime}
                 \left( 
                       \delta({\sigma}) 
                         ({\cal F}_{\mathbf{1}})_{\mu\nu}
                        X_{\text{osc.}}^\nu
                     + \delta({\sigma}-\pi) 
                         ({\cal F}_{\mathbf{2}})_{\mu\nu}
                        X_{\text{osc.}}^\nu
                 \right) \wedge X_{\text{osc.}\, \mu}  \\
 = -\frac{1}{8\pi}
      \left(\delta({\sigma}) ({\cal F}_{\mathbf{1}})_{\mu\nu}
            + \delta({\sigma}-\pi) ({\cal F}_{\mathbf{2}})_{\mu\nu}
      \right)
 e^{i\tau (n_j+m_l)}  \\
  \sum_{j,\, l}  \sideset{}{'}\sum_{n_j\in \mathbb{Z}+\theta_j\atop
                     m_l\in \mathbb{Z}+\theta_l } 
   \big(G e^{in_j\sigma} + V_1 e^{-i n_j\sigma}\big)^\nu
                                        _{\phantom{\nu}\epsilon} C_j^\epsilon
   \big(G e^{i m_l\sigma} + V_1 e^{-i m_l\sigma}\big)^\mu
                                        _{\phantom{\mu}\kappa} C_l^\kappa
  \frac{a^j_{-n_j}\wedge a^l_{-m_l}}{n_j m_l}
 \end{multline} 
We only need to consider the $\tau$-independent terms ($n_j=-m_l$).
Using again $\lambda_l C^T_{-l}= C^T_{-l} (V_2V_1) =C^T_{-l}
 V_2 (V^T_1)^{-1} $  we can split $\Omega_3$ into
${\cal F}_{\parallel\,\mathbf{1}}$ respectively 
${\cal F}_{\parallel\,\mathbf{2}}$ dependent terms:
\begin{equation}
\Omega_3 = -\frac{1}{8\pi} 
   \sum_l  \Sigma_{-l,l}
  \sideset{}{'}\sum_{m_l\in \mathbb{Z}+\theta_l} 
  \frac{a^{-l}_{m_l}\wedge a^l_{-m_l}}{m_l^2} 
\end{equation}
where we have defined:\footnote{We have projected onto parallel 
components of the ${\cal F}_{\mathbf{i}}$-fields, because $G+V_i$
contains the projector ${\cal P}_{\parallel, i}$,
 that removes the perpendicular components.}
\begin{multline}
\Sigma_{-l,l} = 
     C_{-l}^\mu
      \big( 
           \big(G + V_1^{T}){\cal F}_{\parallel\,\mathbf{1}}(G+V_1)
       + 
           (G + \lambda_l V_1^{T}){\cal F}_{\parallel\,\mathbf{2}}
                  (G +V_1\lambda^{-1}_l  )
      \big)_{\mu\nu}
     C_l^\nu  \\
 = C_{-l}^\mu
      \big( 
           \big(G +  V_1^{-1}){\cal F}_{\parallel\,\mathbf{1}}
                                          (G+V_1)
       + 
           (G  + V_2)  {\cal F}_{\parallel\,\mathbf{2}}
                    (G +V_2^{-1})
      \big)_{\mu\nu}
     C_l^\nu
\end{multline}
As $\Omega_2$ and $\Omega_3$ are symmetric w.r.t.\ an exchange 
${\cal F}_{\parallel\,\mathbf{1}}\leftrightarrow 
{\cal F}_{\parallel\,\mathbf{2}}$ we consider only one part, 
e.g.\ the ${\cal F}_{\parallel\,\mathbf{1}}$ dependent
part:
\begin{multline} \label{vanish1}
\left.(\Omega_2+\Omega_3) \right|_{{\cal F}_{\parallel\,\mathbf{1}}}  \\
 =-\frac{1}{8\pi} 
   \sum_l
    C_{-l}^\mu 
     \big(G + V_1^{T})
      \overbrace{\big({\cal F}_{\parallel\,\mathbf{1}}  ( G + V_1) 
            -(G - V_1) 
      \big)_{\mu\nu}}^{(\ast)}
    C_l^\nu
      \sideset{}{'}\sum_{m_l\in \mathbb{Z}+\theta_l} 
  \frac{a^{-l}_{m_l}\wedge a^l_{-m_l}}{m_l^2}                       
\end{multline}
$(\ast)$ vanishes: 
The matrix $R_1$ that determines the Dirichlet conditions on the first brane,
commutes with $V_1$ and  ${\cal F}_{\parallel\,\mathbf{1}}$:
\begin{equation}
 (\ast)=\big({\cal F}_{\parallel\,\mathbf{1}}  ( G + V_1) 
            -(G - V_1) 
      \big)_{\mu\nu} 
  = R \big(2{\cal F}_{\parallel\,\mathbf{1}}
       -2{\cal F}_{\parallel\,\mathbf{1}}\big)
       (G+{\cal F}_{\parallel\,\mathbf{1}})^{-1} =0
\end{equation}
The ${\cal F}_{\parallel\,\mathbf{2}}$ dependent terms cancel analogously. 
Therefore $\Omega_2+\Omega_3$ vanishes and we end up with the Poisson 
brackets \eqref{pbrack1}, \eqref{pbrack2} and \eqref{poissinbrosz} for the
zero-, linear- and oscillator-modes.
The commutators that are obtained by the substitution 
$\{\,.\, ,\,.\, \}\to \tfrac{1}{i}[\,.\, ,\,.\, ]$ are (all others vanishing):
\begin{align}
 \big[a^l_{-m_l} , a^{-l}_{m_l}\big] &= m_l G^{-l\,l}  \\
 \big[H^\jmath , p^\ell\big]         &=  i\frac{\al}{2}
                                        \big(N^{-1}\big)^{\jmath\ell}\\
 \big[H^\jmath , H^\ell\big]              &= i\frac{\al}{2}
      \Big(\big(N^T\big)^{-1}( K^TA_1^{-1 }K  - A_2)N^{-1}\Big)^{\jmath\ell}\\ 
 \big[H^j ,H^k \big]                 &= \frac{\al}{2i}
                                         \big(A_1^{-1 }\big)^{jk} \\
 \big[H^j ,H^\ell \big]              &= i\frac{\al}{2} \big(A_1^{-1 }K
                                                     N^{-1}\big)^{j\ell}
\end{align}

We observe that in contrast to the zero and linear modes
 the quantization of the oscillator 
modes is not affected by the $B$-field. 

As an application we will calculate the commutator 
$\left[X(\tau,\sigma),X(\tau,\sigma^\prime) \right]$ at the string end-points
for the case without Dirichlet conditions.
 It turns out that this 
commutator is ill-defined for $\sigma=\sigma^\prime=0,\pi$ and that it has 
to be regularized.
\subsubsection{\label{toroidaldbranes}
             Quantization of zero and momentum modes in
             toroidal compactifications}

We already noted that the canonical momentum \eqref{canmom} is {\sl not}
a constant of motion, even though the field strengths $F$ do not
depend on space. This is due to the fact that the Lagrangian 
\eqref{bosonicaction}
 contains the vector potential $A$. Therefore $ S_{\text{bos, bdy.}}$ is
 explicitly space dependent
 for any nontrivial  NS field strength $F$:
\begin{equation}
  X_\mu\to X_\mu+\delta X_\mu \qquad \Rightarrow 
   \qquad A_\mu\to  A_\mu+  \delta X^\nu\partial_\nu A_\mu
\end{equation}
In the gauge chosen ($A_\nu = \tfrac{X^\mu}{2}F_{\mu\nu}$) we note however
that the combination of a translation $\delta X$ and a gauge transformation 
$A\to A-\partial\chi$ with $\chi=\tfrac{\delta X^\mu}{2}F_{\mu\nu}X^\nu$
leaves the action invariant.
We consider the following ``generalized momentum'':
\begin{align} \label{constmotionpi}
 \Pi^\mu&=P^\mu+\frac{1}{4\pi\al}\big( 
    \delta({\sigma}) F^{\phantom{\bf{\;}}\mu}_{\bf{1}\phantom{\bf{\;}}\nu}X^\nu
  +\delta({\sigma}-\pi)
                    F^{\phantom{\bf{\;}}\mu}_{\bf{2}\phantom{\bf{\;}}\nu}X^\nu
                   \big)  \\ \label{constmotionpib}
       &=\frac{1}{2\pi\alpha^\prime}\Big( 
                    \Dot{X}^\mu + B^\mu_{\phantom{\mu}\nu}X^{\prime\,\nu}
  +\big( 
    \delta({\sigma}) F^{\phantom{\bf{\;}}\mu}_{\bf{1}\phantom{\bf{\;}}\nu}X^\nu
  +\delta({\sigma}-\pi)
                    F^{\phantom{\bf{\;}}\mu}_{\bf{2}\phantom{\bf{\;}}\nu}X^\nu
                   \big)
                                   \Big) 
\end{align}
In contrast to the canonical momentum \eqref{canmom}, 
$\int d\sigma \Pi^\mu(\tau,\sigma)$ is a constant of motion. It generates
combined translations and gauge transformations. For $F=0$ it reduces to
the ordinary canonical momentum, which is conserved in the  $F=0$ case. 
Therefore we interpret  \eqref{constmotionpi} as a generalization of 
the generators for translations. This is very similar to the magnetic
translation group introduced in condensed matter physics (cf.\ 
\cite{Brown:1964,Zak:1964}). 
In string theory the momentum \eqref{constmotionpi}
already showed up in \cite{Ferrer:1995}. For simplicity, we consider 
the case without Dirichlet conditions. Inserting the solution
\eqref{oscillatorexpansionX} into \eqref{constmotionpib} and integrating 
$\Pi^\mu(\tau,\sigma)$ over $\sigma$ we end up with the $\tau$ independent
expression:
\begin{equation}
  \Pi = \frac{F_1+F_2}{2\pi\al}H +\frac{1}{\al} 
   \big(G+{\cal F}_2{\cal F}_1\big)\big(G+{\cal F}_1\big)^{-1} 
   \sum_{\ell\,:\,\lambda_\ell= 1} C_\ell p^\ell    
\end{equation}
We use some of the results of the following section 
\ref{signullcomm} to further simplify this expression. Especially
the relations \eqref{relationpr2} and \eqref{offdiagvanish}
turn out to be useful. 
 Using in addition 
 $\big(G+{\cal F}_1\big)^{-1}C_\ell=\big(G-{\cal F}_2\big)^{-1}C_\ell$ 
(for $\lambda_\ell =1$),
 which can be deduced from the characteristic polynomial \eqref{charmat},
 we can rewrite the {\sl magnetic translation generator} $\Pi$:\footnote{This
 term is used as $\Pi$ is very similar to the generators
of the group introduced in \cite{Brown:1964,Zak:1964}.} 
\begin{equation}\label{canmomsimpform}
  \Pi = \frac{F_1+F_2}{2\pi\al} \sum_{k\,:\,\lambda_k\neq 1}  C_k H^k
        + \Big(G\mp{\cal F}_{1 \atop 2}\Big) 
           \sum_{\ell\,:\,\lambda_\ell= 1} C_\ell p^\ell 
\end{equation}
We calculate the following commutator of the magnetic translation operators 
$\Pi^\mu$:
\begin{equation}
   \big[\Pi^\mu,\Pi^\nu \big] = \frac{i}{2\pi\al}(F_1+F_2)^{\mu\nu}
\end{equation}
Finite translations $T$ by a vector $R_\mu$ are generated via the exponential
map:
\begin{equation}\label{Tdefin}
   T(R) \equiv e^{i\langle \Pi, R\rangle }
\end{equation}
By using the Campbell-Hausdorff Formula we get the following multiplication
law for the algebra of magnetic translations:
\begin{equation}
 T\big(R^{(1)}\big)T\big(R^{(2)}\big)=  T\big(R^{(2)}\big)T\big(R^{(1)}\big)
                   e^{\frac{i}{2\pi\al}R^{(1)}(F_1+F_2)R^{(2)}}
\end{equation}
Furthermore, the  magnetic translation group is associative. (This justifies
the name ``group''. The inverse of $T(R)$ is $T(-R)$.)

We will now have a closer look to the restrictions that arise if we compactify
the theory on a $d$-torus.  
Strings on a torus generated by a lattice $2\pi\Gamma^d$ (cf.\ section 
\ref{toruscomp}, eq.~\eqref{dtorusdef})
 are described by wave functions that are invariant under all
translations $R^{(i)}\in 2\pi\Gamma^d$. This implies that the commutator 
of two such translations   $T\big(R^{(1)}\big)$ and  $T\big(R^{(2)}\big)$
has to vanish:
\begin{multline} \label{quatizationcondf}
  T\big(R^{(1)}\big)T\big(R^{(2)}\big)-T\big(R^{(1)}\big)T\big(R^{(2)}\big)=0
  \quad\forall \;e^{(1)},e^{(2)}\in \Gamma^d \\
   \Longrightarrow   e^{(1)}\frac{(F_1+F_2)}{\al}e^{(2)}\in \mathbb{Z}
\end{multline}
This condition is  equivalent to the condition that\footnote{$C_2$ denotes
the group of two-cycles on the torus $T^d=\mathbb{R}^d/\Gamma^d$. 
$ c_1(F_1+F_2)$ is the first Chern class w.r.t.\ the $U(1)$ field strength
 $F_1+F_2$.}
\begin{equation}
  \int_{\cal C} c_1\big((F_1+F_2)/\al\big)= 
  \int_{\cal C} \frac{F_1+F_2}{\al} \in 4\pi\mathbb{Z}
                          \quad\forall\, {\cal C}\in C_2
\end{equation}
The normalization in the above formula is unconventional, but one observes
that the result is also a consequence of the mathematical observation 
 that the first Chern class of a $U(1)$-connection on a $d$-torus
takes values in $\mathbb{Z}^{\big({d \atop 2}\big)}$.\footnote{
$\big({d \atop 2}\big)$ is the dimension of $H_2\big(T^d\big)$.}
The matrix $1/\al(F_1+F_2)$ is an antisymmetric map over
the free module $\mathbb{Z}^d$. According to \cite{Lang:1984}, chapter
{\sf XIV}, the corresponding matrix can be brought to a block diagonal form
by a  change of base:
\begin{equation}\label{Fstandardform}
  \frac{1}{\al} S^T (F_1+F_2) S =
  \bigoplus_{j=1}^{\lceil d/2\rceil} 
     \begin{pmatrix}
       0     & f^{(j)} \\
     -f^{(j)} &   0
     \end{pmatrix} \; ,\quad f^{(j)} \in \mathbb{Z},\; S\in SL(d,\mathbb{Z})
\end{equation}
The above sum does {\sl not} need to be an orthogonal one, because in general
there does not exist a change in basis described by $S$ that is
contained in  $SO(d,\mathbb{Z})$ leading at the same time to the
block-diagonal form \eqref{Fstandardform}. 

In section \ref{signullcomm} we will see, that the space spanned
by $C_l$ with $\lambda_l\neq 1$ is perpendicular to the one spanned
by  $C_\ell$  with $\lambda_\ell=1$. Furthermore the projection on these spaces
splits the field strengths $F_i$ according to eq.\ \eqref{fonesplit} and 
  eq.\ \eqref{ftwosplit} (p.\ \pageref{fonesplit}).
We write the torus lattice $\Gamma^d$ as a  direct sum (which is not 
necessarily an orthogonal sum\footnote{This means that two lattice vectors 
$v_1\in\Gamma^p$ and $v_2\in\Gamma^{d-p}$ might have non-vanishing scalar
product.}):
\begin{equation}\label{magntoruslat}
  \Gamma^d =  \Gamma^{d-p}\oplus\Gamma^p
\end{equation}  
$d-p$ is the number of Eigenvectors with
Eigenvalue $\lambda_\ell=1$, that lie in the lattice $\Gamma^d$.
$p$ is the number of Eigenvalues $\lambda_i\neq 1$, with Eigenvectors lying
in the (complexified) space that is isomorphic (though 
not necessarily identical)
 to the $\mathbb{C}$-vector space spanned  $\Gamma^p$. 
Note however that the orthogonal split described by eq.\ \eqref{fonesplit} and 
  eq.\ \eqref{ftwosplit} is in general not compatible with the decomposition
 \eqref{Fstandardform}. Only the space spanned over $\mathbb{R}$
by the Eigenvectors $C_\ell$ is identical with the space spanned  
by the kernel of  \eqref{Fstandardform}. (The kernel of a matrix is unique,
and  $C_\ell$ spans the kernel of $(F_1+F_2)$.)

Requiring that $T\big(2\pi\Gamma^{d-p}\big)$ is represented trivially on all
Eigenfunctions of the Hamiltonian\footnote{As $\Pi$ commutes with the 
Hamiltonian, 
the Eigenfunctions of  the magnetic translations $T$ are Eigenfunctions
of the Hamiltonian as well.} leads via \eqref{Tdefin} to the condition: 
\begin{equation}
 \Pi\in \Gamma^{\ast\,(d-p)}
\end{equation}
It is very convenient, though not necessary, to choose the $ C_\ell$ to
form a basis of lattice $\Gamma^{(d-p)}$. Writing  the vector $\Pi$ in
the dual basis:  $\Pi=  m_\jmath e^\jmath$ with $e^\jmath\in \Gamma^{\ast\,(d-p)}$
we obtain:\footnote{By this we abandon the $C_\ell$ to form an ortho-normalized
set of vectors.}
\begin{equation}\label{singlevaluedcond}
 \sum_\jmath e_\jmath p^\jmath(\vec{m})
    = \Big(G\mp{\cal F}_{1 \atop 2}\Big)^{-1}\sum_\jmath m_\jmath e^\jmath\qquad
           \vec{m} \in \mathbb{Z}^{(d-p)}
\end{equation}
The Hamiltonian \eqref{Hamiltonianmom} for the linear (or momentum) modes
takes the following form:
\begin{equation}
  \label{HamiltonianmomTd}
 { H}_{\text{lin}}(\vec{m}) = \frac{1}{\alpha^\prime} 
              p^\jmath\underbrace{ \langle e_{\jmath},e_\ell\rangle}_{G_{\jmath\ell}} p^\ell 
          =\frac{1}{\alpha^\prime}
         m_\jmath \Big(\big(G-{\cal F}^2_{k}\big)^{-1}\Big)^{\jmath\ell} m_\ell 
           \quad k\in \{1,2\} 
\end{equation}  
The expression in the big parentheses might be called the dual of the
{\it open string metric} in a slight generalization of the open string 
metric introduced 
by Seiberg and Witten \cite{Seiberg:1999vs}. The metric introduced there
does not include the $U(1)$ field strength. Eq.~\eqref{HamiltonianmomTd}
is independent of $k$. Note however, that the $G_{\jmath\ell}$ is the metric
of the lattice $\Gamma^{(d-p)}$, i.e.\ the lattice spanned by 
$e_\jmath$ with $(F_1+F_2)e_\jmath=0$, and not of the full lattice $\Gamma^d$.
 
After solving the linear modes, we are still left with the zero modes,
i.e.\ with the operators $H^i$, $\lambda_i\neq 1$.
Their quantization is dictated by the invariance of the wavefunctions
under translations by $2\pi \Gamma^p$. (The translations
by  $2\pi \Gamma^{p-d}$, are projected out in 
$\langle 2\pi \Gamma^{p-d}, (F_1+F_2)H \rangle$.) 
 As the $H^i$ (in contrast to the $p^\ell$) do not
commute, we can not get simultaneous Eigenfunctions of all $H^i$. 
However we can find a basis, such that $(F_1+F_2)$ becomes block-diagonal
(cf.\ \eqref{Fstandardform}).\footnote{Note that the block-diagonal form
in \eqref{Fstandardform} does of course {\sl not} imply that the (lattice)
 vectors belonging to the individual blocks are perpendicular w.r.t.\ $G$.}
 The respective pairs $\big(H^{(i)}_1 e^{(1)}_i,H^{(i)}_2 e^{(2)}_i\big)$ 
(no sums over $i$),
 then fulfill (cf.\ \eqref{Fstandardform}):
\begin{equation}
   e^{(1)}_i \frac{(F_1+F_2)}{\al} e^{(2)}_j = f^{(i)}\delta_{ij}
\end{equation} 
We want the wavefunctions to be Eigenfunctions of one of the $H^{(i)}_j$.
Without loss of generality we choose the set $H^{(i)}_1$. 
Under the translation $n^i\cdot 2\pi e^{(2)}_i$ an $H_1$-Eigenfunction
 with Eigenvalue $H_1$ acquires a phase:
\begin{equation}\label{translatphase}
 \langle \Pi,n\cdot 2\pi e^{(2)}_i  \rangle =
 2\pi n^i \bigg( \frac{f^{(i)}}{2\pi}H_1^{(i)}
   +  \frac{1}{\al}\vevs{e^{(2)}_{i},(G-{\cal F}_1)e_\ell}p^\ell(\vec{m})
          \bigg)
\end{equation}
Single valuedness of the wavefunction
requires for the Eigenvalue of  $H^{(i)}_1$:
\begin{equation}\label{Heigenvals}
  H^{(i)}_1\big(l^{(i)}\big) =
      2\pi\frac{l^{(i)}}{f^{(i)}} 
     \qquad l^{(i)}\in \mathbb{Z}
\end{equation}  
(as the second term on the r.h.s.\ in \eqref{translatphase} vanishes). 
The inequivalent choices of $ H_1^{(i)}$  are given by  
$ l^{(i)}\in\big\{1\ldots f^{(i)}\big\}$.
The wavefunctions are completely localized in the $H^{(i)}_1$ coordinate.
Translations by $ 2\pi e^{(1)}_i $ map a wavefunction localized at 
$H^{(i)}_1$ to one localized at $H^{(i)}_1+ 2\pi$. Therefore we do not
need to forbid  that the wavefunction  picks up a phase under such a 
translation. Symmetrizing the wavefunction in the $ e^{(1)}_i$ direction
now means to add up all translated wavefunctions.

The algebra 
$\big[H_1^{(i)},H_2^{(2)}\big]=\tfrac{i2\pi}{f^{(i)}}$ is not finitely 
represented by matrices, but admits the usual  infinite-dimensional
 representation: 
\begin{equation}
 \phi\big(H^{(i)}_1\big(l^{(i)})\big)=\exp\big(-i  H^{(i)}_1(l^{(i)}) x^{(i)}\big)
  ,\qquad \Hat{H}_1\simeq  
         i\partial_{x^{(i)}}\, ,\;\hat{H}_2\simeq 
             \tfrac{2\pi}{ f^{(i)}}x^{(i)} 
\end{equation}

The number of inequivalent states (i.e.\ the number that enters partition
functions) is  given by:\footnote{$(F_1+F_2)_{ij}$ denotes
 the restriction of $(F_1+F_2)_{ij}$ to the subspace on which $(F_1+F_2)$ is invertible.}
\begin{equation}
  n_0 = \prod_{j=1}^{p/2} 
       f^{(j)} \qquad\Bigg(=(-1)^{\lceil p/2\rceil}
          \pf\bigg(\frac{(F_1+F_2)_{ij}}{\al}\bigg) \Bigg)
\end{equation}
From the representation of the Pfaffian we see that the multiplicity 
$n_0$ can be rewritten as:
\begin{equation}
  n_0 = \ch_{p/2}\big((F_1+F_2)_{ij}\big)  
\end{equation}
If we are only interested in chiral degrees of freedom, which appear
in super-string compactifications, we do not need to restrict
to the invertible matrix  $(F_1+F_2)_{ij}$. Possible bosonic 
momentum-modes imply
vanishing chiral fermion number. Denoting the multiplicity of chiral
states by $\nu_0$ we can replace Chern character $ \ch_{p/2}$ by
the top Chern character $\ch_{d/2}$ in the following way:\footnote{We assume
the number of compactified dimension to be even.}  
\begin{equation}
  \nu_0 = \ch_{d/2}\big((F_1+F_2)\big) =
          \int_{T^d} \ch  (F_1\otimes F_2) =  \int_{T^d} \ch F_1 \wedge \ch F_2
\end{equation} 
This chiral multiplicity $ \nu_0$ is a special case of the 
{\sl Atiyah-Singer index theorem} for twisted spin-complexes: \footnote{Cf.\
  \cite{Eguchi:1980}, p.~331-334 and \cite{Nakahara}, p.~420-424).}
\begin{equation}
  \nu_0 = 
          \int_{\cal M} \hat{\cal A}({\cal R})\ch E =
            \int_{T^d} \ch F_1 \wedge \ch F_2
\end{equation}  
Like in chapter \ref{orientifolds}, $\hat{\cal A}({\cal R})$ denotes
the A-roof genus of the tangent bundle $T{\cal M}$. $\ch E$ is the
Chern character of the vector-bundle to which the gauge field $F$ is the
curvature two-form.

The multiplicities $\nu_0$  
 can be interpreted as {\it Landau levels} that appear in the case of 
quantized point particles in a (constant) magnetic background field.
In finite systems the degeneracy of Landau levels is finite as well.
If spin is included (Pauli-equation), the Landau levels are split according
to the spin $s=\pm 1/2$. 

We want to conclude with a remark on the quantization condition for 
the $F$-fields \eqref{quatizationcondf}. The torus group $2\pi\Gamma^d$ 
of the open string might
be a proper subgroup of the closed string   torus group (which we denote
for clarity by $2\pi\tilde{\Gamma}^d$). In terms of the 
$\tilde{\Gamma}^d$ basis, $(F_1+F_2)/\al$ might then be $\mathbb{Q}$-valued.
The physical interpretation is as follows: The D-brane wraps the torus
$n$-times ($n\in\mathbb{Z}$). Without $F$-field this would imply that
additional massless modes appear, promoting the gauge symmetry from
the $U(1)$ of the single $D$-brane to a  $U(n)$-symmetry of $n$ $D$-branes.
Due to the  $F$-field, even though the brane is multiply wrapped, no further
massless modes show up, and the gauge symmetry stays a $U(1)$. 

In what follows, we will make some comments on the further quantization
procedure.

\subsection{Hilbert-space, further quantization}
So far we quantized oscillators as well as zero and linear (momentum) modes.
If no electric components are present, and if the $X^0$ (time direction)
has Neumann boundary condition, light-cone quantization might be applied.
This procedure is then completely analogous to the quantization of branes at
angles
\cite{Roy:1986hq,Roy:1987nv,Roy:1988xn,Berkooz:1996km,Breckenridge:1997ar}, 
except for the zero modes. 
 There should exist a description for the
vertex operators as well. For open strings with $B$-field (the 
general ${\cal F}$-field case works the same way with $F_1=-F_2$)
the vertex operators and their OPEs were given explicitly in 
\cite{Schomerus:1999ug,Seiberg:1999vs}. However the situation of
$F_1\neq F_2$ ,  has not yet
been {\sl completely} explored to our knowledge.

If electric components are present, the conditions for applying light-cone
quantization are no longer fulfilled.\footnote{This condition means that
there exist at least one light-cone coordinate whose boundary condition
is Neumann.
 Otherwise this coordinate can not be identified with a 
(transformed) world sheet time because the  world sheet coordinate- and
conformal transformations obey 
$\partial_\sigma\tilde{\tau}(\tau,\sigma=0,\pi)= 
\partial_\sigma\tilde{\sigma}(\tau,\sigma=0,\pi)=0$ reflecting the fact
that the boundary is mapped to itself.}
One could then apply path-integral quantization.\footnote{Even though
the  path integral formalism does not require the very special form
of the mode expansion of light-like coordinates, as the light-cone formalism
does, there are still some obstacles left. For example the path integral
is only defined for euclidian space(-time). Even though one might argue 
that the minkowskian answer is obtained by (a second) Wick rotation, 
it is not clear if this method does not miss some points.}
The boundary conditions
of the ghosts are unaffected by the ${\cal F}$-fields. The partition function
is a product of the bosonic part and the ghost part. Roughly speaking,
the ghost part is the inverse of the partition function of one complex
boson with Neumann boundary conditions. From the Hamiltonian 
\eqref{Hamiltonianosc} and the Poisson bracket \eqref{poissinbrosz} 
 we see that light-like Eigenvectors of $V_2V_1$ with Eigenvalues 
$\xi\neq\pm 1$ would imply the existence of particles with 
{\sl complex} mass\raisebox{1ex}{\tiny 2}. This indicates surely an 
instability. Complex  mass\raisebox{1ex}{\tiny 2} are usually inserted
 into propagators of unstable particles. The instability  encountered
in the presence of electric fields  was observed in string theory by Burgess
\cite{Burgess:1987}, who considered the case of strings with independent 
charges on the end points but with identical $F$-field.
For consistent quantization we will require for the Eigenvalue 
spectrum of $V_2V_1$:
\begin{equation}
|\lambda_i|=1 \quad\forall \quad \text{Eigenvalues }\lambda_i
\end{equation}
As the quantization of the remaining cases is completely analogous to
the known cases cited above from now onwards, we will skip the rest 
of the procedure and turn to some aspect on non-commutativity that arise
in the context of  strings coupled to $B$ and $F$ fields. 
 
\section{\label{commutatorsec}
         The commutator 
    \texorpdfstring{$\left[X(\tau,\sigma),X(\tau,\sigma^\prime) \right]$}
                   {[X(tau,sigma),X(tau,sigma\textgrave\ )]}}
In what follows, we will calculate 
$\left[X(\tau,\sigma),X(\tau,\sigma^\prime)\right]$ and confirm by this the
 disk result of Seiberg and Witten for the 
one-loop case with general constant ${\cal F}$-fluxes on both branes. 
This is of course
expected as the commutator usually only depends on local properties.
For simplicity we restrict to the case without Dirichlet boundary conditions
at some point. To make things simpler, we set $\tau$ to zero. 
However it is easily checked that the expressions that we calculate,
i.e. the oscillator part and the part of the linear modes (the $p_\ell$'s)
 are independent of $\tau$. The commutator part linear in the $p_\ell$'s
cancels out and a  part quadratic in the  $p_\ell$'s does not exist, 
as the $p_\ell$'s commute (cf.\ \eqref{pbrack2}). The $\tau$ dependence
of the oscillator part would drop out explicitly in \eqref{timedropout} anyway
since the commutator $\bigl\{a^j_{-n_j},a^{-l}_{m_l}\bigr\}$ selects 
oscillators with opposite $\tau$ dependence in the exponential.

\subsection{\label{signullcomm}The commutator 
           \texorpdfstring{$\left[X(\tau,0),X(\tau,0) \right]$}
           {[X(tau,0),X(tau,0)]}
           }
We divide the Poisson bracket for the world sheet field $X(\tau,\sigma)$
into an oscillator, a zero-mode and a linear part. For
$\sigma=\sigma^\prime=0$ only the oscillator and zero mode part contribute:
\begin{multline}
 \left\{X^\mu(\tau=0,\sigma),X^\nu(\tau=0,\sigma^\prime)\right\} = \\
 \left\{X^\mu(\tau=0,\sigma)_\text{osc},
                    X^\nu(\tau=0,\sigma^\prime)_\text{osc}\right\}
  + \left\{H^\mu,H^\nu\right\}
\end{multline}
The oscillator part turns out to be:
\begin{multline}\label{timedropout}
  \left\{X^\mu(\tau=0,\sigma=0)_\text{osc},
                    X^\nu(\tau=0,\sigma^\prime=0)_\text{osc}\right\} \\
 =-\frac{\al}{2}\sum_j\sum_l(G+V_1)C_j
         \sideset{}{'}\sum_{n_j\in \mathbb{Z}+\theta_j}
         \sideset{}{'}\sum_{m_l\in \mathbb{Z}+\theta_l}
                 \frac{\bigl\{a^j_{-n_j},a^{-l}_{m_l}\bigr\}}
                                          {n_j(-m_l)}
                                 C_l (G+V_1^T)                      \\
 =-\frac{\al}{2}\sum_j\sum_l(G+V_1)C_j
         \sideset{}{'}\sum_{n_j\in \mathbb{Z}+\theta_j} \frac{i}{n_{-j}}
            C_-j (G+V_1^T)
\end{multline}
The above expression is meaningless unless we do not regularize
the divergent term 
$ \sideset{}{'}\sum_{n_j\in \mathbb{Z}+\theta_j} \frac{i}{n_{-j}}$. We 
regularize by substituting this term:
\begin{equation}
  \sideset{}{'}\sum_{n_j\in \mathbb{Z}+\theta_j} \frac{1}{n_{-j}}=
  \sum_{n\in \mathbb{Z}\atop n-\theta_j\neq 0} \frac{1}{n-\theta_j}
  \rightarrow
   \Biggl(\sum_{n\in \mathbb{Z}^\ast\atop n-\theta_j\neq 0} 
    \frac{1}{n-\theta_j}-\frac{1}{n}\Biggr)-\frac{1}{\theta_j}
    =\pi \cot(-\pi\theta_j)
\end{equation}
Using $\cot(-\pi\theta_j)=i(\lambda_j^{\oh}+\lambda_j^{-\oh})/
                           (\lambda_j^{-\oh}-\lambda_j^{\oh})$
as well as orthogonality of the $C_j$, we can rewrite the regularized Poisson
bracket as 
\begin{multline} \label{oscmodepoisson}
 \left\{X^\mu(\tau,\sigma=0)_\text{osc},   
                    X^\nu(\tau,\sigma^\prime=0)_\text{osc}\right\} \\
 =         \frac{\pi\al}{2}\sum_{j\,:\,\theta_j\neq 0\atop
                          l\,:\,\theta_l\neq 0} 
       \Big( (G+V_1)C_lC_{-l}(V_2V_1-1)^{-1}(V_2V_1+1)C_jC_{-j}(G+V_1^T)\Big)
            ^{\mu\nu}
\end{multline} 
The zero mode expression can be written as:
\begin{multline}\label{zeromodepoisson}
  \left\{H^\mu,H^\nu\right\}  =  \\
  \frac{\al}{2}
  \sum_{i,j}\sum_{a,b}
   \left(\Bigr(e_i-d_a \big(N^{-1}\big)^{T}K^T\Bigr)
    \bigl(A_1^{-1}\bigr)
        \Big(e_j- K\big(N^{-1}d_b\big)\Big)
  \right)^{\mu\nu} 
 \\ + \sum_{a,b}\big(d_aA_2d_b\big)^{\mu\nu}
\end{multline} 
We will now restrict ourselves to the case without Dirichlet conditions.
(The situation where the Dirichlet conditions are in directions which
are perpendicular to {\sl both} ${\cal F}$-fields is a trivial generalization
of the case under consideration. Non-trivial is the calculation of 
the commutator if the Dirichlet conditions of one brane
interfere with the directions in which the ${\cal F}$-field of the other
 brane points.)
If all Eigenvalues  $\lambda$ are equal to one  the oscillator part
vanishes, as well  as $(A_1)^{-1}$ and therefore only the $A_2$ term 
survives. A bit more complicated is the situation where $\lambda\neq 1$ for
all Eigenvalues $\lambda$. However ${\cal F}_1+{\cal F}_2$ is invertible
as it stands in this case and the $e_i$ can be an arbitrary basis of
$\mathbb{R}^n$. $A_2$ vanishes as well as $K$ and the resulting terms are
easily summed up to yield the result stated by Seiberg and Witten in
\cite{Seiberg:1999vs}. A generalization is to allow the Eigenvalues to 
be of both types.  With the $R$-matrices (eq.~\eqref{Rmatrix})
 that determine the Dirichlet  direction now being the identity, the situation
simplifies drastically. The  Eigenvalue equation for $\lambda_i$ is
 given by the vanishing of the determinant of the following matrix:
\begin{multline} \label{charmat}
(V_2V_1-\lambda\id)^\mu_{\phantom{\mu}\nu} = \\
  \Big((G+{\cal F}_2)^{-1}\big[  (1-\lambda) (G+ {\cal F}_2G {\cal F}_1) 
                      - (1+\lambda) ({\cal F}_2 + {\cal F}_1)
                       \big]
       (G+{\cal F}_1)^{-1}G
  \Big)^\mu_{\phantom{\mu}\nu}
\end{multline}
We deduce that the  $(G+{\cal F}_{1})^{-1}GC_i$ or $(G-{\cal F}_{2})^{-1}GC_i$ 
with eigenvalue 
$\lambda_i\neq 1$ can be chosen as the $e_i$ in eq.~\eqref{aonedef}
and the $d_a$ from equation \eqref{KNA2def}
(which are perpendicular to the $e_i$) are then given
by  $(G+{\cal F}_{1})^{-1}GC_\ell=(G-{\cal F}_{2})^{-1}GC_\ell$. We can read of both quantities $(V_2V_1\pm\id)$ from \eqref{charmat}. We will now
rewrite the oscillator part in such a way, that the matrix $A_1$ appears
on one side. This enables us to sum up this term with the 
$e_i(A_1)^{-1}e_j$ term from the zero mode part \eqref{zeromodepoisson} as
the ``denominator'' is then identical and isolated at one side.   
We define the following two (orthogonal) projectors:
\begin{align}
 {\bf Pr}_1 &\equiv \sum_{\lambda_i\neq 1} C_iC_{-i} &
 {\bf Pr}_2 &\equiv \sum_{\lambda_\ell= 1} C_\ell C_{\ell} \\
 \label{orthrel1}
 \Rightarrow {\bf Pr}_1 {\bf Pr}_2&={\bf Pr}_2{\bf Pr}_1=0 &
   {\bf Pr}_1 + {\bf Pr}_2&=\id
 \end{align}
\begin{itemize}
 \item We choose $e_i \equiv (G+{\cal F}_{1})^{-1}C_i$ $d_\ell  
 \equiv (G\pm{\cal F}_{1\atop 2})^{-1}C_\ell$.
 \item The matrices $A_1$ (eq.\ \eqref{aonedef}), $A_2$, $K$ and $N$ 
  (eq.\ \eqref{KNA2def}) now simplify to:\footnote{With the help of relation 
 \eqref{offdiagvanish}
 which will be derived in this section, we can even show, that
$K_{i\ell}$ vanishes completely.}
\begin{equation} 
 \begin{aligned}\label{kna2concrete}
 (A_1)_{ik}    &=
    \frac{1}{4\pi} 
    C_i (G-{\cal F}_{1})^{-1}({\cal F}_1+{\cal F}_2)
        (G+{\cal F}_{1})^{-1}C_k   &\lambda_i,\,\lambda_k\neq 1
              \\
 (K)_{i\ell}   &= \frac{1}{2}
      C_i (G-{\cal F}_{1})^{-1}(G+{\cal F}_{2})C_\ell \quad(=0)
                                    &\lambda_i\neq 1,\,\lambda_\ell=1
               \\
 (N)_{\jmath\ell}&= \frac{1}{2}C_\jmath G C_\ell 
                                     &\lambda_\jmath= \lambda_\ell=1
               \\
 (A_2)_{\jmath\ell} &=
            \pi C_\ell {\cal F}_2 C_\jmath 
 \end{aligned}
\end{equation}
\end{itemize}
and we note that 
\begin{multline}
 \sum_{\lambda_m\neq 1}
  C_{-k} ((C_{-l}(V_2V_2-1)C_{j})^{-1})^{-km}C_m
 C_{-m} (V_2V_2+1)C_n  =\\
 \frac{-1}{4\pi} C_{-k}\big((A_1)^{-1}\big)^{k,-m}
   \overbrace{C_m (G-{\cal F}_{1})^{-1} (G+{\cal F}_{2}) {\bf Pr}_1 
               (G+{\cal F}_{2})^{-1}}
             ^{=C_m (G-{\cal F}_{1})^{-1}}
   (G+{\cal F}_{2}{\cal F}_{1})C_n
\end{multline}
Using this result we sum up the oscillator contribution \eqref{oscmodepoisson}
with the term $e_i(A_1)^{-1}e_j$ from the zero mode contribution 
\eqref{zeromodepoisson}:
\begin{multline}
  \Bigl(
    \bigl\{X^\mu(\tau=0,\sigma)_\text{osc},   
                    X^\nu(\tau=0,\sigma^\prime)_\text{osc}
    \bigr\}  
       +   \frac{\al}{2} e_i\big(A_1^{-1}\big)^{ij}e_j
   \Bigr)^{\mu\nu} \\
  = \frac{\al}{2} 
     \Bigl(
        -(G+{\cal F}_1)C_i\big((A_1)^{-1}\big)^{i,j} \rule{7cm}{0cm}\\
       C_{-j} \big((G-{\cal F}_1)^{-1}
             (G+{\cal F}_{2}{\cal F}_{1})(G+{\cal F}_1)^{-1}
               {\bf Pr}_1(G-{\cal F}_1)^{-1}\\
             +( G-{\cal F}_1 )^{-1}
              \big)
     \Bigr)^{\mu\nu}
\end{multline}
As
\begin{multline}
 C_{-j}( G-{\cal F}_1 )^{-1} \\
  = 
 C_{-j}( G-{\cal F}_1 )^{-1}( G-{\cal F}_1 )
     ( G+{\cal F}_1 )( G+{\cal F}_1 )^{-1}
      ({\bf Pr}_1+{\bf Pr}_2)( G-{\cal F}_1 )^{-1}\\
  = C_{-j}( G-{\cal F}_1 )^{-1}( G-{\cal F}_1{\cal F}_1 )
     ( G+{\cal F}_1 )^{-1}{\bf Pr}_1( G-{\cal F}_1 )^{-1}  
\end{multline}
for $\lambda_j\neq 1$ (as it is the case), the oscillator part simplifies
 to:
\begin{multline}
  \Bigl(
    \bigl\{X^\mu(\tau,0)_\text{osc},   
                    X^\nu(\tau,0)_\text{osc}
    \bigr\}  
       +  \frac{\al}{2} e_i(A_1)^{-1}e_j
   \Bigr)^{\mu\nu} \\
  = -2\pi\al\Bigl(
        (G+{\cal F}_1)^{-1}C_i\big((A_1)^{-1}\big)^{i,j}(A_1)_{i,j}
        {\bf Pr}_1 {\cal F}_1{\bf Pr}_1( G-{\cal F}_1 )^{-1}
        \Bigr)^{\mu\nu} \\
  =  -2\pi\al\Bigl( (G+{\cal F}_1)^{-1} 
      {\bf Pr}_1 {\cal F}_1{\bf Pr}_1( G-{\cal F}_1 )^{-1}
    \Bigr)^{\mu\nu}
\end{multline}
Thus we already obtained a first part of the Poisson bracket.
Before we continue to calculate other parts of the zero-mode contribution,
we will derive some extremely useful relations. As annotated,
we will see that with our choice of the $d_\ell$, the matrix $K$ cf.\
\eqref{kna2concrete} vanishes.
We read of from equation \eqref{charmat} that 
\begin{equation}
 \begin{aligned}
    \bf{Pr}_I &\equiv( G+{\cal F}_{1} )^{-1}{\bf Pr}_2( G+{\cal F}_{1} ) 
    &\text{and }\quad
    \bf{Pr}_{II} &\equiv( G-{\cal F}_{2} )^{-1}{\bf Pr}_2( G-{\cal F}_{2}) 
    \\
    &=( G-{\cal F}_{2} )^{-1}{\bf Pr}_2( G+{\cal F}_{1} )
    & 
    &=( G+{\cal F}_{1} )^{-1}{\bf Pr}_2( G-{\cal F}_{2})
 \end{aligned}
\end{equation}
are projectors.
As $\bf{Pr}_{III}\equiv\bf{Pr}_I\bf{Pr}_{II}$ turns out to be  a projector, 
too, 
$\bf{Pr}_I$ and  $\bf{Pr}_{II}$ commute:  
\begin{align}
  \bf{Pr}_I\bf{Pr}_{II}
  &=\bf{Pr}_{II}\bf{Pr}_{I} \\
  \Rightarrow
  ( G+{\cal F}_{1} )^{-1}{\bf Pr}_2( G-{\cal F}_{2}) 
  &=\Big( G\mp{\cal F}_{2 \atop 1} \Big)^{-1}{\bf Pr}_2( G+{\cal F}_{1})
\end{align}
By multiplying the last result with the (invertible) matrix 
$ ( G+{\cal F}_{1} )$ we note that:
\begin{align}\label{relationpr2}
 {\bf Pr}_2{\cal F}_{1}&=-{\bf Pr}_2{\cal F}_{2} &
  &\text{and} &
   {\cal F}_{1}{\bf Pr}_2&=-{\cal F}_{2}{\bf Pr}_2
\end{align}
with the right equality being the transpose of the left.
We also observe now that $\bf{Pr}_I=\bf{Pr}_{II}$. 
As the projectors
\begin{equation}
 \begin{aligned}
    \bf{Pr}_A &\equiv( G+{\cal F}_{1} )^{-1}{\bf Pr}_1( G+{\cal F}_{1} ) 
    &\text{and }\quad
    \bf{Pr}_{B} &\equiv( G-{\cal F}_{2} )^{-1}{\bf Pr}_1( G-{\cal F}_{2}) 
 \end{aligned}
\end{equation}
fulfill:
\begin{align}
 \bf{Pr}_A{\bf Pr}_I &=0 & \bf{Pr}_B{\bf Pr}_I &=0 &
  \bf{Pr}_A+{\bf Pr}_I = \bf{Pr}_B+{\bf Pr}_I &=\id
\end{align}
we conclude:
\begin{equation}
   \bf{Pr}_A=  \bf{Pr}_B
\end{equation}
Rewriting  $C_j(G-{\cal F}_{1})^{-1}(G+{\cal F}_{2})C_\ell=0$ 
$(\lambda_i\neq 1,\lambda_\ell=1)$ (cf.~eq.\ \eqref{relationpr2} )  we get:
\begin{equation}
  \underbrace{C_i(G-{\cal F}_{1})^{-1}(G-{\cal F}_{1})C_\ell}_{=0} +
  C_i(G-{\cal F}_{1})^{-1}({\cal F}_{1}+{\cal F}_{2})C_\ell=0
\end{equation}
Inserting into this expression the identity 
$(G+{\cal F}_{1})^{-1}({\bf Pr}_1+{\bf Pr}_2)(G+{\cal F}_{1})$ we
obtain (the term  
 $\propto ({\cal F}_{1}+{\cal F}_{2})(G+{\cal F}_{1})^{-1}({\bf Pr}_2)$ 
 vanishes):
\begin{equation}
   C_i (G-{\cal F}_{1})^{-1}({\cal F}_{1}+{\cal F}_{2})
  (G+{\cal F}_{1})^{-1}C_{-k}C_k (G+{\cal F}_{1}) C_\ell = 0 \quad \forall i
\end{equation}
As $(A_1)_{i,-k}= C_i (G-{\cal F}_{1})^{-1}({\cal F}_{1}+{\cal F}_{2})
  (G+{\cal F}_{1})^{-1}C_{-k}$ is a non-singular matrix, we finally arrive at:
\begin{equation}\label{offdiagvanish}
  C_k {\cal F}_{1} C_\jmath = C_k {\cal F}_{2} C_\ell = 0
   \qquad\lambda_k\neq 1,\,\lambda_\ell=1
\end{equation}

As a consequence  $K$ in \eqref{kna2concrete} and
\eqref{zeromodepoisson}  vanish as well.
The remaining term in \eqref{zeromodepoisson} is $\propto 
\sum_{\jmath,\ell} d_\jmath N^{-1}A_2^{-1}N^{-1} d_\ell$. 
It can be rewritten by \eqref{relationpr2}:
\begin{equation}\label{dkakdterm}
 \sum_{\jmath,\ell} d_\jmath N^{-1} A_2 N^{-1} d_\ell 
   =-2\pi\al ( G+{\cal F}_{1} )^{-1}{\bf Pr}_2
              {\cal F}_{1} {\bf Pr}_2( G+{\cal F}_{1})^{-1} 
\end{equation}
The last line in \eqref{dkakdterm} equals 
$-( G-{\cal F}_{2} )^{-1}{\cal F}_2{\bf Pr}_2( G+{\cal F}_{2} )^{-1}$ and
cancels therefore the $A_2$ term in \eqref{zeromodepoisson}. We finally 
arrive at:
\begin{equation}\label{commleft}
 \left[X(\tau,0),X(\tau,0) \right]= i2\pi\al\frac{{\cal F}_1}{G-{\cal F}_1^2}
\end{equation}
\subsection{The commutator  
            \texorpdfstring{$\left[X(\tau,\pi),X(\tau,\pi) \right]$}
            {[X(tau,pi),X(tau,pi)]}}
In this section we will calculate the commutator at the other end of an 
open string stretching between two branes.
The situation is very similar compared to the $\sigma=0$ case. The zero mode
part $\{H^\mu,H^\nu\}$ is unchanged and the linear  (or: momentum modes) do
not contribute for the following reason: The Poisson bracket for
the zero- and momentum-modes  \eqref{pbrack2} does not contain a term 
$\propto \{p^\imath,p^\jmath\}$. The $\tau$-linear term that is
proportional to $\{H^\imath,p^\jmath\}$
is symmetric, and 
does not contribute to the commutator (c.f.\ \eqref{zeromodeexpansion} and 
\eqref{pbrack2}).  We read off from \eqref{pbrack2} 
that there are no terms $\{H^i,p^\jmath\}$ with $H^i$ coupling to  
$(G+{\cal F}_1)^{-1}C_i$ ($C_i$ an Eigenvector with Eigenvalue 
$\lambda_i\neq1$). However  a new term proportional to 
$\sigma\{H^\jmath,p^\ell\}$ (and its transpose) contributes now 
(c.f.\ \eqref{zeromodeexpansion}):
\begin{multline}
  2\pi \big((G+ {\cal F}_1)
    C_\jmath\{H^\jmath,p^\ell\}C_\ell
   (-{\cal F}_1)(G- {\cal F}_1)\big)^{\mu\nu} \\
  =+2\pi\al\big((G+ {\cal F}_1)
     {\bf Pr}_2{\cal F}_1(G- {\cal F}_1)\big)^{\mu\nu} \\
\end{multline}
This term cancels \eqref{dkakdterm}. However there is the ``transposed'' term
as well which gives the final momentum contribution to the commutator:  
\begin{multline}
  2\pi \big((G+ {\cal F}_1){\cal F}_1
    C_\jmath\{p^\jmath,H^\ell\}C_\ell(G- {\cal F}_1)\big)^{\mu\nu} \\
  = -2\pi\al\big((G- {\cal F}_2)
     {\bf Pr}_2{\cal F}_2{\bf Pr}_2(G+ {\cal F}_2)\big)^{\mu\nu} 
\end{multline}

The Poisson-bracket of the oscillator part changes however
slightly: Due  to terms of the type
\begin{equation}
 (G  e^{in_i(\tau+\sigma)} + V_1  e^{in_i(\tau-\sigma)})C_i
\end{equation}
appearing in the mode expansion of $X(\tau,\sigma)$ (eq.\ 
\eqref{oscillatorexpansionX}) the $(G+V_1)$ term in \eqref{timedropout} 
changes to 
\begin{multline}
  \big(G  e^{in_j\pi} + V_1  e^{-in_j(\pi)}\big)C_j
  =(-1)^{n}\Big(
                 \lambda_j^{-\oh}(\lambda _j G
                  C_j+\underbrace{V_1C_j}_{\lambda_jV_2^{-1}C_j})
           \Big) 
  \\
  =(-1)^{n}\lambda_j^{\oh}(G+V_2^{-1})C_j
  =(-1)^{n}\lambda_j^{\oh}(G-{\cal F}_2)^{-1}C_j
\end{multline}
In the same way $C_{-j}(G+V_1^T)$ changes to
\begin{multline}
  C_{-j}\big(G  e^{-in_j\pi} + V_1^T  e^{+in_j(\pi)}\big) \\
  =(-1)^{n}\lambda_{-j}^{\oh}C_{-j}\Big(G+\big(V_2^{-1}\big)^T\Big)
    =(-1)^{n}\lambda_{-j}^{\oh}C_{-j}\big(G+{\cal F}_2\big)^{-1}
\end{multline}
As $\lambda_j^{\oh}\lambda_{-j}^{\oh}=1$ this factor cancels out and the
analog of \eqref{oscmodepoisson} becomes:
\begin{multline} \label{oscmodepoisspi}
 \left\{X^\mu(\tau,\sigma=\pi)_\text{osc},   
                    X^\nu(\tau,\sigma^\prime=\pi)_\text{osc}\right\} \\
 =         \frac{\pi\al}{2}\sum_{j\,:\,\theta_j\neq 0\atop
                          l\,:\,\theta_l\neq 0} 
       \Big( (G+V_2)C_lC_{-l}(V_2V_1+1)(V_2V_1-1)^{-1}C_jC_{-j}(G+V_2^T)\Big)
            ^{\mu\nu}
\end{multline} 
where we interchanged the order of $(V_2V_1+1)$ and $(V_2V_1-1)^{-1}$.
We express the above equation now partially in terms of ${\cal F}_1$ and
${\cal F}_2$:
\begin{multline} \label{oscmodepoisspi2}
 \left\{X^\mu(\tau,\sigma=\pi)_\text{osc},   
                    X^\nu(\tau,\sigma^\prime=\pi)_\text{osc}\right\} \\
 =   -2\pi\al\sum_{i\,:\,\theta_i\neq 0\atop
                          j\,:\,\theta_j\neq 0} 
   \big(( G-{\cal F}_{2} )^{-1}{\bf Pr}_1
          ( G-{\cal F}_{2} )^{-1}(G+{\cal F}_{2}{\cal F}_{1}) 
          ( G+{\cal F}_{1} )^{-1} C_i 
          \\
          (V_2V_1-1)^{-1}_{-i,j}  C_-j
   ( G+{\cal F}_{2} )^{-1}\Big)
            ^{\mu\nu}
\end{multline}  
Next we will rewrite the term $e_i\big(A_1^{-1}\big)^{ij}e_j$ that
stems from the zero mode part:
\begin{multline}\label{A1rewritten}
  4\pi(A_1)_{-i,k}= 
    C_{-i} 
       ( G-{\cal F}_{1} )^{-1}( G+{\cal F}_{2} )( G+{\cal F}_{2} )^{-1}
               ({\cal F}_{1} + {\cal F}_{2})
       ( G+{\cal F}_{1} )^{-1}  
    C_{-k}    \\
   = \sum_{j\,:\,\theta_j\neq 0}
     C_{-i} 
       ( G-{\cal F}_{1} )^{-1}( G+{\cal F}_{2} )
     C_{-j}C_{j} 
      ( G+{\cal F}_{2} )^{-1} ({\cal F}_{1} + {\cal F}_{2})
       ( G+{\cal F}_{1} )^{-1}  
    C_{-k}
\end{multline}
With the help of \eqref{A1rewritten} we
get:
\begin{equation}\label{A1rewritten2}
 e_i\big((A_1)^{-1}\big)^{i,j}e_j = 4\pi
  ( G+{\cal F}_{1} )^{-1} C_{i}
     (C_i \big(V_2V_1-1)C_j)^{-1}\big)_{-i,j} C_{-j}( G+{\cal F}_{2} )^{-1}
\end{equation}
We can now add up the above expression with the oscillator part 
\eqref{oscmodepoisspi2}. Furthermore we multiply \eqref{A1rewritten2}
by $\id= ( G-{\cal F}_{2} )^{-1}({\bf Pr}_1+{\bf Pr}_2)
   ( G+{\cal F}_{2})^{-1}( G-{\cal F}_{2}^2)$:\footnote{Only the
${\bf Pr}_1$ contributes after multiplying with $ e_i(A_1)^{-1}e_j$.}
\begin{multline}
  \Bigl(
    \bigl\{X^\mu(\tau,\pi)_\text{osc},   
                    X^\nu(\tau,\pi)_\text{osc}
    \bigr\}  
       +  \frac{\al}{2} e_i\big(A_1^{-1}\big)^{ij}e_j
   \Bigr)^{\mu\nu} \\
  = -2\pi\al\Bigl(
        (G-{\cal F}_2)^{-1}{\bf Pr}_1( G+{\cal F}_2 )^{-1}
          {\cal F}_2 ({\cal F}_1+{\cal F}_2)
       ( G+{\cal F}_1 )^{-1} C_i
         \\(C_i \big(V_2V_1-1)C_j)^{-1}\big)_{-i,j}
        C_{-j}( G+{\cal F}_2 )^{-1}
        \Bigr)^{\mu\nu} \\
  =  -2\pi\al\Bigl( (G-{\cal F}_2)^{-1} 
      {\bf Pr}_1 {\cal F}_2{\bf Pr}_1( G+{\cal F}_2 )^{-1}
    \Bigr)^{\mu\nu}
\end{multline}
Finally we rewrite the commutator:
\begin{equation}\label{commright}
  \left[X(\tau,\pi),X(\tau,\pi) \right]
   = i2\pi\al\frac{{\cal F}_2}{G-{\cal F}_2^2}
\end{equation}
Besides this result, we found that in the basis 
$(C_i,C_\ell)$ both ${\cal F}$-fields
are of the form:
\begin{align}  \label{fonesplit}
     \biggl( 
     \begin{array}{c}
      C_i \\
     \hline
      C_\jmath
     \end{array} \biggr)
     {\cal F}_1
     \biggl( 
     \begin{array}{c}
      C_k \\
     \hline
      C_\ell
     \end{array} \biggr)
   &=\biggl( 
     \begin{array}{c|c}
       C_i\big({\cal F}_1\big) C_k & 0\\
      \hline
      0 & C_\jmath\big({\cal F}_1\big)C_\ell
   \end{array} \biggr)  \\ \label{ftwosplit}
   \biggl(
     \begin{array}{c}
      C_i \\
     \hline
      C_\jmath
     \end{array} \biggr)
     {\cal F}_2
     \biggl( 
     \begin{array}{c}
      C_k \\
     \hline
      C_\ell
     \end{array} \biggr)
   &=\biggl( 
     \begin{array}{c|c}
       C_i\big({\cal F}_2\big) C_k & 0\\
      \hline
      0 & -C_\jmath\big({\cal F}_1\big)C_\ell
   \end{array} \biggr) 
\end{align}
The lower block can be brought again to block-diagonal form with $2\times 2$
blocks. The upper blocks can of course not be brought to
such a form simultaneously. 
Summarizing, we reproduced and generalized the commutator
relation presented in the cited papers to the one-loop case with constant,
but otherwise completely unrestricted\footnote{Neglecting some
 degenerate cases.}
 electro-magnetic NS-$U(1)$ field strengths $F_1$ and $F_2$ at both 
boundaries, in addition to a constant NSNS two-form flux $B$.

\section{\label{susycondop}Space-time supersymmetry of open strings 
         in constant backgrounds}
Supersymmetry in string-theory (as well as in field theory) implies
in general  
vanishing of  the (complete) partition function.
At one loop-level this is due to Bose-Fermi degeneracy. 
For simplicity we restrict ourselves  to the case without 
Eigenvectors  $C_\lambda$ belonging to an  Eigenvalue $|\lambda|\neq 1$ 
in this section. We assume furthermore, that the we have Neumann
boundary conditions in at least four (or six)  space-time dimensions, while
the remaining  six (or four) dimensions of the superstring might have both
Dirichlet and mixed boundary conditions.\footnote{By ``mixed'' we mean that
an $\cal F$ field can be included.} By the identity \eqref{abstruone}  
(and for orbifolds: \eqref{abstrutwo}) in appendix \ref{riemid} we see
that a necessary condition for a vanishing cylinder partition function 
$\text{A}^{ij}$ in a sector given by the boundary conditions $V_{i}V_{j}$
is a condition on its Eigenvalues:
\begin{gather} \label{susyeq}
 \begin{aligned}
  d&=4 &  d&=6  \\
  0&=\theta_1+\theta_2 \qquad& 0&=\theta_1+\theta_2+\theta_3
 \end{aligned}  \\
 \nonumber
  \text{with Eigenvalues } \{\lambda_i,\bar{\lambda}_i\}\text{ and } 
   \lambda_i= e^{i2\pi\theta_i}
\end{gather}
By an exchange $\lambda_i\to
\bar{\lambda_i}$ the equations \eqref{susyeq} will no longer be fulfilled.
The meaning of \eqref{susyeq} is clear: the rotation $V_{i}V_{j}$ is not
only $\in O(n)$ but even contained in the much smaller group $SU(n/2)$
with $n=4$ (or $n=6$). If two branes should be supersymmetric w.r.t.\ each 
other, it is sufficient,\footnote{Not quite, but we will come to that 
point soon.} that   $V_{2}V_{1}\in SU(n/2)$. 
With several D-branes we will distinguish two cases:
\begin{enumerate}
  \item 
   Given a set of boundary conditions specified by matrices $\{V_i\}$
   some products $V_j V_i$ do not allow to choose Eigenvalues 
   $\lambda^{(j,i)}_k$ s.th.\ $\sum_{k=1}^{n/2}\theta_k$ vanishes (with
   $\theta_k$ in the sum belonging to Eigenvectors that are not obtained
    by complex conjugation). 
   This means that supersymmetry
   is broken in this sector, and $V_j V_i\notin  SU(n/2)$.
  \item 
   For all sectors $V_j V_i$ there  exists a set of 
   Eigenvalues$\{\lambda_i=\exp(i 2\pi \theta_i)|\; 
  \lambda_i\neq \bar{\lambda}_j \}$,  
   s.th.\ $\sum_{i=1}^{n/2}\theta_i$ vanishes. This implies that 
     $V_j V_i\in SU_{j,i}(n/2) $.\footnote{The subscript $j,i$ denotes
    that the $ SU_{j,i}$ belongs to the sector given by 
   $V_j V_i$. Different sectors might lead to different embeddings of
   the associated  $ SU(n/2)$ into $SO(n)$.} 
\end{enumerate}
While it is clear that in the first case all supersymmetry is broken by the
D-branes, the second case is more subtle.
Even though in this case all products $V_jV_i$ lie in an
$SU(n/2)$, the embedding of the individual  $SU(n/2)$ into
$SO(n)$ might differ for different choices of different pairs $(i,j)$.
There is also no warranty that the bulk supersymmetry will be preserved. 
In general, supersymmetry will only be preserved iff there exists a spinor 
$\epsilon_L$ (of definite chirality, namely the one of the left-moving
closed string sector),
 such that the  following spinor equation holds:\footnote{$\Sigma$ denotes the spinor representation.}
\be \label{spinorconda}
 \epsilon_L= \Sigma(V_iV_j)\epsilon_L  \qquad \forall{i,j}
\ee
This  can be seen as 
follows: Writing the left-moving supercharge as: 
$Q_L=\sum_\alpha (\epsilon_L)_\alpha Q_{L,\,\alpha} $ and 
the right moving one as  $Q_R=\sum_\alpha (\epsilon_R)_\alpha Q_{R,\,\alpha}$ 
the combination $Q_L+Q_R$ will be preserved by the D-brane with 
boundary condition $i$, iff:
\be \label{spinorcondb}
  \epsilon_R =\Sigma(V_j) \epsilon_L
\ee 
So far we omitted spinor indices in order to leave the possibility to 
consider either type IIA (with opposite chiralities for left and right movers)
or type IIB (with identical chiralities for left and right movers).
Therefore \eqref{spinorconda} and \eqref{spinorcondb} should
be understood with correct indices (i.e. projections). If for example
eq.\ \eqref{spinorcondb} has no solutions for given chiralities of 
$\epsilon_L$ and $\epsilon_R$, this means that the single brane breaks already
all bulk supersymmetry. On the other hand,  \eqref{spinorcondb} can admit
several solutions.
Combining \eqref{spinorcondb} for two different 
boundary conditions $i$ and $j$ implies equation 
\eqref{spinorconda}.\footnote{Deriving eq.\ 
\eqref{bdy3},p.\ 
\pageref{bdy3} we implicitly assumed, that the $i=2$ boundary has 
$F$-flux $-F_2$ (cf.\ \eqref{bdy2}, p.\ \pageref{bdy2}). This explains why
in \eqref{spinorconda} instead of $V^{-1}_iV_j$ we have to take:
  $V_iV_j$. }

Our analysis is very close to the one described in \cite{Polchinski:1998v2},
 chap.\ 13 and \cite{Witten:2000mf}. See also \cite{Mihailescu:2000dn}.

\subsection{Closed form of the Eigenvalues 
            \texorpdfstring{$\lambda_i$}{lambda(i)} in $d=4,\,6$.}
The (truncated\footnote{By truncated we mean that we
divided by $(1-\lambda)^{(10-d)}$ in order to remove the $\lambda=1$
Eigenvalues of the $10-d$ dimensional space-time.}) characteristic polynomial
for an O(n) rotation is very restricted, as its Eigenvalues are forced to have
modulus one. Furthermore we assume to have even an $SO(n)$ rotation, which
is the case, if the dimension of the brane is the correct one for the
theory under consideration (type IIA or type IIB). From {\it 
Vietas\negphantom{\,}' theorem} on
roots  we conclude that the characteristic polynomial  $\chi(\lambda)$ 
is symmetric and takes the following form in $d=4$ dimensions:\footnote{The
  characteristic polynomial is the determinant of $(V_2V_1-\lambda\id)$,
 cf.~eq.\ \eqref{defV}, 
p.\ \pageref{defV} and eq.\ \eqref{charmat}, p.~\pageref{charmat}.}
\begin{equation}\label{chifour}
 \chi_4(\lambda)\propto a\lambda^4 + b\lambda^3 +c \lambda^2 +b \lambda+a
\end{equation}
In four  dimensions all three coefficients $a,b,c$ can be extracted by 
inserting
$\lambda=0$ and $\lambda=\pm 1$ into $\chi_4(\lambda)$:
\begin{align} 
 a&=\chi_4(0),  & b&= \frac{1}{4}\big(\chi_4(1)-\chi_4(-1)\big),
 & c&= \frac{1}{2}\big(\chi_4(1)+\chi_4(-1)\big)-2\chi_4(0)
\end{align} 
Dividing \eqref{chifour} by $\lambda^2$  and applying a basic  
identity for  $\cos n\phi$,
transforms this equation into a second order polynomial in
 $\cos \pi\theta$.\footnote{Remember that $\lambda$ is related to $\theta$ by
 $\lambda=\exp(-i2\pi \theta)$.}
The two solutions for $\cos(\pi\theta)$ are  given by:
\be \label{explsolfourdim}
 \cos(\pi\theta_{1/2}) = 
 \frac {-b \pm \sqrt {{b}^{2} + 8\, {a}^{2} - 4\, ac}}{4\,a}
\ee 
In $d=6$ two additional terms show up:
\begin{equation}\label{chisix}
  \chi_6(\lambda)\propto  a\lambda^6 + b\lambda^5 
   +c \lambda^4 + d\lambda^3  +c \lambda^2 + b \lambda +a
\end{equation}

To extract all 
four coefficients in $ \chi_6$ one has to insert $\lambda= i$ as well.
Doing so we get:
\be \label{sixdcoeff}
 \begin{aligned} 
  a&=\chi_6(0)  & b&= \frac{\chi_6(1)-\chi_6(-1)}{2}-i\frac{\chi_6(i)}{4}\\
  c&= \frac{\chi_6(1)+\chi_6(-1)}{4}-\chi_6(0)
  & d&= \frac{\chi_6(1)-\chi_6(-1)}{4}+i\frac{\chi_6(i)}{2}
 \end{aligned}
\ee  
By dividing  \eqref{chisix} by 
$\lambda^3$, the Eigenvalue equation
 can be rewritten as a third order polynomial in $x=\cos \pi\theta$:
\be 
 {x}^{3} + \underbrace{\frac{b}{2a}}_{\equiv r}{x}^{2} 
  + \underbrace{\frac{c-3\, a}{4\,a}}_{\equiv s} x 
    +\underbrace{\frac{d- 2\, b}{8\, a}}_{\equiv t} = 0
\ee 
The above equation must admit three real solutions (possibly some of them
coincident) if we express 
the coefficients via \eqref{sixdcoeff}. We can now apply a well known 
formula to solve this equation  in closed form.  
However expressed in terms of $a,b,c,d$ these solutions are quite lengthy.
The explicit solutions blow up even more if we express the result in terms
of $\chi_6(0), \chi_6(\pm 1)$ and $\chi_6(i)$. Therefore we forego without
printing the three solutions. 
To make numerical  calculations simpler, one can also multiply
the characteristic polynomial  by 
$\det (G+{\cal F}_2)\cdot\det (G+{\cal F}_1)$  thereby saving two matrix
inversions. 

It is very simple to impose the {\sl necessary}
 supersymmetry  condition \eqref{susyeq} to the $d=4$ dimensional case.
This leads to one equation:  \eqref{susyeq} implies that
the char.~polynomial $\chi_4(\lambda)$ can be written as:
\be 
 \chi_4(\lambda)= a(\lambda_1-\lambda)^2  (\bar{\lambda}_1-\lambda)^2
\ee 
As two solutions coincide in that case, the root in \eqref{explsolfourdim} 
must vanish:
\begin{align}
   b^2 + 8 a^4 &= 4 a c 
\end{align}
It is an easy exercise to check that this condition is 
indeed fulfilled in the simple cases of tilted D-branes extending
in two directions of the four dimensional space if the sum of their (oriented)
angles vanishes. The same is true for the case of self-dual and anti-self dual
field strengths. 

We conclude this section with a reference to the publications 
\cite{Becker:1995kb} and \cite{Marino:1999af}. In both publications
the condition for membranes and D-branes to preserve supersymmetry were 
investigated. The approach pursued there is to look at the low energy
effective action of the D-brane (membrane). It also captures
the case of curved branes and non-constant NS $F$-fields.
 The classification of
the different branes with their supersymmetry conditions is quite involved,
and one has to consider branes of different dimensions separately.
While \cite{Becker:1995kb} is restricted to vanishing background fields 
$F$ and $B$, \cite{Marino:1999af} considers the case that these
fields are switched on.
In \cite{Marino:1999af} it was also derived that supersymmetric
D-branes of real
dimension three embedded in six-dimensions are not allowed to 
have an ${\cal F}_\parallel$-field living on it. This is not true for
for supersymmetric
D-branes of real dimension two embedded in a complex two-dimensional space.
We will make some further comments in section \ref{slagsec} 
on configurations with
 ${\cal F}_\parallel=0$
and D-branes with half the dimension of the
embedding space in which case  the supersymmetry condition reduces to 
a so called {\it  special La\-gran\-gi\-an submanifold} (short {\it sLag})
 condition.

\addcontentsline{toc}{section}{Concluding remarks}
\section*{Concluding remarks}
In this chapter we quantized the open string for linear boundary conditions,
that arise if two constant $U(1)$ field strengths couple to the string's
boundaries. This quantization is new compared to what has been published in 
literature in several respects:
\begin{itemize}
 \item
 Both field strengths are independent from each other. 
 \item
 We can include Dirichlet conditions as well as so called {\sl mixed boundary
 conditions}. 
 \item
 We have in addition to non equal NS field strengths $F_1$ and $F_2$ 
 the NSNS two-form potential $B$ included.
 \item
 A quantization for the zero and momentum modes in arbitrary
 toroidal compactifications
 is derived from first principles (for the case without Dirichlet 
  boundary conditions). 
\end{itemize}
Some of the results are employed in chapter \ref{ncg} and \ref{magbf}.
However the method developed is applicable for far more general
toroidal orbifold- and orientifold-constructions.


\chapter[Asymmetric Orientifolds]
         {\centerline{\label{ncg} Asymmetric Orientifolds}}

\section{Introduction}
As is known since the work of Connes, Douglas and Schwarz \cite{Connes:1998cr},
matrix theory compactifications on tori with background
three-form flux lead to non-commutative geometry.  
Starting with the early work \cite{Abouelsaood:1987gd} one has  subsequently realized
that open strings moving in backgrounds with non-zero
two-form flux or non-zero gauge fields  have mixed boundary conditions
leading  to a non-commutative geometry on the boundary of the
string world-sheet 
\cite{Cheung:1998nr,Frohlich:1998zm,Ardalan:1998ce,Schomerus:1999ug,Ardalan:1999av,Seiberg:1999vs,Arfaei:1999jt,Chen:1999bf,Lee:1999kj,Chu:1998qz,Chu:1999ta,Chu:1999gi,Banerjee:2002ky}.
We calculated the commutator at the string boundary in the last chapter 
for the one loop case and found agreement with the literature. 

As pointed out in \cite{Seiberg:1999vs},
also the effective theory on the D-branes becomes a non-commutative
Yang-Mills theory. 

We know from the discovery of D-branes, that Dirichlet
branes made their first appearance by studying the realization of T-duality
on a circle in the open string sector \cite{Dai:1989ua}. For instance, 
starting with a D9 brane, the
application of T-duality leads to a D8-brane where the ninth direction
changes from a Neumann boundary condition to a Dirichlet boundary 
condition. Thus, one may pose the question how D-branes with 
mixed Neumann-Dirichlet
boundary conditions fit into this picture. Does there exist
a transformation relating pure Dirichlet or Neumann boundary conditions to
mixed Neumann-Dirichlet boundary conditions?\footnote{Even though we answered this question already partially in 
section \ref{openstringssec} (p.\ \pageref{openstringssec}), 
the discussion presented in this
chapter is very close to our original paper \cite{Blumenhagen:2000fp}.
We think it is illustrative
to adopt main parts of the paper, while being even more specific in
some points. Thereby most parts of this chapter can be read independently 
from the previous chapter.}
At first sight unrelated, there exists the so far unresolved problem of 
what the D-brane content of asymmetric orbifolds is. The simplest asymmetric
orbifold is defined by 
modding out by T-duality itself, which is indeed a symmetry as long
as one chooses the circle at the self-dual radius. Thus,
as was argued in \cite{Dine:1997ji} and applied to Type I compactifications
in \cite{Blumenhagen:1998uf}, in this special case D9- and D5-branes are identified
under the asymmetric orbifold action. However, the general T-duality
group for compactifications on higher dimensional tori contains
more general asymmetric operations. For instance, the 
root lattice of $SU(3)$  allows  an asymmetric $\widehat{\mathbb{Z}}_3$ action.\footnote{A left-right asymmetric $\mathbb{Z}_N$ symmetry is denoted by
  $\widehat{\mathbb{Z}}_N$ (cf.\ section  \ref{enhsym}).}
We made contact with this symmetry already in section \ref{enhsym} of
the introductory chapter on orbifolds. In section \ref{asz3orbi} (p.\ \pageref{asz3orbi}) we considered the
orbifold $T^4/(\mathbb{Z}^{\text{L}}_3 \times\mathbb{Z}^{\text{R}}_3)$.

The closed string sector can very well live with such non-geometric
symmetries \cite{Narain:1987qm} but what about the open string sector? 
Since all Type II
string theories contain open strings in the non-perturbative D-brane sector,
in order for asymmetric orbifolds to be non-perturbatively consistent,
one has to find a realization of such non-geometric symmetries
in the open string sector, as well. Thus, the question arises what
the image of a D9-brane under an asymmetric $\widehat{\mathbb{Z}}_N$ action
is. 

The third motivation for the investigation performed in this chapter is
due to recently introduced orientifolds with D-branes at angles 
\cite{Berkooz:1996km,Blumenhagen:1999md,Angelantonj:1999xf,Blumenhagen:1999ev,Pradisi:1999ii}. 
We investigated orientifold models for which the world-sheet parity 
transformation,
$\Omega$, is combined with a complex conjugation, $\Bar{\sigma}$,
of the compact coordinates. After dividing by a further left-right symmetric
$\mathbb{Z}_N$
space-time symmetry the cancellation of  tadpoles
required the  introduction of  so-called  twisted
open string sectors. These sectors were realized by open strings stretching
between 
D-branes intersecting at non-trivial angles. As was pointed out
in \cite{Blumenhagen:1999md}, these models are related to ordinary $\Omega$ orientifolds
by T-duality. However, under this T-duality the former left-right
symmetric $\mathbb{Z}_N$ action is turned into an asymmetric $\widehat{\mathbb{Z}}_N$
action in the dual model. Thus, we are led to the problem of
describing asymmetric orientifolds in a D-brane language. 
Note, that using pure conformal field theory methods asymmetric
orientifolds were discussed recently in \cite{Bianchi:1999uq}.

In this chapter,  we study the three 
conceptually important problems mentioned
above, for simplicity, in the case of  compactifications on direct 
products of two-dimensional tori.
It turns out that all three problems are deeply related. 
The upshot is that asymmetric rotations turn Neumann boundary conditions
into mixed Neumann-Dirichlet
boundary conditions. This statement is  the solution to the first
problem and allows us to rederive the non-commutative geometry
arising on D-branes with background gauge fields simply by
applying asymmetric rotations to ordinary D-branes.
The solution to the second problem is that asymmetric orbifolds
necessarily contain open strings with mixed boundary conditions. 
In other words: D-branes manage to incorporate left-right asymmetric
 symmetries 
by turning on background gauge fluxes, which renders their world-volume 
geometry non-commutative\footnote{There is an exception
for asymmetric orbifolds with $\widehat{\mathbb{Z}}_2$-action. 
The orientifold in \cite{Blumenhagen:1998uf} of such a model is 
consistent with
D$9$- and D$5$-branes without any fluxes.}. Gauging the left-right asymmetric
 symmetry 
can then lead to an identification of commutative and 
non-commutative geometries. In this sense 
asymmetric Type II orbifolds are deeply related to non-commutative geometry. 
Apparently, the same holds for asymmetric
orientifolds, orbifolds of Type I. Via T-duality the whole plethora of 
$\Omega \Bar{\sigma}$ orientifold models of \cite{Blumenhagen:1999md,Angelantonj:1999xf,Blumenhagen:1999ev} 
is translated into a set of 
asymmetric orientifolds with D-branes of different commutative and 
non-commutative types in the background. We will further  
present a D-brane interpretation
of some of the non-geometric models studied in \cite{Bianchi:1999uq} and
generalizations thereof. 

In section \ref{Dbinasorbs} we describe a special class  of asymmetric 
orbifolds on $T^2$. Employing T-duality we first determine
the tori allowing an asymmetric $\widehat{\mathbb{Z}}_N$ action, where we 
discuss the   $\widehat{\mathbb{Z}}_3$ example in some detail. Afterwards
we study D-branes in such models and also determine the zero-mode
spectrum for some special values of the background gauge 
flux.
In section \ref{asymrotsec}  we apply asymmetric rotations to give an 
alternative
derivation of  the propagator on the disc with mixed Neumann-Dirichlet
boundary conditions. 
In the final section of this chapter
 we apply all our techniques to the explicit
construction of a ${\mathbb{Z}}_3\times \widehat{\mathbb{Z}}_3$ orientifold containing
D-branes with mixed boundary conditions. 

{\it Remark:}
 If the string scale $\al$ is not written explicitly, we have set it to one.

\section{\label{Dbinasorbs}D-branes in asymmetric orbifolds}

In this section we investigate in which way open strings manage
to implement left-right asymmetric symmetries. Naively, one might think
that asymmetric symmetries are an issue only
in the closed string sector, as open strings can be obtained by
projecting onto the left-right symmetric part of the space-time. 
However, historically just requiring  the left-right
asymmetric symmetry under T-duality
on a circle led  to the discovery
of D-branes. This T-duality acts on the space-time coordinates  as
\begin{equation} 
   \label{tdual}
       (X_L,X_R)\to (-X_L,X_R) 
\end{equation}
Thus, the open string sector deals with T-duality by giving rise
to a new kind of boundary condition leading in this case  to 
the well known Dirichlet boundary condition.  
Compactifying on a higher dimensional torus $T^d$, in general with 
non-zero $B$-fields, the T-duality
group gets  enlarged, so that one may ask what the image of Neumann boundary
conditions under these actions actually is. 

In the course of this chapter we restrict ourselves to the two-dimensional
torus $T^2$ and direct products thereof. For concreteness consider
Type IIB compactified on a $T^2$ with complex coordinate
$Z=X_1+iX_2$ allowing a discrete $\mathbb{Z}_N$ symmetry acting as
\begin{equation}
   \label{symm} 
       \Theta: (Z_L,Z_R)\to \left(e^{i\theta} Z_L,
             e^{i\theta} Z_R\right) 
\end{equation}
with $\theta=2\pi /N$.  
The essential observation is that performing a usual T-duality operation
in the $x_1$-direction\footnote{This  T-duality is the same as the 
$D$-duality introduced in section \ref{enhsym} (p.\ \pageref{enhsym}).} 
\begin{equation} \label{xdual}
    T: (Z_L,Z_R)\to (-\Bar{Z}_L,Z_R)
\end{equation}
yields an asymmetric action on the T-dual torus $\hat{T}^2$
\begin{equation}   \label{xdualb}
       \hat{\Theta}=T\Theta T^{-1}: (Z_L,Z_R)\to 
       \left(e^{-i\theta} Z_L,e^{i\theta} Z_R\right) 
\end{equation}
The aim of this chapter is to investigate the properties of 
asymmetric orbifolds defined by actions like \eqref{xdualb}.
The strategy we will follow is  depicted in the following commuting diagram
(figure \ref{diagramm1}):
\begin{figure}[h]
 \begin{center}
 \includegraphics{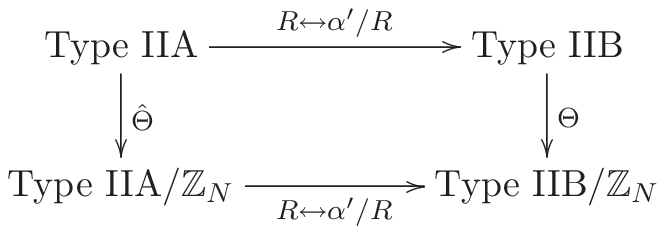}
 \end{center}
\caption{\label{diagramm1} T-duality relation}
\end{figure}
In order to obtain the features of the asymmetric orbifold, 
concerning some questions it is appropriate to directly apply
the asymmetric rotation $\hat{\Theta}$. For other questions
it turns that it is better to first apply a T-duality and then
perform the symmetric rotation $\Theta$ and translate the result back
via a second T-duality. 

\subsection{Definition of the T-dual torus}
The fundamentals underlying this section can be found 
in the review of Giveon, Porrati and Rabinovici \cite{Giveon:1994fu}.
The first step is to define the T-dual torus $\hat{T}^2$ allowing indeed
an asymmetric action \eqref{xdualb}. 
Let the torus $T^2$ be  defined by the following two
vectors
\begin{equation} \label{vect}
    e_1=R_1, \quad\quad e_2=R_2\, e^{i\alpha}
\end{equation}
so that the complex and K\"ahler structures are given by
\begin{equation}  \label{kahlerstruc}
     \begin{aligned}
            \tau &={e_2\over e_1}= {R_2\over R_1}\, e^{i\alpha}, \\
                     \rho&=B_{12}+i R_1 R_2 \sin\alpha
     \end{aligned}
\end{equation}
The left and right moving zero-modes, i.e. Kaluza-Klein and winding
modes, can be written in the following form
\begin{equation}\label{kkw}
      \begin{aligned}  p_L={1\over i\sqrt{\tau_2 \rho_2}} \left[ \tau\, 
      m_1-m_2-\Bar{\rho}(n_1+
   \tau\, n_2) \right] ,\\
       p_R={1\over i\sqrt{\tau_2 \rho_2}} \left[ \tau\, m_1-m_2- {\rho}(n_1+
   \tau\, n_2) \right]
      \end{aligned}
\end{equation} 
Applying T-duality in the $x_1$-direction exchanges the complex-structure
 and 
the K\"ahler modulus yielding the torus $\hat{T}^2$ defined
by the vectors\footnote{A quantity describing the T-dual torus $\hat{T}^2$
 is denoted by a hat ($\,\hat{.}\,$) on it.}
\begin{equation}\label{vectb}
          \hat{e}_1={1\over R_1}, \quad\quad 
           \hat{e}_2={B_{12}\over R_1} + i R_2 \sin\alpha  
\end{equation}
and the two-form flux 
\begin{equation}\label{twoform}
       \hat{B}_{12}={R_2\over R_1} \cos\alpha
\end{equation}
For the Kaluza-Klein and winding modes we get
\begin{equation}\label{kkwb}
    \begin{aligned}  
          p_L&=-{1\over i\sqrt{\hat{\tau}_2 \hat{\rho}_2}} 
                           \left[ \hat{\tau}\, n_1+m_2-\hat{\Bar{\rho}}(m_1+
                              \hat{\tau}\, n_2) \right], \\
          p_R&=-{1\over i\sqrt{\hat{\tau}_2 \hat{\rho}_2}} 
                           \left[ \hat{\tau}\, n_1+m_2-{\hat{\rho}}(m_1
                            + \hat{\tau}\, n_2)
                            \right]
     \end{aligned}
\end{equation}
from which we deduce the relation of the Kaluza-Klein
 and winding quantum numbers
\begin{equation}\label{rel}
         \widehat{m}_1=-n_1,\ \ \widehat{m}_2=m_2,\ \ \widehat{n}_1=-m_1,\ \
             \widehat{n}_2=n_2
\end{equation}
If the original lattice of  $T^2$ allows a crystallographic action of a
$\mathbb{Z}_N$ symmetry, then the T-dual Narain-lattice of
$\hat{T}^2$ does allow
a crystallographic action of the corresponding asymmetric $\widehat{\mathbb{Z}}_N$
symmetry. In view of the orientifold model studied in section 
\ref{asz3orienti}, 
we present the $\mathbb{Z}_3$ case as an easy example. 

\subsection{\label{z3torusncg}The 
     \texorpdfstring{$\widehat{\mathbb{Z}}_3$}
      {hat(Z)(3)} torus}
\noindent
In this section we shortly recall the definition of the $\mathbb{Z}_3$
and its  T-dual $\widehat{\mathbb{Z}}_3$ that was given in section 
\ref{enhsym}.
One starts with the  $\mathbb{Z}_3$ lattice defined by the basis 
vectors
\begin{equation}\label{vectzdrei}
   e^{\bf A}_1=R, \quad\quad 
  e^{\bf A}_2=R\left( {1\over 2}+i{\sqrt{3}\over 2}\right) 
\end{equation}
and arbitrary $B$-field.
The complex-structure and K\"ahler moduli are
\begin{equation}\label{kahlstructtwo}
     \begin{aligned}  \tau^{\bf A}&={1\over 2}+i{\sqrt{3}\over 2}, \\
                     \rho^{\bf A}&=B_{12}+i R^2 {\sqrt{3}\over 2}
     \end{aligned}
\end{equation}
This lattice has the additional property that
it allows a crystallographic action of the reflection at (and consequently: 
along) the $x_2$-axis, 
$\Bar{\sigma}$. This was important for the study of $\Omega\Bar{\sigma}$ orientifolds
in \cite{Blumenhagen:1999md}. We call this lattice a ``lattice of type ${\bf A}$''.
Recall from \cite{Blumenhagen:1999md}, that under $\Omega \Bar{\sigma}$ all three $\mathbb{Z}_3$ fixed points
are left invariant. 
For zero $B$-field one obtains for instance for the T-dual ${\bf A}$ lattice
\begin{equation}\label{vectzdreidual}
       \hat{e}^{\bf A}_1={1\over R}, \quad\quad 
       \hat{e}^{\bf A}_2=iR{\sqrt{3}\over 2}
\end{equation}
and $\hat{b}^{\bf A}={1/2}$.
That this rectangular lattice 
features an asymmetric $\widehat{\mathbb{Z}}_3$ symmetry and that all three 
``fixed points'' of the $\widehat{\mathbb{Z}}_3$ are left invariant under $\Omega$
is not obvious at all. This  shows already how
T-duality can give rise to fairly non-trivial results.

As we have already shown in \cite{Blumenhagen:1999md} (cf.\ section
\ref{wsparityt2}) there exists a 
second $\mathbb{Z}_3$ lattice, 
called type ${\bf B}$, 
allowing a crystallographic action of the reflection $\Bar{\sigma}$, too.
The basis vectors are given by
\begin{equation}\label{vectzdreib}
            e^{\bf B}_1=R, \quad\quad 
            e^{\bf B}_2={R\over 2}+i{R\over 2\sqrt{3}} 
\end{equation}
with arbitrary $B$-field leading to the  complex-structure and K\"ahler moduli
\begin{equation}\label{kahlb}
    \begin{aligned}  \tau^{\bf B}&={1\over 2}+i{1\over 2\sqrt{3}},  \\
                     \rho^{\bf B}&=B_{12}+i {R^2 \over 2 \sqrt{3} }
       \end{aligned}
\end{equation}
For the ${\bf B}$ lattice only one  $\mathbb{Z}_3$ fixed point is invariant
under $\Omega \Bar{\sigma}$, the remaining two are interchanged. 
For $B_{12}=0$ the T-dual lattice is defined by 
\begin{equation}
       \hat{e}^{\bf B}_1={1\over R}, \quad\quad 
       \hat{e}^{\bf B}_2=i{R\over 2\sqrt{3}}
\end{equation}
with  $\hat{b}^{\bf B}={1/2}$.
It is a non-trivial  consequence of T-duality that only one of the three 
$\widehat{\mathbb{Z}}_3$ ``fixed points'' is left invariant under
$\Omega$.

If one requires the lattices to allow simultaneously  a symmetric $\mathbb{Z}_3$
and an asymmetric
$\widehat{\mathbb{Z}}_3$ action one is stuck at the self-dual point $\tau=\rho$ yielding
$R=1$ and $B_{12}=1/2$. Note, that this is precisely  the root lattice of the 
$SU(3)$
Lie algebra. Since now we are equipped with lattices indeed allowing
a crystallographic action of asymmetric $\widehat{\mathbb{Z}}_N$ operations,
we can move forward to discuss their D-brane contents.

\subsection{Asymmetric rotations of D-branes}
In order to divide a string theory by some discrete group we first have
to make sure that the theory is indeed invariant. For the open 
string sector this means that the D-branes also have to be arranged in such a
way that they reflect the discrete symmetry.
Thus, for instance we would like to know what the image of a D0-brane
under an asymmetric
rotation is. In the compact case we can ask this question for
the discrete $\widehat{\mathbb{Z}}_N$ rotations defined in the last subsection,
but we can also pose it quite generally in the non-compact case
using a continuous  asymmetric rotation
\begin{gather}\label{rot}
  \begin{pmatrix}   X'_{1,L} \\ X'_{2,L} 
    \end{pmatrix}
   =\begin{pmatrix} \cos\phi & \sin\phi \\ 
                    -\sin\phi & \cos\phi
    \end{pmatrix}
     \begin{pmatrix}  X_{1,L} \\ X_{2,L}  
     \end{pmatrix}  \quad\quad
      \begin{pmatrix} X'_{1,R} \\ X'_{2,R}
      \end{pmatrix}
    =\begin{pmatrix}  \cos\phi & -\sin\phi \\ 
                    \sin\phi & \cos\phi
     \end{pmatrix}
     \begin{pmatrix}      X_{1,R} \\ X_{2,R} 
     \end{pmatrix}
\end{gather}
As outlined already in the beginning of section \ref{Dbinasorbs}
 (see figure \ref{diagramm1}), 
instead of acting with the asymmetric rotation on the Dirichlet boundary 
conditions of the D0-brane, it is equivalent to go to the T-dual picture,
apply first a symmetric rotation on the branes and then perform a
T-duality transformation in the $x_1$-direction.             
In the T-dual picture the D0-brane becomes a D1-brane filling only the 
$x_1$-direction. 
Thus, the open strings are of Neumann type in the $x_1$-direction and
of Dirichlet type in the $x_2$-direction.
The asymmetric rotation becomes a symmetric rotation, which simply
rotates the D1-brane by an angle $\phi$ in the $x_1$-$x_2$ plane.   
Thus, after the rotation the D1 boundary conditions in these two directions
read 
\begin{equation}\label{bounda}
          \begin{aligned}  \partial_\sigma X_1 + \tan\phi\, 
                    \partial_\sigma X_2 &=0, \\
               \partial_\tau X_2 - \tan\phi\, \partial_\tau X_1 &=0
          \end{aligned}
\end{equation}
If we are on the torus $T^2$ there is a distinction between values
of $\phi$, for which the rotated D1-brane intersects a lattice point, and
values of $\phi$, for which the D1-brane  densely covers the entire
$T^2$. In the first case, one still obtains  
quantized Kaluza-Klein and winding modes
as computed in \cite{Ardalan:1998ce}. 

\begin{figure}
  \setlength{\unitlength}{0.1in}
  \begin{picture}(50,30)
  \SetFigFont{14}{20.4}{\rmdefault}{\mddefault}{\updefault}
  \put(0,0){\scalebox{0.5}{\includegraphics{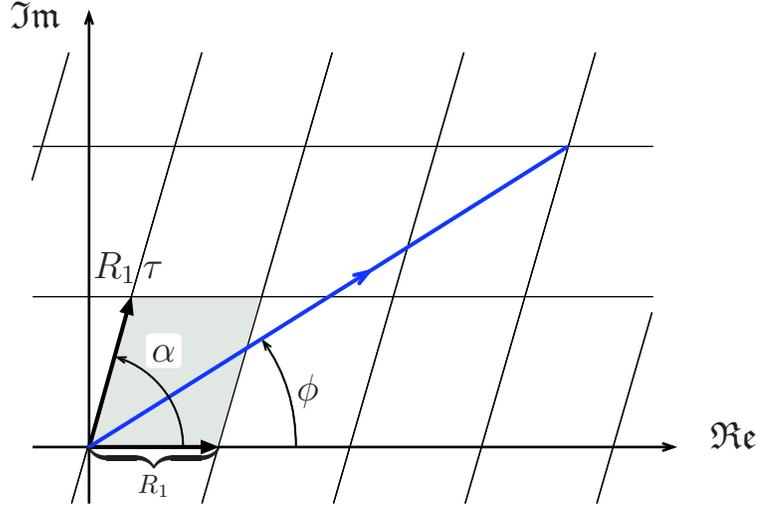}}}
  \put(9,25){$\mathfrak{Im}$}
  \put(45,3){  $\mathfrak{Re}$}
  \put(13.15,3.){\makebox(0,0)[tl]{$\underbrace{\rule{17.1mm}{0mm}}
                  _{R_1}$}}
  \put(13.35,11.9){${\SetFigFont{9}{17.4}{\rmdefault}{\mddefault}{\updefault}R_1\,\tau}$}
 \put(24,5.5){$\phi$}
 \put(16.4,7.5){$\alpha$}
 \end{picture}
\caption[D$1$-brane with 
    wrapping numbers $(n,m)=(3,2)$ on $T^2$]
    {\label{toruswbrane} D$1$-brane 
  ({\color{bl} blue line})   with 
    wrapping numbers $(n,m)=(3,2)$ on $T^2$.
    Note that
    the minimal distance of two such D-branes is smaller than the torus spacing 
   $R_1$ and $R_2 \equiv R_1|\tau |$ while the length is larger.}
\end{figure}

If the D1-brane runs $n$-times around the $e_1$ circle
and $m$ times around the $e_2$ circle until it intersects a lattice point, 
the relation\footnote{$m$ and  $n$ are 
{\sl co-prime}. If $m$ and  $n$ would have a greatest common devisor $p$, this
would mean that the brane is wrapped $p$ times around the one-cycle defined
by $(m/p, n/p)$.}
\begin{equation}\label{inte} 
           \cot\phi=\cot\alpha+{n\over m \tau_2}
\end{equation}
holds. As an example we show in figure \ref{toruswbrane} a rotated D1-brane 
with $n=3$ and $m=2$.
In the following we will mostly consider D-branes of the
first kind, which we will call rational D-branes.  
Finally, T-duality in the $x_1$-direction has the effect of exchanging 
$\partial_\sigma X_1\leftrightarrow  - \partial_\tau X_1$, leading to the
boundary conditions \cite{Sheikh-Jabbari:1998yi}
\begin{equation}\label{boundb}
      \begin{aligned}  \partial_\sigma X_1 + \cot\phi\, 
              \partial_\tau X_2 &=0 ,\\
             \partial_\sigma X_2 - \cot\phi\, \partial_\tau X_1 &=0 
      \end{aligned}
\end{equation}
As emphasized already, 
one could also perform the asymmetric rotation directly on the Dirichlet
boundary conditions for the D0-brane and derive the same result. Thus, 
we conclude that an asymmetric rotation turns a D0-brane into 
a D2-brane with mixed boundary conditions. The last  statement is the main
result of this chapter. As has been discussed intensively after
the talks of Witten and
Seiberg at the {\it Strings} conference in 1999
(and their related paper \cite{Seiberg:1999vs}),
 mixed boundary conditions arise from open strings
traveling in a background with non-trivial two-form flux, $B$,  or
non-trivial gauge flux, $F$,
\begin{equation}\label{boundc}
     \begin{aligned}  \partial_\sigma X_1+ (B+F)_{12}\partial_\tau X_2 &=0, \\
                  \partial_\sigma X_2 - (B+F)_{12}\partial_\tau X_1 &=0 
     \end{aligned}
\end{equation}
Thus, we can generally identify 
\begin{equation}
    \cot\phi={\cal F}={B+F}
\end{equation}
which in the rational case becomes (note that $\cot\phi$ is not necessarily
rational)
\begin{equation}\label{identb}
    \cot\phi=\cot\alpha + {n\over m \tau_2} ={B+F} 
\end{equation}
Since the $B$ field is related to the shape of the torus $T^2$ and
the $F$ field to the D-branes, from  \eqref{identb} we extract the 
following identifications\footnote{The $F$ given in the following formula
is written in the euclidian basis. In terms of a lattice basis for
 $T^2$ it is simply:
$F=n/m$. $m$ and $n$ are co-prime.} 
\begin{equation}\label{schnee}
         B=\cot\alpha, \quad\quad F={n\over m \tau_2}
\end{equation}
In section \ref{asymrotsec}  we will further elaborate the relation between
asymmetric rotations and D-branes with mixed boundary conditions. We
will present an alternative derivation of some of the non-commutativity 
properties known  for such boundary conditions.
In the remainder of this section we will focus our attention on  the momentum-
 and zero-mode spectrum for open strings stretched between D-branes with mixed 
boundary conditions.
In particular, we will demonstrate  that in the compact case   
open strings stretched between identical rational D-branes  do  
have a non-trivial momentum-mode spectrum.
This is in sharp contrast to some statements in the literature 
\cite{Kakushadze:1998cd}
saying that Neumann boundary conditions allow Kaluza-Klein 
 momentum, Dirichlet boundary
conditions allow non-trivial winding but general mixed D-branes do have
neither of them.  

\subsection{Kaluza-Klein and winding modes, zero mode degeneracy}
In this section we will calculate the spectrum of the linear modes
for two dimensional D-branes with $U(1)$-$F$ fluxes and for the 
T-dual configuration i.e. one dimensional D-branes on a $T^2$.
The results are easily generalized to torus compactifications of the more
general form:
\be
 {\cal M}_{10}=\mathbb{R}^{1,9-2d}\times\prod_{j=1}^d T^2_{(j)}  
\end{equation}
We will also explain the zero mode degeneracy that appears if
the $F$ fluxes are different at the two ends of a string. 
In the T-dual picture this degeneracy corresponds to the
topological intersection number of two branes on a torus.

\subsubsection{D$2$-branes with $F$-flux on 
                  \texorpdfstring{$T^2$}{T\texttwosuperior}}
Since we can not easily visualize a D-brane with mixed boundary conditions,
we first determine the zero-mode spectrum in the closed string
tree channel and then 
transform the result into the open string loop channel.
In contrast to the underlying publication \cite{Blumenhagen:2000fp},
 we have now a direct method
to quantize open strings with mixed boundary conditions in toroidal 
compactifications (cf.\ chap.\ \ref{strbg}). However, we first calculate 
the spectrum by the use of boundary state formalism. After this, we use the
results of the canonical quantization of the preceeding chapter, to derive
an independent mass formula. These mass formul\ae\; perfectly agree for the 
case $m=1$ (cf.\ eq.\ \eqref{identb}) without further explanation.

However for $m\neq 1$ (or fractional valued $F$-flux) there is a slight
mismatch if we do not make further assumptions. In section
\ref{toroidaldbranes} it was argued, that $F/\al$ has to be integer-valued
(in a basis of the torus lattice $\Gamma^d$). It was however mentioned
that fractional valued $F/\al$ corresponds to multiply wrapped branes,
i.e.\ the lattice  $\Gamma^d$ of the closed string sector should be
substituted by a bigger lattice $\tilde{\Gamma}^d$, s.th.\ $F/\al$ is
integer valued in a basis of $\tilde{\Gamma}^d$.
Alternatively, we can demand the string wave function to be invariant 
only under shifts $\in m2\pi\Gamma^d$. By this we get perfect agreement
even in the case $m\neq 1$.

We are looking for boundary states (see also \cite{Brunner:1999fj})
in the closed string theory
satisfying the following boundary state conditions
\begin{equation}\label{boundd}
   \begin{aligned} 
 \left[\partial_\tau X_{1,cl} + \cot\phi\, \partial_\sigma
   X_{2,cl}\right]|B\rangle 
    &=0 ,\\
  \left[\partial_\tau X_{2,cl} - \cot\phi\, \partial_\sigma
    X_{1,cl}\right]|B\rangle 
      &=0
   \end{aligned}
\end{equation}
Rewriting \eqref{boundd} in terms of the complex coordinate  
the boundary condition reads
\begin{equation}\label{bounde}
     \left[\partial_\tau Z_\text{cl} -i \cot\phi\, \partial_\sigma
   Z_\text{cl}\right]|B\rangle =0
\end{equation}
There is an analogous condition for the  hermitian conjugate field $\Bar{Z}_\text{cl}$.
Using the mode expansion
\begin{equation}\label{mode}
     \begin{aligned} Z_\text{cl}&={z_0\over 2} + {1\over 2}(p_L+p_R)\tau + 
{1\over 2} (p_L-p_R)\sigma    \\
   &\rule{4cm}{0cm} + {i\over \sqrt{2}} 
      \sum_{n\ne 0} \left( {\alpha_n\over n} e^{-in (\tau+\sigma)} +
              {\tilde\alpha_n\over n} e^{-in (\tau-\sigma)} \right), \\
   \Bar{Z}_\text{cl}&={\Bar{z}_0\over 2} + {1\over 2}(\Bar{p}_L+\Bar{p}_R)\tau + 
{1\over 2} (\Bar{p}_L-\Bar{p}_R)\sigma \\
  &\rule{4cm}{0cm} - {i\over \sqrt{2}}
    \sum_{n\ne 0} \left( {\Bar{\alpha}_n\over n} 
     e^{-in (\tau+\sigma)}
   +
      {\Bar{\tilde\alpha}_n\over n} e^{-in (\tau-\sigma)} \right)
      \end{aligned}
\end{equation}
one obtains 
\begin{equation}\label{modeb}
     \begin{aligned}\left[ (p_L+p_R)-i\cot\phi\,(p_L-p_R)\right]
       |B\rangle&=0 ,\\
      \big[ \alpha_n+e^{2i\phi} \tilde{\alpha}_{-n} \big]
      |B\rangle&=0
     \end{aligned}
\end{equation}
with similar conditions for the fermionic modes. 
Inserting \eqref{kkwb} and \eqref{rel} into the first equation of 
\eqref{modeb} one can solve
for the Kaluza-Klein and winding modes
\begin{equation}\label{solve}
            \widehat{m}_1=-{n\over m} \widehat{n}_2,\quad\quad 
            \widehat{m}_2={n\over m} \widehat{n}_1
\end{equation}
giving rise to the following momentum mode spectrum:
\begin{equation}\label{speccl}
                    M^2_\text{cl}(\vec{r})=
               { |r_1+r_2\,\hat{\tau}|^2\over \hat{\tau}_2}\, 
                     { |n+m\, \hat{\rho}|^2\over \hat{\rho}_2}, 
                    \qquad \vec{r}\in \mathbb{Z}^2
\end{equation}
with $r,s\in\mathbb{Z}$.
We observe that this agrees with the spectrum derived in \cite{Ardalan:1998ce} by 
employing T-duality. 
The oscillator part of a bosonic boundary state satisfying 
\eqref{modeb} is given by
\begin{equation}\label{boundsol}
    |B\rangle_{(n,m)} =\sum_{r,s\in\mathbb{Z}} {\rm exp}
    \bigg(\sum_{n\in\mathbb{Z}}
    {1\over n} e^{2 i\phi} \alpha_{-n} {\tilde{\alpha}}_{-n} \bigg) 
     |r,s\rangle_{(n,m)}
\end{equation}
Using this boundary state we compute the tree channel annulus partition 
function. 
Transforming  the result via a modular transformation into loop channel,
we can extract the momentum-mode contribution and conclude that open
strings stretching between identical rational  D-branes 
carry non-vanishing linear modes giving rise to masses
\begin{equation}\label{specb}
                     M^2_{\text{op}}(\vec{s}) =
            { |s_1+s_2\,\hat{\tau}|^2\over \hat{\tau}_2}\, 
                     {  \hat{\rho}_2 \over |n+m\,\hat{\rho}|^2}, 
                    \qquad \vec{s} \in\mathbb{Z}^2
\end{equation}
As announced, due to the results of the preceding chapter, we have an
alternative way at hand to derive this formula.
The mass formula for an open string on a lattice $\Gamma^d$ (without
Dirichlet bdy.-conditions) is given by formula \eqref{HamiltonianmomTd} (p.\ 
\pageref{HamiltonianmomTd}). 
However we interpret $m$ as the wrapping number of the D$2$ brane
and require invariance only under $m 2\pi\Gamma^2$ which is
only a subgroup of the closed string torus  $2\pi\Gamma^2$.
This means that we consider the $m^2$-fold cover of the torus
defined by $2\pi\Gamma^2$.
The single valuedness condition of the wavefunction under
 $m 2\pi\Gamma^2$ now reads (cf.\ eq.\ \eqref{singlevaluedcond},
 p.\ \pageref{singlevaluedcond}): 
\begin{equation}
  \label{singlevaluedcondncg}
 \sum_{i=1}^2 e_i p^i(\vec{s})
    = \Big(G\mp{\cal F}_{1 \atop 2}\Big)^{-1}\frac{1}{m}\sum_{i=1}^2 s_i e^i\qquad
           \vec{s} \in \mathbb{Z}^{2}
\end{equation}
This leads via \eqref{Hamiltonianmom} (p.\ \pageref{Hamiltonianmom}) to 
the mass-formula for open strings in $F$-field backgrounds:
\begin{equation}\label{canopstrmass}
 M^2_{\text{op}}(\vec{s})=  
    \frac{\vec{s}^T\,G\vec{s}}{\det(m G) + (n+m B_{12})^2}          
\end{equation}
Using the definitions \eqref{taudef} and \eqref{rhodef} (p.\ 
 \pageref{taudef}), we note that \eqref{specb} and \eqref{canopstrmass} agree.

Summarizing, we now have the means to compute annulus amplitudes
for open strings stretched between different kinds of D-branes with
rational mixed boundary conditions. 

There is still the question of the number of  {\it Landau levels}  that occur
if both field strength $F_1$ and $F_2$ do not agree.
By the means of section \eqref{toroidaldbranes} we have a
 concrete formula to calculate this degeneracy. However, 
we have to take into account, that the two branes appearing in
$n_{i,j}$ have in general different wrapping numbers $m_i$ and $m_j$.
Therefore we expect that wavefunctions can only be invariant
  under translations $m_i m_j 2\pi\Gamma^2$ instead of being
invariant under $2\pi\Gamma^2$. Taking this into account in the
derivation of \eqref{Heigenvals} (p.\ \pageref{Heigenvals}) we see
that the degeneracy on each $T^2$ is given by:
\begin{equation} \label{landaudeg}
   n_{i,j} =  (-1) \pf \big(m_i m_j(F_i+F_j)\big) = ( n_j m_i- n_i m_j)
\end{equation}
The Landau degeneracy \eqref{landaudeg} equals the {\it topological 
intersection} number in the T-dual picture where the two dimensional 
$F$-flux branes correspond to one-dimensional branes with non-trivial
intersection. 

We observed in chapter \ref{orbifolds} that the NSNS  $B$ field has to 
be quantized, if we want to gauge the world sheet parity $\Omega$ (cf.\ eq.\ 
\eqref{quantizedB}, p.\ \pageref{quantizedB}). For the two-torus this means
that $B_{12}/\al=1/2$.  $\Omega$ exchanges the $\sigma=0$ with the $\sigma=\pi$
boundary of the string. Thereby the $\tau$-derivative changes sign. This
can be absorbed in redefining $F_i\to -F_i$. (We have to include boundary
conditions with reversed $F$-field.) For  $\Omega$ to be a symmetry, the 
open string mass spectrum has to be invariant. From eq.\ \eqref{canopstrmass}
we deduce that the action of $\Omega$ for backgrounds with $B_{12}/\al=1/2$ is:
\begin{gather}  
\Omega_{B_{12}=1/2} :\;
 \begin{array}{rl}
      n&\to\; n^\prime= n+m  \\
      m&\to\; m^\prime=-m
  \\ 
  F=\frac{1}{m}\begin{pmatrix}
      0 & n  \\
       -n & 0
    \end{pmatrix}  
    &\to \; F^\prime=
   \frac{1}{m}\begin{pmatrix}
      0 & -(n+m)  \\
       n+m & 0
    \end{pmatrix}  
 \end{array}        
\end{gather}
For  vanishing $B$-field we simply get:
\begin{gather}
 \Omega_{B_{12}=0} :\;
 \begin{array}{rl}
      n&\to\; n^\prime= n  \\
      m&\to\; m^\prime=-m
 \end{array}    
\end{gather}
This has an important consequence (especially for model-building, as
we will see in the next chapter): 
the difference between the multiplicity of states 
\begin{align} 
 n_{ij}-  n_{i\Omega(j)} &\in 2\mathbb{Z}\qquad\text{for } B_{12}=0 \\
 n_{ij}-  n_{i\Omega(j)} &\in \phantom{2}\mathbb{Z}
                  \qquad\text{for } B_{12}=\al/2
\end{align}
has always to be an even number for vanishing $B$ field, while it can be both
odd and even for $B_{12}=\al/2$.

As an example, we discuss the $\widehat{\mathbb{Z}}_3$ case
in some more detail.

\paragraph{D-branes in the asymmetric 
           \texorpdfstring{$\widehat{\mathbb{Z}}_3$}{hat(Z)(3)} orbifold}
Consider the $\widehat{\mathbb{Z}}_3$ lattice of type ${\bf A}$ and start with
a D$_1$-brane with pure Dirichlet boundary conditions ($\phi=0$)
\begin{equation}\label{branea}
        \begin{aligned}  \partial_\tau X_1 &=0 ,\\
                          \partial_\tau X_2 &=0 
        \end{aligned}
\end{equation}
Successively applying the asymmetric $\widehat{\mathbb{Z}}_3$ this D-brane is 
mapped to 
a mixed D$_2$-brane with boundary conditions ($\phi=2\pi/3$)
\begin{equation}\label{braneb}
       \begin{aligned}  \partial_\sigma X_1 - 
                 {1\over \sqrt{3}}\partial_\tau X_2 &=0, \\
                      \partial_\sigma X_2 + 
                    {1\over \sqrt{3}} \partial_\tau X_1 &=0 
       \end{aligned}
\end{equation}
and a mixed D$_3$-brane with boundary conditions ($\phi=-2\pi/3$)
\begin{equation}
    \begin{aligned}  \partial_\sigma X_1 + {1\over \sqrt{3}}
             \partial_\tau X_2 &=0, \\
                      \partial_\sigma X_2 - 
            {1\over \sqrt{3}} \partial_\tau X_1 &=0 
    \end{aligned}
\end{equation}
In the orbifold theory these three kinds of D-branes are identified. 
This reflects that their background fields are being identified according to 
\begin{equation}\label{ident}
       {\cal F} \sim {\cal F} + {1 \over \sqrt{3}} 
\end{equation}
or equivalently 
\begin{equation}\label{identbranes}
        \phi \sim \phi + {2\pi \over 3}
\end{equation}
The two coordinates $X_1$ and $X_2$ yield the following 
contribution to the annulus partition function for open strings 
stretched between identical D-branes
\begin{equation}\label{anna}
    A^{\alpha\beta}_{ii}= {\thef{\alpha}{\beta}\over \eta^3}
     \left(\sum_{r\in\mathbb{Z}} e^{-2\pi t{r^2\over R^2}}\right) 
     \left(\sum_{s\in\mathbb{Z}} e^{-2\pi t{3s^2\over 4R^2}}\right) 
\end{equation}
independent of  $i\in\{1,2,3\}$. 
Open strings stretched between different kinds of
D-branes give rise to shifted moding  and yield the partition
function
\begin{equation}\label{annb}
        A_{i,i+1}^{\alpha\beta}=n_{i,i+1} {\thef{{1\over 3}+\alpha}{\beta}
        \over \thef{{1\over 3}+\alpha}{{1\over 2}} } 
\end{equation}
which looks like a twisted open string sector.
As we know from 
\cite{Blumenhagen:1999md,Angelantonj:1999xf,Blumenhagen:1999ev} 
(and the preceding discussion)  we have to take into
account extra multiplicities, $n_{i,i+1}$, which have a natural
geometric interpretation as  multiple intersection
points of D-branes at angles  in the T-dual picture.
By this reasoning we find that
for the {\bf A} type lattice the extra factor is one. 
However, for the three D-branes generated by $\widehat{\mathbb{Z}}_3$ 
when one starts with a D-brane with pure Neumann boundary conditions,
$\phi\in\{\pi/2,\pi/6,-\pi/6\}$, 
T-duality tells us that there must appear an extra factor of three
in front of the corresponding annulus amplitude 
\eqref{annb}. In the orientifold construction presented in section 
\ref{asz3orienti}
these multiplicities are important to give consistent models. 

\subsubsection{\label{onedimbsec}D$1$-branes on 
               \texorpdfstring{$T^2$}{T\texttwosuperior}}
In this section we will calculate the masses of the linear (i.e.~momentum)
modes of a D-brane that has Dirichlet bdy.-cond.~in one direction
and Neumann conditions in the perpendicular direction. This 
situation is depicted in figure \ref{toruswbrane} (p.~\pageref{toruswbrane}).
We denote the vector tangential to the brane by $\vec{t}$ and the 
one normal to the brane by $\vec{n}$:
\begin{align}
  \vec{t}&\equiv n e_1 +m e_2 &
  \vec{n}&\equiv (m e_1^\ast -n e_2^\ast) 
    \left(\big(\begin{smallmatrix}m \\  -n \end{smallmatrix}\big)^T 
    G^\ast\big(\begin{smallmatrix}m \\ -n\end{smallmatrix}\big) \right)^{-1/2}
\end{align}
With these conventions, we see that the linear part of the field 
$X$ becomes (cf.\ \eqref{zeromodeexpansion}, p.~\pageref{zeromodeexpansion}):
\begin{equation}
   X_\text{lin} = \tau \,\vec{t}\, p_{\vec{t}} + \sigma \,\vec{n} \,p_{\vec{n}}
\end{equation} 
$\vec{n}p_{\vec{n}}$ points perpendicular to the D-brane. Therfore $\pi p_{\vec{n}}$
must be an integer multiple of the minimal distance of two branes on the
torus. Therfore we get:\footnote{To calculate the minimal distance between
two D$1$-branes we use an elementary theorem of number theory: {\sl
The units in the ring $\mathbb{Z}/n\mathbb{Z}$ consist of those 
residue classes $\bmod\, n\mathbb{Z}$ which are represented by integers
$m\neq 0$ and prime to $n$} (cf.\ \cite{Lang:1984}, chapter
{\sf II}, \S 2 ). This implies that given two integers $n$ and $m$ which
are prime w.r.t.\ each other there exist two integers $a$, $b$ s.th.:
$an+bm=1$.}
\begin{equation}
    p_{\vec{n}}(s_2) = s_2  \left(\big(\begin{smallmatrix}m \\  -n \end{smallmatrix}\big)^T 
    G^\ast\big(\begin{smallmatrix}m \\ -n\end{smallmatrix}\big) \right)^{-1/2} 
    \qquad s_2\in\mathbb{Z}
\end{equation} 
The momentum in the $\tau$-direction is conserved. It reads (c.f.\ 
eq.~\eqref{constmotionpib}, p.~\pageref{constmotionpib}): 
\begin{equation}
  \Pi_{\vec{t}} =\frac{1}{2\pi} \bigg(\|\vec{t}\| p_{\vec{t}}
    + \overbrace{\frac{\vec{t}}{\|\vec{t}\|} B_{12}\vec{n}}^{=B_{12}} p_{\vec{n}}
    (s_2)
                 \bigg)
\end{equation} 
Requiring invariance of the wavefunction under a translation by 
$2\pi  \|\vec{t}\|$ in  the $\vec{t}$ direction leads to
\begin{equation}
   p_{\vec{t}}(s_1,s_2) =
     \left(\big(\begin{smallmatrix}n \\  m \end{smallmatrix}\big)^T 
     G\big(\begin{smallmatrix} n \\ m\end{smallmatrix} \big)\right)^{-1/2}
    \big(s_1  - B_{12}\, p_{\vec{n}}(s_2) \big) 
   \qquad s_1,s_2\in\mathbb{Z}
\end{equation} 
The resulting   mass\raisebox{1ex}{\tiny 2} formula takes the form:
\begin{equation}\label{canopstrmassangles}
 M^2_{\text{op}}(\vec{s})=  
   \frac{ \vec{s}^T\,
          \Big(\begin{smallmatrix}
             1         &  -B_{12}             \\
             -B_{12}   &   \det (G) + B_{12}^2
             \end{smallmatrix}
           \Big)
          \vec{s}}
         {\big(\begin{smallmatrix}n \\  m \end{smallmatrix}\big) 
          G\big(\begin{smallmatrix}n \\ m\end{smallmatrix}\big)},
              \qquad \vec{s} \in\mathbb{Z}^2  
\end{equation}
This formula is T-dual (under the duality denoted by ``$D$''
in section \ref{enhsym}, eq.\ \eqref{Dduality}) to the mass formula 
\eqref{canopstrmass} and \eqref{specb}.
Of course we can rewrite \eqref{canopstrmassangles} in terms of the 
complex-structure $\tau$ and the K\"ahler structure $\rho$:
\begin{equation}\label{canopstrmassanglesb}
                     M^2_{\text{op}}(\vec{s}) =
            { |s_1+s_2\,\rho|^2\over \rho_2}\, 
                     {  \tau_2 \over |n+m\,\tau|^2}, \qquad \vec{s} \in\mathbb{Z}^2
\end{equation}

\section{\label{asymrotsec}Asymmetric rotations and non-commutative geometry}

In section \ref{Dbinasorbs} we have pointed out that on $T^2$ or $\mathbb{R}^2$ 
D-branes with mixed boundary
conditions can be generated by simply applying an asymmetric
rotation to an ordinary D-brane with pure Neumann or Dirichlet
boundary conditions. Thus, it should be possible to rederive earlier
results for the two-point function on the disc
\begin{equation}\label{twopoint}
          \langle X_i(z)\, X_j(z') \rangle
\end{equation}
for the operator product expansion (OPE) between   vertex operators
on the boundary
\begin{equation}\label{vertb}
       e^{i p X}(\tau)\,   e^{i q X}(\tau') 
\end{equation}
by applying an asymmetric rotation on the corresponding quantities
for open strings ending on D0-branes in flat space-time.

\subsection{Two-point function on the disc}

The two-point function on the disc for both $X_1$ and $X_2$ of Dirichlet type
reads
\begin{equation}\label{twopointb}
       \begin{aligned}  \langle X_i(z)\, X_j(z') \rangle&=
                 -\alpha'\delta_{ij}\left( \ln |z-z'| -\ln |z-\Bar{z}'| \right)
                 \\
                &=-\alpha'\delta_{ij}{1\over 2}
                \left( \ln (z-z') + \ln (\Bar{z}-\Bar{z}')
             -\ln (z-\Bar{z}') - \ln (\Bar{z}-{z}')\right) 
       \end{aligned}
\end{equation}
from which, formally using 
\begin{equation}\label{move}
           X_i(z)=X_{i,L}(z) + X_{i,R}(\Bar{z})
\end{equation}
we can directly read off the individual contributions from
the left- and right-movers.
Performing the asymmetric rotation
\begin{equation}\label{asymrota}
         X_L\to A\, X_L, \quad\quad X_R\to A^T X_R 
\end{equation}
where $A$ denotes an element of $SO(2)$, leads to the 
following expression
for the propagator in the rotated coordinates
\begin{equation}\label{twopointc}
  \begin{aligned} 
    \langle X_i(z)\, X_j(z') \rangle=&
                 -\alpha'\delta_{ij} \ln |z-z'| - \alpha'\delta_{ij}
               \left( \sin^2\phi-\cos^2\phi\right) \ln |z-\Bar{z}'| \\
               &-\alpha' \epsilon_{ij} \sin\phi\, \cos\phi\,
               \ln\left({z-\Bar{z}'\over \Bar{z}-z' } \right)
  \end{aligned}
\end{equation}
This expression agrees precisely  with the propagator derived in \cite{Abouelsaood:1987gd}
with the identification
\begin{equation}\label{nappi}
      {\cal F}=\begin{pmatrix} 0 & \cot\phi \\
                 -\cot\phi & 0
               \end{pmatrix} 
\end{equation}
Thus, by applying an asymmetric rotation we have found an elegant and short
way of deriving this  propagator without explicit reference to the
boundary conditions or the background fields. Moreover, since the
commutative D0-brane is related in this
smooth way to a non-commutative  D2-brane, it is suggesting that
also both effective theories arising on such branes are related
by some smooth 
transformation. Such an  explicit map between the commuting and
the non-commuting effective gauge theories  has been determined in 
\cite{Seiberg:1999vs}.

\subsection{The OPE of vertex operators}

In this subsection we apply an asymmetric rotation also
to the operator product expansion of tachyon vertex operators 
${\cal O}(z)= e^{ipX}(z)$ on the 
boundary. Of course this OPE is a direct consequence of the
correlator \eqref{twopointc} restricted to the boundary, but nevertheless
we would like to see whether we can generate  the non-commutative
$\ast$-product directly
via an asymmetric rotation. 
Taking care of the left- and right-moving contributions in the
OPE between vertex operators living on a pure Dirichlet boundary
we can write for $|z|>|z'|$
\begin{equation}\label{verta} 
      e^{i p X}(z)\,   e^{i q X}(z') =
        {(z-z')^{{\alpha'\over 2} p_L q_L }  \,
        (\Bar{z}-\Bar{z}')^{{\alpha'\over 2} p_R q_R }
       \over (z-\Bar{z}')^{{\alpha'\over 2} p_L q_R }\,
             (\Bar{z}-{z}')^{{\alpha'\over 2} p_R q_L }}\
             e^{i (p+q) X}(z')+\ldots 
\end{equation}
Now we apply an asymmetric rotation \eqref{asymrota} together with 
\begin{equation}\label{asymrot}
     \begin{aligned} &p_L\to A\, p_L, \quad\quad p_R\to A^T p_R, \\
                       &q_L\to A\, q_L, \quad\quad q_R\to A^T q_R
     \end{aligned}
\end{equation}
and, after all, identifying
$p_L=p_R$, $q_L=q_R$ we obtain
\begin{multline}\label{vertba}  
         e^{i p X}(z)\,   e^{i q X}(z') = \\
           {\bigl[(z-z')  
        (\Bar{z}-\Bar{z}')\bigr]^{{\alpha'\over 2} p q } \over
        \bigl[ (z-\Bar{z}')  
        \times  (\Bar{z}-{z}')\bigr]^{{\alpha'\over 2} \cos(2\phi)\, p q }}
         \bigl( {z-\Bar{z}'\over \Bar{z}-z'} \bigr)^{-{\alpha'\over 2}
        \epsilon_{ij} p_i q_j \sin(2\phi) } \
             e^{i (p+q) X}(z')+\ldots 
\end{multline}
Restricting \eqref{vertba} to the boundary and choosing the same branch cut as 
in \cite{Seiberg:1999vs} we finally arrive at 
\begin{multline} \nonumber
            e^{i p X}(\tau)\,   e^{i q X}(\tau') = \\
           (\tau-\tau')^{\alpha' p q (1+\sin^2\phi-\cos^2\phi)}\,
            {\rm exp}\bigl(-i \pi \alpha' \sin\phi \cos\phi  
           \epsilon_{ij} p_i q_j\bigr) \ 
             e^{i (p+q) X}(\tau')+\ldots 
\end{multline}
This is precisely the OPE derived in 
\cite{Schomerus:1999ug,Seiberg:1999vs}. It shows that it is indeed possible to derive the 
$\ast$-product $e^{ipX}(\tau)e^{iqX}(\tau')\sim e^{ipX}\ast e^{iqX}(\tau')$
directly via an asymmetric rotation, 
where the non-commutative algebra ${\cal A}$ of functions $f$ and $g$
is defined as
\begin{equation}   \label{starprod}
   f\ast g=fg-i\pi \alpha' \sin\phi\, \cos\phi\, \epsilon_{ij}\,
   \partial_if\partial _jg+ \dots
\end{equation}

\subsection{\label{coordcommutesec}The commutator of the coordinates}
\begin{figure}
\begin{minipage}{\textwidth}
 \begin{minipage}[b]{6.0cm}
 \begin{center}
\begin{picture}(0,0)%
\includegraphics{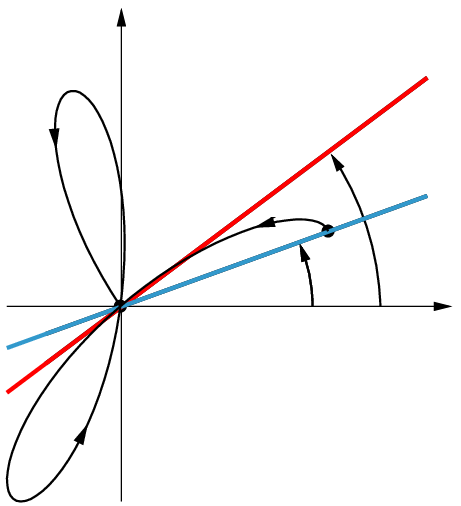}%
\end{picture}%
\setlength{\unitlength}{2486sp}%
\begingroup\makeatletter\ifx\SetFigFont\undefined%
\gdef\SetFigFont#1#2#3#4#5{%
  \reset@font\fontsize{#1}{#2pt}%
  \fontfamily{#3}\fontseries{#4}\fontshape{#5}%
  \selectfont}%
\fi\endgroup%
\begin{picture}(3437,3794)(3513,-3143)
\put(6353,-2150){\makebox(0,0)[lb]{\smash{\SetFigFont{14}{16.8}{\rmdefault}{\mddefault}{\updefault}{\color[rgb]{0,0,0}$X_1$}%
}}}
\put(3698,315){\makebox(0,0)[lb]{\smash{\SetFigFont{14}{16.8}{\rmdefault}{\mddefault}{\updefault}{\color[rgb]{0,0,0}$X_2$}%
}}}
\put(6791,-890){\makebox(0,0)[lb]{\smash{\SetFigFont{12}{14.4}{\rmdefault}{\mddefault}{\updefault}{\color[rgb]{0,0,0}$\sigma=0$}%
}}}
\put(6791, 40){\makebox(0,0)[lb]{\smash{\SetFigFont{12}{14.4}{\rmdefault}{\mddefault}{\updefault}{\color[rgb]{0,0,0}$\sigma=\pi$}%
}}}
\put(6286,-576){\makebox(0,0)[lb]{\smash{\SetFigFont{14}{16.8}{\rmdefault}{\mddefault}{\updefault}{\color[rgb]{0,0,0}$\phi_2$}%
}}}
\put(6791,-890){\makebox(0,0)[lb]{\smash{\SetFigFont{12}{14.4}{\rmdefault}{\mddefault}{\updefault}{\color[rgb]{0,0,0}$\sigma=0$}%
}}}
\put(6791, 40){\makebox(0,0)[lb]{\smash{\SetFigFont{12}{14.4}{\rmdefault}{\mddefault}{\updefault}{\color[rgb]{0,0,0}$\sigma=\pi$}%
}}}
\put(5963,-1469){\makebox(0,0)[lb]{\smash{\SetFigFont{14}{16.8}{\rmdefault}{\mddefault}{\updefault}{\color[rgb]{0,0,0}$\phi_1$}%
}}}
\end{picture}
  \end{center}
 \end{minipage} \hfill
  \raisebox{2.4cm}{
   \begin{minipage}[b]{5.5cm}
    \caption[Branes at angles]{\label{braneanglepic}
                               String attached to branes at angles.
                               The relative angle is $\phi_2-\phi_1$.
                               The boundary conditions are given by eq.\
                               \eqref{sigmanull} and \eqref{sigmanullb}
                               (Cf.\ fig.\ \ref{angledbranesfig},
                                p.\ \pageref{angledbranesfig}).}
   \end{minipage}
 }
\end{minipage}
\end{figure}

While the two-point function derived above already implies that the commutator 
of the coordinate fields is non-vanishing, i.e.\ the geometry on the D-brane 
non-commutative, we would like to rederive this result directly via 
studying D-branes with mixed boundary conditions, as well. 
This is done by the quantization of the bosonic coordinate fields of 
the open string. We start with the T-dual situation
with two D-branes intersecting at an arbitrary angle $\phi_2-\phi_1$ 
(see figure \ref{braneanglepic}).
The open string boundary conditions at $\sigma=0$ are 
\begin{equation}\label{sigmanull}
    \begin{aligned}  \partial_\sigma X_1 + \tan\phi_1\, 
                   \partial_\sigma X_2 &=0, \\
                      \partial_\tau X_2 - \tan\phi_1\, \partial_\tau X_1
                      &=0
    \end{aligned}
\end{equation}
and at $\sigma=\pi$  we require
\begin{equation}\label{sigmanullb}
    \begin{aligned}  \partial_\sigma X_1 + \tan\phi_2\, 
                   \partial_\sigma X_2, &=0 ,\\
                      \partial_\tau X_2 - \tan\phi_2\, \partial_\tau X_1
                      &=0
    \end{aligned}
\end{equation}
The mode expansion satisfying these two boundary conditions looks like
\begin{equation}\label{modeg}
    \begin{aligned} X_1=x_1 +&
    i \sqrt{\alpha'} \sum_{n\in\mathbb{Z}}  {\alpha_{n+\nu}\over n+\nu} 
           e^{-i(n+\nu) \tau} \cos[(n+\nu)\sigma +\phi_1] + \\
    &i \sqrt{\alpha'} \sum_{m\in\mathbb{Z}}  {\alpha_{m-\nu}\over m-\nu} 
           e^{-i(m-\nu) \tau} \cos[(m-\nu)\sigma -\phi_1], \\
    X_2=x_2 +&
    i \sqrt{\alpha'} \sum_{n\in\mathbb{Z}}  {\alpha_{n+\nu}\over n+\nu} 
           e^{-i(n+\nu) \tau} \sin[(n+\nu)\sigma +\phi_1] -\\
    &i \sqrt{\alpha'} \sum_{m\in\mathbb{Z}}  {\alpha_{m-\nu}\over m-\nu} 
           e^{-i(m-\nu) \tau} \sin[(m-\nu)\sigma -\phi_1]
    \end{aligned}
\end{equation}
with $\nu=(\phi_2-\phi_1)/\pi$. Using the usual commutation relation 
\begin{equation}\label{osci}
     [\alpha_{n+\nu}, \alpha_{m-\nu}]=(n+\nu)\, \delta_{m+n,0} 
\end{equation}
and the vanishing of the commutator of the center of mass coordinates $x_1$ 
and $x_2$ one can easily show that for D-branes at angles the 
general equal time commutator vanishes
\begin{equation}\label{commc}
         [ X_i(\tau,\sigma), X_j(\tau,\sigma') ]=0 
\end{equation}
Therefore, the geometry of D-branes at angles, but without background 
gauge fields, is always commutative.  

Performing  T-duality in the $x_1$ direction one gets the two 
mixed boundary conditions for the open strings
\begin{equation}\label{sigmanullc}
    \begin{aligned}  \partial_\sigma X_1 + \cot\phi_1\, 
                   \partial_\tau X_2 &=0 ,\\
                      \partial_\sigma X_2 - \cot\phi_1\, \partial_\tau X_1
                      &=0 
    \end{aligned}
\end{equation}
at $\sigma=0$ and 
\begin{equation}\label{sigmanulld}
    \begin{aligned}  \partial_\sigma X_1 + \cot\phi_2\, 
                      \partial_\tau X_2 &=0, \\
                      \partial_\sigma X_2 - \cot\phi_2\, \partial_\tau X_1
                      &=0 
    \end{aligned}
\end{equation}
at $\sigma=\pi$.
This is a very special case of what we considered in chapter \ref{strbg}. 
It corresponds to the following field strengths:
\begin{align}
   {\cal F}_1&= \begin{pmatrix} 
                0 & -\cot\phi_1  \\ \cot\phi_1 & 0
               \end{pmatrix} 
   & {\cal F}_2&= \begin{pmatrix} 
                0 & \cot\phi_2  \\ -\cot\phi_2 & 0
               \end{pmatrix} 
\end{align}
In section \ref{commutatorsec} we proved that the commutator of the
 coordinate fields $X$ takes the following values at its boundaries:
\begin{equation} 
 \left[X(\tau,\sigma),X(\tau,\sigma) \right]\big|_{
             \sigma\in\partial {\cal M}_j}
   = i2\pi\al\frac{{\cal F}_j}{G-{\cal F}_j^2} \qquad j=1,2
\end{equation}  
This result was obtained for the two-torus  
\cite{Chu:1998qz,Chu:1999ta} as well.

At the end of this section let us briefly comment on the algebraic  
structure of the non-commutative torus we have obtained by the asymmetric
rotation on the D-branes. As shown in the previous section, the tachyon
vertex operator ${\cal O}=e^{ipX}(\tau)$ leads to a non-commutative
algebra ${\cal A}$, defined in eq.\ \eqref{starprod}.
As explained in \cite{Seiberg:1999vs}, the algebra ${\cal A}$ of tachyon vertex operators
can be taken at either end of the open string.
Therefore the open string states form a bi-module ${\cal A}\times {\cal A}'$,
where ${\cal A}$ is acting on the boundary $\sigma=0$ and
${\cal A}'$ on the boundary $\sigma=\pi$ of the open string.
Specifically, for an open string whose first boundary $\sigma =0$ is 
related to
a D-brane with parameter $\phi_1$ and whose second boundary $\sigma=\pi$
is attached to a D-brane with parameter $\phi_2$, the algebra ${\cal A}$
of functions on the non-commutative torus
is generated
by 
\begin{equation}\label{ua}
    \begin{aligned}
    &U_1=\exp\Big(iy_1-{2\pi^2 \alpha' 
{\cal F}_1\over 1+{\cal F}_1^2}(\partial /\partial y_2)\Big),
\\ 
     &U_2=\exp\Big(iy_2+{2\pi^2 \alpha'
{\cal F}_1\over 1+{\cal F}_1^2}(\partial /\partial y_1)\Big)
    \end{aligned}
\end{equation}
which obey 
\begin{equation}\label{uu}
       U_1U_2=\exp\Big(-2\pi i{2\pi\alpha'{\cal F}_1\over 1
       +{\cal F}_1^2}\Big)U_2U_1
\end{equation}
On the other hand, the algebra 
 ${\cal A}'$
is generated
by 
\begin{equation}\label{utilde}
     \begin{aligned}
&\tilde U_1
=\exp\Big(iy_1+{2\pi^2\alpha'{\cal F}_2\over 1+{\cal F}_2^2}
(\partial /\partial y_2)\Big),\\
&\tilde U_2
=\exp\Big(iy_2-{2\pi^2\alpha'{\cal F}_2\over 1+{\cal F}_2^2}
(\partial /\partial y_1)\Big)
     \end{aligned}
\end{equation}
obeying 
\begin{equation}\label{uutilde}
      \tilde U_1\tilde U_2
     =\exp\Big(2\pi i{2\pi\alpha'{\cal F}_2\over 1+{\cal F}_2^2}\Big)
       \tilde U_2\tilde U_1
\end{equation}

\section{\label{asz3orienti}Asymmetric orientifolds}
Another motivation for studying such asymmetric orbifolds arises in 
the construction of Type I vacua. In \cite{Blumenhagen:1999md,Blumenhagen:1999ev}
we have 
considered so-called supersymmetric 
orientifolds with D-branes at angles in six and four space-time dimensions 
which in the six-dimensional case were defined as
\begin{equation}\label{oriend}
        {{\rm Type\, IIB}\ {\rm on}\ T^4\over 
       \{\Omega\Bar{\sigma}, \Theta \}}
\end{equation}
with $\Bar{\sigma}:z_i\to -\Bar{z}_i$, the $z_i$ being the complex coordinates of 
the $T^4$. 
Upon T-dualities in the directions of their real parts 
one obtains an ordinary orientifold where, however,
the space-time symmetry becomes asymmetric
\begin{equation}\label{orienc}
       {{\rm Type\, IIB}\ {\rm on}\ \hat{T}^4\over 
       \{\Omega, \Hat{\Theta} \} }
\end{equation}
In the entire derivation in section \ref{Dbinasorbs} we 
have identified the two constructions explicitly via T-duality, relating 
branes with background fields to branes at angles. 
While in the $\Omega \Bar{\sigma}$ orientifolds $\Theta$ identified branes at different 
locations on the tori, $\hat{\Theta}$ now maps branes with different values 
of their background gauge flux upon each other. As the background fields determine 
the parameter which rules the non-commutative geometry, branes with different 
geometries are identified according to \eqref{ident}. However the
non-commutativity
is solely restricted to the compactification space, i.e.\: 
\begin{equation} 
 \Big(\prod_{i}\Hat{T}_{(i)}^2\Big)\Big/\widehat{\mathbb{Z}}_N
\end{equation}
Each $\Hat{T}_{(i)}^2$ has to admit the respective $\widehat{\mathbb{Z}}_N$ symmetry.
  The $\widehat{\mathbb{Z}}_3$ tori 
\eqref{z3hatmetric}, p.~\pageref{z3hatmetric} ({\bf A}-torus)
and  \eqref{z3hatmetricB}, p.~\pageref{z3hatmetricB}
 ({\bf B}-torus) both have a $\rho_2$ component of
order one. As $\rho_2$ is the volume of the two-torus measured in terms of
$\al$, the compactification size is roughly speaking the string
scale. No large volume of the asymmetric ${\widehat{\mathbb{Z}}_3}$ (and more
general:  ${\widehat{\mathbb{Z}}_N}$) orbifold exists. Even though we will get
$SO(N)$ gauge groups in the $\mathbb{Z}_3^L\times\mathbb{Z}_3^R$ orientifold
on $T^4$ this is not in contradiction with the observation that no
non-commutative gauge theories with  gauge-groups other than $U(n)$
 seem to  exist (c.f. \cite{Seiberg:1999vs,Matsubara:2000gr,Dorn:2002ah}): The
$\mathbb{Z}_3^L\times\mathbb{Z}_3^R$ orientifold 
does not admit a decompactification limit.  

From the above mentioned identification it is now clear that the 
${\cal N}=(0,1)$ supersymmetric asymmetric $\widehat{\mathbb{Z}}_N$ orientifolds
\eqref{orienc} have the same one loop partition functions as the corresponding 
symmetric  $\mathbb{Z}_N$ orientifolds \eqref{oriend}.
The only difference is that instead of D7-branes at angles, we introduce
D9-branes with appropriate background fields. 
Thus, a whole class of asymmetric orientifolds has already been
studied in the T-dual picture involving D-branes at angles. 
One could repeat the whole computation for
the asymmetric orientifolds \eqref{orienc},
 getting of course identical results. 
Note, the model \eqref{orienc} is really a Type I vacuum, as $\Omega$
itself is gauged. Thus, in principle there exist the possibility
that heterotic dual models exist. Of course, in six dimensions most models
have more than one tensor-multiplet so that no perturbative heterotic
dual model can exist. It would be interesting to look for 
heterotic duals for the four dimensional models discussed in \cite{Blumenhagen:1999ev}.

\subsection{\label{z3z3asymorienti} Orientifolds on the  
            \texorpdfstring{$\big(T^2\times T^2\big)\big/
                           \big(\mathbb{Z}_3^L\times\mathbb{Z}_3^R\big)$}
                         {(T\texttwosuperior \textmultiply T\texttwosuperior)/
                          (Z(3)L\textmultiply Z(3)R)}
            orbifold background} 
In the following we will construct the even more general 
six-dimensional $\mathbb{Z}_3\times
\widehat{\mathbb{Z}}_3$ orientifold
\begin{equation}  \nonumber 
         {{\rm Type\, IIB}\ {\rm on}\ \hat{T}^4\over 
            \{\Omega, \Theta,\hat\Theta \}} 
\end{equation}
which is T-dual to 
\begin{equation} \nonumber 
      {{\rm Type\, IIB}\ {\rm on}\ T^4\over \{\Omega\Bar{\sigma},\hat\Theta,
        \Theta \} }
\end{equation}
where in fact, as shown in section \ref{z3torusncg}, 
the two tori are identical
$T^4=\hat{T}^4=SU(3)^2$. The freedom to choose their complex-structures 
gives rise 
to a variety of three distinct models, which are denoted by ${\bf AA,AB,BB}$
 as in \cite{Blumenhagen:1999md}. 
Note, that the same orbifold group is generated
by a pure left-moving $\mathbb{Z}_{3L}$, $\Theta_L=\hat\Theta\Theta^{-1}$, and
a  pure right-moving $\mathbb{Z}_{3R}$, $\Theta_R=\hat\Theta\Theta$.
As was also shown in \cite{Bianchi:1999uq} this model actually
has ${\cal N}=(1,1)$ supersymmetry, but one can get 
${\cal N}=(0,1)$ supersymmetry by turning on non-trivial discrete
torsion. 
\subsubsection{\label{tadpcancz3z3}Tadpole cancellation}
\noindent
The computation of the various one-loop amplitudes is straightforward.
For the loop channel Klein bottle amplitude  we obtain
\begin{equation}\label{kleinncg}
    \begin{aligned} K^{(ab)}={16c\over 12} \int_0^\infty {dt\over t^4} 
 \frac{1}{\eta^4} \Bigl[ &\rho_{00}\, \Lambda^a \Lambda^b 
                + \rho_{01} + \rho_{02}                                  \\
            &+ n^{(ab)}_{\hat\Theta,\Omega}\, \rho_{10}+
              n^{(ab)}_{\hat\Theta,\Omega\Theta^2}\,\epsilon \rho_{11}+
             n^{(ab)}_{\hat\Theta,\Omega\Theta}\,\Bar{\epsilon} \rho_{12}\\
            &+ n^{(ab)}_{\hat\Theta^2,\Omega}\, \rho_{20}+
              n^{(ab)}_{\hat\Theta^2,\Omega\Theta^2}\,\Bar{\epsilon} \rho_{21}+
          n^{(ab)}_{\hat\Theta^2,\Omega\Theta}\,\epsilon \rho_{22} \Bigr]
    \end{aligned}
\end{equation}
where $c\equiv {\rm V}_6/\left( 8\pi^2 \alpha^\prime \right)^3$ and $\epsilon$ is 
a phase factor defining the discrete torsion. Furthermore we use 
functions $\rho_{gh}$ that are adopted 
from  \cite{Bianchi:1999uq}. We already expressed the torus partition
function of the orbifold theory in terms of these functions 
(section \ref{asz3orbi}). They are given in appendix \ref{confblocks} 
(p.~\pageref{confblocks}) together with their modular properties.  
$g$ and $h$ denote the twists resp.\ the projections in the 
partition function
$\rho_{gh}$: $g,h\in\{(0,1/3,-1/3),(0,2/3,-2/3)\}$ for which we use the shorter
notation $g,h\in\{0,1,2\}$.
The index $(ab)$ denotes the three possible choices of lattices,
${\bf AA}$, ${\bf AB}$ and ${\bf BB}$, and $\Lambda^a$ are the zero mode
contributions \eqref{specb} to the partition function
\begin{equation}\label{zeros}
    \begin{aligned} 
   \Lambda^{\bf A}
           &=\frac{1}{\eta^2}\sum_{m_1,m_2} e^{-\pi t \left[
               m_1^2+{4\over 3}\left({m_1\over 2}-m_2\right)^2 \right]} 
              =\sum_{i=0}^2\chi_i 
               \\
   \Lambda^{\bf B}
           &=\frac{1}{\eta^2}\sum_{m_1,m_2} e^{-\pi t \left[
               m_1^2+12\left({m_1\over 2}-m_2\right)^2 \right]}
             =\chi_0 
    \end{aligned}
\end{equation}
$\chi_i$ are  $SU(3)$ characters with argument $q=\exp(-4\pi t)$
(cf.\ eq.\ \eqref{su3char}, p.\ \pageref{su3char}).
Finally, $n^{(ab)}_{\Sigma_1,\Sigma_2}$ denotes the trace of the action of
$\Sigma_2$ on  the fixed points in the $\Sigma_1$ twisted sector.
Taking into account that the origin is the only common fixed point
of $\mathbb{Z}_3$ and $\widehat{\mathbb{Z}}_3$, they can be determined to be
\begin{gather}\label{anzahl}
              n^{(ab)}_{\hat\Theta,\Omega}=
             \begin{cases}
                        9& \text{for }  $({\bf AA})$ \\
                        3& \text{for }  $({\bf AB})$ \\
                        1& \text{for }  $({\bf BB})$
             \end{cases}
\end{gather}
and 
\begin{gather}\label{anzahlb}
               n^{(ab)}_{\hat\Theta,\Omega\Theta}=
              n^{(ab)}_{\hat\Theta^2,\Omega\Theta^2}=
            \begin{cases}
                        -3& \text{for } $({\bf AA})$ \\
                 i\sqrt{3}& \text{for } $({\bf AB})$ \\
                         1& \text{for } $({\bf BB})$ 
            \end{cases}
\end{gather}
The remaining numbers are given by complex conjugation of \eqref{anzahlb}.
Applying a modular transformation to \eqref{kleinncg} yields the tree channel 
Klein bottle amplitude
\begin{equation}\label{kleintree}
    \begin{aligned} \widetilde{K}^{(ab)}=2{32c\over 3} 
       \int_0^\infty {dl} 
   \frac{1}{\eta^4} \Bigl[ & \rho_{00}\, 
               \tilde\Lambda^a \tilde\Lambda^b + 
            {1\over 3} n^{(ab)}_{\hat\Theta^2,\Omega}\, \rho_{01} + 
            {1\over 3} n^{(ab)}_{\hat\Theta,\Omega}\, \rho_{02}      \\
         &+ 3 \rho_{10} - 
             n^{(ab)}_{\hat\Theta^2,\Omega\Theta^2}\, \Bar{\epsilon}\rho_{11}-
             n^{(ab)}_{\hat\Theta,\Omega\Theta^2}\, \epsilon\rho_{12}\\
         &+ 3 \rho_{20} - 
             n^{(ab)}_{\hat\Theta^2,\Omega\Theta}\, \epsilon\rho_{21}-
             n^{(ab)}_{\hat\Theta,\Omega\Theta}\, \Bar{\epsilon}\rho_{22} 
             \Bigr]
    \end{aligned} 
\end{equation}
The lattice contributions are\footnote{The argument of the
 $SU(3)$ characters is $q=\exp (-4\pi l)$.}
 \begin{equation}\label{zerosb}
    \begin{aligned} \tilde\Lambda^{\bf A}&=\frac{\sqrt{3}}{\eta^2}
                        \sum_{m_1,m_2} e^{-3\pi l 
            \left[
               m_1^2+{4\over 3}\left({m_1\over 2}-m_2\right)^2 \right]}
                =\sqrt{3}\chi_0
               \\
                \tilde\Lambda^{\bf B}&=\frac{1}{\sqrt{3}\eta^2}
                \sum_{m_1,m_2} e^{-\pi l \left[
               {1\over 3}m_1^2+4\left({m_1\over 2}-m_2\right)^2 \right]}
               =\frac{1}{\sqrt{3}}\sum_{i=0}^2\chi_i
    \end{aligned}
\end{equation}
In order to cancel these tadpoles we now introduce D-branes with
mixed boundary conditions. For both the ${\bf A}$ and the ${\bf B}$
lattice we choose three kinds of D-branes with 
$\theta\in\{\pi/2,\pi/6,-\pi/6\}$. The asymmetric $\widehat{\mathbb{Z}}_3$ cyclically
permutes these three branes, whereas the symmetric ${\mathbb{Z}}_3$ leaves
every brane invariant and acts with a $\gamma_{\Theta,i}$ matrix on  the
Chan-Paton factors of each brane. Since $\widehat{\mathbb{Z}}_3$ permutes
the branes, all three $\gamma_{\Theta,i}$ actions must be the same.
The computation of the annulus amplitude gives 
\begin{equation}\label{ann}
    \begin{aligned} A^{(ab)}=\rule{2.1cm}{0.cm}&
    \\{2c\over 12} \int_0^\infty {dt\over t^4} 
     \frac{1}{\eta^4} \Bigl[ &{\rm M^2}\rho_{00}\, \Lambda^a \Lambda^b + 
               ({\rm Tr}\, \gamma_\Theta )^2\,  \rho_{01} + 
               ({\rm Tr}\, \gamma_{\Theta^2} )^2\, \rho_{02} +\\
            & {\rm M^2}\, n^{(ab)}_{\hat\Theta,1}\, \rho_{10}+
              ({\rm Tr}\, \gamma_\Theta )^2\,
               n^{(ab)}_{\hat\Theta,\Theta}\,\epsilon \rho_{11}+
             ({\rm Tr}\, \gamma_{\Theta^2} )^2\,
             n^{(ab)}_{\hat\Theta,\Theta^2}\,\Bar{\epsilon} \rho_{12}+\\
            & {\rm M^2} \, n^{(ab)}_{\hat\Theta^2,1}\, \rho_{20}+
            ({\rm Tr}\, \gamma_\Theta )^2\,
            n^{(ab)}_{\hat\Theta^2,\Theta}\,\Bar{\epsilon} \rho_{21}+
           ({\rm Tr}\, \gamma_{\Theta^2} )^2\,
          n^{(ab)}_{\hat\Theta^2,\Theta^2}\,\epsilon \rho_{22} \Bigr]
    \end{aligned}
\end{equation}
where the $\hat\theta$ twisted sector is given by open strings stretched
between D-branes with $\theta_i$ and $\theta_{i+1}$. 
Thus, $n^{(ab)}_{\hat\Theta,1}$ denotes the intersection number
of two such branes and $n^{(ab)}_{\hat\Theta,\Theta}$ the number
of intersection points invariant under $\Theta$.
The actual numbers turn out to be the same as the multiplicities of 
the closed string twisted sectors in \eqref{anzahl} and \eqref{anzahlb}.
For the tree channel amplitude we obtain
\begin{equation}\label{anntreencg}
   \begin{aligned} \tilde{A}^{(ab)}=2{c\over 6} 
       \int_0^\infty {dl} 
      \frac{1}{\eta^4}\Bigl[ &{\rm M}^2\, \Bigl( \rho_{00}\, 
           \tilde\Lambda^a \tilde\Lambda^b + 
            {1\over 3} n^{(ab)}_{\hat\Theta^2,1}\, \rho_{01} + 
            {1\over 3} n^{(ab)}_{\hat\Theta,1}\, \rho_{02}\Bigr)     \\
         &+ ({\rm Tr}\, \gamma_\Theta )^2 \left( 3 \rho_{10} - 
             n^{(ab)}_{\hat\Theta^2,\Theta}\, \Bar{\epsilon}\rho_{11}-
             n^{(ab)}_{\hat\Theta,\Theta}\, \epsilon\rho_{12}\right) \\
         &+ ({\rm Tr}\, \gamma_{\Theta^2} )^2 \left( 3 \rho_{20} - 
             n^{(ab)}_{\hat\Theta^2,\Theta^2}\, \epsilon\rho_{21}-
             n^{(ab)}_{\hat\Theta,\Theta^2}\, \Bar{\epsilon}\rho_{22}\right)
       \Bigr]
    \end{aligned}
\end{equation}
Finally, one has to compute the M\"obius amplitude 
\begin{equation}\label{moeb}
    \begin{aligned} M^{(ab)}= 
      \\-{2c\over 12} \int_0^\infty &{dt\over t^4}
      \frac{1}{\eta^4}\Bigl[ {\rm M}\, \rho_{00}\, \Lambda^a \Lambda^b + 
       {\rm Tr}( \gamma_{\Omega\Theta}^T\gamma_{\Omega\Theta}^{-1} )\, 
            \rho_{01} + 
             {\rm Tr}( \gamma_{\Omega\Theta^2}^T\gamma_{\Omega\Theta^2}^{-1} )
               \, \rho_{02} \\
           + & {\rm M}\, n^{(ab)}_{\hat\Theta,\Omega}\, \rho_{11}+
              {\rm Tr}( \gamma_{\Omega\Theta}^T\gamma_{\Omega\Theta}^{-1} )\,
               n^{(ab)}_{\hat\Theta,\Omega\Theta}\,\epsilon \rho_{12}+
             {\rm Tr}( \gamma_{\Omega\Theta^2}^T\gamma_{\Omega\Theta^2}^{-1} )
        \,   n^{(ab)}_{\hat\Theta,\Omega\Theta^2}\,\Bar{\epsilon} \rho_{10}
     \\+
     & {\rm M}\, n^{(ab)}_{\hat\Theta^2,\Omega} \rho_{22}+
              {\rm Tr}( \gamma_{\Omega\Theta}^T\gamma_{\Omega\Theta}^{-1} )\,
               n^{(ab)}_{\hat\Theta^2,\Omega\Theta}\,\Bar{\epsilon} \rho_{20}+
             {\rm Tr}( \gamma_{\Omega\Theta^2}^T\gamma_{\Omega\Theta^2}^{-1} )
          n^{(ab)}_{\hat\Theta^2,\Omega\Theta^2}\epsilon \rho_{21} 
          \Bigr]\rule[-2.5ex]{0.0ex}{2ex}
    \end{aligned}
\end{equation}
with argument $q=-{\rm exp}(-2\pi t)$.
Transformation into tree channel leads to the expression
\begin{equation}\label{moebtree}
   \begin{aligned} \widetilde{M}^{(ab)}=-2{8c\over 3} 
       \int_0^\infty {dl} 
      \frac{1}{\eta^4}\Bigl[ &{\rm M}\, \Bigl( \rho_{00}\, 
           \tilde\Lambda^a \tilde\Lambda^b + 
            {1\over 3} n^{(ab)}_{\hat\Theta,\Omega}\, \rho_{01} + 
            {1\over 3} n^{(ab)}_{\hat\Theta^2,\Omega}\, \rho_{02}\Bigr) \\
         &+  {\rm Tr}( \gamma_{\Omega\Theta^2}^T\gamma_{\Omega\Theta^2}^{-1} )
           \left( 3 \rho_{11} - 
          n^{(ab)}_{\hat\Theta,\Omega\Theta^2}\, \Bar{\epsilon}\rho_{12}-
        n^{(ab)}_{\hat\Theta^2,\Omega\Theta^2}\, \epsilon\rho_{10}\right) \\
         &+  {\rm Tr}( \gamma_{\Omega\Theta}^T\gamma_{\Omega\Theta}^{-1} )
            \left( 3 \rho_{22} - 
             n^{(ab)}_{\hat\Theta,\Omega\Theta}\, \epsilon\rho_{20}-
             n^{(ab)}_{\hat\Theta^2,\Omega\Theta}\, \Bar{\epsilon}\rho_{21}
            \right)
       \Bigr]
   \end{aligned}
\end{equation}
The three tree channel amplitudes give rise to two 
tadpole cancellation conditions
\begin{equation}\label{tadpole}
    \begin{aligned}  &{\rm M}^2-16\, {\rm M}+64=0, \\
                        &({\rm Tr}\, \gamma_\Theta )^2-16 
            {\rm Tr}( \gamma_{\Omega\Theta^2}^T\gamma_{\Omega\Theta^2}^{-1} )
                 +64=0
    \end{aligned}
\end{equation}
Thus, we have ${\rm M}=8$ D9-branes of each kind and the action
of $\mathbb{Z}_3$ on the Chan-Paton labels has to satisfy ${\rm Tr} \gamma_\Theta=8$
implying that we have the simple solution that $\gamma_\Theta$ is 
the identity matrix. 

\subsubsection{The massless spectrum}

Having solved the tadpole cancellation conditions we can move forward
and compute the massless spectrum of the effective commutative field theory 
in the non-compact six-dimensional space-time.
In computing the massless spectra we have to take into account
the actions of the operations on the various fixed points. 
In  the closed string sector  we find the spectra shown in 
table  \ref{asZ3orclspec}. The computation of the massless spectra in the open string sector
is also straightforward and yields the result in table \ref{asZ3oropspec}.
\begin{table}
 \renewcommand{\arraystretch}{1.2}
  \begin{center}
 \begin{tabular}{|c|c|c|}
 \hline
 $\epsilon$ & (ab) & spectrum       \\
 \hline\hline
 $1$ & $-$ & $(1,1)\,  {\rm Sugra}+4\times V_{1,1} $ \\
 \hline
          & ${\bf AA}$ & $(0,1)\, {\rm Sugra} +6\times T +15\times H$ \\
 $e^{\pm 2\pi i/3}$ 
          & ${\bf AB}$ & $(0,1)\, {\rm Sugra}+9\times T +12\times H$ \\
          & ${\bf BB}$ & $(0,1)\, {\rm Sugra}+10\times T +11\times H$ \\
 \hline
 \end{tabular}
 \caption{\label{asZ3orclspec}Closed string spectra of the
          $\big(T^2\times T^2\big)
          \big/\big(\mathbb{Z}_3^L\times\mathbb{Z}_3^R\big)$-orientifold
         }
 \end{center}
\end{table}

\begin{table}
 \renewcommand{\arraystretch}{1.2}
 \begin{center}
 \begin{tabular}{|c|c|c|}
 \hline
 $\epsilon$ & (ab) & spectrum       \\
\hline\hline
 $1$ & $-$ & $V_{1,1}$ in $SO(8)$ \\
\hline
                    & ${\bf AA}$ & $V$ in $SO(8)$ +$4\times H$ in ${\bf 28}$ \\
 $e^{\pm 2\pi i/3}$ & ${\bf AB}$ & $V$ in $SO(8)$ +$1\times H$ in ${\bf 28}$ \\
                    & ${\bf BB}$ & $V$ in $SO(8)$  \\
\hline
 \end{tabular}
 \caption{\label{asZ3oropspec} Open string spectra of the
          $\big(T^2\times T^2\big)
          \big/\big(\mathbb{Z}_3^L\times\mathbb{Z}_3^R\big)$-orientifold}
 \end{center}
\end{table}
All the spectra shown in table \ref{asZ3orclspec} and table \ref{asZ3oropspec}
 satisfy the cancellation
of the non-factorizable anomaly. Note, that the configurations ${\bf AB}$ 
and ${\bf BB}$
were not analyzed in \cite{Bianchi:1999uq}. Thus, we have successfully applied
the techniques derived in section \ref{Dbinasorbs} and section 
\ref{asymrotsec} to the construction of asymmetric orientifolds.

\addcontentsline{toc}{section}{Concluding remarks}
\section*{Concluding remarks}
In this chapter  we have pointed out a relationship between the
realization of asymmetric operations in the open string sector and 
non-commutative geometry arising at the boundaries of open string
world-sheets. More concretely, we have shown that  a left-right 
asymmetric rotation transforms an ordinary Neumann or 
Dirichlet boundary condition into a mixed Neumann-Dirichlet boundary condition. 
We have employed this observation to rederive the non-commutativity
relations for the open string. Moreover, we have solved the problem
of how the open string sector manages to incorporate asymmetric
symmetries. It simply turns on background gauge fluxes.
Finally, we have considered a concrete asymmetric Type I vacuum,
where D-branes with mixed boundary conditions were introduced
to cancel all tadpoles. 

We have restricted ourselves to the case of products of two-dimensional
tori. With the insights gained in the preceeding chapter it is very suggestive
how to  generalize these ideas to more general asymmetric elements of the
T-duality group. 
As it is known that the
T-duality group $SO(d,d,\mathbb{Z})$ is generated by only three classes of 
generators,\footnote{I.e.\ integer shifts in the NSNS $B$-field, change of the
  torus-basis ($\in SL(d,\mathbb{Z})$) and a so called factorized duality, 
which is
the generalization of the $R\al\to\al/R$ T-duality. A nice review on 
T-duality is \cite{Giveon:1994fu}.
 There also the original references (which are quite
a lot) concerning the generators of $SO(d,d,\mathbb{Z})$ can be found.}
 we can map each
 kind of constant open string  boundary condition under T-duality in addtion
 to the closed string background fields.

It would be also interisting to discuss the dual heterotic description. 

Furthermore, it would be interesting to see whether via the asymmetric
rotation one can gain further insight into the relation between  
the effective non-commutative and commutative gauge theories 
on the branes.


\chapter[Toroidal orientifolds with magnetized  versus intersecting D-Branes]
        { \label{magbf}\centerline{Toroidal orientifolds with}
                        \centerline{magnetized  versus}
                       \centerline{intersecting D-Branes}
        }

In this chapter we will mainly present the results published
in  publication \cite{Blumenhagen:2000wh,Blumenhagen:2000vk}. 
We investigated strings on toroidal orientifolds with D$9$-branes
which are allowed to carry arbitrary magnetic background fluxes. 
We restricted ourselves  to block diagonal NS $U(1)$ fields $F$.
Therefore the pure $\Omega$-orientifold is T-dual to a $\bar{\sigma}\Omega$-orientifold
with $\bar{\sigma}$ being a reflection which has a fixed point locus of half
the dimension of the compact space. As a consequence the T-dual picture leads 
 to an orientifold $6$-plane (O$6$-plane) in compactifications on 
 six-tori (and to an O$7$-plane in compactifications on 
 four-tori). The O$6$-plane is charged under 
 a RR $7$-form while the O$7$-plane carries RR $8$-form  charge. The RR charge
will be canceled by D$6$- (resp.\ D$7$-) branes without 
any $U(1)$ background field. The T-dual picture has the advantage to admit
a purely geometric interpretation, with multiplicities of open string 
states given  by the intersection numbers of the corresponding D$p$ branes.
Because the interpretation of the intersection numbers in the $F$-field
picture is less direct though possible (cf.\ section \ref{toroidaldbranes},
 p.\ \pageref{toroidaldbranes}), 
we will from time to time switch between the description in terms of fluxes
and the  purely geometrical description in terms of intersecting D$p$ branes.
We will discuss both the technical description as well as applications to
phenomenology.

\section{Introduction}   
The search for realistic string vacua is one of the burning open problems
within superstring theory.
A phenomenologically viable string compactification should contain at least
three chiral fermion generations, the Standard Model gauge group and
broken space-time supersymmetry. In the context of `conventional'
string compactifications the requirement of getting chiral fermions
is usually achieved by considering compact, internal background spaces
with nontrivial topology rather than simple tori.
In particular, when analyzing the Kaluza-Klein fermion spectra 
\cite{Witten:1981me}
a net-fermion generation number arises if the internal Dirac operator
has zero modes. For example, considering heterotic string compactifications
on Calabi-Yau threefolds \cite{Candelas:1985en} , 
the net-generation number is equal to $|\chi|/2$,
where $\chi$ is the Euler number of the Calabi-Yau space. 
Chiral fermions are also present in a large
class of heterotic orbifold compactifications \cite{Dixon:1985jw}, 
as well as in free bosonic \cite{Lerche:1987cx} 
and fermionic \cite{Kawai:1987ah,Antoniadis:1987rn} constructions. 
Type II string models with chiral fermions can be constructed by locating
D-branes at transversal orbifold or conifold singularities 
\cite{Douglas:1996sw,Klebanov:1998hh},
or by considering intersections of D-branes and NS-branes 
\cite{Landsteiner:1998gh,Brunner:1998jr,Hanany:1998tb};
chiral Type I models were first proposed in \cite{Angelantonj:1996uy}.
Moreover 
orbifold compactifications of eleven-dimensional M-theory
can lead to chiral fermions, as discussed e.g. in
\cite{Horava:1996qa,Witten:1996mz,Faux:1999hm,Kaplunovsky:1999ia,Faux:2000dv}.

The phenomenological requirement of breaking space-time supersymmetry can
be met in various ways. In the context of heterotic string compactifications
gaugino condensation \cite{Derendinger:1985kk,Dine:1985rz} or the
 Scherk-Schwarz mechanism \cite{Scherk:1979ta,Cremmer:1979uq,Ferrara:1989jx,
Antoniadis:1998ki,Antoniadis:1998ep,Antoniadis:1999ux,Cotrone:1999xs}
lead to potentially interesting models with supersymmetry broken at low
energies. In addition, as it was realized more recently, Type II models
on nontrivial background spaces with certain D-brane configurations 
possess broken space-time supersymmetry. Especially, when changing the
GSO-projections tachyon free Type 0 orientifolds 
in four dimensions can be constructed \cite{Angelantonj:1998gj,Blumenhagen:1999bd,Blumenhagen:1999ns,Blumenhagen:1999ad}. 
Alternatively, orientifolds on six-dimensional orbifolds with brane-antibrane
configurations provide interesting scenarios \cite{Antoniadis:1999xk,Aldazabal:1999jr,Aldazabal:1999tw,Aldazabal:2000sk,Angelantonj:1999ms,Angelantonj:2000xf}, where supersymmetry is
left unbroken in the gravity bulk, but broken in the open string sector
living on the brane-antibrane system.

Finally the quest for a realistic gauge group with sufficiently low rank
is met in heterotic strings by choosing appropriate gauge vector bundles
on the Calabi-Yau spaces \cite{Witten:1986bz}, 
which can be alternatively described by turning on {\it Wilson lines}
in Calabi-Yau or also in orbifold compactifications \cite{Ibanez:1987xa}.
\footnote{A Wilson line defines an embedding of the fundamental group
$\pi_1({\cal X})$ into the gauge bundle of the theory. Wilson lines
on compactification spaces ${\cal X}$ with abelian fundamental group only
lead to rank conserving symmetry breaking. However in  manifolds
with {\sl nonabelian} $\pi_1({\cal X})$  one  can achieve rank reducing
gauge-symmetry breaking via Wilson lines.}
On the Type II side  the gauge group can be  reduced  
by Wilson lines or, in the T-dual picture, by placing the branes at
different positions inside the internal space.

As it should have become clear from the previous discussion, `standard'
heterotic, Type I or Type 
II compactifications on simple 6-tori do not meet any of the three above
requirements. However, as we will discuss in \ref{d9baneswflux}, turning on 
magnetic fluxes in the internal directions of the D-branes, thereby inducing 
mixed Neumann-Dirichlet boundary conditions for open strings equivalent to a 
non-commutative internal geometry \cite{Connes:1998cr,Douglas:1998fm}
on the branes, all three goals can be
achieved in one single stroke.\footnote{In chapter \ref{ncg} which is mainly 
based on \cite{Blumenhagen:2000fp}
we have discussed Type I string compactifications on non-commutative
asymmetric orbifold spaces.} 
Specifically, we will discuss Type I string compactifications on a product
of $d$ non-commutative two-tori to $10-2d$ non-compact Minkowski dimensions
($d=2,3$),
i.e. the ten-dimensional
background spaces ${\cal M}_{10}$ we are considering have the following form:
\begin{equation}
 \label{backgr}
  {\cal M}_{10}=\mathbb{R}^{1,9-2d}\times {\cal X}_{2d},\qquad
   {\cal X}_{2d}=\prod_{j=1}^d T^2_{(j)}
\end{equation}
Since we assume that the purely internal magnetic $F$-field is block diagonal
and constant for all D$9$-branes:
\begin{equation}\label{blockdiag}
 F_{ab}= \bigoplus_{j=1}^{d} 
     \begin{pmatrix}
       0     & F^{(j)} \\
     -F^{(j)} &   0
     \end{pmatrix}                   
\end{equation}
these boundary conditions are T-dual to D$(9-d)$-branes which have an
angle of 
\begin{equation}
 \phi^{(j)}= \arctan (F^{(j)}+B^{(j)})
\end{equation}
wrt.\ the $X^{(j)}$ axis. (T-duality is performed along the $Y$-direction:
  $R_{Y}/\sqrt{\al}\to \sqrt{\al} R_{Y}$.) Applying the 
 results of chapter \ref{strbg} we find that the open string coordinates 
of the D$9$-branes
fulfill  the following equal-$\tau$ commutation relation in the $F$ (= {\it flux})
-picture:
\begin{equation}
 \label{commutator} 
  [ X_{10-2j}(\tau,\sigma) ,X_{11-2j}(\tau,\sigma) ]
  \bigr|_{\sigma\in\partial{\cal M}}=i\Theta^{(j)}\, ,
   \qquad j=1,\dots , d
\end{equation}
The non-commutative deformation parameter $\Theta^{(j)}$ 
in eq.\ \eqref{commutator} is defined by:
\begin{equation}
 \label{defpar}
 \Theta^{(j)}\equiv-2\pi\al\frac{  F^{(j)}+B^{(j)}}{1+(F^{(j)}+B^{(j)})^2}
\end{equation} 
The entire internal non-commutative torus will actually consist out of 
different
sectors with different non-commutative deformation parameters, because
we will introduce several D9-branes  with different magnetic fluxes.
We will show that the spectrum of open strings,
with mixed boundary
conditions in the
internal directions is generically chiral, breaks space-time supersymmetry and
leads to gauge groups of lower rank.
It is however important to stress
that
the effective gauge theories in the uncompactified part of space-time
are still commutative, and therefore are Lorentz invariant and local
field theories.

This construction is the D-brane extended version of 
\cite{Bachas:1995ik}, where
it was already observed that turning on magnetic flux in a toroidal
Type I compactification leads to supersymmetry breaking and 
chiral massless spectra in four space-time dimensions. 
However, the consistency conditions for such models were derived
in the effective non-supersymmetric gauge theories, leaving the
actual string theoretic conditions an open issue. 
We will show that, with all the insights gained  
in the description of D-branes with magnetic flux, 
we are now able to achieve a complete string theoretic understanding, 
giving rise to certain extensions and modifications of the purely 
field theoretical analysis.  
As a solution to the tadpole cancellation conditions we can get
different sectors of D-branes with different magnetic fluxes, corresponding
to different non-commutative boundary conditions.  
Chirality then arises in sectors of open strings which have ends on branes 
with different gauge flux, while the presence of any solitary flux is 
not sufficient. The gauge groups that act on the D-branes with non-vanishing 
flux are unitary instead of orthogonal or symplectic in accord with the 
general statement that only these are compatible with a non-commutative 
deformation of the coordinate algebra.

As already mentioned, it is sometimes very helpful to employ an  
equivalent T-dual
description, where the background fields vanish and the 
torus is entirely commutative, but
the D$(9-d)$-branes intersect at various different angles 
\cite{Blumenhagen:1999md,Blumenhagen:1999ev,Blumenhagen:1999db}. 
This description allows to
present  a more intuitive picture of the open string sector
involved in such models.  
Chiral fermions then arise
due to the nontrivial geometric boundary conditions of the intersecting 
D-branes,\footnote{The appearance of {\sl chiral} fermions
at intersections of angled D-branes
was discovered in \cite{Berkooz:1996km}.}
 which at the same time generically break space-time supersymmetry
\footnote{However if the angles $\Delta \phi_j$ 
of the branes fulfill special conditions
 e.g.\ $\sum_{j=1}^d\Delta \phi_j=0$ supersymmetry is preserved. It will turn
 out that requiring RR-tadpole cancellation in purely toroidal 
$\bar{\sigma}\Omega$ orientifolds has only supersymmetric solutions with
 $\phi_j=0\;\forall  j$. This excludes chirality.}
and lower the rank of the gauge group. 

The chapter is organized as follows. In the next section we analyze the
one-loop amplitudes and the resulting tadpole cancellation conditions
for D9-branes with mixed Neumann-Dirichlet boundary conditions 
moving in the background of $d$
two-dimensional tori ($d=2,3$). In section \ref{sixdimsec} we discuss specific
six-dimensional models ($d=2$) working out the non-supersymmetric,
chiral spectrum. We also point out some subtleties 
involving the mechanisms of supersymmetry breaking in `nearly' supersymmetric 
brane configurations. 
In section \ref{fourdimsec} we move on to chiral, non-supersymmetric
four-dimensional models ($d=3$), reconsider in particular the model 
presented in \cite{Bachas:1995ik} with GUT-like gauge group 
$G=U(5)\times U(3)\times U(4) \times U(4)$ and display another 
4 generation model with `Standard Model' gauge group
$G=U(3)\times U(2)\times U(1)^r$.\footnote{For other recent 
bottom up attempts
to obtain GUTs and the Standard Model from branes see \cite{Antoniadis:2000en,Aldazabal:2000sa,Krause:2000gp}.}
Some phenomenological problems of this model are stressed at the end. 
This chapter is organized as follows. In the next section we analyze the
one-loop amplitudes and the resulting tadpole cancellation conditions
for D9-branes with mixed Neumann-Dirichlet boundary conditions 
moving in the background of $d$
two-dimensional tori ($d=2,3$). 

\section{One loop amplitudes}
In \cite{Bachas:1995ik} it was observed that turning on magnetic flux in a toroidal
Type I compactification leads to supersymmetry breaking and in general
to chiral massless spectra in four space-time dimensions. 
The consistency conditions for such models were derived
in the effective non-supersymmetric gauge theory but not in the full 
string theory. 
In this section we will show that, with the inclusion of
D-branes with magnetic flux, respectively D-branes
at angles, we are now able to derive the string theoretic tadpole
cancellation conditions. 

\subsection{\label{d9baneswflux}D9-branes with magnetic fluxes}

As our starting point we consider
the orientifold
\begin{equation}
 \label{oria}   
    \frac{\text{ Type IIB  on } T^{2d}}{ \Omega }
\end{equation}
In the following we will assume that $T^{2d}$ splits into
a direct product of $d$ two-dimensional tori $T^2_{(j)}$ 
with coordinates $X^{(j)}_1$, $X^{(j)}_2$ and radii $R^{(j)}_1$, 
$R^{(j)}_2,\ j=1, \dots ,d$.  
We restrict ourselves to purely imaginary
 complex structures 
and 
vanishing antisymmetric NSNS tensor field $B$.\footnote{As noted 
in chapter \ref{toruscomp}, eq.\ \eqref{quantizedB} a quantized $B$-field
would allow  $\Omega$ to be a symmetry. The case of non-vanishing NS
 $B$-fields  on some $T^2$ and its T-dual interpretation was considered 
in \cite{Blumenhagen:2000ea}.}
Turning on magnetic flux $F_{ab}$ on 
a D9-brane changes the pure Neumann 
boundary conditions into mixed Neumann-Dirichlet conditions.
This case was investigated in the preceding chapter(s).
Especially we gave formul\ae\, for the masses of the momentum modes
(eq.\ \eqref{canopstrmass}, p.\ \pageref{canopstrmass}). In addition we
gave a formula for the zero mode degeneracy (eq.\ \eqref{landaudeg}). 
In chapter \ref{strbg} we related the modings of the oscillator modes
to the Eigenvalues of a matrix constructed form the two $U(1)$ fields
coupling to the endpoints of the string. However here we will use the
``classical'' derivation of the solution given by \cite{Abouelsaood:1987gd}.
The boundary conditions for strings with $F$-field are given by 
(cf.\ \eqref{sigmanulld}, p. \pageref{sigmanulld}):
\begin{equation}
    \begin{array}{cc}  \partial_\sigma X_1 + \cot\phi_i\, 
                      \partial_\tau X_2 &=0                             \\
                      \partial_\sigma X_2 - \cot\phi_i\, \partial_\tau X_1
                      &=0 
    \end{array}\quad\text{ with }\quad
    \bigg\{\begin{array}{cc}
          i=1&\text{ and }\sigma=0 \\
          i=2 &\text{ and }\sigma=\pi
    \end{array}
\end{equation}
The angles $\phi_i$ are related to the $F$-fields that couple to the 
string via:
\begin{align}
   {\cal F}_1&= \begin{pmatrix} 
                0 & -\cot\phi_1  \\ \cot\phi_1 & 0
               \end{pmatrix} 
   & {\cal F}_2&= \begin{pmatrix} 
                0 & \cot\phi_2  \\ -\cot\phi_2 & 0
               \end{pmatrix} 
\end{align}
The mode expansion that solves these boundary conditions is
\begin{equation}\label{modeh}
    \begin{aligned} X_1=x_1 -&
    \sqrt{\alpha'} \sum_{n\in\mathbb{Z}}  {\alpha_{n+\nu}\over n+\nu} 
           e^{-i(n+\nu) \tau} \sin[(n+\nu)\sigma +\phi_1] - \\
    &\sqrt{\alpha'} \sum_{m\in\mathbb{Z}}  {\alpha_{m-\nu}\over m-\nu} 
           e^{-i(m-\nu) \tau} \sin[(m-\nu)\sigma -\phi_1], \\
    X_2=x_2 +&
    i \sqrt{\alpha'} \sum_{n\in\mathbb{Z}}  {\alpha_{n+\nu}\over n+\nu} 
           e^{-i(n+\nu) \tau} \sin[(n+\nu)\sigma +\phi_1] -\\
    &i \sqrt{\alpha'} \sum_{m\in\mathbb{Z}}  {\alpha_{m-\nu}\over m-\nu} 
           e^{-i(m-\nu) \tau} \sin[(m-\nu)\sigma -\phi_1]
    \end{aligned}
\end{equation}
We will not review the mass formul\ae. Instead of working with D9-branes 
with  various magnetic fluxes, we will now use 
the T-dual description in terms of D-branes at angles
\cite{Blumenhagen:1999md,Blumenhagen:1999ev,Blumenhagen:1999db}, 
which  allows to
present a more intuitive picture of the open string 
sector involved in such models: 

\subsection{D$(9-d)$-branes at angles}   
Applying a T-duality in all $X^{(j)}_2$ 
directions (which is a special version of the $D$-duality \eqref{z2duality},
  p.~\pageref{z2duality})
\begin{equation}
  \nonumber 
   R^{(j)}_2 \rightarrow R^{(j)'}_2 = 1/ R^{(j)}_2 
\end{equation}
leads to boundary conditions
for D$(9-d)$-branes intersecting 
at angles, where the angle of the D$(9-d)$-brane relative to
the $X^{(j)}_1$ axes is given by 
\begin{equation}
 \label{angles}
 \tan\phi^{(j)}=F^{(j)}
\end{equation} 
(In the following we will omit the prime on
the dual radii.)
This T-duality also maps $\Omega$ onto $\bar{\sigma}\Omega$, 
where $\bar{\sigma}$ acts as complex conjugation on all the $d$  
complex coordinates along the $T^2_{(j)}$ tori. 
Thus, instead of \eqref{oria} we are considering the orientifold
\begin{equation}\label{orib}
    \frac{{\rm Type\, II}\ {\rm on}\ T^{2d}}{\bar{\sigma}\Omega}  
\end{equation}
For $d$ even we have to take Type IIB
and for $d$ odd Type IIA, as explained in section \ref{wsparityt2}. 
Note that after performing this T-duality
transformation the internal coordinates  are completely commutative.

Let $j\in\{1,\ldots,d\}$ again 
label the $d$ different two-dimensional tori
and $a\in\{1,\ldots,K\}$ the different kinds of 
D$(9-d)$-branes, which are distinguished by different angles 
on at least one torus. Moreover,
we are only considering branes which do not densely cover any of the
two-dimensional tori. Thus, 
the position of a D$(9-d)$-brane is described 
by two sets of integers $(n^{(j)}_a,m^{(j)}_a)$, 
labeling how often
the D-branes are wound around the two fundamental cycles of each $T^2_{(j)}$. 
The angles of such a brane with the axes $X_1^{(j)}$ are given by
(cf.~figure \ref{toruswbrane}, p.\pageref{toruswbrane},
were we assumed the lattice vector $e_1$ to
be parallel to the $\mathfrak{Re}$-axis.): 
\begin{equation}
 \label{angle} 
   \cot(\phi^{(j)})=\frac{n^{(j)}+m^{(j)}\tau_1^{(j)}}{m^{(j)}
   \tau_2^{(j)}}
                 = \cot(\alpha)+\frac
                    {n^{(j)}_a R_1^{(j)}}{m^{(j)} \sin(\alpha) R_2^{(j)} }
\end{equation}
$\alpha$ is the angle between the two generating lattice vectors:
$\mathfrak{Re}(\tau)=\cos(\alpha) R_2^{(j)}$, $\mathfrak{Im}(\tau)
=\sin(\alpha) R_2^{(j)}$.
These conventions are shown in figure \ref{toruswbrane},
p. \pageref{toruswbrane}, where we have omitted the index $(j)$ which labels
the different two-tori $T^2_{(j)}$. 
However
in order for $\bar{\sigma}$ to be a symmetry the complex structure is 
fixed either to be purely imaginary or to have real part $\tau_1=1/2$ 
(cf.~section \ref{wsparityt2}, p. \pageref{wsparityt2}).
The open string boundary condition is easily derived:
\begin{align}\label{bdycondsangles}
   \partial_- X(\tau,\sigma)&={\cal R}\bigl(\vec{\phi}\bigr)\partial_+
                       X (\tau,\sigma)\bigr|_{\sigma\in\partial{\cal M}} &
   {\cal R}&=D\bigl(\vec{\phi}\bigr)^T \bar{\sigma}D\bigl(\vec{\phi}\bigr)
\end{align}  
$D(\vec{\phi})$ is a rotation described by a set of  angles 
$\vec{\phi}=(\phi^{(1)},\ldots,\phi^{(d)})$. In our case $D(\vec{\phi})$ is
block-diagonal, each $2\times 2$ block acting on a $T^2_{(j)}$.
In all cases considered in this thesis ${\cal R}$ is also block diagonal 
(except for chapter \ref{strbg}, where our considerations are more general). 
In orthogonal coordinates each block is of the form:
\begin{equation}
 {\cal R}^{(j)} \bigl(\phi^{(j)}\bigr) 
  = \begin{pmatrix}
   \cos \bigl(2\,\phi^{(j)}\bigr) &- \sin \bigl(2\,\phi^{(j)}\bigr) \\
   -\sin \bigl(2\,\phi^{(j)}\bigr)
    & \rule[0ex]{0.ex}{2.8ex}-\cos \bigl(2\,\phi^{(j)}\bigr)
    \end{pmatrix}
\end{equation}
In complex coordinates (i.e.\ $Z=1/\sqrt{2}(X_1+iX_2)$) 
${\cal R}^{(j)}\bigl(\phi^{(j)}\bigr)$ this
 combination of a reflection and a subsequent rotation by
$-2\,\phi^{(j)}$ looks very simple:
\begin{align}
 {\cal R}^{(j)} \bigl(\phi^{(j)}\bigr)  \bigl(Z^{(j)}\bigr) 
  &= e^{-2\phi^{(j)}} \bar{Z}^{(j)} 
  & 
 {\cal R}^{(j)} \bigl(\phi^{(j)}\bigr)  \bigl(\bar{Z}^{(j)}\bigr) 
  &= e^{2\phi^{(j)}} Z^{(j)} 
\end{align} 
Since $\bar{\sigma}\Omega$ reflects the D-branes at the axis $X^{(j)}_1$,
for each brane labeled by
$\bigl(n^{(j)}_a,m^{(j)}_a\bigr)$ we must also introduce the mirror brane
with $\bigl(n^{(j)}_{a'},m^{(j)}_{a'}\bigr)=$\\
 $\bigl(n^{(j)}_a,-m^{(j)}_a\bigr)$. 
The values $m^{(j)}_a =0 \neq n^{(j)}_a$ 
and $n^{(j)}_a =0 \neq m^{(j)}_a$ correspond to branes located along
one of the axis. The horizontal D-branes  translate via T-duality 
into D9-branes with vanishing 
flux and the vertical ones into branes of lower dimension with pure
Dirichlet boundary conditions.     
A solution to the boundary conditions \eqref{bdycondsangles} was given in 
section \ref{onedimbsec}, p.\ \pageref{onedimbsec} and 
section  \ref{coordcommutesec}, p.\ \pageref{coordcommutesec}.
   
The questions we are going to deal with in the following 
are: Is it possible to cancel 
all or at least the RR tadpoles originating from the Klein bottle amplitude 
by D9-branes with non-vanishing magnetic fluxes $F^{(j)}$,
or equivalently by D$(9-d)$-branes at nontrivial angles $\phi^{(j)}$?  
Taken that supersymmetry is broken generically by such a background, are 
there configuration which still preserve some amount of supersymmetry? 
This would provide a string scenario with partial supersymmetry 
breaking. 
Finally, what are the phenomenological properties of such compactifications?
Concerning the first question we find a positive answer in the sense that 
the RR tadpole can 
be canceled, while supersymmetry is always broken entirely.
This shows up both in the spectrum and a  non-vanishing NSNS tadpole.
Tachyons are always present in compactifications
on $T^4$ and for some region in parameter space 
of the four dimensional compactifications as well.
However it was shown after our publication appeared
(\cite{Aldazabal:2000dg,Ibanez:2001nd,Rabadan:2001mt}) 
that there exist toroidal
models, which are  free from tachyons (but still suffering from
NSNS tadpoles).
 Interestingly,   models with non-trivially intersecting 
D-branes generically contain
chiral fermions  motivating us to study how far one can get 
in deriving the Standard Model in this setting.  
However, later we will mention an 
obstacle to construct phenomenologically 
realistic models in this simple approach on a $T^6=(T^2)^3$ which
is a product of only $\bf A$-type two-tori.

Technically we first have to compute all contributions 
to the massless RR tadpole. The cancellation conditions will then imply 
relations for the number of D9-branes and their respective background fluxes. 
This computation will be performed in the T-dual picture, where 
D9-branes with background fields are mapped to D$(9-d)$-branes, and the 
background fields translate into relative angles. This picture 
allows to visualize the D-branes  easily and gives a much better 
intuition than dealing with sets of D9-branes, all filling the same space 
but differing by background fields.

\subsection{Klein bottle amplitude}

The loop channel Klein bottle amplitude for \eqref{orib} can be computed
straightforwardly
\begin{equation}
 \label{klein}
   {K}=2^{(5-d)} c\, (1-1) \int_0^\infty {dt\over t^{(6-d)}} 
         {1\over 4}
          {\thef{0}{1/2}^4 \over \eta^{12} } \prod_{j=1}^d 
         \Biggl(
         \sum_{r,s\in\mathbb{Z}} e^{-\pi t \left( 
               {r^2\big/\big(R^{(j)}_1\big)^2 } +
              s^2 \big(R^{(j)}_2\big)^2 \right) } 
         \Biggr) 
\end{equation}
with $c= {V_{10-2d}}/\left( 8\pi^2 \alpha' \right)^{5-d}$. 
Transforming \eqref{klein} into tree channel, one obtains the
following massless RR tadpole
\begin{equation}
 \label{tadkl}
   \int_0^\infty  dl\, 2^{(13-d)}  \prod_{j=1}^d \left(
             {R^{(j)}_1\over R^{(j)}_2} \right) .
\end{equation}
The tree channel Klein bottle amplitude allows to determine the 
normalization of the corresponding cross-cap states
\begin{equation}
 \label{crosscap} |C\rangle={ 2^{({d/2}-4)} }  
        \Bigg(\prod_{j=1}^d {R^{(j)}_1\over R^{(j)}_2}\Bigg)^{\frac{1}{2}}
           \left(
            |C_{\rm NS}\rangle+ |C_{\rm R}\rangle\right) 
\end{equation}
\subsection{Annulus amplitude}

Next we calculate all contributions of open strings stretching between 
the various D$(9-d)$-branes, generically located at nontrivial relative 
angles. We will both  include the case, where the relative angle 
is vanishing, i.e. the background gauge flux is equal on both branes, 
and the case, where the angle is $\pi/2$ and 
the field gets infinitely large on, say, $p$ of the tori. 
 
We start with the contributions of strings with both ends on the same 
brane. 
The T-dual of the  
Kaluza-Klein and winding spectrum in eq.\ \eqref{specb} (p.\ \pageref{specb})
is given by  \eqref{canopstrmassanglesb} and \eqref{canopstrmassanglesb}
(p.\ \pageref{canopstrmassanglesb}). With our simplifying assumptions that
$B=0$ it reads:\footnote{$B\neq 0$ does not modify the the RR tadpole
conditions, as they are topological. It also does not change the moding of
the oscillators, as it does not enter the boundary conditions. Therefore
$B=0$ is a very mild simplification.}
\begin{equation}
 \label{kwind}
 M_a^2=\sum_{j=1}^d \left( \Bigg( {r^{(j)}_a 
        \over V^{(j)}_a} \Bigg)^2 
        + \big(s^{(j)}_a \big)^2 \left( {R^{(j)}_1 R^{(j)}_2  
        \over V^{(j)}_a} \right)^2 \right)
\end{equation}
with
\begin{equation} 
 \label{lenght}
    V^{(j)}_a=\sqrt{ \Big(R^{(j)}_1 n^{(j)}_a\Big)^2 
                 +\Big(R^{(j)}_2 m^{(j)}_a\Big)^2 }
\end{equation}
denoting the volume of the brane on $T^2_{(j)}$.
It is now straightforward to compute the loop channel annulus amplitude
for open strings starting and ending on the same brane 
and transform it to the tree channel
\begin{equation}
 \label{anntree} \tilde{A}_{aa}=c\, N_a^2 (1-1) \int_0^\infty dl\, 
      {1\over 2^{(d+1)}} \prod_{j=1}^d {\Big(V^{(j)}_a\Big)^2 
\over R^{(j)}_1 R^{(j)}_2}\ 
            {\thef{1/2}{0}^4\over \eta^{12}}
        \sum_{r,s} e^{-\pi l \widetilde{M}^2_a} 
\end{equation}
with 
\begin{equation}
 \label{mtild} 
   \widetilde{M}_a^2=\sum_{j=1}^d 
      \left( {\big(r^{(j)}_a\big)^2 \Big(V^{(j)}_a\Big)^2}
        + \big(s^{(j)}_a \big)^2
             \left( {V^{(j)}_a \over  R^{(j)}_1 R^{(j)}_2 }
     \right)^2  \right)
\end{equation}
$N_a$ counts the numbers of different kinds of branes.      
Using \eqref{anntree} one can determine the normalization of the boundary
state, which has the schematic form 
\begin{equation}
 \label{boundmb} 
   \ket{D_a}={2^{-({d/2}+1)}}  
        \left(\prod_{j=1}^d {V^{(j)}_a
              \over \sqrt{R^{(j)}_1 R^{(j)}_2}}\right)\left(
     \ket{D_{a,{\rm NS}}}+ \ket{D_{a,{\rm R}}}\right) 
\end{equation}
Reflecting the brane on a single $T^2_{(j)}$ by a $\pi$ rotation onto itself 
corresponds to $(n^{(j)}_a,m^{(j)}_a)\to (-n^{(j)}_a,-m^{(j)}_a)$ 
and, as can be determined in the boundary state approach, 
changes the sign of the RR charge, thus exchanging branes and anti-branes. 

Using the  boundary state \eqref{boundmb} we can compute the tree channel
annulus amplitude for an open string stretched between two different
D-branes                                
\begin{multline}
 \label{anndiv}
   \tilde{A}_{ab}=\int_0^\infty  dl\,
      \langle D_a| e^{-lH_{cl}}| D_b\rangle \\
     =\frac{c}{2} N_a N_b  I_{ab} \int_0^\infty dl 
        (-1)^d\sum_{\alpha,\beta\atop\in\{0,\oh\}} 
             (-1)^{2(\alpha+\beta)}  
            { \thef{\alpha}{\beta}^{4-d}\,   \prod_{j=1}^{d}
                \thef{\alpha}{\Delta(\phi^{(j)})_{ab}+\beta}
                \rule[-2ex]{0ex}{0ex}\over
                  \eta^{12-3d}\,  \prod_{j=1}^{d}
                \thef{1/2}{\Delta(\phi^{(j)})_{ab}+1/2} } 
\end{multline}
where the coefficient
\begin{equation}
 \label{coeff}
     I_{ab}=\prod_{j=1}^d
           \left( n^{(j)}_a m^{(j)}_b- m^{(j)}_a n^{(j)}_b\right) 
\end{equation}
is the (oriented) {\it intersection number} of the two branes.
We have defined the oriented angle between brane $a$ and $b$ on the
torus $T^2_{(j)}$ by:
\be
 \Delta(\phi^{(j)})_{ab}\equiv \bigl(\phi^{(j)}_b-\phi^{(j)}_a\bigr)/\pi
\ee
It gives rise to an extra multiplicity in the annulus loop channel, which
we have to take into account, when we compute the massless spectrum. 
In order to properly include the case where some 
$\phi^{(j)}_a=\phi^{(j)}_b$, 
one needs to employ the relation 
\begin{equation}
 \label{limit}
    \lim\limits_{\psi \rightarrow 0}
             {2 \sin (\pi \psi)  
             \over \thef{1/2}{1/2+\psi}} = -{1 \over \eta^3}
\end{equation}
and include a sum over KK momenta and windings as in \eqref{anntree}.
The contribution to the massless RR tadpole due to \eqref{anntree} and 
\eqref{anndiv} is
\begin{equation}
 \label{tadann} \int_0^\infty dl\, N_a N_b\, 2^{(3-d)} 
           \,\prod_{j=1}^d
                 \frac{ \left(R^{(j)}_1 \right)^2 n^{(j)}_a n^{(j)}_b + 
                       \left(R^{(j)}_2\right)^2 m^{(j)}_a m^{(j)}_b} 
                       {R^{(j)}_1 R^{(j)}_2 } 
\end{equation}
The loop channel annulus can be obtained by a modular transformation:
\begin{multline}
  \label{annloop}
      {A}_{ab}=c\ N_a N_b\, I_{ab}
      \int_0^\infty  {dt\over t^{(6-d)}}\, \frac{1}{ 4}\cdot  
       \frac{1}{\eta^{12-3d}}  \\
       \cdot \sum_{\alpha,\beta\in\{0,1/2\}} (-1)^{2(\alpha+\beta)}
        \frac{
            \thef{\beta}{\alpha}^{4-d}\,   
          \prod_{j=1}^{d}e^{i2\pi(\oh-\alpha) \Delta(\phi^{(j)})_{ab}}
               \thef{\Delta(\phi^{(j)})_{ab}+\beta}
                      {{\alpha }}\rule[-2ex]{0ex}{0ex}}
            {  \prod_{j=1}^{d}
               \thef{\Delta(\phi^{(j)})_{ab}+1/2}{1/2} } 
\end{multline} 

\subsection{M\"obius  amplitude}

Computing the overlap between the crosscap state \eqref{crosscap}
and a boundary
state \eqref{boundmb} yields  the contribution of the brane D$(9-p)_a$
to the M\"obius amplitude 
\begin{multline} 
 \label{moediv}
    \widetilde{M}_{a}=
            \mp c\, N_a\, 2^5\, (-1)^{d}\, 
                \int_0^\infty dl\,
                  {\prod_{j=1}^d  m^{(j)}_a}  \\
          \cdot \sum_{\alpha,\beta\in\{0,1/2\}}  (-1)^{2(\alpha+\beta)}  
                 { \thef{\alpha}{\beta}^{4-d}\,   \prod_{j=1}^{d}
               \thef{\alpha}{\phi^{(j)}_a/ \pi+\beta}
                \rule[-2ex]{0ex}{0ex}\over
                 \eta^{12-3d}\,  \prod_{j=1}^{d}
               \thef{1/2}{\phi^{(j)}_a/\pi+1/2} }
\end{multline} 
with argument $q=-\exp (-4\pi l)$.
Therefore the contribution to the RR tadpole is
\begin{equation}
  \label{tadmoe}
    \mp \int_0^\infty dl\, N_a\,  2^{(9-d)} 
           \,\prod_{j=1}^d \left( {R^{(j)}_1\over R^{(j)}_2} n^{(j)}_a 
           \right)
\end{equation}
The overall sign in \eqref{moediv} and \eqref{tadmoe} is fixed by the tadpole
cancellation condition.
In the loop channel the contribution of the M\"obius strip results from 
strings starting on one brane and ending on its mirror partner. 
The extra multiplicity given by the numbers $m^{(j)}_a$ of intersection 
points invariant under $\bar{\sigma}$ needs to be regarded as before.  
Now we have all the ingredients to study the relations which derive from 
the cancellation of massless RR tadpoles.

\section{\label{sixdimsec} Compactifications to six dimensions}

\begin{figure}[t]
 \begin{center}
 \begin{picture}(0,0)%
\includegraphics{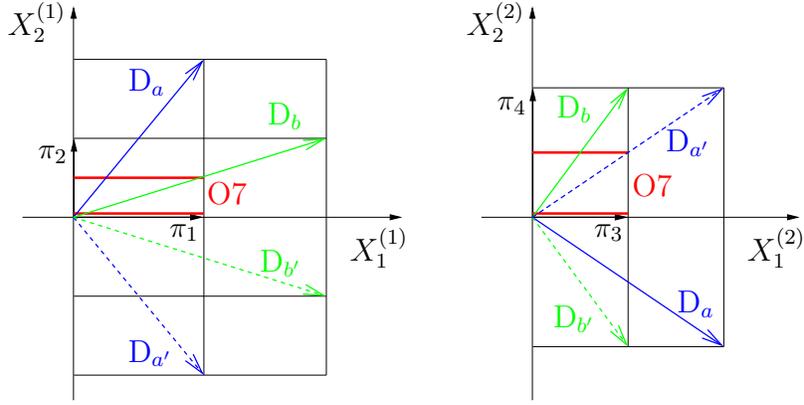}%
\end{picture}%
\setlength{\unitlength}{2072sp}%
\begingroup\makeatletter\ifx\SetFigFont\undefined%
\gdef\SetFigFont#1#2#3#4#5{%
  \reset@font\fontsize{#1}{#2pt}%
  \fontfamily{#3}\fontseries{#4}\fontshape{#5}%
  \selectfont}%
\fi\endgroup%
\begin{picture}(9457,4734)(-899,-7618)
\put(-899,-3211){\makebox(0,0)[lb]{\smash{\SetFigFont{12}{14.4}{\rmdefault}{\mddefault}{\updefault}{\color[rgb]{0,0,0}$X_2^{(1)}$}%
}}}
\put(4591,-3211){\makebox(0,0)[lb]{\smash{\SetFigFont{12}{14.4}{\rmdefault}{\mddefault}{\updefault}{\color[rgb]{0,0,0}$X_2^{(2)}$}%
}}}
\put(6121,-5631){\makebox(0,0)[lb]{\smash{\SetFigFont{11}{13.2}{\rmdefault}{\mddefault}{\updefault}{\color[rgb]{0,0,0}{$\pi_3$}}%
}}}
\put(4956,-4111){\makebox(0,0)[lb]{\smash{\SetFigFont{11}{13.2}{\rmdefault}{\mddefault}{\updefault}{\color[rgb]{0,0,0}{$\pi_4$}}%
}}}
\put(1036,-5631){\makebox(0,0)[lb]{\smash{\SetFigFont{11}{13.2}{\rmdefault}{\mddefault}{\updefault}{\color[rgb]{0,0,0}{$\pi_1$}}%
}}}
\put(-534,-4696){\makebox(0,0)[lb]{\smash{\SetFigFont{11}{13.2}{\rmdefault}{\mddefault}{\updefault}{\color[rgb]{0,0,0}{$\pi_2$}}%
}}}
\put(7951,-5911){\makebox(0,0)[lb]{\smash{\SetFigFont{12}{14.4}{\rmdefault}{\mddefault}{\updefault}{\color[rgb]{0,0,0}$X_1^{(2)}$}%
}}}
\put(3196,-5911){\makebox(0,0)[lb]{\smash{\SetFigFont{12}{14.4}{\rmdefault}{\mddefault}{\updefault}{\color[rgb]{0,0,0}$X_1^{(1)}$}%
}}}
\put(541,-3886){\makebox(0,0)[lb]{\smash{\SetFigFont{12}{14.4}{\rmdefault}{\mddefault}{\updefault}{\color[rgb]{0,0,1}$\text{D}_a$}%
}}}
\put(1486,-5236){\makebox(0,0)[lb]{\smash{\SetFigFont{12}{14.4}{\rmdefault}{\mddefault}{\updefault}{\color[rgb]{1,0,0}O$7$}%
}}}
\put(6571,-5146){\makebox(0,0)[lb]{\smash{\SetFigFont{12}{14.4}{\rmdefault}{\mddefault}{\updefault}{\color[rgb]{1,0,0}O$7$}%
}}}
\put(541,-7126){\makebox(0,0)[lb]{\smash{\SetFigFont{12}{14.4}{\rmdefault}{\mddefault}{\updefault}{\color[rgb]{0,0,1}$\text{D}_{a'}$}%
}}}
\put(2116,-6046){\makebox(0,0)[lb]{\smash{\SetFigFont{12}{14.4}{\rmdefault}{\mddefault}{\updefault}{\color[rgb]{0,1,0}$\text{D}_{b'}$}%
}}}
\put(2206,-4291){\makebox(0,0)[lb]{\smash{\SetFigFont{12}{14.4}{\rmdefault}{\mddefault}{\updefault}{\color[rgb]{0,1,0}$\text{D}_b$}%
}}}
\put(5671,-4201){\makebox(0,0)[lb]{\smash{\SetFigFont{12}{14.4}{\rmdefault}{\mddefault}{\updefault}{\color[rgb]{0,1,0}$\text{D}_b$}%
}}}
\put(5626,-6721){\makebox(0,0)[lb]{\smash{\SetFigFont{12}{14.4}{\rmdefault}{\mddefault}{\updefault}{\color[rgb]{0,1,0}$\text{D}_{b'}$}%
}}}
\put(6976,-4606){\makebox(0,0)[lb]{\smash{\SetFigFont{12}{14.4}{\rmdefault}{\mddefault}{\updefault}{\color[rgb]{0,0,1}$\text{D}_{a'}$}%
}}}
\put(7111,-6541){\makebox(0,0)[lb]{\smash{\SetFigFont{12}{14.4}{\rmdefault}{\mddefault}{\updefault}{\color[rgb]{0,0,1}$\text{D}_a$}%
}}}
\end{picture}
 \caption[D$7$-Brane configuration on  $T^4$.]
        {\label{dsevenbranes}
           D$7$-brane configuration with $\Bar{\sigma}$ images on the $T^4$.
         The orientifold 7 planes are painted in \textcolor{r}{red}. The 
         fundamental one cycles of the torus are denoted by
         $\pi_1\ldots\pi_4$.}
 \end{center}
\end{figure}

We are compactifying Type I strings on a four-dimensional torus and
cancel the tadpoles by introducing stacks 
of D9-branes with magnetic fluxes. 
The T-dual arrangement of D7-branes at angles 
looks like the situation depicted
in figure \ref{dsevenbranes}, where we have drawn only two 
types of D7-branes labeled by 
$a$ and $b$ and their mirror partners $a'$ and $b'$, 
the angles being chosen arbitrary.
\subsection{Six-dimensional models}
The complete annulus amplitude is a sum over all open strings stretched 
between the various D7-branes
\begin{multline}
  \label{anncom}
        \tilde{A}_{\text{tot}}=\sum_{a=1}^K 
        \left( \tilde{A}_{aa} +
         \tilde{A}_{a'a'}+  \tilde{A}_{aa'} +
          \tilde{A}_{a'a}\right) + \\
    \sum_{a<b}  \left( \tilde{A}_{ab} +\tilde{A}_{ba} + 
         \tilde{A}_{a'b'}+ \tilde{A}_{b'a'} + 
    \tilde{A}_{ab'} +\tilde{A}_{ba'}+
          \tilde{A}_{a'b}+\tilde{A}_{b'a} \right) 
\end{multline}  
Using \eqref{anntree} and \eqref{tadann} and adding up all these various 
contributions yields the  following two RR tadpoles: 
\begin{equation}
  \label{tadannsix}
  \int_{0}^\infty dl\, 8\, 
       \frac{R^{(1)}_1 R^{(2)}_1}{R^{(1)}_2 R^{(2)}_2} \left(
            \sum_{a=1}^K N_a\,  n^{(1)}_a n^{(2)}_a
              \right)^2   
      + \int_{0}^\infty dl\, 8\, 
      \frac{R^{(1)}_2 R^{(2)}_2}{R^{(1)}_1 R^{(2)}_1}  \left(
            \sum_{a=1}^K N_a\,  m^{(1)}_a m^{(2)}_a
              \right)^2 \rule[-4.3ex]{0ex}{0ex}
\end{equation}
For the total M\"obius amplitude we obtain the RR tadpole
\begin{equation}
  \label{tadmoeb}
   \mp \int_0^\infty dl\, 2^8\, {R^{(1)}_1 R^{(2)}_1
      \over R^{(1)}_2 R^{(2)}_2}\,
         \sum_{a=1}^K N_a\,   n^{(1)}_a n^{(2)}_a 
\end{equation}
Note, the two special cases of $N_9$ horizontal and  $N_5$ vertical
D7-branes are contained in \eqref{tadannsix} and \eqref{tadmoeb}
 by  setting $N_a=N_9/2$ respectively  $N_a=N_5/2$.
Choosing the minus sign in \eqref{tadmoeb} we get the two RR tadpole 
cancellation conditions
\begin{equation}
 \begin{aligned}
  \label{tadcancel}
   \Biggl(\prod_{j=1}^{2} \tau_2^{(j)}\Biggr)^{-1}
      = {R^{(1)}_1 R^{(2)}_1\over R^{(1)}_2 R^{(2)}_2} 
                &:\quad\quad \quad  &
            \sum_{a=1}^K N_a\, n^{(1)}_a n^{(2)}_a &=16,  \\
     \prod_{j=1}^{2} \tau_2^{(j)}
      =    {R^{(1)}_2 R^{(2)}_2\over R^{(1)}_1 R^{(2)}_1} 
                &: \quad\quad\quad  &
            \sum_{a=1}^K N_a\,  m^{(1)}_a m^{(2)}_a &=0
 \end{aligned}
\end{equation}
As one might have expected, 
pure D9-branes with $m_a^{(j)}=0$ only contribute to the tadpole 
proportional to the product of the inverse imaginary parts
of the complex structures of the two-tori.
 The D5-branes with 
$n^{(j)}_a=0$ are responsible only for the tadpole that
is  proportional to the product of the imaginary parts
of the complex structures.  
Remarkably, by choosing multiple winding numbers, $n^{(j)}_a>1$, 
one can  reduce the rank of
the gauge group. As usual in non-supersymmetric models, 
there remains an uncanceled NSNS tadpole, which
needs to be canceled by a Fischler-Susskind mechanism.
However it is also possible that the equations of motion lead to a degenerate
compactification space (i.e. a degenerate torus). 

In the section \ref{slagsec} we shall show that except  for the trivial 
case, when $m^{(1)}_a=m^{(2)}_a=0$ for all $a$, i.e. vanishing 
gauge flux on all the D9-branes, supersymmetry is
broken and tachyons develop for open strings stretched between different 
branes.  In contrast to the breaking of supersymmetry 
in a brane-antibrane system
these tachyons cannot be removed by turning on Wilson-lines, which is related 
via T-duality to shifting the position of 
the branes by some constant vector. At any 
non trivial angle there always remains an intersection point of two 
D7-branes where the tachyons can localize.    
Also the lowest lying bosonic spectrum depends
on the radii of the torus, which determine the relative angles. 
The zero point energy in the NS sector of a string stretching 
between two different branes is shifted by 
\begin{equation}
  \label{vacenergy} \Delta E_{0,\text{NS}} = {1 \over 2} 
                \sum_{j=1}^d { \phi_a^{(j)} -\phi_b^{(j)} \over \pi}
\end{equation}
using the convention $\phi_a^{(j)} -\phi_b^{(j)} \in (0,\pi/2]$. 
Even assuming a standard GSO projection, the lightest physical state 
can easily be seen to be tachyonic except for the supersymmetric situation 
with $\phi_a^{(1)} -\phi_b^{(1)} = \phi_a^{(2)} -\phi_b^{(2)}$. 
We shall find in the next section that tadpole cancellation prohibits this 
solution, except when all fluxes vanish.   
 
On the contrary, 
the chiral fermionic massless spectrum is independent of the moduli and 
we display it in table \ref{spectab6d}.
\begin{table} 
  \renewcommand{\arraystretch}{1.2}
 \begin{center}
 \begin{tabular}{|c|c|l|}
 \hline
 Spin & Representation & \multicolumn{1}{c|}{Multiplicity} \\
      & (gauge group)  &              \\
 \hline\hline
  &   &  
   $2 m^{(1)}_a m^{(2)}_a (n^{(1)}_a n^{(2)}_a +1)$ 
     \rule[2.8ex]{0ex}{0ex}\\
   \raisebox{1.5ex}[-1.5ex]{$(1,2)$} 
    & \raisebox{1.5ex}[-1.5ex]{${\bf A}_a+{\bf \Bar{A}}_a$}
     &  $=    \tfrac{1}{2} (I_{a\,a'}+ I_{a\,\text{O}7}$)
        \rule[2.8ex]{0ex}{0ex}\\
   \hline
   &   &
   $2 m^{(1)}_a m^{(2)}_a (n^{(1)}_a n^{(2)}_a -1)$ 
    \rule[2.8ex]{0ex}{0ex}
    \\  
   \raisebox{1.5ex}[-1.5ex]{$(1,2)$} & 
     \raisebox{1.5ex}[-1.5ex]{${\bf S}_a+{\bf \Bar{S}}_a$}
     &  $=   \tfrac{1}{2} (I_{a\,a'}- I_{a\,\text{O}7})$  \\
  \hline 
   $(1,2)$ & $({\bf N}_a,\overline{\bf N}_b)
     +(\overline{{\bf N}}_a,{\bf N}_b)$ & 
   $ I_{a\,b} $\\
   $(1,2)$ & $({\bf N}_a,{\bf N}_b)+
   (\overline{\bf N}_a,\overline{\bf N}_b)$ & $I_{a\,b'}$
             \\ 
 \hline
 \end{tabular}
 \caption[ Chiral 6D massless open string spectrum]
         {\label{spectab6d}  Chiral 6D massless open string spectrum.
          The intersection form $I$ is introduced in section \ref{slagsec}.}
 \end{center}
\end{table} 
(${\bf A}_a$ and ${\bf S}_a$ denote the antisymmetric resp.\ symmetric 
tensor representations with respect to $U(N_a)$, $SO(N_a)$ or $Sp(N_a)$.)
Since $\Bar{\sigma}\Omega$ exchanges 
a brane with its mirror brane, the Chan-Paton indices of strings ending on 
a stack of branes 
with non-vanishing gauge flux have no $\Omega$ projection and the gauge 
group is $U(N_a)$. If $\Bar{\sigma}\Omega$ leaves branes invariant,
 i.e. the 
flux vanishes or is infinite, corresponding to pure D9- or D5-branes, 
the gauge factor is $SO(N_a)$ or $Sp(N_a)$, respectively.

The degeneracy of states stated in the third column of table \ref{spectab6d} 
is essentially given by the intersection numbers
of the D7-branes. 
Whenever it is formally negative, one has to pick the $(2,1)$ spinor 
of opposite 
chirality taking into  account the opposite orientation of the branes at 
the intersection. As was pointed out earlier, a change of the orientation 
switches the RR charge in the tree channel translating  into the 
opposite GSO projection in the loop channel.
Therefore  the other chirality survives the GSO projection in the 
R sector. 
If the multiplicity is zero, this does not mean
that there are no massless open string states in this sector, it only
means that the spectrum is not chiral. This happens precisely  
when  two branes lie on top of each other in one of the two $T^2_{(j)}$ tori. 
Then the extra zero modes give rise to an extra spinor state of opposite 
chirality.
The chiral spectrum shown in table \ref{spectab6d}  does indeed cancel the irreducible
$R^4$ and $F^4$ anomalies.

We have also considered a $\mathbb{Z}_2$ orbifold background, together 
with non-vanishing magnetic flux, which changes the second 
condition in \eqref{tadcancel} to 
\begin{equation}
 \begin{aligned}
 \label{tadcancelorb}
     \frac{R^{(1)}_2 R^{(2)}_2}{R^{(1)}_1 R^{(2)}_1} &:& \qquad
       \sum_{a=1}^K N_a\,  m^{(1)}_a m^{(2)}_a &=16 
 \end{aligned}
\end{equation}
and leads to a projection $SO(N_a),\ Sp(N_a) \rightarrow U(N_a /2)$ 
on pure D9- and D5-branes but no further changes on D9-branes 
with non-vanishing flux. 
In this background it appears to be possible to construct also 
supersymmetric models \cite{Angelantonj:2000hi}.

\section{\label{fourdimsec}Four dimensional models}

The completely analogous computation as in six dimensions 
can be performed for the compactification of Type I strings on a 6-torus 
in the presence of additional gauge fields. 
Now we cancel the tadpoles by D9-branes with magnetic fluxes
on all three 2-tori respectively, in the T-dual picture, 
by D6-branes at angles. 
One obtains four independent tadpole cancellation conditions
\begin{equation}
  \label{tadcancelf}
 \begin{aligned} 
  \Biggl(\prod_{j=1}^{3} \tau_2^{(j)}\Biggr)^{-1}
      = {R^{(1)}_1 R^{(2)}_1 R^{(3)}_1\over R^{(1)}_2 R^{(2)}_2 R^{(3)}_2}
            &: &\qquad
       \sum_{a=1}^K N_a\, n^{(1)}_a n^{(2)}_a n^{(3)}_a &=16  \\
   \Bigl(\tau_2^{(1)}\Bigr)^{-1}\prod_{j=2}^{3} \tau_2^{(j)}
   ={R^{(1)}_1 R^{(2)}_2 R^{(3)}_2\over R^{(1)}_2 R^{(2)}_1 R^{(3)}_1} 
            &: &\qquad
       \sum_{a=1}^K N_a\,  n^{(1)}_a m^{(2)}_a m^{(3)}_a &=0  \\
   \tau_2^{(1)}\Bigl(\tau_2^{(2)}\Bigr)^{-1}\tau_2^{(3)}
   ={R^{(1)}_2 R^{(2)}_1 R^{(3)}_2\over R^{(1)}_1 R^{(2)}_2 R^{(3)}_1} 
            &: &\qquad
       \sum_{a=1}^K N_a\,  m^{(1)}_a n^{(2)}_a m^{(3)}_a &=0 \\
    \prod_{j=1}^{3} \tau_2^{(j)}
   ={R^{(1)}_2 R^{(2)}_2 R^{(3)}_1\over R^{(1)}_1 R^{(2)}_1 R^{(3)}_2} 
             &: &\qquad
       \sum_{a=1}^K N_a\,  m^{(1)}_a m^{(2)}_a n^{(3)}_a &=0
  \end{aligned} 
\end{equation} 
(For convenience they are  given in the  picture with D6-branes at
angles.) 
Again the gauge group contains a $U(N_a)$ factor for each stack of 
D9-branes with non-vanishing flux, an $SO(N_a)$ gauge factor for 
a stack with vanishing flux and an $Sp(N_a)$ factor for a stack of 
D5-branes. 
The general spectrum of chiral fermions with respect to
the gauge group factors is presented in table \ref{spect4D}.
\begin{table}
  \renewcommand{\arraystretch}{1.2}
 \begin{center}
 \begin{tabular}{|c|l|l|}
 \hline
  Representation & \multicolumn{1}{c|}{Multiplicity}       \\
\hline
    &   $4 m^{(1)}_a m^{(2)}_a m^{(3)}_a 
   (n^{(1)}_a n^{(2)}_a n^{(3)}_a +1)$
    \rule[2.8ex]{0ex}{0ex}                                 \\
     \raisebox{1.5ex}[-1.5ex]{$({\bf A}_a)_L$}  & 
     $=   \tfrac{1}{2} (I_{a'\,a}+ I_{\text{O}6\,a})$   \\
   \hline \rule[2.8ex]{0ex}{0ex}   
    &  $4 m^{(1)}_a m^{(2)}_a m^{(3)}_a 
   (n^{(1)}_a n^{(2)}_a n^{(3)}_a -1)$               \\  
  \raisebox{1.5ex}[-1.5ex]{$({\bf S}_a)_L$} &  
      $=   \tfrac{1}{2} (I_{a'\,a}- I_{\text{O}6\,a})$      \\
  \hline
  $(\overline{{\bf N}}_a,{\bf N}_b)_L$ &   $I_{a\,b}$   \\ 
  $({\bf N}_a,{\bf N}_b)_L$ & $I_{a' b}$          \\
 \hline
 \end{tabular}  
 \caption{\label{spect4D}  Chiral 4D massless open string spectrum.}
 \end{center}
\end{table}
Whenever the intersection number in the second column is formally negative, 
one again has to take the conjugate representation. The 
spectrum in table 2 is free of non-abelian gauge anomalies.

In the next subsections we discuss some examples and point out some 
phenomenological issues  for these models.

\subsection{A 24 generation $SU(5)$ model}

\begin{figure}
  \setlength{\unitlength}{0.1in}
  \begin{picture}(25,30)
  \SetFigFont{14}{20.4}{\rmdefault}{\mddefault}{\updefault}
  \put(0,0){
   \scalebox{0.5}{\includegraphics{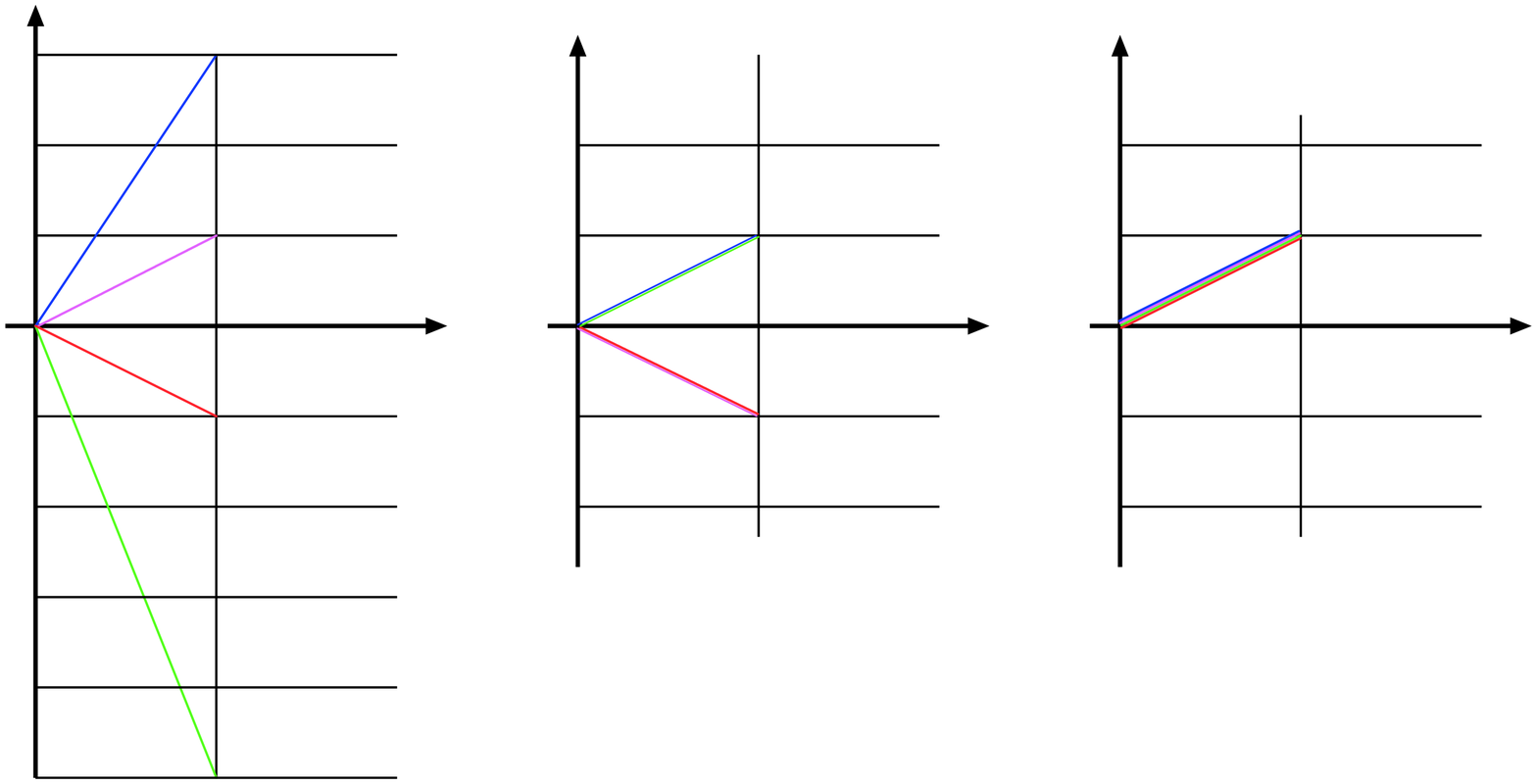}}}
  \put(0,27){$X_2^{(1)}$} \put(14.,16){$X_1^{(1)}$} 
  \put(17.75,27){$X_2^{(2)}$} \put(31.75,16){$X_1^{(2)}$} 
  \put(35.5,27){$X_2^{(3)}$}  \put(49.5,16){$X_1^{(3)}$}
  \SetFigFont{10}{20.4}{\rmdefault}{\mddefault}{\updefault}
  \put(8,24.4){$U(5)_1$}\put(8,18.5){$U(4)_3$}
  \put(8,12.6){$U(4)_4$}\put(8,0.8){$U(3)_2$}
  \put(25.75,18.5){$U(5)_1,\,U(3)_2$}
  \put(25.75,12.6){$U(4)_3,\,U(4)_4$} 
  \end{picture} 
\caption[D$6$-brane configuration of the 24 generation 
             model]
       {\label{24genpic}D$6$-brane configuration of the 24 generation 
             model ($\bar{\sigma}$-pictures of D$6$ branes  omitted).}
\end{figure}

Having found a way to break supersymmetry, 
to reduce the rank of the gauge group and to produce chiral spectra
in four space-time dimensions, it is tempting to search in a compact
bottom-up approach for brane
configurations producing  massless spectra close to the 
Standard Model. The tachyons are not that dangerous from the effective
field theory point of view, as they simply may serve as Higgs-bosons 
for spontaneous gauge symmetry breaking, anticipating a mechanism to generate 
a suitable potential keeping their vacuum expectation values finite. 
In \cite{Bachas:1995ik} a three generation
GUT model was presented, which we shall  revisit 
in the following. 
The gauge group of the model is $G=U(5)\times U(3)\times U(4)\times U(4)$
with maximal rank, so that we have to choose all $n^{(j)}_a=1$. 
The following choice of $m^{(j)}_a$ then 
satisfies all tadpole cancellation conditions \eqref{tadcancelf}:
\begin{equation}
  \label{mchoice}
   m^{(j)}_a=\begin{pmatrix} 3 & -5 & 1 & -1 \\
                               1 &  1 & -1 & -1\\
                               1 &  1 & 1 & 1 
                \end{pmatrix}
\end{equation}
This configuration of D6-branes is displayed in figure 
\ref{24genpic}, where the mirror 
branes have been omitted. 
The chiral part of the fermionic massless spectrum is shown in table 
\ref{spectab2}. 
No chiral fermions transform under both the $U(5)\times U(3)$ gauge 
group and the $U(4)\times U(4)$ gauge group, but there will of course also be
non-chiral bifundamentals. If we think of the $SU(5)$ factor as a 
GUT gauge group, then this model has 24 
generations\footnote{In \cite{Bachas:1995ik} this
model was advocated as a three generation model. We can formally
reproduce the model in \cite{Bachas:1995ik} by dividing the matrix \eqref{mchoice}
 by a factor of two. However, this is inconsistent as it 
would violate the condition that the $m^{(j)}_a$'s
have to be integers. Thus, we conclude that in string theory only
the choice \eqref{mchoice} is correct and the model is 
actually a 24 generation model.}. We shall see in the following that it 
is actually impossible
to get a model with three or any odd number of generations if we restrict
all tori to admit a purely imaginary complex structure $\tau$. 
\begin{table}
  \renewcommand{\arraystretch}{1.2}
 \begin{minipage}[b]{6.0cm}
 \begin{center}
 \begin{tabular}{|l|c|}
 \hline
  $U(5)\times U(3)\times U(4)^2$\rule[2.2ex]{0ex}{0ex} & Multipl.\ \\ 
  \hline\hline
  $({\bf 10},{\bf 1},{\bf 1},{\bf 1})$ &  $24$  \\
 $({\bf 1},{\bf 3},{\bf 1},{\bf 1})$   &  $40$  \\
 $(\Bar{\bf 5},\Bar{\bf 3},{\bf 1},{\bf 1})$ &  $8$  \\
  \hline
 $({\bf 1},{\bf 1},\Bar{\bf 6},{\bf 1})$ &  $8$  \\
 $({\bf 1},{\bf 1},{\bf 1},{\bf 6})$ &  $8$  \\
 \hline  \multicolumn{2}{c}{}
 \\ \multicolumn{2}{c}{}
 \\  \multicolumn{2}{c}{}
 \\ 
 \end{tabular}
  
 \caption{\label{spectab2} 
          Chiral left-handed fermions for 24 generation  model}
 \end{center}
 \end{minipage}\hfill
 \begin{minipage}[b]{6.5cm}
 \begin{center}
 \begin{tabular}{|l|c|}
 \hline
  $SU(3)\times SU(2)\times U(1)^4$ & Multipl. \rule[2.2ex]{0ex}{0ex}\\
  \hline\hline
  $({\bf 3},{\bf 2})_{(1,1,0,0)}$ \rule[2.2ex]{0ex}{0ex}&  $2$  \\
  $({\bf 3},{\bf 2})_{(1,-1,0,0)}$ &  $2$  \\
  $(\Bar{\bf 3},{\bf 1})_{(-1,0,-1,0)}$ &  $4$  \\
  $(\Bar{\bf 3},{\bf 1})_{(-1,0, 1,0)}$ \rule[-1ex]{0ex}{0ex}&  $4$  \\
  \hline
  $({\bf 1},{\bf 2})_{(0,1,0,1)}$ &  $2$  \\
  $({\bf 1},{\bf 2})_{(0,-1,0,1)}$ &  $2$ \\
  $({\bf 1},{\bf 1})_{(0,0,-1,-1)}$ &  $4$  \\
  $({\bf 1},{\bf 1})_{(0,0, 1,-1)}$\rule[-1ex]{0ex}{0ex} &  $4$  \\
  \hline
 \end{tabular}
 \caption{\label{spectab4d} Chiral left-handed fermions for  4 generation 
            model}
 \end{center}
\end{minipage} 
\end{table}

\subsection{\label{fourgenmodel}A four generation model}
The tadpole cancellation condition 
\begin{equation}
  \label{tadcancelfb}
  \sum_{a=1}^K N_a\, n^{(1)}_a n^{(2)}_a n^{(3)}_a =16
\end{equation} 
tells us that we can reduce the rank of the gauge group 
right from the beginning by choosing some $n^{(j)}_a>1$. Therefore, 
we can envision a model where we start with the gauge group
$U(3)\times U(2)\times U(1)^r$ at the string scale.
In order to have three quark generations in the $({\bf 3},{\bf 2})$ 
representation of $SU(3)\times U(2)$,
we necessarily need $I_{12}=3$ and $I_{12'}=0$. 
It turns out that
this is not possible if all three tori are of $\bf A$-type as in this case
$I_{ab}-I_{ab'}$ is always an even number. 
After publishing the paper, the $\bf B$-torus 
as well as products of $\bf A$ and $\bf B$-type tori were considered
in \cite{Blumenhagen:2000ea}. It was shown that the obstruction 
of  $I_{ab}-I_{ab'}=$even can be evaded in these compactifications.
There it was shown as well, that it
is {\sl not} possible to have $I_{12}=3$ and $I_{12'}=0$\,\footnote{This
would imply that
all three quark generations transform in the $({\bf 3},{\bf 2})$ of
$U(3)\times U(2)$.} as well as RR-tadpole cancellation. However
it is possible that {\sl two}   quark generations transform in 
the  $({\bf 3},{\bf 2})$ and {\sl one} generation in the $({\bf 3},\bar{{\bf
    2}})$. This means that one quark generation has opposite $U(1)$-charge 
w.r.t.\
the $U(2)$ stack.\footnote{The fundamental representation of 
$SU(2)\simeq SO(3)$ is (pseudo-)real.
 By the bar over the ${\bf 2}$ we mean that the 
$U(2)$-representation has opposite $U(1)$ charge w.r.t.\ to the unbarred 
${\bf 2}$.}  

Ibanez \& al.\ used combinations of $\bf A$ and $\bf B$-type tori
 in \cite{Ibanez:2001nd} to construct a class of
(toroidal) models with chiral spectrum extremely close to the Standard Model.

\begin{figure}
 \setlength{\unitlength}{0.1in}
  \begin{picture}(45,26)
  \SetFigFont{14}{20.4}{\rmdefault}{\mddefault}{\updefault}
  \put(0,0){
   \scalebox{0.5}{\includegraphics{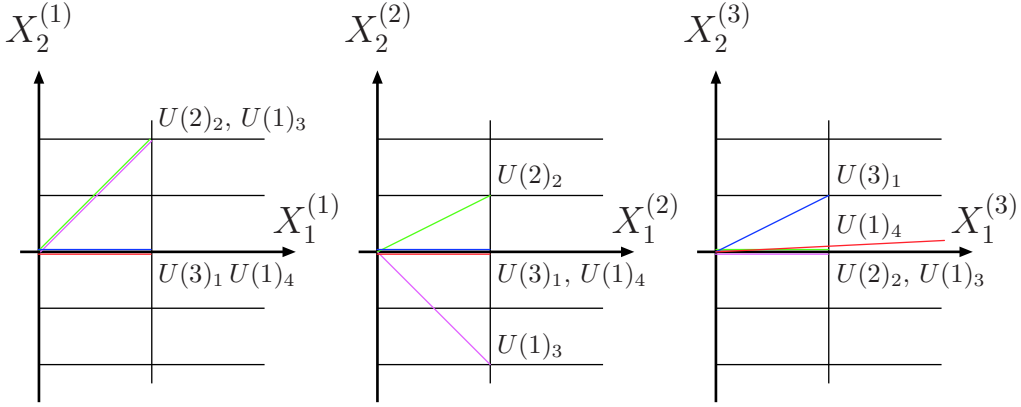}}}
  \put(0,19){$X_2^{(1)}$} \put(14.,9){$X_1^{(1)}$} 
  \put(17.75,19){$X_2^{(2)}$} \put(31.75,9){$X_1^{(2)}$} 
  \put(35.5,19){$X_2^{(3)}$}  \put(49.5,9){$X_1^{(3)}$} 
  \SetFigFont{10}{20.4}{\rmdefault}{\mddefault}{\updefault}
  \put(8,14.6){$U(2)_2,\, U(1)_3$}\put(8,6.4){$U(3)_1\, U(1)_4$}
  \put(25.75,11.6){$U(2)_2$} 
  \put(25.75,6.4){$U(3)_1,\,U(1)_4$}
  \put(25.75,2.7){$U(1)_3$}
  \put(43.5,11.6){$U(3)_1$}
  \put(43.5,9){$U(1)_4$}
  \put(43.5,6.4){$U(2)_2,\, U(1)_3$}
  \end{picture}
 \caption[D$6$-brane configuration of the 4 generation model]
    {\label{4genpic}D$6$-brane configuration of the 4 generation 
             $SU(3)\times SU(2)\times U(1)_Y\times U(1)^2$ model 
             ($\bar{\sigma}$-pictures of D$6$ branes  omitted).}

\end{figure}

The model we found in \cite{Blumenhagen:2000wh}
closest to the 4 generation Standard Model is presented in the following.
We choose the gauge group $U(3)\times U(2)\times U(1)^2$ and the 
following configuration of four stacks of D-branes: 
\begin{align}
  \label{nchoiceb}  
      n^{(j)}_a&=\begin{pmatrix} 1 & 1 & 1 & 1 \\
                                  1 & 1 & 1 & 1 \\
                                  1 &  1 & 1 & 10 
                   \end{pmatrix}\quad\quad
                 &
      m^{(j)}_a&=\begin{pmatrix} 0 & 2 & 2 & 0 \\
                                  0 & 1 & -2 & 0 \\
                                  1 &  0 & 0 & 1
                  \end{pmatrix}
\end{align} 
The configuration has been illustrated in figure \ref{4genpic}.
The resulting chiral massless spectrum is shown in table \ref{spectab4d2}.
Computing the mixed $G^2-U(1)$ anomalies one realizes that one of the
abelian gauge factors is anomalous, which needs to be cured by the
Green-Schwarz mechanism. The other three anomaly-free abelian gauge groups
include a suitable hypercharge $U(1)$  
\begin{equation}
    U(1)_Y={1\over 3} U(1)_1 + U(1)_3 - U(1)_4 
\end{equation}
so that the spectrum finally looks like the one in table \ref{spectab4d2}.
\begin{table}[b]
 \begin{center}
 \begin{tabular}{|l|c|}
 \hline
 $SU(3)\times SU(2)\times U(1)_Y\times U(1)^2$ \rule[2.2ex]{0ex}{0ex}& 
  Multiplicity
  \\
 \hline\hline
 $({\bf 3},{\bf 2})_{({1\over 3},1,0)}$  \rule[2.2ex]{0ex}{0ex}  & $2$  \\
 $({\bf 3},{\bf 2})_{({1\over 3},-1,0)}$   &  $2$  \\
 $(\Bar{\bf 3},{\bf 1})_{(-{4\over 3},0,-1)}$ &  $4$  \\
 $(\Bar{\bf 3},{\bf 1})_{({2\over 3},0, 1)}$ \rule[-1.55ex]{0ex}{0ex} &$4$\\ 
 \hline
 $({\bf 1},{\bf 2})_{(-1,1,0)}$ &    $2$  \\
 $({\bf 1},{\bf 2})_{(-1,-1,0,)}$ &  $2$  \\
 $({\bf 1},{\bf 1})_{(0,0,-1)}$ &    $4$  \\
 $({\bf 1},{\bf 1})_{(2,0,1)}$\rule[-1.2ex]{0ex}{0ex}  &     $4$  \\
 \hline
\end{tabular}
 \caption{\label{spectab4d2} Chiral left-handed fermions for 4 generation 
            model including anomaly-free $U(1)$-charges}
 \end{center}
\end{table} 
We found a semi-realistic, non-supersymmetric, 
four generation Standard Model like spectrum 
with two gauged flavor symmetries and right-handed neutrinos. 
In order to determine the Higgs sector, we would have to investigate the bosonic
part of the spectrum. However, this is not universal but depends
on the radii of the six-dimensional torus. We will not elaborate
this further but instead discuss another important issue concerning the 
possible phenomenological relevance of these models. 

Since we break supersymmetry already at the string scale $M_s$, in order
to solve the gauge hierarchy problem we must choose 
$M_s$  in the TeV region. Let us employ the T-dual picture 
of D6-branes at angles again to analyze the situation in more detail. 
Using the relations
\begin{align}
   M_{\text{Pl}}^2 &\sim {M_s^8 V_6\over g_s^2} & 
        {1\over (g^{(a)}_{\text{YM}})^2} &\sim {M_s^3 V_a\over g_s}
\end{align}
where $V_a$ denotes the volume of some D6-brane in the internal
directions
\begin{equation}
     V_a=\prod_{j=1}^3 V^{(j)}_a 
\end{equation}
and $g^{(a)}_{\rm YM}$ the gauge coupling on this brane. 
They imply
\begin{equation}
    M_s\sim \alpha_{\text{YM}}^{(a)} M_{\text{Pl}} 
         {V_a\over \sqrt{V_6} } 
\end{equation}
Therefore, for the TeV scenario to work one needs
\begin{equation}
 \label{condi}   {V_a\over \sqrt{V_6} } \ll 1 
\end{equation}
for all D6-branes. However, chirality for the fermionic spectrum of 
an open string stretched between
any two D6-branes implies that the two branes in question do not lie on 
top of each other on any of the three $T^2_{(j)}$ tori. 
In other words the two branes already span the entire torus 
and the condition \eqref{condi} cannot be realized. 

\section{(In-) Stability of purely toroidal orientifolds}
In this chapter we will make some comments about the stability of 
purely toroidal orientifolds. Stability usually demands the vanishing of the
partition function, as the partition function in string theory 
is interpreted as
a dilaton potential. To do consistent string perturbation theory,
no tadpoles are allowed at any order of the string perturbation theory.
(Otherwise one could hope that higher order contributions could ``repair''
a tadpole that originates from lower order terms.) Actually the dilaton does
not couple to $\chi=0$ amplitudes, however other string-excitations do.
Supersymmetry guarantees the vanishing of the partition function\footnote{In
  the light-cone quantized string on the pp-wave the partition function does
  not vanish, but this is believed to be connected to the fact that the
  light-cone gauge does not cover all string states (i.e.\ $p^+=0$).
  (Private communication with Matthias Gaberdiel.)}.
Therefore imposing supersymmetry is a very general method to get stable 
models. Compactifications on flat tori, which are the starting point of
our construction, have a supersymmetric closed string sector. 
The three unoriented  $\chi=0$ diagrams (Klein-bottle,
Cylinder and M\"obius strip) are divergent  only due to closed string tadpoles.
However also in the open string sector we can often isolate a single 
excitation that is responsible for an instability: the tachyon. 
Due to the  tachyon relation $M^2_\text{tach}<0$ it has been suggested 
that the tachyon can condense like a Higgs field $\phi$  whose
potential $V(\phi)\propto -\mu^2|\phi|^2+\lambda^2 |\phi|^4$
 if expanded around  $\phi=0$ contains a negative
 mass\raisebox{1ex}{\tiny 2}-term
as well. Tachyon condensation in terms of {\it string field theory} 
has been investigated in a variety of papers. We only mention
two of the first \cite{Moeller:2000jy,Berkovits:2000hf}, 
a complete list would require more than
hundred entries.  However we will mention the conditions for absence of 
tachyons, which where published after the paper this chapter is 
based on, was finished. First however we show that {\sl chiral} supersymmetric
$\Omega\Bar{\sigma}$ orientifolds of the torus are impossible. This restriction
does not apply for all toroidal orbifolds.\footnote{However the 
$\mathbb{Z}_3$ orientifold investigated in \cite{Blumenhagen:2001te} is 
always non-supersymmetric for chiral models.} The first example 
of a four dimensional, chiral supersymmetric
 $\Omega\Bar{\sigma}$-orientifold was the 
$\mathbb{Z}_2\times\mathbb{Z}_2$ orientifold
 investigated by Cvetic \& al.\ which
in addition has the nice feature to admit quasi-realistic models. 
The second example is the $\mathbb{Z}_4$  $\Omega\Bar{\sigma}$-orientifold
 on a six torus 
\cite{Blumenhagen:1999ev} which admits chiral supersymmetric
models as well \cite{Blumenhagen:2002wn,Blumenhagen:2002gw}. Some of
them are even phenomenological interesting.\footnote{Meanwhile similar features have been found in the
$\Omega\Bar{\sigma}$-orientifold of the
$\mathbb{Z}_4\times\mathbb{Z}_2$-orbifold \cite{Honecker:2003vq}.}
We will postpone this discussion
to the next chapter.   

\subsection{\label{slagsec}Supersymmetric brane configurations and special La\-gran\-gi\-an submanifolds
(sLags)}
The mathematical foundations of calibrations and sLags are described for
example  in the classic publication \cite{Harvey:1982a} and a newer 
article by Joyce \cite{Joyce2002}.
Here we will state some important definitions and consequences.
Without going into the details, we state the common fact that D-branes
that are wrapped around so called {\it special La\-gran\-gi\-an submanifolds}
(sLags) preserve some amount of the bulk (or closed string)
supersymmetry \cite{Becker:1995kb}, if no NS $F$ field is switched on
and if no NSNS $B$-field component is along the D-brane. 
There are many ways to introduce the notion of a sLag. One is to start 
by defining what a {\it calibration} is.
 Roughly speaking a calibration is a {\sl closed} 
$k$-form $\phi$ on a manifold ${\cal M}$ (with volume form  $\text{vol}$) 
such that given
any oriented  $k$ plane $V$ in the tangent bundle  $T{\cal M}$  the induced  
$k$-volume $\text{vol}_k|_V$
form is always bigger or equal than the restriction of $\phi$:
\be
   \phi|_V \le\text{vol}_k|_V
\ee
In this sense a $\phi$-calibrated submanifold ${\cal N}\subset {\cal M}$ is 
  a submanifold of dimension $k$ s.th.\ for all oriented tangent planes 
 $T_x{\cal N}$  of ${\cal N}$ the
  restriction of $\phi$ equals the restriction of the volume form:
\be
   \phi|_{T_x{\cal N}} = \text{vol}_k|_{T_x{\cal N}} \qquad 
   \forall x\in {\cal N}
\ee  
The striking result (of a theorem) is that a calibrated submanifold is always
volume-mini\-miz\-ing in its homology class.

CY$n$-spaces  ${\cal M}$ 
(cf.\ \ref{susyorbs}, p.\ \pageref{susyorbs}) consist of
a quadruple\footnote{This follows either from the definition of 
the CY space, or if the  CY space is introduced differently, by further
theorems.}  $({\cal M},J,g, \Omega)$. ${\cal M}$ is
the complex compact manifold (of complex dimension $n$)
 itself, $J$ the corresponding complex structure. 
$g$ is the K\"ahler metric with the {\it Levi-Civita connection} 
leading to $SU(n)$ holonomy.
(The K\"ahler form $\omega$ is is obtained from $J$ and $g$). 
$\Omega$ is a non-zero covariantly constant $(n,0)$-form s.th.:
\be
 \frac{\omega^m}{m!} = (-1)^{m(m-1)/2} \Big(\frac{i}{2}\Big)^m 
   \Omega \wedge \bar{\Omega}
\ee
By the above definition of $\Omega$ the real part $\mathfrak{Re} (\Omega)$ 
is automatically a calibration on ${\cal M}$.\footnote{${\mathfrak Re} (\exp(i\theta)\Omega)$ is another calibration form.
         In what follows, we will consider only calibrations w.r.t.\
         $\exp(i\theta)=1$. }
A  $\mathfrak{Re} (\Omega)$-calibrated submanifold of a CY$n$-fold is called a
``special Lagrangian submanifold'' (sLag). SLags have real dimension $n$.

A theorem states that given a CY$n$-fold ${\cal M}$
(defined as above) and
a real $n$-dimensional submanifold ${\cal N}$, then ${\cal N}$ admits
an orientation making it into a sLag iff both $\omega|_{\cal N}=0$ and
 $\mathfrak{Im}(\Omega|_{\cal N})=0$ are fulfilled.\footnote{If only
 the condition $\omega|_{\cal N}=0$ is obeyed, ${\cal N}$ is called a
 ``Lagrangian submanifold''. The supplement ``special'' means the additional
 property $\mathfrak{Im}(\Omega|_{\cal N})=0$. Langrangian submanifolds
 are defined more generally in symplectic geometry, where $\omega$ denotes the
 symplectic form.}

In $\mathbb{C}^2$ the sLags are given by holomorphic curves.

The case of $\bar{\sigma}\Omega$ orientifolds with sLags has been 
studied in great generality in \cite{Blumenhagen:2002wn}. 
We will now take advantage of some of the facts inherited by sLags.
Applying these properties to the four dimensional case of
compactification on $T^2\times T^2$, $\Omega$ is given by $dz_1\wedge dz_2$.
The D$6$-branes projected on this four-torus are real two-dimensional.
We have restricted ourselves to flat branes, each a product of two 
 one-cycles, a one-cycle on each two-torus $T^2_{(j)}$.  
 The sLag condition then gives rise to the condition that
the sum of the two oriented angles vanishes:
\be 
   \phi^{(1)}+\phi^{(2)}=0
\ee 
The O$7$-plane, i.e. the fixed locus of $\bar{\sigma}$ has 
$\phi_1=\phi_2=0$.\footnote{The fact that $O(9-d)$-planes can be interpreted as fixed loci of
          $\bar{\sigma}$ is explained in \cite{Blumenhagen:2002wn}.}
In other words: the  O$7$-plane is parallel to the $X^{(i)}_{1}$-axis.
Canceling the RR-tadpole means that the complete holonomy-class of the cycle
associated with the D-branes cancels exactly 
the holonomy  of the  O$7$-plane:\footnote{The factor of $8$ in front of
 $\pi_{\text{O}7}$ is determined by the tadpole cancellation conditions 
\eqref{tadcancel}. $\pi_{a'}$ denotes the $\bar{\sigma}$-image of  $\pi_a$.}
\be 
  \sum_a N_a (\pi_a+\pi_{a'}) = 8\pi_{\text{O}7}
\ee 
The $\pi_a$ are elements of the homology generated by the basis (cf.\ figure
\ref{dsevenbranes}):\footnote{We have split the basis. Element of the second
 summand do not appear in our models.} 
\begin{multline} 
 \overline{\left<p_1\equiv\pi_1\otimes\pi_3, p_2\equiv\pi_2\otimes\pi_4,
                  p_3\equiv\pi_2\otimes\pi_3, p_4\equiv\pi_1\otimes\pi_4
           \right>}
 \\ \oplus 
 \overline{\left<\pi_1\otimes \pi_2, \pi_3\otimes \pi_4\right>}
    = H_2\big(T^4\big)
\end{multline} 
In the $p$-basis the cycle wrapped by the D$7$-brane $a$ is expressed
by:\footnote{$I$ denotes the intersection matrix. It can be used to calculate
              the intersection number as well (cf.\ table \ref{spectab6d}).}
\be 
 \pi_a = \begin{pmatrix} n_a^{(1)}n_a^{(2)} \\
                         m_a^{(1)}m_a^{(2)} \\
                         m_a^{(1)}n_a^{(2)} \\
                         n_a^{(1)}m_a^{(2)} 
         \end{pmatrix}
   \qquad
   I =\left(\begin{array}{cc|cc}    0 &  1 &    &       \\
                          1 &  0 &    \multicolumn{2}{c}{\raisebox{1.5ex}[-1.5ex]{0}}      \\
                          \hline& &  0 &  1     \\
                 \multicolumn{2}{c|}{\raisebox{1.5ex}[-1.5ex]{0}}    & 1 &  0  
            \end{array}    
      \right)
\ee 
In the above basis the homology class of  the  O$7$-plane
 (i.e.\ the $\bar{\sigma}$-invariant locus) is given by:
\be
  \pi_{\text{O}7}=4\, p_1 = (4,0,0,0)^T
\ee 
It is now clear that the configuration with the smallest volume that 
lies in the same Homology class as the O$7$ plane, is the one in which 
all D-branes are parallel to the $X^{(i)}_1$ axis. Any flat  tilted brane
with identical 
$\pi_{\text{O}7}$-component and 
non-zero angle wrt.\ the $X^{(i)}_1$-axis will have larger volume. Such a
brane
configuration can not be a sLag since sLags are volume minimizing in their
homology class. 
 Thus, in the absence of NSNS $B$-fields with components parallel to the 
branes and in the absence of an NS $F$-flux, the only supersymmetry preserving 
D$7$-brane configurations that cancel the RR-tadpole, are the ones where
all branes are parallel to the O$7$-plane.\footnote{However the brane
might be deformed, s.th. it is no longer flat: a theorem by McLean 
states that the dimension of the moduli space of a sLag ${\cal N}$ equals
 the first Betti number $b^1({\cal N})$ (cf. \cite{McLean1998,Joyce2002}).
 In the case of the flat tow-torus however, $b^1({\cal N})$ just equals the
number of independent translations of the D-brane in its normal direction,
 which is two. Therefore we conclude that a deformation that transforms a
flat D-brane to a non-flat one, will spoil the sLag condition and as a
consequence will break supersymmetry. Similar to the CY-case where the
geometric K\"ahler cone gets complexified by the NSNS $B$-field, we can 
add Wilson lines, i.e.\ flat gauge connections, without breaking
supersymmetry. The number of independent  Wilson lines
equals $\dim (\pi_1)$ which in turn again equals $b^1({\cal N})$ 
(cf.\ \cite{Douglas2001:tr}).}
 These configurations are however the ones  without chiral fermions.

The same arguments go through for the six-torus (i.e. the four-dimensional
models). 
In contrast to the four compact dimensions, the six-dimensional torus forces
$B$-field components along the D$6$-brane to vanish, as well as $F=0$ for
the NS $U(1)$-fields. This was shown in \cite{Marino:1999af}.
Therefore we conclude that pure toroidal compactifications can not
reconcile  both supersymmetry and chirality w.r.t.\ the gauge group. 

In the next section, we summarize what has been found out 
on the existence and non-existence of (open-) string tachyons in
toroidal compactifications.

\subsection{Tachyons in toroidal orientifolds}
As we already noted, tachyons can only appear in the open
string sector, as the closed string sector is supersymmetric. In our
original publication we concluded that in six-dimensional models open string
tachyons generically appear due to the fact that two D$7$-branes that are rotated
by an  angle  $(\phi_1,\phi_2)$ always imply negative ground state energy, except
for the supersymmetric case $\phi_1=-\phi_2$. 
Thus in $\Omega\bar{\sigma}$-orientifolds on $T^4$
we always encounter one or several open string tachyons, iff supersymmetry is
broken.
\begin{figure} 
 \begin{minipage}[t]{6.0cm}
 \begin{center}
 \setlength{\unitlength}{0.1in}
  \begin{picture}(25,26.5)
  \SetFigFont{14}{20.4}{\rmdefault}{\mddefault}{\updefault}
  \put(0,0){\scalebox{0.25}{\includegraphics{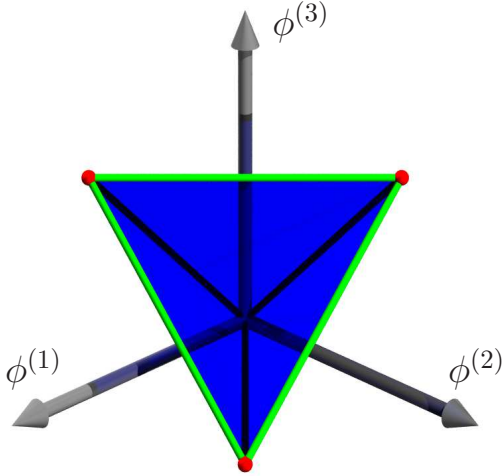}}}
  \put(0,5){$\phi^{(1)}$} 
  \put(23.2,5){$\phi^{(2)}$} 
  \put(14,23.2){$\phi^{(3)}$}
  \end{picture}
\end{center}
 \end{minipage} \hfill
  \raisebox{2ex}{\begin{minipage}[t]{6.0cm}
    \caption[Tachyon region]
      {\label{tachregion}
      Tachyon region:
      Tachyons develop outside the tetrahedron. The vertices (red)
      preserve ${\cal N}=4$, and the edges (green) ${\cal N}=2$ supersymmetry.
      The faces (transparent blue) only preserve ${\cal N}=1$. 
      While inside the tetrahedron 
      (${\cal N}=0$) no open string tachyons are present,
      the brane configuration
      generically destabilizes due to closed string tadpoles in this case.
      The four vertices sit at 
      $(\phi^{(1)},\phi^{(2)},\phi^{(3)})$=$(0,0,0)$, $(\pi,\pi,0)$,
      $(0,\pi,\pi)$, $(\pi,0,\pi)$.}
   \end{minipage}}
\end{figure}
However for open strings stretching between two D$6$-branes with 
angles $(\phi_1\neq 0,\phi_2\neq 0,\phi_3\neq 0)$ tachyon free regions
exist in the parameter space 
(\cite{Aldazabal:2000dg,Ibanez:2001nd,Rabadan:2001mt}).
We represent the distinct cases in figure
\ref{tachregion} (cf.\ \cite{Ibanez:2001nd}). The tachyon free region
is inside the tetrahedron. The vertices, edges and faces are not only tachyon
free, but carry also different amounts of supersymmetry. The vertices preserve
${\cal N}=4$ , the edges ${\cal N}=2$ and the faces  ${\cal N}=1$. However
different faces carry different kinds of ${\cal N}=1$ supersymmetry which
is determined by the signs of the supersymmetry condition: 
$\phi_1\pm\phi_2\pm'\phi_3=0$. Similarly, two adjacent faces carry only one
common supersymmetry. The vertices are however equivalent, as the situation 
is invariant under shifting the tetrahedron by lattice-vectors of the type:
\begin{equation}
  \Gamma = \left\{l\cdot(\pi,\pi,0)+m\cdot(0,\pi,\pi)+n\cdot(\pi,0,\pi)\,
                  |l,n,m\in \mathbb{Z}
           \right\}
\end{equation}
In other words, branes rotated in exactly two complex planes by angles
$\pi$ preserve the 
same supersymmetry. (A rotation by $\pi$ in a single plane would transform the brane
to an antibrane.) Actually the tetrahedron in figure \ref{tachregion} 
should be repeated each lattice vector, which we did not do for clarity
of the picture.  Outside the tetrahedron tachyons are present.
It turned out to be possible to build tachyon-free toroidal models 
with a chiral 
spectrum extremely close to the Standard Model \cite{Ibanez:2001nd}. 
Even thought tachyons  might be absent, a non-supersymmetric 
orientifold-model is unstable if there are uncanceled NSNS-tadpoles. 
If the closed string moduli corresponding to the tadpoles are related
to the angles of the branes, the model can be driven to point where open
string tachyons appear.
 In the toroidal models  complex-structure  moduli  influence
the angles of the branes, and these are exactly the ones which develop
 NSNS tadpoles (besides the dilaton). 

\addcontentsline{toc}{section}{Concluding remarks}
\section*{Concluding remarks}
In this paper we have investigated Type I string compactifications
on non-commutative tori, which are due to constant magnetic fields along the 
world volumes of the D$9$-branes being wrapped around the internal space.
The number of chiral fermions (arising
from strings ``stretched'' between two branes with different 
NS $F$-field-strength on each $T^2$) can be interpreted as a Landau degeneracy.
It is given by the {\sl Atiyah-Singer index theorem} for twisted 
spin-complexes.

In the T-dual picture the magnetized D$9$-branes become lower-dimensional
branes, while the $\Omega$-parity projection now includes 
a complex conjugation on all $T^2$'s: 
$\Omega\xrightarrow{\text{T-duality}}\bar{\sigma}\Omega$.
In this picture, the chiral fermions are located at the brane 
intersection-points. The analog of the Landau degeneracy is the
topological intersection number of two D-branes.  

In the given setting, we found a four-dimensional model with
Standard Model gauge group $SU(3)\times SU(2)\times U(1)_Y$
(times some abelian  flavor  gauge groups) with four generations
of Standard Model fermions and also a four-dimensional 24 generation
$SU(5)$ GUT-like model. 
However it turned out, that the Ansatz $T^6=\prod_j T^2_{(j)\,\mathbf A}$  involving
only ${\bf A}$-type tori leads always to an even number 
of left handed quark generation, as the difference of the intersection number
$I_{ab}-I_{ab'}$ is always even. As shown 
in a subsequent publication \cite{Blumenhagen:2000ea},
this problem can be evaded by
considering more general $T^6$-tori involving ${\bf B}$-type two-tori, as well.

We found that in chiral toroidal orientifolds, supersymmetry is always broken.
This could lead to open string tachyons. However many models have 
been found after our publication, in which tachyons are absent (cf.\ 
\cite{Ibanez:2001nd}). Nevertheless, closed string tadpoles tend to
destabilize
the model and drive it to a singular limit of the $T^6$. Dilaton tadpoles
can in general push a non-supersymmetric  model to either a strongly-coupled
 or free regime. 
Investigations if a non-supersymmetric string model could be stabilized
by the  {\it Fischler-Susskind mechanism} 
\cite{Fischler:1986ci, Fischler:1986tb} have been undertaken as well (cf.\
\cite{Dudas:2000ff,Blumenhagen:2000dc}).

The stability problems would be absent, if one can construct
supersymmetric models. That it is indeed possible to reconcile both
supersymmetry and chirality has been shown by Cvetic \& al.\
\cite{Cvetic:2001nr} who considered a (left-right symmetric)
$T^6/(\mathbb{Z}_2\times\mathbb{Z}_2)$ Type IIA orientifold. 
This orientifold has phenomenologically appealing  solutions. 
The second example of a   Type IIA orientifold involving
branes at angles and admitting chiral supersymmetric solutions was published
by us some time later (cf.\ \cite{Blumenhagen:2002gw}).
We found a solution with a gauge-group that can be broken down to
the SM gauge group in a supersymmetry preserving way by giving 
VEVs to fields in the low energy effective action. The matter
content of the resulting model is extremely close to the MSSM.
We will review the  $T^6/\mathbb{Z}_4$ 
$\Bar{\sigma}\Omega$-orientifold in the next chapter. 


\chapter[The  \texorpdfstring{$\Bar{\sigma}\Omega$}
              { sigma Omega }-Orientifold
          on  \texorpdfstring{$(T^2\times T^2\times T^2)/\mathbb{Z}_4$}
              {(T\texttwosuperior \textmultiply T\texttwosuperior
                 \textmultiply T\texttwosuperior)/Z(4)}]
  { \label{z4}\centerline{The $\Bar{\sigma}\Omega$-Orientifold } 
              \centerline{on $(T^2\times T^2\times T^2)/\mathbb{Z}_4$}
  }
Intersecting brane world models have been the subject of elaborate
string model building for several years
\cite{Blumenhagen:2000wh,Angelantonj:2000hi,Blumenhagen:2000vk,Angelantonj:2000rw,Aldazabal:2000dg,Aldazabal:2000cn,Blumenhagen:2000ea,Ibanez:2001nd,
Forste:2001gb,Rabadan:2001mt,Blumenhagen:2001te,Cvetic:2001tj,Cvetic:2001nr,
 Bailin:2001ie,Ibanez:2001dj,Blumenhagen:2001mb,Honecker:2001dj,Honecker:2002hp,Cremades:2002te,Blumenhagen:2002ua,Cremades:2002cs,Kokorelis:2002ip,Garcia-Bellido:2002zu,Cremades:2002dh,Kokorelis:2002zz,Serna:2002ux,Cvetic:2002qa,Cvetic:2002wh,Klein:2002vu,Blumenhagen:2002wn,Uranga:2002pg,Bailin:2002gg,Burgess:2003hx,Kokorelis:2002ns,Blumenhagen:2002vp,Pradisi:2002vu,Kokorelis:2002tf,Bailin:2002tv}.
The main new ingredient in these models is that they contain
intersecting D-branes and open strings in a consistent manner
providing  simple mechanisms to generate chiral fermions and  to
break supersymmetry \cite{Bachas:1995ik,Berkooz:1996km}. Most attempts for
constructing realistic models were dealing with non-supersymmetric
configurations of D-branes, mainly because non-trivial, chiral
supersymmetric 
intersecting brane world models are not easy to find. It is known
for instance that flat factorizing D-branes on the six-dimensional
torus as well as on the $T^6/\mathbb{Z}_3$ orbifold can never give rise
to supersymmetric models except for the trivial
non-chiral configuration where all D$6$-branes are located on top of
the orientifold plane \cite{Blumenhagen:2001te}. Supersymmetric models clearly
have some advantages over the non supersymmetric ones. From the
stringy point of view such models are stable, as not only the
Ramond-Ramond (R-R) tadpoles cancel but also the
Neveu-Schwarz-Neveu-Schwarz (NS-NS) tadpoles. From the
phenomenological point of view, since the gauge hierarchy problem
is solved by supersymmetry, one can work in  the conventional
scenarios with a large string scale close to the Planck  scale or
in an intermediate regime \cite{Burgess:1998px}.
For an overview on other Type I constructions see \cite{Angelantonj:2002ct}.

The only semi-realistic supersymmetric models that have been found
so far are defined in the $T^6/\mathbb{Z}_2\times \mathbb{Z}_2$ orientifold
background and were studied in a series of papers
\cite{Cvetic:2001tj,Cvetic:2001nr,Cvetic:2002qa,Cvetic:2002wh,Pradisi:2002vu}.\footnote{Meanwhile
  the $\mathbb{Z}_4\times\mathbb{Z}_2$ orientifold with 
  $\bar{\sigma}\Omega$-projection has been studied. Interesting
 non-chiral supersymmetric solutions have been found as well (cf.\ \cite{Honecker:2003vq}).}
 Besides their
phenomenological impact, Type IIA supersymmetric intersecting
brane worlds with orientifold six-planes and D$6$-branes are also
interesting from the stringy point of view, as they are expected
to lift to M-theory on singular $G_2$ manifolds \cite{Kachru:2001je}.

The aim of this chapter (and the underlying publication 
\cite{Blumenhagen:2002gw})
is to pursue the study of intersecting brane
worlds on orientifolds with a particular emphasis on the
systematic construction  of semi-realistic supersymmetric
configurations. Note, that without the orientifold projection
supersymmetric intersecting brane configurations do not exist, as
the overall tension always would be positive. Interestingly, from
the technical point of view, the $\mathbb{Z}_4$ orbifold involves some
new insights, as not all 3-cycles are inherited from the torus. In
fact, a couple of 3-cycles arise in the $\mathbb{Z}_2$ twisted sector
implying that this model contains so-called fractional D$6$-branes,
which have been absent in the  $\mathbb{Z}_2\times \mathbb{Z}_2$ and $\mathbb{Z}_3$
orbifolds. To treat these exceptional cycles accordingly, we will
make extensive use of the formalism developed in \cite{Blumenhagen:2002wn}.

It will turn out that supersymmetric models in general can be
constructed in a straightforward way. But as in other model
building approaches, finding semi-realistic three generation
models turns out to be quite difficult. Fortunately, we will
finally succeed in constructing a globally  supersymmetric three
generation Pati-Salam model with gauge group $SU(4)\times
SU(2)_L\times SU(2)_R$ and the Standard Model matter in addition
to some exotic matter in the symmetric and antisymmetric
representation of the two $SU(2)$ gauge groups. In this chapter, we
will mainly focus on the new and interesting string model building
aspects and leave a detailed investigation of the phenomenological
implications of the discussed models for future work.

This chapter is organized as follows. In section \ref{intcysec} we review some of
the material presented in \cite{Blumenhagen:2002wn}  about the 
general structure of
intersecting  brane worlds on Calabi-Yau manifolds. We will review
those formul\ae\, which will be extensively used in the rest of the
paper. In section \ref{3cyclsec} we start to investigate the ${\cal X}=T^6/\mathbb{Z}_4$
orbifold and in particular derive an integral basis for the
homology group $H_3({\cal X},\mathbb{Z})$, for which the intersection form
involves the Cartan-matrix of the Lie-algebra $E_8$. The main
ingredient in the construction of such an integral basis will be
the physical motivated introduction of fractional D-branes which
also wrap around exceptional (twisted) 3-cycles in ${\cal X}$. In section
\ref{z4orientifoldsec} we construct the orientifold models of Type IIA on the orbifold
${\cal X}$ and discuss the orientifold planes, the action of the
orientifold projection on the homology and the additional
conditions arising for supersymmetric configurations. In section \ref{pssec} 
we construct as a first example a globally supersymmetric four
generation Pati-Salam model. Finally, in section \ref{3genpssec} we elaborate 
on a supersymmetric model with initial gauge symmetry $U(4)\times
U(2)^3\times U(2)^3$ and argue that by brane recombination it
becomes a supersymmetric three generation Pati-Salam model. By
using conformal field theory methods, for this model we determine
the chiral and also the massless non-chiral spectrum, which turns
out to provide Higgs fields in just the right representations in
order to break the model down to the Standard Model. At the end of
the paper we describe both the GUT breaking and the electroweak
breaking via brane recombination processes. We also make a
prediction for the Weinberg angle at the string scale.

\section{\label{intcysec}Intersecting Brane Worlds on Calabi-Yau spaces}

Before we present our new model, we would like to briefly
summarize some of the results presented in \cite{Blumenhagen:2002wn}
 about Type IIA
orientifolds on smooth Calabi-Yau spaces. If the manifold admits
an anti-holomorphic involution $\Bar{\sigma}$, the combination
$\Omega\Bar{\sigma}$ is indeed a symmetry of the Type IIA model.
Taking the quotient with respect to this symmetry introduces an
orientifold six-plane into the background, which wraps a special
Lagrangian 3-cycle of the Calabi-Yau.\footnote{By ``special
Lagrangian 3-cycle'' we mean the real three-dimensional special
Lagrangian submanifold, that lies in the same homology class as the 
 3-cycle.} In order to cancel  the
induced RR-charge, one introduces stacks of $N_a$ D$6$-branes which
are wrapped on 3-cycles $\pi_a$. Since under the action of
$\Bar{\sigma}$ such a 3-cycle, $\pi_a$,  is in general mapped to a
different 3-cycle, $\pi_a'$, one has to wrap the same number of
D$6$-branes on the latter cycle, too. The equation of motion for the
RR 7-form implies the RR-tadpole cancellation condition,
\begin{equation}
 \label{tadhom}
  \sum_a  N_a\, (\pi_a + \pi'_a)-4\, \pi_{O6}=0. 
\end{equation}
If it is possible to wrap a connected smooth D-brane on such a
homology class,  the  stack of D$6$-branes supports a $U(N_a)$ gauge
factor. Note, that it is not a trivial question if in a given
homology class such a connected smooth manifold does exist.
However, as we will see in section \ref{3genpssec} for special cases, there are
physical arguments ensuring that such smooth D-branes exist.

The Born-Infeld action provides an expression for the open string
tree-level scalar potential which by differentiation leads to an
equation for the NS-NS tadpoles 
\begin{equation}
 \label{susy}
  {V}=T_6\,
 {e^{-\phi_{4}} \over M_s^3\sqrt{{\text{ Vol}({\cal X})}}}
            \Bigl( \sum_a  N_a \left( {\rm Vol}({\rm D}6_a) +
              {\rm Vol}({\rm D}6'_a) \right) -4\, {\rm Vol}({\rm O}6)
            \Bigr) 
\end{equation}
with the four-dimensional dilaton given by
$e^{-\phi_{4}}=M_s^3\sqrt{{\text{Vol}({\cal X})}}e^{-\phi_{10}}$ and $T_6$
denoting the tension of the D$6$-branes. By 
${\rm Vol}({\rm D}6_a)$ we mean the three dimensional internal volume of the
D$6$-branes. Generically, this scalar potential is non-vanishing
reflecting the fact that intersecting branes do break
supersymmetry. If the cycles (more precisely: the corresponding
submanifolds) are special Lagrangian (sLag) but
calibrated with respect to $3$-forms $\mathfrak{Re}(e^{i\theta}\Omega_{3,0})$
with different constant phase factors $\exp (i\theta)$, the expression gets
simplified to\footnote{$\widehat\Omega_{3,0}=e^{i\theta}\Omega_{3,0}$} 
\begin{equation}
 \label{dbi}
  {V}=T_6\, e^{-\phi_{4}} \left( \sum_a{N_a
   \left| \int_{\pi_a} \widehat\Omega_{3,0} \right|} +
 \sum_a{N_a \Biggl| \int_{\pi'_a} \widehat\Omega_{3,0} \Biggr|}-
 4 \left| \int_{\pi_{{\rm O}6}} \widehat\Omega_{3,0} \right |\right) 
\end{equation}
In this case, all D$6$-branes preserve some supersymmetry but not
all of them the same. Models of this type have been discussed in
\cite{Cremades:2002te,Cremades:2002cs}. In the case of a completely
 supersymmetric
model, all 3-cycles are calibrated with respect to the same 3-form
as the O6-plane implying that the disc level scalar potential
vanishes due to the RR-tadpole condition \eqref{tadhom}.

In \cite{Blumenhagen:2002wn}  it was argued and confirmed by many
 examples that the
chiral massless spectrum charged under the $U(N_1)\times
\ldots\times U(N_k)$ gauge group of a configuration of $k$
intersecting stacks of D$6$-branes can be computed from the
topological intersection numbers as shown in table \ref{chiralspecgen}. 
\begin{table}
 \renewcommand{\arraystretch}{1.2}
 \begin{center}
 \begin{tabular}{|c|c|}
 \hline
  Representation  & Multiplicity \\
 \hline\hline
  $[{\bf A_a}]_{L}$  &
  ${1\over 2}\left(\pi'_a\circ \pi_a+\pi_{{\rm O}6} \circ \pi_a\right)$ \\
  $[{\bf S_a}]_{L}$
   & ${1\over 2}\left(\pi'_a\circ \pi_a-\pi_{{\rm O}6} \circ\pi_a\right)$\\ 
  \hline
  $[{\bf (\Bar{N}_a,N_b)}]_{L}$  & $\pi_a\circ \pi_{b}$ \\
  $[{\bf (N_a, N_b)}]_{L}$ & $\pi'_a\circ \pi_{b}$   \\
  \hline 
 \end{tabular}
 \caption{\label{chiralspecgen} Chiral spectrum in $d=4$ }
 \end{center}
\end{table}
Since in six dimensions the intersection number between two
3-cycles is anti-symmetric, the self intersection numbers do vanish
implying the absence of chiral fermions in the adjoint
representation. Negative intersection numbers correspond to chiral
fermions in the conjugate representations. Note, that if we want
to apply these formul\ae\, to orientifolds on singular toroidal
quotient spaces, the intersection numbers have to be computed in
the orbifold space and not simply in the ambient toroidal space.
After these preliminaries, we will discuss the $\mathbb{Z}_4$ orientifold in
the following sections.
\section{\label{3cyclsec}3-cycles in the \texorpdfstring{$\mathbb{Z}_4$}{Z(4)}
          orbifold}

We consider Type IIA string theory compactified on the orbifold
background $T^6/\mathbb{Z}_4$, where the action of the $\mathbb{Z}_4$ symmetry,
$\Theta$, on the internal three complex coordinates reads
\begin{equation}
 \label{actio}   z_1\to e^{\pi i \over 2}\, z_1, \quad
              z_2\to e^{\pi i \over 2}\, z_2, \quad
              z_3\to e^{-\pi i}\, z_3 
\end{equation}
with $z_1=x_1+i x_2$,\, $z_2=x_3+i x_4$ and $z_3=x_5+i x_6$.
This action preserves ${\cal N}=2$ supersymmetry in four dimensions
so that the orbifold describes a singular limit of a Calabi-Yau
threefold. The Hodge numbers of this threefold are given by
$h_{21}=7$ and $h_{11}=31$, where
1 complex- and 5 K{\"a}hler-moduli arise in the untwisted
sector. The $\Theta$ and $\Theta^3$ twisted sectors contain 16
$\mathbb{Z}_4$ fixed points giving rise to 16 additional K\"ahler moduli.
In the $\Theta^2$ twisted sector, there are 16 $\mathbb{Z}_2$ fixed points from
which 4 are also $\mathbb{Z}_4$ fixed points. The latter ones
contain 4 K\"ahler moduli whereas the remaining twelve $\mathbb{Z}_2$ fixed
points are organized in pairs under the $\mathbb{Z}_4$ action giving
rise to 6 complex- and 6 K\"ahler-moduli.
The fact that
the $\mathbb{Z}_2$ twisted sector contributes $h^{\text{tw}}_{21}=6$ elements to
the number of complex structure deformations and therefore contains
what might be called twisted 3-cycles, is the salient new feature
of this $\mathbb{Z}_4$ orbifold model as compared to the intersecting brane
world models studied so far.

Given this supersymmetric closed string background, we take
the quotient by the orientifold projection $\Omega\Bar{\sigma}$,
where $\Bar{\sigma}$ is an anti-holomorphic involution
$z_i\to  e^{i \phi_i} \Bar{z}_i$ of the manifold.
Note, that this orientifold model is not T-dual to
the $\mathbb{Z}_4$ Type IIB orientifold model studied first in \cite{Aldazabal:1998mr}.
In the latter model there did not exist any
supersymmetric brane configurations canceling all
tadpoles induced by the orientifold planes.
In fact, as was pointed out in \cite{Blumenhagen:2000fp} our model is 
T-dual to a Type IIB orientifold on
an asymmetric $\mathbb{Z}_4$ orbifold space. 
Slightly different $\mathbb{Z}_4$ Type IIB orientifold models
were studied in \cite{Angelantonj:1999ms,Klein:2000hf}.

Our orientifold projection breaks supersymmetry in the bulk to 
${\cal N}=1$ and introduces
an orientifold O$6$-plane located at the fixed point locus of
the anti-holomorphic involution.
The question arises if one can introduce D$6$-branes, generically not aligned
to the orientifold plane, in order to cancel the tadpoles induced by the
presence of the O$6$ plane.
The simplest such model where the D$6$-branes lie on top of the orientifold plane
has been investigated in \cite{Blumenhagen:1999ev}.

\subsection{Crystallographic actions}

\begin{figure}
  \begin{minipage}[b]{7.5cm}
   \begin{center}
\begin{picture}(0,0)%
\includegraphics{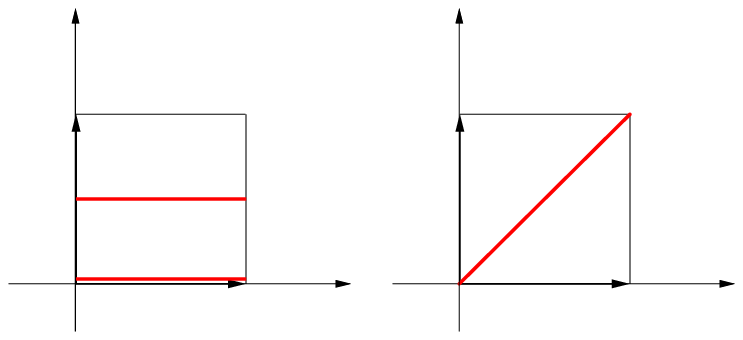}%
\end{picture}%
\setlength{\unitlength}{2072sp}%
\begingroup\makeatletter\ifx\SetFigFont\undefined%
\gdef\SetFigFont#1#2#3#4#5{%
  \reset@font\fontsize{#1}{#2pt}%
  \fontfamily{#3}\fontseries{#4}\fontshape{#5}%
  \selectfont}%
\fi\endgroup%
\begin{picture}(6842,3085)(-934,-2234)
\put(5311,-2196){\makebox(0,0)[lb]{\smash{\SetFigFont{12}{14.4}{\rmdefault}{\mddefault}{\updefault}{\color[rgb]{0,0,0}$X_{1,3}$}%
}}}
\put(3776,524){\makebox(0,0)[lb]{\smash{\SetFigFont{17}{20.4}{\rmdefault}{\mddefault}{\updefault}{\color[rgb]{0,0,0}{\bf B}}%
}}}
\put(266,524){\makebox(0,0)[lb]{\smash{\SetFigFont{17}{20.4}{\rmdefault}{\mddefault}{\updefault}{\color[rgb]{0,0,0}{\bf A}}%
}}}
\put(1801,-2196){\makebox(0,0)[lb]{\smash{\SetFigFont{12}{14.4}{\rmdefault}{\mddefault}{\updefault}{\color[rgb]{0,0,0}$X_{1,3}$}%
}}}
\put(2811,-511){\makebox(0,0)[lb]{\smash{\SetFigFont{12}{14.4}{\rmdefault}{\mddefault}{\updefault}{\color[rgb]{0,0,0}{$\pi_{2,4}$}}%
}}}
\put(4591,-2176){\makebox(0,0)[lb]{\smash{\SetFigFont{12}{14.4}{\rmdefault}{\mddefault}{\updefault}{\color[rgb]{0,0,0}{$\pi_{1,3}$}}%
}}}
\put(-719,-511){\makebox(0,0)[lb]{\smash{\SetFigFont{12}{14.4}{\rmdefault}{\mddefault}{\updefault}{\color[rgb]{0,0,0}{$\pi_{2,4}$}}%
}}}
\put(1081,-2176){\makebox(0,0)[lb]{\smash{\SetFigFont{12}{14.4}{\rmdefault}{\mddefault}{\updefault}{\color[rgb]{0,0,0}{$\pi_{1,3}$}}%
}}}
\put(2576,569){\makebox(0,0)[lb]{\smash{\SetFigFont{12}{14.4}{\rmdefault}{\mddefault}{\updefault}{\color[rgb]{0,0,0}{$X_{2,4}$}}%
}}}
\put(-934,589){\makebox(0,0)[lb]{\smash{\SetFigFont{12}{14.4}{\rmdefault}{\mddefault}{\updefault}{\color[rgb]{0,0,0}{$X_{2,4}$}}%
}}}
\end{picture}
    \rule[-2.5cm]{0ex}{0ex}
    \caption[Anti-holomorphic involutions]
      {\label{involpic}Anti-holomorphic involutions. The O$6$-planes, i.e.\ the
       fixed loci under $\Bar{\sigma}$ are painted in \textcolor{r}{red}.}
         \end{center}
  \end{minipage} \hfill
  \begin{minipage}[b]{5cm}
   \begin{center}
    \begin{picture}(0,0)%
\includegraphics{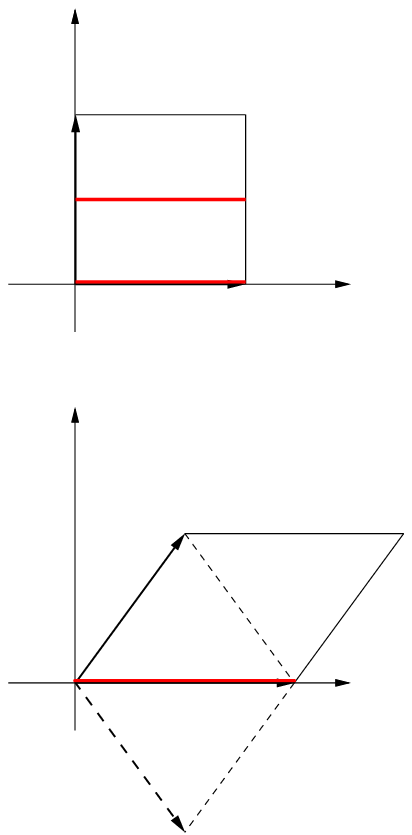}%
\end{picture}%
\setlength{\unitlength}{2072sp}%
\begingroup\makeatletter\ifx\SetFigFont\undefined%
\gdef\SetFigFont#1#2#3#4#5{%
  \reset@font\fontsize{#1}{#2pt}%
  \fontfamily{#3}\fontseries{#4}\fontshape{#5}%
  \selectfont}%
\fi\endgroup%
\begin{picture}(3717,7564)(-834,-6713)
\put(1801,-5796){\makebox(0,0)[lb]{\smash{\SetFigFont{12}{14.4}{\rmdefault}{\mddefault}{\updefault}{\color[rgb]{0,0,0}$X_{5}$}%
}}}
\put(1801,-2196){\makebox(0,0)[lb]{\smash{\SetFigFont{12}{14.4}{\rmdefault}{\mddefault}{\updefault}{\color[rgb]{0,0,0}$X_{5}$}%
}}}
\put(266,524){\makebox(0,0)[lb]{\smash{\SetFigFont{17}{20.4}{\rmdefault}{\mddefault}{\updefault}{\color[rgb]{0,0,0}{\bf A}}%
}}}
\put(991,-3121){\makebox(0,0)[lb]{\smash{\SetFigFont{17}{20.4}{\rmdefault}{\mddefault}{\updefault}{\color[rgb]{0,0,0}{\bf B}}%
}}}
\put(1081,-5776){\makebox(0,0)[lb]{\smash{\SetFigFont{12}{14.4}{\rmdefault}{\mddefault}{\updefault}{\color[rgb]{0,0,0}{$\pi_{5}$}}%
}}}
\put(1081,-2176){\makebox(0,0)[lb]{\smash{\SetFigFont{12}{14.4}{\rmdefault}{\mddefault}{\updefault}{\color[rgb]{0,0,0}{$\pi_{5}$}}%
}}}
\put(-719,-511){\makebox(0,0)[lb]{\smash{\SetFigFont{12}{14.4}{\rmdefault}{\mddefault}{\updefault}{\color[rgb]{0,0,0}{$\pi_{6}$}}%
}}}
\put( 46,-4156){\makebox(0,0)[lb]{\smash{\SetFigFont{12}{14.4}{\rmdefault}{\mddefault}{\updefault}{\color[rgb]{0,0,0}{$\pi_6$}}%
}}}
\put(-834,589){\makebox(0,0)[lb]{\smash{\SetFigFont{12}{14.4}{\rmdefault}{\mddefault}{\updefault}{\color[rgb]{0,0,0}{$X_{6}$}}%
}}}
\put(-834,-3056){\makebox(0,0)[lb]{\smash{\SetFigFont{12}{14.4}{\rmdefault}{\mddefault}{\updefault}{\color[rgb]{0,0,0}{$X^{6}$}}%
}}}
\end{picture}
     \caption{\label{orientationpic}Orientations of the third
       $T^2$}
       \phantom{space}
   \end{center}
  \end{minipage}
\end{figure}

Before dividing Type IIA string theory by the discrete symmetries
$\mathbb{Z}_4$ and $\Omega\Bar{\sigma}$, we have to ensure that the torus
$T^6$ does indeed allow crystallographic actions of these symmetries.
For simplicity, we assume that $T^6$ factorizes as
$T^6=T^2\times T^2\times T^2$.
On the first two $T^2$s the $\mathbb{Z}_4$ symmetry enforces a rectangular
torus with complex structure $U=1$.
On each torus two different anti-holomorphic
involutions
\begin{equation}
 \label{invol}
  \begin{aligned}   {\bf A}&: z_i\to  \Bar{z}_i \\
                    {\bf B}&: z_i\to  e^{i {\pi\over 2}} \Bar{z}_i  
  \end{aligned} 
\end{equation}
do exist. These two cases are shown in figure \ref{involpic}, where we have indicated the fixed point
set of the orientifold projection $\Omega\Bar{\sigma}$.\footnote{The same
distinction between the involutions ${\bf A}$ and ${\bf B}$ occurred
for the first time in the papers
\cite{Blumenhagen:1999md,Angelantonj:1999xf,Blumenhagen:1999ev,Forste:2000hx}.}

Since on the third torus the $\mathbb{Z}_4$ acts like a reflection, its
complex structure is unconstrained. But again there exist two different
kinds of involutions, which  equivalently correspond to  the two possible
choices of the orientation of the torus as shown in 
figure \ref{orientationpic}.
For the ${\bf A}$-torus its complex structure is given by $U=i U_2$
with $U_2$ unconstrained and for the ${\bf B}$-torus the complex structure
is given by $U={1\over 2}+i U_2$.
Therefore, by combining all possible choices of complex conjugations
we get eight possible orientifold models.
However, taking into account that the orientifold model on the $\mathbb{Z}_4$
orbifold does not only contain the orientifold planes related
to $\Omega\Bar{\sigma}$ but also the orientifold planes
related to $\Omega\Bar{\sigma} \Theta$, $\Omega\Bar{\sigma} \Theta^2$ and
$\Omega\Bar{\sigma} \Theta^3$, only four models
$\{${\bf AAA,ABA,AAB,ABB}$\}$ are actually different.

\subsection{A non-integral basis of 3-cycles}
In order to utilize  the formul\ae\, from section \ref{intcysec}, we have to find
the independent 3-cycles on the $\mathbb{Z}_4$ orbifold space.
Since we already know that the third Betti number, $b_3=2+2 h_{21}$,
is equal to sixteen, we expect to find precisely this number
of independent 3-cycles.

\begin{figure}
 \begin{minipage}[t]{7.0cm}
  \begin{center}
 \begin{picture}(0,0)%
\includegraphics{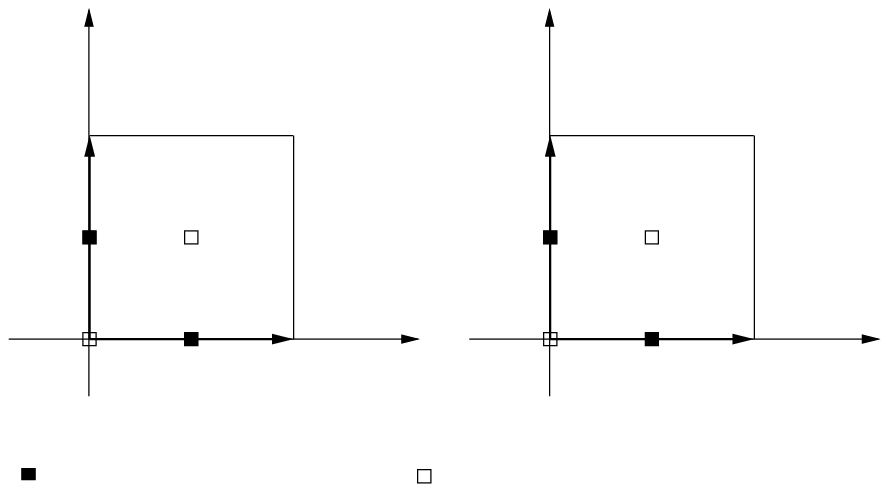}%
\end{picture}%
\setlength{\unitlength}{2486sp}%
\begingroup\makeatletter\ifx\SetFigFont\undefined%
\gdef\SetFigFont#1#2#3#4#5{%
  \reset@font\fontsize{#1}{#2pt}%
  \fontfamily{#3}\fontseries{#4}\fontshape{#5}%
  \selectfont}%
\fi\endgroup%
\begin{picture}(6762,3670)(-854,-2819)
\put(2656,569){\makebox(0,0)[lb]{\smash{\SetFigFont{14}{16.8}{\rmdefault}{\mddefault}{\updefault}{\color[rgb]{0,0,0}$X_4$}%
}}}
\put(-854,569){\makebox(0,0)[lb]{\smash{\SetFigFont{14}{16.8}{\rmdefault}{\mddefault}{\updefault}{\color[rgb]{0,0,0}$X_2$}%
}}}
\put(-359,-2739){\makebox(0,0)[lb]{\smash{\SetFigFont{12}{14.4}{\rmdefault}{\mddefault}{\updefault}{\color[rgb]{0,0,0}$:\;\mathbb{Z}_2$-fixed point}%
}}}
\put(2656,-2761){\makebox(0,0)[lb]{\smash{\SetFigFont{12}{14.4}{\rmdefault}{\mddefault}{\updefault}{\color[rgb]{0,0,0}$:\;\mathbb{Z}_4$-fixed point $\big(\in \left\{\mathbb{Z}_2\mbox{-fixed points}\right\}\big)$}%
}}}
\put(3466,-1131){\makebox(0,0)[lb]{\smash{\SetFigFont{10}{12.0}{\rmdefault}{\mddefault}{\updefault}{\color[rgb]{0,0,0}{$3$}}%
}}}
\put(4241,-1131){\makebox(0,0)[lb]{\smash{\SetFigFont{10}{12.0}{\rmdefault}{\mddefault}{\updefault}{\color[rgb]{0,0,0}{4}}%
}}}
\put(5311,-2196){\makebox(0,0)[lb]{\smash{\SetFigFont{14}{16.8}{\rmdefault}{\mddefault}{\updefault}{\color[rgb]{0,0,0}$X_3$}%
}}}
\put(4241,-1906){\makebox(0,0)[lb]{\smash{\SetFigFont{10}{12.0}{\rmdefault}{\mddefault}{\updefault}{\color[rgb]{0,0,0}{$2$}}%
}}}
\put(3466,-1906){\makebox(0,0)[lb]{\smash{\SetFigFont{10}{12.0}{\rmdefault}{\mddefault}{\updefault}{\color[rgb]{0,0,0}{$1$}}%
}}}
\put(-44,-1131){\makebox(0,0)[lb]{\smash{\SetFigFont{10}{12.0}{\rmdefault}{\mddefault}{\updefault}{\color[rgb]{0,0,0}{$3$}}%
}}}
\put(731,-1131){\makebox(0,0)[lb]{\smash{\SetFigFont{10}{12.0}{\rmdefault}{\mddefault}{\updefault}{\color[rgb]{0,0,0}{4}}%
}}}
\put(1801,-2196){\makebox(0,0)[lb]{\smash{\SetFigFont{14}{16.8}{\rmdefault}{\mddefault}{\updefault}{\color[rgb]{0,0,0}$X_1$}%
}}}
\put(-44,-1906){\makebox(0,0)[lb]{\smash{\SetFigFont{10}{12.0}{\rmdefault}{\mddefault}{\updefault}{\color[rgb]{0,0,0}{$1$}}%
}}}
\put(3776,524){\makebox(0,0)[lb]{\smash{\SetFigFont{17}{20.4}{\rmdefault}{\mddefault}{\updefault}{\color[rgb]{0,0,0}{\bf A}}%
}}}
\put(266,524){\makebox(0,0)[lb]{\smash{\SetFigFont{17}{20.4}{\rmdefault}{\mddefault}{\updefault}{\color[rgb]{0,0,0}{\bf A}}%
}}}
\put(731,-1906){\makebox(0,0)[lb]{\smash{\SetFigFont{10}{12.0}{\rmdefault}{\mddefault}{\updefault}{\color[rgb]{0,0,0}{$2$}}%
}}}
\put(2811,-511){\makebox(0,0)[lb]{\smash{\SetFigFont{12}{14.4}{\rmdefault}{\mddefault}{\updefault}{\color[rgb]{0,0,0}{$\pi_4$}}%
}}}
\put(1081,-2176){\makebox(0,0)[lb]{\smash{\SetFigFont{12}{14.4}{\rmdefault}{\mddefault}{\updefault}{\color[rgb]{0,0,0}{$\pi_1$}}%
}}}
\put(4591,-2176){\makebox(0,0)[lb]{\smash{\SetFigFont{12}{14.4}{\rmdefault}{\mddefault}{\updefault}{\color[rgb]{0,0,0}{$\pi_3$}}%
}}}
\put(-719,-511){\makebox(0,0)[lb]{\smash{\SetFigFont{12}{14.4}{\rmdefault}{\mddefault}{\updefault}{\color[rgb]{0,0,0}{$\pi_2$}}%
}}}
\end{picture}
  \end{center}
  \end{minipage} \hfill
  \begin{minipage}[t]{4.0cm}
    \begin{center}
    \caption[Orbifold fixed points]{\label{orbifixpic} 
               Orbifold fixed points. The second $\bf A$-torus
               can also be interpreted as a $\bf B$-torus
               (cf.\ fig.\ \ref{involpic} and 
                sect.\ \ref{wsparityt2}, p.\ \pageref{wsparityt2}-
                \pageref{complexrotation}).}
    \end{center}
   \end{minipage}
\end{figure}

One set of 3-cycles we get for free as they descend from the ambient space.
Consider the three-cycles  inherited from the torus $T^6$.
We call the two fundamental cycles on the torus $T^2_I$ ($I=1,2,3$)
$\pi_{2I-1}$ and $\pi_{2I}$ and moreover we define
the toroidal 3-cycles
\begin{equation}
 \label{picyc}
 \pi_{ijk} \equiv \pi_i\otimes\pi_j\otimes\pi_k.
\end{equation}
Taking orbits under the $\mathbb{Z}_4$ action, one can deduce the following
four $\mathbb{Z}_4$ invariant 3-cycles
\begin{equation}
 \label{invcyc}
 \begin{aligned}
        \rho_1 &\equiv 2 (\pi_{135}-\pi_{245} ),   \qquad
                   \bar{\rho}_1   \equiv 2 (\pi_{136}-\pi_{246} ) \\
        \rho_2 &\equiv 2 (\pi_{145}+\pi_{235} ), \qquad
                   \bar{\rho}_2  \equiv 2 (\pi_{146}+\pi_{236} )  
 \end{aligned}
\end{equation}
The factor of two in \eqref{invcyc} is due to the fact that
$\Theta^2$ acts trivially on the toroidal 3-cycles.
In order to compute the intersection form, we make use of the following
fact: if the 3-cycles $\pi^t_a$
on the torus are arranged in orbits of length $N$
under some  $\mathbb{Z}_N$ orbifold group, i.e.
\begin{equation}
 \label{orbit}
\pi_a \equiv  \sum_{i=0}^{N-1} \Theta^i \pi^t_a 
\end{equation}
the intersection number between two such 3-cycles on the orbifold space
is given by
\begin{equation}
 \label{inta}
     \pi_a\circ\pi_b={1\over N} \left(\sum_{i=0}^{N-1} \Theta^i \pi^t_a
                                \right)
           \circ \left(\sum_{j=0}^{N-1} \Theta^j \pi^t_b \right) 
\end{equation}
Therefore, the intersection form for the four 3-cycles \eqref{invcyc} reads
\begin{equation}
 \label{form} 
    I_{\rho}=\bigoplus_{i=1}^2  \begin{pmatrix} 0 & -2 \\
                                                2 & 0  \\
                                \end{pmatrix}
\end{equation}
The remaining twelve 3-cycles arise in the $\mathbb{Z}_2$ twisted
sector of the orbifold.
Since $\Theta^2$ acts non-trivially only onto the first two $T^2$,
in the $\mathbb{Z}_2$ twisted sector the sixteen $\mathbb{Z}_2$ fixed
points do appear as shown in figure \ref{orbifixpic}.

The unfilled boxes in the figure indicate the $\mathbb{Z}_2$ fixed points
 which are also fixed under the $\mathbb{Z}_4$ symmetry.
After blowing up the
orbifold singularities, each of these fixed points gives rise an exceptional 2-cycle $e_{ij}$ with
the topology of $S^2$. These exceptional 2-cycles can be combined with
the two fundamental 1-cycles on the third torus to form what might
be called exceptional 3-cycles with the topology $S^2\times S^1$.
However, we have to take into account the $\mathbb{Z}_4$ action, which leaves
four fixed points invariant and arranges the remaining twelve in six
pairs.
Since the $\mathbb{Z}_4$ acts by reflection on the third torus,
its action on the exceptional cycles $e_{ij}\otimes \pi_{5,6}$
is
\begin{equation}
 \label{actionthe} 
       \Theta\left(e_{ij}\otimes \pi_{5,6}\right)=-
                    e_{\theta(i)\theta(j)} \otimes \pi_{5,6} 
\end{equation}
with
\begin{equation}
 \label{actiontheb}
      \theta(1)=1,\quad  \theta(2)=3,\quad \theta(3)=2,\quad
                      \theta(4)=4
\end{equation}
Due to the minus sign in \eqref{actionthe} the
invariant $\mathbb{Z}_4$ fixed points drop out and what remains are precisely
the twelve 3-cycles
\begin{equation}
 \label{excyc}
 \begin{aligned}
  \varepsilon_{1}       & \equiv (e_{12}-e_{13})\otimes\pi_5, &  \qquad
  \bar{\varepsilon}_{1}  \equiv (e_{12}-e_{13})\otimes\pi_6 \\
  \varepsilon_{2}       & \equiv (e_{42}-e_{43})\otimes\pi_5, &  \qquad
  \bar{\varepsilon}_{2}  \equiv (e_{42}-e_{43})\otimes\pi_6 \\
  \varepsilon_{3}       & \equiv (e_{21}-e_{31})\otimes\pi_5, &  \qquad
  \bar{\varepsilon}_{3}  \equiv (e_{21}-e_{31})\otimes\pi_6 \\
  \varepsilon_{4}       & \equiv (e_{24}-e_{34})\otimes\pi_5, &  \qquad
  \bar{\varepsilon}_{4}  \equiv (e_{24}-e_{34})\otimes\pi_6 \\
  \varepsilon_{5}       & \equiv (e_{22}-e_{33})\otimes\pi_5, &  \qquad
  \bar{\varepsilon}_{5}  \equiv (e_{22}-e_{33})\otimes\pi_6 \\
  \varepsilon_{6}       & \equiv (e_{23}-e_{32})\otimes\pi_5, &  \qquad
  \bar{\varepsilon}_{6}  \equiv (e_{23}-e_{32})\otimes\pi_6 
 \end{aligned}
\end{equation}
Utilizing \eqref{inta} the resulting intersection form is simply
\begin{equation}
  \label{formex}
    I_{\varepsilon}=\bigoplus_{i=1}^6  \begin{pmatrix} 0 & -2 \\
                                                       2 &  0  
                                       \end{pmatrix}
\end{equation}
These 3-cycles  lie in $H_3({\cal X},\mathbb{Z})$ but do not form an
integral basis of the free module since their intersection form is not
uni-modular.

\subsection{An integral basis of 3-cycles}

The cycles which are missing so far are the ones corresponding
to what is called fractional D-branes \cite{Douglas:1996sw,Douglas:1997xg}.
In our context these are
D-branes wrapping only one-half times around the toroidal
cycles $\{\rho_1,\bar\rho_1,\rho_2,\bar\rho_2\}$ while
wrapping simultaneously  around some of the exceptional 3-cycles.
Therefore in the orbifold limit such branes are stuck at the
fixed points and one needs at least two such fractional
D-branes in order to form a brane which can be moved into the bulk.

To proceed, we need a rule of what combinations of toroidal and
exceptional cycles are allowed  for a fractional D-brane.
Such a rule can be easily gained from our physical intuition.
A D-brane wrapping for instance the toroidal cycle
${1\over 2}\rho_1$ can only wrap around those exceptional
3-cycles that correspond to the $\mathbb{Z}_2$ fixed points the flat D-brane
is passing through.  In our case, when the brane is lying along
the $X_{1,3,5}$-axis on the three $T^2$s, the allowed exceptional
cycles are $\{\varepsilon_1,\varepsilon_3,\varepsilon_5\}$.
Therefore, the total homological cycle the D-brane is wrapping on  can be
for instance
\begin{equation}
  \label{frac1} \pi_a={1\over 2}\rho_1+{1\over 2}(
                \varepsilon_1+\varepsilon_3+\varepsilon_5) 
\end{equation}
The relative signs for the four different terms in \eqref{frac1}
are still free parameters and
at the orbifold point do correspond to turning on a discrete Wilson line
along a longitudinal internal direction of the D-brane.
 Note, that this construction is completely analogous to
the construction of boundary states for fractional D-branes
\cite{Sen:1998ii,Diaconescu:1999dt,Gaberdiel:2000jr}
carrying also a charge under some $\mathbb{Z}_2$ twisted sector states.

As an immediate consequences of this rule, only unbarred
respectively  barred cycles can be combined into fractional cycles,
as they wrap the same fundamental 1-cycle on the third $T^2$.
Apparently, the only non-vanishing intersection numbers are
between barred and unbarred cycles. Any unbarred fractional D-brane
can be expanded as 
\begin{equation}
 \label{expa}
     \pi_a=v_{a,1} \rho_1 + v_{a,2}
            \rho_2 + \sum_{i=1}^6 v_{a,i+2}\, \varepsilon_i 
\end{equation}
with half-integer valued coefficients $v_{a,i}$.
By exchanging the two fundamental cycles
on the third $T^2$, we can associate to it  a barred brane
\begin{equation}
  \label{expb}
    \Bar{\pi}_a=v_{a,1} \Bar{\rho}_1 + v_{a,2} \Bar{\rho}_2 
     + \sum_{i=1}^6 v_{a,i+2}\, \Bar{\varepsilon}_i
\end{equation}
with the same coefficients $v_{a,i+8}=v_{a,i}$ for $i\in\{1,\ldots,8\}$.
Using our rule we can construct all linear combinations
with ``self''-intersection number $\pi\circ \Bar{\pi}=-2$, where
we also have to keep in mind that the cycles form a lattice, i.e.
integer linear combinations of cycles are again cycles.\\
In the following we list all the fractional
3-cycles with  ``self''-intersection number $\pi\circ \Bar{\pi}=-2$.
These cycles can be divided into 3 sets:
\renewcommand{\labelenumi}{\bf\alph{enumi})}
\begin{enumerate}
\item $\{(v_1,v_2;v_3,v_4;v_5,v_6;v_7,v_8)\;|\; v_1+v_2=\pm 1/2,\,
                                   v_3+v_4=\pm 1/2,\,
                                  v_5+v_6=\pm 1/2,\,
                                  v_7+v_8=\pm 1/2;\,
       v_1+v_3+v_5+v_7 = 0\ {\rm mod}\ 1\}$. These combinations are
      obtained by observing which fixed points
      the flat branes parallel to the fundamental cycles do
      intersect. These define $8\cdot 16=128$ different fractional 3-cycles.

\item $\{(v_1,v_2 ;v_3,v_4;0,0;0,0),\,(v_1,v_2 ;0,0;v_5,v_6;0,0),\,
     (v_1,v_2 ;0,0;0,0;v_7,v_8),$\\ $ (0,0;v_3,v_4;v_5,v_6 ;0,0)
     ,\, (0,0;v_3,v_4;0,0;v_7,v_8),\, (0,0;0,0;v_5,v_6;v_7,v_8)$\\$ \;|
     \; v_i \in {\pm 1/2} \} \}$. The first three kinds of cycles  are
     again constructed from branes lying parallel to the x,y-axis on one
     $T^2$ and stretching along the diagonal on the other $T^2$.
     The remaining  three kinds of cycles  arise from integer
     linear combinations of the cycles introduced so far.
     Thus, in total this yields $6\cdot 16 = 96 $ 3-cycles
     in the second  set.

\item $\{(v_1,v_2 ;v_3,v_4;v_5,v_6;v_7,v_8)\; |
         \; {\rm exactly\ one }\ v_i=\pm
     1\ {\rm , rest\ zero} \}$. Only the vectors with $v_1=\pm
     1$ or $v_2\pm1$ can be derived from untwisted branes. They are purely
     untwisted. The purely twisted ones again arise from linear
     combinations. This third set contains $2\cdot 8= 16$ 3-cycles.
\end{enumerate}
Altogether there are 240 of such 3-cycles with
``self''-intersection number $-2$,  which intriguingly just
corresponds to the number of roots of the $E_8$ Lie algebra. Now,
it is easy to write a computer program searching for a basis among
these 240 cycles, so that the intersection form takes the
following form 
\begin{equation}
 \label{eacht}
       I=\begin{pmatrix} 0 & C_{E_8} \\
                       -C_{E_8} & 0 
         \end{pmatrix}
\end{equation}
where $C_{E_8}$ denotes the Cartan matrix of $E_8$
\begin{equation}
 \label{cartan}   C_{E_8}=\left( 
  \begin{array}{rrrrrrrr}
   -2  &   1 &   0 &  0 &  0 &  0 &  0 &   0 \\
    1  &  -2 &   1 &  0 &  0 &  0 &  0 &   0 \\
    0  &   1 &  -2 &  1 &  0 &  0 &  0 &   0 \\
    0  &   0 &   1 & -2 &  1 &  0 &  0 &   0 \\
    0  &   0 &   0 &  1 & -2 &  1 &  0 &   1 \\
    0  &   0 &   0 &  0 &  1 & -2 &  1 &   0 \\
    0  &   0 &   0 &  0 &  0 &  1 & -2 &   0 \\
    0  &   0 &   0 &  0 &  1 &  0 &  0 &  -2 
  \end{array}
 \right)
\end{equation}
One possible choice for the ``simple roots'' is
\begin{equation}
 \label{simple}
 \begin{aligned}
  \vec{w}_1 &= 
   {1\over 2 }(-1,\phantom{-}0, -1,\phantom{-}0,-1,\phantom{-}0,-1,
            \phantom{-}0) \\
  \vec{w}_2 &= {1\over 2 } (\phantom{-}1,\phantom{-}0,\phantom{-}1,
            \phantom{-}0,\phantom{-}1,\phantom{-}0,-1,\phantom{-}0) \\
  \vec{w}_3 &= {1\over 2 } (\phantom{-}1,\phantom{-}0, -1,\phantom{-}0,-1,
            \phantom{-}0,\phantom{-}1,\phantom{-}0) \\
  \vec{w}_4 &= {1\over 2 } (-1,\phantom{-}0,\phantom{-}1,\phantom{-}0,
            \phantom{-}0,\phantom{-}1,\phantom{-}0,\phantom{-}1) \\ 
  \vec{w}_5 &= {1\over 2 } (\phantom{-}0,\phantom{-}1,-1,
            \phantom{-}0,\phantom{-}1,\phantom{-}0,\phantom{-}0,-1) \\
  \vec{w}_6 &= {1\over 2 } (\phantom{-}0,-1,\phantom{-}1,\phantom{-}0,-1,
            \phantom{-}0,\phantom{-}0,-1) \\
  \vec{w}_7 &= {1\over 2 } (\phantom{-}0,\phantom{-}1,\phantom{-}0,
            \phantom{-}1,\phantom{-}0,-1,\phantom{-}0,\phantom{-}1) \\
  \vec{w}_8 &= {1\over 2 } (\phantom{-}0,-1, \phantom{-}0,-1,\phantom{-}0, -1,
            \phantom{-}0,\phantom{-}1)
 \end{aligned}
\end{equation}
Since the Cartan matrix is unimodular, we indeed have constructed
an integral basis for the homology lattice $H_3({\cal X},\mathbb{Z})$. In the
following, it turns out to be more convenient to work with the
non-integral orbifold basis allowing also half-integer
coefficients. However, as we have explained not all such cycles
are part of $H_3({\cal X},\mathbb{Z})$, so we have to ensure each time we use
such fractional 3-cycles that they are indeed contained in the
unimodular lattice $H_3({\cal X},\mathbb{Z})$, i.e. that they are integer linear
combinations of the basis \eqref{simple}.

\section{\label{z4orientifoldsec}Orientifolds of the  
                \texorpdfstring{$\mathbb{Z}_4$}{Z(4)} Type IIA orbifold}

Equipped with the necessary information about the 3-cycles in the
$\mathbb{Z}_4$ toroidal orbifold, we can move forward and consider the
four inequivalent orientifold models in more detail.

\subsection{O6-planes in the  \texorpdfstring{$\mathbb{Z}_4$}{Z(4)}
             orientifold}

First, we have to determine the 3-cycle of the O6-planes.
Let us discuss this computation for the ${\bf ABB}$
model in some more detail, as this orientifold will
be of main interest for its potential to provide
semi-realistic standard-like models.

We have to determine the fixed point sets of the four
relevant orientifold projections $\{ \Omega\Bar{\sigma},
 \Omega\Bar{\sigma}\Theta,
\Omega\Bar{\sigma}\Theta^2,\Omega\Bar{\sigma}\Theta^3\}$.
The results are listed in table \ref{o6planeabb}.
\begin{table}
 \renewcommand{\arraystretch}{1.2}
 \begin{center}
 \begin{tabular}{|c|c|}
 \hline
  Projection  & fixed point set  \\
\hline\hline
 $\Omega\,\Bar{\sigma}$  & $2\, \pi_{135} + 2\, \pi_{145}  $  \\
  $\Omega\,\Bar{\sigma}\,\Theta$  
   & $2\, \pi_{145} + 2\, \pi_{245}- 4\, \pi_{146}- 4\, \pi_{246}$  \\
 $\Omega\, \Bar{\sigma}\, \Theta^2$  & $2\, \pi_{235} - 2\, \pi_{245} $ \\
 $\Omega\, \Bar{\sigma}\, \Theta^3$  
   & $-2\, \pi_{135} + 2\, \pi_{235}+ 4\, \pi_{136}- 4\, \pi_{236}$   \\ 
 \hline
 \end{tabular}
 \caption{\label{o6planeabb} O6-planes for {\bf ABB} model}
 \end{center}
\end{table}
Adding up all contributions we get
\begin{equation}
  \label{orid}
  \begin{aligned}  \pi_{\text{O}6}&=4\, \pi_{145} +4\, \pi_{235} +4\, \pi_{136}
                       -4\,\pi_{246} -4\, \pi_{146} -4\, \pi_{236} \\
                      &=2\, \rho_2+2\, \Bar{\rho}_1 -2\, \Bar{\rho}_2 .\\ 
  \end{aligned}
\end{equation}
Thus, only bulk cycles appear in $\pi_{\text{O}6}$ reflecting the fact
that in the conformal field theory the orientifold planes
carry only charge under untwisted R-R fields 
\cite{Blumenhagen:2000wh,Blumenhagen:1999ev}.
The next step is to determine the action of $\Omega\Bar{\sigma}$ on the
homological cycles. This can easily be done for the orbifold
basis. We find for the toroidal 3-cycles
\begin{equation}
  \label{actbas}
  \begin{aligned}   \rho_1\to \rho_2,& \qquad 
                       \Bar{\rho}_1\to \rho_2- \Bar{\rho}_2 \\
                      \rho_2\to \rho_1,& \qquad  
                   \Bar{\rho}_2\to \rho_1- \Bar{\rho}_1 
 \end{aligned}
\end{equation}
and for the exceptional cycles
\begin{equation}
 \label{actbasex}
 \begin{aligned} &\varepsilon_1\to -\varepsilon_1 \quad\quad
                 \Bar{\varepsilon}_1\to -\varepsilon_1+\Bar{\varepsilon}_1 \\
                        &\varepsilon_2\to -\varepsilon_2 \quad\quad
                 \Bar{\varepsilon}_2\to -\varepsilon_2+\Bar{\varepsilon}_2 \\
                        &\varepsilon_3\to \varepsilon_3\phantom{-} \quad\quad
                 \Bar{\varepsilon}_3\to \varepsilon_3-\Bar{\varepsilon}_3 \\
                        &\varepsilon_4\to \varepsilon_4 \phantom{-}\quad\quad
                 \Bar{\varepsilon}_4\to  \varepsilon_4-\Bar{\varepsilon}_4 \\
                        &\varepsilon_5\to \varepsilon_6 \phantom{-} \quad\quad
                 \Bar{\varepsilon}_5\to \varepsilon_6-\Bar{\varepsilon}_6 \\
                        &\varepsilon_6\to \varepsilon_5\phantom{-} \quad\quad
                 \Bar{\varepsilon}_6\to \varepsilon_5-\Bar{\varepsilon}_5 
 \end{aligned}
\end{equation}
Consistently, the orientifold plane \eqref{orid} is invariant under the 
$\Omega\Bar{\sigma}$ action.
For the other three orientifold models,
the results for the O$6$ planes and the action of $\Omega\Bar{\sigma}$ on the
homology lattice can be found in appendix \ref{oplaneszfour}.
In principle, we have now provided all the information that is necessary to
build intersecting brane world models on the $\mathbb{Z}_4$ orientifold.
However, since we are particularly interested in supersymmetric models we need
to have control not only over topological data of the D$6$-branes but
over the nature of the sLag cycles as well.

\subsection{\label{susycyclsec}Supersymmetric cycles}

The metric at the orbifold point is flat up to some isolated
orbifold singularities. Therefore, flat D$6$-branes in a given
homology class are definitely special Lagrangian.
We restrict our D$6$-branes to be flat and factorizable
in the sense that they can be described by six wrapping numbers,
$(n_I,m_I)$ with $I=1,2,3$, along the fundamental toroidal cycles,
where for each $I$ the integers $(n_I,m_I)$ are relatively co-prime.
Given such a bulk brane, one can compute the homology class
that it wraps expressed in the $\mathbb{Z}_4$ basis
\begin{equation}
   \label{bulki}
  \begin{aligned}
    \pi^{\text{bulk}}_a=&\left[ (n_{a,1}\,n_{a,2} -
                    m_{a,1}\,m_{a,2}) n_{a,3}\right]\, \rho_1 +
                    \left[(n_{a,1}\,m_{a,2} + m_{a,1}\,n_{a,2})
                   n_{a,3}\right]\, \rho_2  \\
                  +&\left[(n_{a,1}\,n_{a,2} - m_{a,1}\,m_{a,2})
                    m_{a,3}\right]\, \Bar{\rho}_1 +
                    \left[(n_{a,1}\,m_{a,2} + m_{a,1}\,n_{a,2})
                m_{a,3}\right]\, \Bar{\rho}_2 
 \end{aligned}
\end{equation}
For the ${\bf ABB}$ orientifold,
 the condition that such a D$6$-brane preserves the same supersymmetry
as the orientifold plane is simply
\begin{equation}
 \label{susygh}
         \varphi_{a,1}+ \varphi_{a,2}+ \varphi_{a,3}={\pi\over 4}
                 \;\mod\ 2\pi
\end{equation}
with
\begin{equation}
  \label{tangi} 
            \tan\varphi_{a,1}={m_{a,1}\over n_{a,1}}, \quad
              \tan\varphi_{a,2}={m_{a,2}\over n_{a,2}}, \quad
           \tan\varphi_{a,3}={U_2\, m_{a,3}\over n_{a,3}+{1\over 2} m_{a,3} } 
\end{equation}
Taking the $\tan(...)$  on both sides of equation \eqref{susygh} we can
 reformulate the supersymmetry condition in terms of wrapping numbers
(Note, that this only yields a necessary condition as  $\tan(...)$ is just
 periodic mod $\pi$.)
\begin{equation}
 \label{susywar}
    U_2={ \left(n_{a,3}+{1\over 2} m_{a,3}\right) \over m_{a,3}}
          { \left(n_{a,1}\,n_{a,2} - m_{a,1}\,m_{a,2} - n_{a,1}\,m_{a,2} -
               m_{a,1}\,n_{a,2} \right)
           \over \left(n_{a,1}\,n_{a,2} - m_{a,1}\,m_{a,2} + n_{a,1}\,m_{a,2} +
            m_{a,1}\,n_{a,2} \right)}
\end{equation}
Therefore, the complex structure of the third torus in general is
already fixed by one supersymmetric D-brane. In case one introduces
more D$6$-branes, one gets non-trivial conditions on the wrapping numbers
of these D-branes.
The supersymmetry conditions for the other three orientifold models
are summarized in  appendix \ref{susyz4} (p. \pageref{susyz4}).

Working only with the bulk branes \eqref{bulki},
 the model building possibilities
are  very restricted. In particular, it seems to be impossible
to get large enough gauge groups to accommodate the Standard Model
gauge symmetry, $U(3)\times U(2)\times U(1)$,  of at least rank six.
One such supersymmetric model with only bulk branes and rank four
has been constructed in \cite{Blumenhagen:2002wn}.
Now, to enlarge the number of possibilities, we also allow such
flat, factorizable branes to pass through $\mathbb{Z}_2$ fixed points and
split into fractional D-branes. Thus, according to our rule
we allow fractional D-branes wrapping the cycle
\begin{equation}
  \label{fractio}
   \pi_a^{\text{frac}}={1\over 2} \pi_a^{\text{bulk}} + {n_{a,3} \over 2}
                        \left[ \sum_{j=1}^6  w_{a,j} \varepsilon_j\right]
                  +  {m_{a,3 }\over 2}
                     \left[ \sum_{j=1}^6  w_{a,j} \Bar{\varepsilon}_j\right] 
\end{equation}
with $w_{a,j}\in\{ 0,\pm 1\}$.  To make contact with the formerly introduced
coefficients $v_{a,j}$ , we define
\begin{equation}
 \label{formercoef}
                 v_{a,j}={n_{a,3} \over 2} \, w_{a,j},\quad\quad
                 v_{a,j+8}={m_{a,3} \over 2} \, w_{a,j}
\end{equation}
for $j\in\{1,\ldots,8\}$.
In \eqref{fractio} we have taken into account that the $\mathbb{Z}_2$ fixed points
all lie on the first two two-dimensional tori and that on the third torus
fractional D-branes do have  winding numbers along the two fundamental
1-cycles. Moreover, since $\varepsilon_j$
and $\Bar{\varepsilon}_j$ only differ by the cycle on the third torus,
their coefficients in \eqref{fractio} must indeed be equal.

These fractional D$6$-branes do correspond to the following boundary states
in the conformal field theory of the $T^6/\mathbb{Z}_4$ orbifold model
\begin{multline}
  \label{bound}
       \ket{D^\text{frac};(n_I,m_I),\alpha_{ij}}= 
   \\
       {1\over 4\sqrt 2} \left(\prod_{j=1}^2 \sqrt{n_j^2+m_j^2}\right)\,
           \sqrt{n_3^2+n_3 m_3 +{\textstyle{m_3^2\over 2}}} 
          \biggl(
     \bigl|D;(n_I,m_I)\bigr\rangle_U +  
             \bigl|D;\Theta(n_I,m_I)\bigr\rangle_U\biggr)
   \\
      +{1\over 2\sqrt 2} \,
           \sqrt{n_3^2+n_3 m_3 +\textstyle{{m_3^2\over 2}}}
       \phantom{dfssssssssdsdfadssdddsssssadsadssdfsdfsdffsd}
   \\
    \times \biggl(
     \sum_{i,j=1}^{4} \alpha_{ij} \bigl|D;(n_I,m_I),e_{ij}\bigr\rangle_T + 
     \sum_{i,j=1}^{4} \alpha_{ij} \bigl|D;\Theta(n_I,m_I),\Theta(e_{ij})
                  \bigr\rangle_T \biggr)
\end{multline}
In the schematic form of the boundary state \eqref{bound} there are
 contributions
from both the untwisted and the $\mathbb{Z}_2$ twisted sector and we have taken
the orbit under the $\mathbb{Z}_4$ symmetry $\Theta$ with the following action
on the winding numbers
\begin{equation}
  \label{windact}
    \Theta(n_{1,2},m_{1,2})=(-m_{1,2}, n_{1,2}),\quad\quad
               \Theta(n_{3},m_{3})=-(n_3, m_{3}) 
\end{equation}
implying that $\Theta^2$ acts like the identity on the boundary
states. This explains why only two and not four untwisted boundary
states do appear in \eqref{bound}. Note, that in the sum over the $\mathbb{Z}_2$
fixed points, for each D$6$-brane precisely  four coefficients take
values $\alpha_{ij}\in\{-1,+1\}$ and the remaining ones are
vanishing. The $\alpha_{ij}$ are of course directly related to the
coefficients $w_i$ appearing in the description of the
corresponding fractional 3-cycles. For the interpretation of these
coefficients $\alpha_{ij}$, one has to remember that changing the
sign of $\alpha_{ij}$ corresponds to turning on a discrete $\mathbb{Z}_2$
Wilson line along one internal direction of the brane
\cite{Sen:1998ii,Gaberdiel:2000jr}. The action of $\Theta$ on the twisted sector
ground states $e_{ij}$ is the same as in \eqref{actionthe}. The
elementary boundary states like $|D;(n_I,m_I)\rangle_U$ are the
usual ones for a flat D$6$ brane with wrapping numbers $(n_I,m_I)$
on $T^6=T^2\times  T^2\times T^2$ and can be found in Appendix C.
The important normalization factors in \eqref{bound} are fixed by the
Cardy condition (cf.\ \cite{Cardy1989}), stating that the result for the 
annulus partition function must coincide for  the loop- and the tree-channel
computation.

Since  the brane and its $\mathbb{Z}_4$ image only break the
supersymmetry down to ${\cal N}=2$ , one gets a ${\cal N}=2$
$U(N)$ vector multiplet on each stack of fractional D-branes. The
scalars in these vector multiplets correspond to the position of
the D$6$-brane on the third $T^2$ torus, which is still an open
string modulus.

Coming back to the homology cycles, following our general rule for
fractional branes imposes further constraints on the
coefficients because only those exceptional cycles
are allowed to contribute which are intersected by the flat
D-brane. The only allowed exceptional 3-cycles are summarized in
table \ref{exceptcycles}, depending on the wrapping numbers of the 
first two tori $T^2$. 
\begin{table}
 \renewcommand{\arraystretch}{1.2}
 \begin{center}
 \begin{tabular}{|c|c|c|c|}
   \cline{2-4}
   \multicolumn{1}{c|}{}
   & $n_1$ odd, $m_1$ odd & $n_1$ odd, $m_1$ even & $n_1$ even, $m_1$ odd  \\
   \hline
   $n_2$ odd &    & $\varepsilon_3$,  $\varepsilon_4$ 
         & $\varepsilon_3$, $\varepsilon_4$  \\
   $m_2$ odd &    & $\varepsilon_5$,  $\varepsilon_6$ 
         &   $\varepsilon_5$, $\varepsilon_6$ \\
   \hline 
         &  $\varepsilon_1$,  $\varepsilon_2$  
         & $\varepsilon_1$, $\varepsilon_3$, $\varepsilon_5$  
           & $\varepsilon_1$,$\varepsilon_3$, $\varepsilon_6$   \\
    $n_2$ odd  & $\varepsilon_5$,  $\varepsilon_6$  
         & $\varepsilon_1$,$\varepsilon_4$, $\varepsilon_6$  
           & $\varepsilon_1$,$\varepsilon_4$, $\varepsilon_5$   \\
    $m_2$ even &  & $\varepsilon_2$,$\varepsilon_3$, $\varepsilon_6$  
            & $\varepsilon_2$,$\varepsilon_3$, $\varepsilon_5$  \\
         &  & $\varepsilon_2$,$\varepsilon_4$, $\varepsilon_5$  
            & $\varepsilon_2$,$\varepsilon_4$, $\varepsilon_6$  \\
   \hline
   &  $\varepsilon_1$,  $\varepsilon_2$  & $\varepsilon_1$,
       $\varepsilon_3$, $\varepsilon_6$  & $\varepsilon_1$,
       $\varepsilon_3$, $\varepsilon_5$   \\
   $n_2$ even & $\varepsilon_5$,  $\varepsilon_6$  & $\varepsilon_1$,
       $\varepsilon_4$, $\varepsilon_5$  & $\varepsilon_1$,
       $\varepsilon_4$, $\varepsilon_6$  \\
   $m_2$ odd &  & $\varepsilon_2$,
       $\varepsilon_3$, $\varepsilon_5$  & $\varepsilon_2$,
       $\varepsilon_3$, $\varepsilon_6$  \\
  &  & $\varepsilon_2$,
       $\varepsilon_4$, $\varepsilon_6$  & $\varepsilon_2$,
       $\varepsilon_4$, $\varepsilon_5$  \\
   \hline
 \end{tabular}
 \caption{\label{exceptcycles}Allowed exceptional cycles}
 \end{center}
\end{table}
At first glance, there is a
mismatch between the number of parameters describing a 3-cycle and
the corresponding  boundary state. For each D$6$-brane there are
three non-vanishing parameters $w_i$ but four $\alpha_{ij}$.
However, a flat fractional brane and its $\mathbb{Z}_4$ image always
intersect in precisely one $\mathbb{Z}_4$ fixed point times a circle on
the third $T^2$.

\begin{figure}
 \begin{center}
  \begin{minipage}[b]{7.0cm}
   \setlength{\unitlength}{0.1in}
   \begin{picture}(50,30)
   \SetFigFont{14}{20.4}{\rmdefault}{\mddefault}{\updefault}
   \put(0,0){\scalebox{0.5}{\includegraphics{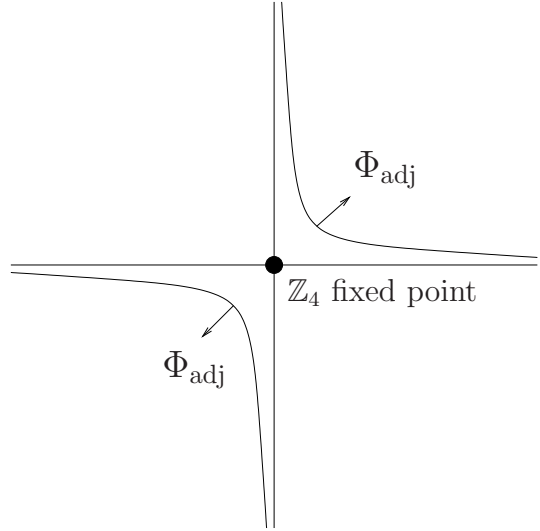}}}
   \put(8,8){$\Phi_\text{adj}$}
   \put(14,12){\SetFigFont{12}{20.4}{\rmdefault}{\mddefault}{\updefault}
              $\mathbb{Z}_4$ fixed point}
   \put(18,18.5){$\Phi_\text{adj}$}
   \end{picture}
  \end{minipage}
 \end{center}  
 \begin{center} 
   \caption{\label{branerecpic}Recombined branes}
 \end{center}
\end{figure}

Since $\Theta$ acts on this fixed locus  with a minus sign, this
twisted sector effectively drops out of the boundary state \eqref{bound}.
A different way of saying this is that at the intersection between
the brane and its $\mathbb{Z}_4$ image, there lives a hyper-multiplet,
$\Phi_\text{adj}$, in the adjoint representation. Since it is an
${\cal N}=2$ super-multiplet, there exists a flat direction in the
D-term potential corresponding to the recombination of the
two branes into a single brane. This single brane of course no longer
runs to the $\mathbb{Z}_4$ invariant fixed point.
This brane recombination process is depicted in figure \ref{branerecpic}.
\noindent

A non-trivial test for our
considerations is the condition that a fractional brane \eqref{fractio}
transformed to the $E_8$-basis must have  integer
coefficients. To see this, we write the $8\times8$ matrix \eqref{cartan}
and a second identical copy as the two diagonal blocks of a
$16\times16$ matrix, and then act with the inverse of the
transposed matrix onto a general vector \eqref{fractio}. Then we have
to investigate the different cases according to table \ref{exceptcycles}
separately. For instance for the case $n_1$ odd, $n_2$ odd, $m_1$
even, $m_2$ odd and fractional cycles $\varepsilon_3$,
$\varepsilon_4$ with signs $w_{3}$, $w_{4}$ respectively, we
substitute $m_1=2k_1$
and  obtain the following vector in the $E_8$-basis: 
\begin{equation}
  \label{proofinteger}
  \bigg[
    \Big({1\over 2 }(n_{1} m_{2}-w_{3})+k_{1} n_{2}
    \Big)n_{3},\
    \Big({1 \over 2 }(n_{1} n_{2}-w_{3}) -k_{1}m_{2}+n_{1} m_{2}+2
         k_{1} n_{2}
    \Big)n_{3},\ldots
    \bigg]
\end{equation}
 Already for the first two
components we can see what generally happens for all cases and
components: since $n_1$, $n_2$, $m_2$ and $w_{3}$ are
non-vanishing and because products of odd numbers are also odd, just sums
and differences of two odd numbers occur and these are always even or
zero and therefore can be divided by 2 and still lead to
integer coefficients.
Having defined  a well understood set of supersymmetric
fractional D$6$-branes, we are now in
the position to search  for phenomenologically
interesting supersymmetric intersecting brane worlds.

\section[A four generation supersymmetric Pati-Salam model]
        {\label{pssec}\centerline{A four generation supersymmetric}
          \centerline{Pati-Salam model}
        }
In this section we present the construction of a semi-realistic
supersymmetric intersecting brane world model. This provides  an
application of the formalism developed in the previous sections.
It turns out that the ${\bf ABB}$ orientifold model is the most
appropriate one for doing this. Using the fractional D$6$-branes
introduced in the last section, one finds that by requiring that
no (anti-) symmetric representations of the
$U(N_a)$ gauge groups do appear, only very few sufficiently small mutual
intersection numbers arise. For the ${\bf ABB}$ model with the
complex structure of the last torus being $U_2=1$, an extensive
computer search reveals that essentially only mutual intersection
numbers $(\pi_a\circ \pi_b,\pi'_a\circ \pi_b)=(0,0),(\pm 2,\mp
2)$ are possible. Even with these intersection numbers it is
possible to construct a four generation
supersymmetric Pati-Salam model with initial gauge group
$U(4)\times U(2)\times U(2)$. A typical model of this sort can be
realized by the three stacks of D$6$-branes presented in table \ref{branes4genmod}.

\begin{table}[h]
 \renewcommand{\arraystretch}{1.2}
 \begin{center}
 \begin{tabular}{|c|c|l|}
   \hline
   Stack  & $(n_I,m_I)$  & \multicolumn{1}{c|}{Homology cycle} \\
   \hline\hline
    &  & $\pi_1={1\over 2}\left(
          \rho_1+\rho_2-\varepsilon_5+\varepsilon_6\right)$  \\
   \raisebox{1.5ex}{ U(4)} & \raisebox{1.5ex}{$(-1,0;1,1;-1,0)$} 
         & $\pi'_1={1\over 2}\left(
          \rho_1+\rho_2+\varepsilon_5-\varepsilon_6\right)$  \\
   \hline
     &    & $\pi_2={1\over 2}(-\rho_1+\rho_2+ 2\Bar{\rho}_1
                                   -2\Bar{\rho}_2$ \\ 
     &    &  
           \multicolumn{1}{r|}{$+\varepsilon_5-\varepsilon_6
                                   -2\Bar{\varepsilon}_5+2\Bar{\varepsilon}_6
                                     )$}    \\
   \raisebox{1.5ex}{ U(2)}       & \raisebox{1.5ex}{$(0,1;-1,-1;-1,2)$ }
         & $\pi'_2={1\over 2}(-\rho_1+\rho_2+2\Bar{\rho}_1-2\Bar{\rho}_2
                             \phantom{)}$ \\
         &                       & 
           \multicolumn{1}{r|}  {$ +\varepsilon_5-\varepsilon_6
                               -2\Bar{\varepsilon}_5+2\Bar{\varepsilon}_6)$} \\
   \hline
   &  
         & $\pi_3= {1\over 2} ( -\rho_1+\rho_2+2\Bar{\rho}_1
                                   -2\Bar{\rho}_2$ \\
         &        &
           \multicolumn{1}{r|}{$-\varepsilon_5+\varepsilon_6+
                                   2\Bar{\varepsilon}_5-2\Bar{\varepsilon}_6 
                                   )$}              \\
 \raisebox{1.5ex}{ U(2)}   & \raisebox{1.5ex}{$(-1,0;1,-1;1,-2)$}
         & $\pi'_3={1\over 2}(-\rho_1+\rho_2+2\Bar{\rho}_1
                                   -2\Bar{\rho}_2  $  \\
         &                 &
           \multicolumn{1}{r|}{$-\varepsilon_5+\varepsilon_6
                             +2\Bar{\varepsilon}_5-2\Bar{\varepsilon}_6)$} \\
 \hline
 \end{tabular}
 \caption{\label{branes4genmod}D$6$-branes for a 4 generation PS-model}
 \end{center}
\end{table}
Computing the intersection numbers for these D$6$-branes and using
the general formula for the chiral massless spectrum, one gets
the massless modes shown in table \ref{chiralspec4genmod}.
\begin{table}
 \renewcommand{\arraystretch}{1.2}
 \begin{center}
 \begin{tabular}{|c|clc|}
  \hline   
       n  &  \multicolumn{3}{c|}{$SU(4)\times SU(2)\times SU(2)\times U(1)^3$}  \\
  \hline\hline 
       2 &\rule{6.5ex}{0ex} &$(4,2,1)_{(1,-1,0)}$        & \\
       2 & &$(4,2,1)_{(1,1,0)}$         & \\
  \hline 
       2 & &$(\Bar{4},1,2)_{(-1,0,1)}$  & \\  
       2 & &$(\Bar{4},1,2)_{(-1,0,-1)}$ & \\
  \hline
\end{tabular}
 \caption{\label{chiralspec4genmod}Chiral spectrum for  4
              generation PS-model}
 \end{center}
\end{table}
Here we have normalized  as usual the gauge
fields in  the diagonal $U(1)_a\subset U(N_a)$ sub-algebras as
\begin{equation}
  \label{normi} 
          A^\mu_{U(1)_a}={1\over N_a}\,{\rm Tr}
          \left(A^\mu_{U(N_a)}\right) 
\end{equation}
Note, that all non-abelian gauge
anomalies are canceled. Adding up all homological cycles, one
finds that the RR-tadpole cancellation condition \eqref{tadhom} is
indeed satisfied. A nice check is whether the NS-NS tadpole
cancellation condition \eqref{susy} is also satisfied, as it should be
for a globally supersymmetric configuration. For the contribution
of the O6-plane to the scalar potential, one finds 
\begin{equation}
 \label{tens}
        V_{\text{O}6}=-T_6\, e^{-\phi_4} 16 \sqrt{2}
              \left(
                    {1\over \sqrt{U_2}} +2 \sqrt{U_2} 
              \right) 
\end{equation}
whereas the three stacks of D$6$-branes give
\begin{equation}
  \label{tenso}
  \begin{aligned}
         V_{1}&=T_6\, e^{-\phi_4} 16 \sqrt{2} {1\over \sqrt{U_2}} \\
         V_{2,3}&=T_6\, e^{-\phi_4} 16 \sqrt{2} \sqrt{U_2} 
  \end{aligned}
\end{equation}
We see that the scalar potential vanishes for all values
of the complex structure $U_2$ of the third torus.
Thus, the disc level scalar potential indeed vanishes and
we have constructed a globally supersymmetric
intersecting brane world model with gauge group $U(4)\times
U(2)\times U(2)$.

\subsection{Green-Schwarz mechanism}
Computing in the usual way the mixed $U(1)_a-SU(N_b)^2$ anomalies,
one confirms the general result derived in \cite{Blumenhagen:2002wn} 
\begin{equation}
   \label{mixed}
    A_{ab}={N_a\over 4}(-\pi_a+\pi'_a)\circ(\pi_b+\pi'_b) 
\end{equation}
In our example there is only one anomalous $U(1)$ while $U(1)_2$ and
$U(1)_3$ are anomaly-free. This anomaly is canceled by some
generalized Green-Schwarz mechanism involving the axionic
couplings from the Chern-Simons terms in the effective action on
the D$6$-branes\;\footnote{The sign in front of $v'_{a,i}$ is due to the fact
 that $ F_{a'}=- F_{a}$. Similarly this sign cancels in \eqref{gsmechb}.} 
\begin{equation}
   \label{gsmech}
       S^F_{CS}=  \sum_{i=1}^{b_3}    
         \int d^4 x\,
              N_a\,  (v_{a,i} - v'_{a,i} )\, B_i\wedge F_a  
\end{equation}
and
\begin{equation}
   \label{gsmechb}
            S^{F\wedge F}_{CS}=  \sum_{i=1}^{b_3}    
        \int d^4 x\,
              (v_{b,i} + v'_{b,i} )\, \Phi_i\, {\rm Tr}(F_b\wedge F_b) 
\end{equation}
where $B_i$ is defined as the integral of the RR 5-form over the
corresponding 3-cycle and similarly $\Phi_i$ is defined
as the integral of the RR 3-form over the
corresponding 3-cycle.
Taking into account the Hodge duality between the fields $B_i$ and
$\Phi_{i+8}$ these axionic couplings indeed cancel the mixed anomalies.
For more details we refer the reader to the general discussion
in \cite{Blumenhagen:2002wn}.

As was pointed out in \cite{Ibanez:2001nd} the couplings \eqref{gsmech}
can generate a mass term  for $U(1)$ gauge fields
even if they are not anomalous.
The massless $U(1)$s are given by the kernel
of the matrix
\begin{equation} 
   \label{matric}
       M_{ai}= N_a (v_{a,i} - v'_{a,i} )  
\end{equation}
In our model it can be easily seen that $U(1)_2$ and $U(1)_3$
remain massless, so that the final gauge symmetry
is $SU(4)\times SU(2)\times SU(2)\times U(1)^2$.
We will not discuss this model any further but move forward
to the construction of a more realistic model with three generations.

\section[Three generation supersymmetric Pati-Salam model]
        {\label{3genpssec}\centerline{Three generation supersymmetric}
           \\ \centerline{Pati-Salam model}}

For the ${\bf ABB}$ model with the complex structure of the last
torus fixed at $U_2=1/2$, a computer search shows that only sufficiently small
mutual intersection numbers $(\pi_a\circ \pi_b,\pi'_a\circ
\pi_b)=(0,0),(\pm 1,0),(0,\pm 1)$ are possible. These numbers
allow the construction of a three generation model in the
following way. First, we start with seven stacks of D$6$-branes with
an initial gauge symmetry $U(4)\times U(2)^6$ and choose  the
wrapping numbers as shown in table \ref{d63genmod}.
\begin{table}[h]
 \renewcommand{\arraystretch}{1.2}
 \begin{center}
 \begin{tabular}{|c|l|}
\hline
  Stack    & \multicolumn{1}{c|}{Homology cycle ($\rho,\,\epsilon$-basis)} \\
  \hline\hline
         & $\pi_1={1\over 2}
           \left(\Bar{\rho}_1-\Bar{\varepsilon}_1-\Bar{\varepsilon}_3
                 -\Bar{\varepsilon}_5\right)$ \\
    \raisebox{1.5ex}{$U(4)_1$}  
          & $\pi'_1={1\over 2}
           \left(\rho_2-\Bar{\rho}_2+\varepsilon_1-\varepsilon_3
                 -\varepsilon_6-\Bar{\varepsilon}_1+\Bar{\varepsilon}_3
                 +\Bar{\varepsilon}_6 \right)$ \\
  \hline
  \hline
           & $\pi_2={1\over 2}
              \left(\Bar{\rho}_1-\Bar{\varepsilon}_1+\Bar{\varepsilon}_3
                    +\Bar{\varepsilon}_5\right)$ \\
     \raisebox{1.5ex}{$U(2)_2$} 
           & $\pi'_2={1\over 2}
              \left(\rho_2-\Bar{\rho}_2+\varepsilon_1+\varepsilon_3
                    +\varepsilon_6-\Bar{\varepsilon}_1
                    -\Bar{\varepsilon}_3-\Bar{\varepsilon}_6 \right)$ \\
  \hline
           & $\pi_3={1\over 2}
              \left(\Bar{\rho}_1-\Bar{\varepsilon}_2
                    +\Bar{\varepsilon}_3+\Bar{\varepsilon}_6\right)$ \\
     \raisebox{1.5ex}{$U(2)_3$}  
           & $\pi'_3={1\over 2}
              \left(\rho_2-\Bar{\rho}_2+\varepsilon_2+\varepsilon_3
                    +\varepsilon_5-\Bar{\varepsilon}_2
                    -\Bar{\varepsilon}_3-\Bar{\varepsilon}_5 \right)$ \\
  \hline
           & $\pi_4={1\over 2}
             \left(\Bar{\rho}_1+\Bar{\varepsilon}_2+\Bar{\varepsilon}_3
                   +\Bar{\varepsilon}_6\right)$ \\
        \raisebox{1.5ex}{$U(2)_4$}   
           & $\pi'_4={1\over 2}
             \left(\rho_2-\Bar{\rho}_2-\varepsilon_2+\varepsilon_3
                   +\varepsilon_5+\Bar{\varepsilon}_2
                   -\Bar{\varepsilon}_3-\Bar{\varepsilon}_5 \right)$ \\
  \hline
  \hline
           & $\pi_5={1\over 2}
             \left(\Bar{\rho}_1+\Bar{\varepsilon}_1-\Bar{\varepsilon}_3
                   +\Bar{\varepsilon}_5\right)$ \\
   \raisebox{1.5ex}{$U(2)_5$} 
           & $\pi'_5={1\over 2}
             \left(\rho_2-\Bar{\rho}_2-\varepsilon_1-\varepsilon_3
                   +\varepsilon_6+\Bar{\varepsilon}_1+\Bar{\varepsilon}_3
                   -\Bar{\varepsilon}_6 \right)$ \\
\hline
           & $\pi_6={1\over 2}
             \left(\Bar{\rho}_1+\Bar{\varepsilon}_1+\Bar{\varepsilon}_4
                   -\Bar{\varepsilon}_6\right)$ \\
   \raisebox{1.5ex}{$U(2)_6$}
           & $\pi'_6={1\over 2}
             \left(\rho_2-\Bar{\rho}_2-\varepsilon_1+\varepsilon_4
                   -\varepsilon_5+\Bar{\varepsilon}_1
                   -\Bar{\varepsilon}_3+\Bar{\varepsilon}_5 \right)$ \\
  \hline
           & $\pi_7={1\over 2}
             \left(\Bar{\rho}_1+\Bar{\varepsilon}_1-\Bar{\varepsilon}_4
                   -\Bar{\varepsilon}_6\right)$ \\
   \raisebox{1.5ex}{$U(2)_7$}
           & $\pi'_7={1\over 2}
             \left(\rho_2-\Bar{\rho}_2-\varepsilon_1-\varepsilon_4
                   -\varepsilon_5+\Bar{\varepsilon}_1
                   +\Bar{\varepsilon}_3+\Bar{\varepsilon}_5 \right)$ \\ 
\hline
 \end{tabular}
 \caption[D$6$-branes for 3 generation PS-model]
  {\label{d63genmod} D$6$-branes for 3 generation PS-model.
             All the branes have wrapping numbers $(n_I,m_I)=(1,0;1,0;0,1)$
             for the untwisted part.}
 \end{center}
\end{table} 

Adding up all
homological 3-cycles, one realizes that the RR-tadpole
cancellation condition is satisfied. The contribution of the
O6-plane tension to the scalar potential is 
\begin{equation}
  \label{tensa}
    {V}_{\text{O}6}=-T_6\, e^{-\phi_4} 16 \sqrt{2}\left(
                    {1\over \sqrt{U_2}} +2 \sqrt{U_2} \right) 
\end{equation}
whereas the seven stacks of D$6$-branes give
\begin{equation}
  \label{tensob}
    \begin{aligned}
      {V}_{1}&=T_6\, e^{-\phi_4} 16 \sqrt{ {1\over 4\, U_2}+ U_2}       \\
      {V}_{2,\ldots,7}&=T_6\, e^{-\phi_4} 8 \sqrt{{1\over 4\, U_2}+ U_2}
    \end{aligned}
\end{equation}
Adding up all terms, one finds that indeed the NS-NS tadpole
vanishes just for $U_2={1\over 2}$. This means that in contrast to
the four generation model, here supersymmetry really fixes the
complex structure of the third torus (if we assume that no curved
 sLags exist in a neighborhood of $U_2={1\over 2}$).
 This freezing of moduli for
supersymmetric backgrounds is very similar to what happens for
instance in recently discussed compactifications with
non-vanishing R-R fluxes \cite{Giddings:2001yu,Giddings:2001yu}.

In terms of ${\cal N}=2$ super-multiplets, the model contains
vector multiplets in the gauge group $U(4)\times U(2)^3\times
U(2)^3$ and in addition two hyper-multiplets in the adjoint
representation of each unitary gauge factor. The complex scalar in
the vector multiplet corresponds to the unconstrained position of
each stack of D$6$-branes on the third $T^2$. As described at the end of
section \ref{susycyclsec}, the hyper-multiplet appears on the intersection
between a stack of branes and its $\mathbb{Z}_4$ image. By computing the
intersection numbers, we derive the chiral spectrum as shown in
table \ref{chiralspec7stack}, where $n$ denotes the number of 
chiral multiplets in the
respective representation as given by the intersection number.

\begin{table}
 \renewcommand{\arraystretch}{1.2}
 \begin{center}
 \begin{tabular}{|c|c|c|}
 \hline
 Field & n  & $U(4)\times U(2)^3\times U(2)^3$   \\
 \hline\hline
    $\Phi_{1'2}$ & 1 & $(4;2,1,1;1,1,1)$ \\
    $\Phi_{1'3}$ & 1 & $(4;1,{2},1;1,1,1)$ \\
    $\Phi_{1'4}$ & 1 & $(4;1,1,{2};1,1,1)$ \\
   \hline    
    $\Phi_{1'5}$ & 1 & $(\Bar{4};1,1,1;\Bar{2},1,1)$ \\
    $\Phi_{1'6}$ & 1 & $(\Bar{4};1,1,1;1,\Bar{2},1)$ \\
    $\Phi_{1'7}$ & 1 & $(\Bar{4};1,1,1;1,1,\Bar{2})$ \\
   \hline\hline
    $\Phi_{2'3}$ & 1 & $(1;\Bar{2},\Bar{2},1;1,1,1)$ \\
    $\Phi_{2'4}$ & 1 & $(1;\Bar{2},1,\Bar{2};1,1,1)$ \\
    $\Phi_{3'4}$ & 1 & $(1;1,\Bar{2},\Bar{2};1,1,1)$ \\
   \hline
    $\Phi_{5'6}$ & 1 & $(1;1,1,1;2,{2},1)$ \\
    $\Phi_{5'7}$ & 1 & $(1;1,1,1;2,1,{2})$ \\
    $\Phi_{6'7}$ & 1 & $(1;1,1,1;1,{2},{2})$ \\
\hline
\end{tabular}
 \caption{\label{chiralspec7stack}Chiral spectrum for a 7-stack model}
 \end{center}
\end{table}
First, we notice that all non-abelian anomalies cancel including
formally also the $U(2)$ anomalies.

In order to proceed and really get a three generation model, it is
necessary to break the two triplets $U(2)^3$ down to their
diagonal subgroups. Potential gauge symmetry breaking candidates
in this way are the chiral fields
$\{\Phi_{2'3},\Phi_{2'4},\Phi_{3'4}\}$ and
$\{\Phi_{5'6},\Phi_{5'7},\Phi_{6'7}\}$ from table \ref{chiralspec7stack}. However, one
has to remember that these are chiral ${\cal N}=1$ super-multiplets
living on the intersection of two D-branes in every case. Let us
review what massless bosons localized on intersecting D-branes
indicate.

\subsection{Brane recombination}
If two stacks of D-branes preserve a common ${\cal N}=2$ supersymmetry, then
a massless hyper-multiplet, $H$, localized on the intersection,
signals a possible deformation of the two stacks of D-branes into 
recombined D-branes which wrap a complex cycle. Note, that two
factorizable branes can only preserve ${\cal N}=2$ supersymmetry
if they are parallel on one of the three $T^2_I$ tori.  The
complex cycle has the same volume as the sum of volumes of the two
D-branes before the recombination process occurs. In the effective
low energy theory, this recombination can be understood as a Higgs
effect where a flat direction $\langle h_1\rangle=\langle
h_2\rangle$ in the  D-term potential 
\begin{equation}
  \label{dtermpo}
     V_D={1\over 2 g^2}\left( h_1 \Bar{h}_1 - h_2 \Bar{h}_2\right)^2 
\end{equation}
exists, along
which the $U(N)\times U(N)$ gauge symmetry is broken to the
diagonal subgroup.\footnote{If on one of the two stacks there sits
 only a single D$6$-brane, the F-term potential
$\phi h_1 h_2$ forbids the existence of a flat direction
with $\langle h_1\rangle=\langle
h_2\rangle$. This is the field theoretic correspondence  of the fact
that there do not exist large instantons in the $U(1)$ 
gauge group. We thank A. Uranga for pointing this out to us.} 
Here $h_1$ and $h_2$ denote the two complex
bosons inside the hyper-multiplet. Thus, in this case without
changing the closed string background, there exists an open string
modulus, which has the interpretation of a Higgs field in the low
energy effective theory. Note, that in the T-dual picture, this is
just the deformation of
a small instanton into an instanton of finite size.
In our concrete models such ${\cal N}=2$
Higgs sectors are coupled at brane intersections to chiral ${\cal
N}=1$ sectors. Note, that the brane recombination in the effective
gauge theory cannot simply be described by the renormalizable
couplings. In order to get the correct light spectrum, one also
has to take into account stringy higher dimensional couplings.

When the two D-branes only preserve ${\cal N}=1$ supersymmetry and
support a massless chiral super-multiplet $\Phi$ on the
intersection \cite{Kachru:1999vj,Witten:2000mf}, the situation gets a little bit
more involved. In this case, the analogous D-term potential is of
the form 
\begin{equation}
    \label{dtermpoc} V_D={1\over 2 g^2}\left( \phi  \Bar{\phi}\right)^2
\end{equation}
 which tells us that, unless there are more chiral fields involved, 
simply by giving a VEV to the massless boson
 $\phi$,
we do not obtain a flat direction of the D-term potential and
therefore break supersymmetry. Nevertheless, the massless modes
indicate that the intersecting brane configuration lies on a line
of marginal stability in the complex structure moduli space. By a
small variation of the complex structure, a Fayet-Iliopoulos (FI)
term, $r$, is introduced that changes the D-term potential  to
\begin{equation}
  \label{dtermpod}
       V_D={1\over 2 g^2}\left( \phi \Bar{\phi} +r \right)^2 
\end{equation}
 Therefore, for $r<0$ the field $\phi$ becomes
tachyonic and there exists a new stable supersymmetric minimum of
the D-term potential. The intersecting branes then have combined
into one D-brane wrapping a special Lagrangian 3-cycle in the
underlying Calabi-Yau. For a finite FI-term $r$, this 3-cycle has
smaller volume than the two intersecting branes. However, the two
volumes are precisely equal on the line of marginal stability.
This means that on the line of marginal stability, there exists a
different configuration with only a single brane which also
preserves the same ${\cal N}=1$ supersymmetry and has the same
volume as the former pair of intersecting D-branes. Again the
gauge symmetry is broken to the diagonal subgroup. It has to be
emphasized that in this case the two configurations are not simply
linked by a Higgs mechanism in the effective low energy gauge
theory. As mentioned before, in order to deform the intersecting
brane configuration into the non-flat D-brane wrapping a special
Lagrangian 3-cycle, one first has to deform the closed string
background and then let the tachyonic mode condense. Therefore,
the description of this process is intrinsically stringy and
should be better described by string field theory rather than the
effective low  energy gauge theory\footnote{In the context
of so-called quasi-supersymmetric intersecting brane world models
\cite{Cremades:2002cs} , it has been observed that indeed the brane
recombination of ${\cal N}=1$ supersymmetric intersections cannot
simply be described by a Higgs mechanism of massless modes. It was
suggested there that the stringy nature of this transition has the
meaning that also some massive, necessarily non-chiral, fields are
condensing during the brane recombination. At least from the
effective gauge theory point of view, this could induce the right
mass terms which are necessary for an understanding of the new
massless modes after the recombination. We leave it for future
work to find the right effective description of this transition,
but we can definitely state that it must involve some stringy
aspects as the complex structure changes, i.e. the closed string
background.}. For $r>0$, the non-supersymmetric intersecting
branes are stable and have a smaller volume than the recombined
brane. The lift of these brane recombination
processes to M-theory was discussed in \cite{Uranga:2002ag}.

After this little excursion, we come back to our model. We have
seen that the condensation of hyper-multiplets is under much better
control than the condensation of chiral multiplets. Therefore, we
have to determine the Higgs fields in our model as well, meaning
to compute the non-chiral spectrum. This cannot be done by a
simple homology computation, but fortunately we do know the exact
conformal field theory at the orbifold point. Using the boundary
states \eqref{bound}, we can determine the non-chiral matter living on
intersections of the various stacks of D-branes. One first
computes the overlap between two such boundary states and then
transforms the result to the open string channel to get the
annulus partition function, from which one can read off the
massless states. This is a straightforward but tedious
computation, which also confirms the chiral spectrum in table \ref{chiralspec7stack}.
Thus the conformal field theory result agrees completely with the
purely topological computation of the intersection numbers.

\begin{figure}[h]
 \begin{center}
 \includegraphics{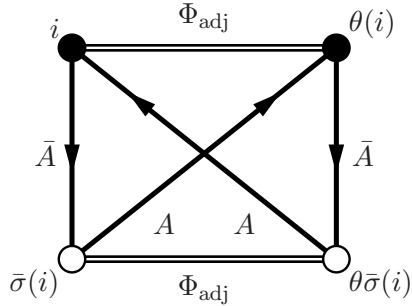}
 \end{center}
 \caption{\label{adjhiggsquiver} Adjoint higgsing}
\end{figure}

Computing the non-chiral spectrum just for one stack of $U(2)$
branes and their $\mathbb{Z}_4$- and $\Omega\bar{\sigma}$-images, one first
finds the  well known hyper-multiplet,
$\Phi_\text{adj}=(\phi_\text{adj},\tilde{\phi}_\text{adj})$, in the adjoint
representation of $U(2)$ localized on the intersection of a brane
and its $\mathbb{Z}_4$ image. Moreover, there are two chiral multiplets,
$\Psi_A$ and $\Psi_{\bar{A}}$, in the ${\bf A}$ respectively ${\bf
\bar{A}}$ representation arising from the $(\pi_i,\pi'_i)$ sector.
Since the two chiral fields carry conjugate representations of the
gauge group, they cannot be seen by the topological intersection
number which in fact vanishes, $\pi_i\circ\pi^\prime_i=0$.
 We have depicted the resulting  quiver diagram for these three fields in
figure \ref{adjhiggsquiver}.
For each closed polygon in the quiver diagram (respecting the directions of
the arrows), the
associated product of fields can occur in the holomorphic
super-potential. In our case, the following two terms can appear
\begin{equation}
 \label{massterm} 
  W=\phi_{\text{adj}} \Psi_A\Psi_{\bar{A}} + \tilde{\phi}_{\text{adj}}
    \Psi_A\Psi_{\bar{A}} 
\end{equation}
which generate  a mass for  the
anti-symmetric fields when the adjoint multiplet gets a VEV.
As we have mentioned already in the last
section, giving a VEV to this adjoint field localized on the intersection
between a brane and its $\mathbb{Z}_4$ image, leads to the recombination
of these two branes. The recombined brane no longer passes through the
$\mathbb{Z}_4$ invariant intersection points.
After computing all annulus partition functions for
pairs of D-branes from table \ref{d63genmod}, we find the total non-chiral spectrum
listed in table \ref{nonchiralspec7stack}.
\begin{table}
 \renewcommand{\arraystretch}{1.2}
 \begin{center}
 \begin{tabular}{|c|c|c|}
\hline
 field &  n  & $U(4)\times U(2)^3\times U(2)^3$   \\   
\hline\hline
 $H_{12}$ & 1 & $(4;\Bar{2},1,1;1,1,1)+c.c.$ \\
 $H_{13}$ & 1 & $(4;1,\Bar{2},1;1,1,1)+c.c.$ \\
 $H_{14}$ & 1 & $(4;1,1,\Bar{2};1,1,1)+c.c.$ \\
\hline
 $H_{15}$ & 1 & $(\Bar{4};1,1,1;{2},1,1)+c.c.$ \\
 $H_{16}$ & 1 & $(\Bar{4};1,1,1;1,{2},1)+c.c.$ \\
 $H_{17}$ & 1 & $(\Bar{4};1,1,1;1,1,{2})+c.c.$ \\
\hline\hline
 $H_{25}$ & 1 & $(1;{2},1,1;\Bar{2},1,1)+c.c.$ \\
 $H_{26}$ & 1 & $(1;{2},1,1;1,\Bar{2},1)+c.c.$ \\
 $H_{27}$ & 1 & $(1;{2},1,1;1,1,\Bar{2})+c.c.$ \\
 $H_{35}$ & 1 & $(1;1,{2},1;\Bar{2},1,1)+c.c.$ \\
 $H_{36}$ &  1 & $(1;1,{2},1;1,\Bar{2},1)+c.c.$ \\
 $H_{37}$ & 1 & $(1;1,{2},1;1,1,\Bar{2})+c.c.$ \\
 $H_{45}$ & 1 & $(1;1,1,{2};\Bar{2},1,1)+c.c.$ \\
 $H_{46}$ & 1 & $(1;1,1,{2};1,\Bar{2},1)+c.c.$ \\
 $H_{47}$ & 1 & $(1;1,1,{2};1,1,\Bar{2})+c.c.$ \\
\hline
\end{tabular}
 \caption{\label{nonchiralspec7stack}Non-chiral spectrum (Higgs fields)}
 \end{center}
\end{table}

It is interesting that we find  Higgs fields  which might break a
$SU(4)\times SU(2)\times SU(2)$ gauge symmetry in a first step
down to the  Standard Model and in a second step down to
$SU(3)\times U(1)_{em}$. However, the Higgs fields which would
allow us to break the product groups $U(2)^3$ down to their
diagonal subgroup are not present in the non-chiral spectrum.

\subsection{D-flatness}

However, we do have the massless chiral bifundamental fields
$\{\Phi_{2'3},\ldots,\Phi_{6'7}\}$ living on  intersections
preserving ${\cal N}=1$ supersymmetry.
As we have already mentioned, for isolated brane intersections
these massless fields indicate
that the complex structure moduli are chosen such that
one sits on a line of marginal stability. On one side
of this line, the intersecting branes break supersymmetry
without developing a tachyonic mode. This indicates that
the intersecting brane configuration is stable.
But on the other side of the line, the former massless
chiral field becomes tachyonic and after condensation
leads to a new in general non-flat supersymmetric brane
wrapping a special Lagrangian 3-cycle. Since the tachyon
transforms in the bifundamental representation, on this brane
the gauge symmetry is broken to its diagonal subgroup.

We therefore expect for our compact situation that at least
locally these bifundamental chiral multiplets indicate the
existence of a recombined brane of the same volume but with the
gauge group broken to the diagonal subgroup. In order to make our
argument save, we need to show that the D-terms allow, that for
certain continuous deformations of the complex structure moduli,
just the  fields
$\{\Phi_{2'3},\Phi_{2'4},\Phi_{3'4},\Phi_{5'6},\Phi_{5'7},
\Phi_{6'7}\}$  become
tachyonic, leaving the VEVs of the remaining 
$\{\Phi_{1'2},\Phi_{1'3},\Phi_{1'4},\Phi_{1'5},\Phi_{1'6},
\Phi_{1'7}\}$  vanishing.
 Then the fields $\{\Phi_{2'3},\Phi_{2'4},\Phi_{3'4},\Phi_{5'6},\Phi_{5'7},
\Phi_{6'7}\}$ condense to a new supersymmetric ground state
and the gauge symmetries $U(2)^3$ are broken to the diagonal
$U(2)$s. From general arguments for open string models
with ${\cal N}=1$ supersymmetry, it is known that the complex-structure 
moduli only
appear in the D-term potential, whereas the K\"ahler moduli only
appear in the F-term potential \cite{Brunner:1999jq,Kachru:1999vj,Kachru:2000an,Aganagic:2000gs}.

Remember that the Green-Schwarz mechanism requires the Chern-Simons
couplings to be of the form
\begin{equation}
    \label{gsmechc}  S_{\text{CS}}
    =   \sum_{i=1}^{b_3} \sum_{a=1}^k \int d^4 x\, M_{ai}\, B_i\wedge
               {1\over N_a}\,{\rm tr}(F_a) 
\end{equation}
The supersymmetric completion involves a coupling of the
auxiliary field $D_a$
\begin{equation}
   \label{fib}  
           S_{\text{FI}}
            =\sum_{i=1}^{b_3} \sum_{a=1}^k \int d^4 x\,  M_{ai}\,
              {\partial{\cal K}\over \partial \phi_i} \,
               {1\over N_a}\,{\rm tr}(D_a) 
\end{equation}
where $\phi_i$ are the super-partners of the Hodge duals of
the RR 2-forms and ${\cal K}$ denotes the K\"ahler potential.
Thus, these couplings give rise to FI-terms depending
on the complex structure moduli which we parameterize
simply by ${\cal B}^i\equiv{\partial{\cal K}/\partial \phi_i}$
\footnote{For our purposes we do not need the precise form of
the K\"ahler potential as long as the map from the complex structure
moduli $\phi_i$ to the new parameters ${\cal B}^i$ is one to one. But this
is the case, at least for a  sufficiently small open set $U\ni {\cal B}^i$,
as the functional determinant for the map between
these two sets of variables is equal to
$\det\left({\partial^2 {\cal K}\over \partial \phi_i\partial \phi_j}\right)$,
which is non-vanishing for a positive definite metric
on the complex structure moduli space ($i$ and $j$ including both holomorphic
and antiholomorphic indices).}.

Let us now discuss the D-term potential for the $U(4)\times U(2)^3\times
U(2)^3$ gauge fields
in our model and see whether it allows supersymmetric
ground states of the type described above.
The D-term potential including only the chiral matter and
the FI-terms  in general reads
\begin{equation}
     \label{dtermp}
     \begin{aligned} 
            V_D&=\sum_{a=1}^k
              \sum_{r,s=1}^{N_a} {1\over 2 g_a^2}(D^{rs}_a)^2 \\
                 &=\sum_{a=1}^k \sum_{r,s=1}^{N_a}
       {1\over 2 g_a^2}
   \Biggl(\sum_{b=1}^k\sum_{p=1}^{N_b} q_{ab}
              \, \Phi^{rp}_{ab} \,    \Bar{\Phi}^{sp}_{ab}+
                g_a^2\sum_{i=1}^{b_3}  {M_{ai}\over N_a}\, {\cal B}^i\,\delta^{rs}
   \Biggr)^2 
      \end{aligned}
\end{equation}
where the indices $(r,s)$ numerate the $N_a^2$ gauge fields in the adjoint
representation
of the gauge factor $U(N_a)$ and the sum over $b$ is over all
chiral fields charged under $U(N_a)$.
The gauge coupling constants $g_a$ depend on the complex
structure moduli as well, but since we are only interested in the leading order
effects, we can set them to constant values on the line of marginal
stability.  Since all branes {\sl in our particular model} have the same
volume there, in the following we set them to the same constant:
\begin{equation}
  g_a = g \qquad\forall a\in\{1,\ldots7\}
\end{equation}
To make things simpler we will set them to one.\footnote{This simplification
  will not affect the following conclusions.} For the time
being, we are only interested 
in breaking the $U(2)^3\times U(2)^3$ part of the gauge group
down to  the diagonal $U(2)\times U(2)$ by deforming the closed 
string background. We achieve this by giving appropriate VEVs to the
fields ${\cal B}^i$ in the D-term potential \eqref{dtermp}. 
    
In our case, the charges $q_{ab}$ can be read off from table
 \ref{chiralspec7stack}\,\footnote{The charge is $q_a=1$ if the 
representation of the $U(N_a)$ is $N_a$ and $q_a=-1$ if it is
$\Bar{N}_a$.}
and
the Green-Schwarz  couplings $M_{ai}$ from table \ref{d63genmod} using the 
definition 
\eqref{matric}.

With the explicit D-term potential at hand, we will sketch the argument, 
why expectation values for the fields $\Phi_{ab}$ with $a$ and $b$ 
belonging to two different gauge groups, will break the gauge symmetry
from $U(N)_a\times U(N)_b$  to a diagonal subgroup   $U(N)_\alpha$. First, we
will make a comment on the possible  form of the $(\Phi_{ab})^{rp}$.
We note that $\sum_{p=1}^{N_b} \Phi_{ab}^{rp}\Bar{\Phi}_{ab}^{sp}=
(\Phi_{ab}\Phi_{ab}^\dagger)^{rs}$ is a hermitian matrix. If only
one such matrix is present, then
$\delta^{rs}$ in \eqref{dtermp} always forces $\Phi_{ab}\Phi_{ab}^\dagger$
to be a diagonal matrix. 
Depending on the exact form of the D-term, and especially in the example at
hand, the hermitian matrices $\Phi_{ab}\Phi_{ab}^\dagger$ might be forced to
 diagonal form, even if several of such bifundamental fields are present. 
This means that the vectors $C^{(i)}_{ab}$, $i=1\ldots N_a$, that form 
$\Phi_{ab}$ as row- or column-vectors, are orthogonal and 
normalized to $||C^{(i)}_{ab}||=\chi_{ab}\;\forall i$.  Consequently $\Phi_{ab}$ is a $U(N_a)$ matrix up to a 
constant factor $\chi_{ab}$.
  With a further unitary transformation,
this time involving both gauge groups, we can achieve:\footnote{The diagonal
form is obtained by $\Phi_{ab}\to U_a\Phi_{ab}U_b$ with $U_b=U_a^{-1}$. Such
a $U_a$ always exists for one matrix $\Phi_{ab}$. 
Potential phases can be eliminated, because 
one can transform $\Phi_{ab}$ by independent  $U(N)_a$ and $U(N)_b$ 
matrices on both indices. However, with too many fields $\Phi_{ab}$
involved in the D-term a
simultaneous diagonalization of all VEVs $\Phi_{ab}$ might fail for
generic cases. In the case at hand the small number of fields allows
for simultaneous diagonalization.}
\be\label{stdformphi}
  \Phi_{ab}= \chi_{ab}\mathds{1}_{N_a}, \qquad \chi_{ab}\in \mathbb{R}_+
\ee
Assuming for the moment that we have only one field $\Phi_{ab}$ acquiring a
VEV, we will  show that it gives masses to 
a special linear combination of the vector potentials $A^\mu_a$ and
$A^\mu_b$, while leaving another combination massless, thereby breaking
the $U(N_a)\times U(N_b)\to U(N_\alpha)$ (with $N_a=N_b=N_\alpha$).
What we will present is a generalization of the usual $U(1)$
gauge symmetry breaking
by a suitable D-term  (i.e.\ a D-term with the right sign) to
the case where the chiral fields transform in the bifundamental 
representation of  $N_1\times \bar{N}_2$. 
We split $\Phi_{ab}$ into its (diagonal) VEV 
$\vev{\Phi_{ab}}=\chi_{ab}\mathds{1}_{N_a}$ and a field
 $C_{ab}(x)$ that describes the fluctuations around the
 VEV:\footnote{$U^{-1}_{a,b}(x)$ are $x$-dependent unitary matrices.}
\be \label{phiatvev}
 \Phi_{ab}= U_a(x)\big(\chi_{ab} + C_{ab}\big) U^{-1}_b(x)
\ee 
As we are just interested in the (gauge-invariance breaking)
 mass term of the gauge field we need  to consider only the  
 covariant derivative of $ \Phi_{ab}$:
\be  \label{covderiv}
  \nabla_\mu \Phi_{ab} = \partial_\mu  \Phi_{ab}
   -i A^a_\mu \Phi_{ab} + i \Phi_{ab} A^b_\mu
\ee
 Next we will make a finite gauge change in the gauge potentials:
\be
  \begin{aligned}\label{guaugetr}
    A^a_\mu &\to A^{\prime\,a}_\mu \equiv U^{-1}_a(x) A^a_\mu U_a(x)  
                 +  i(\partial_\mu U_a(x))  U^{-1}_a(x) \\
    A^b_\mu &\to  A^{\prime\,b}_\mu\equiv U^{-1}_b(x) A^b_\mu U_b(x)
               +  i(\partial_\mu U_b(x))  U^{-1}_b(x)
  \end{aligned}
\ee  
Expressed in terms of the gauge transformed fields \eqref{guaugetr}, the
covariant derivative \eqref{covderiv} becomes:
\be  \label{covderprime}
  \nabla_\mu \Phi_{ab} = U_a(x)\big( \nabla^\prime_\mu C_{ab} +
                                  i\chi_{ab}(A^{\prime\,b}_\mu - 
                                             A^{\prime\,a}_\mu )
                               \big)U^{-1}_b(x)
\ee
The kinetic term  $\tr\big((\nabla \Phi_{ab})^2\big)$ leads to a mass term:
\be \label{massmatrixdt}
 \tr\big((\nabla \Phi_{ab})^2\big) 
  = \tr (\nabla^\prime C_{ab})^2 + \chi_{ab}^2
   \tr \bigg(\begin{array}{c} A^\prime_a \\ \hline A^\prime_b 
            \end{array}   
       \bigg)^\dagger
       \bigg(\begin{array}{c|c}
                  \mathds{1}_{N_a}  & - \mathds{1}_{N_a}      \\  
                  \hline 
                 - \mathds{1}_{N_a}  &  \mathds{1}_{N_a}
             \end{array}
       \bigg)           
       \bigg(\begin{array}{c} A^\prime_a \\ \hline A^\prime_b 
            \end{array}   
       \bigg)     
\ee
The mass matrix appearing in \eqref{massmatrixdt} has $\rk N_a$, leaving
the diagonal combination $ A^\prime_a+A^\prime_b$ massless, while giving 
$ A^\prime_a-A^\prime_b$ a mass of order $\chi_{ab}$\;\footnote{If the gauge
kinetic term is canonically normalized.}. Thereby the gauge 
symmetry gets broken: 
$U(N_a)\times U(N_b)\to U(N_\alpha),\,(N_a=N_b=N_\alpha)$.
 The kinetic term
of the gauge field is invariant under the gauge 
transformation.\footnote{Of course the gauge kinetic function 
$f(\phi)$ that multiplies the term $F\wedge\ast F$ has to be expressed
in terms of \eqref{phiatvev}, but it does {\sl not} produce a mass term or even
worse, a term
compensating the mass term in \eqref{massmatrixdt}.}
The generalization to an additional field $\Phi_{ac}$ with VEV  
$\chi_{ac}\mathds{1}_{N_a}$ is straightforward
and leads again to a mass term:
\be \label{massmattwovev}
   \tr \left(\begin{array}{c} A^\prime_a \\ \hline A^\prime_b 
                            \\ \hline A^\prime_c
            \end{array}   
       \right)^\dagger
       \left(\begin{array}{c|c|c}
           (\chi_{ab}^2+\chi_{ac}^2)\mathds{1}_{N_a}  
                 & - \chi_{ab}^2\mathds{1}_{N_a} 
             & -\chi_{ac}^2 \mathds{1}_{N_a}
             \\  
             \hline \rule[-1ex]{0ex}{3.2ex}
             - \chi_{ab}^2\mathds{1}_{N_a}  
             & \chi_{ab}^2 \mathds{1}_{N_a} &   0   \\ 
              \hline \rule{0ex}{2.2ex}
              -\chi_{ac}^2\mathds{1}_{N_a}  
               & 0 & \chi_{ac}^2\mathds{1}_{N_a} 
             \end{array}
        \right)
       \left(\begin{array}{c} A^\prime_a \\ \hline A^\prime_b 
                            \\ \hline A^\prime_c
            \end{array}   
        \right)
\ee
The linear combination $ A^\prime_a+ A^\prime_b+A^\prime_c$ 
stays massless, while the two other (orthogonal) linear combinations have
 mass\raisebox{1ex}{\tiny 2} of the order $\chi_{ab}^2+\chi_{ac}^2\pm
 \sqrt{\chi_{ab}^4-\chi_{ab}^2\chi_{ac}^2\chi_{ac}^4}$. It might be
 even possible
to give a VEV $\chi_{bc}\mathds{1}_{N_a}$
 to a field $\Phi_{bc}$ transforming in the
bifundamental $N_b\times \bar{N}_c$ of $U(N_b)\times U(N_c)$ (if it exists).
The generalization of the mass matrix \eqref{massmattwovev} is 
straightforward:
\begin{multline}
      \left(\begin{array}{c|c|c}
            (\chi_{ab}^2+\chi_{ac}^2)\mathds{1}  
                  & - \chi_{ab}^2\mathds{1} 
              & -\chi_{ac}^2 \mathds{1}
             \\  
              \hline \rule[-1ex]{0ex}{3.2ex}
              - \chi_{ab}^2\mathds{1}  
              & \chi_{ab}^2 \mathds{1} &   0   \\ 
               \hline \rule{0ex}{2.2ex}
               -\chi_{ac}^2\mathds{1}  
                & 0 & \chi_{ac}^2\mathds{1} 
             \end{array}
      \right)   
           \\    
    \longrightarrow
     \left(\begin{array}{c|c|c}
           (\chi_{ab}^2+\chi_{ac}^2)\mathds{1}  
                 & - \chi_{ab}^2\mathds{1} 
             & -\chi_{ac}^2 \mathds{1}
             \\  
             \hline \rule[-1ex]{0ex}{3.2ex}
             - \chi_{ab}^2\mathds{1}  
             & (\chi_{ab}^2 +\chi_{bc}^2) \mathds{1}
                  & -\chi_{bc}^2\mathds{1} 
                \\ 
              \hline \rule{0ex}{2.2ex}
              -\chi_{ac}^2\mathds{1}  
               & -\chi_{bc}^2\mathds{1} 
            & (\chi_{ac}^2+\chi_{bc}^2)\mathds{1} 
             \end{array}
     \right)
\end{multline}
leaving again   $ A^\prime_a+ A^\prime_b+A^\prime_c$ massless
while giving masses to the other two orthogonal combinations.

From the fact that $(N_a)^{-1}M_{ai}$ 
(cf.\ \eqref{matric}, p.\ \pageref{matric})
can be written as 
\be
 \frac{M_{ai}}{N_a}=  (\mathds{1}-\bar{\sigma})_{i}^{\phantom{i}j} v_{a,j}
\ee
we deduce that the components of ${\cal B}^i$ that
can contribute to the FI-term \eqref{dtermp}  belong to a vector space of 
dimension $b_3 /2$.\footnote{The rank of $\mathds{1}-\bar{\sigma}$ is of this 
dimension.}
It is the Eigenspace of   
$\big(\bar{\sigma}^T\big)_{i}^{\phantom{i}j}$ with Eigenvalue $-1$.
 Actually in our seven stack example the rank of $M_{ai}$ is only
seven.  We are now searching for vectors $ {\cal B}$, that do not
couple to the first gauge group which is the $U(4)$:
\be \label{u4unbrok}
  {\cal B}^i (\mathds{1}-\bar{\sigma})_{i}^{\phantom{i}j} v_{1,j} 
    \stackrel{!}{=}0
\ee
This condition ensures  the $U(4)$ to be unbroken as it forces
vanishing VEVs for  the fields $\Phi_{a\in\{1,1'\},b}$. 
(The  $U(4)$ is an 
essential part of the Pati-Salam model.) 
Giving a VEV $\chi_{ab}$ to one of the remaining fields is equivalent
to imposing that  
\be \label{vevrelation}
  V^{(ab)}_c=  \frac{M_{ci}}{N_c} {\cal B}^i_{(ab)}
\ee
equals a certain vector $V^{(ab)}$. In this way we can associate to each
 VEV $\chi_{ab}$ a vector ${\cal B}^i_{(ab)}$. 
Linear combinations of $\sum_{ab}
\l_{ab}V^{ab}$, $l_{ab}>0$, will then lead to VEVs $\l_{ab}\chi_{ab}$ for
the fields $\Phi_{ab}$. In other words: the classical moduli space
of the fields  
\be \label{sixvevs}
 \big\{\Phi_{2',3}, \Phi_{2',4} , \Phi_{3',4}, \Phi_{5',6}, \Phi_{5',7}
       , \Phi_{6',7} 
 \big\}
\ee
is a cone of real dimension six.\footnote{It is not a vector space, as
negative coefficients $l_{ab}$ multiplying the basis vectors would break supersymmetry
 as they hinder the D-term \eqref{dtermp} to vanish.}
This is only possible as the matrix $M_{ai}$ is of rank seven, i.e.\ of
maximal rank. As a consequence, the relation \eqref{vevrelation} gets 
invertible. Furthermore the chiral field-content of our model (c.f.\ table
\ref{d63genmod}, p. \pageref{d63genmod}) leads to an unambiguous
 mapping between VEVs of 
the six fields \eqref{sixvevs} and the vectors $ V^{(ab)}$ 
in \eqref{vevrelation} (iff we impose the condition of unbroken
$U(4)_1$, eq.\ \eqref{u4unbrok}).  
Therefore we can give the fields in \eqref{sixvevs} arbitrary VEVs  that
are proportional to the unit matrix $\mathds{1}_{N_a}$.
As explained above, giving non-vanishing VEVs to at least two of the fields
 $ \big\{\Phi_{2',3}, \Phi_{2',4} , \Phi_{3',4}\big\}$ will lead to 
the diagonal gauge breaking:
\be
  U(2)_2\times U(2)_3\times U(2)_4 \to U(2)_b
\ee 
Analogously giving VEVs to two or three of the fields
 $ \big\{\Phi_{5',6}, \Phi_{5',7} , \Phi_{6',7} \big\}$ leads to the
 gauge breaking:
\be
  U(2)_5\times U(2)_6\times U(2)_7 \to U(2)_c
\ee 
The string theoretic interpretation of this low energy description is 
as follows:
There exists a supersymmetric configuration  where
the branes $\{\pi_2,\pi'_3,\pi'_4\}$ and similarly the branes
$\{\pi_5,\pi'_6,\pi'_7\}$ have recombined into a single
 brane within the same homology class, thereby breaking the gauge
symmetry from 
\be
   U(4)_1\times \big(U(2)_2\times U(2)_3\times U(2)_4 \big) \times
                \big( U(2)_5\times U(2)_6\times U(2)_7 \big) 
\ee
down to 
\be
   U(4)_a\times  U(2)_b\times U(2)_c 
\ee
which is a three stack model.

In what follows we will give non-trivial VEVS only to the fields 
\be \label{compd}
 \begin{aligned}
  \vevs{\Phi_{2',3}}> 0 &,  \vevs{\Phi_{2',4}} > 0 
  &\text{ and }\quad 
  \vevs{\Phi_{5',6}}> 0 &, \vevs{\Phi_{5',7}}  > 0
 \end{aligned}
\ee
 while leaving
$\vevs{\Phi_{3',4}}=\vevs{\Phi_{6',7}}=0$.
This simplifies some of the following mass formul\ae\, 
(cf.\ eq.~\eqref{massmatrixsmallquiv} to \eqref{massmatrixinmatrixhyp}).
We will make some comments about the changes that occur for non-vanishing 
VEVs $\vevs{\Phi_{3',4}}>0$ and $\vevs{\Phi_{6',7}} >0$.

\subsection{Gauge symmetry breaking}
After this recombination process we are left with only three stacks of
D$6$-branes wrapping the homology cycles
\begin{equation}
   \label{recom}
          \pi_a=\pi_1, \quad \pi_b=\pi_2+\pi'_3+\pi'_4,
              \quad \pi_c=\pi_5+\pi'_6+\pi'_7 
\end{equation}
 These branes are
not factorizable but we have presented arguments ensuring
that they preserve the same supersymmetry as the closed string
sector and the former intersecting brane 
configuration.\footnote{Since we get chiral fields in the (anti-)symmetric
representations after brane recombination, one might check
if those intersection numbers can also be obtained by flat
factorizable D-branes. Remember that we had the first assumption
that there are no such chiral fields in the (anti-)symmetric
representations. In fact, after an extensive computer search we
have not been able to find a model with just factorizable D-branes
generating the chiral spectrum of table \ref{chiralspecps}.} The chiral spectrum
for this now 3 stack model is shown in table \ref{chiralspecps}.
\begin{table}
 \renewcommand{\arraystretch}{1.2}
 \begin{center}
 \begin{tabular}{|c|c|clc|}
 \hline
 field & n  & \multicolumn{3}{c|}{$SU(4)\times SU(2)\times SU(2)\times U(1)^3$}   \\
 \hline\hline
 $\Phi_{ab}$   &  2  & \rule{5ex}{0ex}& $(4,2,1)_{(1,-1,0)}$               & \\
 $\Phi_{a'b}$  &  1  & & $(4,2,1)_{(1,1,0)}$                & \\
 \hline
 $\Phi_{ac}$   &  2  & & $(\Bar{4},1,2)_{(-1,0,1)}$         & \\
 $\Phi_{a'c}$  &  1  & & $(\Bar{4},1,2)_{(-1,0,-1)}$        & \\
 \hline
 $\Phi_{b'b}$  &  1  & & $(1,S+A,1)_{(0,2,0)}$              & \\
 $\Phi_{c'c}$  &  1  & & $(1,1,\Bar{S}+\Bar{A})_{(0,0,-2)}$ & \\
 \hline
\end{tabular}
 \caption{\label{chiralspecps}Chiral spectrum for 3 stack PS-model}
 \end{center}
\end{table}
The intersection numbers $\pi'_{b,c}\circ \pi_{b,c}$ do
not vanish any longer, therefore giving  rise to chiral
multiplets in the symmetric and anti-symmetric representation of the $U(2)$
gauge factors. Clearly, these chiral fields are needed in order to cancel the
formal non-abelian $U(2)$ anomalies.
Computing the mixed anomalies for this model, one finds that
two $U(1)$ gauge factors are anomalous and that the  only anomaly free
combination is
\begin{equation}
  \label{freean} 
    U(1)=U(1)_a-3\, U(1)_b -3\, U(1)_c 
\end{equation}
The quadratic axionic couplings reveal
that the matrix $M_{ai}$ in \eqref{matric}
has a trivial kernel and therefore
all three $U(1)$ gauge groups become massive and survive
as global symmetries.
To summarize, after the recombination of some of the $U(2)$ branes
we have found  a supersymmetric 3 generation
Pati-Salam model with gauge group $SU(4)\times SU(2)_L\times SU(2)_R$
which accommodates the Standard Model matter in addition
to some exotic matter in the (anti-)symmetric representation
of the $SU(2)$ gauge groups.

\begin{figure}[h]
 \rule{0.25cm}{0cm}
 \begin{minipage}[b]{7.0cm}
 \begin{center}
  \raisebox{-1ex}{
                  \includegraphics{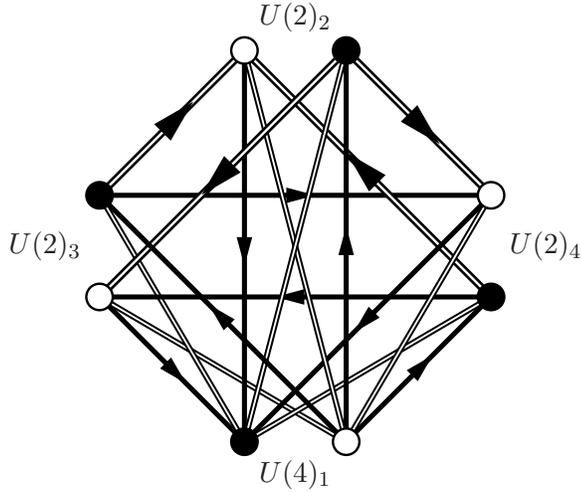}
                 }
 \end{center}
 \end{minipage}\hfill
  \raisebox{1ex}{\begin{minipage}[b]{4.6cm} 
  \caption[Quiver diagram for the branes $\{1,2,3,4\}$]
          {\label{smallquiver} Quiver diagram for the branes $\{1,2,3,4\}$.
           The arrows denote a chiral multiplet. Double lines with
           arrows denote chiral multiplets where the scalar field gets
           a VEV due to a complex structure deformation (c.f.\ eq.\ 
           \eqref{compd}).
           Double lines without an arrow are hyper-multiplets. The filled
           dots describe a stack of branes and the unfilled dots are the
           $\Omega\Bar{\sigma}$ images.}
  \end{minipage}}
 \end{figure}

To compute the massless non-chiral spectrum after the recombination,
we have to determine which Higgs fields receive a mass from  couplings
with the condensing chiral bifundamental fields.
As we have explained earlier, the applicability of the low energy
effective field theory is limited but still is the only
information we have. So, we will see how far we can get.
We first consider the sector of the branes $\{\pi_1, \ldots, \pi_4\}$ in
figure \ref{smallquiver}.
The chiral fields are indicated by an arrow and non-chiral fields
by a double line without an arrow.
 The (chiral) fields which receive a VEV after small complex
structure deformations are depicted by a double line with an arrow.

Let us decompose  the Higgs fields inside one hyper-multiplet into its
two chiral components $H_{1j}=\big(h^{(1)}_{1j}, h^{(2)}_{1j}\big)$ for
 $j=2,3,4$.
We observe a couple of closed  triangles in the quiver diagram in 
figure \ref{smallquiver}
that give rise to the following Yukawa couplings in the super-potential:
\begin{equation}
  \label{eq:supercoup1}
 \begin{aligned} 
  \includegraphics{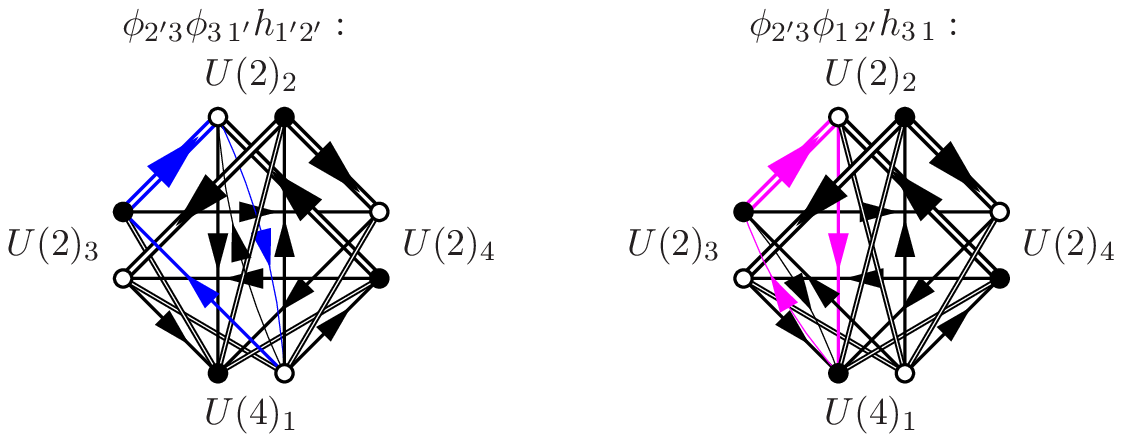}
 \end{aligned}
\end{equation}
and
\begin{equation}
  \label{eq:supercoup2}
 \begin{aligned} 
  \includegraphics{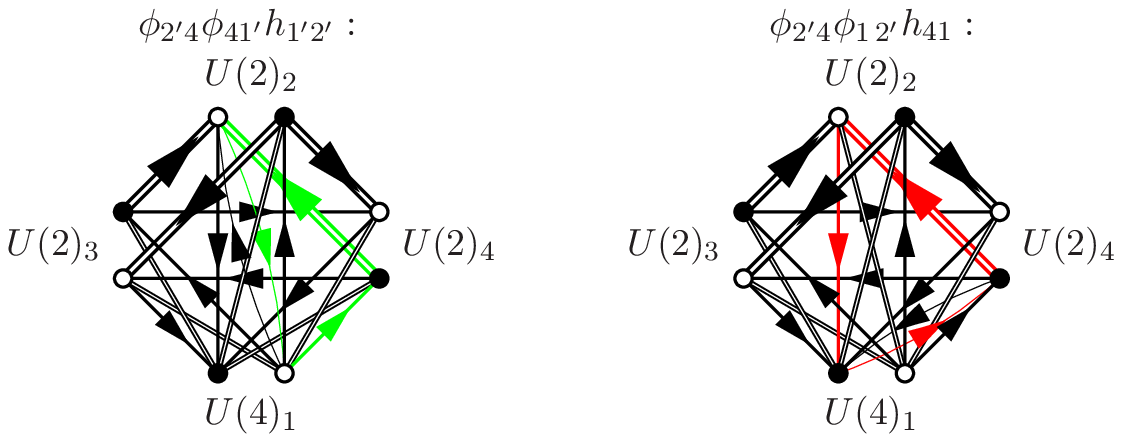}
 \end{aligned}
\end{equation}

Condensation of the chiral fields $\Phi_{2'3}$ and $\Phi_{2'4}$
leads to a mass matrix for the six fields $\{\Phi_{1'2},
\Phi_{1'3}, \Phi_{1'4},h^{(2)}_{12},h^{(2)}_{13}, h^{(2)}_{14}\}$
of rank four. The mass terms look schematically:\footnote{As strings
 with ends on $(a\,b)$ 
  and $\Bar{\sigma}$-pictures $(\bar{\sigma}(a)\bar{\sigma}(b))=(a'b')$
   (primes) are identified, we loosely
  identify the corresponding fields.}\raisebox{0.8ex}{\tiny , }\footnote{We
  have included the possibility of a non-vanishing VEV $\vevs{\Phi_{3'4}}>0$ 
in parentheses. If all three VEVs are non-zero, the mass\raisebox{1ex}{\tiny
  2} matrix will be of rank six. In what follows, we assume 
$\vevs{\Phi_{3'4}}=0$.}
\begin{equation} \label{massmatrixsmallquiv}
   \left( 
     \begin{smallmatrix}
      h^{(2)}_{12}  \\  h^{(2)}_{13}  \\  h^{(2)}_{14} \\
      \Phi_{1'2}    \\  \Phi_{1'3}    \\   \Phi_{1'4} 
     \end{smallmatrix}
   \right) ^\dagger
   \left( 
     \begin{array}{c|c}
     \renewcommand{\arraystretch}{2}
      0 
    &\begin{smallmatrix} 
         0      & \vevs{\Phi_{2'3}} & \vevs{\Phi_{2'4}} \\
     \vevs{\Phi_{2'3}} & 0          & \big(\vevs{\Phi_{3'4}}\big)   \\
     \vevs{\Phi_{2'4}}
     \rule[-.8ex]{0.ex}{0.1ex} & \big(\vevs{\Phi_{3'4}}\big)
           \rule[-1.15ex]{0.ex}{0.ex}             & 0    
     \end{smallmatrix} 
    \\ \hline
      \begin{smallmatrix} 
         0      & \vevs{\Phi_{2'3}} & \vevs{\Phi_{2'4}}
        \rule[0ex]{0.ex}{1.6ex} \\
     \vevs{\Phi_{2'3}} & 0          &     \big(\vevs{\Phi_{3'4}}\big)  \\
     \vevs{\Phi_{2'4}}  & \big(\vevs{\Phi_{3'4}}\big) &     0     
     \end{smallmatrix}
       & 0   
   \end{array} \right)
    \left( 
     \begin{smallmatrix}
      h^{(2)}_{12}  \\  h^{(2)}_{13}  \\  h^{(2)}_{14} \\
      \Phi_{1'2}    \\  \Phi_{1'3}    \\   \Phi_{1'4} 
     \end{smallmatrix}
   \right)
\end{equation}
Thus, one combination of the three fields $\Phi$,
one combination of the three fields $h^{(2)}$ and furthermore the three fields
$h^{(1)}$ remain massless. These modes just fit into the three
chiral fields in table \ref{chiralspecps} in addition to one further 
hyper-multiplet
in the $(4,{2},1)$ representation of the Pati-Salam gauge group
$U(4)\times U(2)\times U(2)$.\footnote{This hyper-multiplet would gain a mass
  if we give a VEV to the field $\Phi_{3'4}$.} The condensation for the second
triplet of $U(2)$s is completely analogous and leads to a
massless hyper-multiplet in the $(4,1,{2})$ representation.\footnote{The
  hyper-multiplet in the $(4,1,{2})$ representation would become massive if we
  give  $\Phi_{6',7}$ a VEV.}
\begin{figure}
 \rule{0.025cm}{0cm}
 \begin{minipage}[b]{6.8cm}
 \raisebox{0.5cm}{\includegraphics{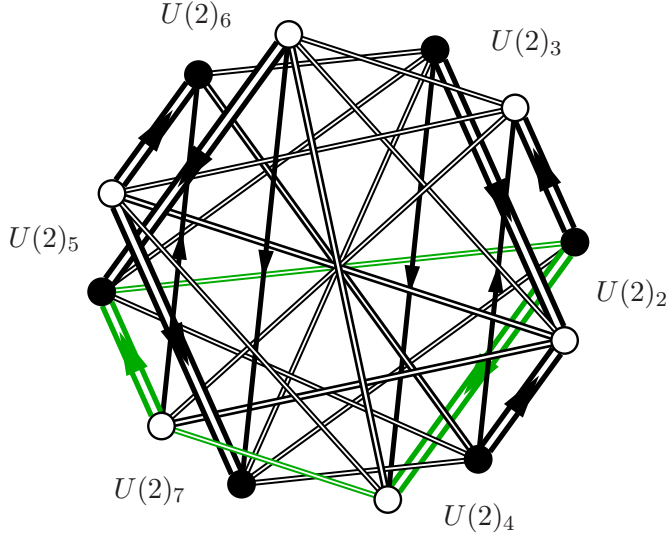}}
 \end{minipage}\hfill
  \begin{minipage}[b]{4.025cm} 
  \caption[Quiver diagram for the branes $\{2,3,4,5,6,7\}$]{
   \label{bigquiver}Quiver diagram for the branes $\{2,3,4,5,6,7\}$. The right
   stacks ($U(2)_2\ldots U(2)_4$) and the left branes ($U(2)_2\ldots U(2)_4$)
   will combine into the right and left $U(2)$s of the left-right symmetric
   Pati-Salam Model after the scalar fields \eqref{compd} get VEVs. The
   chiral multiplets that accommodate these condensed scalars are indicated
   by arrows (chirality) and double lines (VEV). }
  \end{minipage}
\end{figure} 
The quiver diagram involving the six $U(2)$ gauge groups is shown in figure
\ref{bigquiver}. In this quiver diagram the closed polygon
$(2-4'-7'-5-2)$ (marked in \textcolor[rgb]{0,0.67,0}{green}) 
 generates a mass term after condensation of $\Phi_{2'3}$ and $\Phi_{5'7}$ for
one chiral component inside 
$\{H_{25},H_{47}\}$ and the sub-quiver $(2-4'-5'-7-2)$ (not marked )
generates a mass term for one chiral component in
$\{H_{46},H_{47}\}$ (Remember that a
hyper-multiplet consists of two chiral multiplets of opposite
charge, $H=(h^{(1)},h^{(2)})$). In analogy the mass terms for all
 nine Hyper-multiplets
 are obtained from fig.~\ref{bigquiver}, too. They have the form:
\begin{equation}\label{massmatrixhyp}
  \Psi^\dagger
    \left( 
     \begin{array}{c|c|c}
        0      &  M_1 & M_2\\
     \hline
     M_1& 0    &     0     \\
     \hline
     M_2& 0   & 0\rule[-.45ex]{0.ex}{0.1ex}  
     \end{array} 
    \right)
  \Psi
\end{equation}
Here we have defined:
\begin{equation}\label{massmatrixinmatrixhyp}
 \begin{aligned}
   M_1&= 
    \left( 
     \begin{array}{ccc}
        0                & \Phi_{2'3}\Phi_{5'6} & \Phi_{2'4}\Phi_{5'6} \\
    \Phi_{2'3}\Phi_{5'6} & 0                    &     0     \\
    \Phi_{2'4} \Phi_{5'6} & 0                    & 0\rule[-.45ex]{0.ex}{0.1ex}  
     \end{array} 
    \right) 
    \\
  M_2&=  
    \left( 
     \begin{array}{ccc}
        0                & \Phi_{2'3}\Phi_{5'7} & \Phi_{2'4}\Phi_{5'7} \\
    \Phi_{2'3}\Phi_{5'7} & 0                    &     0     \\
    \Phi_{2'4} \Phi_{5'7} & 0                    & 0\rule[-.45ex]{0.ex}{0.1ex}  
     \end{array} 
    \right) 
   \\   
  \Psi^T&=
   \left( 
     \begin{array}{ccc|ccc|ccc}
       h_{25} & h_{35} & h_{45} &
       h_{26} & h_{36} & h_{46} &
       h_{27} & h_{37} & h_{47} 
     \end{array}
   \right)  
 \end{aligned}
\end{equation}
The mass matrix \eqref{massmatrixhyp} for the
chiral fields has rank six, so that three combinations of the four
chiral fields, $h^{(1)}$, in $\{H_{36},H_{37}, H_{46},H_{47}\}$
remain massless. Since the intersection numbers in table 
\ref{chiralspecps} tell us
that there are no chiral fields in the $(1,2,2)$ representation of
the $U(4)\times U(2)\times U(2)$ gauge group, the other chiral
components, $h^{(2)}$, of the hyper-multiplets must also gain a mass
during brane recombination. A very similar behavior was found in 
\cite{Cremades:2002cs},
and it was pointed out that this might involve the
condensation of massive string modes, as well. These
would at least allow the correct mass terms in the quiver diagram.
We expect that the quiver diagram really tells us half of the complete
story, so that the non-chiral spectrum of the
three generation Pati-Salam model is as listed in table \ref{nchiral3stack}.
\begin{table}
 \renewcommand{\arraystretch}{1.2}
 \begin{center}
 \begin{tabular}{|c|c|c|}
 \hline
 field & n  & $U(4)\times U(2)\times U(2)$   \\
 \hline \hline
 $H_{aa}$  & 1 & $({\rm Adj},1,1)+c.c.$ \\
 $H_{bb}$  & 1 & $(1,{\rm Adj},1)+c.c.$ \\
 $H_{cc}$  & 1 & $(1,1,{\rm Adj})+c.c.$ \\
 \hline
 $H_{a'b}$  & 1 & $(4,{2},1)+c.c.$ \\
 \hline
 $H_{a'c}$  & 1 & $(4,1,{2})+c.c.$ \\
 \hline
 $H_{bc}$  & 3 & $(1,2,\Bar{2})+c.c.$ \\
\hline
 \end{tabular}
 \caption{\label{nchiral3stack}Non-chiral spectrum for 3 stack PS-model}
 \end{center}
\end{table}
Intriguingly, these are just appropriate Higgs fields to break
the Pati-Salam gauge group down to the Standard Model.

\subsection{Getting the Standard Model}
It is beyond the scope of this chapter to discuss all the
phenomenological consequences of this 3 generation Pati-Salam model.
However, we would like to present two possible ways of breaking
the GUT Pati-Salam model down to
the Standard Model.

\subsubsection{Adjoint Pati-Salam breaking}
There are still the adjoint scalars related
to the unconstrained positions of the branes on the third
$T^2$. By moving one of the four D$6$-branes away from the $U(4)$ stack,
or in other words by giving VEVs to appropriate fields in the adjoint
of $U(4)$, we can break the gauge group down to
$U(3)\times U(2)\times U(2)\times U(1)$. Indeed the resulting spectrum
as shown in table \ref{chiral4stlrspec} looks like a three generation left-right
symmetric extension of the Standard Model.
\begin{table}
 \renewcommand{\arraystretch}{1.2}
 \begin{center}
 \begin{tabular}{|c|clc|c|}
 \hline
 n  & \multicolumn{3}{c|}{$SU(3)_c\times SU(2)_L\times SU(2)_R\times U(1)^4$}  &
             $U(1)_{B-L}$ \\
 \hline \hline
 1 & \rule{7ex}{0ex}& $(3,2,1)_{(1,1,0,0)}$  & & $\phantom{-}{1\over 3}$ \\
 2 & & $(3,2,1)_{(1,-1,0,0)}$ & & $\phantom{-}{1\over 3}$\\
 \hline
 1 & & $(\Bar{3},1,2)_{(-1,0,-1,0)}$ & & $-{1\over 3}$ \\
 2 & & $(\Bar{3},1,2)_{(-1,0,1,0)}$  & & $-{1\over 3}$ \\
 \hline
 1 & & $(1,2,1)_{(0,1,0,1)}$  & & ${-1}$ \\
 2 & & $(1,2,1)_{(0,-1,0,1)}$ & & ${-1}$ \\
 \hline
 1 & & $(1,1,2)_{(0,0,-1,-1)}$ & & $\phantom{-}{1}$ \\
 2 & & $(1,1,2)_{(0,0,1,-1)}$  & & $\phantom{-}{1}$ \\
 \hline
 1 & & $(1,S+A,1)_{(0,2,0,0)}$ & & $\phantom{-}0$ \\
 1 & & $(1,1,\Bar{S}+\Bar{A})_{(0,0,-2,0)}$ & & $\phantom{-}0$ \\
 \hline
\end{tabular}
 \caption{\label{chiral4stlrspec}Chiral spectrum for 4 stack 
                  left-right symmetric SM}
 \end{center}
\end{table}
Performing the anomaly analysis, one finds
two anomaly free $U(1)$s, of which
the combination ${1\over 3}(U(1)_1-3U(1)_4)$ remains massless
even after the Green-Schwarz mechanism. This linear combination
in fact is the $U(1)_{B-L}$ symmetry, which is expected to be anomaly-free
in a model with right-handed neutrinos.

By giving a VEV to fields
in the adjoint of $U(2)_R$, one obtains the next symmetry breaking, where the two $U(2)_R$
branes split into two $U(1)$ branes. This gives rise to the gauge symmetry
$U(3)\times U(2)_L\times U(1)_R\times U(1)_R\times U(1)$.
In this case the following two $U(1)$ gauge factors remain
massless after checking the
Green-Schwarz couplings
\begin{equation}
  \label{massu}
  \begin{aligned}    U(1)_{B-L}&={1\over 3}(U(1)_1-3U(1)_5) \\
                        U(1)_Y&={1\over 3}U(1)_1 +U(1)_3-U(1)_4-U(1)_5 
  \end{aligned}
\end{equation}
It is very assuring that we indeed
obtain a massless hypercharge. The final supersymmetric chiral
spectrum is listed in table \ref{chiral5stspec} with respect to the unbroken 
gauge symmetries.

\begin{table}
 \renewcommand{\arraystretch}{1.2}
 \begin{center}
 \begin{tabular}{|c|c|clc|c|}
 \hline
 n  & field & \multicolumn{3}{c|}{$SU(3)\times SU(2)\times U(1)^3$} & $U(1)_Y \times U(1)_{B-L}$  \\
\hline \hline
 1 &   $q_L$ & \rule{1.5ex}{0ex}& $(3,2)_{(1,1,0,0,0)}$ & & $\left({1\over 3},{1\over
    3}\right)$ \\
 2 &   $q_L$ & & $(3,2)_{(1,-1,0,0,0)}$ & & $\left({1\over 3},{1\over
    3}\right)$ \\
\hline
 1 &   $u_R$ & & $(\Bar{3},1)_{(-1,0,-1,0,0)}$ & & $\left(-{4\over 3},-{1\over
    3}\right)$ \\
 2 &   $u_R$ & & $(\Bar{3},1)_{(-1,0,0,1,0)}$ & & $\left(-{4\over 3},-{1\over
    3}\right)$ \\
 2 &   $d_R$ & & $(\Bar{3},1)_{(-1,0,1,0,0)}$ & & $\left({2\over 3},-{1\over
    3}\right)$ \\
 1 &   $d_R$ & & $(\Bar{3},1)_{(-1,0,0,-1,0)}$ & & $\left({2\over 3},-{1\over
    3}\right)$ \\
\hline
 1 &   $l_L$ & & $(1,2)_{(0,1,0,0,1)}$     & & $\left(-{1},-{1}\right)$ \\
 2 &   $l_L$ & & $(1,2)_{(0,-1,0,0,1)}$    & &  $\left(-{1},-{1}\right)$ \\
\hline
 2 &   $e_R$   & & $(1,1)_{(0,0,1,0,-1)}$    & & $\left({2},{1}\right)$ \\
 1 &   $e_R$   & & $(1,1)_{(0,0,0,-1,-1)}$   & & $\left({2},{1}\right)$ \\
 1 &   $\nu_R$ & & $(1,1)_{(0,0,-1,0,-1)}$ & & $\left({0},{1}\right)$ \\
 2 &   $\nu_R$ & & $(1,1)_{(0,0,0,1,-1)}$  & & $\left({0},{1}\right)$ \\
\hline
 1 &   $$ & & $(1,S+A)_{(0,2,0,0,0)}$      & & $\left({0},{0}\right)$  \\
 1 &   $$ & & $(1,1)_{(0,0,-2,0,0)}$       & & $\left(-{2},{0}\right)$  \\
 1 &   $$ & & $(1,1)_{(0,0,0,-2,0)}$       & & $\left({2},{0}\right)$  \\
 2 &   $$ & & $(1,1)_{(0,0,-1,-1,0)}$      & & $\left({0},{0}\right)$  \\
  \hline
\end{tabular}
 \caption{\label{chiral5stspec}Chiral spectrum for 5 stack SM}
 \end{center}
\end{table}

The anomalous $U(1)_1$ can be identified with the baryon number
operator and survives the Green-Schwarz mechanism as a global
symmetry. Therefore, in this model the baryon number
is conserved and the proton is stable.
Similarly, $U(1)_5$ can be identified with the lepton number
and also survives as a global symmetry.
To break the gauge symmetry $U(1)_{B-L}$, one can recombine
the third and the fifth stack of D$6$ branes, which is expected to
 correspond  to giving
a VEV to the Higgs field $H_{3'5}$. We will see in section
\ref{bifupsbreaksec} that this brane recombination gives a mass 
to the right-handed neutrino.

To proceed, let us compute the relation between the Standard Model
gauge couplings at the PS-breaking   scale at string tree
level. The $U(N_a)$ gauge couplings for D$6$-branes are given by
\begin{equation}
   \label{gaugec} {4\pi\over g_a^2}={M^3_s\over g_s } {\text{Vol}(\text{D}6_a)}
\end{equation}
where Vol$(\text{D}6_a)$ denotes the internal volume of the 3-cycle the
D$6$-branes are  wrapping on. During the brane recombination process the
volume of the recombined brane is equal to the sum of the volumes
of the two intersecting branes. Therefore, we have the
following ratios for the volumes of the five stacks of D$6$-branes in our model
\begin{equation}
   \label{ratioa}
      { \text{Vol}(\text{D}6_2)}={\text{Vol}(\text{D}6_3)}={\text{Vol}(\text{D}6_4)}=3{\text{Vol}(\text{D}6_1)}, \quad
     {\text{Vol}(\text{D}6_5)}={ \text{Vol}(\text{D}6_1)} 
\end{equation}
This allows us at string tree level to determine the ratio of the 
Standard Model gauge couplings at the PS-breaking scale to be
\begin{equation}
   \label{ratiob}
      {\alpha_{s}\over \alpha_Y}={11\over 3}, \quad\quad
                {\alpha_{w}\over \alpha_Y}={11\over 9}
\end{equation}
leading to a Weinberg angle $\sin^2 \theta_w=9/20$ which differs
from the usual $SU(5)$ GUT prediction
$\sin^2(\theta_w)=3/8$. Encouragingly, from \eqref{ratiob} we get the
right order for the sizes of the Standard Model gauge couplings
constants, $\alpha_{s}> \alpha_{w}>\alpha_Y$.
It would be interesting to analyze
whether this GUT value is
consistent with the low energy data at the weak scale. A potential
problem is the  appearance of colored Higgs fields in table \ref{nchiral3stack},
which would spoil the asymptotic freedom of the $SU(3)$. In  order to
improve this situation one needs a model with less
non-chiral matter, i.e. a model where not so many open string sectors
actually preserve ${\cal N}=2$ supersymmetry.

\subsubsection{\label{bifupsbreaksec}Bifundamental  Pati-Salam breaking}
\begin{figure}
\begin{center}
\includegraphics{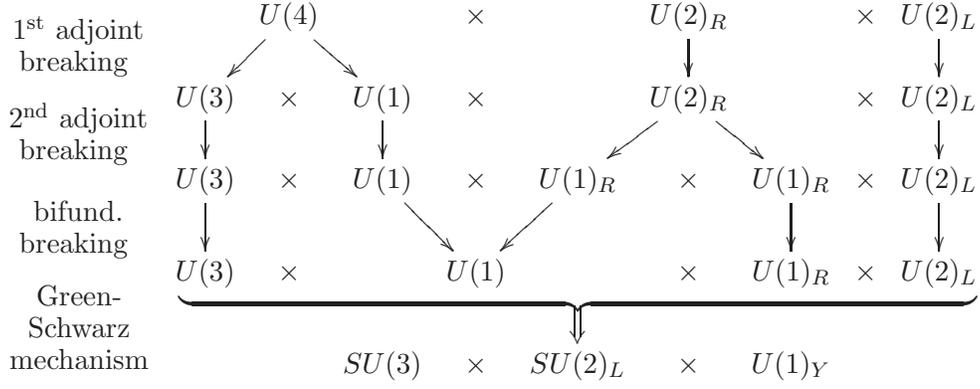}
\caption{\label{gaugesbreakdiag} Gauge symmetry breaking of
              $U(4)\times U(2)_L\times U(2)_R$}
\end{center}
\end{figure}

We can also use directly the bifundamental Higgs fields like
$H_{a'c}$ to break the model down to the Standard Model gauge
group. This higgsing in string theory should correspond to a
recombination of one of the four D$6$-branes wrapping $\pi_a$ with one of
the branes wrapping $\pi'_c$. Thus, we get the following four
stacks of D$6$-branes
\begin{equation}
   \label{recomp}
                 \pi_A=\pi_a, \quad
         \pi_B=\pi_b, \quad \pi_C=\pi_a+\pi'_c, \quad \pi_D=\pi_c
\end{equation}
supporting  the initial gauge group $U(3)\times U(2)\times
U(1)^2$. The tadpole cancellation conditions are still satisfied.
One gets the chiral spectrum by computing the homological
intersection numbers as shown in table \ref{chiral4stspec}.
\begin{table}
 \renewcommand{\arraystretch}{1.2}
 \begin{center}
 \begin{tabular}{|c|c|c|clc|c|}
 \hline
 n & field  & sector &  \multicolumn{3}{c|}{$SU(3)_c\times SU(2)_L\times U(1)^4$} & $U(1)_Y$ \\
\hline \hline
 2 & $q_L$ &  $(AB)$ & \rule{3.5ex}{0ex} & $(3,2)_{(1,-1,0,0)}$ & & $\phantom{-}{1\over 3}$ \\
 1 & $q_L$ &  $(A'B)$ & & $(3,2)_{(1,1,0,0)}$  & & $\phantom{-}{1\over 3}$ \\
\hline
 1 & $u_R$ &  $(AC)$ & & $(\Bar{3},1)_{(-1,0,1,0)}$  & & $-{4\over 3}$ \\
 2 & $d_R$ &  $(A'C)$ & & $(\Bar{3},1)_{(-1,0,-1,0)}$ & & $\phantom{-}{2\over 3}$ \\
\hline
 2 & $u_R$ &  $(AD)$ & & $(\Bar{3},1)_{(-1,0,0,1)}$  & & $-{4\over 3}$ \\
 1 & $d_R$ &  $(A'D)$ & & $(\Bar{3},1)_{(-1,0,0,-1)}$ & & $\phantom{-}{2\over 3}$ \\
\hline
 2 & $l_L$ &  $(BC)$ & & $(1,2)_{(0,-1,1,0)}$  & & $-{1}$  \\
 1 & $l_L$ &  $(B'C)$ & & $(1,2)_{(0,1,1,0)}$  & & $-{1}$  \\
\hline
 1 & $e_R$ &  $(C'D)$ & & $(1,1)_{(0,0,-1,-1)}$ & & $\phantom{-}{2}$ \\
 1 & $e_R$ & $(C'C)$   & & $(1,1)_{(0,0,-2,0)}$ & & $\phantom{-}2$ \\
 1 & $e_R$ & $(D'D)$   & & $(1,1)_{(0,0,0,-2)}$ & & $\phantom{-}2 $ \\
\hline
 1 & $S$  & $(B'B)$   & & $(1,S+A)_{(0,2,0,0)}$ & & $\phantom{-}0$ \\
\hline
\end{tabular}
 \caption{\label{chiral4stspec}Chiral spectrum for 4 stack SM}
 \end{center}
\end{table}
By computing the mixed anomalies, one finds that there are
two anomalous $U(1)$ gauge factors and two anomaly free ones
\begin{equation}
    \label{massub}
    \begin{aligned}
                   U(1)_Y&={1\over 3}U(1)_A-U(1)_C - U(1)_D \\
                        U(1)_{K}&=U(1)_A-9\, U(1)_B +9\, U(1)_C-9\,U(1)_D
    \end{aligned}
\end{equation}
Remarkably, the axionic couplings just leave the hypercharge massless,
so that we finally get the Standard Model gauge group
$SU(3)_C\times SU(2)_L\times U(1)_Y$.
In this model only the baryon number generator can be identified with
 $U(1)_1$, whereas
the lepton number is broken. Therefore, the proton is stable and lepton number
violating couplings as Majorana mass terms are possible.
Note, that there are no  massless right-handed neutrinos
in this model. As we have mentioned already, this model is related
to the model discussed in the last section by a further brane recombination
process, affecting the mass of the right-handed neutrinos.
This brane recombination can be considered as a stringy mechanism
to generate GUT scale masses for the right-handed neutrinos \cite{Cremades:2002cs}.
The different ways of gauge symmetry breaking that have been discussed so 
far are depicted in figure \ref{gaugesbreakdiag}.

It is evident
from table \ref{chiral4stspec} that there is also something unusually going on
with the right-handed leptons. Only one of them
is realized as a bifundamental field, the remaining
two are given by symmetric representations of $U(1)$.
This behavior surely will have consequences for the allowed
couplings, in particular for the Yukawa couplings and the electroweak Higgs
mechanism.

Computing the gauge couplings, we find the following
ratios for the internal volumes of the four 3-cycles
\begin{equation}
   \label{ratioc}
         {\text{Vol}(\text{D}6_2)}={\text{Vol}(\text{D}6_4)}
          =3{\text{Vol}(\text{D}6_1)}, \quad
             {\text{Vol}(\text{D}6_3)}=4{\text{Vol}(\text{D}6_1)} 
\end{equation}
This allows us to determine the ratio of the Standard Model
gauge couplings at the GUT scale to be again
\begin{equation}
   \label{ratiod}
        {\alpha_{s}\over \alpha_Y}={11\over 3}, \quad\quad
                {\alpha_{w}\over \alpha_Y}={11\over 9} 
\end{equation}
leading to a Weinberg angle $\sin^2 \theta_w={9/20}$.
Thus, both models provide  the same prediction for the Weinberg-angle
at the GUT scale.

\subsubsection{Electroweak symmetry breaking}
Finally, we would like to make some comments on electroweak
symmetry breaking in this model.
From the quiver diagram of the $U(4)\times U(2)\times U(2)$ Pati-Salam
model we do not expect that the three Higgs fields in the $(1,\Bar{2},2)$
representation
get a mass during the brane recombination process. Therefore,
our model does contain appropriate Higgs fields to participate in the
 electroweak
symmetry breaking.
The three Higgs fields, $H_{bc}$,  in the Pati-Salam model in table \ref{nchiral3stack}
give rise to the Higgs fields
\begin{equation}
   \label{higgsfield}
             H_{BD}=(1,2)_{(0,1,0,-1)} +c.c.\, ,
                  \quad\quad H_{B'C}=(1,2)_{(0,1,1,0)} +c.c.
\end{equation}
for the $SU(3)_c\times SU(2)_L\times U(1)_Y$ model above.

Of course supersymmetry should already be broken by some mechanism
above the electroweak symmetry breaking scale, but nevertheless
we can safely discuss the expectations  from the purely
topological data of the corresponding brane recombination process.
Since we do not want to break the color $SU(3)$, we still take a
stack of three D$6$-branes which are wrapped on the cycle $\pi_\alpha=\pi_A$.
Giving a VEV to the fields $H_{BD}$ is expected to correspond
to the brane recombination
\begin{equation}
   \label{bewbrane}  \pi_\beta=\pi_B+\pi_D
\end{equation}
However, for the brane recombination
\begin{equation}
   \label{newbrane}
         \pi_\gamma=\pi'_B+\pi_C
\end{equation}
the identification with the corresponding
field theory deformation is slightly more subtle, as the
intersections between these two branes  support both the massless
chiral  multiplet $l_L^{B'C}$ as listed in table \ref{chiral4stspec} and the Higgs
field $H_{B'C}$. Thus, the intersection preserves only ${\cal N}=1$ 
supersymmetry and one might expect that some combination of
$l_L^{B'C}$ and  $H_{B'C}$ are involved in the brane recombination
process. Even without knowing all the details, in the following we
can safely compute the chiral spectrum via intersection numbers.

After the brane recombination we have
a naive gauge group $U(3)\times U(1)\times U(1)$, which however is
broken by the Green-Schwarz couplings to $SU(3)_c\times
U(1)_{\text{em}}$ with
\begin{equation}
   \label{elektro} 
                U(1)_{\text{em}}={1\over 6}
                   U(1)_\alpha-{1\over 2} U(1)_\beta +
                  {1\over 2} U(1)_\gamma
\end{equation}
Interestingly, just $U(1)_{\text{em}}$ survives this brane recombination
process. Moreover, all intersection numbers vanish, so that there
are no  chiral massless fields, i.e. all quarks and leptons
in table \ref{chiral4stspec} have gained a mass including the left-handed neutrinos
and the exotic matter.
Looking at the charges in table \ref{chiral4stspec}, one realizes that in the
leptonic sector this Higgs effect cannot be the usual one, where simply
$l_L$ and $e_R$ receive a mass via some Yukawa couplings.
Here also higher dimensional couplings, like the dimension five coupling
\begin{equation}
   \label{higherc}
       W\sim  {1\over M_s} \Bar{H}_{BD}\,\Bar{H}_{BD}\,  S\, e^{D'D}_R 
\end{equation}
are relevant. These couplings induce a mixing of the Standard Model
matter with the exotic field, $S$. Thus we can state, that
by realizing some of the right-handed leptons in the (anti-)symmetric
representation, the exotic field is needed to give
all leptons a mass during electroweak symmetry breaking.
It remains to be seen whether the induced masses can be consistent
with the low-energy data.

\addcontentsline{toc}{section}{Concluding remarks}
\section*{Concluding remarks}

In this chapter we have studied intersecting brane worlds for the
$T^6/\mathbb{Z}_4$ orientifold background with special emphasis on
supersymmetric configurations. We have found as a first
non-trivial result a supersymmetric three generation
Pati-Salam type extension of the Standard Model with some exotic
matter. The chiral matter content is only slightly extended by one
chiral multiplet in the (anti-)symmetric representation of
$SU(2)_L$. The presence of this exotic matter can be traced back
to the fact that we were starting with a Pati-Salam gauge group,
where the anomaly constraints forced us to introduce additional
matter. Issues which arose for non-supersymmetric models will also
appear in the supersymmetric setting. Since the Green-Schwarz
mechanism produces global $U(1)$ symmetries, the allowed couplings
in the effective gauge theory are usually much more constrained
than for the Standard Model.

With such model at hand, many
phenomenological issues deserve to be studied, as for instance
mechanisms for supersymmetry breaking, the generation of soft
breaking terms, Yukawa and higher dimensional couplings, the
generation of $\mu$-terms and gauge coupling unification.\footnote{Yukawa 
couplings for  toroidal $\bar{\sigma}\Omega$-orientifolds have been
 calculated in
\cite{Cremades:2003qj} and \cite{Cvetic:2003ch} (also four-point couplings).
Abel and Owen have investigated  three- and four-point tree-level amplitudes
for intersecting branes in \cite{Abel:2003vv} and extended their analysis in a
more recent paper to N-point disk amplitudes \cite{Abel:2003yx}. 
 However it is not clear if these calculations 
might be generalized to the case of recombined D-branes that 
appear in the models discussed in this chapter. The evolution of 
gauge-couplings including  threshold corrections 
was investigated for  
supersymmetric $\bar{\sigma}\Omega$-orientifolds in \cite{Lust:2003ky}.} 
 It also
remains to be seen whether the electroweak Higgs effect indeed
produces the correct masses for all quarks and leptons. Moreover,
one should check whether the renormalization of the gauge
couplings from the string respectively the PS-breaking scale down
to the weak scale can lead to acceptable values for the Weinberg
angle.\footnote{In \cite{Blumenhagen:2003jy} issues concerning gauge-coupling
unification were addressed in the context of supersymmetric 
$\bar{\sigma}\Omega$-orientifolds with D$6$-branes.}

The motivation  for this analysis was to start a
systematic search for realistic supersymmetric intersecting brane
world models. We have worked out some of the technical model
building aspects when one is dealing with more complicated
orbifold backgrounds containing in particular twisted sector
3-cycles. These techniques can be directly generalized to, for
instance, the $\mathbb{Z}_6$ orientifolds  \cite{Blumenhagen:1999ev} or 
the $\mathbb{Z}_N\times\mathbb{Z}_M$  orientifold  models
\cite{Forste:2000hx}.\footnote{Meanwhile
  the $\mathbb{Z}_4\times\mathbb{Z}_2$ $\bar{\sigma}\Omega$-orientifold with 
  projection has been studied \cite{Honecker:2003vq}.} It could be worthwhile to
undertake a similar study for these orbifold models, too.

The final goal would be to find a realization of the
MSSM in some simple intersecting brane world model.
As should have become clear from our analysis, while
phenomenologically interesting non-supersymmetric models are fairly
easy to get, the same is not true for the supersymmetric
ones. Requiring supersymmetry imposes very strong constraints
on the possible configurations and as we have observed in the
$\mathbb{Z}_4$ example, also the supply of possible  intersection numbers
is very limited.
These obstructions appear to be less surprising, when one contemplates that
for smooth backgrounds, by lifting to M-theory,
the construction of an ${\cal N}=1$
chiral intersecting brane world background with O$6$ planes and
D$6$ branes is equivalent
to the construction of a compact singular $G_2$ manifold.
In this respect it would be interesting whether certain M-theory
orbifold constructions like the one discussed in \cite{Doran:2002iz} are 
dual to the kind of models discussed in this chapter.

At a certain scale close to the TeV scale supersymmetry has to be
broken. For the intersecting brane world scenario one might
envision different mechanisms for such a breaking. First, we might
use the conventional mechanism of gaugino condensation  via some
non-perturbative effect.
Alternatively, one could build models where the MSSM is localized on a
number of D-branes, but where the RR-tadpole cancellation
conditions requires the introduction of hidden sector branes, on
which supersymmetry might be broken. This breaking could be
mediated gravitationally to the Standard Model branes. A third
possibility is to get D-term supersymmetry breaking by generating
effective Fayet-Iliopoulos terms via complex structure
deformations. We think that these issues and other
phenomenological questions deserve to be studied in the future.

\newpage
\chapter*{Conclusions}
\addcontentsline{toc}{chapter}{Conclusions}
In this thesis we have investigated specific kinds of open-string theories.
A striking feature of all constructions we considered is that they
potentially contain
chiral fermions. From the string theoretic point of view, chiral fermions
arise due to non-trivial boundary conditions of the open-string. 
In the case of world-sheet supersymmetry the Ramond sector yields a reduced
number of zero-modes.
To be more specific, this could lead to a single zero-mode after GSO projection.
By compactification to four space-time dimensions this results in a single 
Weyl fermion.
 However the number of  world-sheet bosonic zero modes
(e.g. intersection points or Landau levels) can be increased by the boundary
conditions. 
As the Hilbert space of the string states is a
product of world-sheet bosons and fermions: 
${\cal H}_\text{bos}\otimes{\cal H}_\text{ferm}$, the degeneracy 
 of world-sheet bosonic zero modes is inherited by the Ramond-sector.
Furthermore one has to take into account, that there might appear 
space-time fermions of both chiralities. Therefore the multiplicity
of world-sheet bosons encounters possible signs. As a result the total number
of fermions with definite chirality is given by purely {\sl topological}
quantities like the intersection number of D-branes or the index of the twisted
spin complex. 
In model-building this degeneracy is (roughly speaking) interpreted as
the number of generations of a specific particle type. Therefore the 
degenerate states should be split by some mechanism\footnote{Such a mechanism
should have a string-theoretic interpretation. For example
 different masses could
arise due to Yukawa-couplings which are associated to some area of the
world-sheet in intersecting brane-world models (cf.\
\cite{Cremades:2003qj,Cvetic:2003ch}).}
 into states of different
masses, but otherwise identical quantum numbers.   
 By adjusting the topological data in a bottom-up approach, we could generate
many phenomenologically appealing spectra. 

As a basis for the subsequent chapters  we quantized the
open string with linear, but independent boundary conditions that are induced
by D-branes of arbitrary dimension with constant NSNS $B$-  and NS $F$-field(s)
 in chapter \ref{strbg}. 
We confirmed the disk result of Seiberg and Witten on the non-commutativity
of open-string boundaries for the one-loop case (with arbitrary, constant
${\cal F}$-fluxes on the string-endpoints, but without Dirichlet conditions).
Furthermore we investigated the zero- and momentum-mode spectrum in toroidal
compactifications. It was shown that a kind of Landau degeneracy shows up, if
the string end-points couple to different NS $F$-fields. This is an example
of the degeneracy  mentioned above. 

Chapter \ref{ncg} mainly reviews the content of our publication 
\cite{Blumenhagen:2000fp}. 
In this article we investigated the D-brane spectrum of asymmetric
orbifolds and orientifolds. It turned out that left-right asymmetric twists 
imply in many cases the presence of D-branes with magnetic background fluxes,
since the D-brane configuration has to be symmetrized under the (asymmetric)
orbifold Group $G$. As an example we presented a space-time six-dimensional
model that was obtained from orientifolding a 
$T^4/\mathbb{Z}^\text{L}_3\times\mathbb{Z}^\text{R}_3$-orbifold.\footnote{We
  considered three variations of this orientifold. The main difference  in
  these  three different orientifolds is due to two
  alternative  actions of the world sheet
  parity  $\Omega$ on the zero- and momentum-modes. This is 
  equivalent to different choices for the 
  complex- and K\"ahler-structure of each of the two two-tori 
 ($T^4=T^2\times T^2$). The two different choices 
  result in three inequivalent orientifold models.
  One of these models was investigated before in \cite{Bianchi:1999uq} by
  means of conformal field theory. However a D-brane interpretation of the
  open-string sector could not be given in this former publication.}
This article was finished before the work
presented in chapter \ref{strbg} was done. With some new insights gained in 
this chapter it was now possible to answer some so far open questions, like the
quantization of the open-string momentum modes on toroidally compactified
D-branes both for magnetized branes and lower dimensional branes from first principles.\footnote{The
  quantization for open strings on lower dimensional branes was done earlier
for vanishing NSNS $B$-field.} This quantization was derived up to now only
indirectly via the open-closed string correspondence (i.e.\ boundary states,
cf.\ \cite{Arfaei:1999af,Arfaei:1999kr}).  

Chapter \ref{magbf} is devoted to purely toroidal orientifolds. We considered
both space-time six- and four-dimensional compactifications, i.e.\
compactifications on $T^6$ resp.\ $T^4$. Computations can be either done in
the pure $\Omega$-orientifold with D$9$-branes carrying NS $U(1)$-fluxes in
the compact directions, or in the T-dual picture, where $\Omega$ is combined
with complex conjugation $\bar{\sigma}$. Here the O-plane fills only a real
subspace of the torus and its RR charge is canceled by D-branes of the same 
dimensionality. This means that we introduce D$7$-branes for the
$T^4$-compactification and  D$6$-branes on the six-torus $T^6$. Chiral
fermions arise in the  $\bar{\sigma}\Omega$-orientifold at the intersection points of the
 D$7$- resp.\ D$6$-branes. The number of chiral fermion generations is then
 given by the topological intersection numbers, while in the flux picture it
 is due to Landau degeneracies, which can be calculated by an index-theorem.\footnote{There are some subtleties, like the
   impossibility of obtaining the complete left-handed quark sector just from
   two kinds of branes (cf.\ section \ref{fourgenmodel}).} 
 In order to get an interesting spectrum, one tries to distribute D-branes in
 such a way that 
 a) they cancel the RR tadpole (thereby ensuring anomaly cancellation) and 
 b) they yield the desired intersection numbers (resp.\ index).
 We were able to construct a four generation model with SM-like spectrum
 and gauge group $SU(3)\times SU(2)\times U(1)_Y\times U(1)^2$. The
 obstruction that the number of  generations has to be even can be overcome
 by including so called $\mathbf{B}$-type tori as shown in the subsequent
 publication \cite{Blumenhagen:2000ea}.\footnote{$\mathbf{B}$-type means on the
  in the ``flux'' picture that the NSNS $B$-field is set to $\al/2$, 
  which is still an $\Omega$-symmetric background. 
  In the T-dual ``branes at angles'' picture, the real part of the
  complex-structure $\tau$ of the $\mathbf{B}$-type two-torus is fixed to
   $\tau_1=1/2$.} 
However chiral configurations in purely toroidal constructions always break
supersymmetry. Therefore the solutions might be unstable (i.e.\ divergent to
a singular limit). Besides the Fischler-Susskind mechanism\footnote{It is
  however not clear, if the Fischler-Susskind mechanism does lead to a
  non-degenerate and non-supersymmetric limit.} there might exist
further   (yet unknown) mechanisms to stabilize the non-supersymmetric vacua.

Chapter \ref{z4} deals with the  $\bar{\sigma}\Omega$-orientifold of an  ${\cal N}=2$
supersymmetric $T^6/\mathbb{Z}_4$ orbifold.
It is the second example besides the $\mathbb{Z}_2\times\mathbb{Z}_2$
orientifold models (cf.\ \cite{Cvetic:2001nr}) that a
$\bar{\sigma}\Omega$-orientifold admits chiral supersymmetric solutions.
This chapter is mainly
based on our publication \cite{Blumenhagen:2002gw}, however more detailed in
some points.
We concentrate in the second half on an $U(4)\times U(2)^3_\text{L}\times
U(2)^3_\text{R}$-model which we can break down to a three generation 
Pati Salam  $U(4)\times U(2)_\text{L}\times U(2)_\text{R}$-model, while
preserving supersymmetry. This is done by giving VEVs to fields in the low
energy effective action. The chiral fermion spectrum in the latter model is
given again by topological intersection numbers, while the non-chiral spectrum
(i.e.\ hypermultiplets) is obtained by field theoretic considerations. Fields
in the non-chiral spectrum serve as Higgs-particles for  breaking the model
down to an MSSM like spectrum.

While non-supersymmetric models have to deal with instabilities due to NSNS
tadpoles (and potentially with open string tachyons, too), the supersymmetric
models are stable.\footnote{However instabilities might be induced by quantum
  or instanton corrections or so far unknown mechanisms.}
However it is still an open issue how to break supersymmetry in a way that is
manifestly compatible with string-theory.\footnote{Field-theoretic
  considerations in related models do exist, some of them involving
  non-perturbative effects.}
The brane recombination process deserves a microscopic
explanation.\footnote{This topic has recently been addressed by Hashimoto at 
  the {\it  Strings 2003}  conference, however in a more general context.}
Since special Lagrangian submanifolds play a prominent role in the construction
of supersymmetric intersecting brane worlds, a richer knowledge about these
objects is desirable, also in the more general context than 
Calabi-Yau orbifolds.

As another aim one could consider other toroidal orientifolds, searching again
for a stringy realization of the MSSM.

We want to conclude with these few suggestions, even though many other 
questions related to this kind of orientifold constructions 
should be  addressed, too.

\newpage
\chapter*{Acknowledgments}
\addcontentsline{toc}{chapter}{Acknowledgments}
First of all I would like to thank  Professor Dieter L\"ust who offered
me the possibility  to do my Ph.D.\ under his supervision. 
I am very grateful to Ralph Blumenhagen, who taught me a lot of things
in string theory. I would also like to thank Boris K\"ors who was Ph.D.\
student in Berlin during my Diploma and early Ph.D.\ time. With all
of these three gentlemen I published several papers, two of them during 
the time of my Ph.D.~thesis. I would also like to thank Tassilo Ott.
With him I had an extremely fruitful collaboration at the end of my thesis 
resulting in a publication, too. 

I thank the colleagues who shared the  office with me for providing
a pleasant atmosphere:
Matthias Br\"andle, Volker Braun, Stefano Chiantese, Claus Jeschek and
 Christoph Sieg.

I am indebted to Stephan Stieberger for proof-reading my thesis.

With many other current and former members of our group I had extended and 
fruitful discussions.
I would like to thank:
Gianguido Dall'Agata,
Oleg Andreev,
Klaus Behrndt,
Gabriel Cardoso,
Gottfried Curio,
Bernd-Dietrich D\"orfel,
Harald Dorn,
Johanna Erdmenger,
Andrea Gregori,
Johannes Grosse,
Zachary Guralnik,
Robert Helling, 
Albrecht Klemm,
Georgios Kraniotis,
Ingo Kirsch,
Axel Krause,
Karl Landsteiner,
Calin Lazaroiu,
Andr{\'e} Miemiec,
Aalok Misra,
Hans-J\"org Otto,
Nikolaos Prezas,
Ingo Runkel,
Mario Salizzoni
and 
Radu Tatar.

I am also grateful to our secretary Sylvia Richter and our former 
secretary Susanne Preisser for helping me with administrative problems. 

I would like to express my gratitude to my parents, who  supported me
during  my University studies.

My work was funded by the ``Deutsche Forschungsgemeinschaft'' (DFG) 
under the ``Schwerpunktprogramm'' (Priority Programme) 1096 with project  
number DFG Lu 419/7-2.

The Berlin group was supported by the EEC contract ERBFMRXCT96-0045 as well.


\begin{appendix}
\chapter{\label{thetafunctions}Theta-functions and related functions}
\section{\texorpdfstring{$\eta$ and $\vartheta$-functions}
         {Eta and theta-functions}, identities and 
transformation under $SL (2,\mathbb{Z})$}
The $\vartheta$-functions are  defined as follows
\beqn
\thef{a}{b}(\tau) = \sum_{n\in \mathbb{Z}} q^{\oh(n+a)^2} \,
{\rm e}^{2i\pi (n+a) b}
\label{vts}
\eeqn
$q$ is defined by  $q={\rm e}^{2\pi i\tau }$ ($\mathfrak{Im} \tau > 0$).
The  $\vartheta$-functions admit the following representation as an
infinite product:
\beqn
\frac{\thef{a}{b}}{\eta} = {\rm e}^{2i\pi a b} \, q^{\oh a^2 - \frac1{24}} \,
          \prod_{n=1}^\infty \big(1 + q^{n+a -\oh} {\rm e}^{2i\pi b} \big) \,
          \big(1 + q^{n-a -\oh} {\rm e}^{-2i\pi b} \big),
\label{vtp}
\eeqn
with  $\eta$ being the Dedekind $\eta$-function:
\beqn
\eta =  q^{\frac1{24}} \,
\prod_{n=1}^\infty (1 - q^n)
\label{deta}
\eeqn

\subsection{Transformation under $SL(2,\mathbb{Z})$:}
The $\vartheta$- and $\eta$-functions
transform under the generators $S$, $T$ of the modular group
$SL(2,\mathbb{Z})$  and under  $P=TST^2S$ as follows:
\newline
\rule{0.ex}{3ex}$S$-transformation:
\begin{align} 
&\tau\stackrel{S}{\rightarrow}-1/\tau  \\ \label{Stransf}
\frac{1}{\eta}\thef{a}{b} 
(\tau)&=e^{2\pi i\,ab}\frac{1}{\eta}\thef{b}{-a}(-1/\tau)  &
\eta(\tau)&=\sqrt{\frac{i}{\tau}}\eta(-1/\tau)
\end{align} 
\rule{0.ex}{3ex}$T$-transformation:
\begin{align} \label{Ttransf}
&\tau\stackrel{T}{\rightarrow}\tau+1  \\
\thef{a}{b} 
(\tau)&=  e^{i\pi \left(a^2-a\right)}\thef{a}{b-a+1/2}(\tau+1) \\
\eta(\tau)&=e^{-i\pi/12}\eta(\tau+1)
\end{align} 
The following $P$-transformation prove useful  in transforming
 M\"obius-strip amplitudes
from loop- to tree-channel. We have already inserted
the M\"obius-strip relation $t = \tfrac{1}{8l}$. The parameters $a$ and $b$
are  not independent in the M\"obius amplitude s.th.\ the phases
get a simpler form. \\ 
\rule{0.ex}{3ex}$P$-Transformation:
\begin{equation}\label{Ptransf}
 P=T\circ S\circ T^2\circ S \qquad
 \tau=it+\frac{1}{2}=\frac{i}{8l}+\frac{1}{2}\stackrel{P}{\rightarrow}
 \tau=i2l+\frac{1}{2} \qquad P\equiv T\circ S\circ T^2\circ S 
\end{equation}
\begin{align} \label{Ptransf2}
  \frac{1}{\eta}{\thef{a}{b}}\left(\tau = it+\tfrac{1}{2}\right) 
    &= e^{-i\pi \,\left( \frac{3}{2} + a^2 + 
        2 \left(2a +b\right)\left( b-1 \right)\right)   }
    \frac{1}{\eta}\thef{ 2 b-a}{  b-a-3/2}\left(\tau=i2l+\tfrac{1}{2}\right)
  \\
 \label{Ptransf3}
 \eta\left(\tau = it+\tfrac{1}{2}\right)&= 
    \sqrt{l}\, \eta\left(\tau = i2l+ \tfrac{1}{2}\right)
 \end{align} 
\subsection{\label{riemid}Identities between 
            \texorpdfstring{$\vt$}{theta}-functions}
The $\vt$-functions obey several Riemannian identities 
\cite{mum}. Supersymmetry shows up in the partition functions by
the vanishing of the vacuum-amplitudes. The phases of the different
sectors ($(NS,+)$, $(NS,-)$,
$(R,+)$),  which implicitly determine the GSO projection can be 
determined by these identities. For $u_1+u_2+u_3=0$ we have:  
\begin{align}
 \label{abstruone}
 \sum_{\alpha,\beta\in\{0,1/2\}} \epsilon_{\alpha,\beta} \ 
  \thef{\alpha}{\beta} \prod_{i=1}^3 \thef{\alpha}{\beta+u_i} = 0 \\
 \label{abstrutwo}
 \sum_{\alpha,\beta\in\{0,1/2\}} \epsilon_{\alpha,\beta} \ \thef{\alpha}{\beta}
\thef{\alpha}{\beta+u_3}
\prod_{i=1}^2 \thef{\alpha+\oh}{\beta+u_i} = 0\\ \non
\epsilon_{0,0}=1,\quad\epsilon_{0,1/2}=\epsilon_{1/2,0}=-1
\end{align}
We set $u_3=0$ in the six dimensional models of chapter \ref{magbf}.

\section{Poisson resummation formula for lattice sums}
Sums of the following type are involved in traces over Kaluza Klein and
winding contributions (for $\vec{a},\vec{b}\in\mathbb{R}^d$, 
  $S$ a real symmetric and 
non degenerate $d\times d$ matrix):
\begin{multline}\label{poissonf1}
\sum_{\vec{v}\in\mathbb{Z}^d} 
 e^{i2\pi(\vec{v}+\vec{a})\cdot\vec{b}}
 e^{-\pi t\left((\vec{v}+\vec{a})^T\, S\,(\vec{v}+\vec{a})\right)}
 \\
 =t^{-d/2}\frac{1}{\sqrt{\det S}}\sum_{\vec{w}\in\mathbb{Z}^d} 
 e^{-i2\pi\vec{w}\cdot\vec{a}}
 e^{-\frac{\pi}{t}\left( (\vec{w}+\vec{b})^T\,S^{-1}\,(\vec{w}+\vec{b})\right)}
\end{multline}

\section{\label{confblocks}Conformal blocks in $D=6$}
In this section we summarize the conformal blocks that we use to shorten the
notation for the asymmetric  $\mathbb{Z}^L_3\times\mathbb{Z}^R_3$ orbifold
on $T^2\times T^2$
in section \ref{asz3orbi} (p.\ \pageref{asz3orbi}) and
its orientifolded version (or open descendants) in 
section \ref{asz3orienti} (p.\ \pageref{asz3orienti}).
We define:\footnote{We adopted our notation 
                    from \cite{Bianchi:1999uq} and restricted it to the
                    case $d=4$, i.e.\ compactification on a $T^4$.}
\begin{equation}\label{rhos}
    \begin{aligned} \rho_{00}&=\frac{1}{2}\sum_{\alpha,\beta=0,{1\over 2}}
           (-1)^{2\alpha+2\beta+4\alpha\beta}{\thef{\alpha}{\beta}^4\over 
                \eta^{4}}, & & \\ 
            \rho_{0h}&=\frac{1}{2}\sum_{\alpha,\beta=0,{1\over 2}}
           (-1)^{2\alpha+2\beta+4\alpha\beta}{\thef{\alpha}{\beta}^2\over 
                \eta^2} \prod_{i=1}^2 2\sin(\pi h_i)
                {\thef{\alpha}{\beta+h_i}\over 
                \thef{{1\over 2}}{{1\over 2}+h_i}}  & h&\ne 0  \\
            \rho_{gh}&=\frac{1}{2}\sum_{\alpha,\beta=0,{1\over 2}}
           (-1)^{2\alpha+2\beta+4\alpha\beta}{\thef{\alpha}{\beta}^2\over 
                \eta^2} \prod_{i=1}^2 
                {\thef{\alpha+g_i}{\beta+h_i}\over 
                \thef{{1\over 2}+g_i}{{1\over 2}+h_i}}   & g,h&\neq 0 
    \end{aligned}
\end{equation}
The functions \eqref{rhos} transform under $\tau\xrightarrow{S}-1/\tau$ like:
\be
 \begin{aligned}\label{rhosS}
  \rho_{00} &\rightarrow & \rho_{00} & \\
  \rho_{0h} &\rightarrow
   & (2 \sin{\pi h})^{2}\, \rho_{h0}& &h\neq 0 \\
  \rho_{h0} &\rightarrow
   & (2 \sin{\pi h})^{-2}\, \rho_{0,-h}& &h\neq 0 \\
  \rho_{gg} &\rightarrow&
   -\rho_{g,-g}                 & &g\neq 0 \\
  \rho_{g,-g} &\rightarrow
   & -\rho_{-g,-g}            & &\qquad g\neq 0 
 \end{aligned}
\end{equation} 
The modular $T$ transformation $\tau\xrightarrow{T}\tau+1$ acts by:
\begin{equation} \label{rhosT}
 \frac{\rho_{gh}}{\eta^{2}} \xrightarrow{ \tau\rightarrow\tau+1}
 \frac{\rho_{g,g+h}}{\eta^{2}}
\end{equation}
As abbreviations for  lattice sums we define (cf.~\cite{Bianchi:1999uq}):
\begin{equation}
 \begin{aligned}
  \Lambda_{SU(3)^2}    &\equiv \big(|\chi_0|^2+|\chi_1|+
                                  |\chi_2|^2
                             \big)^2
\nonumber\\
  \Lambda_{R}        &\equiv \chi_0^{2}\nonumber\\
  \Lambda^\omega_{W} &\equiv \big(
                                  \chi_0+e^{i2\pi\omega/3}\chi_1
                                   +e^{-i2\pi\omega/3}\chi_2
                              \big)^{2}  \\
  \Lambda_{W}        &\equiv\Lambda^0_{W}
 \end{aligned}
\end{equation}
where $\chi_0,\chi_1$ and $\chi_2$  are
the $SU(3)$ characters at level one defined by formula \eqref{su3char} 
(p. \pageref{su3char}).

\chapter{\label{omegstform}  Equivalence classes of unitary 
  symmetric and anti-symmetric matrices.}
In this appendix we show that any symmetric or anti-symmetric
 unitary $n$-dimensional matrix $U$ can be brought to the form \eqref{symform}
 (or \eqref{asymform}) (page \pageref{symform}) via a transformation:
\begin{equation}\label{basestransfa}
 V^TUV\qquad V\in U(n)
\end{equation}
Any unitary matrix $U$ can be brought to diagonal form by conjugation:
\begin{equation} \label{udiag}
 \begin{gathered}
  \exists \,W,\;W\in U(n):\qquad\widetilde{U}= W^{-1} UW,\quad W\in U(n) \\
  \widetilde{U}=\diag \bigl(e^{i{\lambda_1}}\ldots e^{i{\lambda_n}}\bigr),
  \,\quad\lambda_i\in\mathbb{R}
 \end{gathered}
\end{equation}
 In any case we have for arbitrary vectors $e_i,f_j\in\mathbb{C}^n$
   ($\left<\cdot,\cdot\right> $ denotes the {\sl hermitian}
 inner product):
 \begin{align}\label{scal1}
  \bigl<\overline{U} e_i,U f_j\big> 
  =\bigl< e_i,U^T U f_j\big> 
\end{align}
We also note that a basis change \eqref{basestransfa} corresponds
to (note the complex conjugation in $\bar{d}_i$):
\begin{equation}\label{basetransf2}
 \begin{gathered}
  U_{ik}
   = \sum_{j,k=1}^n\bigl< \bar{d}_i,c_j\big> 
   \bigl< c_j,U c_k\big> 
    \bigl< c_k,d_l\big> = 
   \bigl< \bar{d}_i, U d_k\big>   
   \\
  \bigl< c_i,c_j\big> =\bigl< d_i,d_j\big> =\delta_{ij}
\end{gathered} 
\end{equation}
We choose the $d_i$ to be Eigenvectors of $U$: $U d_i= \exp (i\lambda_i)d_i$.
 To incorporate
both the symmetric and antisymmetric case we leave the phase arbitrary:
\begin{equation}
 U=e^{i\phi}U^T
\end{equation}
Inserting this result and $e_i=\bar{d}_i$, $f_j=d_j$ into \eqref{scal1}
we get:
\begin{equation}
  e^{i(\lambda_i-\lambda_j)}\bigl< \bar{d}_i,d_j\big>  
  =e^{i\phi}\bigl< \bar{d}_i,d_j\big> 
\end{equation}
For  $\phi\in\{0,\pi\}$ we see that 
$U_{ij}=\exp{i\lambda_j}\cdot\bigl< \bar{d}_i, d_j\big> =0$ iff
$(\lambda_i-\lambda_j)\neq\phi\mod 2\pi$.  
This means that $\widetilde{U}$ is block-diagonal in this basis. Each block
 is associated with an Eigenvalue $\lambda_j$ and the individual blocks
 $\Lambda_i(\lambda_i)$ are symmetric (resp.\ anti-symmetric) :
\begin{equation}
 U= \left(\begin{array}{c|c|c|c|c|c|c}
    \phantom{\rule{0ex}{1.8ex}} \Lambda_1(\lambda_1)  
        & \multicolumn{1}{c}{0}  &\hdotsfor{4}  & 0 
    \\ \cline{1-2}
     0 & \phantom{\rule[-0.3ex]{0ex}{2.1ex}} \Lambda_2(\lambda_2) & 
         \multicolumn{1}{c}{0}& \hdotsfor{3}   & 0 
    \\
   \cline{2-2}
     \multicolumn{1}{c}{\vdots} &\multicolumn{1}{c}{0} & 
           \multicolumn{1}{c}{\ddots}  
        & \multicolumn{3}{c}{}  & \multicolumn{1}{c}{\vdots}
    \\
    \multicolumn{1}{c}{\vdots} & \multicolumn{1}{c}{\vdots}
    & \multicolumn{1}{c}{} &  \multicolumn{1}{c}{\ddots} 
    &  \multicolumn{2}{c}{} & \multicolumn{1}{c}{\vdots}
    \\
    \cline{7-7}
     \multicolumn{1}{c}{0}  &\multicolumn{1}{c}{0} & \hdotsfor{4}  
      & 
     \multicolumn{1}{|c}{\phantom{\rule[-0.3ex]{0ex}{1.8ex}} 
     \Lambda_s(\lambda_s)} 
   \end{array}
  \right)
\end{equation}
However the blocks $\Lambda_i$ are in general neither proportional
to the identity matrix nor
to the standard symplectic form. However we can achieve this
by finitely many repetitions of the described procedure:
\begin{enumerate}
 \item As $U$ is still unitary  it can be diagonalized
       again by conjugation with a block-diagonal unitary matrix
       (cf.\ \eqref{udiag}). 
 \item We then transform $U$ with the 
       modified transformation \eqref{basestransfa} into this basis
       \eqref{basetransf2}. 
\item  By induction we will reduce the size of 
       the blocks $\Lambda_i$ (Obviously the size can not grow since
       the unitary base transformations do not mix the different blocks
       $\Lambda_i$).
\end{enumerate}
In the symmetric case $U=U^T$ this size reduction stops iff   
$\Lambda_i$ is a matrix with Eigenvalue $\exp (i\lambda_j)$ for
all of the vectors on which $\Lambda_i$ acts non-trivially. This
means that $\Lambda_i$  is proportional to the identity matrix $\id_{n_i}$.
By a transformation
\eqref{basestransfa} with $V$  acting block-diagonally on the separate blocks
by $\exp(-i\lambda_i/2)\id_{n_i}$, we transform $\Lambda_i$   to a matrix
with all vectors having Eigenvalue one. (That means: $\Lambda_i$ is
transformed to the identity). Applying this procedure to
all blocks $\Lambda_i$  gives the identity matrix $U=\id_n$.

In the anti-symmetric case $U=-U^T$ this size reduction stops iff   
$\Lambda_i$ is a block matrix acting non-trivially only on vectors
with Eigenvalue $\lambda_i$ and $-\lambda_i$. By a reordering of the
basis (which is actually an $SO(n_i,\mathbb{Z})$ transformation.) we obtain:
\begin{equation}
 \Lambda_i=
 e^{i\lambda_i}\begin{pmatrix} 
            0 & \id_{n_i/2}\\ 
            -\id_{n_i/2} & 0
           \end{pmatrix}
\end{equation}
By a similar  rescaling as in the symmetric case (i.e.\ by 
$V_i=\exp(-i\lambda_i/2)\id_{n_i}$) $U_i$ is seen to be of the standard
symplectic form. Applying this procedure for all blocks $\Lambda_i$
and reordering the basis,  $U$ can be transformed to standard symplectic from:
 \begin{equation}\label{standsympl}
 U=\begin{pmatrix} 
            0 & \id_{n/2}\\ 
            -\id_{n/2} & 0
           \end{pmatrix}
\end{equation}
By a last transformation with $V=\exp (i\pi/2)$ we can transform U to the
commonly used form \eqref{asymform} (p.\ \pageref{asymform}).

We have proven that there exists only one 
  equivalence class with respect to the transformation 
\eqref{basestransfa} for either a symmetric or a antisymmetric unitary matrix
$U$. In the symmetric case this class can be represented by the identity
 matrix. In the anti-symmetric case this class can be represented by 
the standard symplectic form \eqref{standsympl}. 

\chapter{\label{loerentzform} Spectrum and Eigenvectors of Lorentz
                       transformations}
It is rather well known that orthogonal matrices (which are a subset of
the space of unitary matrices) can be diagonalized by unitary matrices.
Their Eigenvectors make an orthogonal system, that might be normalized.
The Eigenvalues have modulus one. We will now do the analogous classification
 for Lorentz transformations, i.e.\ those linear maps 
$\Lambda\in GL(n+1,\mathbb{R})$ that preserve the {\sl minkowskian}
 metric $G$. 
By minkowskian we mean: one time and $n$ space directions such that
we have a light-cone.
Surprisingly (or not) it turns out that the number of 
Eigenvectors might be lower than the dimension of space time. 
 Our analysis is not valid for a metric with more
time directions because this would ruin the  structure of a (light)-cone which
is essential in our proof.
\begin{theorem}\label{lorentzdecomp}A finite dimensional Lorentz 
transformation $\Lambda\in SO(1,n)$ preserving the corresponding metric $G$
admits   
$n+1$ Eigenvectors, if there are no single light-like Eigenvectors
with Eigenvalue $\pm1$. (This is a sufficient but, not a necessary condition).
 In this case   $n-1$ of the Eigenvalues and Eigenvectors
 are {\sl complex}
(denoted by $\lambda_2\ldots\lambda_n$) with $|\lambda_i|=1$
 (including real as a subset) and two {\sl light-like and  real} with  
$\lambda_0=\lambda_1^{-1}$  or one time-like Eigenvector with 
$\lambda_0=\pm 1$
and $n$    non-time-like Eigenvectors with Eigenvalue $|\lambda_i|=1$.
\end{theorem}
For the $SO(n)$ case the proof makes use of the fact that every Eigenvector
has non-vanishing norm. Starting from one Eigenvector $v\in V$
(which always exists
for non-degenerate maps in the {\sl complexified} vector space $V^\mathbb{C}$)
one can then build
the orthogonal complement $W$ of this Eigenvector. In our case 
we can build the orthogonal complement $V_v\equiv\{w\in V|
 \langle v,w\rangle=0\}$ as long as $\|v \|\neq 0$:\footnote{
As $v\in V^\mathbb{C}$ we also have to complexify the scalar product.
By $\langle \, .\, ,\, .\,\rangle$ we denote the hermitian version (i.e.
complex conjugation on the first vector) of the real Minkowski
 scalar product.}  
\begin{equation}
  V_v = P_v (V)
\end{equation}
with  $P_v$ being the projector defined by: 
\begin{align} \label{projlor}
 P_v(w) \equiv  w- v\frac{\langle v,w\rangle}{\|v \|} 
\end{align} 
We can now proceed by induction starting from one Eigenvector $v$:
\begin{enumerate}
\item If $v$ is not light-like we note that $\lambda_v$ has modulus one.
      As we have assumed $\Lambda$ to be real (appropriate to our 
      application in chapter \ref{strbg}) we also deduce that $\bar{v}$ is an
      Eigenvector with Eigenvalue $\lambda_{\Bar{v}}=\Bar{\lambda}_v$.
      We will then project onto  $W=P_v (V)$ which is left invariant by
      $\Lambda$ ($\langle v,\Lambda{w\rangle}=\lambda_v
       \langle\Lambda{ v},\Lambda{w\rangle}=\langle v,w\rangle=0$).
\item  If $v$ is light-like its Eigenvalue is assumed to be real:
       $\lambda$ complex would imply that $\Bar{v}$ is also Eigenvector with
      Eigenvalue $\Bar{\lambda}$. If $\langle v,\Bar{v}\rangle\neq 0$
      this implies $\lambda^2=1$. This implies that $v$ is real 
      (up to a phase). This contradicts  $\langle v,\Bar{v}\rangle\neq0$.
     The second possibility, $\langle v,\Bar{v}\rangle=0$  
     implies that  $v$ and $\Bar{v}$
     are linear dependent and  up to a phase: real. (One can see this
     by explicitly writing down the scalar product for two light-like
     vectors. If one normalizes the time component $v_0=w_0=a\in\mathbb{R}$
     one discovers for the space component that the vectors should be
     perpendicular wrt.\ the minkowskian scalar product: $\langle
     \vec{v},\vec{v}\rangle_\text{herm.}
      =\langle\vec{w},\vec{w}\rangle_\text{herm.}
      =\langle\vec{w},\vec{v}\rangle_\text{herm.}$. This implies
     that $\vec{v}=\vec{w}$. By  
     $\langle\, . \, ,\, . \,\rangle_\text{herm.}$ we mean the hermitian, 
     positive definite scalar product of the space components.) 
     We denote the Eigenvalue of the light-like $v$ by $\lambda_0$.
     If $\lambda_0\neq\pm 1$ we note that there is another light-like
     Eigenvector:
      As all non-light-like Eigenvectors only contribute Eigenvalues
     with $|\lambda|=1$, in order for $\det\Lambda=\pm 1$ 
       there has to exist at least one other 
       light-like Eigenvector $w$ with
       $\lambda_1\equiv\lambda_w\neq\lambda_0$.
      Due to the structure of the light cone,
      $w$ has non-vanishing scalar product with $v$. This implies 
       $\lambda_1= \lambda_0^{-1}\,(\in\mathbb{R}$). We can now linearly
      combine $v$ and $w$ into non-light-like vectors 
      and project (by \eqref{projlor} ) on the orthogonal complement which has
      two dimensions less. The orthogonal complement  $W_{v,w}$ fulfills of 
      course $\Lambda\big(W_{v,w}\big)\subset W_{v,w}$.
\item If $v$ is light-like with Eigenvalue $\lambda=\pm 1$ the determinant 
      argument does not apply. There might or might not be additional
      time-like Eigenvectors. If another linear indep.\ light-like
      Eigenvector exists, we could project out the 
      space spanned by the two light-like Eigenvectors.
      If not, we can make no further general and simple statement,
      if other non-light-like Eigenvectors exist.\footnote{One could think
      to project by the projector  
      ${\cal P}_\parallel$ or ${\cal P}_\perp$ (cf.\ eq.\ \eqref{lightlikeproj}
      ,p.\ \pageref{lightlikeproj} ) defined
      in chapter \ref{strbg}. The resulting two spaces have however
      non-vanishing scalar-products. The so defined complement of
      the light-like Eigenvector $v$: $W_v\equiv\ker (\mathds{1}-{\cal P}_v)$
      is not invariant under the Lorentz-transformation $\Lambda$:
       $\Lambda\big(W_{v}\big)\not\subset W_{v}$ }
      We simply have to exclude this
      latter case for our solution in chapter \ref{strbg}.
\end{enumerate}
The procedure can be applied on the remaining subspace $W$ until all
Eigenvectors are found. There is one comment in order about the time-like
Eigenvector which might exist. In principle it could have Eigenvalue 
$\lambda=\exp(i\phi), \phi\in\mathbb{R}$. However only $\lambda=\pm 1$ is
actually possible: $v$ time-like implies $\Bar{v}$ time-like. Because there
are no two orthogonal time-like directions, $v$ is real up to a phase.
This restricts the Eigenvalue of time-like Eigenvectors to be either 
plus or minus one.

Of course there can  be maximally two lin.\ indep.\ light-like 
Eigenvectors $v$ and $w$.
This follows from considerations on the different scalar products 
and from the fact that linear independent light-like vectors have
non-vanishing scalar products.

In the case that $\Lambda\in O(1,n+1)$ (and not only $SO(1,n+1)$)
 it can happen that 
a time-like Eigenvector with $\lambda=-1$ exists.  This would correspond 
to a time reversal. Applied to the discussion on boundary conditions
 in chapter \ref{strbg} this is interpreted  as  a brane  localized
in time, i.e.\ an instanton (of possibly higher space dimension). 

We will now make some comments about the situation with only one light-like
Eigenvector with Eigenvalue $\lambda=\pm 1$. \\
We consider the case $\lambda= +1$ first:\\
Investigations seem 
      to exclude the degenerate case for $SO(1,n)$ rotations with 
      $n<4$. A light-like Eigenvector with $\lambda= +1$ implies 
      (probably) always a second, lin.~indep.~light-like 
      Eigenvector with $\lambda= +1$.\footnote{We can not yet
      completely exclude this case as we made the Ansatz  
      $\Lambda(F)=(G+F)^{-1}(G-F)$ with an antisymmetric $F$ for
      a Lorentz rotation which is not the most general one. Also
      the Ansatz $\Lambda(\widetilde{F})=\exp F$ 
      (each $SO(1,n)^+$ transformation might
      be written that way \cite{Nishikawa:1983}, but not
      general $O(1,n)$ transformation) does not necessarily imply that
      each light-like Eigenvector of $\Lambda$ is a Nullvector of 
      $\widetilde{F}$.
      This one observes already from the fact 
       that there exist $\widetilde{F}\neq 0$ which imply 
      $\mathds{1}=\exp\widetilde{F}$. In fact these are exactly those
       $\widetilde{F}$ with Eigenvalues $\lambda\in i2\pi\mathbb{Z}$ 
      \cite{Gottlieb:2001fl}. 
      The statement we could make is that no antisymmetric
      matrices $F$ exist below five dimensions which admit exactly
      one light-like  Nullvector but  no further Nullvector.}
      However in five space time dimensions one can construct Lorentz
      transformations by $\Lambda(F)=(G+F)^{-1}(G-F)$ with an antisymmetric $F$
      that has precisely one light-like Null-vector. The Null-vectors of
      $F$ are  the Eigenvectors of $\Lambda(F)$  with Eigenvalue
      one and vice versa. Numerical analysis  gives some hints that in the 
      case (i.e.\ $n=4$) of precisely one light like Eigenvector
      the number of Eigenvectors decreases, but we can not
      yet make a definite statement in terms of a proof.
 As a quite general example we choose for $n=4$: 
\begin{equation}
 F_{\mu\nu}=
 \left(
  \begin{array}{ccccc}
     0 &  0 &  a &  b & c  \\
     0 &  0 & -a & -b & -c   \\ 
    -a &  a &  0 &  d & e   \\ 
    -b &  b & -d &  0 & h   \\
    -c &  c & -e & -h & 0
  \end{array}
 \right)
\end{equation}
$F$ has a one dimensional light-like Null-space $\lambda\cdot v$ with
$v=(1,1,0,0,0)$. This leads to the following Jordan decomposition of
the associated Lorentz transformation $\Lambda$:
\begin{equation}\label{jordandecomp}
  \left(
  \begin{array}{c|cc}
   \begin{smallmatrix}
     1 & 1 & 0 \\
     0 & 1 & 1 \\
     0 & 0 & 1 
   \end{smallmatrix}     
   &          0  
    \\
   \hline   
    0 & 
   \begin{smallmatrix} \rule{0ex}{2.65ex}
    -\left( \frac{-1 + d^2 + e^2 + h^2 + 2\,
      {\sqrt{-d^2 - e^2 - h^2}}}{1 + d^2 + e^2 + h^2} \right) & 0 \\ 
     0 & -\left( \frac{-1 + d^2 + e^2 + h^2 - 2\,
      {\sqrt{-d^2 - e^2 -h^2}}}{1 + d^2 + e^2 + h^2} \right)
   \end{smallmatrix}      
  \end{array}
  \right)
\end{equation}
\end{appendix}
Of course the base change in \eqref{jordandecomp} 
destroys the property of preserving 
$G=(-1,1,1,1,1)$. However it shows that the number of Eigenvectors 
is smaller than the dimension of  space-time. It is also interesting
that this effect only occurs above four space-time dimensions. We do not yet
know if this
degenerate case has physical consequences. At least our method for generating
a general solution to the boundary conditions misses some dofs. It seems
to be complicated to find the general solution and if possible to quantize it.
 We will leave this as a purpose for future work.\\
The case with exactly one light-like Eigenvector with Eigenvalue
$\lambda=-1$ occurs
if we consider a D-brane  of the type described above at
the $\sigma=0$ end-point, and a light-like D-brane without any 
${\cal F}$-field at the $\sigma=\pi$ end-point of the string,
with the light-like 
Eigenvector of $(G+{\cal F})^{-1}  (G-{\cal F})$ being perpendicular to the 
second brane. The $R$-matrix (cf.~eq.~\eqref{Rmatrix}, p.~\pageref{Rmatrix})
would then reflect the light-like Eigenvector of the first matrix, resulting in
exactly one light-like Eigenvector with Eigenvalue minus one. Therefore both
the light-like $\lambda=1$ and the light-like $\lambda=-1$ case are 
connected.

\chapter{Quantities of the 
            $ \texorpdfstring{\big(T^2\times T^2\times 
                          T^2\big)/\mathbb{Z}_4$}
       {(T\texttwosuperior \textmultiply T\texttwosuperior
                 \textmultiply T\texttwosuperior)/Z(4)}-Orientifold}
 In this appendix we present some useful quantities of the 
 $(T^2\times T^2\times T^2)/\mathbb{Z}_4$-orientifold that we discussed
 in chapter \ref{z4}. 

\section{\label{oplaneszfour}Orientifold planes}
We present the results for the O6-planes
and the action of $\Omega\Bar{\sigma}$ on the homology lattice
for the other three orientifold models.
The result is summarized in table \ref{taba1}.
\begin{table}
 \renewcommand{\arraystretch}{1.2}
 \begin{center}
 \begin{tabular}{|c|c|}
 \hline
  model         & O6-plane \\
  \hline
  ${\bf AAA}$   & $4\, \rho_1 -2\, \Bar{\rho}_2 $   \\
  \hline
  ${\bf AAB}$   & $2\, \rho_1 + \rho_2 -2\, \Bar{\rho}_2 $         \\
  \hline
  ${\bf ABA}$   & $2\, \rho_1 + 2\, \rho_2+  2\, \Bar{\rho}_1  -2\,
                  \Bar{\rho}_2 $   \\
  \hline
  ${\bf ABB}$   & $ 2\, \rho_2+2\, \Bar{\rho}_1 -2\, \Bar{\rho}_2 $ \\ 
 \hline 
 \end{tabular}
 \caption{\label{taba1}O6-planes of the $T^6/\mathbb{Z}_4$ orientifold}
 \end{center}
\end{table}
For the action of $\Omega\Bar{\sigma}$ on the orbifold basis we find:
\begin{itemize}
\item{{\bf AAA}:}
For the toroidal 3-cycles we get
\begin{equation}\label{actbasa}
 \begin{aligned} &\rho_1\to \rho_1, \quad  \Bar{\rho}_1\to -\Bar{\rho}_1 \\
                 &\rho_2\to -\rho_2, \quad  \Bar{\rho}_2\to \Bar{\rho}_2 
 \end{aligned}
\end{equation}
and for the exceptional cycles
\begin{equation}\label{actbasexa}
 \begin{aligned}   &\varepsilon_i\to \varepsilon_i \quad\quad
                   \Bar{\varepsilon}_i\to -\Bar{\varepsilon}_i 
                         &\quad\forall i\in\{1,\ldots,6\}
 \end{aligned}
\end{equation}
\item{{\bf AAB}:}
  For the toroidal 3-cycles we get
  \be\label{actbasb}
   \begin{aligned} 
        &\rho_1\to \rho_1, \quad  \Bar{\rho}_1\to \rho_1 -\Bar{\rho}_1 \\
        &\rho_2\to -\rho_2, \quad  \Bar{\rho}_2\to -\rho_2+\Bar{\rho}_2 
   \end{aligned}
  \end{equation} 
  and for the exceptional cycles
  \be\label{actbasexb}
    \begin{aligned} &\varepsilon_i\to \varepsilon_i \quad\quad
                   \Bar{\varepsilon}_i\to \varepsilon_i-\Bar{\varepsilon}_i 
                   &\quad\forall i\in\{1,\ldots,6\}
   \end{aligned}
  \end{equation}
\item{{\bf ABA}:}
  For the toroidal 3-cycles we get
  \be\label{actbasc}
   \begin{aligned}
       &\rho_1\to \rho_2, \quad  \Bar{\rho}_1\to -\Bar{\rho}_2 \\
       &\rho_2\to \rho_1, \quad  \Bar{\rho}_2\to -\Bar{\rho}_1 
    \end{aligned}
  \end{equation} 
  and for the exceptional cycles
  \be\label{actbasexc}
   \begin{aligned}     &\varepsilon_1\to -\varepsilon_1 \quad\quad
                        \Bar{\varepsilon}_1\to \Bar{\varepsilon}_1 \\
                        &\varepsilon_2\to -\varepsilon_2 \quad\quad
                        \Bar{\varepsilon}_2\to \Bar{\varepsilon}_2 \\
                        &\varepsilon_3\to \varepsilon_3\phantom{-} \quad\quad
                        \Bar{\varepsilon}_3\to -\Bar{\varepsilon}_3 \\
                        &\varepsilon_4\to \varepsilon_4 \phantom{-}\quad\quad
                        \Bar{\varepsilon}_4\to -\Bar{\varepsilon}_4 \\
                        &\varepsilon_5\to \varepsilon_6 \phantom{-} \quad\quad
                        \Bar{\varepsilon}_5\to -\Bar{\varepsilon}_6 \\
                        &\varepsilon_6\to \varepsilon_5\phantom{-} \quad\quad
                        \Bar{\varepsilon}_6\to -\Bar{\varepsilon}_5 
   \end{aligned}
  \end{equation}
\end{itemize}

\section{\label{susyz4}Supersymmetry conditions}
\noindent
In this appendix we list the supersymmetry conditions for the remaining
three orientifold models.
\begin{itemize}
 \item{{\bf AAA}:} The condition that such a D6-brane preserves the 
                   same supersymmetry
                   as the orientifold plane is simply
  \begin{equation}\label{susya}    
             \varphi_{a,1}+ \varphi_{a,2}+ \varphi_{a,3}=0 \mod 2\pi
  \end{equation}
  with
 \be\label{tangia}  \tan\varphi_{a,1}={m_{a,1}\over n_{a,1}}, \quad
              \tan\varphi_{a,2}={m_{a,2}\over n_{a,2}}, \quad
               \tan\varphi_{a,3}=\frac{U_2\, m_{a,3}}{n_{a,3}} 
 \end{equation}
 This implies the following necessary condition in terms of the wrapping
 numbers
 \be\label{susywara}
     U_2=-{ n_{a,3}  \over m_{a,3}}\cdot
                  { \left(n_{a,1}\,m_{a,2} +
               m_{a,1}\,n_{a,2} \right)
           \over \left(n_{a,1}\,n_{a,2} - m_{a,1}\,m_{a,2}  \right)}
 \end{equation}

\item{{\bf AAB}:} The condition that such a D6-brane preserves the same 
                  supersymmetry as the orientifold plane is simply
 \be\label{susyb} 
     \varphi_{a,1}+ \varphi_{a,2}+ \varphi_{a,3}=0 \mod  2\pi
 \end{equation}
 with
 \begin{equation}\label{tangib} 
          \tan\varphi_{a,1}={m_{a,1}\over n_{a,1}}, \quad
          \tan\varphi_{a,2}={m_{a,2}\over n_{a,2}}, \quad
          \tan\varphi_{a,3}={U_2\, m_{a,3}\over n_{a,3}+{1\over 2} m_{a,3} } 
 \end{equation}
 This implies the following necessary condition in terms of the wrapping numbers
 \begin{equation}\label{susywarb}
           U_2=-{ \left(n_{a,3}+{1\over 2} m_{a,3}\right)  \over m_{a,3}}\cdot
                  { \left(n_{a,1}\,m_{a,2} +
               m_{a,1}\,n_{a,2} \right)
                 \over \left(n_{a,1}\,n_{a,2} - m_{a,1}\,m_{a,2}  \right)}
 \end{equation}
\item{{\bf ABA}:} The condition that such a D6-brane preserves the same 
                  supersymmetry as the orientifold plane is simply
 \begin{equation}\label{susyc}      \varphi_{a,1}+ \varphi_{a,2}+ \varphi_{a,3}={\pi\over 4}
                 \mod 2\pi
 \end{equation}
with
 \begin{equation}\label{tangic}
              \tan\varphi_{a,1}={m_{a,1}\over n_{a,1}}, \quad
              \tan\varphi_{a,2}={m_{a,2}\over n_{a,2}}, \quad
              \tan\varphi_{a,3}={U_2\, m_{a,3}\over n_{a,3} } 
 \end{equation}
This implies the following necessary condition in terms of the wrapping
numbers
 \begin{equation}\label{susywarc} 
           U_2={ n_{a,3} \over m_{a,3}}\cdot
               { \left(n_{a,1}\,n_{a,2} - m_{a,1}\,m_{a,2} - n_{a,1}\,m_{a,2} -
                       m_{a,1}\,n_{a,2} 
                \right)
                   \over 
                \left(n_{a,1}\, n_{a,2} - m_{a,1}\,m_{a,2} + n_{a,1}\,m_{a,2} +
                      m_{a,1}\,n_{a,2} 
                \right)
               }
 \end{equation}
\end{itemize}

\section{Fractional boundary states}
\noindent
The unnormalized boundary states in light cone gauge for D6-branes at angles 
in the untwisted sector are given by
\begin{equation}\label{boundaz}
  \begin{aligned} |D;(n_I,m_I)\rangle_{U}=
                &|D;(n_I,m_I),\text{NSNS},\eta=1\rangle_{U}+
                   |D;(n_I,m_I),\text{NSNS},\eta=-1\rangle_{U} \\
                   +&|D;(n_I,m_I),\text{RR},\eta=1\rangle_{U}+
                   |D;(n_I,m_I),\text{RR},\eta=-1\rangle_{U} 
  \end{aligned}
\end{equation}
with the coherent state
\begin{equation}\label{boundbz}
  \begin{aligned}
       &|D;(n_I,m_I),\eta\rangle
      \\&= \int dk_2 dk_3 \sum_{\vec r,\vec s} {\rm exp}
          \biggl(-\sum_{\mu=2}^3 \sum_{n>0} {1\over n} \alpha^\mu_{-n}
               \tilde \alpha^\mu_{-n} \\ 
     &\rule{3.7cm}{0cm}-\sum_{I=1}^3 \sum_{n>0} {1\over 2n} \Bigr(e^{2i\varphi_I}
          \zeta^I_{-n} \tilde{\zeta}^I_{-n}  +
        e^{-2i\varphi_I} \Bar{\zeta}^I_{-n} \tilde{\Bar{\zeta}}^I_{-n} \Bigr)\\
        &\rule{7.cm}{0cm}+ i \eta \bigl[ {\rm fermions} \bigr] \biggr)
             |\vec r,\vec s,\vec k, \eta \rangle
   \end{aligned}
\end{equation}
Here $\alpha^\mu$ denotes the two real non-compact directions and
$\zeta^I$  the three complex compact directions.
The angles $\varphi_I$ of the D6-brane relative to the horizontal
axis on each of the three internal tori $T^2$ can be expressed
by the wrapping numbers $(n_I,m_I)$ as listed in appendix \ref{susyz4}.
The boundary state \eqref{boundbz} involves a sum over the internal
Kaluza-Klein and winding ground states parameterized by
$(\vec r,\vec s)$.
The mass of  these KK and winding modes on each $T^2$ in general reads
\begin{equation}\label{spec} 
          M^2_{I}={ |r_I+s_I\,{U_I}|^2\over U_{I,2} }\,
                  { |n_I+m_I\, {T_{I}}|^2\over {T_{I,2}} }
\end{equation}
with $r_I,s_I\in\mathbb{Z}$ as above and $U_I$ and $T_I$ denote the
complex and K\"ahler structure on the torus \cite{Blumenhagen:2000wh}.
 If the brane
carries some discrete Wilson lines, $\vartheta=1/2$,  appropriate
factors of the form $e^{i s R \vartheta}$ have to be introduced
into the winding sum in \eqref{boundb}.

\noindent In the $\Theta^2$ twisted sector, the boundary state
involves the analogous sum over the fermionic spin structures
\eqref{bounda} with 
\begin{equation}\label{boundcz}
  \begin{aligned}
   &|D;(n_I,m_I),e_{ij}, \eta
                  \bigr\rangle_T \\
    &=\int dk_2 dk_3  \sum_{r_3,s_3} {\exp}
          \biggl(-\sum_{\mu=2}^3 \sum_{n>0} {1\over n} \alpha^\mu_{-n}
               \tilde \alpha^\mu_{-n}  \\
          &\rule{3.7cm}{0cm}-\sum_{I=1}^2\sum_{r\in\mathbb{Z}^+_0 
      +{1\over 2}} {1\over 2r} 
                    \left(e^{2i\varphi_I}\zeta^I_{-r} \tilde{\zeta}^I_{-r} +
    e^{-2i\varphi_I} \Bar{\zeta}^I_{-r} \tilde{\Bar{\zeta}}^I_{-r} \right)\\
    &\rule{3.7cm}{0cm}-\sum_{n>0 } {1\over 2n} \left (e^{2i\varphi_3}
    \zeta^3_{-n} \tilde{\zeta}^3_{-n}  +
   e^{-2i\varphi_3} \Bar{\zeta}^3_{-n} \tilde{\Bar{\zeta}}^3_{-n} \right)\\
       &\rule{6.2cm}{0cm}+ i \eta \bigl[ {\rm fermions} \bigr] \biggr)
             |r_3,s_3,\vec k,e_{ij},\eta \rangle
 \end{aligned}
\end{equation}
where $e_{ij}$ denote the 16 $\mathbb{Z}_2$ fixed points. Here, we have
taken into account that the twisted boundary state can only have
KK and winding modes on the third $T^2$ torus and that the bosonic
modes on the two other $T^2$ tori carry half-integer modes.

\newpage

\bibliography{thesis_l_goerlich}
\bibliographystyle{JHEP}
\end{document}